\documentclass[twocolumn]{aa}

\usepackage{txfonts}

\usepackage{graphicx}

\usepackage{aalongtable}
\usepackage{subfig,subtab}

\def\etal{{et\,al.}}

\def\degr{\ifmmode ^{\circ}\else$^{\circ}$\fi}
\def\amin{\ifmmode ^{\prime}\else$^{\prime}$\fi}
\def\asec{\ifmmode ^{\prime\prime}\else$^{\prime\prime}$\fi}
\def\fm{\hbox{$.\!\!^{\rm m}$}}            % Fractions of magnitudes
\def\fss{\hbox{$.\!\!^{\rm s}$}}        % Fractions of seconds
\def\fdg{\hbox{$.\!\!^\circ$}}          % Fractions of degrees
\def\farcs{\hbox{$.\!\!^{\prime\prime}$}}  % Fractions of arcseconds
\def\farcm{\hbox{$.\!\!^{\prime}$}}  % Fractions of arcmin
\newbox\grsign \setbox\grsign=\hbox{$>$}
\newdimen\grdimen \grdimen=\ht\grsign
\newbox\laxbox \newbox\gaxbox
\setbox\gaxbox=\hbox{\raise.5ex\hbox{$>$}\llap
     {\lower.5ex\hbox{$\sim$}}}\ht1=\grdimen\dp1=0pt
\setbox\laxbox=\hbox{\raise.5ex\hbox{$<$}\llap
     {\lower.5ex\hbox{$\sim$}}}\ht2=\grdimen\dp2=0pt
\def\gax{$\mathrel{\copy\gaxbox}$}
\def\lax{$\mathrel{\copy\laxbox}$}

\def\h{$^{\rm h}$}\def\m{$^{\rm m}$}
\def\s{$^{\rm s}$}
\def\fxo{f$_{\rm X}/f_{\rm opt}$}
\def\fxv{f$_{\rm X}/f_{\rm V}$}

\begin{document}

\title{Optical counterparts of ROSAT X-ray sources in two selected fields
at low vs. high Galactic latitudes}

\titlerunning{ROSAT counterparts in Com and Sge}

\author{J. Greiner \inst{1}, G.A. Richter \inst{2}}

\offprints{J. Greiner, jcg@mpe.mpg.de}

\institute{
  Max-Planck-Institut f\"ur extraterrestrische Physik, 85740 Garching, Germany
  \and
  Sternwarte Sonneberg, 96515 Sonneberg, Germany}

\date{Received 14 October 2013 / Accepted 12 August 2014}

%\abstract{Context}{Aim}{Method}{Results}{Conclusions}

\abstract{The optical identification of large number of X-ray sources
such as those from the ROSAT All-Sky Survey is challenging with
conventional spectroscopic follow-up observations.}
%Aim
{ We investigate two ROSAT All-Sky Survey fields of 
size 10\degr $\times$ 10\degr\ each, one at galactic latitude 
b = 83\degr\ (26 Com), the other
at b = --5\degr\ ($\gamma$ Sge), in order to 
optically identify the majority of sources.}
%Method
{We used optical variability, among other more standard methods, 
as a means of
  identifying a large number of ROSAT All-Sky Survey sources.
All objects fainter 
than about 12 mag and brighter than about 17 mag, in or near the error 
circle of the ROSAT positions, were tested for optical
variability on hundreds 
of archival plates of the Sonneberg field patrol.} 
% Results
{The present paper contains probable optical identifications of 
altogether 256 of the 370 ROSAT sources analysed. In particular,
 we found 126 AGN (some of them may be 
misclassified  CVs), 17 likely clusters of galaxies,
16 eruptive double stars (mostly CVs), 43 chromospherically active stars, 
65 stars brighter than about 13 mag, 7 UV Cet stars, 
3 semiregular resp. slow irregular 
variable stars of late spectral type, 
2 DA white dwarfs, 1 Am star, 1 supernova remnant and 1 planetary nebula.
As expected, nearly all AGN are found in the high-galactic latitude field,
while the majority of CVs is located at low galactic latitudes.
We identify in total 72 new variable objects. }
{X-ray emission is, expectedly, tightly correlated with
optical variability, and thus our new method for optically identifying 
X-ray sources is demonstrated to be feasible. Given the large number of
optical plates used, this method was most likely not more efficient
than e.g. optical spectroscopy. However, it required no telescope time,
only access to archival data.}

\keywords{X-rays:  stars   --  stars: variables -- 
techniques: photometric -- surveys }

\maketitle

\section{Introduction}

In the past, many attempts were made to investigate 
the stellar content in the Solar neighbourhood, as well as in the 
whole Galaxy, and from that deduce
the structure of our 
Galaxy by studying the spatial distribution and the dynamics of different
types of objects within the Galaxy. An early summary of such analyses
based on optical data is given by Hoffmeister \etal\ (1985). 
Variable stars played an important role right from the beginning; e.g.,
 Richter (1968) investigated the structure of our Galaxy by means 
of the statistics of the variable stars of the Sonneberg field patrol.

Since that time, many objects have been found in other spectral regions by new 
ground-based (e.g. 2MASS, Pan-STARRS, PTF, SDSS) or 
space-based (ROSAT, GALEX, WISE, Fermi) surveys,
and their spatial distribution is of interest. In this 
connection the X-ray sources found by ROSAT are very important 
because of its large number of about 200 000 sources, 
and their
all-sky distribution (note that pointed Chandra, XMM-Newton or Swift/XRT
observations together cover only about 7--8\% of the sky).
Beyond the sheer number, these surveys at other wavelengths have the
additional advantage of suffering from completely different 
selection biases, thus
improving our understanding of the limits of optical surveys.
 Examples include not only source populations dominated by X-ray
emission like single neutron stars (Haberl 2005), but also classical
optical populations like chromospherically active stars 
(Schmitt \& Liefke 2004).

Before these surveys can be used for this kind of study, rather complete
optical identifications are necessary. In general, standard practice of
optical identification of X-ray sources is via optical spectroscopy.
Crude identifications can also be made based on optical colours and on
the ratio of optical-to-X-ray flux (Maccacaro \etal\ 1982, Stocke \etal\ 1991,
Beuermann \etal\ 1999), since different object classes
have different efficiencies in producing X-ray emission.
For the selection of blazars, the correlation of X-ray and
radio surveys have proven very efficient (e.g. Brinkmann \etal\ 1995).
However, for the purpose of the present paper we have chosen a
different approach, namely to use archival data available from
the Sonneberg optical sky survey. The underlying idea was that the majority 
of X-ray-emitting objects is expected to be variable and that the chance 
coincidence of a variable inside the ROSAT error box is small.

The Sonneberg sky survey consists of two major parts: the sky patrol and the
field patrol (Br\"auer \etal\ 1999). The sky patrol records the entire 
northern sky since 1926 with 14 short-focus cameras in two colours down 
to a limiting magnitude of $\sim$14\m.
The field patrol monitored 80 selected 10\degr$\times$10\degr\ fields
between 1926 and 1995 in one colour down to a limiting magnitude of 
$\sim$ 18\m\ at the plate centre.

Here we present the results of the classification of selected ROSAT 
X-ray point sources 
and their optical counterparts based on archival photographic plates, 
complemented by the standard identification methods
 of a low- ($\gamma$ Sge, b = --5\degr) and a high- (26 Com, b = 83\degr) 
galactic latitude field, 
respectively, in order to make statements about the population 
characteristics of various types of these objects and, after all, about the 
population differences at different galactic latitudes. 

\section{The data}

\subsection{The X-ray data from the ROSAT all-sky survey}
\label{sec:RASS_Xray}

The ROSAT all-sky-survey was performed between August 1990 and January 1991
with the position-sensitive proportional counter (PSPC), sensitive in the 
energy range 0.1--2.4 keV (Tr\"umper 1983). The two fields were scanned
by ROSAT over a time period of nearly three to four weeks each: the Com field
during Dec. 4--22, 1990 (TJD 48229--48247), and the Sge field during
Oct. 7 -- Nov. 3, 1990 (TJD 48171--48198. The exposure times
in the two fields are 250--500 sec (Com) and 300--590 sec (Sge), respectively,
and are typically distributed over 20--30 individual scans. 
There is an exposure gradient over the fields according to ecliptic
latitude.

After extracting the data of the two 10\degr $\times$ 10\degr\ fields,
the EXSAS package (Zimmermann et al. 1994) was used for the data 
reduction. The adapted source detection technique consists of several steps:
first, all possible sources are
identified by means of a "sliding window" technique and removed from the
data. This procedure was applied twice with two different sizes
of the sliding window in order to account for different coverage
of sky regions at different off-axis angles, which leads to varying
merged point-spread functions across the fields. Also, the likelihood
threshold was set at a very low value to allow many source candidates
to join the list.
Second, a background map is produced with a bi-cubic spline fit to the
resulting image. Finally, a maximum likelihood algorithm is applied
to the background-subtracted data (e.g., Cruddace, Hasinger \& Schmitt 1988) 
in three separate pulse-height channel ranges. Each candidate source of the 
sliding-window list is tested in this way with a likelihood threshold of 10. 
If a  source is detected in more 
than one energy band, the detection corresponding to a higher likelihood  
value is considered. 
As a result, we have detected 238 (26 Com) and 132 ($\gamma$ Sge) 
X-ray point sources in the above two fields.
The ROSAT source positions provided here are typically accurate to 
less than a 30\asec\ radius. The error is dominated by
systematic effects, hence independent of the brightness
of the sources.

\begin{figure}[t]
 \includegraphics[width=0.85\columnwidth,angle=-90, bb=55 70 560 610,clip]{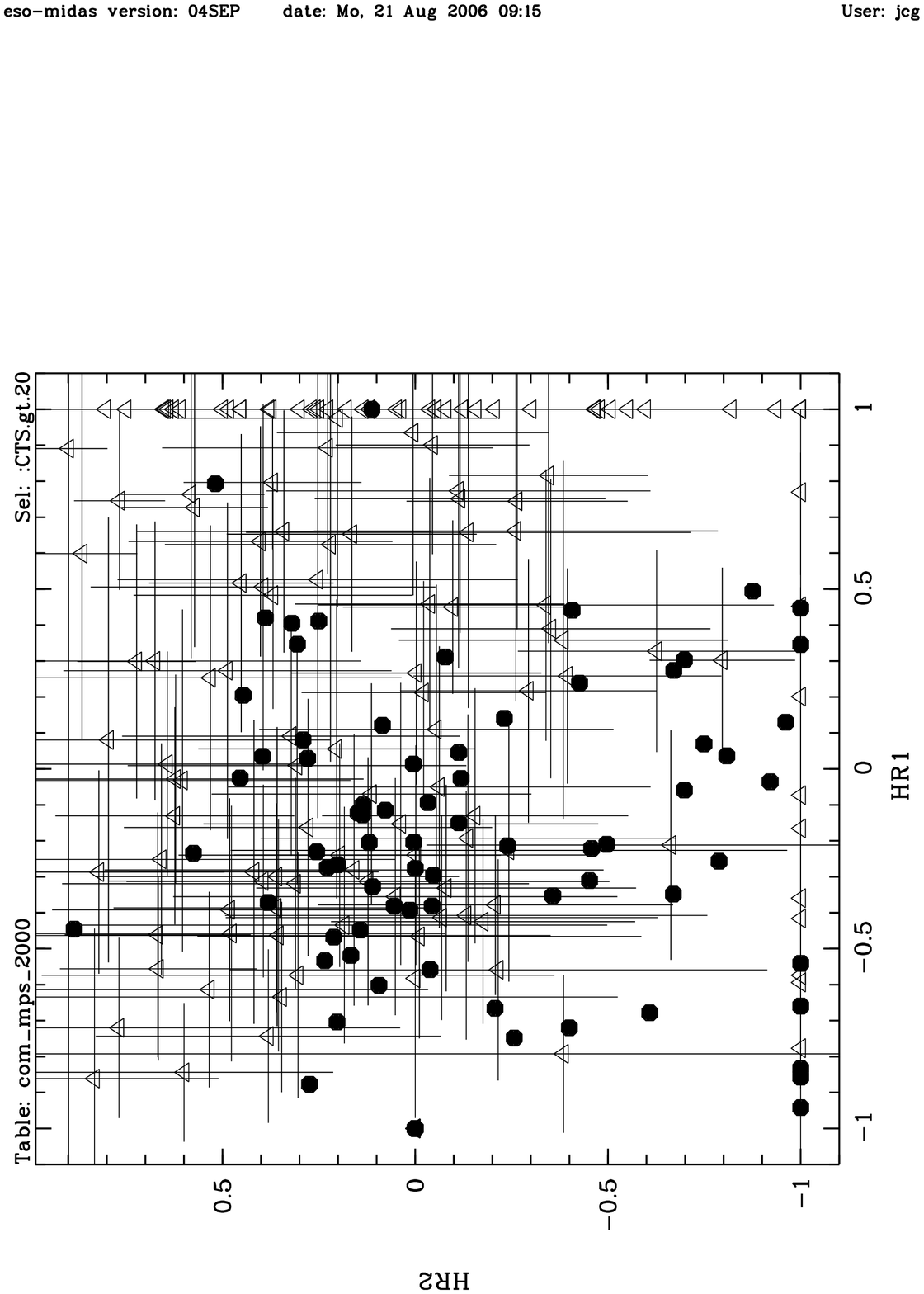}
 \includegraphics[width=0.85\columnwidth,angle=-90, bb=55 70 560 610,clip]{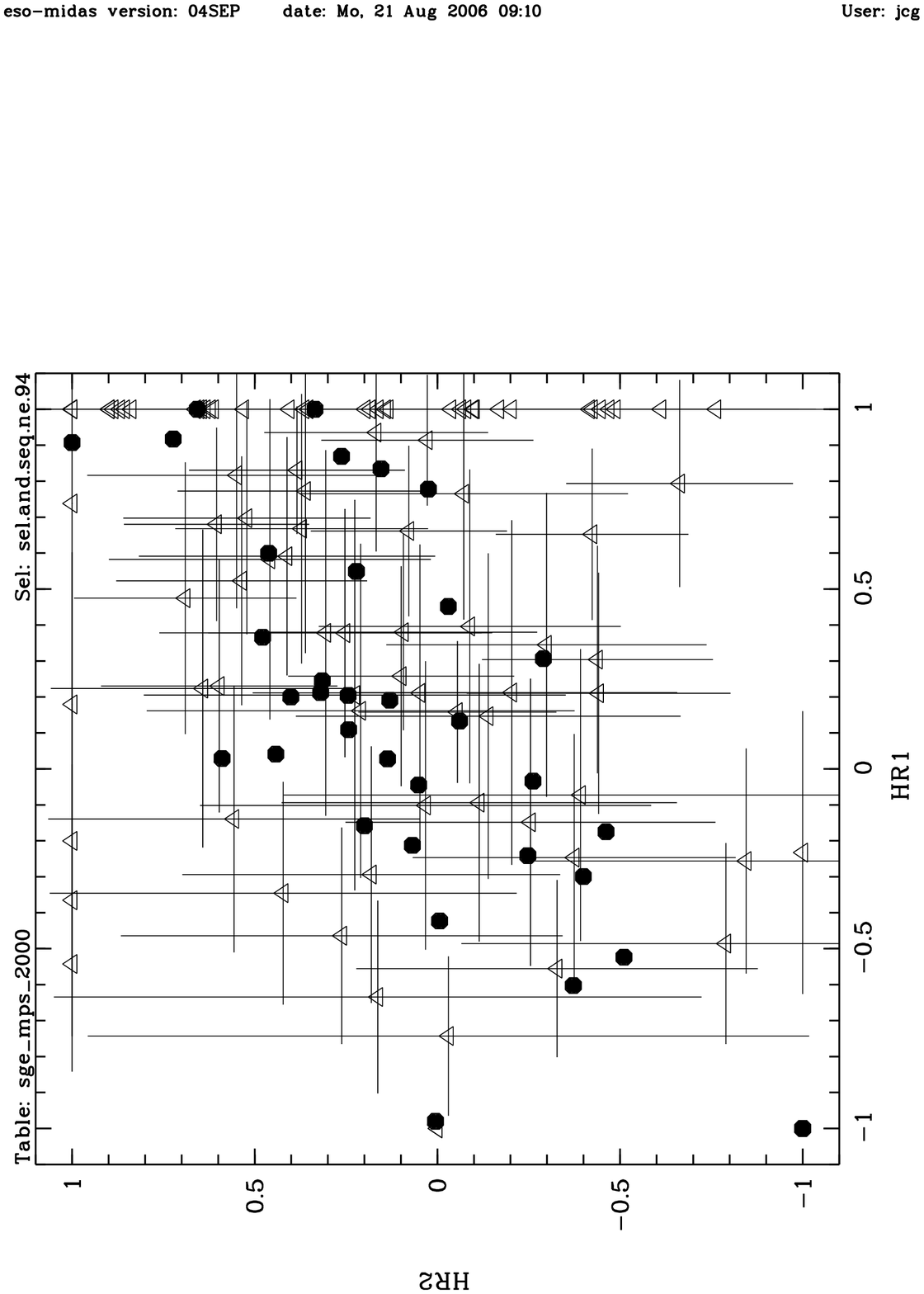}
 \caption{Distribution of the X-ray hardness ratios of all detected
  sources in the two fields (top: Com; bottom: Sge). Sources with
  more than 20 cts are shown with filled symbols, and spectral parameters
  are given in Tables \ref{specparcom},\ref{specparsge}, while sources with  
  fewer than 20 cts are shown with open triangles and their error bars.
  In the bottom panel note the substantial difference in the distribution
  which is the effect of larger absorption in the Sge field
  due to its low galactic latitude,  i.e.  scarcity of sources with 
  negative HR1 and/or with negative HR2. 
 See the text for the definition of HR1 and HR2.}
\label{hr}
\end{figure}

Tables \ref{comX},\ref{sgeX} contain the main X-ray data of the ROSAT 
sources in 26 Com and 
$\gamma$ Sge, respectively.  Subitem ``{\it a}'' refers to the Com field
and subitem ``{\it b}'' to the Sge field.
\begin{itemize}
\vspace{-0.22cm}
\item[] $\!\!\!\!${\it Column 1}: Running number. 
\item[] $\!\!\!\!${\it Column 2}: Coordinates (2000.0) of the ROSAT position 
in right ascension (\h, \m, \s) and declination (\degr, \amin, \asec). 
\item[] $\!\!\!\!${\it Column 3}: Name of source when contained in the 
ROSAT all-sky survey 
catalogue (RXS; Voges \etal\ 1999).
\item[] $\!\!\!\!${\it Column 4}: Statistical error of the RXS position 
(from the RXS catalogue).
\item[] $\!\!\!\!${\it Column 5}: Vignetting corrected mean X-ray intensity 
in PSPC counts/sec. 
\item[] $\!\!\!\!${\it Column 6+7}: X-ray spectral shape, expressed in 
terms of two hardness ratios
HR1 and HR2. The hardness ratio HR1 is defined as the normalized count 
difference
(N$_{\rm 50-200}$ -- N$_{\rm 10-40}$)/(N$_{\rm 10-40}$ + N$_{\rm 50-200}$),
where N$_{\rm a-b}$
denotes the number of counts in the PSPC between channels a  and b.
Similarly, the hardness ratio HR2 is defined as
(N$_{\rm 91-200}$ -- N$_{\rm 50-90}$)/N$_{\rm 50-200}$.
HR1 is sensitive to the Galactic foreground absorbing column.
\item[] $\!\!\!\!${\it Column 8}: The observed (not extinction corrected!)
 X-ray flux in the 0.1--2.4 keV band in units of 10$^{-13}$ erg/cm$^2$/s.
 This is a gross underestimate for objects identified as clusters of 
 galaxies (or candidates), as this is derived from point-source
 PSF-fitting.
\item[] $\!\!\!\!${\it Column 9}: 
Most likely optical identification 
according to various criteria
(see text). The symbol is that of the corresponding object in
column 1 of Table \ref{comopt},\ref{sgeopt}, 
except for the Sge ``sources'' 87-91,93-94 which are detections
of flux enhancements of an extended supernova remnant (SNR).
A dash means that no optical identification can be proposed, and a
question mark after the symbol denotes some doubts due to inconclusive
data or other alternatives. 
``Cluster'' denotes cases where the X-ray emission
is more likely associated with the galaxy cluster gas emission rather than
individual galaxies in that cluster.
\end{itemize}

We note that there is not a one-to-one correspondence of sources
detected with the above procedure and those published in the
ROSAT all-sky survey (RXS) catalogue. This is due to the fact that
this program, and in particular the source detection in the ROSAT
all-sky survey, had been started in 1993 with the processing state
of the data of that time. In contrast, the RXS survey was
prepared several years later with various processing improvements.
We have no indication that our version suffers any systematic problems
which post-facto is proven by the very similar identification rates
(see Table \ref{tab:IDstat}).
The  major difference in the source detection is, though, that we
used a slightly lower maximum likelihood threshold (8) as compared
to that of the RXS catalogue (10), and therefore the present source
list is more extensive than the RXS catalogue in these two areas.
A noteworthy difference in approach is that the RXS catalogue lists
the statistical error for each source, some of which are as small
as 8\asec. At this size, the positional error is affected by
systematic effects of at least a similar level. We therefore start
out identification process with a generic 30\asec\ error circle for
all sources.

The distribution of hardness ratios of all sources
is shown in Fig. \ref{hr},
and shows the significant differences caused by the different galactic
foreground absorbing column in the two fields:
The fraction of soft sources is much lower in the Sge field.

For the brighter X-ray sources (more than 20 counts) in both fields, 
we have performed spectral fitting using three different models:
a power law, a thermal bremsstrahlung and a blackbody model.
Together with the absorption by neutral hydrogen which was always
left as a free variable, a total of three parameters were fit to
the X-ray spectrum. The selection of the ``most appropriate'' 
spectral fit was based on the goodness of the fit (reduced $\chi^2$)
and the consistency with the expected spectral shape for the
type of object according to the most likely identification - thus
it was an iterative process. The fit parameters of the selected
``most appropriate''  model for each X-ray source are reported in
Table \ref{specparcom},\ref{specparsge}, and the unfolded
photon spectra are shown in Fig. \ref{comspec},\ref{sgespec}.

Rough distance estimates can be made for the brighter 
X-ray sources based on the measured hydrogen column density.
For the Com field, at galactic latitude $b=+83$\degr,
the foreground column density is in the range 
(0.10--0.26)$\times$10$^{21}$ cm$^{-2}$ 
(corresponding to $A_{\rm V}$=0.06--0.15 mag), and sources with
less than this column should be at less than 100 pc distance.
In contrast, sources in the Sge field suffer an absorbing column
in the range of (1.3--12.7)$\times$10$^{21}$ cm$^{-2}$
(corresponding to $A_{\rm V}$=0.7--7 mag). While this
larger dynamic range allows a rough distance estimate, it also implies
that more distant sources are easily absorbed, thus falling below
the sensitivity threshold of the ROSAT telescope during the survey
exposure.

\begin{figure}[t]
\hspace{-0.3cm}
 \includegraphics[width=0.75\columnwidth,angle=-90]{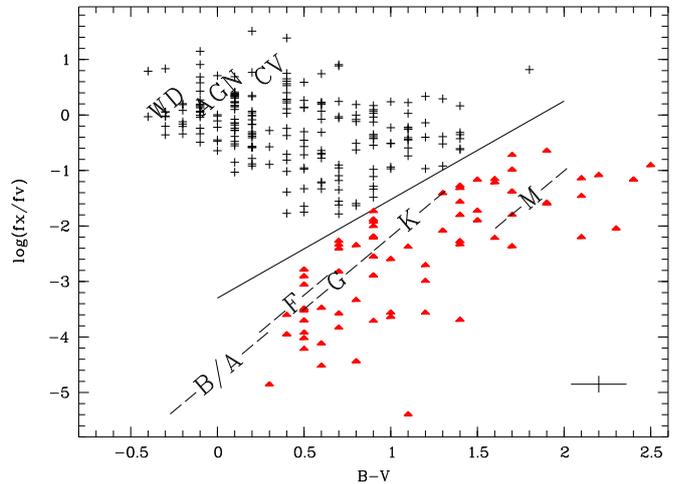}
 \caption{The X-ray to optical flux is plotted over the $B-V$ colour for all
% 316
  objects of the Com field (Table \ref{comopt}). The straight line marks 
  the upper boundary of the region populated by stars (Stocke et al. 1991),
  and the areas populated by various spectral types are taken
  from Beuermann \etal\ (1999).
  Thus, triangles denote secure star identifications.
  Accretion-powered systems (CVs, AGN) can reach high f$_{\rm X}/f_{\rm opt}$ 
  values. Since most of those systems have blue optical colours, they 
  populate the upper left corner. The high number of objects
  reflects the fact that the Com field is at high galactic latitude
  where the fraction of AGN is high.  
  No extinction correction has been applied, since the correction factor
  is smaller than a factor of two (Fig. \ref{lxloptAv}).
  This and the error in $B-V$ are visualized by the cross in the lower
  right corner. }
\label{stocke}
\end{figure}

\subsection{The optical plate material}
\label{sec:plates}

For the 
field $\gamma$ Sge we have altogether 239 plates of the  400/1600 mm 
quadruplet astrographs with 10\degr $\times$ 10\degr\ and 
87 plates taken with the 400/1950 mm 
quadruplet astrograph with 8\fdg5 by 8\fdg5. They cover the time interval 
1935 -- 1995 more or less continuously. Moreover, for some overlapping 
zones we have 227 plates of the neighbouring field $\gamma$ Aql, 192 plates 
for $\beta$ Del. Furthermore, for stars brighter than 16 mag we have 
193 plates 
of the Zeiss triplet 170/1400 mm. 

For the field 26 Com, we have 
226 plates taken with the 400/1600 mm astrographs, 204 plates taken with 
the 400/1950mm astrograph, partly with displaced centre, and 294 plates 
(1400/1600 mm) of the overlapping neighbouring field 5 Com,
covering the time interval 1960--1995.  
The best Sonneberg plates 
reach a sensitivity of nearly 18 mag at the plate centre.

Finally, we 
used the POSS Sky Survey prints, and in some cases plates of the 
Tautenburg 2\,m Schmidt telescope with a limiting magnitude of 
$\sim$21 mag, at the best.

\subsection{Swift X-ray observations}
\label{sec:Swiftobs}

In the process of this study it became clear that there was
some fraction of sources for which a unique identification
was not possible. This was particularly true for the Sge field, where
often more than one optically variable object was located within
the ROSAT error circle. We therefore proposed and obtained short
Swift (Gehrels \etal\ 2004) X-ray observations to obtain a more precise
X-ray position. Typically, the X-ray telescope (XRT; Burrows \etal\ 2005) 
onboard Swift provides
positions at the 5\asec\ accuracy level for bright sources, which
degrades somewhat for faint sources.
The exposure time has been chosen to provide a secure detection
even if the source dropped by a factor of three in 
X-ray intensity.
In general, however, the number of photons detected for these
sources is not sufficient for deriving a proper X-ray spectrum.

\subsection{SDSS-III spectra}
\label{sec:SDSSspec}

At a very late stage of this paper, when practically all analysis 
was already finished, the DR9 (IIIrd phase of the Sloan Digital Sky Survey,
SDSS) release contained spectra of about 130 sources in the Coma field.
Obviously, this resource has been used.

\begin{subtables}
\begin{longtable}{rccccrrrc}
\caption{\label{comX} ROSAT X-ray sources in the Com field. All positions
  have an error radius of 30\asec. The ID column provides cross-reference to
  tables \ref{comopt},\ref{sgeopt} and Figs. \ref{comfc},\ref{sgefc}. } \\
\hline\hline
  No & Coordinates (2000.0) & RXS J Name & RXS Err & CR   & HR1~~~ & HR2~~~ & F$_{\rm X}$/10$^{-13}$ & ID \\
     & RA (\h \m \s) Decl (\degr\ \amin\ \asec)  &  & (\asec)& (c/s)&  & &erg/cm$^2$/s & \\
\hline
\endfirsthead
\caption{continued.}\\
\hline\hline
  No & Coordinates (2000.0) & RXS J Name & RXS Err & CR   & HR1~~~ & HR2~~~ & F$_{\rm X}$/10$^{-13}$ & ID\\
     & RA (\h \m \s) Decl (\degr\ \amin\ \asec)  &  & (\asec)& (c/s)&  & &erg/cm$^2$/s  & \\
\hline
\endhead
\hline
\endfoot
  1 & 12 33 55.9 +26 02 41 & 123356.4+260249 & 16 & 0.040 & -0.31$\pm$0.34 &  0.12$\pm$0.63 &  3.15 & a? \\ 
  2 & 12 36 17.8 +25 59 00 &                - &  - & 0.042 &  0.80$\pm$0.19 &  0.37$\pm$0.23 &  4.64 & a \\ 
  3 & 12 54 09.1 +25 55 26 &                - &  - & 0.137 & -0.60$\pm$0.12 &  0.09$\pm$0.33 & 11.50 & a\\ 
  4 & 12 23 08.9 +25 50 53 & 122308.9+255058 & 18 & 0.051 &  0.42$\pm$0.27 &  0.39$\pm$0.27 & 12.50 & a \\ 
  5 & 12 24 02.4 +25 51 05 &                - &  - & 0.012 &  0.27$\pm$0.47 &  0.49$\pm$0.43 &  1.28 & a\\ 
  6 & 12 22 30.7 +25 50 25 &                - &  - & 0.680 &  0.12$\pm$0.06 &  0.09$\pm$0.08 & 57.47 & a\\ 
  7 & 12 41 36.0 +25 52 44 & 124136.3+255244 & 15 & 0.050 & -0.86$\pm$0.14 & -1.00$\pm$0.00 &  3.24 & a\\ 
  8 & 12 28 12.6 +25 49 38 &                - &  - & 0.068 & -0.39$\pm$0.19 &  0.01$\pm$0.37 &  1.97 & a\\
\end{longtable}

\begin{longtable}{rccccrrrc}
\caption{\label{sgeX} ROSAT X-ray sources in the Sge field. All positions
 have an error radius of 30\asec. } \\
\hline\hline
  No & Coordinates (2000.0) & RXS J Name & RXS Err & CR   & HR1~~~ & HR2~~~ &  F$_{\rm X}$/10$^{-13}$ & ID \\
     & RA (\h \m \s) Decl (\degr\ \amin\ \asec)  &  & (\asec)& (c/s)&  & &erg/cm$^2$/s & \\
\hline
\endfirsthead
\caption{continued.}\\
\hline\hline
  No & Coordinates (2000.0) & RXS J Name & RXS Err & CR   & HR1~~~ & HR2~~~ &  F$_{\rm X}$/10$^{-13}$ & ID\\
     & RA (\h \m \s) Decl (\degr\ \amin\ \asec)  &  & (\asec)& (c/s)&  &  &erg/cm$^2$/s & \\
\hline
\endhead
\hline
\endfoot
  1 & 19 37 42.7 +24 15 13 & 193742.4+241452 & 11 & 0.037 &  0.65$\pm$0.24 & -0.42$\pm$0.26 &  4.09 & b\\ 
  2 & 19 56 28.8 +23 54 57 & 195628.7+235445 & 17 & 0.014 &  1.00$\pm$0.00 &  0.87$\pm$0.41 &  1.57 & a\\ 
  3 & 19 58 00.1 +23 54 05 & 195759.6+235356 & 17 & 0.060 &  0.37$\pm$0.18 &  0.48$\pm$0.17 &  4.10 & a?\\ 
  4 & 20 07 02.5 +23 52 46 & 200702.4+235242 & 13 & 0.010 &  0.21$\pm$0.41 &  0.05$\pm$0.46 &  1.16 &a\\ 
  5 & 19 53 13.8 +23 51 08 & 195314.1+235110 & 18 & 0.023 &  1.00$\pm$0.00 &  0.16$\pm$0.36 &  2.59 &a\\ 
  6 & 20 20 03.8 +23 38 11 & 202003.8+233818 & 10 & 0.135 & -0.30$\pm$0.12 & -0.40$\pm$0.20 & 11.55 & a\\ 
  7 & 19 45 51.9 +23 30 27 & 194552.5+233015 & 16 & 0.048 &  0.45$\pm$0.20 & -0.03$\pm$0.26 &  8.37 &a\\ 
  8 & 19 57 26.3 +23 28 43 & 195726.6+232844 & 10 & 0.027 &  0.26$\pm$0.31 &  0.10$\pm$0.31 &  2.96 & --\\
\end{longtable}
\end{subtables}
\nopagebreak

\twocolumn

\section{Optical identifications of the ROSAT sources}

\subsection{Optical variability}
\label{sec:ID_var}

As mentioned above, we will use the detection of optical variability 
and the course of the brightness changes
as an additional essential criterion 
for the identification of the ROSAT sources.
Given the number of ``only'' about 100 000 known optically variable 
objects on the sky,
i.e. 2.5 per square degree,
a crude likelyhood of finding an optically variable object within a
30\asec\ ROSAT position is 2$\times$10$^{-4}$.
A more thorough estimate which takes into account the brightness
and amplitude of variables, has been made by Richter (1968), according
to whom the percentage of variables with amplitude $>$0.3 mag among all
stars up to 17\m\ is about 0.23\%. The total number of stars
($<$17\m) in the Sge area is estimated to be 7.25$\times$10$^5$.
This implies a likelihood of finding any optically variable object within a
30\asec\ ROSAT position of about 3.8$\times$10$^{-2}$ for the Sge field.
Concerning the other field (26 Com), we deduce from Richter \& Greiner
(1999) that it contains about 1/20 of the number of stars existing
in the Sge field. If we assume that the percentage of variables
is also 0.23\% (which may not be the case due to the different galactic
latitude), we arrive at a 2$\times$10$^{-3}$ chance coincidence for a
variable object being within a 30\asec\ ROSAT position.
We note in passing that in a more recent paper Vogt et al. (2004) estimated
the percentage of variable stars among all stars to be as large as 7.9\% --
a factor 34 more than the above mentioned value. This estimate is based on
two selection effects: (i) the sample of Richter includes stars between
12 and 17 mag, while that of Vogt contains stars brighter than about 11.5 mag.
The brighter the stars, the larger the percentage of the (more frequent
variable) giant stars. (ii) the sample by Richter is limited to brightness
amplitudes $>$0.3 mag, while that of Vogt includes everything down to 
\gax 0.1 mag. It is known that the number of variables increases steeply
with smaller amplitude, though a quantitative estimate covering different
types of variables is still missing.

\begin{figure}[t]
 \hspace{-0.5cm}
 \includegraphics[width=0.99\columnwidth,angle=-90]{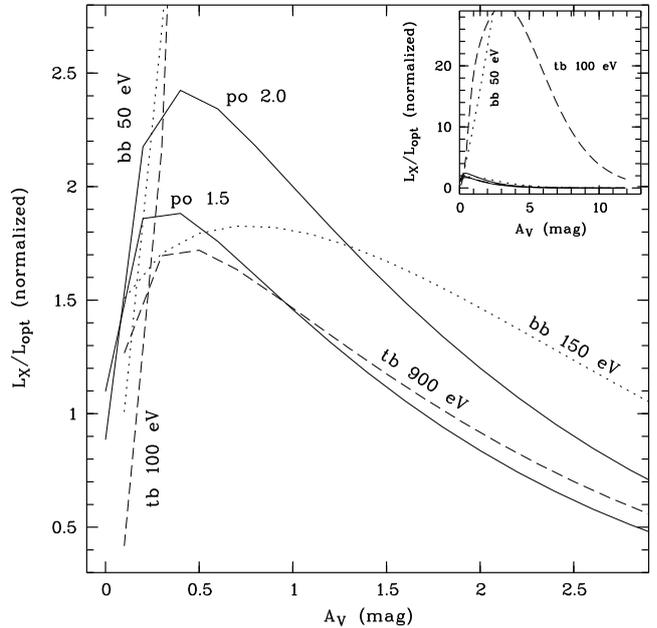}
 \caption{The ratio of absorption-corrected X-ray (0.1-2.4 keV) to 
   extinction-corrected optical (V band) luminosity as a function
   of visual extinction, with the insert showing the range up to
   $A_{\rm V}$=12 mag. The ordinate corresponds to the correction
   which has to be applied to the X-ray to optical luminosity ratio
   if the observed values without extinction correction are used.
   Shown are two blackbody (bb) models with 50 and 150 eV, 
   two thermal bremsstrahlung (tb) models with 100 eV and 900 eV,
   and two power law (po) models with photon index 1.5 and 2.0, 
   respectively. Except for very soft X-ray spectra (50 eV blackbody
   and 100 eV thermal bremsstrahlung) this ratio is within a factor
   of two with respect to the uncorrected luminosity values.
   This justifies the decision to use observed, not extinction-corrected
   values to determine the  X-ray to optical luminosity ratio (see text).
        }
\label{lxloptAv}
\end{figure}

Four additional criteria were applied in order to support the
optical identification of the X-ray sources:
\begin{enumerate}
\item Using the
objective prism spectra taken with the Hamburg Schmidt
telescope on Calar Alto, which mainly selected stars and AGN.
In the Hamburg objective prism survey (Hagen \etal\ 1995, 
Bade \etal\ 1998) spectra are
taken in the 3400--5400 \AA\, range with a dispersion of 1390 \AA/mm down to
17--18th mag covering the whole northern hemisphere except the galactic
plane ($\mid$ b $\mid >$20\degr). A systematic identification of
X-ray sources from the Bright Source Catalog of the ROSAT all-sky survey has 
been published by Zickgraf \etal\ (2003). 
\item Including the positional correlation with the X-ray positions,
  i.e. giving lower weight to more distant sources.
\item For the brighter sources
 and multiple optical counterpart candidates: 
 Obeying consistency between X-ray
 absorption as determined from the X-ray spectra and visual extinction.
\item Evaluating the X-ray to optical intensity ratio for known populations.
 It is long known (Maccacaro et al. 1982, Stocke et al. 1991)
 that stars populate only some sub-phase-space in the 
 log(f$_{\rm X}/f_{\rm opt}$) vs. $B-V$ diagram, while accretion-powered
 systems can reach much larger f$_{\rm X}/f_{\rm opt}$ values.
 We have used these properties to select the more likely optical
 counterpart (see Fig. \ref{stocke} which shows, for the Com field, 
 the stars (filled triangles) according
 to Stocke et al. (1991) together with the upper boundary of
 values reachable by stars (straight line)). We used the X-ray fluxes
  in the 0.1--2.4 keV band as determined either from spectral fitting
 (bright sources) or count-to-flux conversion (for faint sources;
more specifically: for sources with HR1$\ge$0 we use 
f$_{\rm X} = 0.9\times 10^{11}$ cts cm$^2$ erg$^{-1}$ matching a power law 
spectrum of photon index 2, while for sources with HR1$<$0 we use 
f$_{\rm X} = 1.27\times 10^{11}$ cts cm$^2$ erg$^{-1}$ matching a blackbody
spectrum with temperature of a few hundred eV)
and  the  $V$  magnitudes 
(with f$_{\rm opt}$ = 4.26$\times 10^{-6 - 0.4\times V}$)
from USNO-A2 (if not available, then $B$). 
We did not apply extinction
  correction, neither to the X-ray fluxes nor to the visual magnitudes,
 as these corrections are uncertain and the net effect is within
 a factor of 2 for most sources (Fig. \ref{lxloptAv}).
 This allows to select the most probable stellar counterpart, but not 
 to discriminate between e.g. star and quasar 
(since then a large \fxo\ range is allowed).
\end{enumerate}

An unequivocal optical identification of the ROSAT sources was not possible 
in some cases because the accuracy of the ROSAT X-ray positions is not better 
than 30\asec, and in 
most cases there are several possible optical counterpart candidates within
the error circle 
and therefore Swift/XRT observations were obtained. 
This is particularly true in the low galactic latitude field 
$\gamma$ Sge. Moreover, the real counterpart may be invisible in some cases,
i.e. fainter than our limiting magnitude. Therefore we can only estimate
the reliability of a supposed optical identification by statistical methods
(see, e.g., Richter \& Greiner 1999). 
But anticipating the
results, we can say that the probability of a positive optical identification 
is very high if we are dealing either with a bright object (as a rule an 
object with an HD or a NGC number), or an active galactic nucleus (quasar, 
BL Lac object, Seyfert galaxy). As a new element for optical identification
we also consider an object with brightness variability as a very likely
optical counterpart 
since the chance probability of a variable object to be inside the error 
circle of the ROSAT position is rather small (see above).

Thus, we have tested the objects in or near the error circles of the ROSAT 
positions for variability with the following exceptions: 
1. Very bright objects, mostly HD stars, which are too bright to be tested 
on astrographic plates. 
Empirically, most of them will be BY Dra stars 
with amplitudes rarely more than 0.1 mag and therefore their variability
is hard to discover on photographic plates. 
2. Objects fainter than about 17.5 mag near the plate centre and about 
16.5 mag near the edge of a plate. 
3. galaxies (with exception of some AGN). 

All magnitudes, if not noticed otherwise, are photographic and not 
far from the $B$ magnitude in the system of Johnson and Morgan
(typically $B$ = $m_{\rm Phot}$ + 0.1 mag).

The results are given in Tables \ref{comopt},\ref{sgeopt} and \ref{IDsum}, 
and Figures \ref{comfc},\ref{sgefc} contain the finding charts of all 
ROSAT sources, 
with each
chart having a size of 2\farcm5$\times$2\farcm5. The labelling is the same as 
in column 2 of Table  \ref{comopt},\ref{sgeopt}.

Tables \ref{comopt},\ref{sgeopt} contain the data of optical objects 
inside or near the error 
circles of the ROSAT positions. As a rule, all objects brighter than 
about 18 mag are listed. In some cases, if of interest, also fainter 
objects are added. The columns contain the following informations: 
\begin{itemize}
\vspace{-0.22cm}
\item[] $\!\!\!\!${\it Column 1}: Running number of the ROSAT objects 
from Table  \ref{comX},\ref{sgeX}, appended with
an alphabetical specification of all optical objects 
inside or near the error circle of about 30\asec\ of the ROSAT position.
These letters are also used as labels in Fig. \ref{comfc},\ref{sgefc}.
\item[] $\!\!\!\!${\it Column 2}: Coordinates (2000.0) of the optical objects 
in right ascension
(\h,\m,\s) and declination (\degr, \amin, \asec). A dash means that no object 
visible on our plates (with some exceptions)
is within the error circle of the ROSAT source.
\item[] $\!\!\!\!${\it Column 3}: Distance $D$ between ROSAT position and 
optical position in
arcseconds, unless the optical object is extended.
\item[] $\!\!\!\!${\it Column 4}: Number in usual star or nebular object 
catalogues.
\item[] $\!\!\!\!${\it Column 5}: If the object is variable: 
Name from the General Catalogue of 
Variable Stars and its supplements (GCVS, Moscow). If not named, the 
preliminary designation of newly discovered Sonneberg variables is given 
by the usual S-number (some prominent cases among the
of order 70 new variables  discovered in this work have already been
published by us separately, and already received an IAU variable star name).
In few cases, the number in the New Catalogue of Suspected Variable Stars 
(NSV catalogue, Moscow 1982) is given. 
If the object was found to be constant on all plates, a "C" is given 
while a "C?" means that there could be
small amplitude variations which are marginal in our data.
A void place 
means that the object was not tested for variability, for example bright stars 
(mostly HD stars), and very faint stars below the plate limit.
\item[] $\!\!\!\!${\it Column 6}: Type of object and type of variability 
corresponding to the 
nomenclature of the GCVS. G = galaxy. GCl = cluster of galaxies. 
AGN = active galactic nucleus. 
AGN? = supposed AGN only by reason of its blue colour, though in single 
cases it may be a white dwarf or a cataclysmic variable. If possible, further 
sub-classification of AGN: QSO = quasistellar object, 
BLL = BL Lacertae object, SY = Seyfert galaxy. 
ULX = ultra-luminous X-ray source.
CV = cataclysmic variable. 
Further specification: UGSU = SU UMa type, AM = AM Her type, 
NC = very slow nova.
% NL = novalike. 
E = eclipsing variable,
EA = Algol type, 
EB = Beta Lyrae type. CA = chromospherically active star. 
If possible, further subclasses are specified: 
RS = RS CVn, BY = BY Dra. UV = UV Cet type. 
LB resp. SRB = slowly  irregular resp. semiregular variable of late 
spectral type. PN = planetary nebula. 
\item[] $\!\!\!\!${\it Column 7}: Spectral type. FG means F or G star 
according to the objective prism spectra taken with the Hamburg Schmidt
telescope on Calar Alto (Bade \etal\ 1998).
\item[] $\!\!\!\!${\it Columns 8-10}: RBV magnitudes; if existing, mostly
from the USNO-A2 catalogue.
\item[] $\!\!\!\!${\it Column 11}: Brightness amplitude, generally in the blue.
\item[] $\!\!\!\!${\it Column 12}: Logarithm of the ratio of X-ray to 
visual flux. This ratio is arbitrary for variable sources, since
the catalogued $V$ (or ($B$) band value is taken 
which is not contemporaneous to the X-ray observation.
\end{itemize}

Dates or light curves of already known variables can be found in the 
literature 
(see notes on individual objects). For newly discovered variable objects 
we give light curves either for the whole time of observation
(Figs. \ref{comolc},\ref{sgeolc}), or for interesting time intervals 
(Figs. \ref{com27lc} - \ref{sge131lc}; see above section). 

The {\it Simbad} and {\it NED} catalogues were used to search for coincidences,
and matches are listed either in 
Table \ref{comopt}, \ref{sgeopt}, or mentioned in sections 
\ref{sec:IDComNotes} and \ref{sec:IDSgeNotes}.

\subsection{Pointed X-ray observations}
\label{sec:ID_PX}

In addition to the ROSAT all-sky survey and optical data, we have 
used the ROSAT pointed observations with both the position-sensitive
proportional counter (PSPC) and the high-resolution imager (HRI), 
the {\it Chandra} and {\it XMM-Newton} archives, and
dedicated Swift/XRT observations. This provided more accurate
source positions which were particularly helpful in the Sge field
due to the heavy crowding in the galactic plane. It also provided
a second epoch X-ray observation which allows us to assess X-ray
variability for a number of sources. The results of the dedicated
Swift/XRT observations are summarized in Tables \ref{comxrt} and 
\ref{sgexrt}. Improved positions and variability are always 
mentioned in the notes for the individual objects, sections 
\ref{sec:IDComNotes} and \ref{sec:IDSgeNotes}.

\begin{subfigures}
\begin{figure*}[ht]
\includegraphics[width=3.9cm, bb=76 410 385 760, angle=-90,clip]{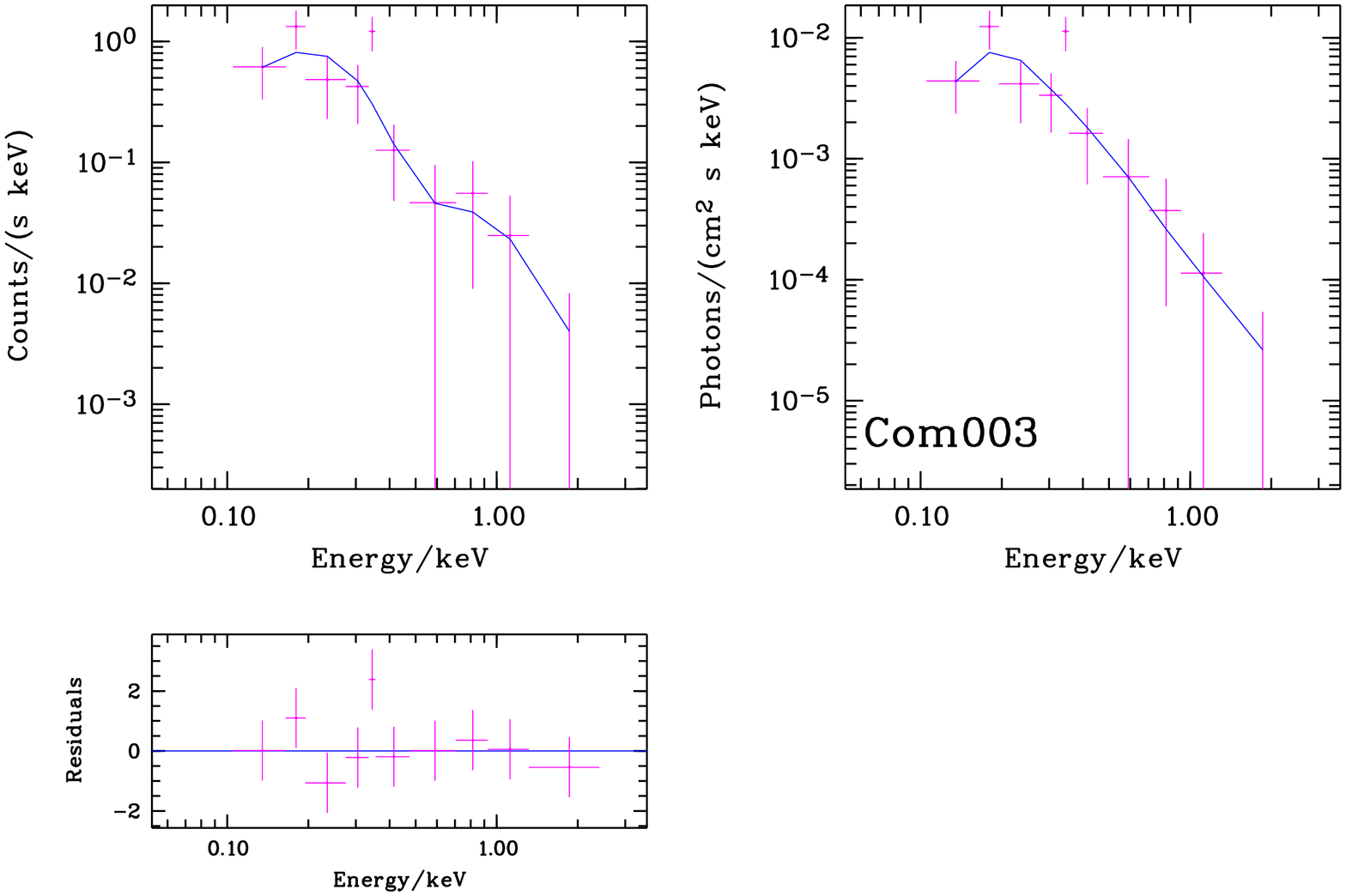}
\includegraphics[width=3.9cm, bb=76 410 385 760, angle=-90,clip]{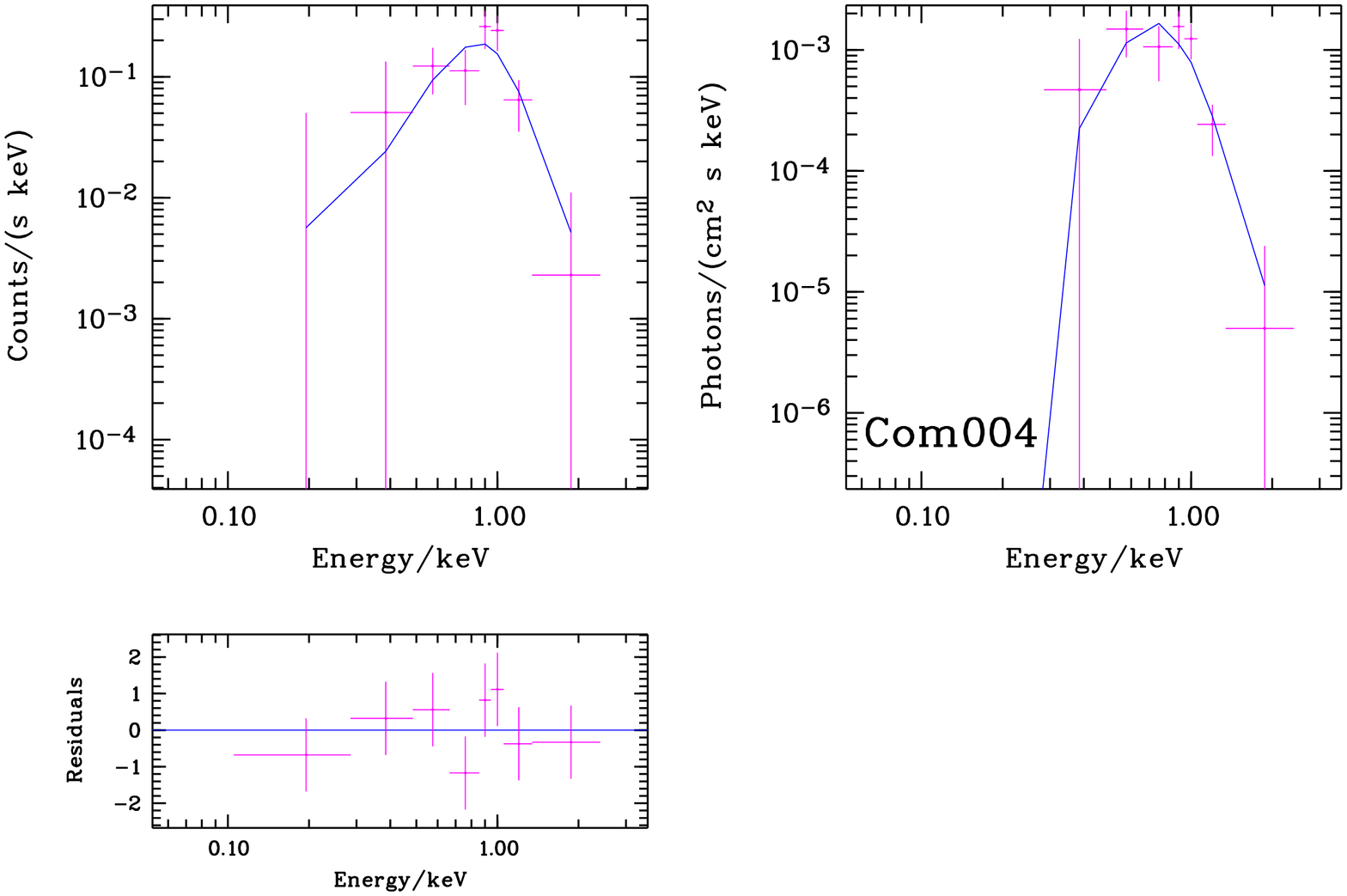}
\includegraphics[width=3.9cm, bb=76 410 385 760, angle=-90,clip]{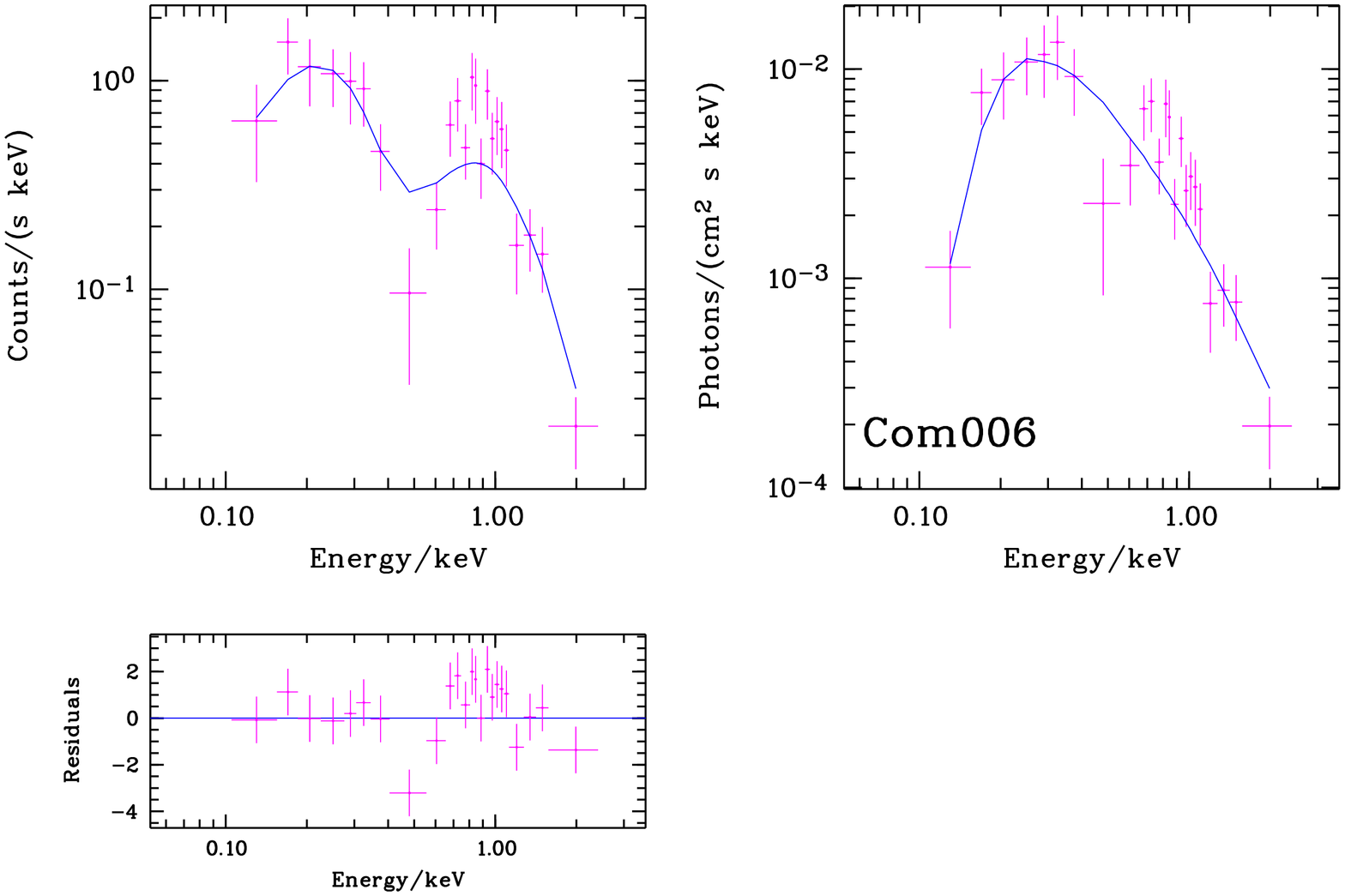}
\includegraphics[width=3.9cm, bb=76 410 385 760, angle=-90,clip]{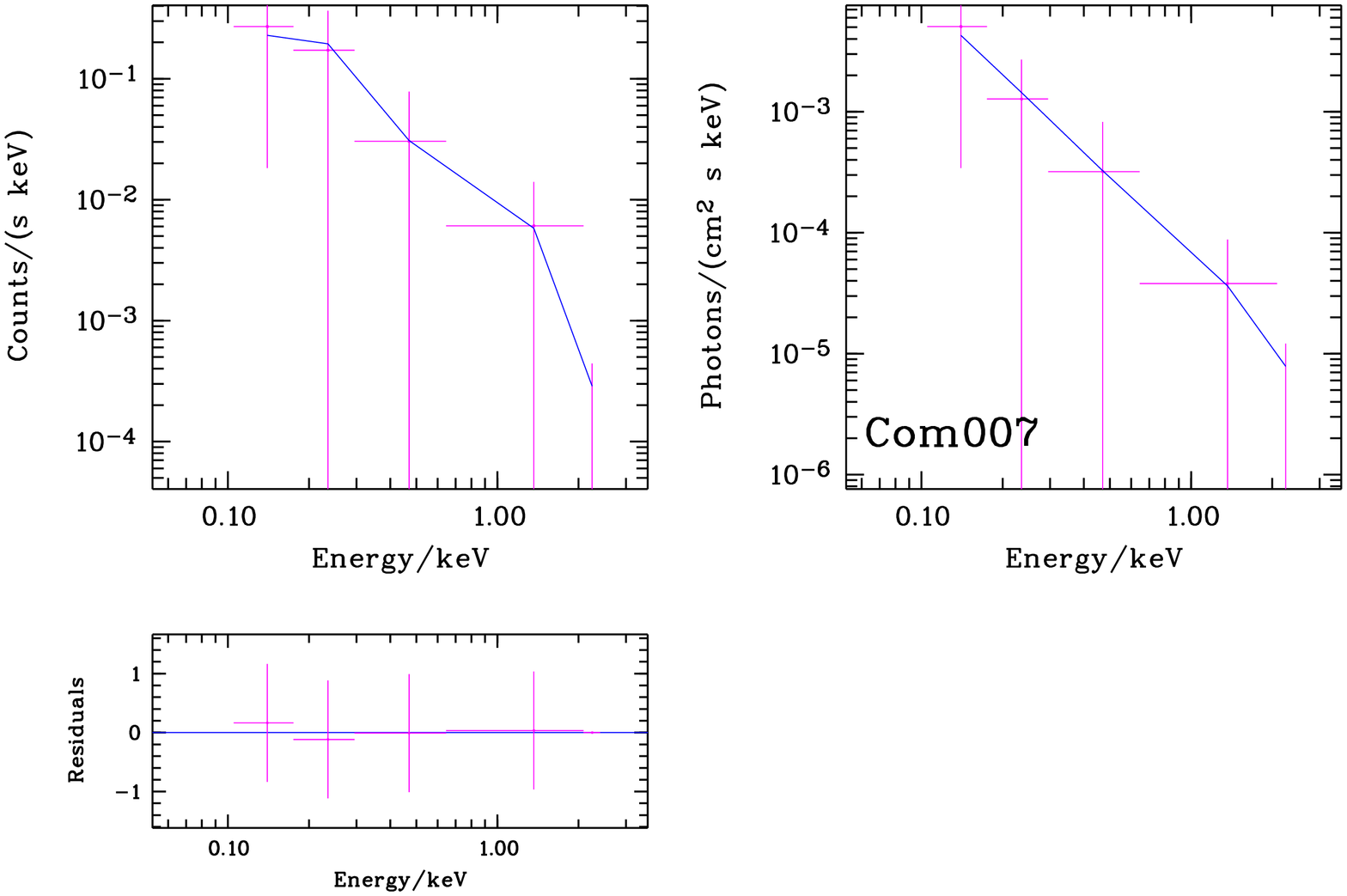}

\includegraphics[width=3.9cm, bb=76 410 385 760, angle=-90,clip]{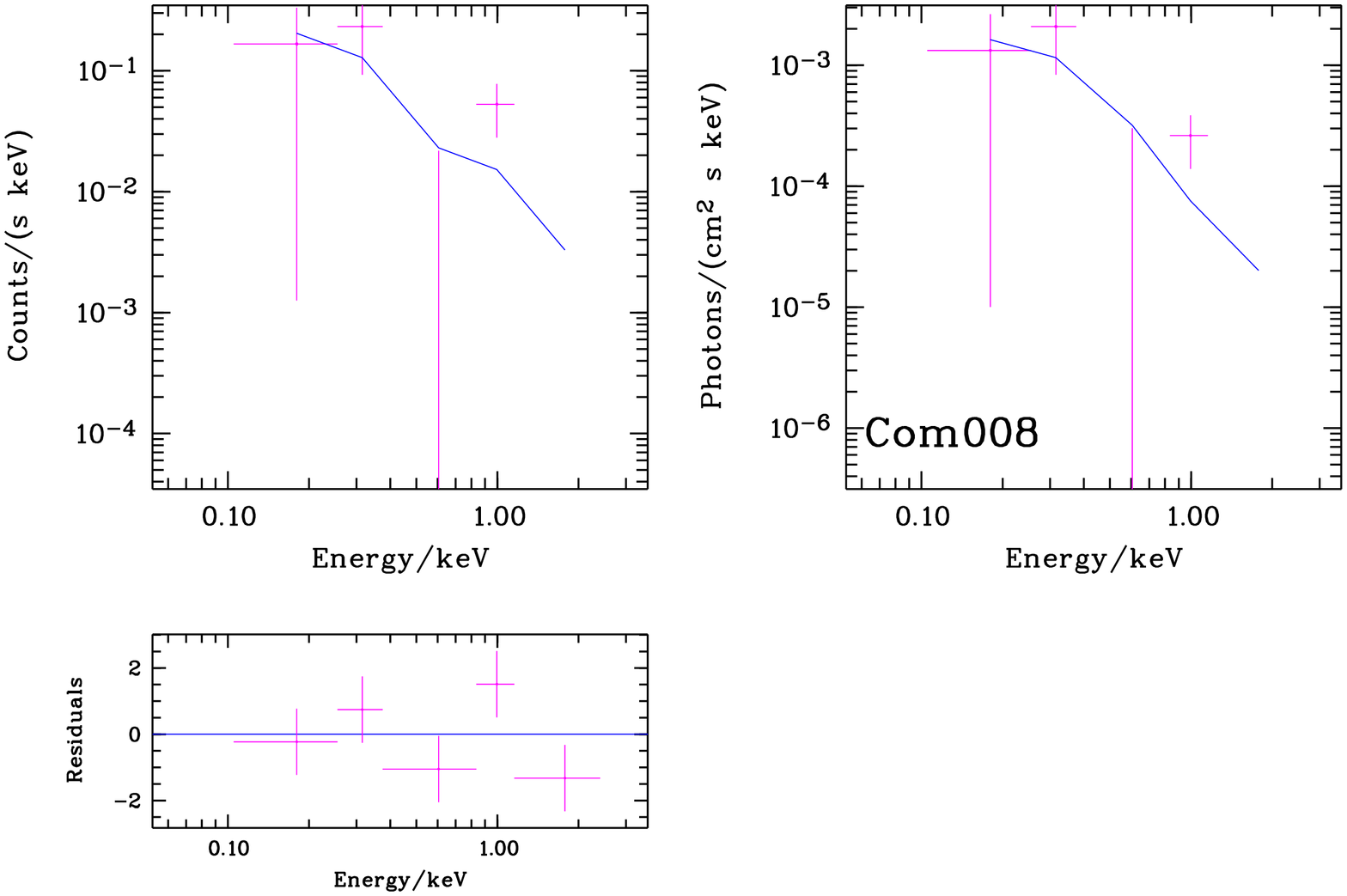}
\includegraphics[width=3.9cm, bb=76 410 385 760, angle=-90,clip]{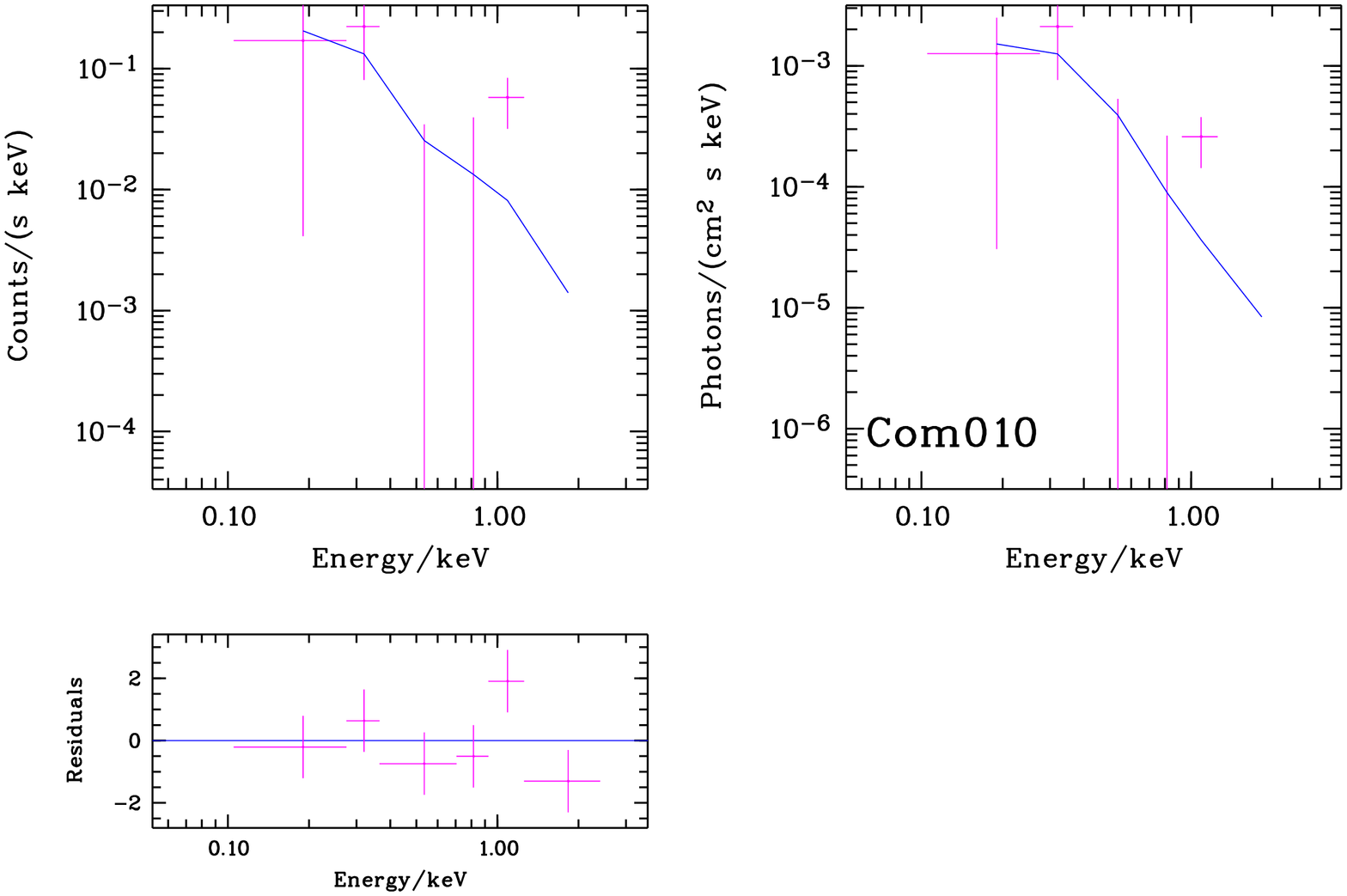}
\includegraphics[width=3.9cm, bb=76 410 385 760, angle=-90,clip]{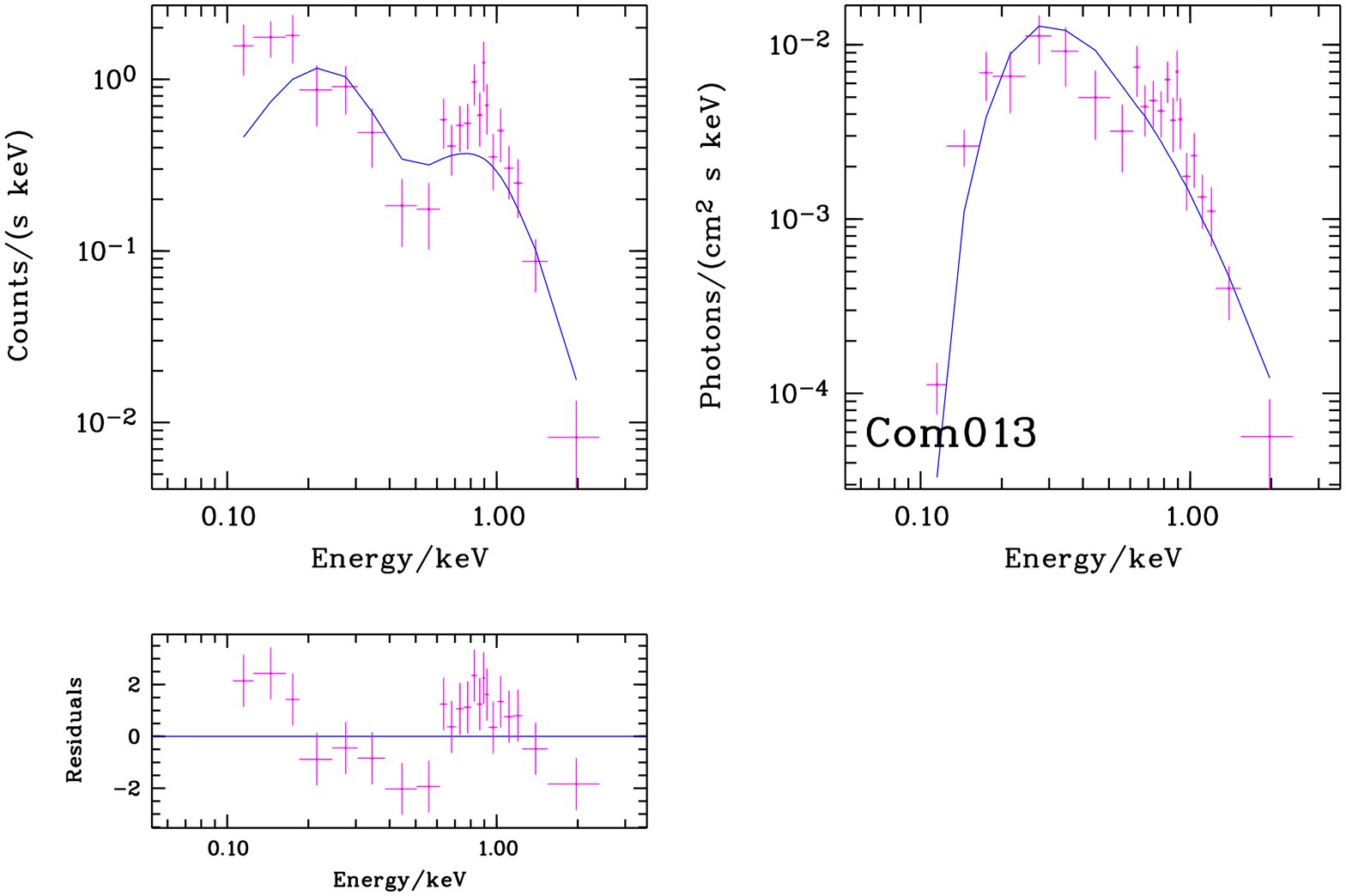}
\includegraphics[width=3.9cm, bb=76 410 385 760, angle=-90,clip]{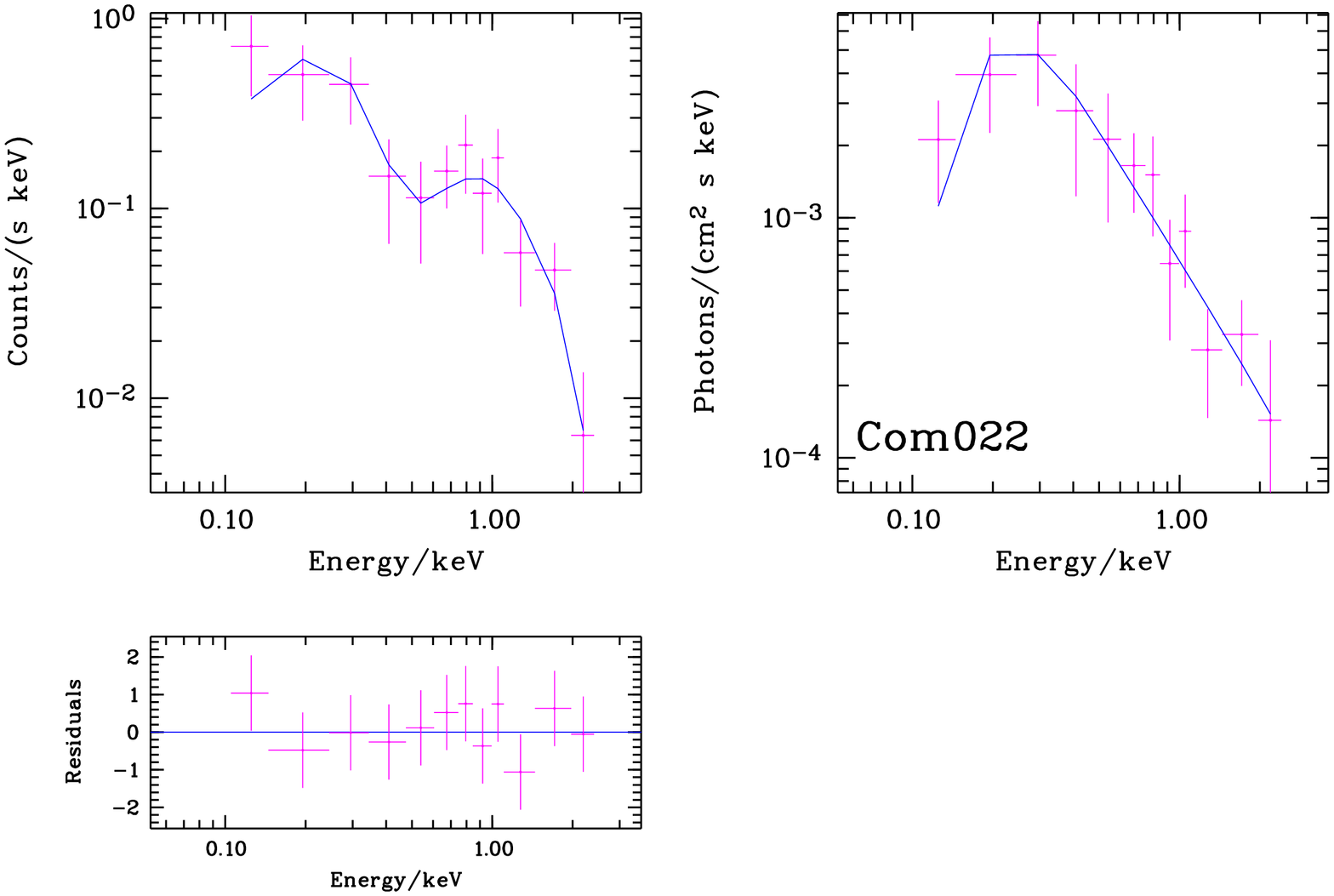}

\includegraphics[width=3.9cm, bb=76 410 385 760, angle=-90,clip]{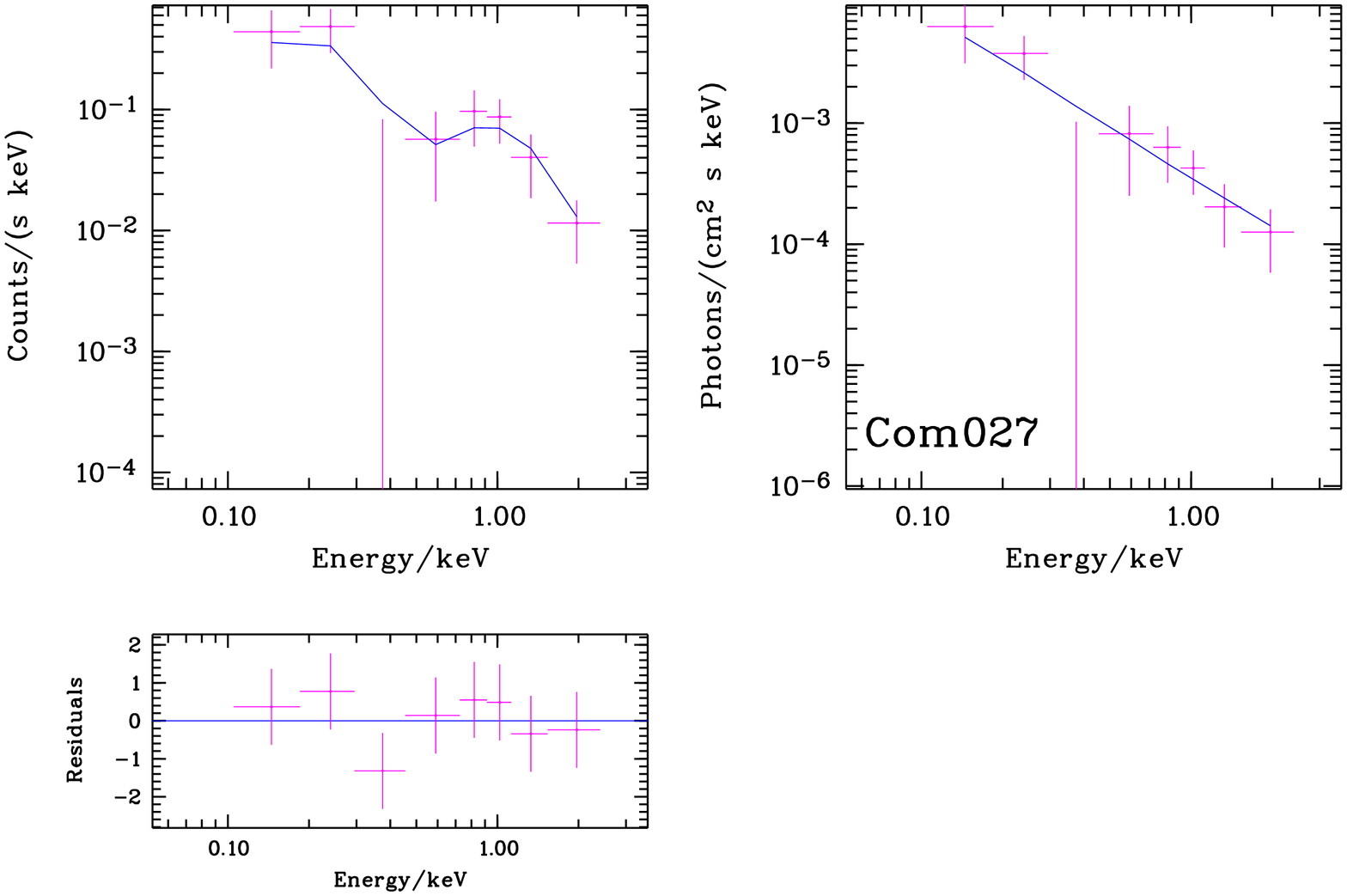}
\includegraphics[width=3.9cm, bb=76 410 385 760, angle=-90,clip]{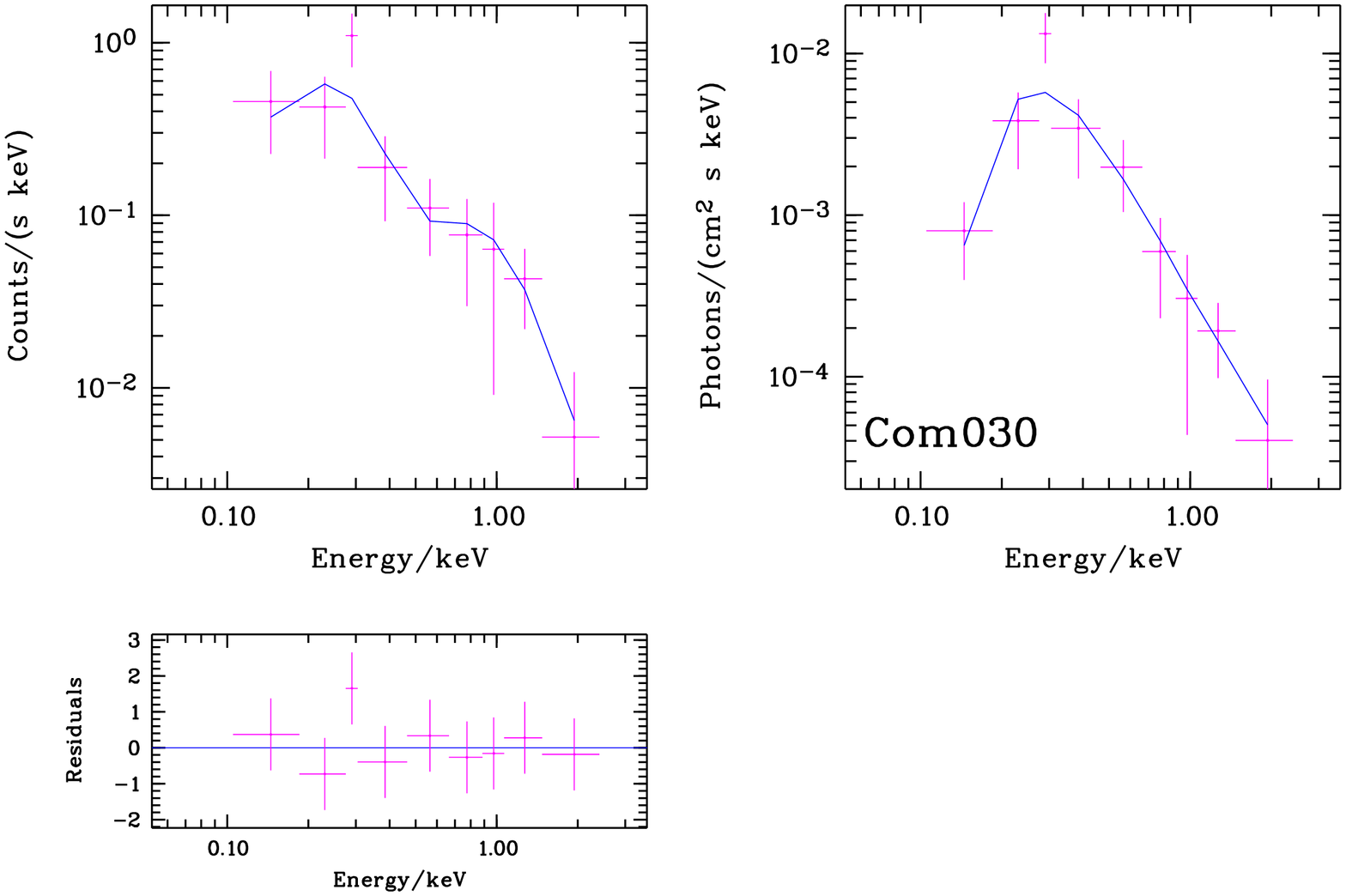}
\includegraphics[width=3.9cm, bb=76 410 385 760, angle=-90,clip]{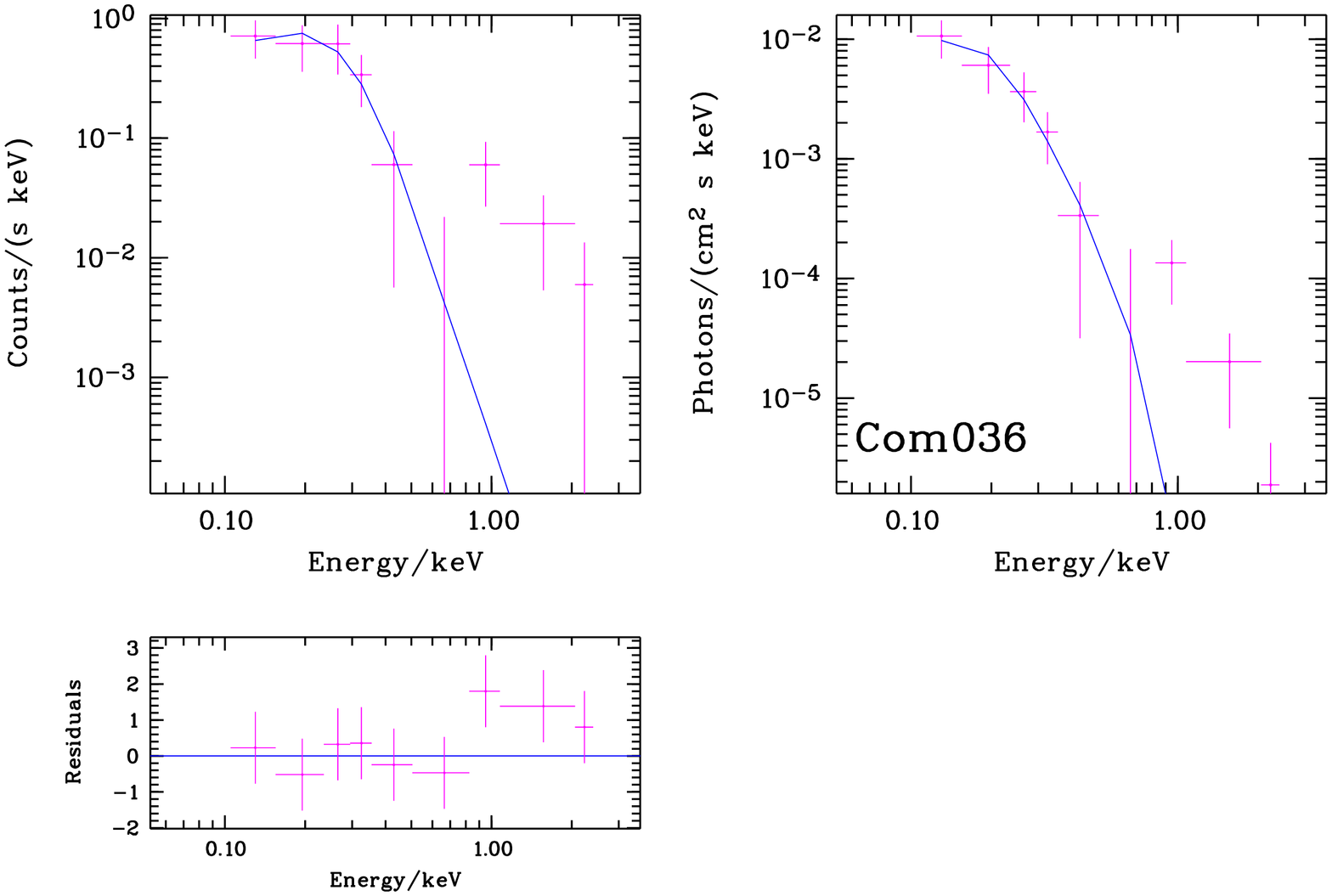}
\includegraphics[width=3.9cm, bb=76 410 385 760, angle=-90,clip]{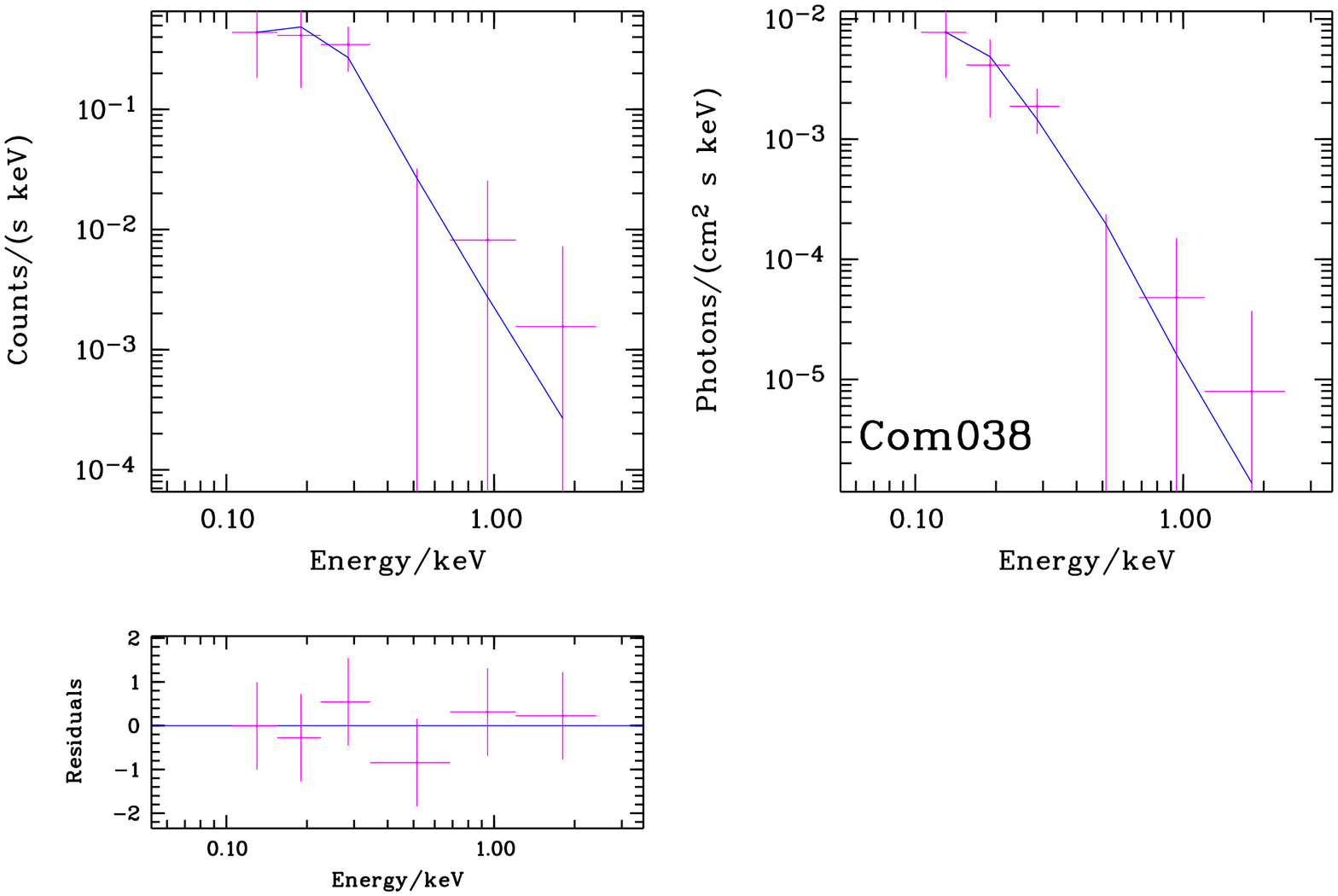}

\includegraphics[width=3.9cm, bb=76 410 385 760, angle=-90,clip]{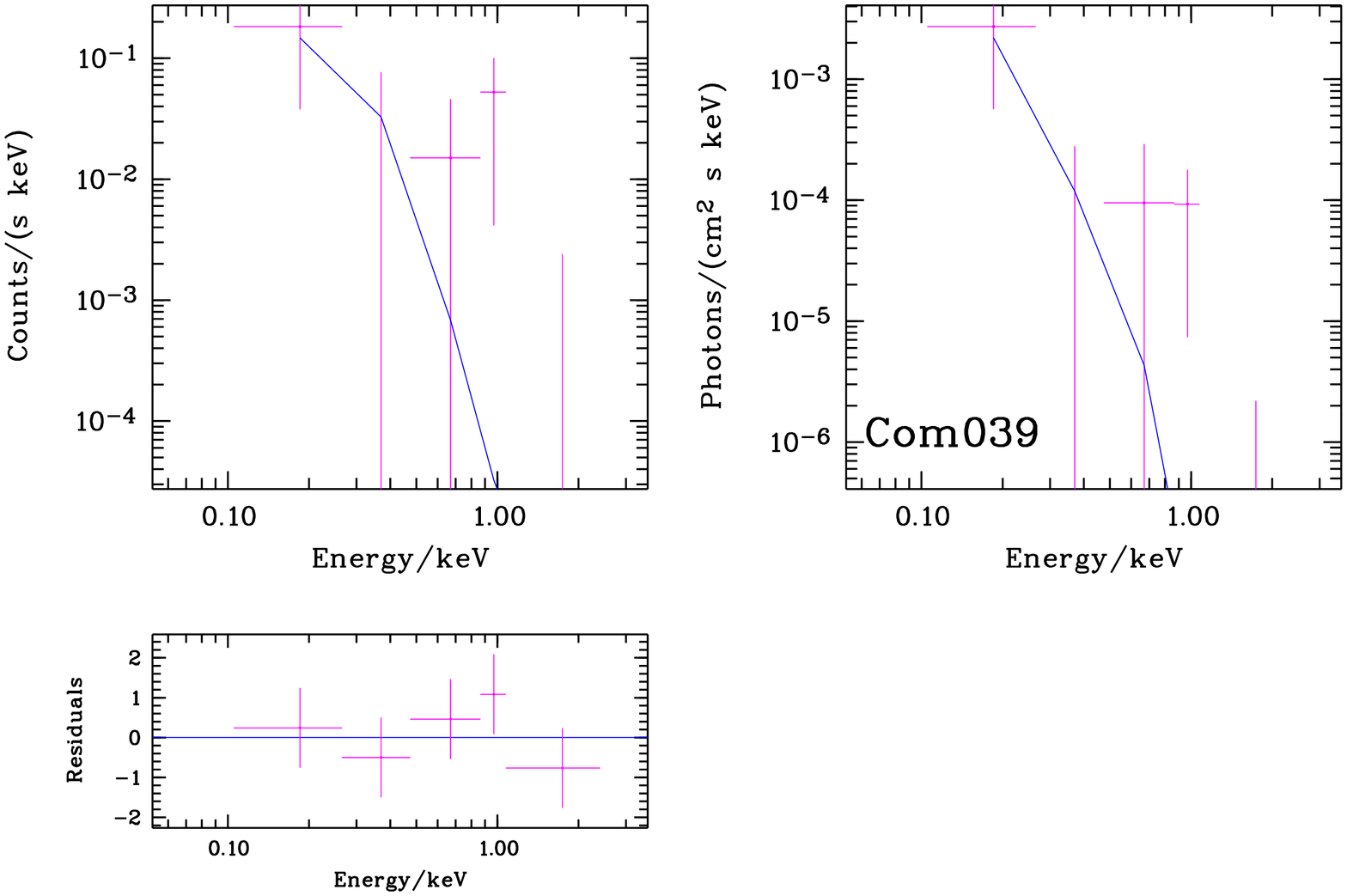}
\includegraphics[width=3.9cm, bb=76 410 385 760, angle=-90,clip]{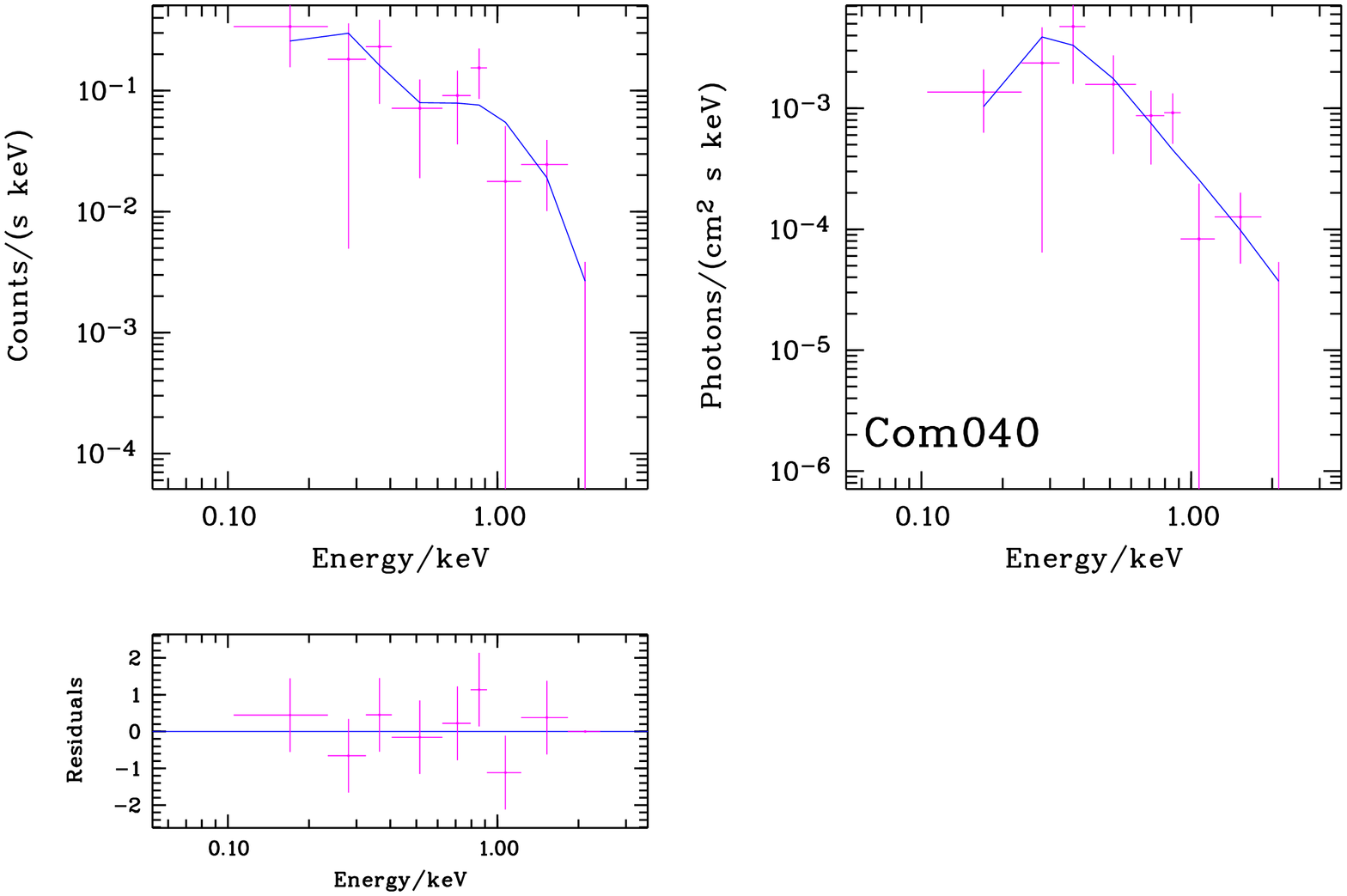}
\includegraphics[width=3.9cm, bb=76 410 385 760, angle=-90,clip]{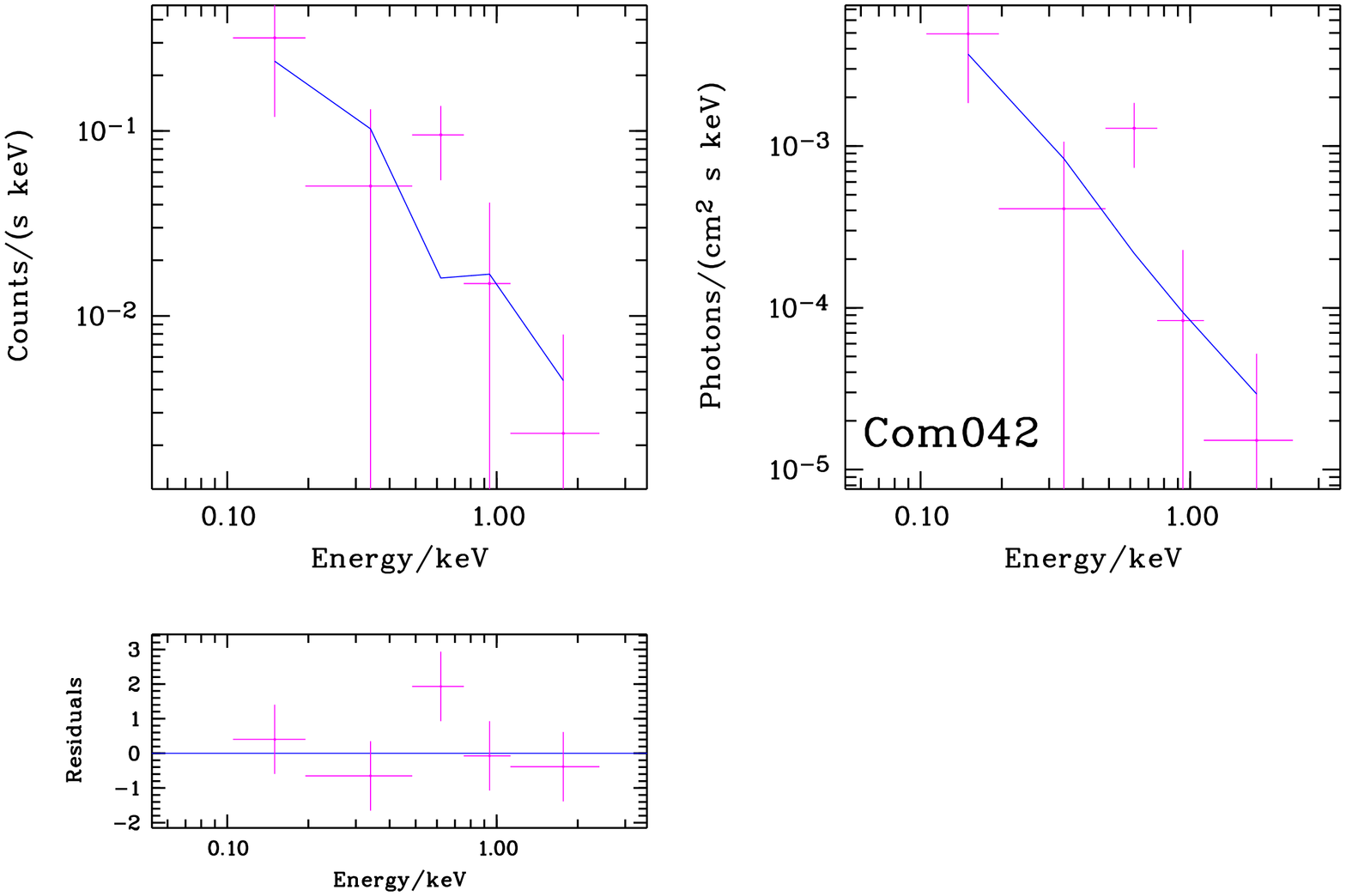}
\includegraphics[width=3.9cm, bb=76 410 385 760, angle=-90,clip]{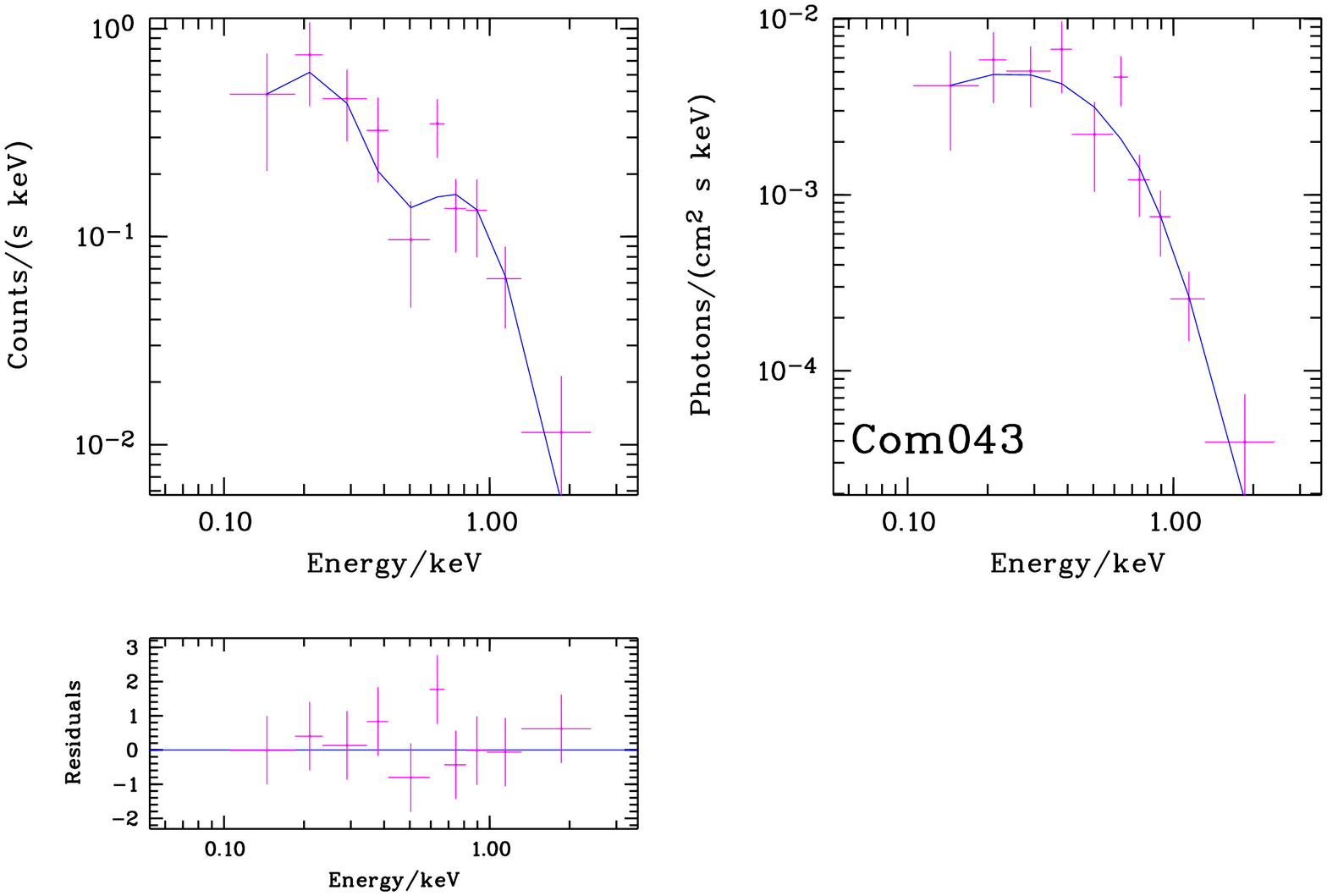}

\includegraphics[width=3.9cm, bb=76 410 385 760, angle=-90,clip]{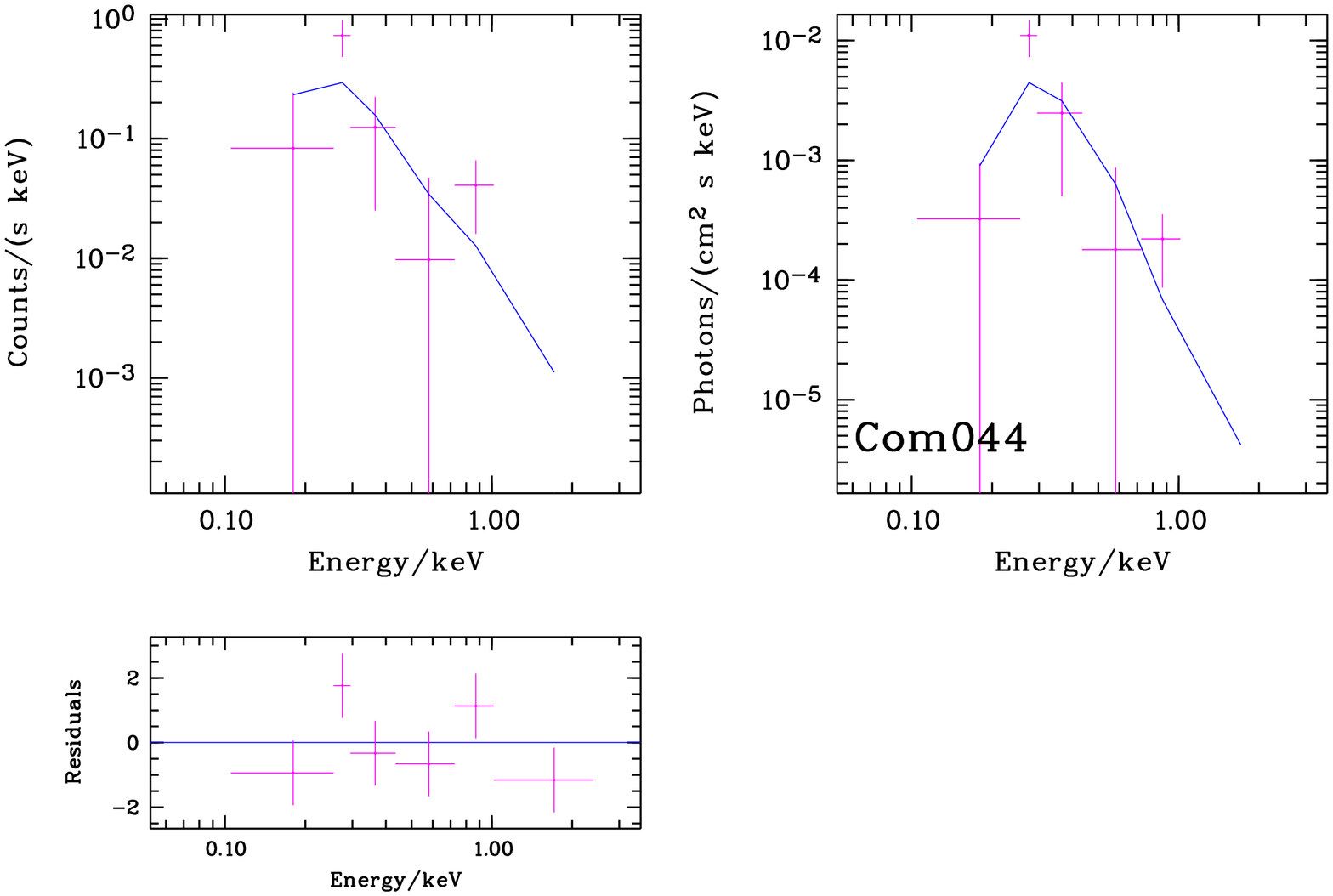}
\includegraphics[width=3.9cm, bb=76 410 385 760, angle=-90,clip]{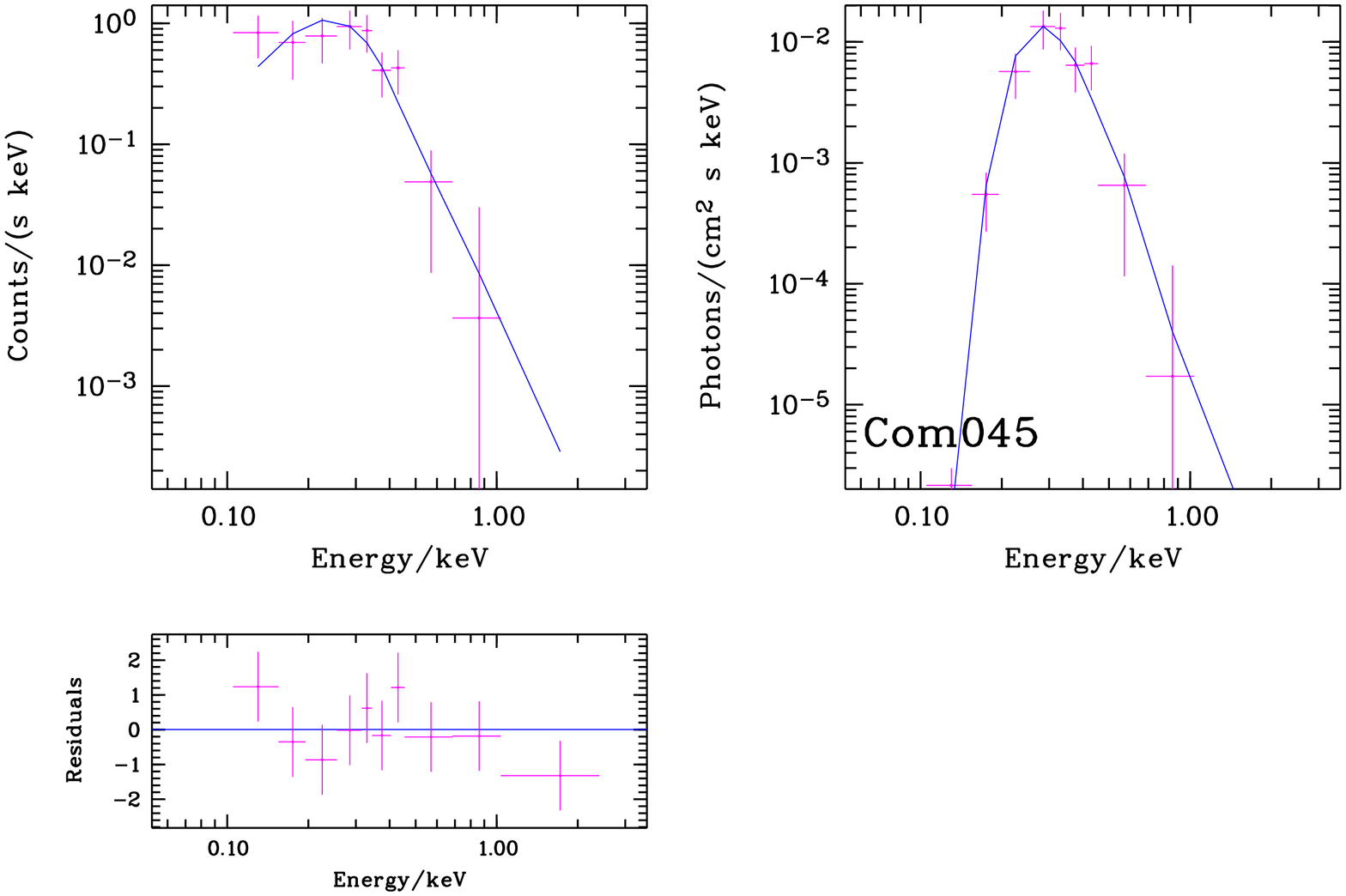}
\includegraphics[width=3.9cm, bb=76 410 385 760, angle=-90,clip]{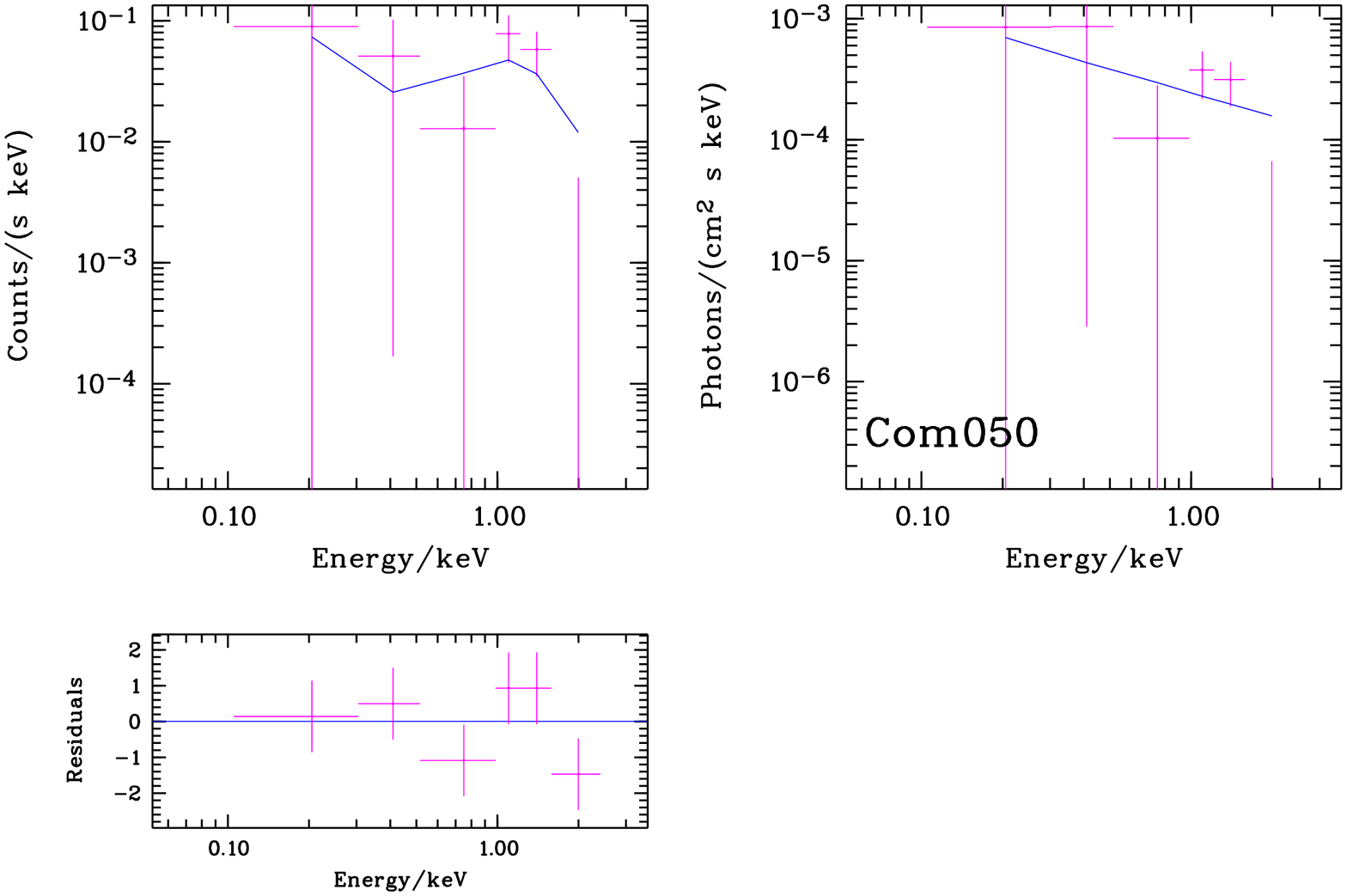}
\includegraphics[width=3.9cm, bb=76 410 385 760, angle=-90,clip]{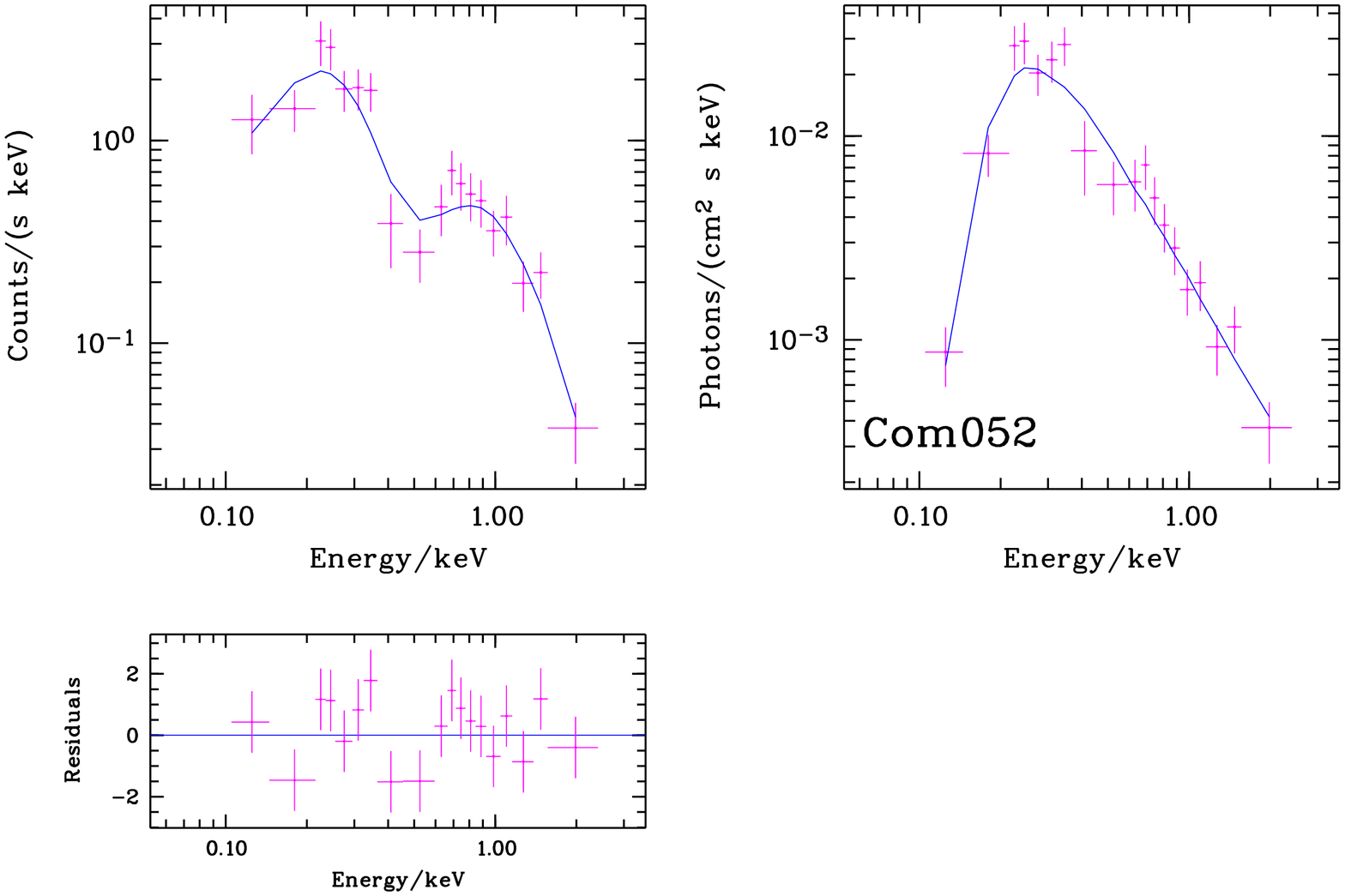}

\includegraphics[width=3.9cm, bb=76 410 385 760, angle=-90,clip]{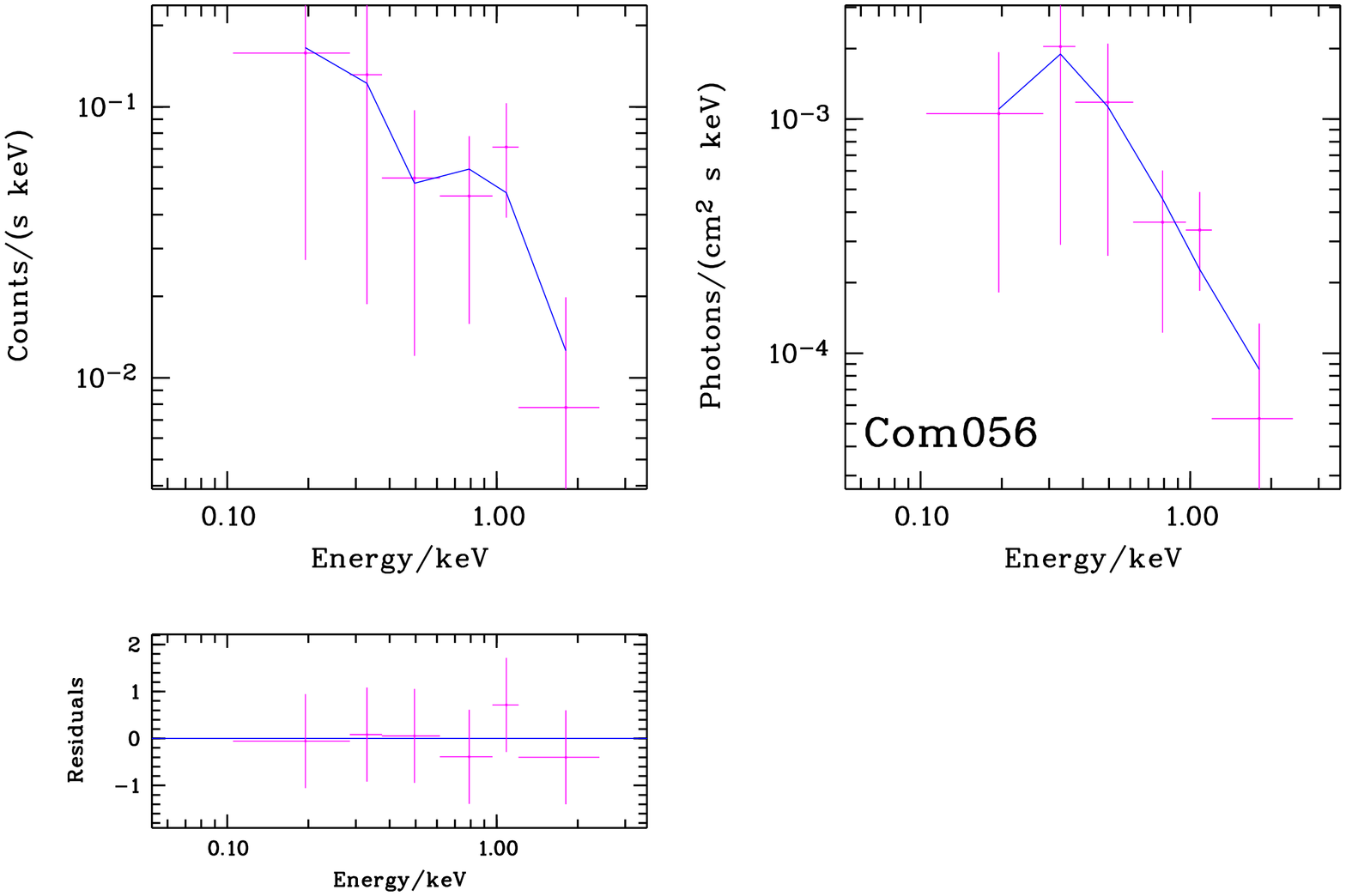}
\includegraphics[width=3.9cm, bb=76 410 385 760, angle=-90,clip]{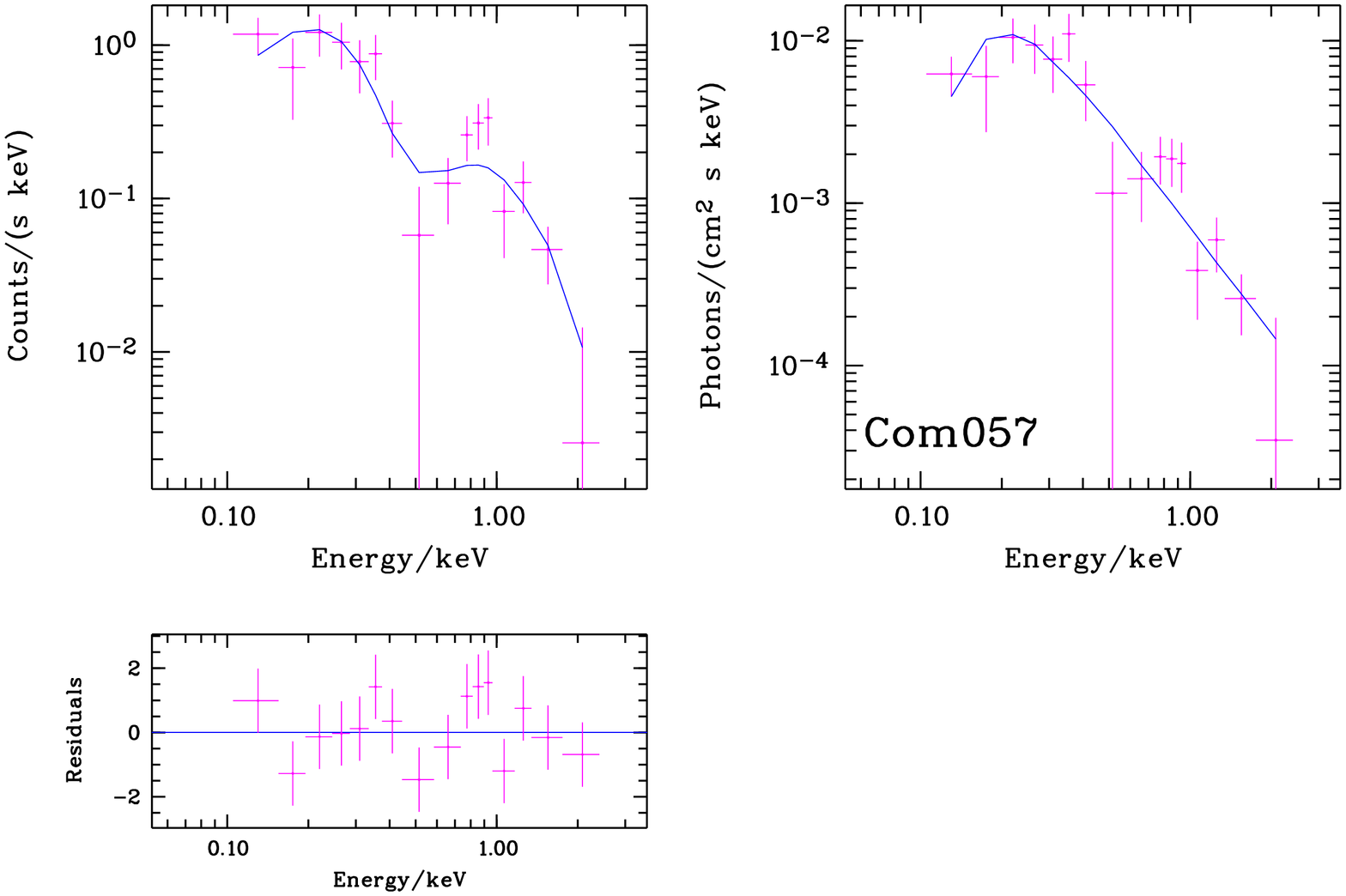}
\includegraphics[width=3.9cm, bb=76 410 385 760, angle=-90,clip]{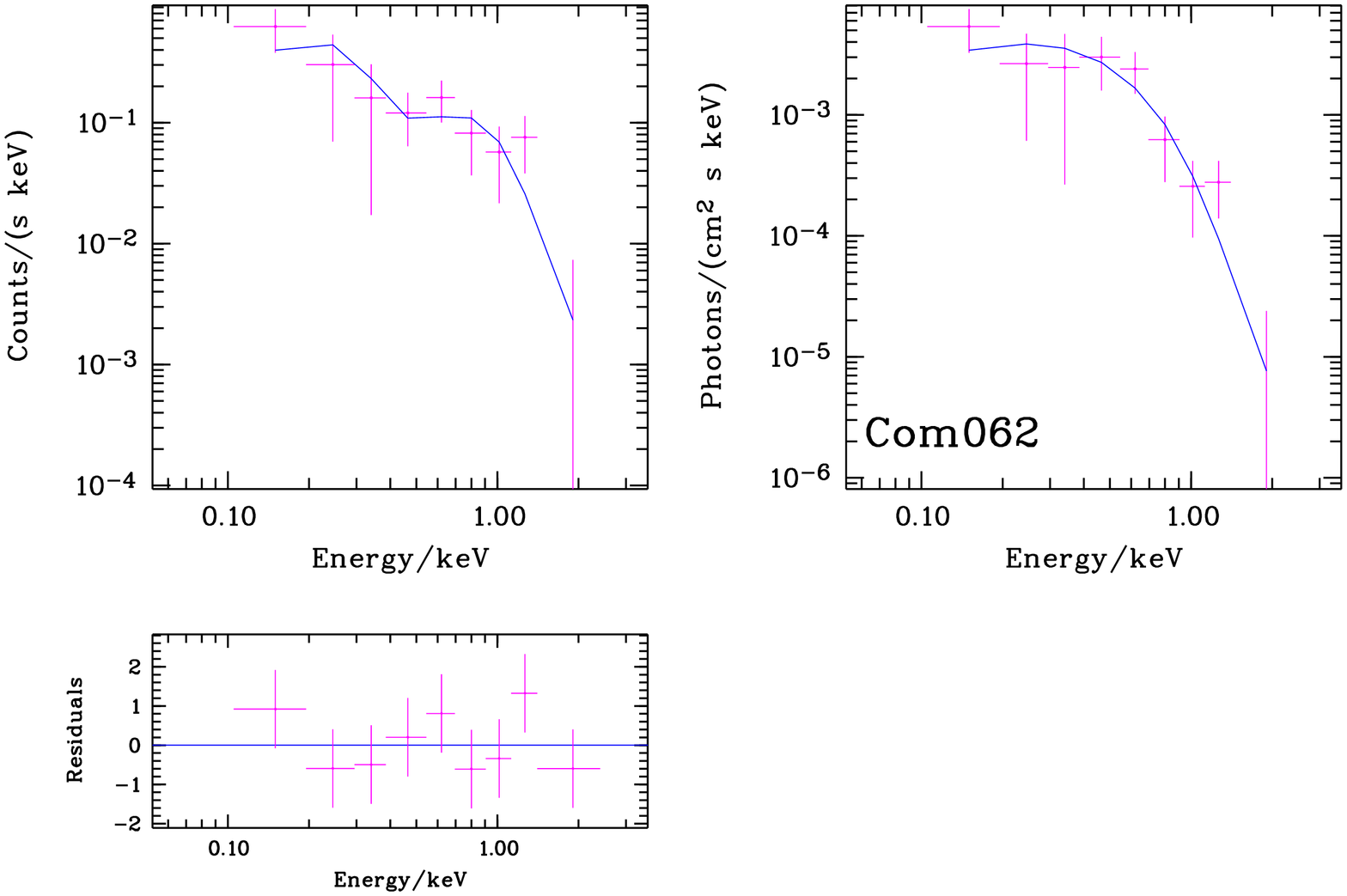}
\includegraphics[width=3.9cm, bb=76 410 385 760, angle=-90,clip]{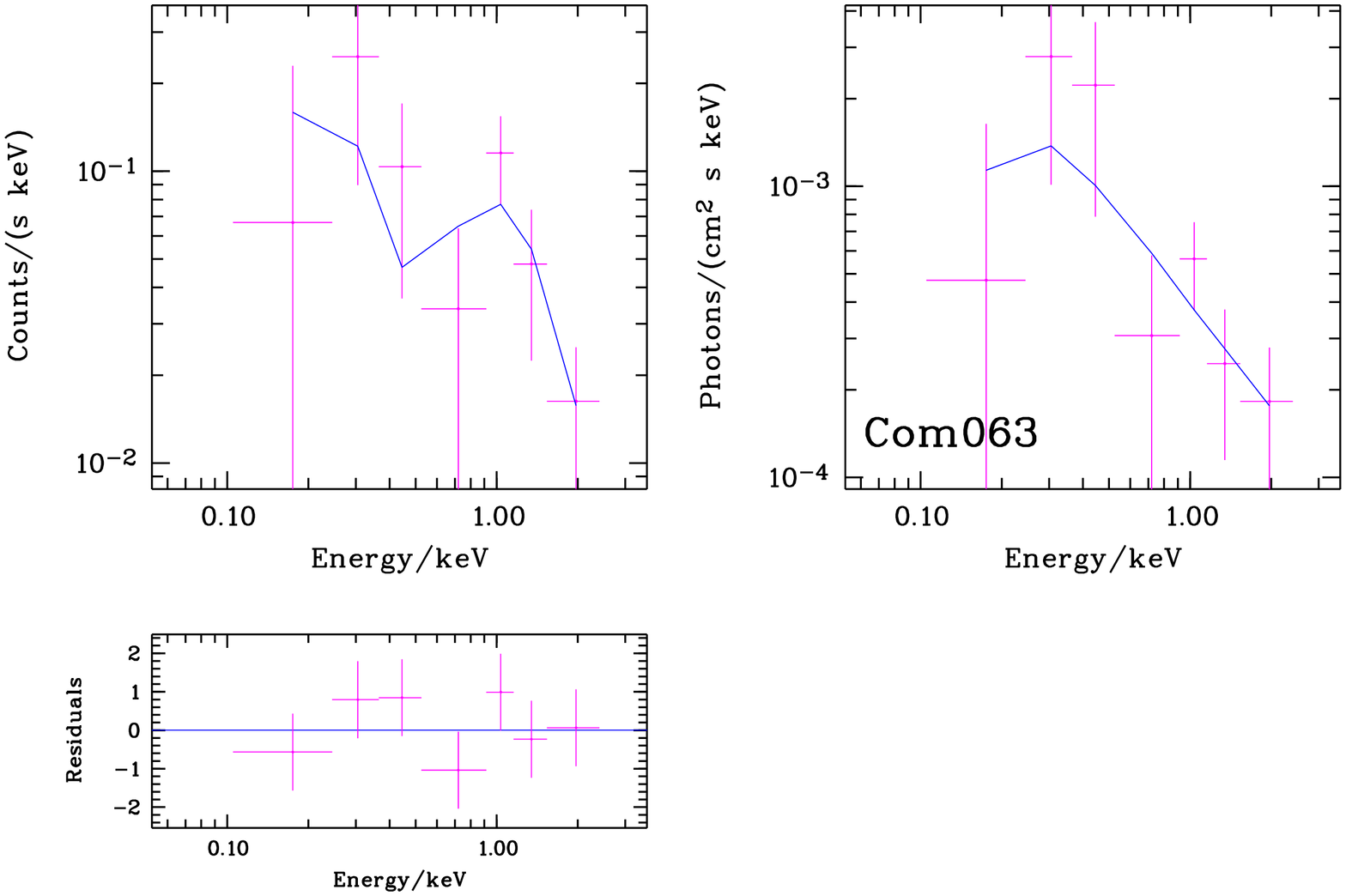}

\caption[comXspec]{\label{comspec} ROSAT survey spectra of Com sources. 
For each source, the
photon spectrum is shown. The choice of the model for a given source
was made based on the lowest reduced $\chi^2$ and consistency with the
optical properties (see Table \ref{specparcom} for model 
and spectral parameters).}
\end{figure*}

\setcounter{figure}{0}

\begin{figure*}[ht]

\includegraphics[width=3.9cm, bb=76 410 385 760, angle=-90,clip]{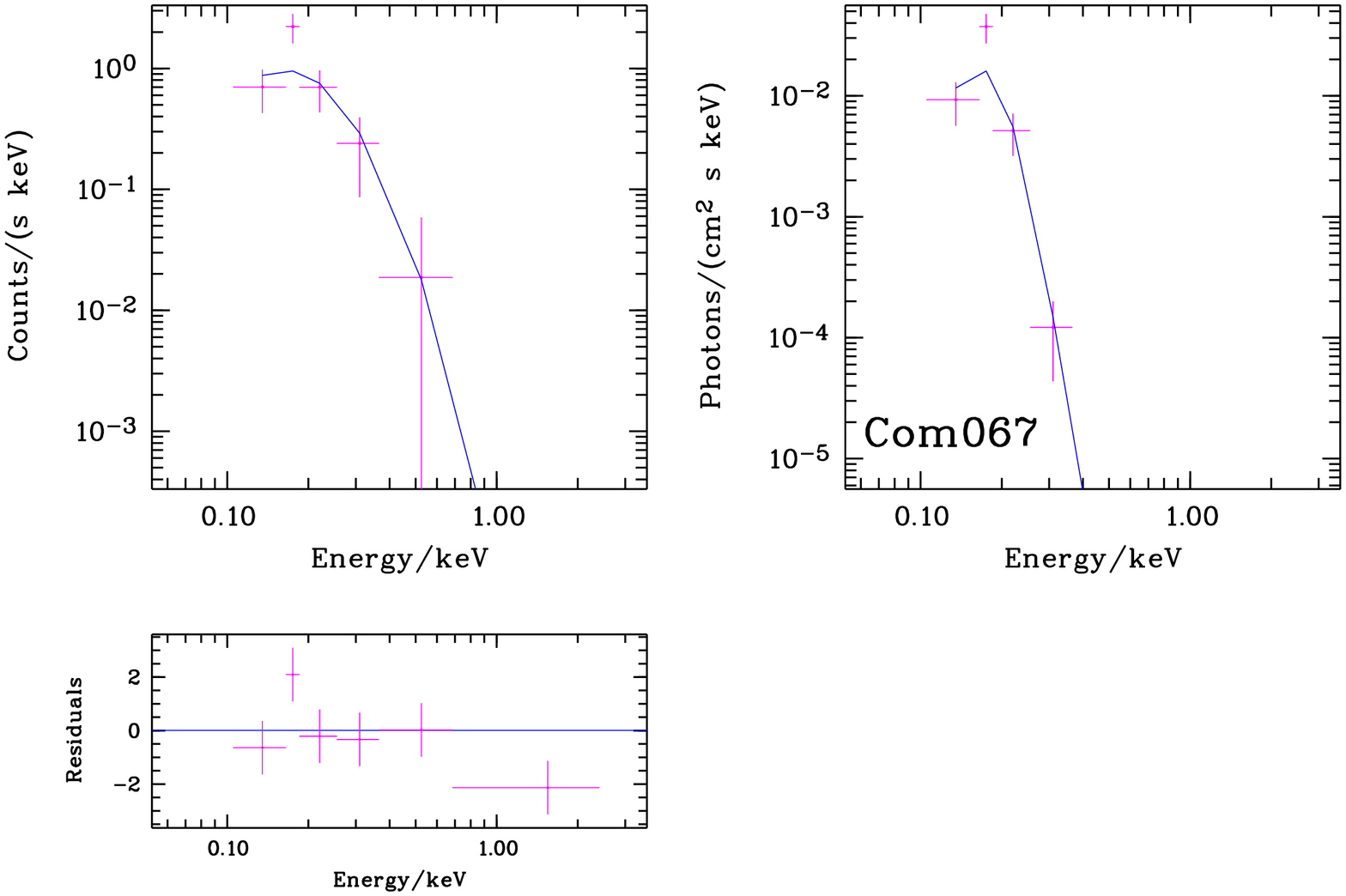}
\includegraphics[width=3.9cm, bb=76 410 385 760, angle=-90,clip]{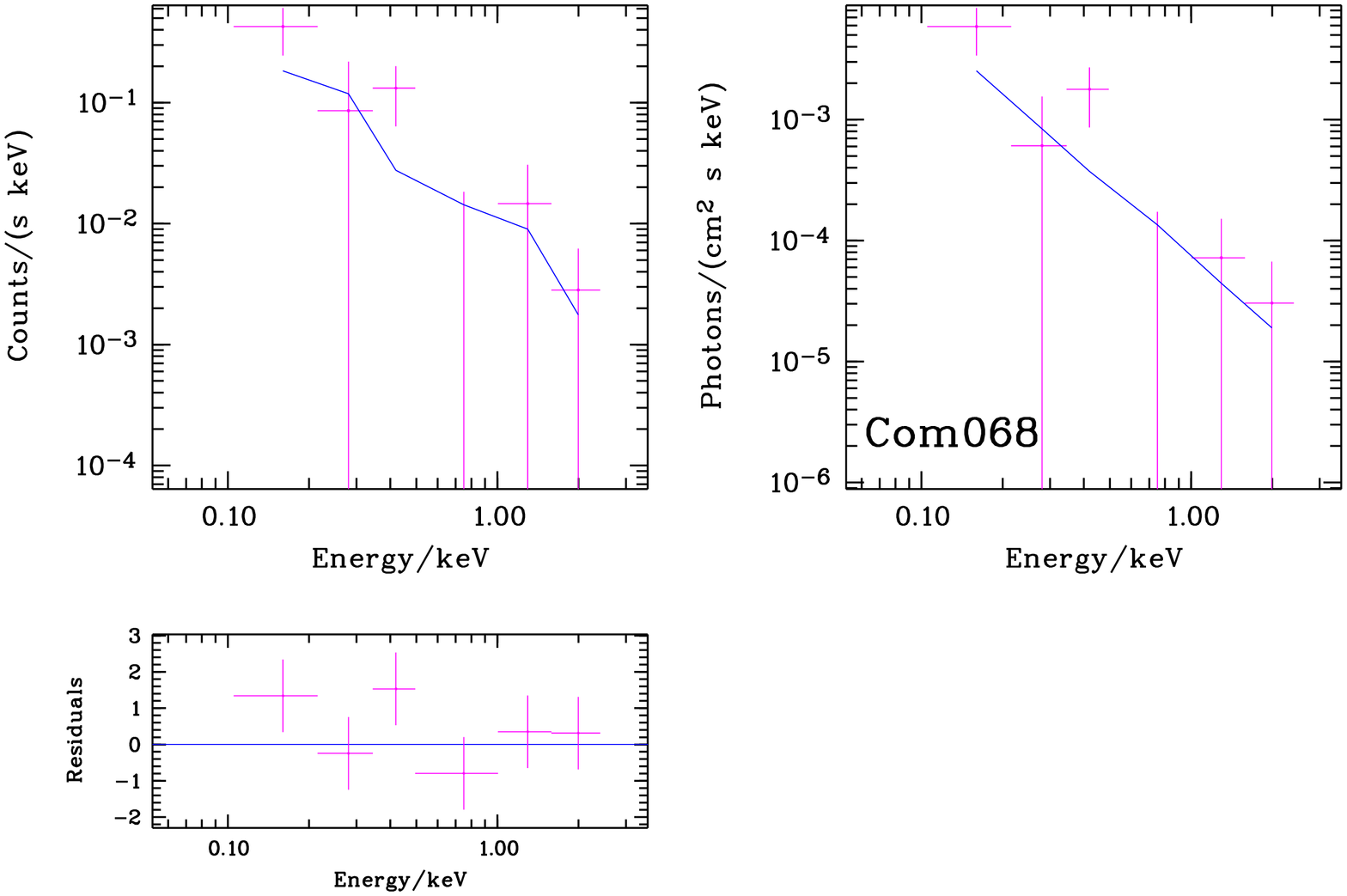}
\includegraphics[width=3.9cm, bb=76 410 385 760, angle=-90,clip]{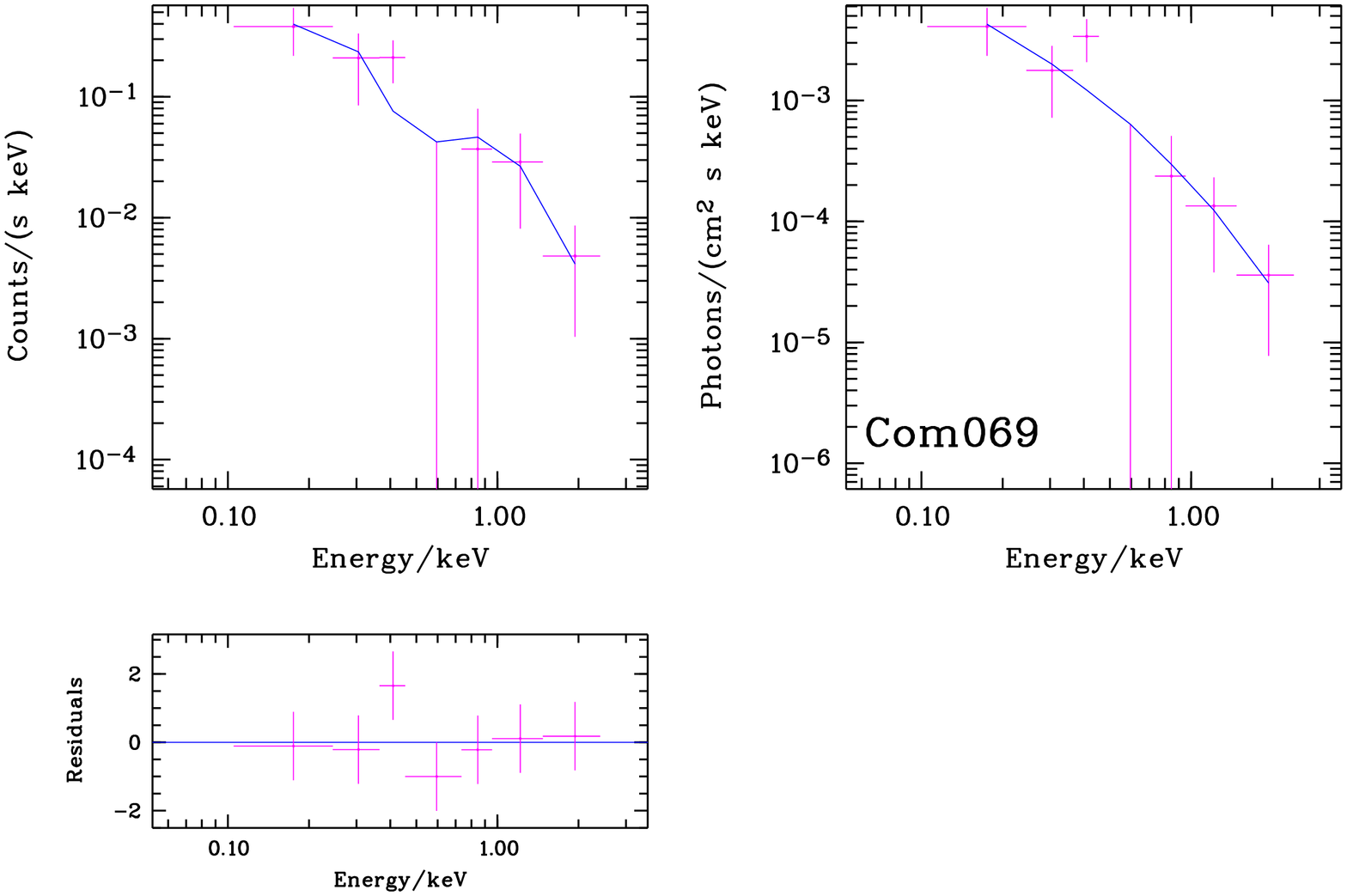}
\includegraphics[width=3.9cm, bb=76 410 385 760, angle=-90,clip]{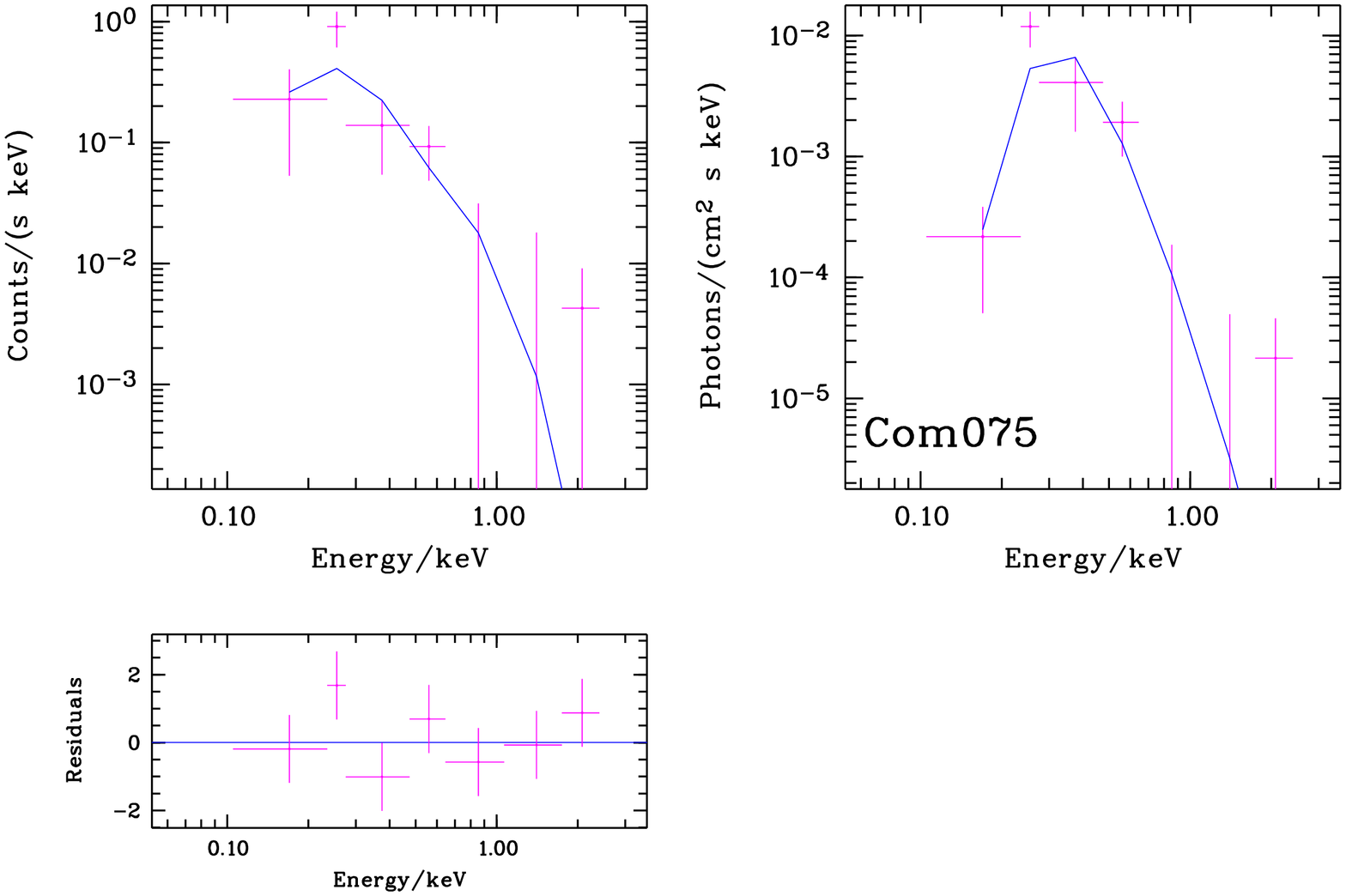}

\includegraphics[width=3.9cm, bb=76 410 385 760, angle=-90,clip]{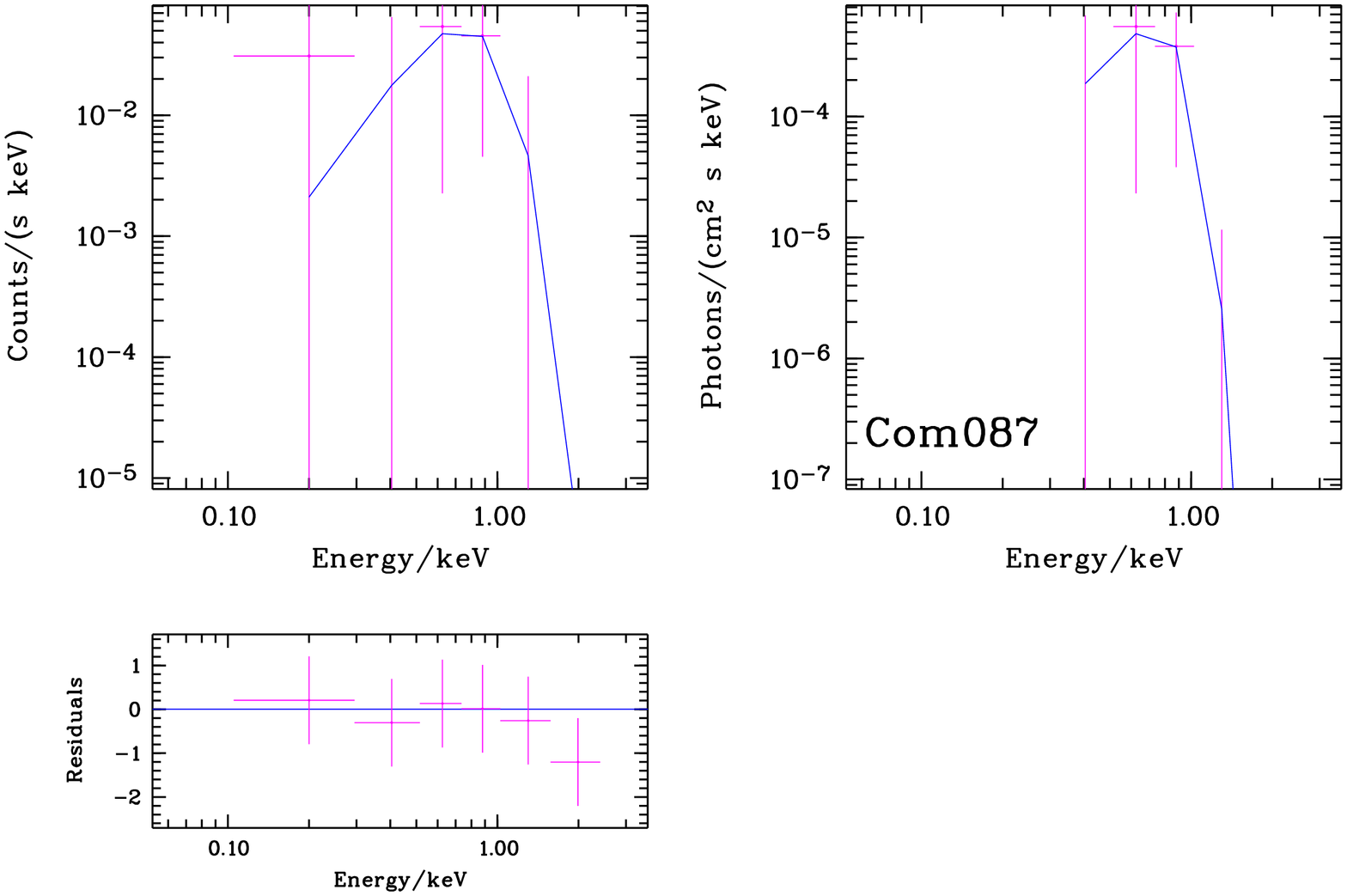}
\includegraphics[width=3.9cm, bb=76 410 385 760, angle=-90,clip]{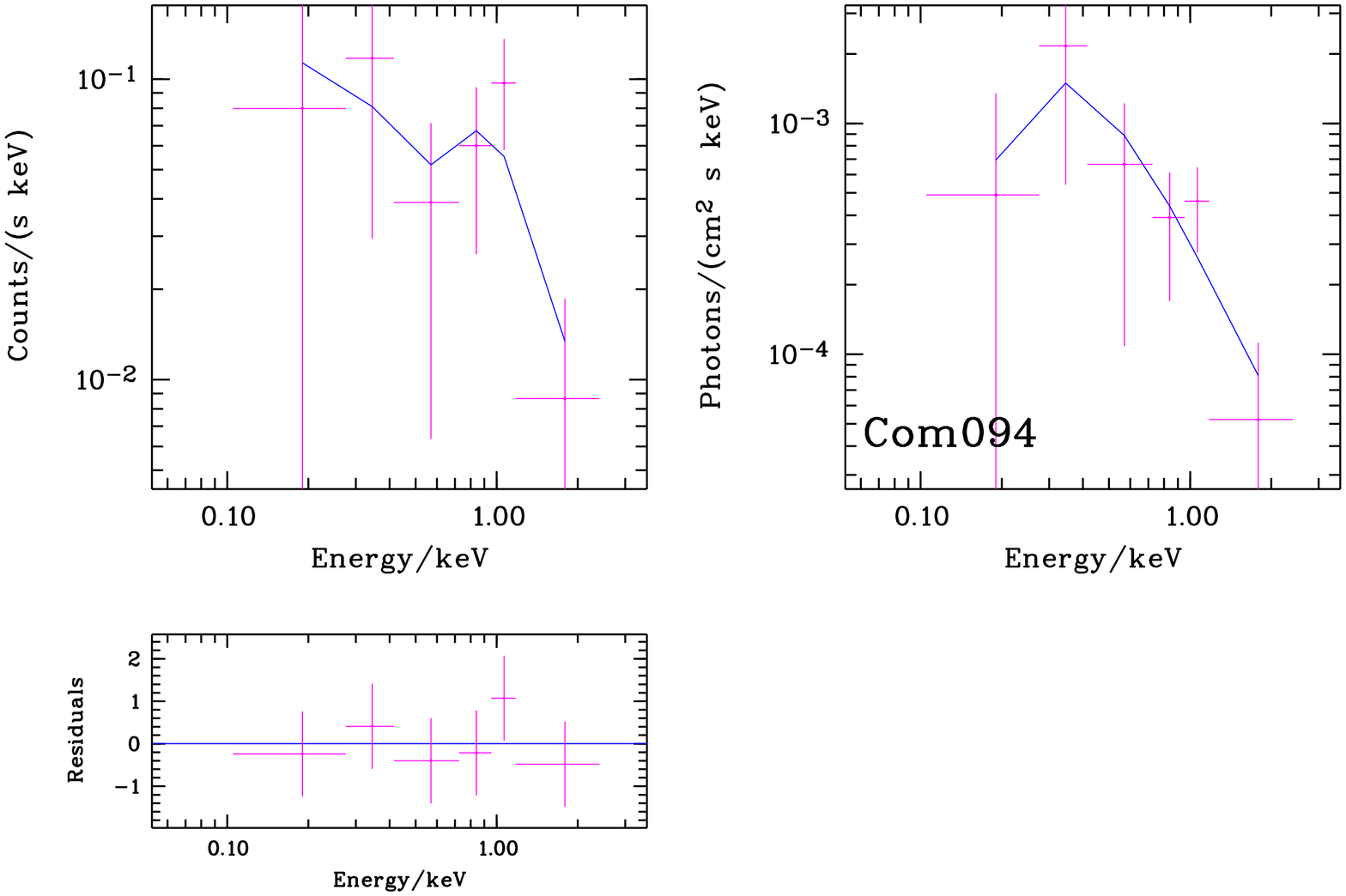}
\includegraphics[width=3.9cm, bb=76 410 385 760, angle=-90,clip]{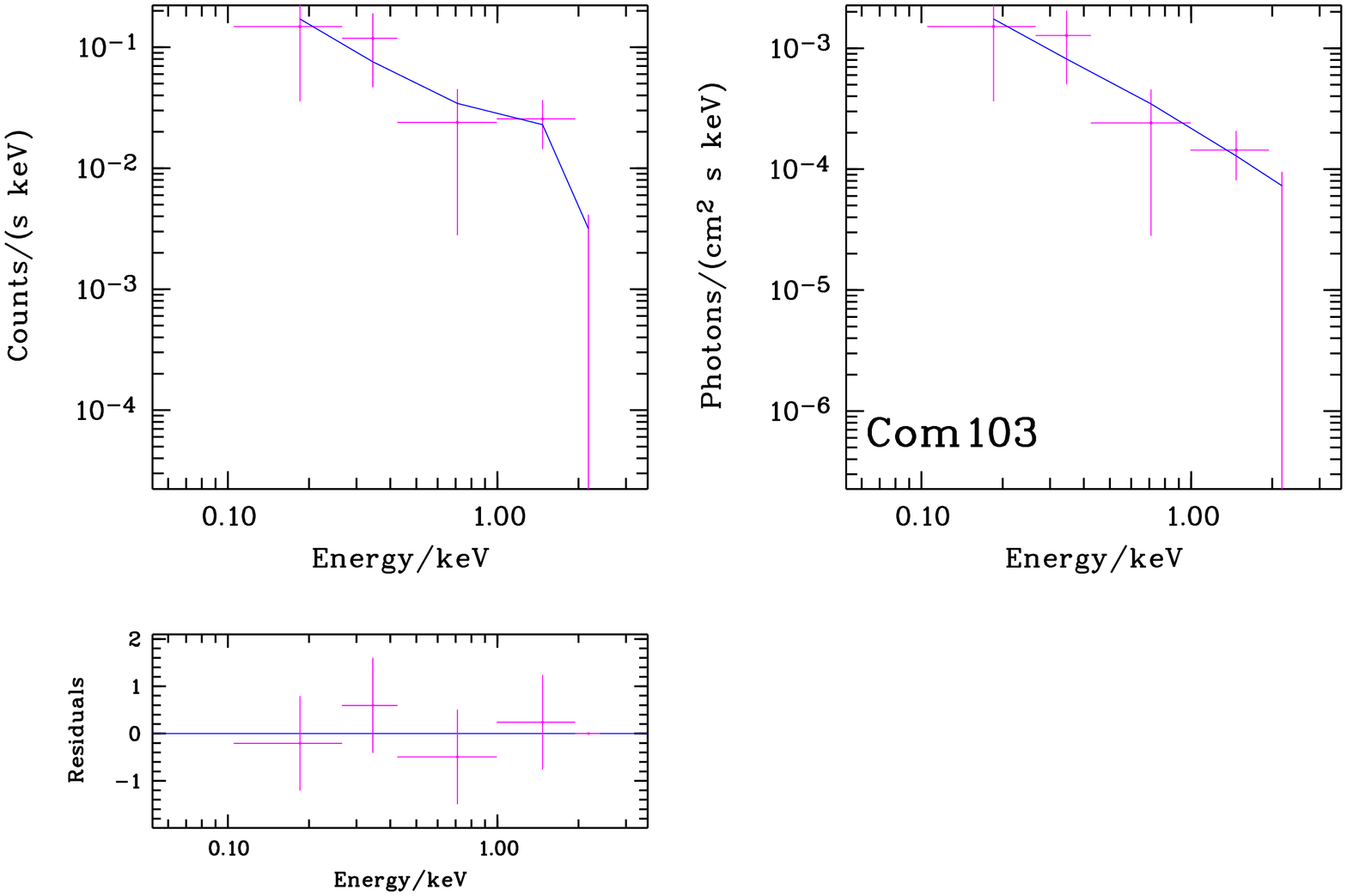}
\includegraphics[width=3.9cm, bb=76 410 385 760, angle=-90,clip]{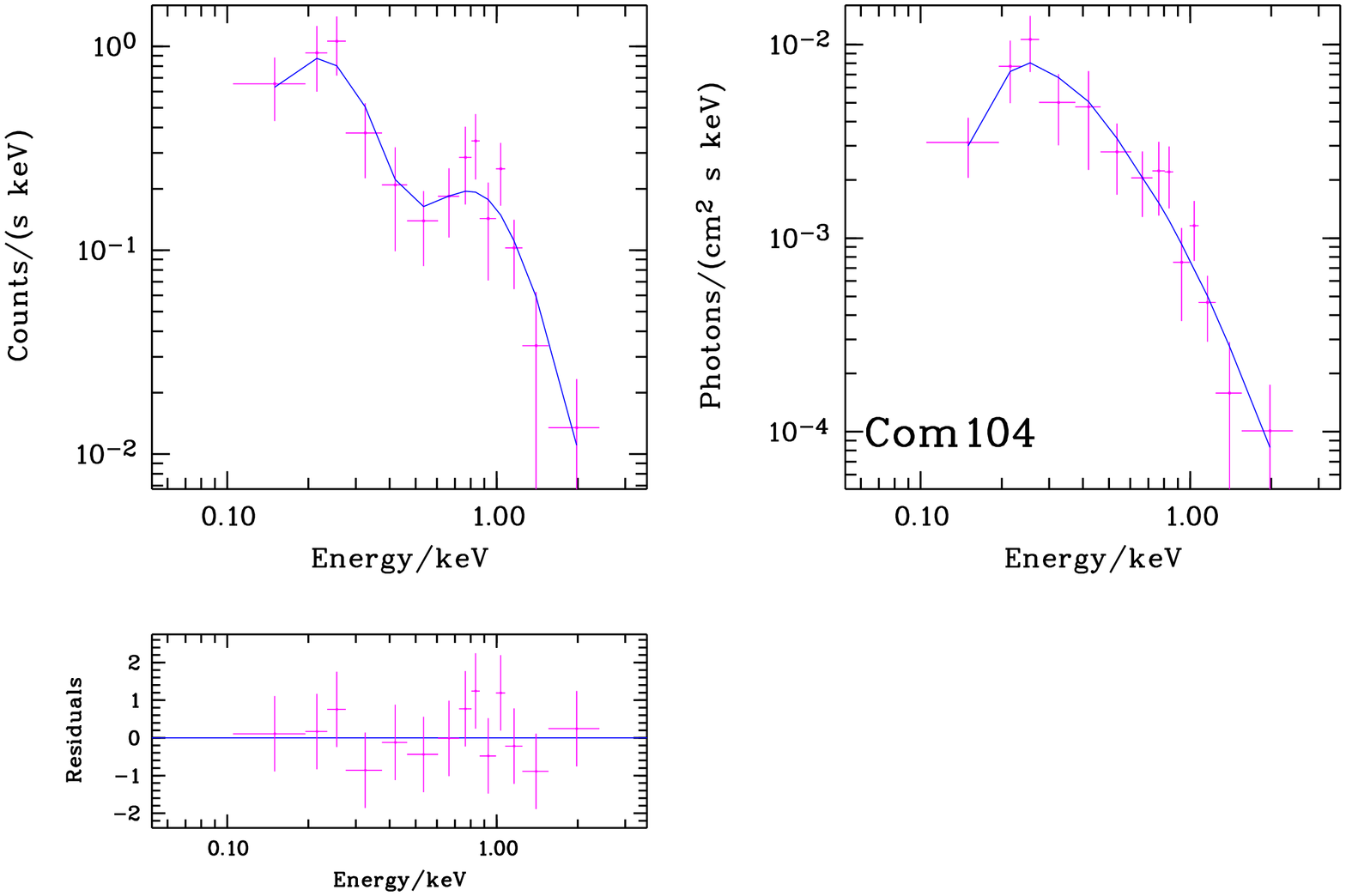}

\includegraphics[width=3.9cm, bb=76 410 385 760, angle=-90,clip]{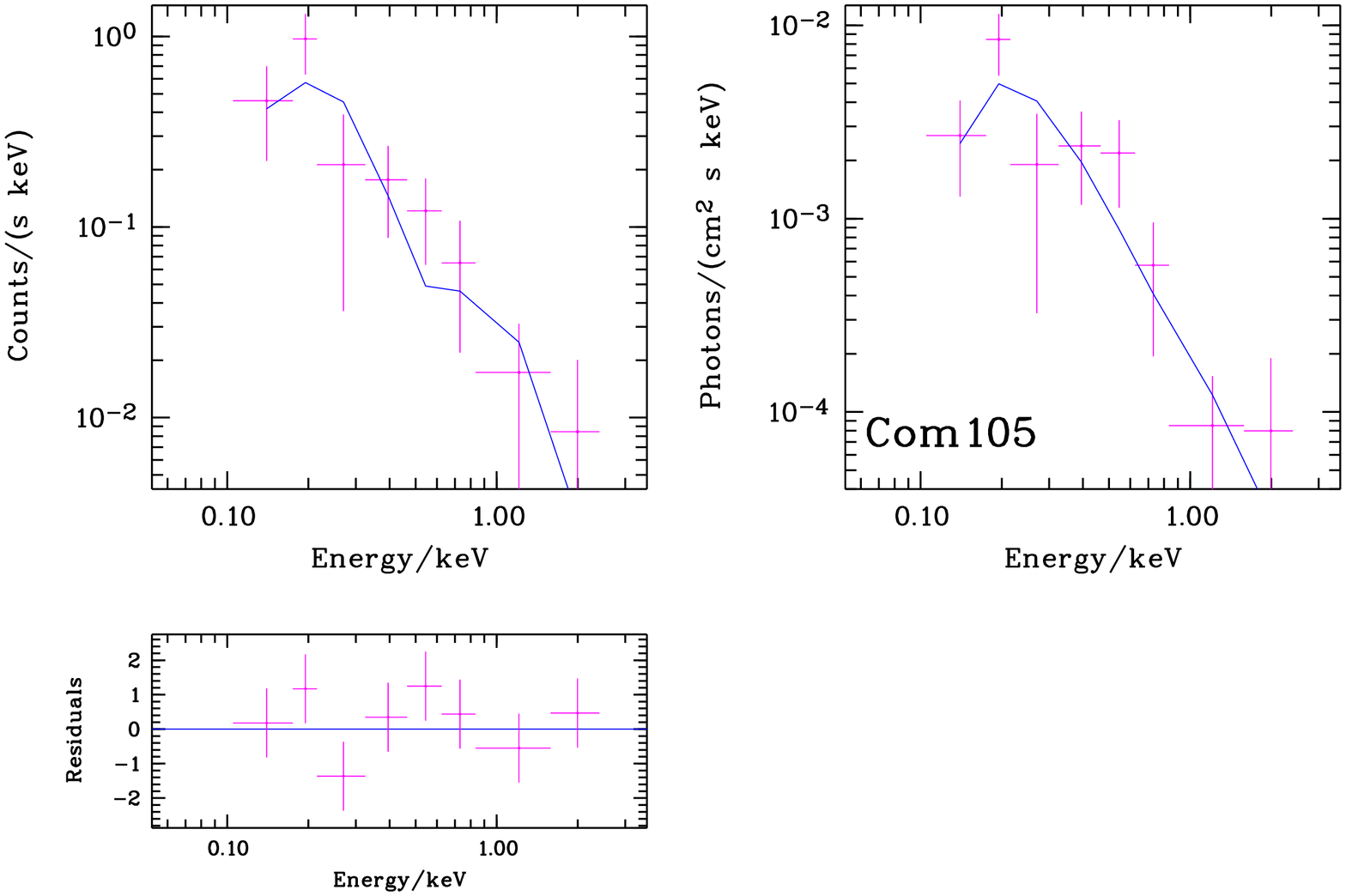}
\includegraphics[width=3.9cm, bb=76 410 385 760, angle=-90,clip]{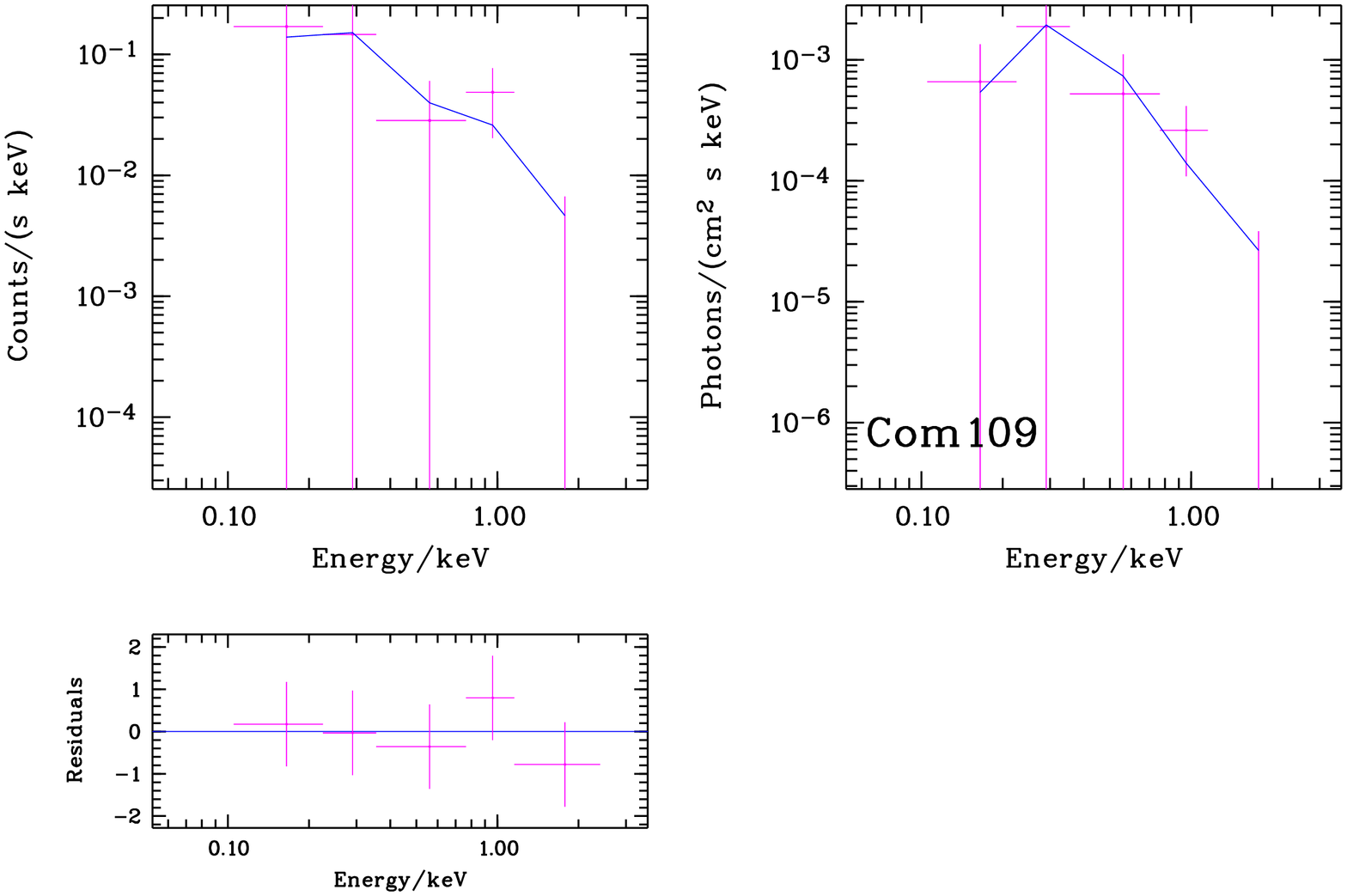}
\includegraphics[width=3.9cm, bb=76 410 385 760, angle=-90,clip]{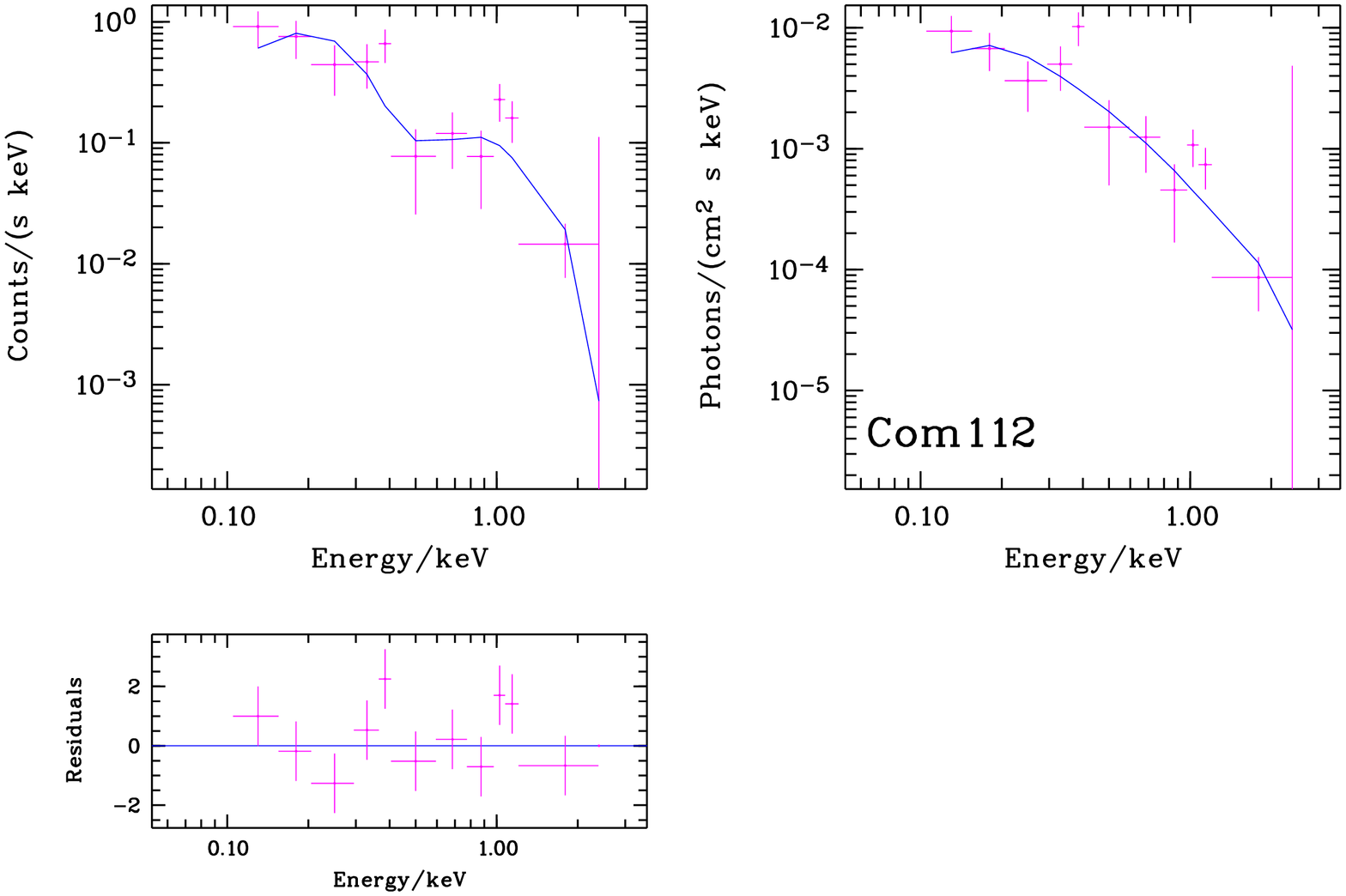}
\includegraphics[width=3.9cm, bb=76 410 385 760, angle=-90,clip]{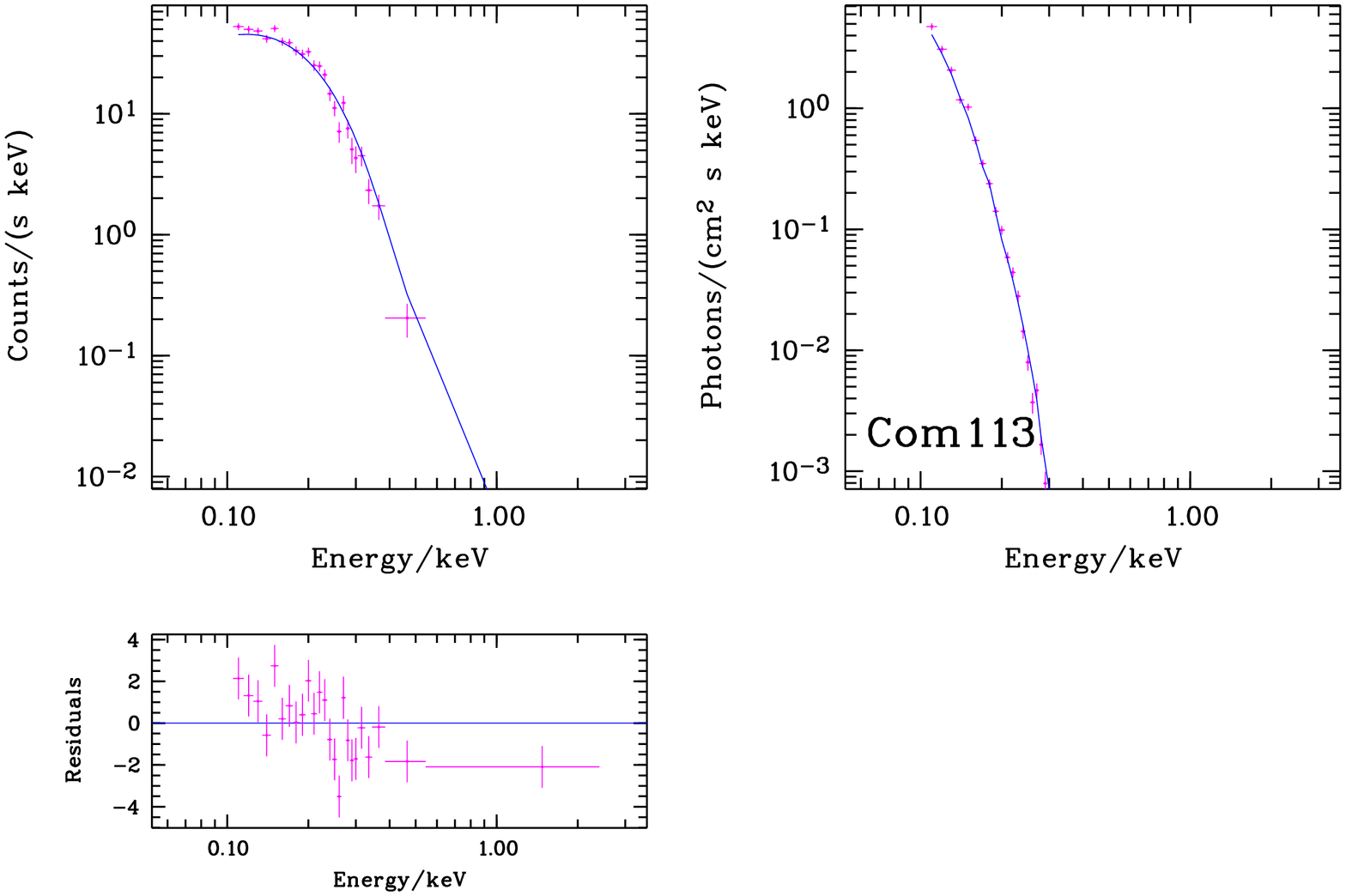}

\includegraphics[width=3.9cm, bb=76 410 385 760, angle=-90,clip]{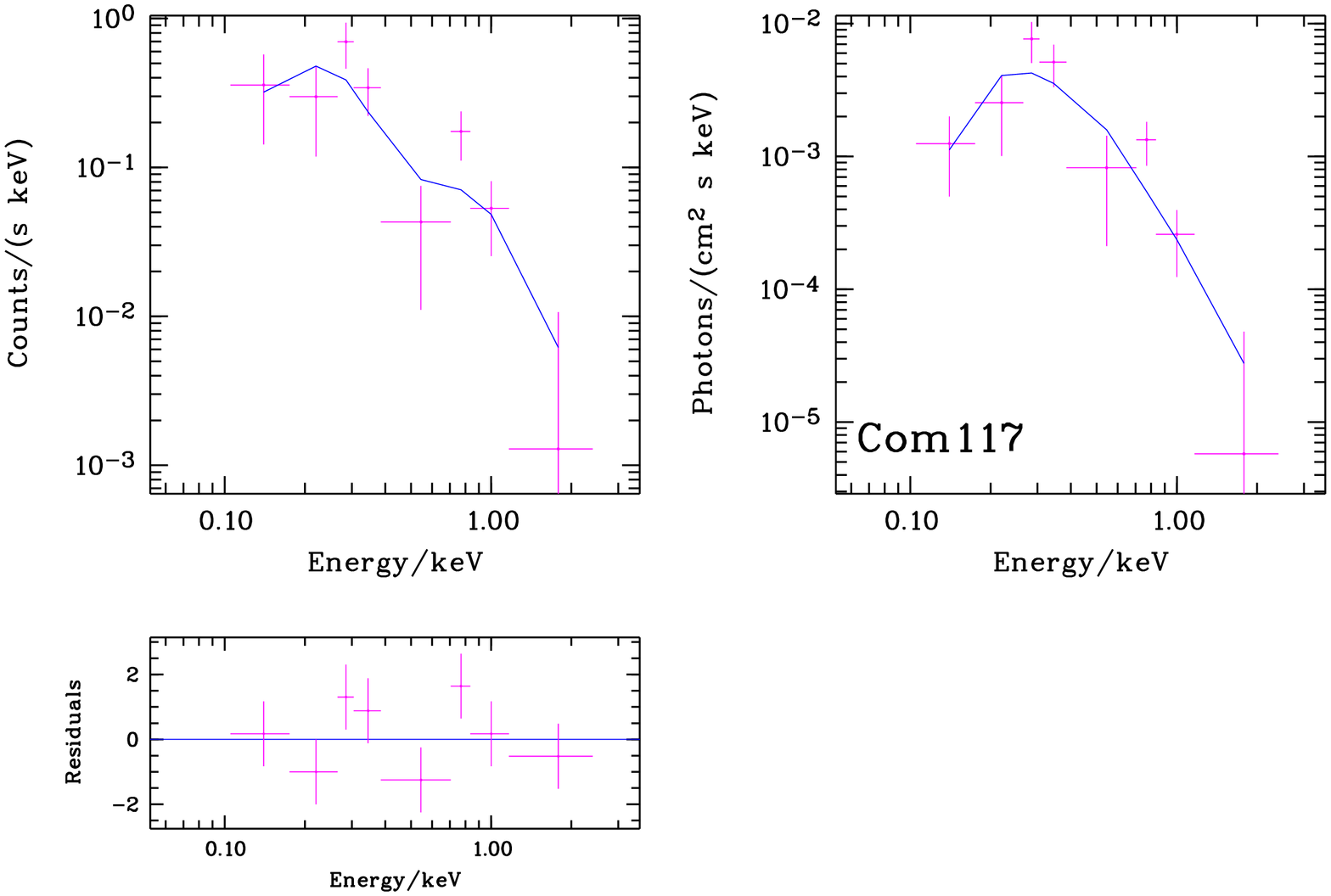}
\includegraphics[width=3.9cm, bb=76 410 385 760, angle=-90,clip]{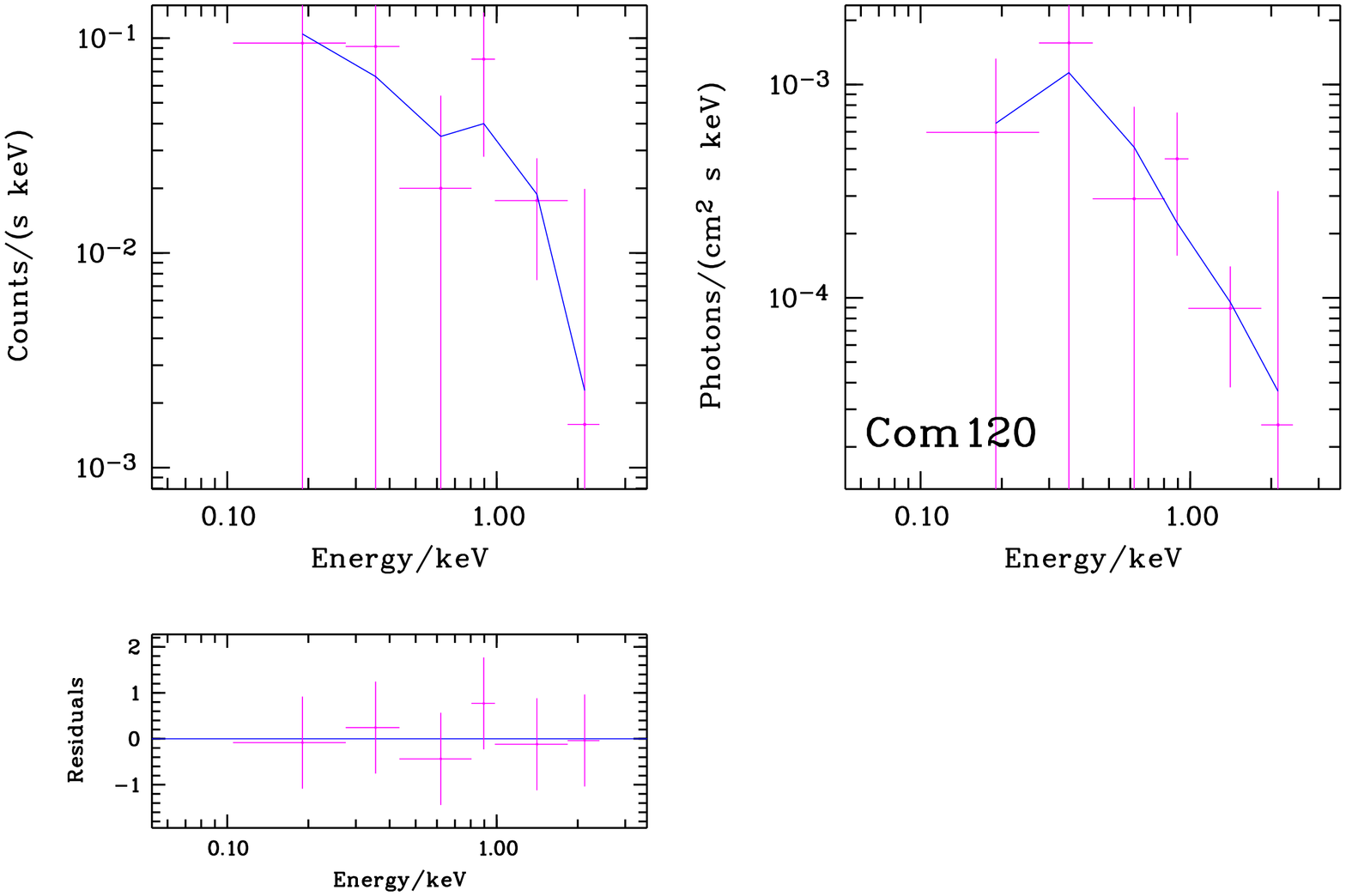}
\includegraphics[width=3.9cm, bb=76 410 385 760, angle=-90,clip]{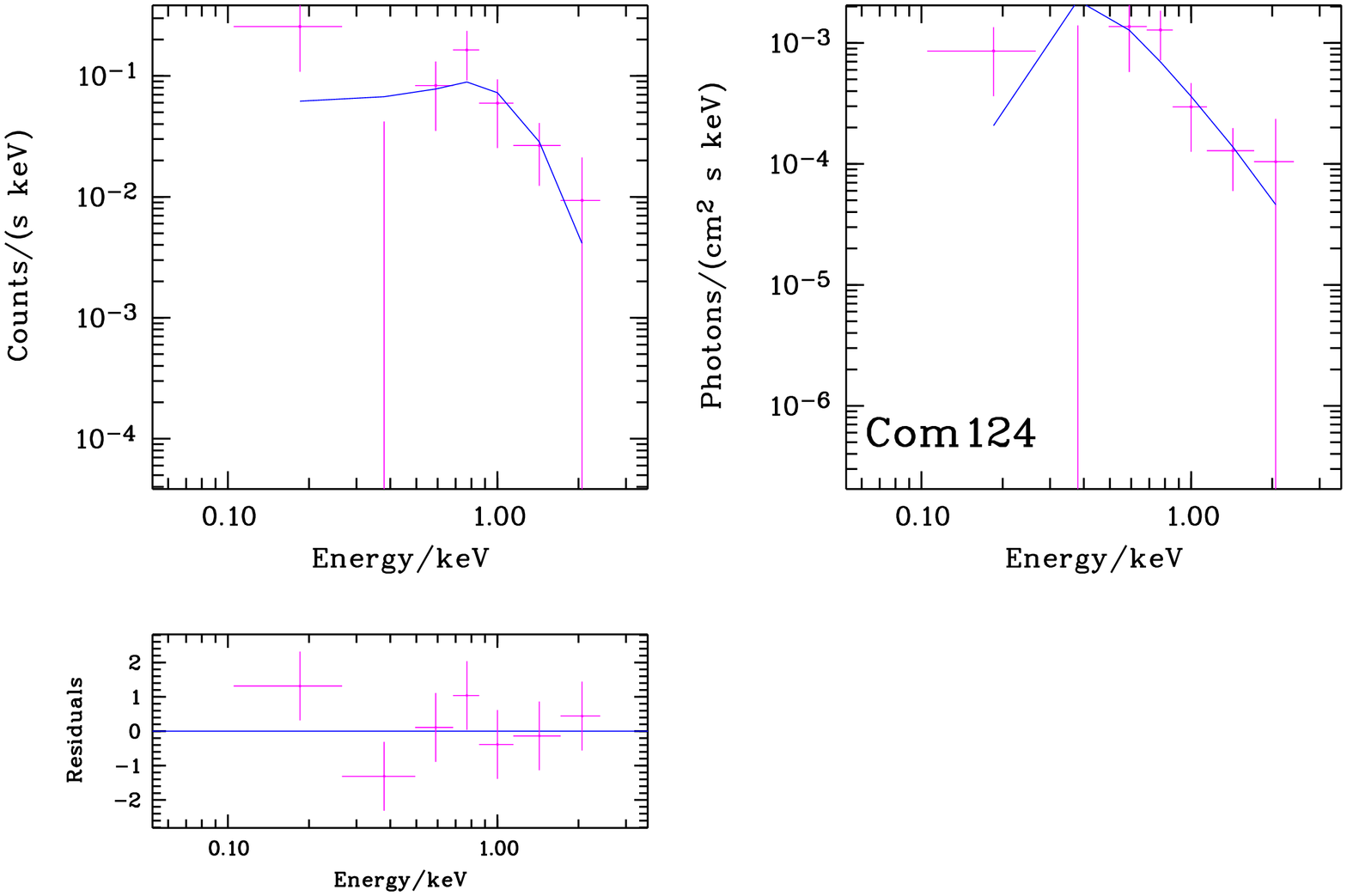}
\includegraphics[width=3.9cm, bb=76 410 385 760, angle=-90,clip]{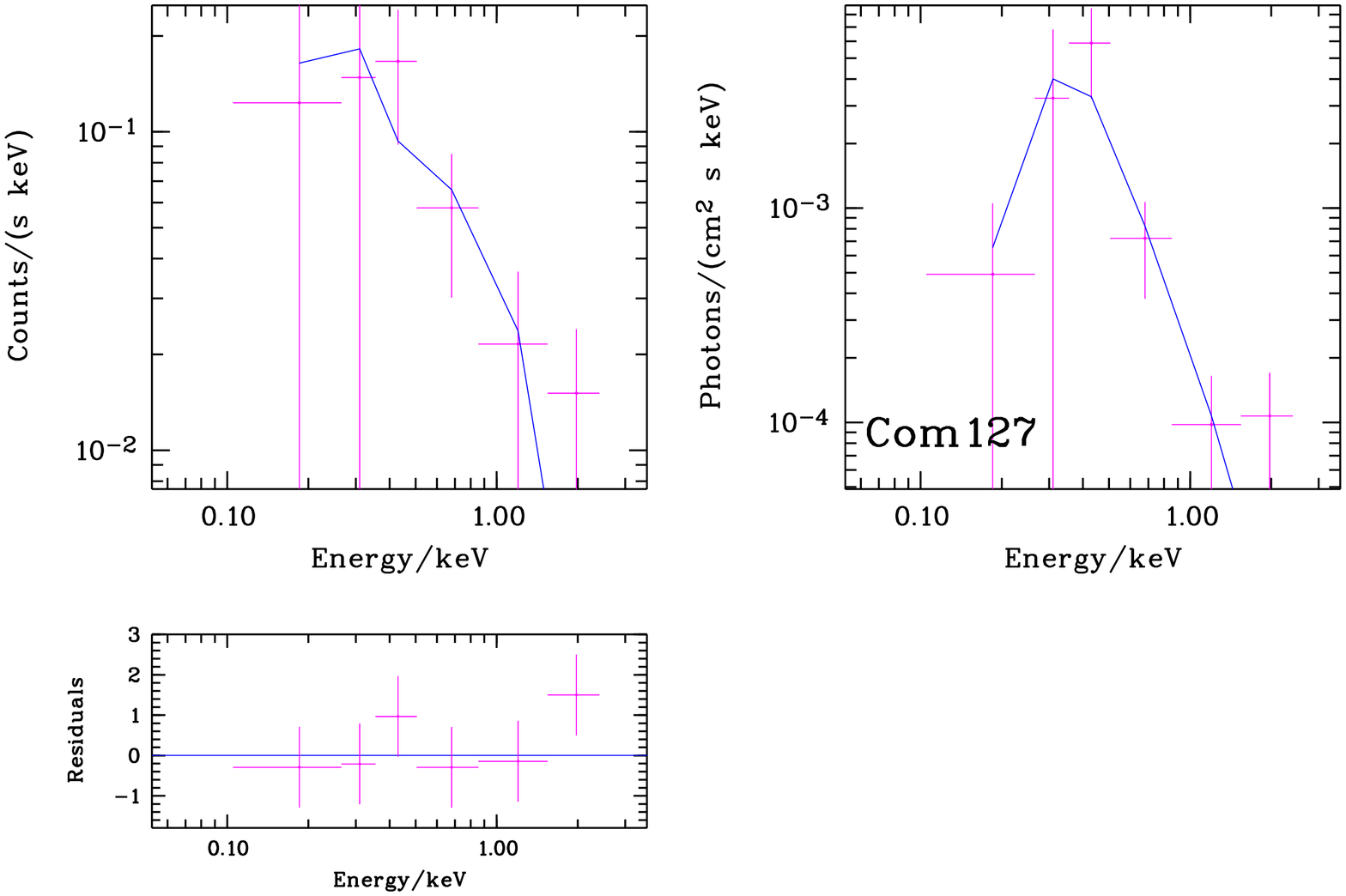}

\includegraphics[width=3.9cm, bb=76 410 385 760, angle=-90,clip]{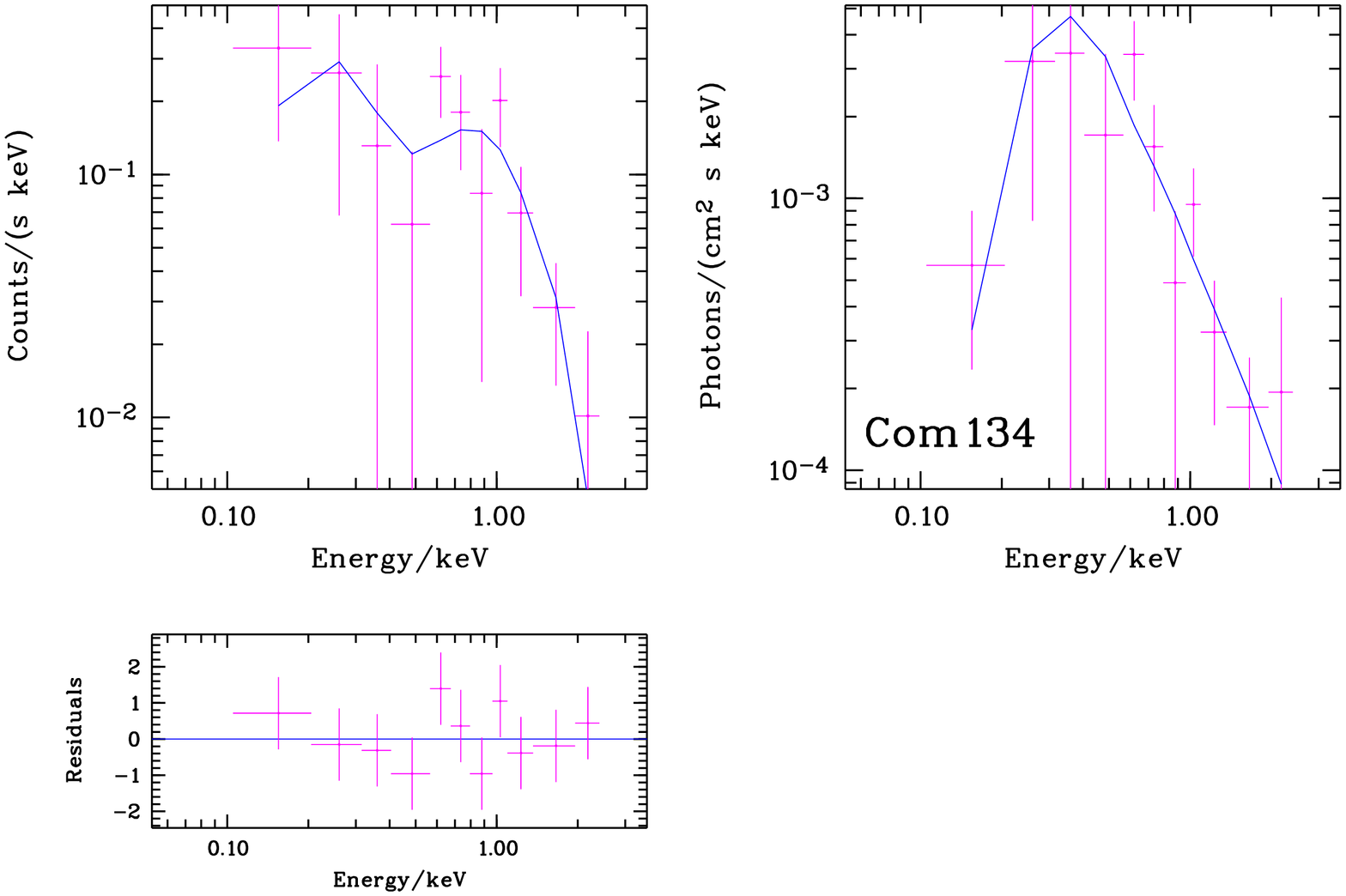}
\includegraphics[width=3.9cm, bb=76 410 385 760, angle=-90,clip]{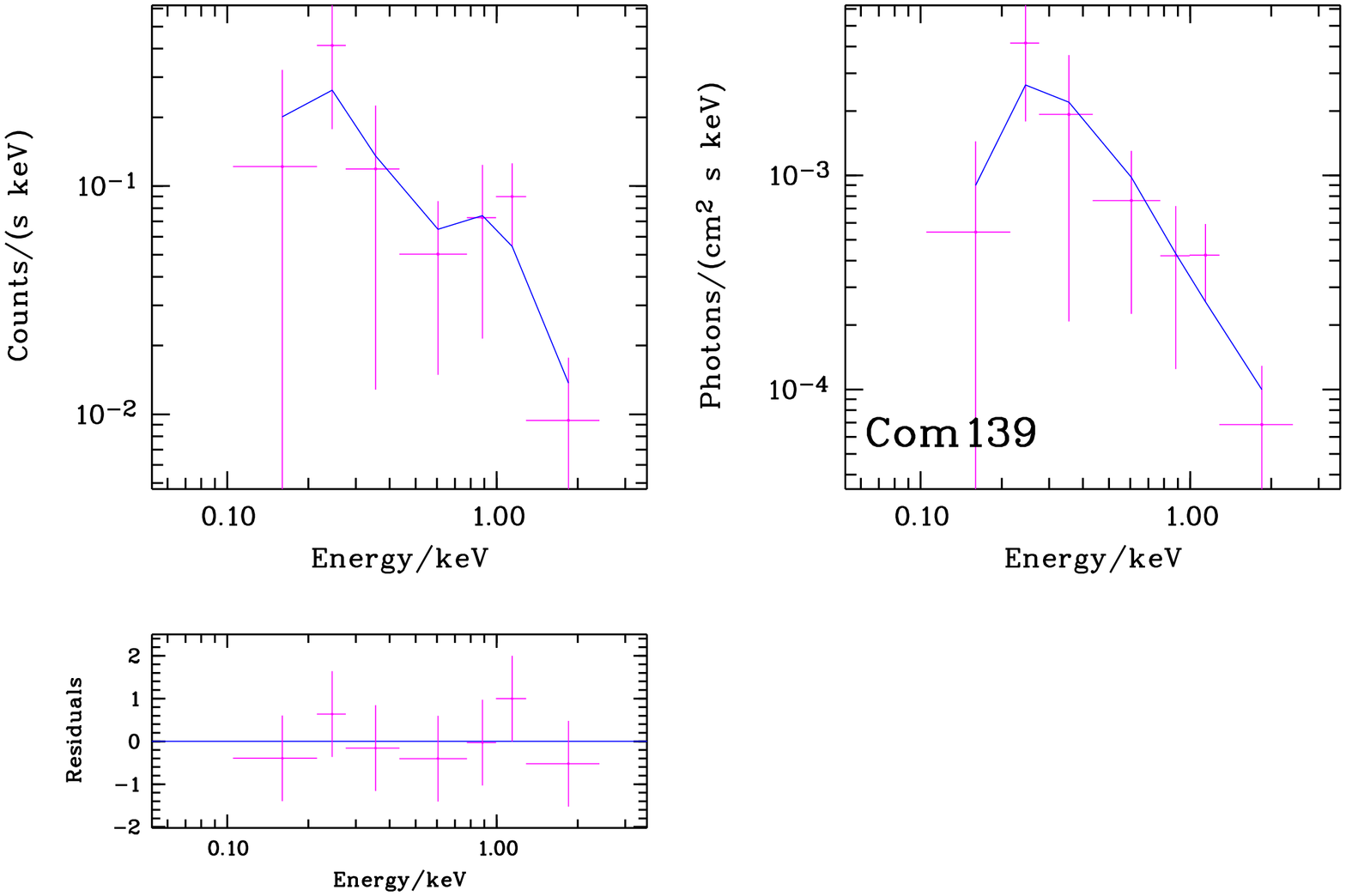}
\includegraphics[width=3.9cm, bb=76 410 385 760, angle=-90,clip]{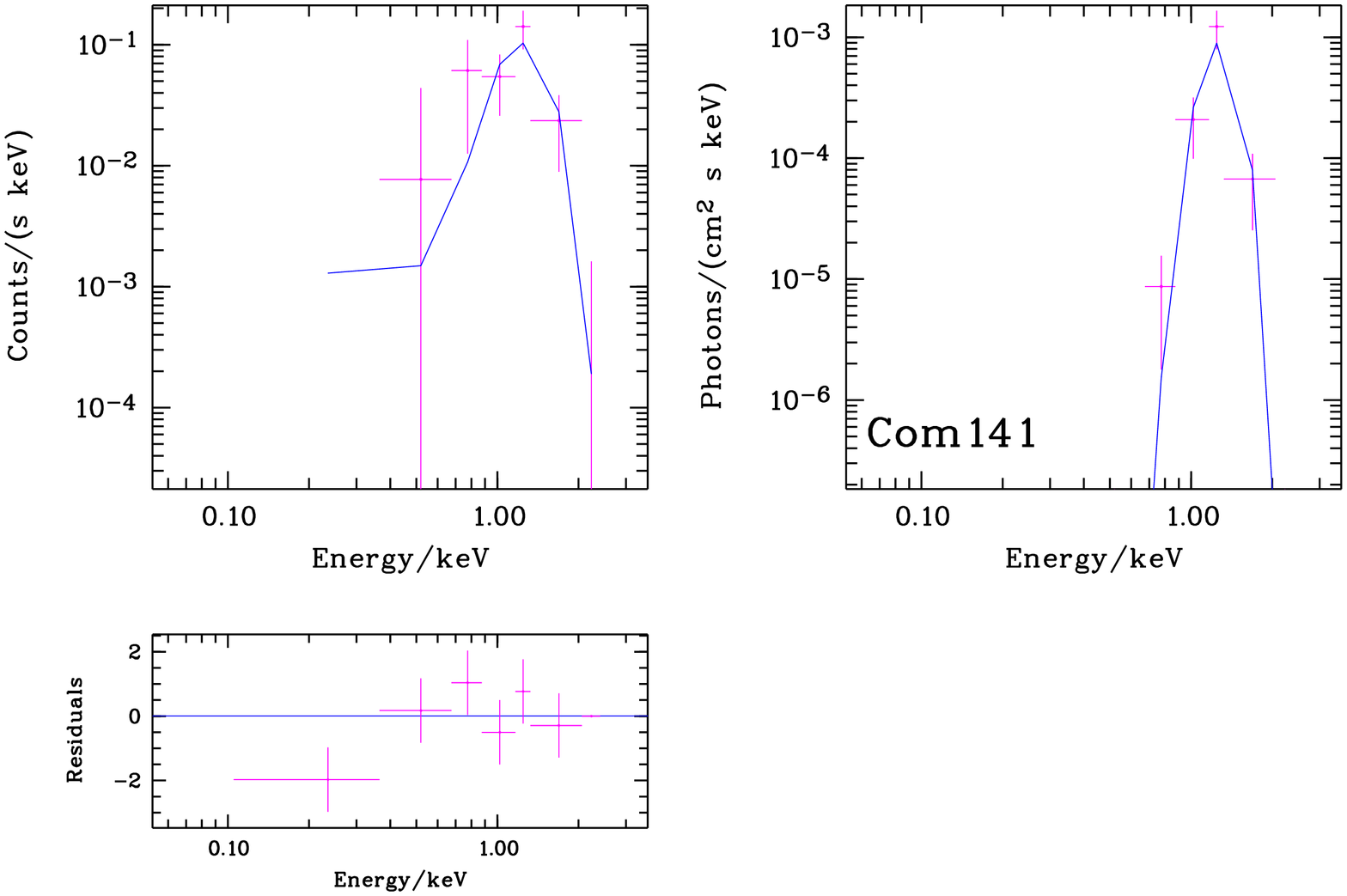}
\includegraphics[width=3.9cm, bb=76 410 385 760, angle=-90,clip]{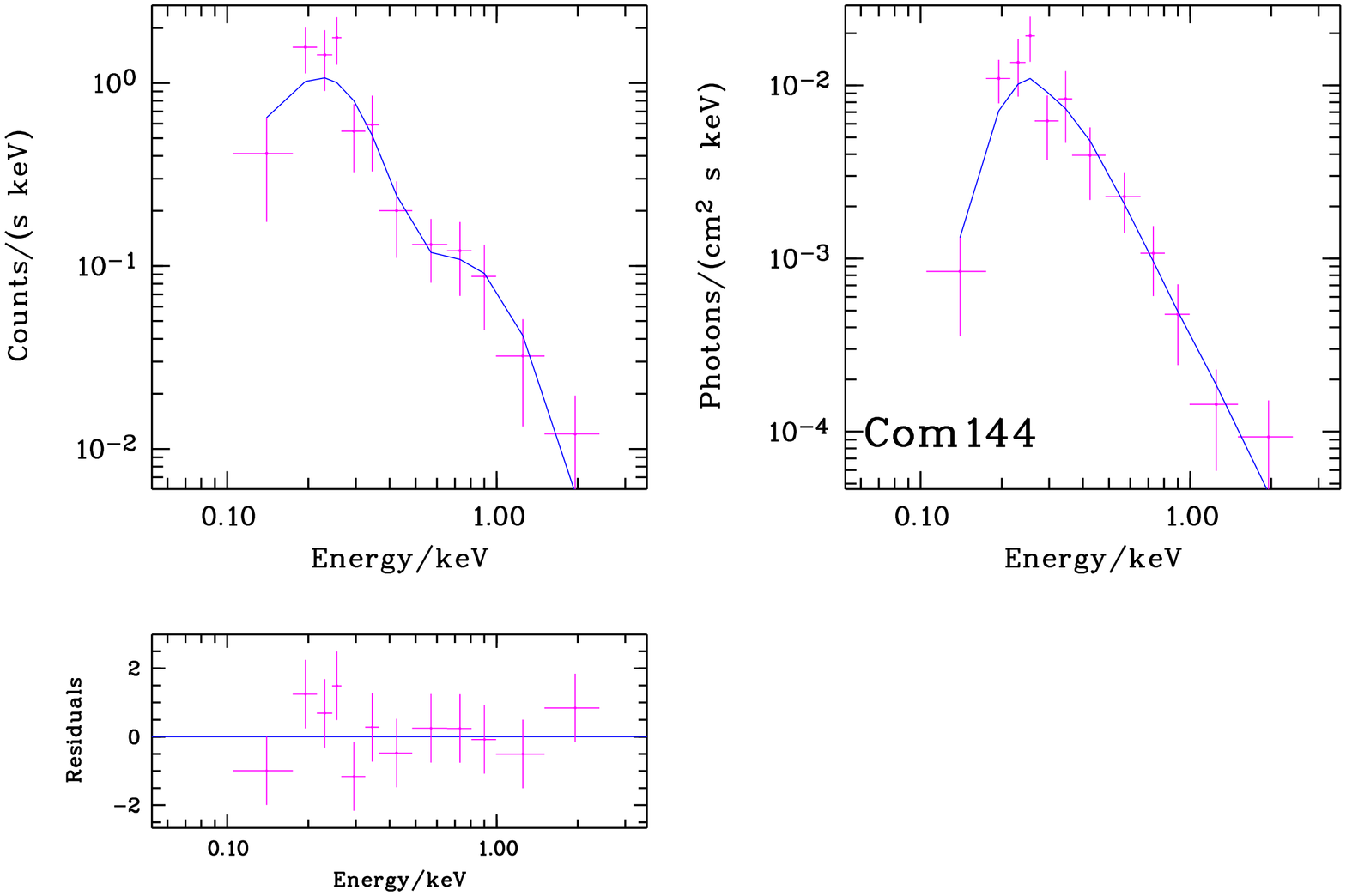}

\includegraphics[width=3.9cm, bb=76 410 385 760, angle=-90,clip]{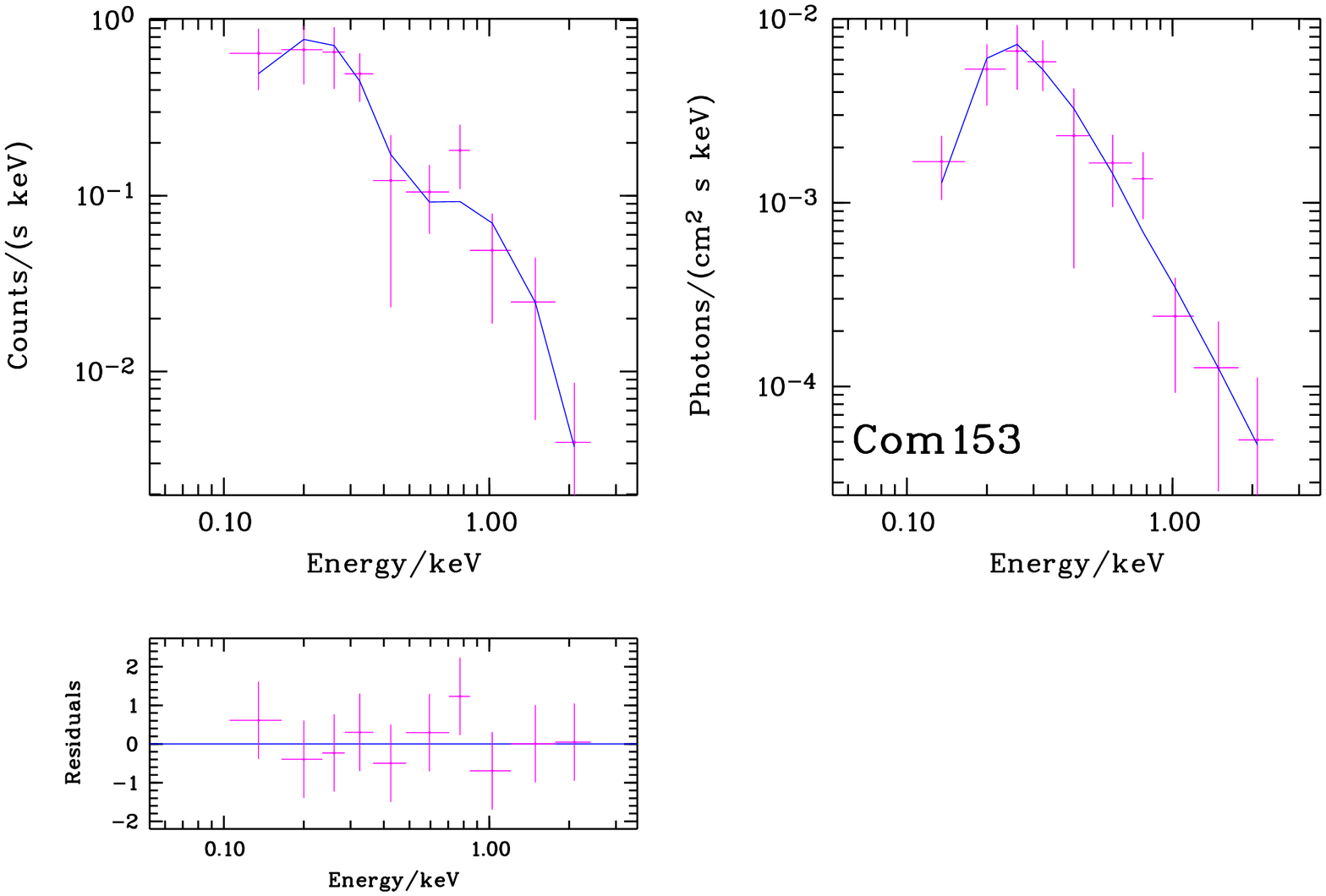}
\includegraphics[width=3.9cm, bb=76 410 385 760, angle=-90,clip]{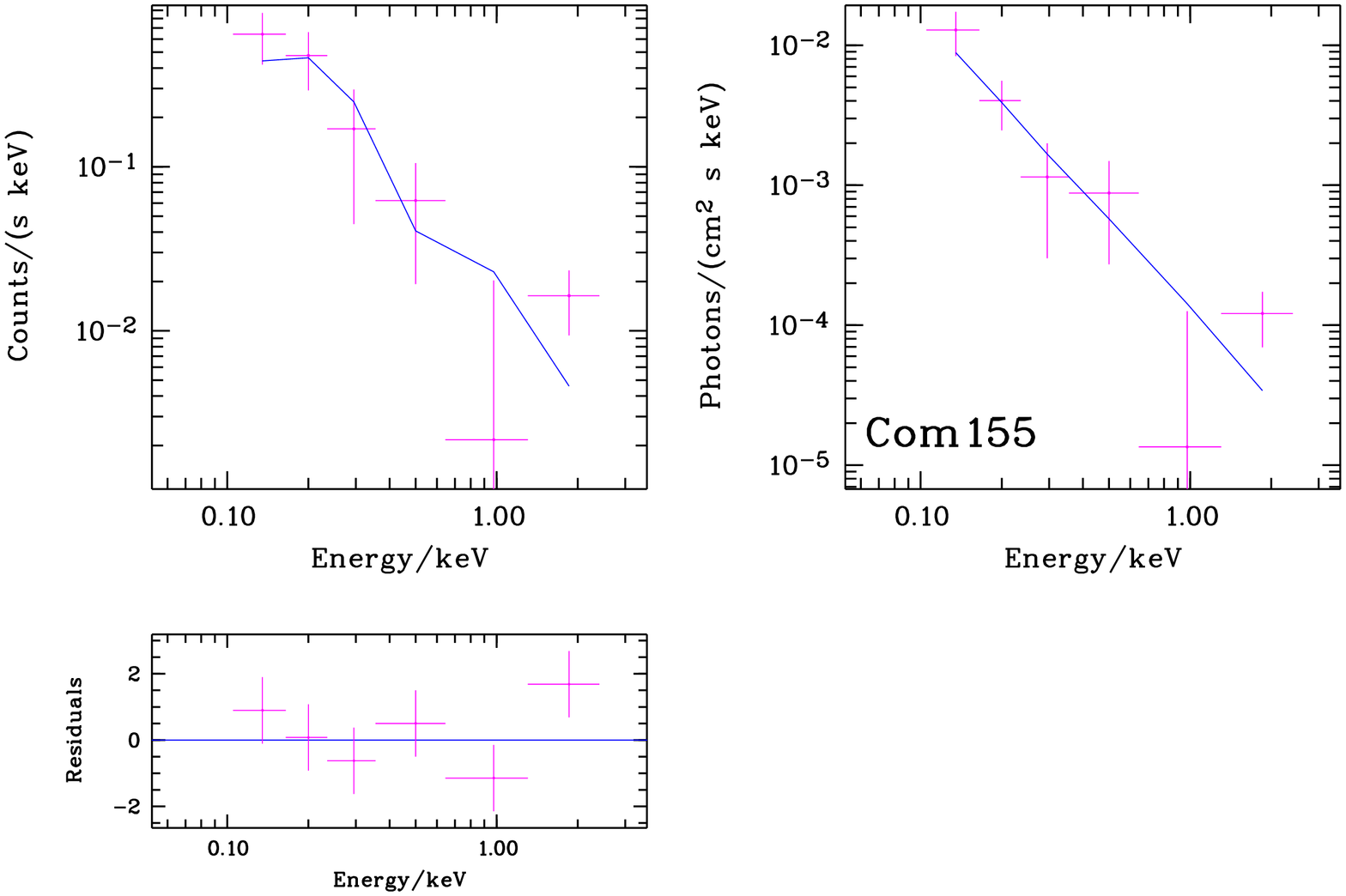}
\includegraphics[width=3.9cm, bb=76 410 385 760, angle=-90,clip]{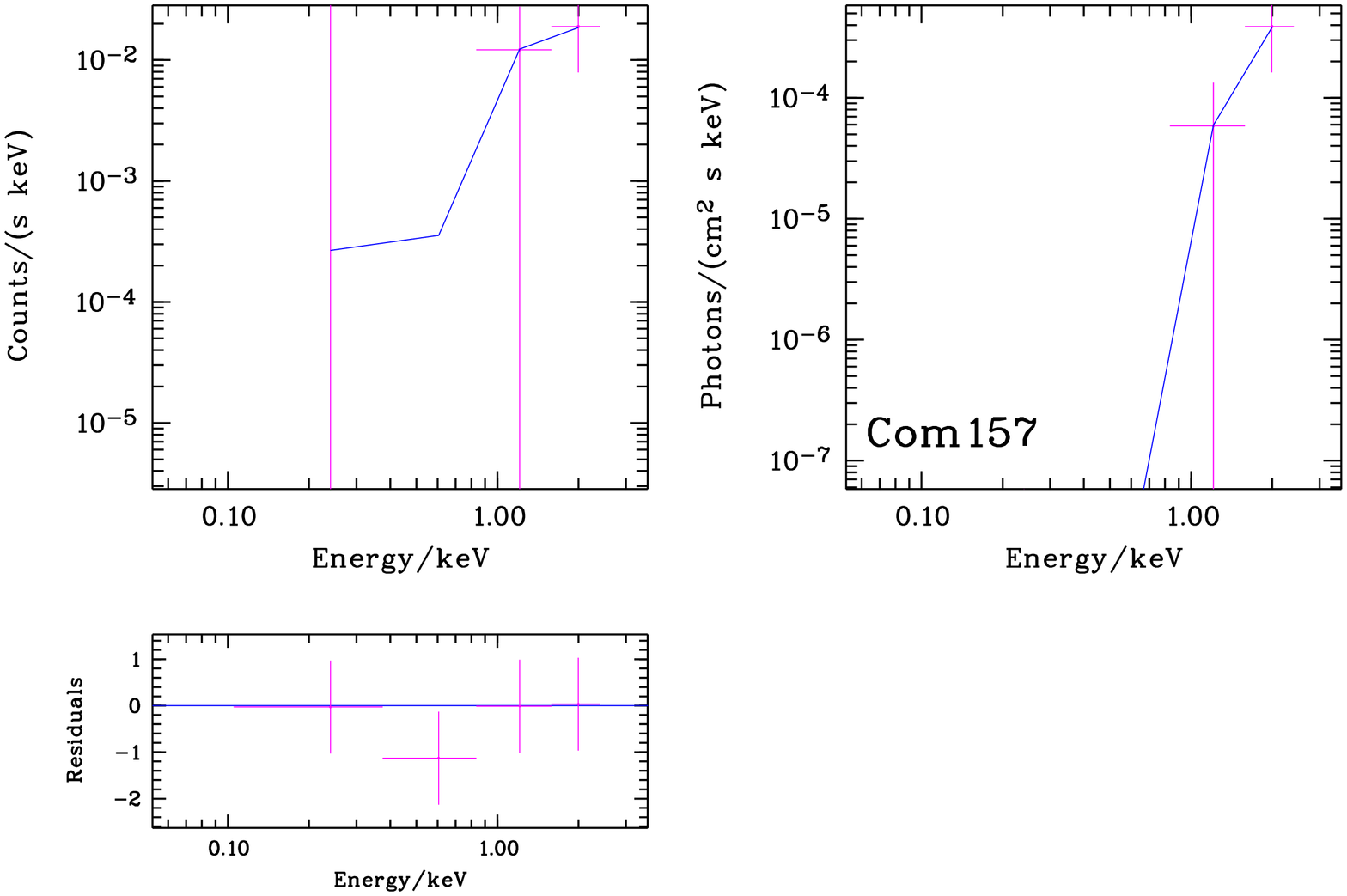}
\includegraphics[width=3.9cm, bb=76 410 385 760, angle=-90,clip]{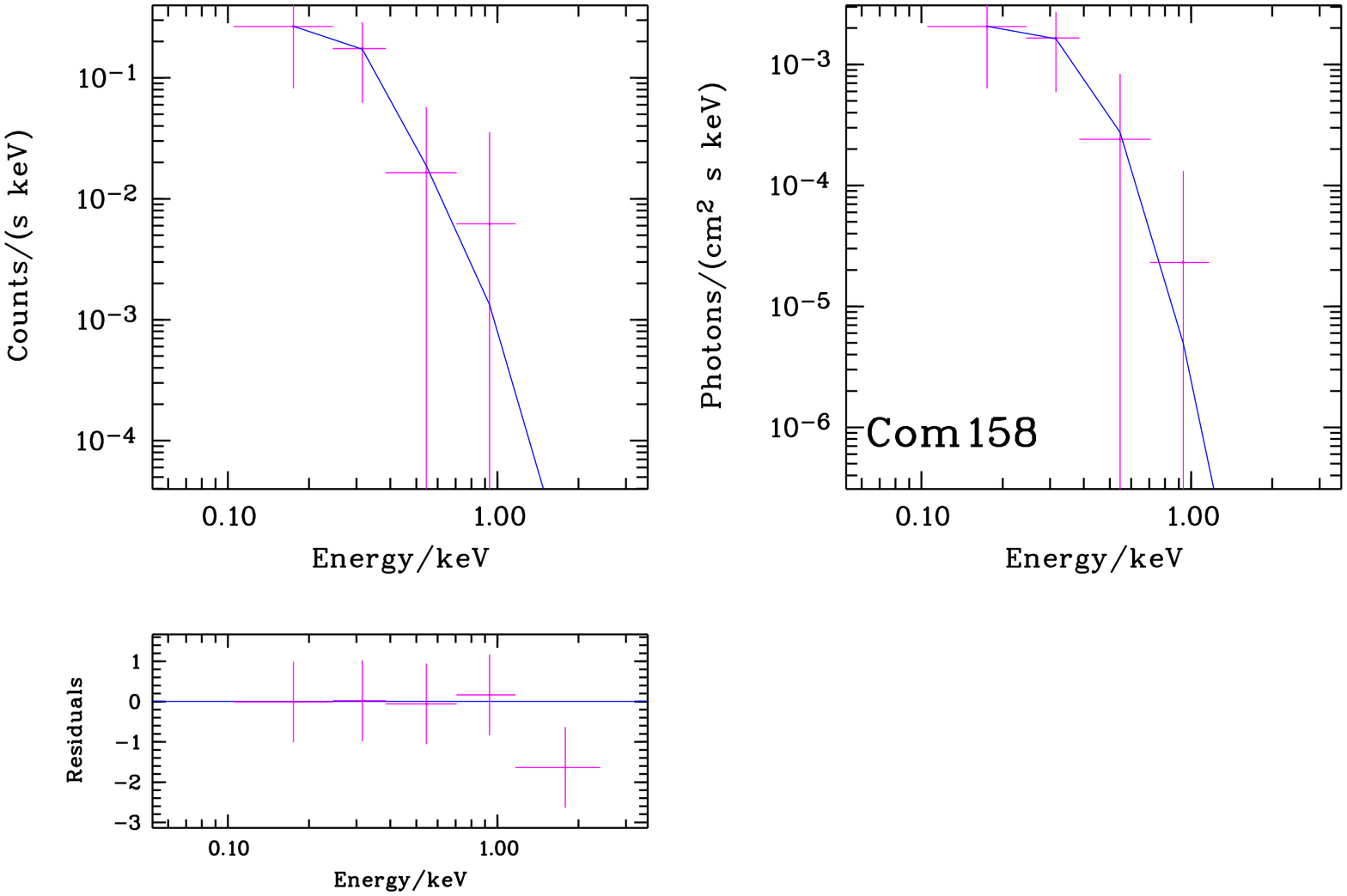}

 \caption[fsp]{{\bf contd.} ROSAT survey spectra of Com sources.}
 \end{figure*}

\setcounter{figure}{0}

\begin{figure*}[ht]

\includegraphics[width=3.9cm, bb=76 410 385 760, angle=-90,clip]{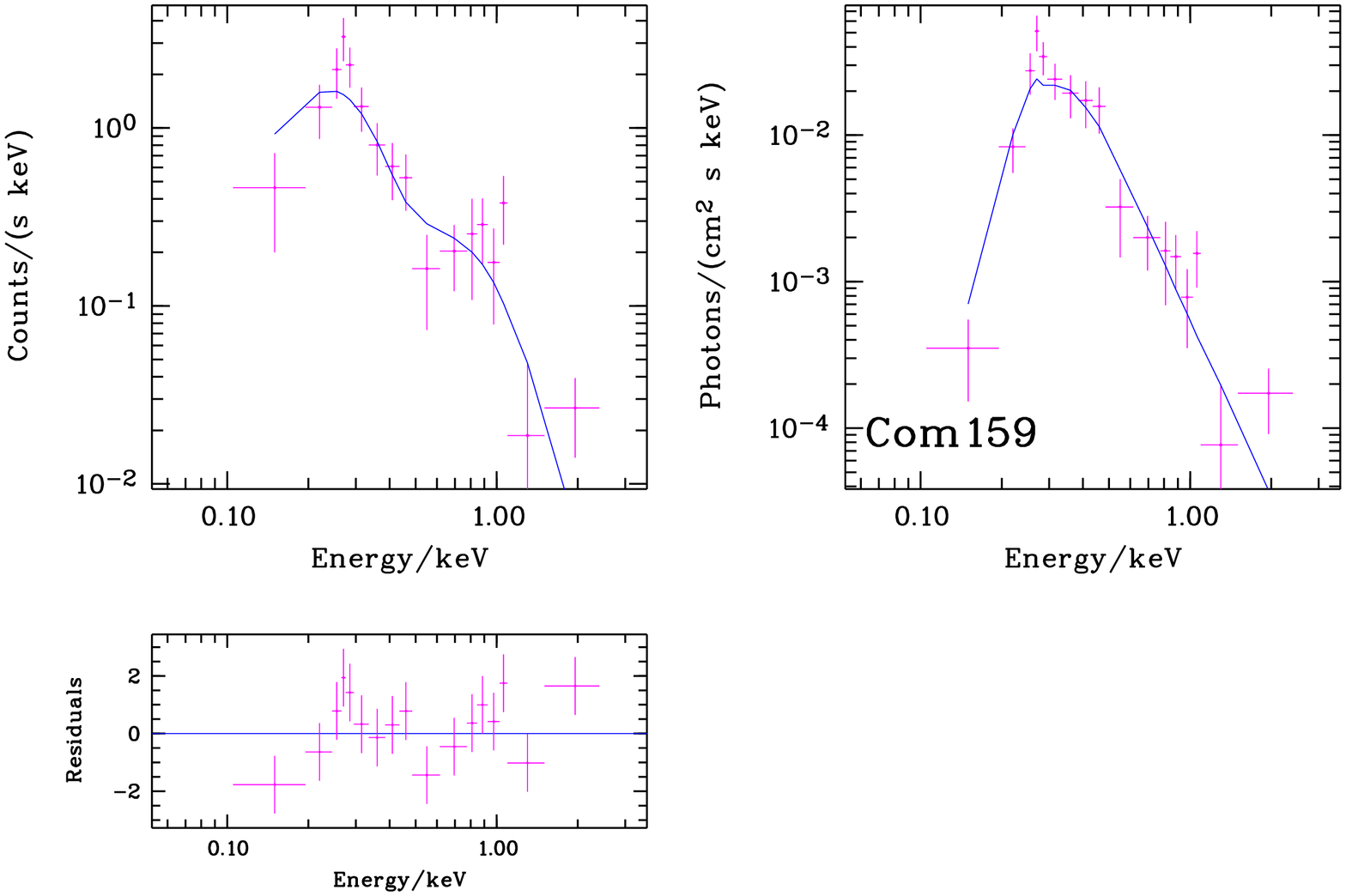}
\includegraphics[width=3.9cm, bb=76 410 385 760, angle=-90,clip]{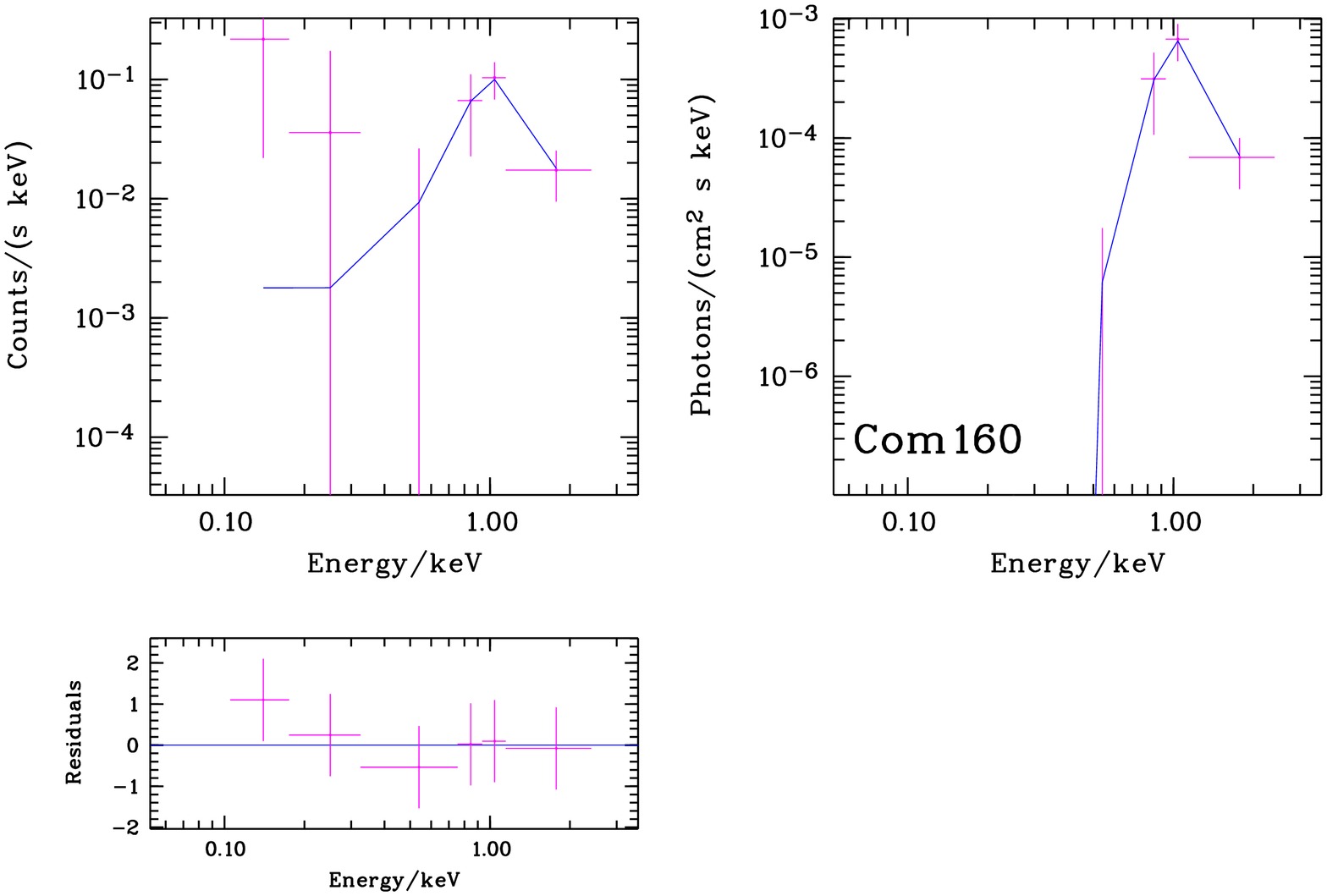}
\includegraphics[width=3.9cm, bb=76 410 385 760, angle=-90,clip]{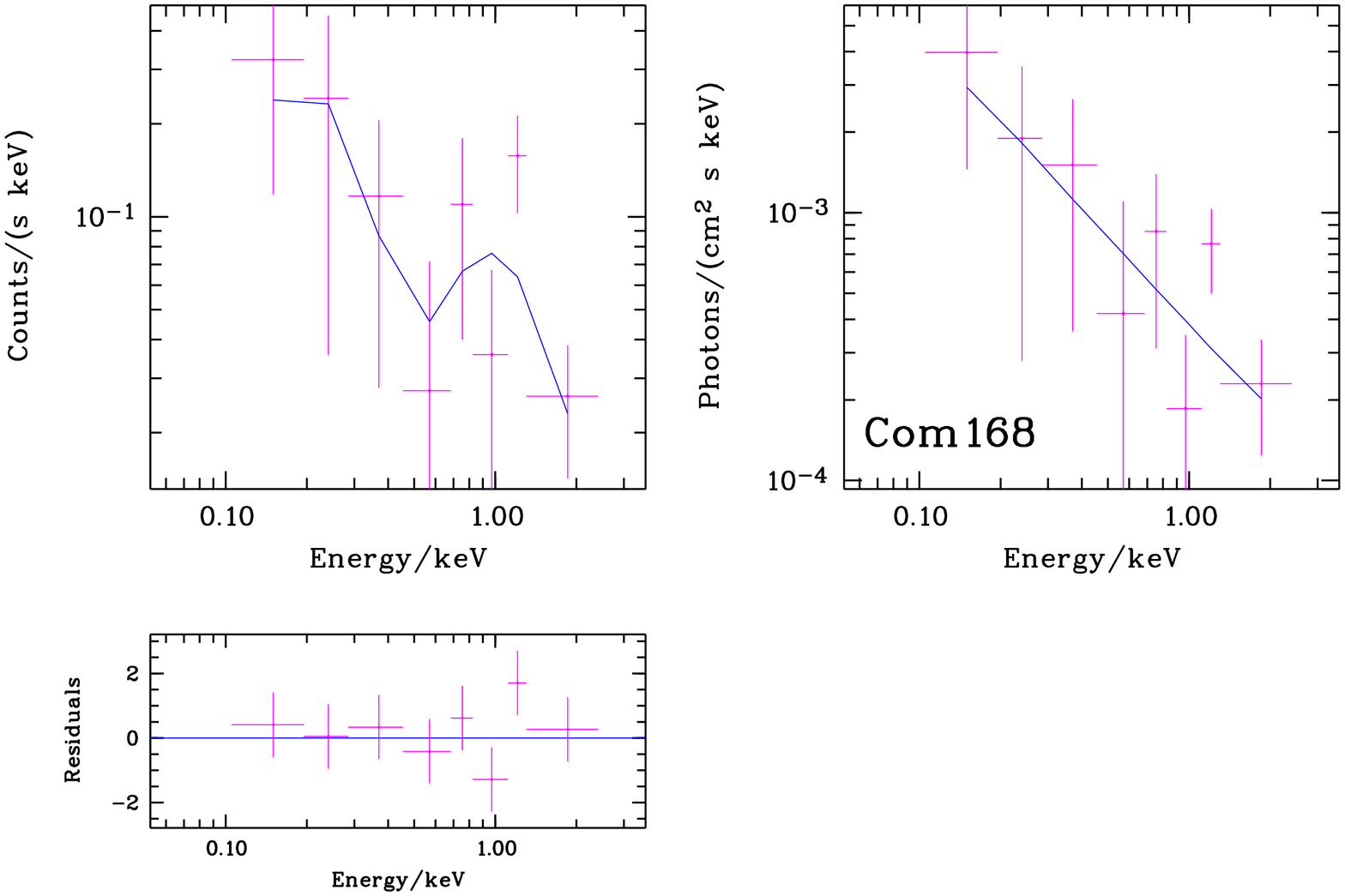}
\includegraphics[width=3.9cm, bb=76 410 385 760, angle=-90,clip]{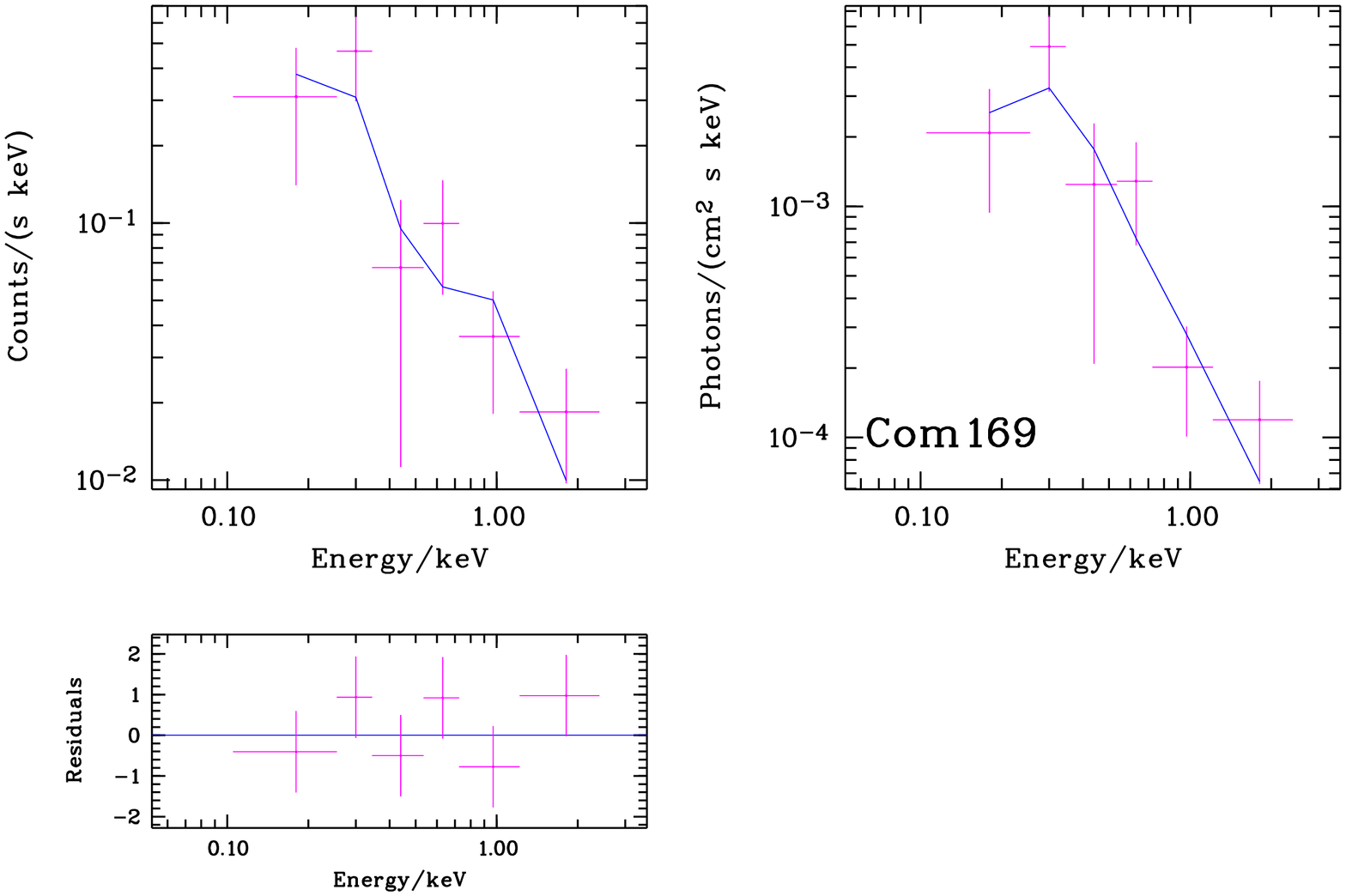}

\includegraphics[width=3.9cm, bb=76 410 385 760, angle=-90,clip]{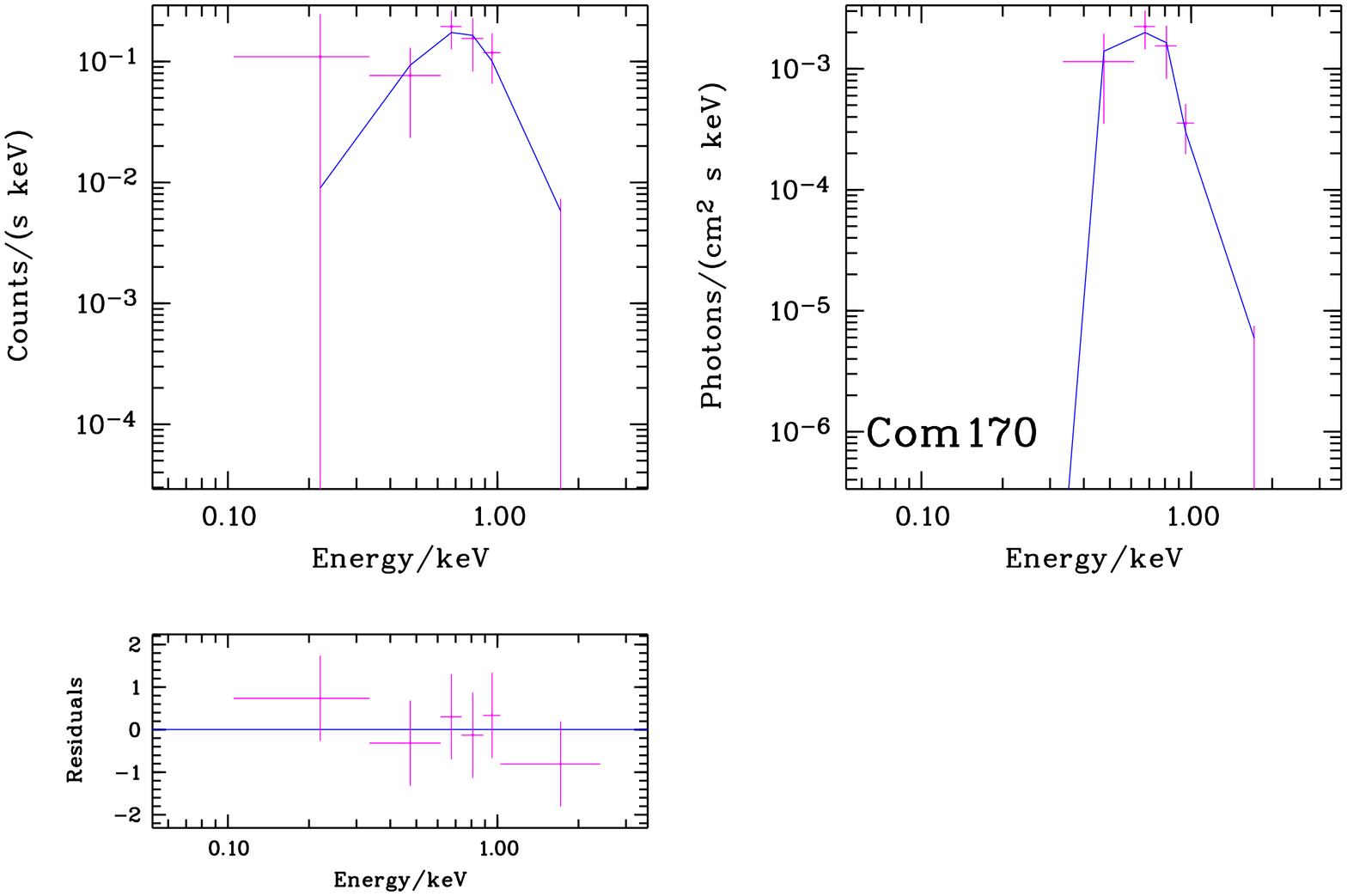}
\includegraphics[width=3.9cm, bb=76 410 385 760, angle=-90,clip]{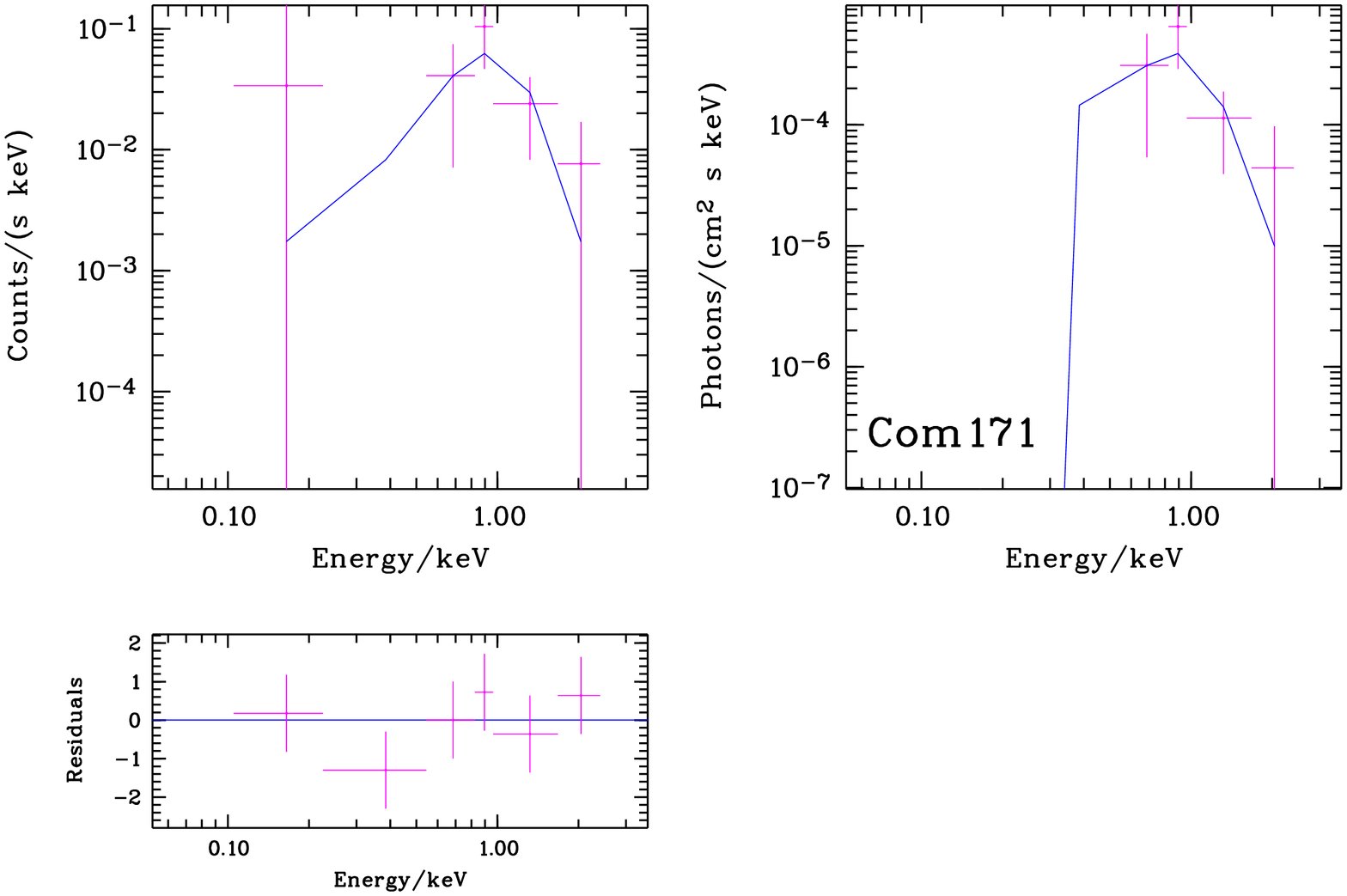}
\includegraphics[width=3.9cm, bb=76 410 385 760, angle=-90,clip]{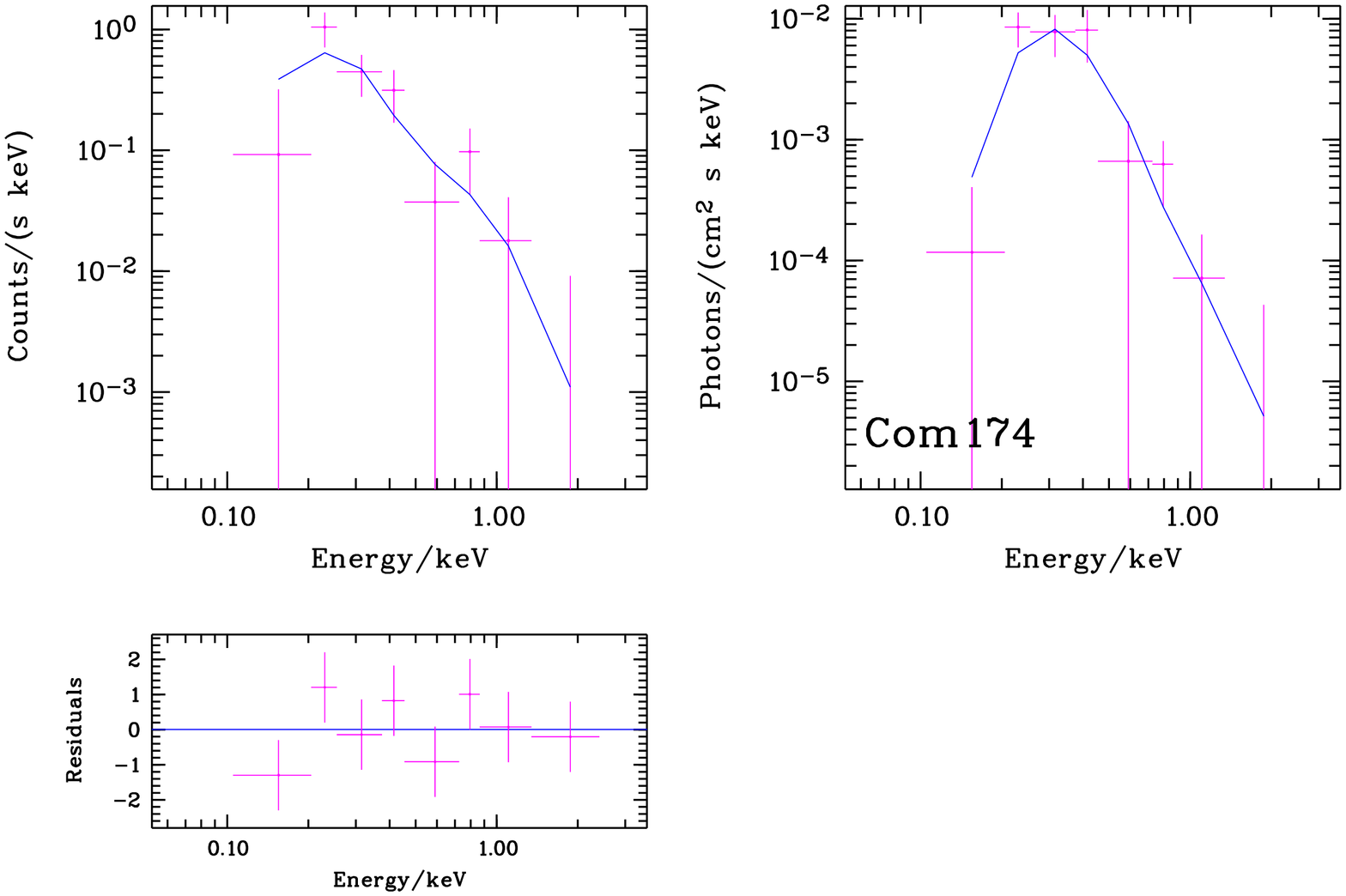}
\includegraphics[width=3.9cm, bb=76 410 385 760, angle=-90,clip]{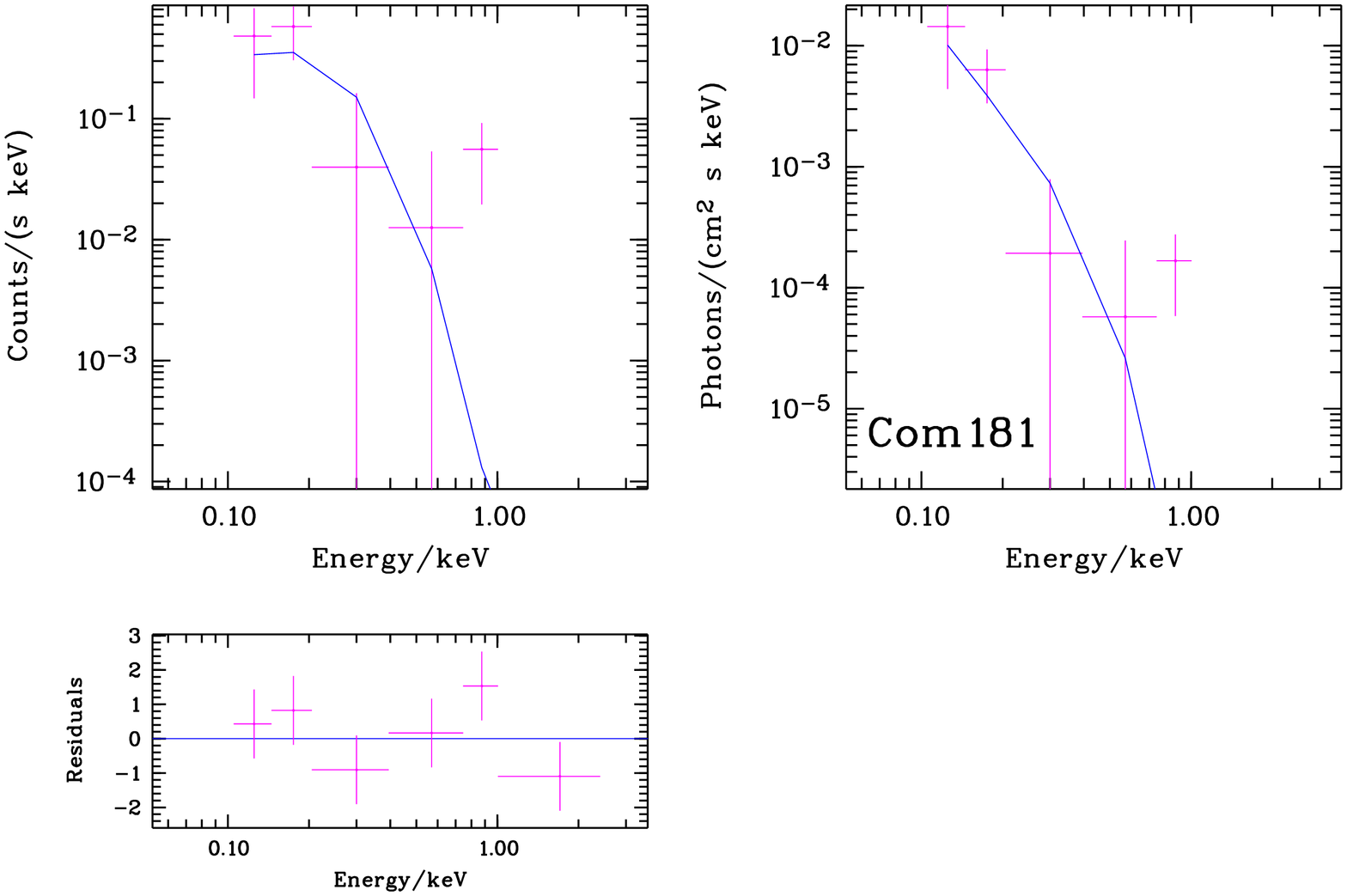}

\includegraphics[width=3.9cm, bb=76 410 385 760, angle=-90,clip]{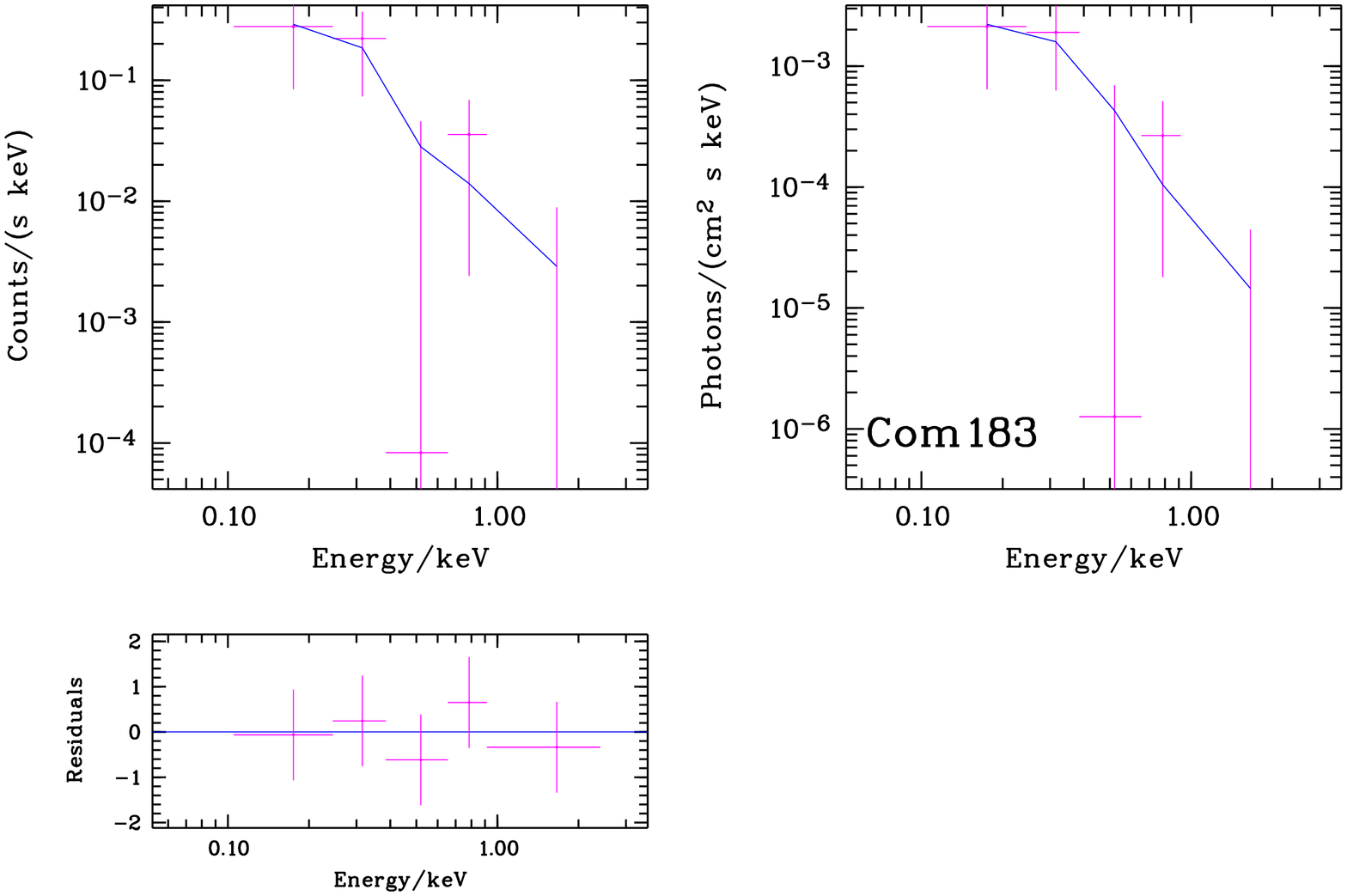}
\includegraphics[width=3.9cm, bb=76 410 385 760, angle=-90,clip]{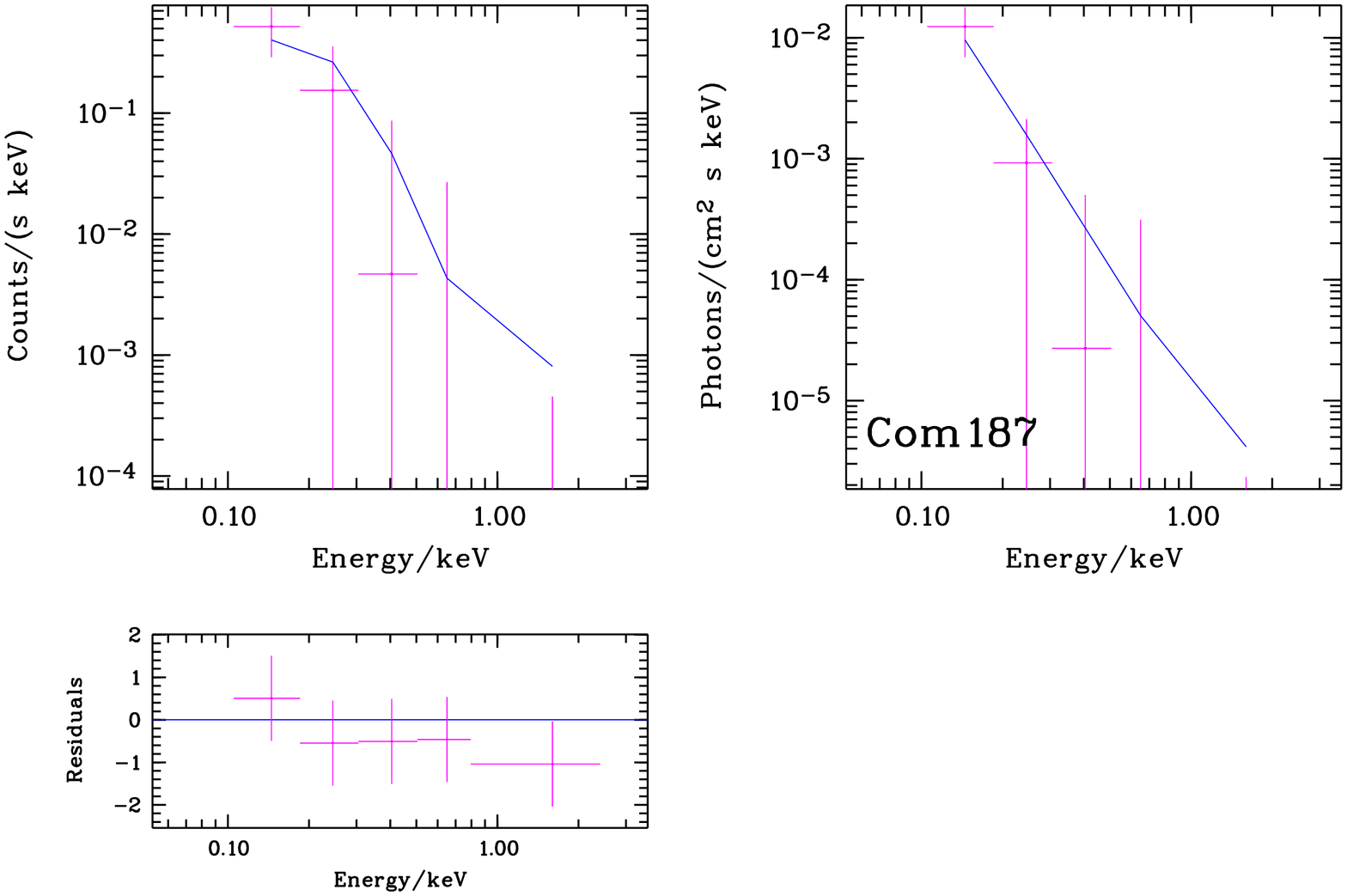}
\includegraphics[width=3.9cm, bb=76 410 385 760, angle=-90,clip]{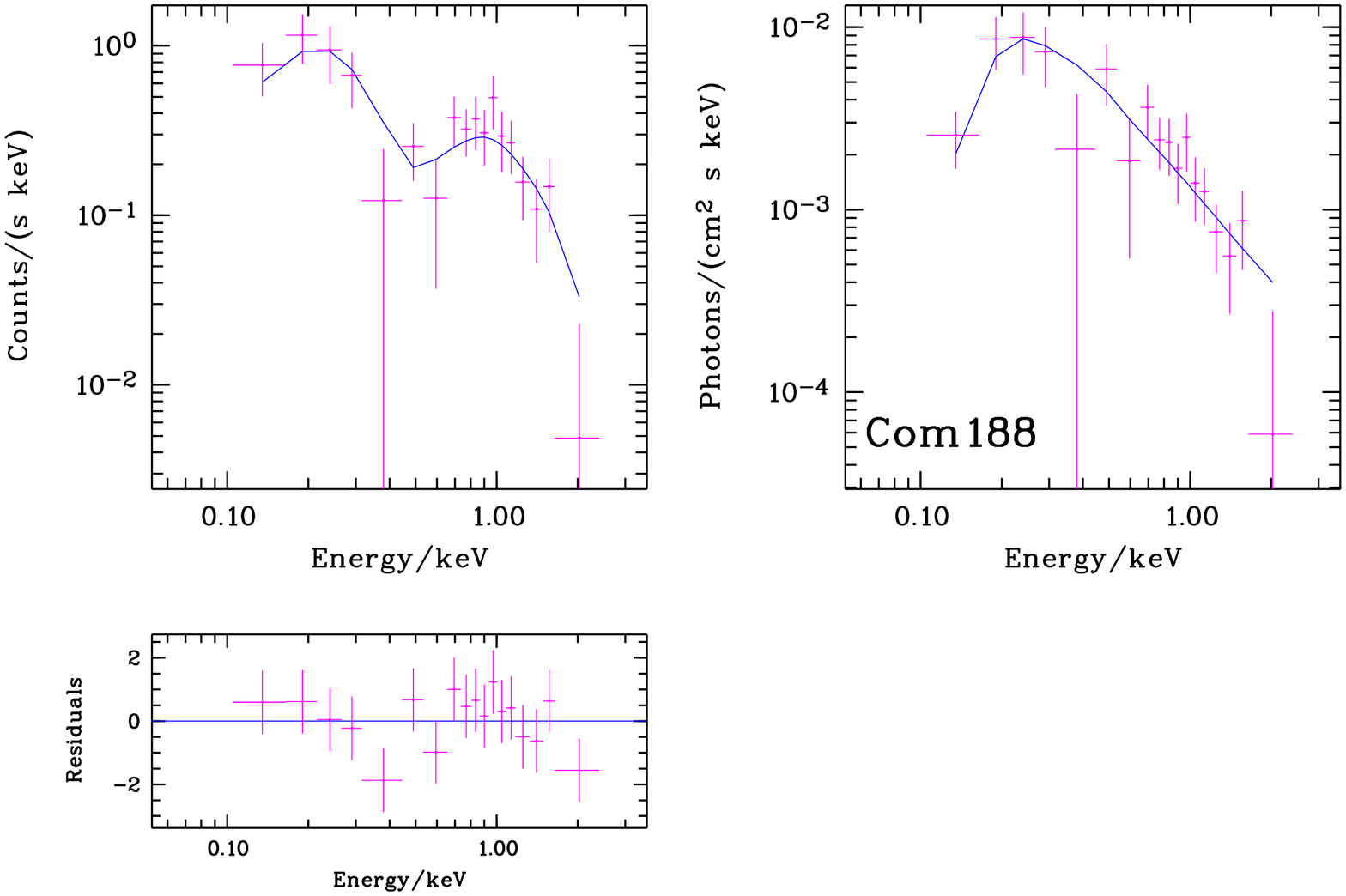}
\includegraphics[width=3.9cm, bb=76 410 385 760, angle=-90,clip]{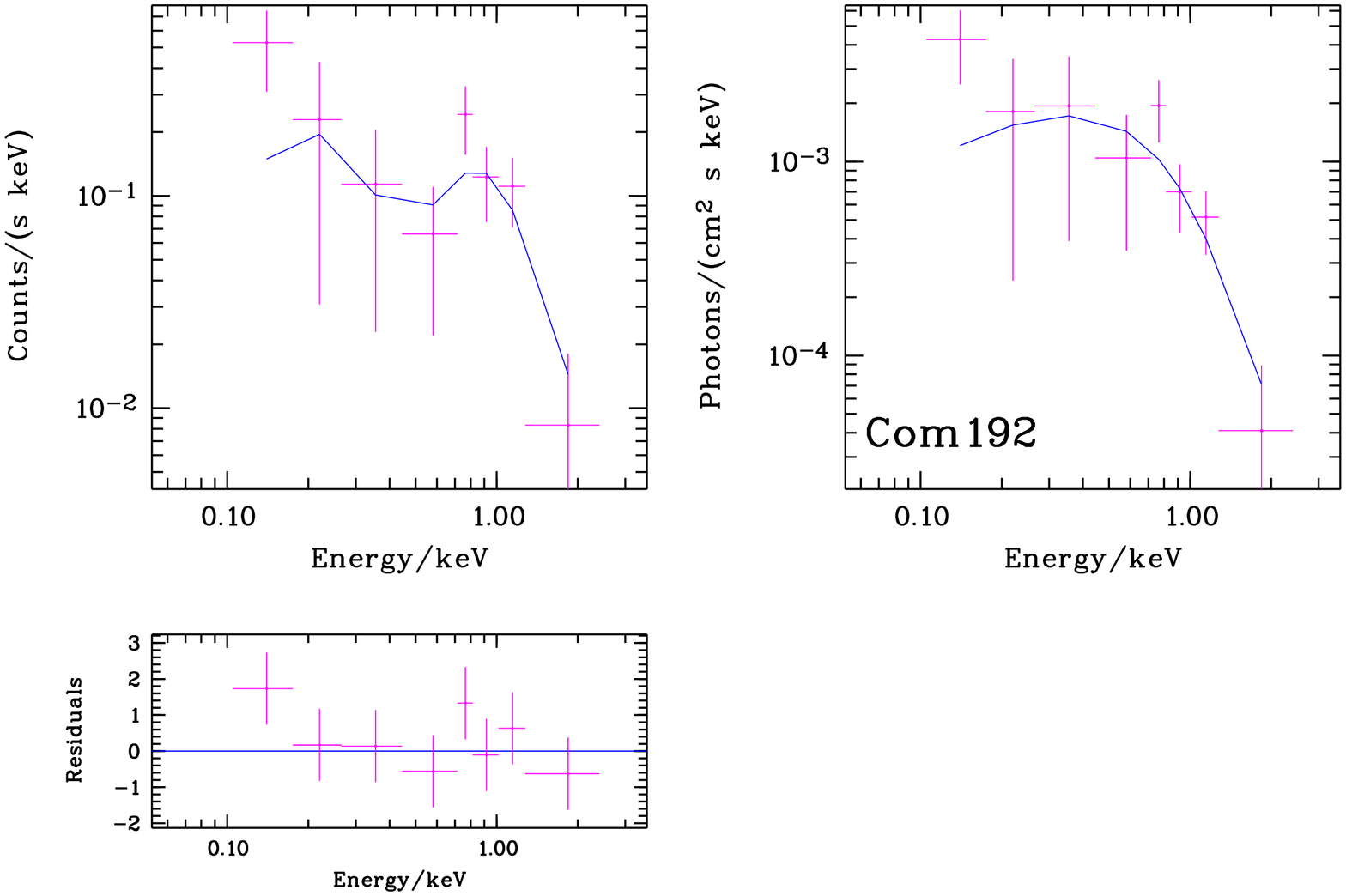}

\includegraphics[width=3.9cm, bb=76 410 385 760, angle=-90,clip]{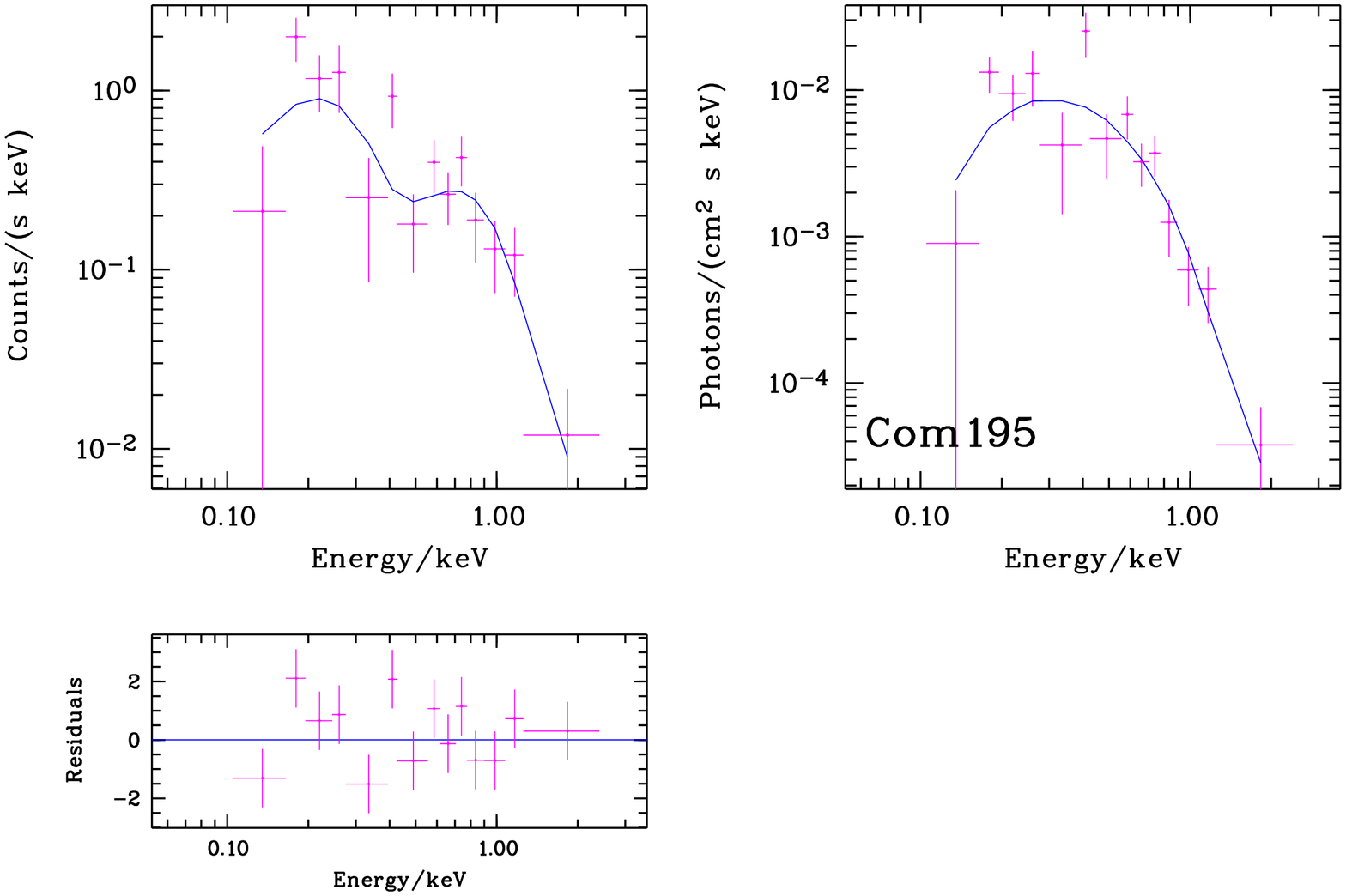}
\includegraphics[width=3.9cm, bb=76 410 385 760, angle=-90,clip]{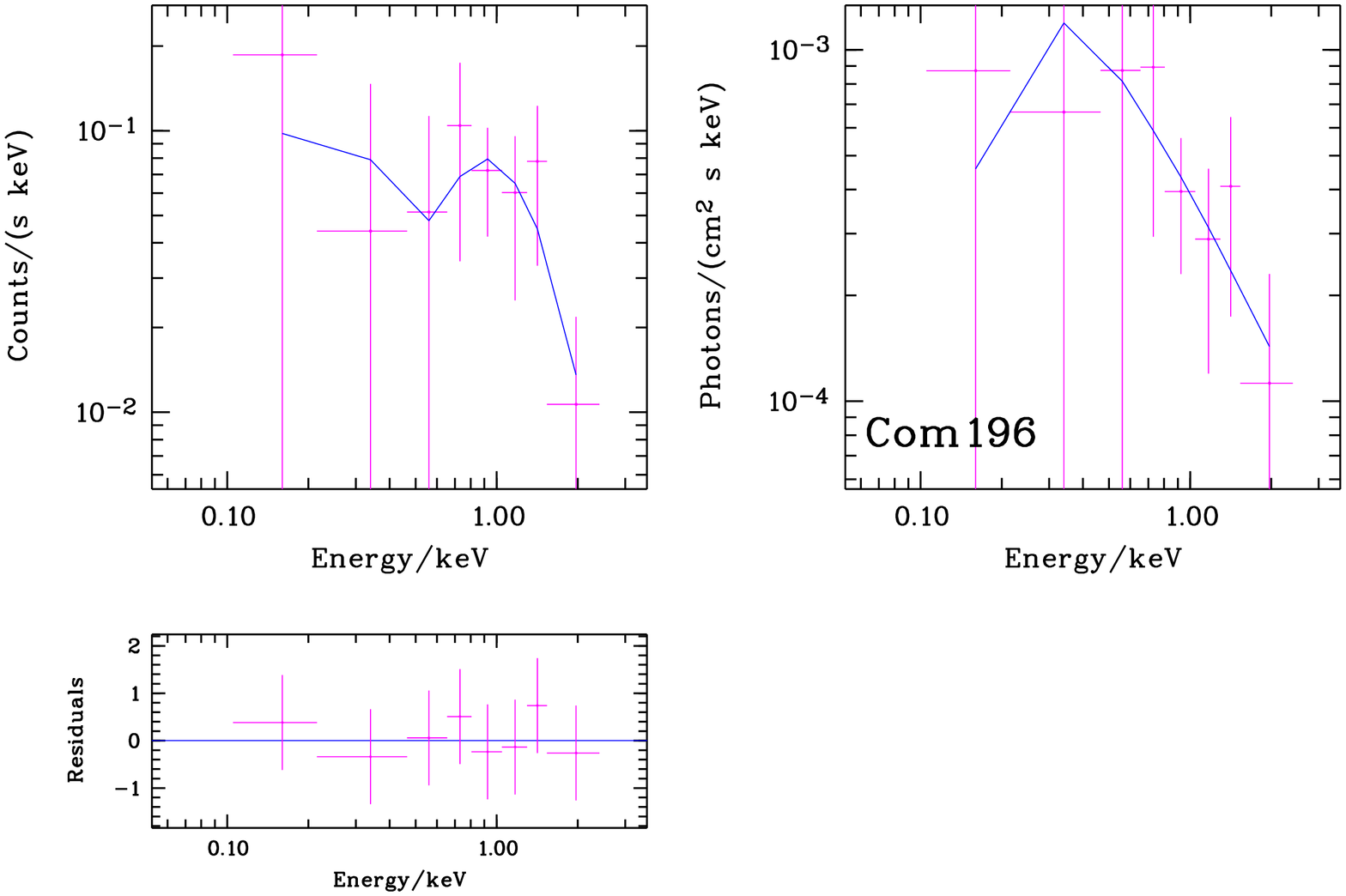}
\includegraphics[width=3.9cm, bb=76 410 385 760, angle=-90,clip]{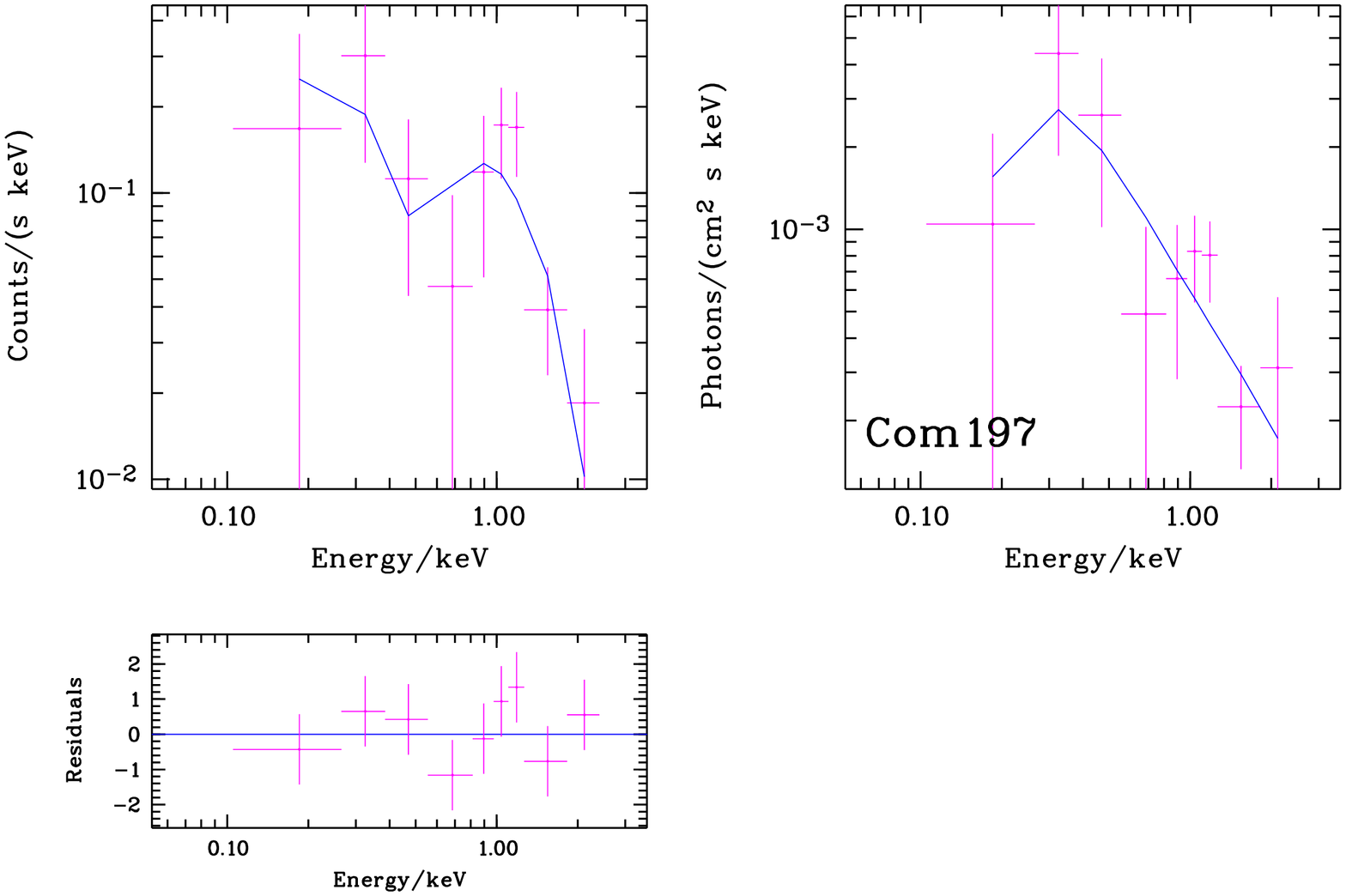}
\includegraphics[width=3.9cm, bb=76 410 385 760, angle=-90,clip]{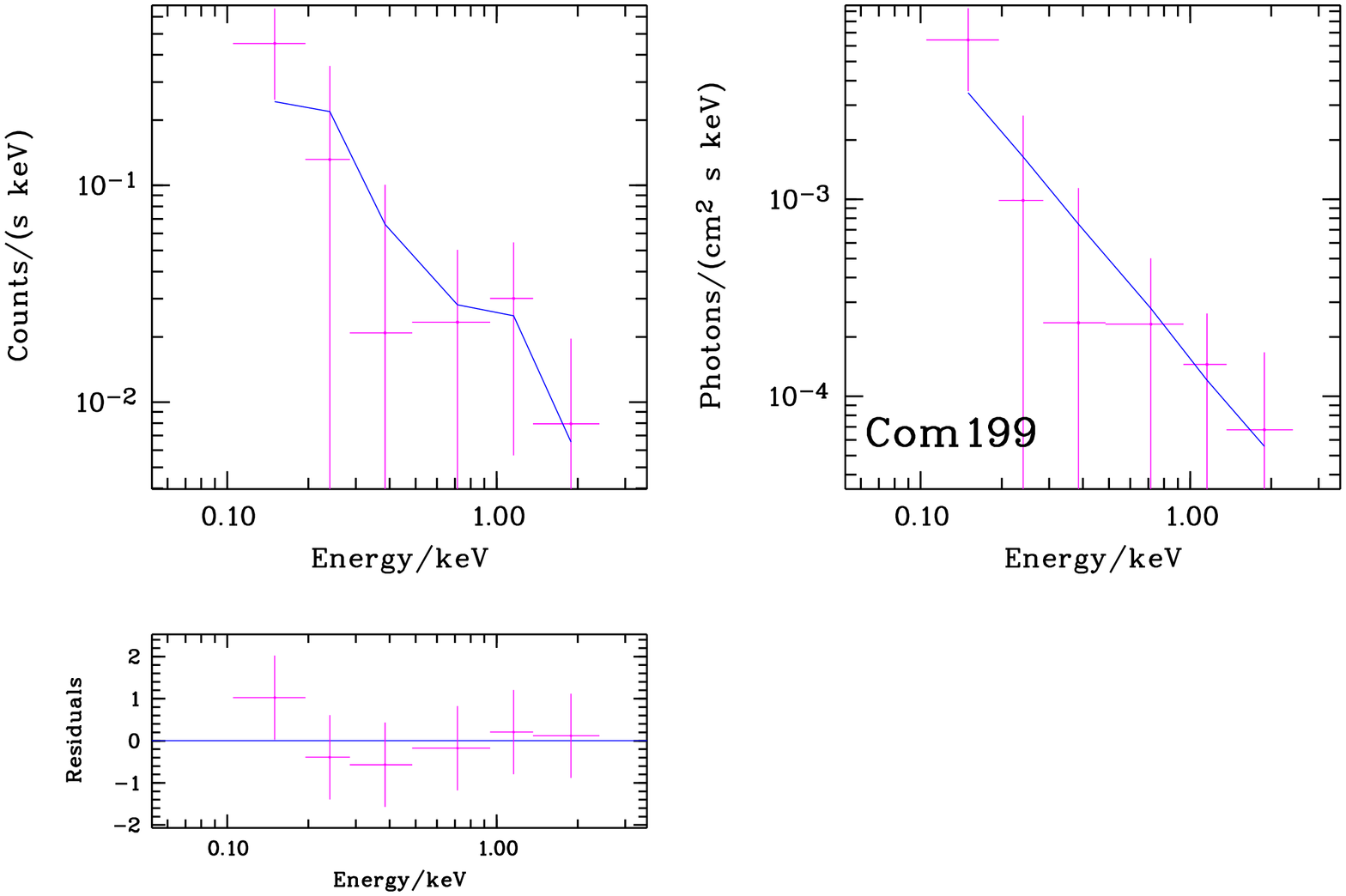}

\includegraphics[width=3.9cm, bb=76 410 385 760, angle=-90,clip]{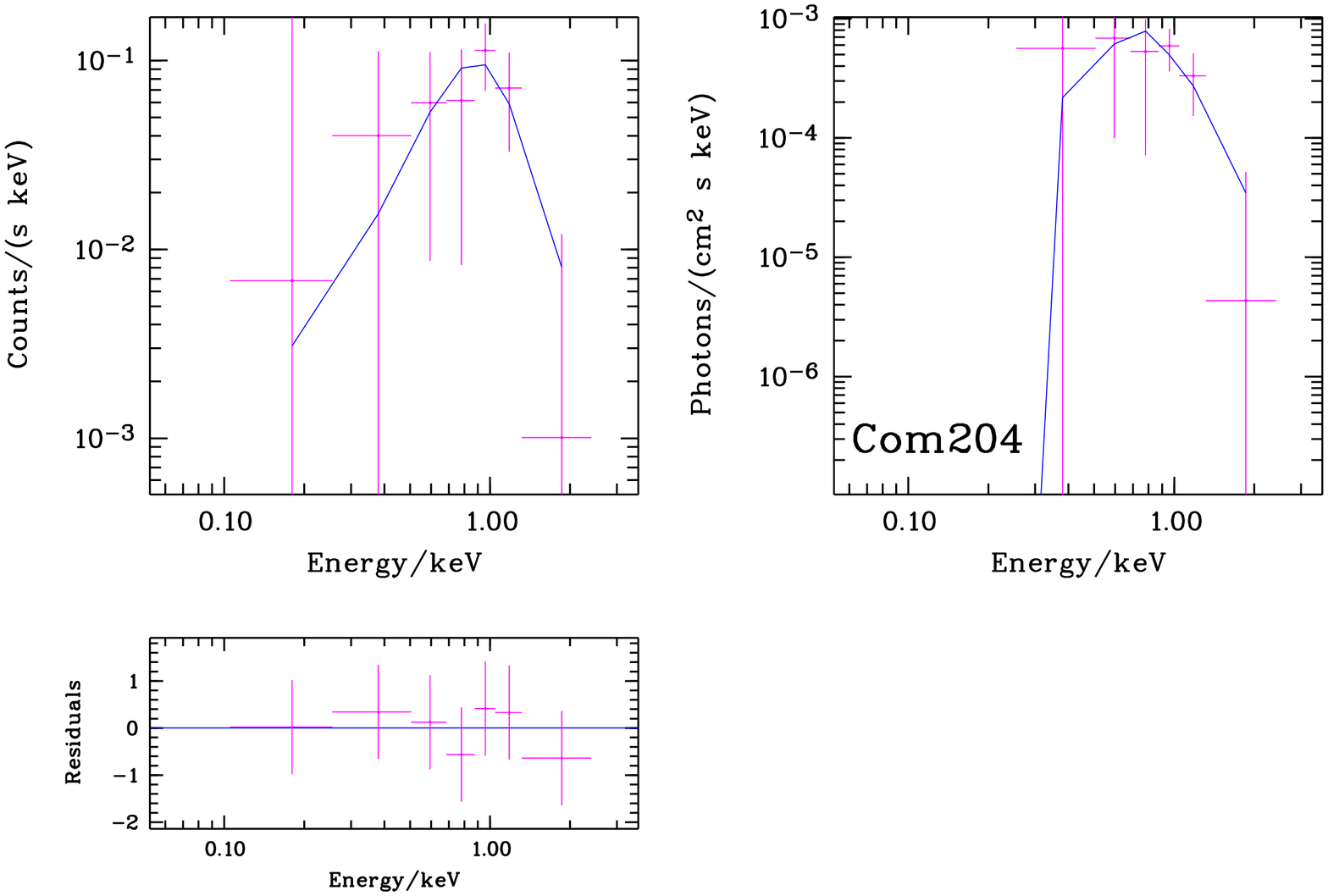}
\includegraphics[width=3.9cm, bb=76 410 385 760, angle=-90,clip]{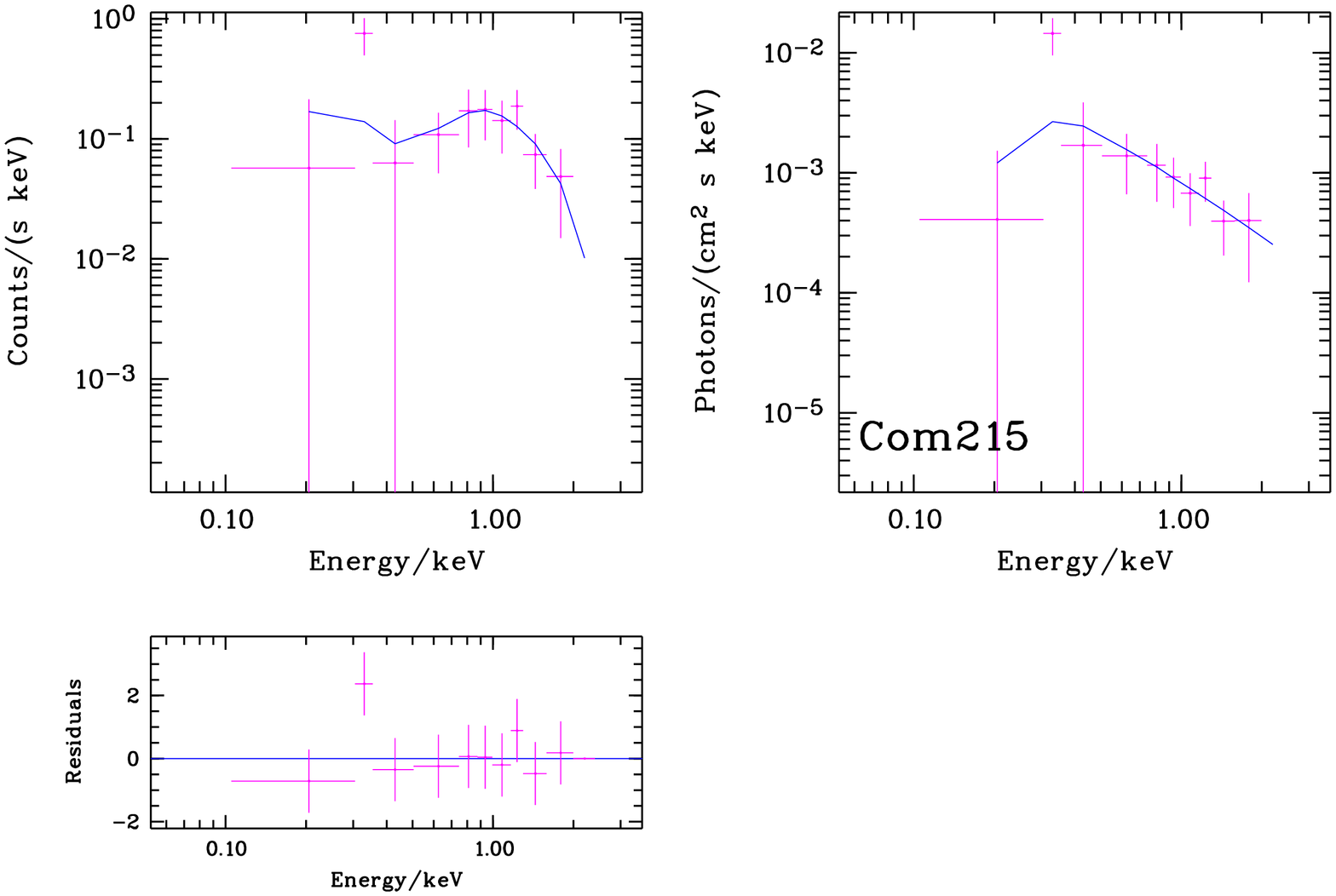}
\includegraphics[width=3.9cm, bb=76 410 385 760, angle=-90,clip]{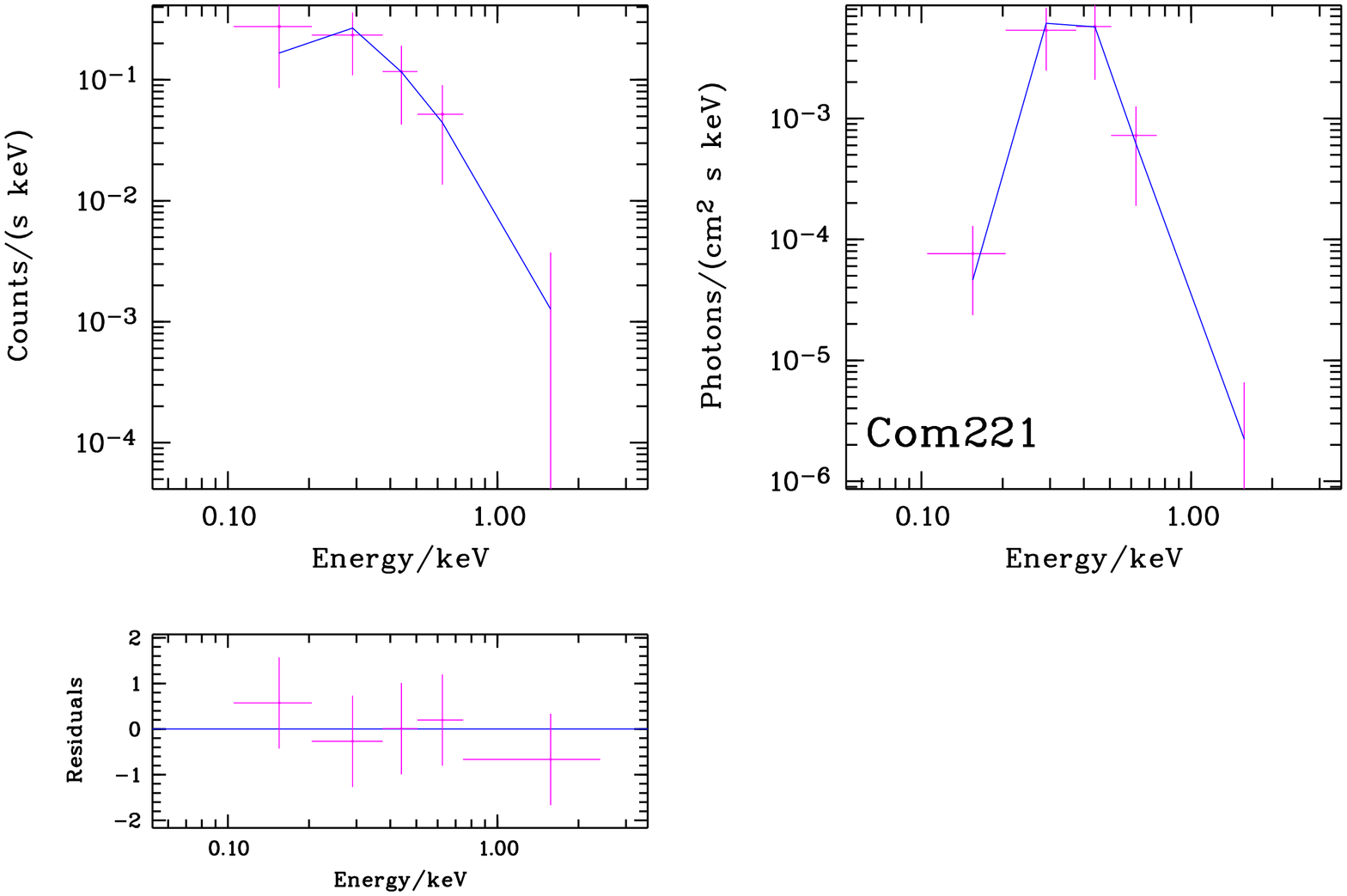}
\includegraphics[width=3.9cm, bb=76 410 385 760, angle=-90,clip]{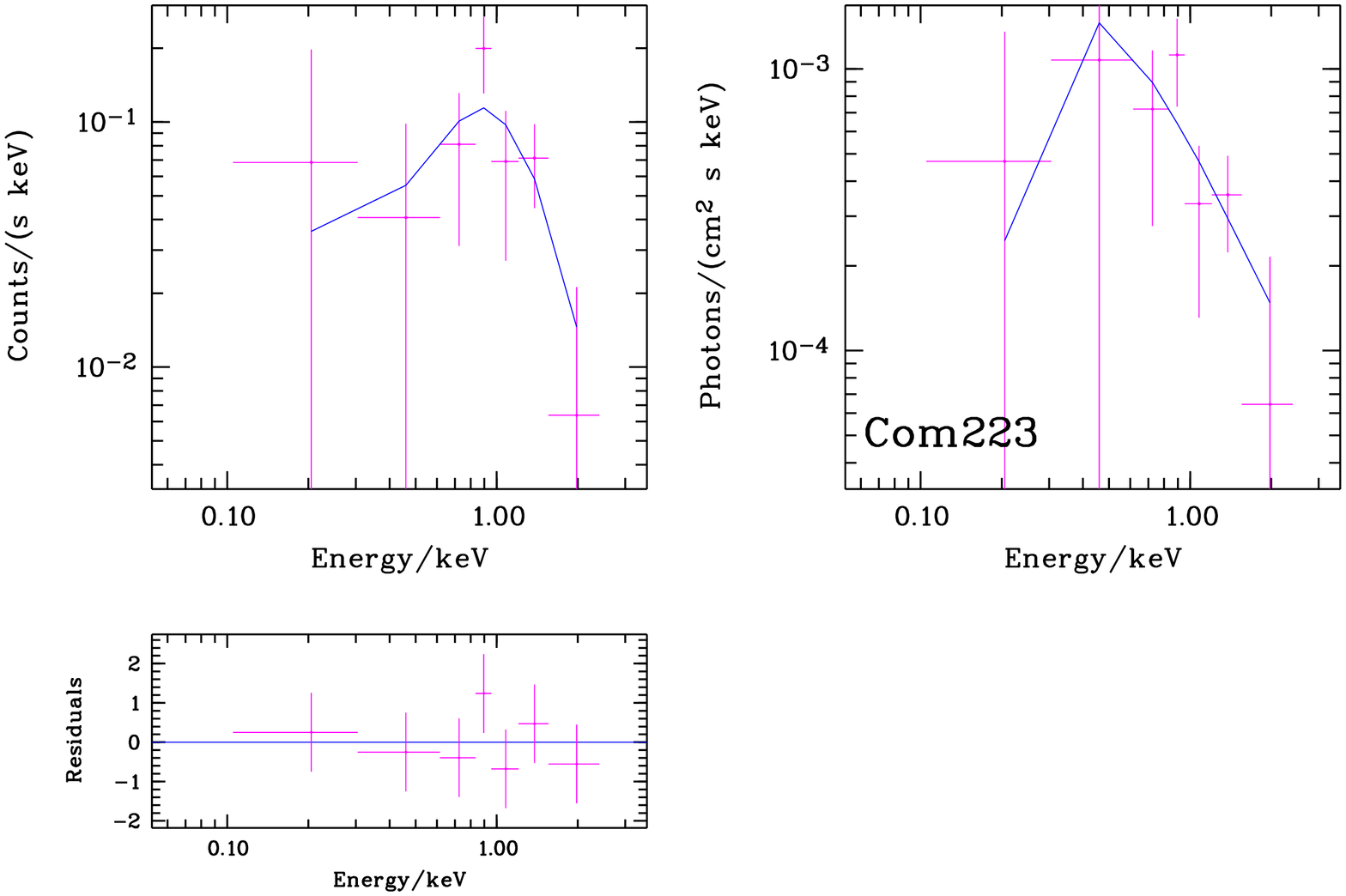}

\includegraphics[width=3.9cm, bb=76 410 385 760, angle=-90,clip]{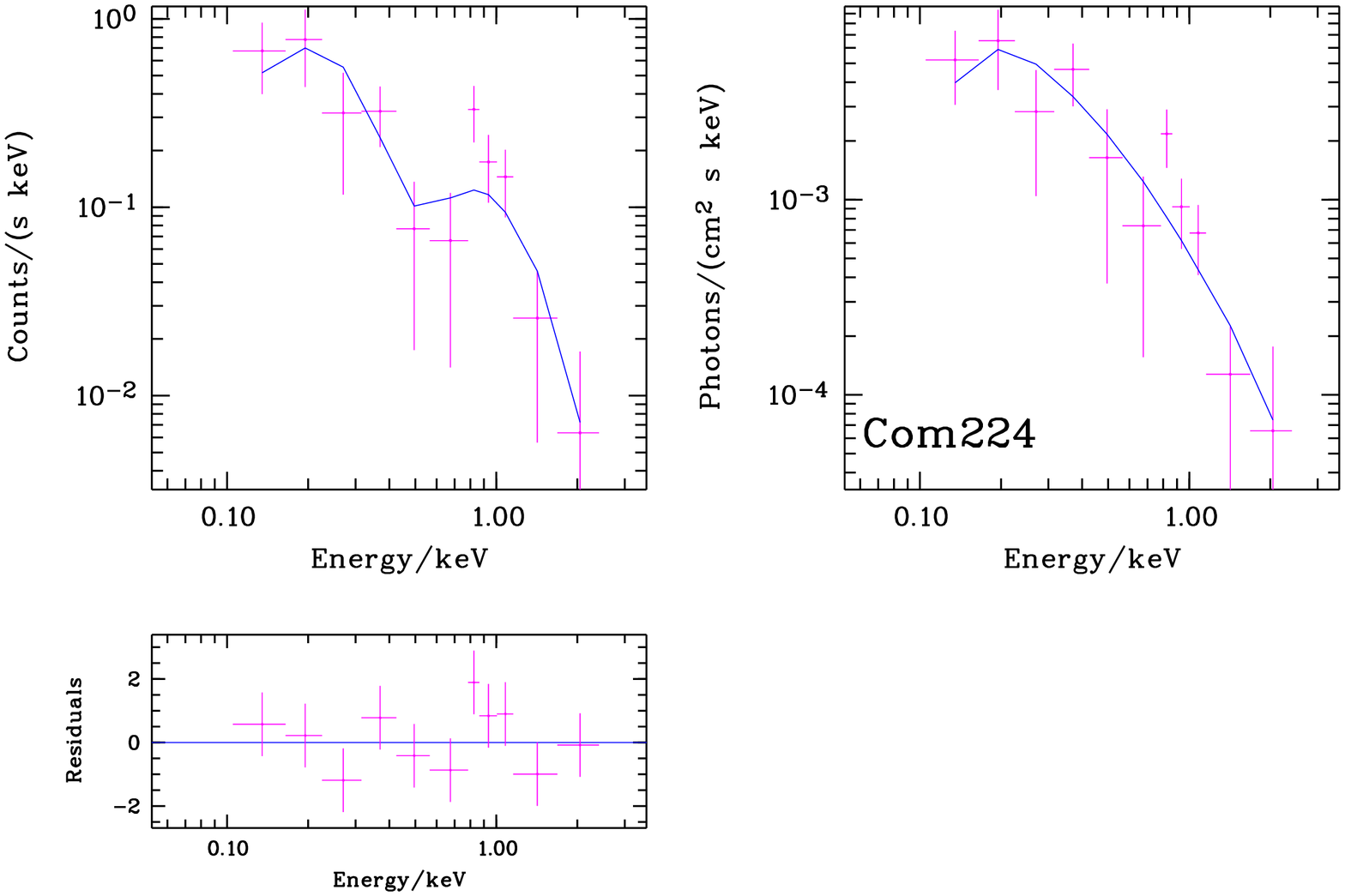}
\includegraphics[width=3.9cm, bb=76 410 385 760, angle=-90,clip]{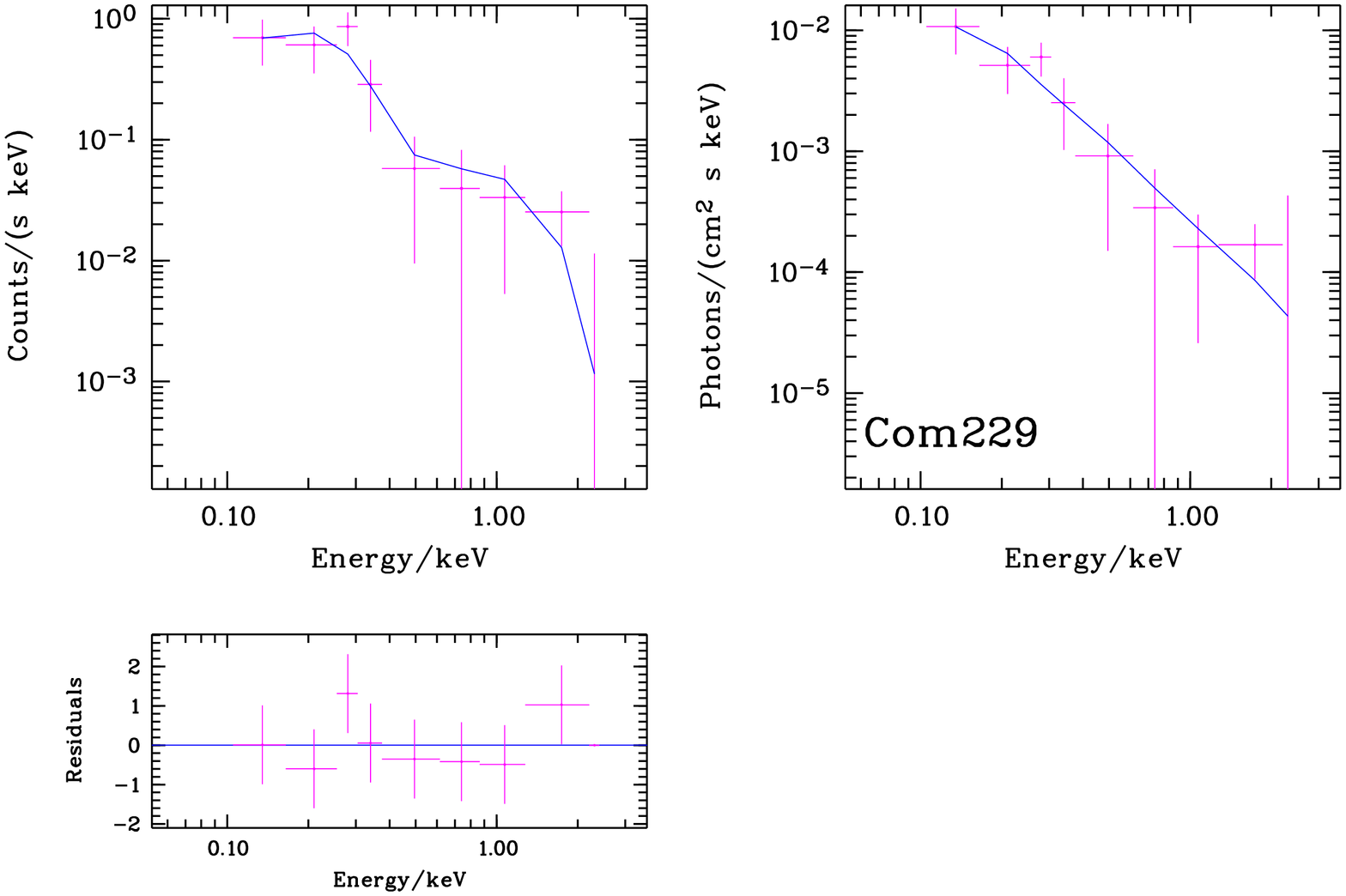}
\includegraphics[width=3.9cm, bb=76 410 385 760, angle=-90,clip]{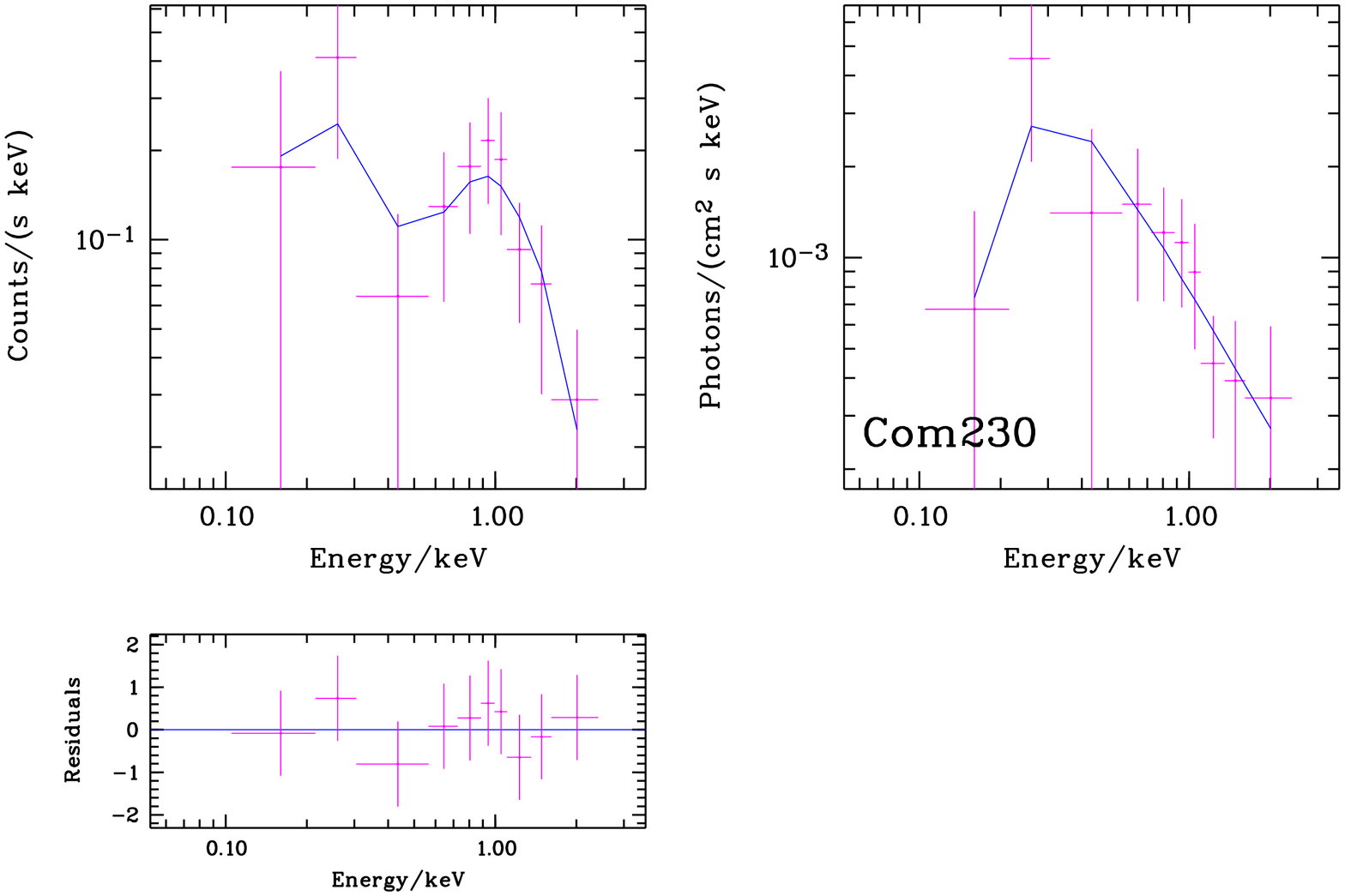}
\includegraphics[width=3.9cm, bb=76 410 385 760, angle=-90,clip]{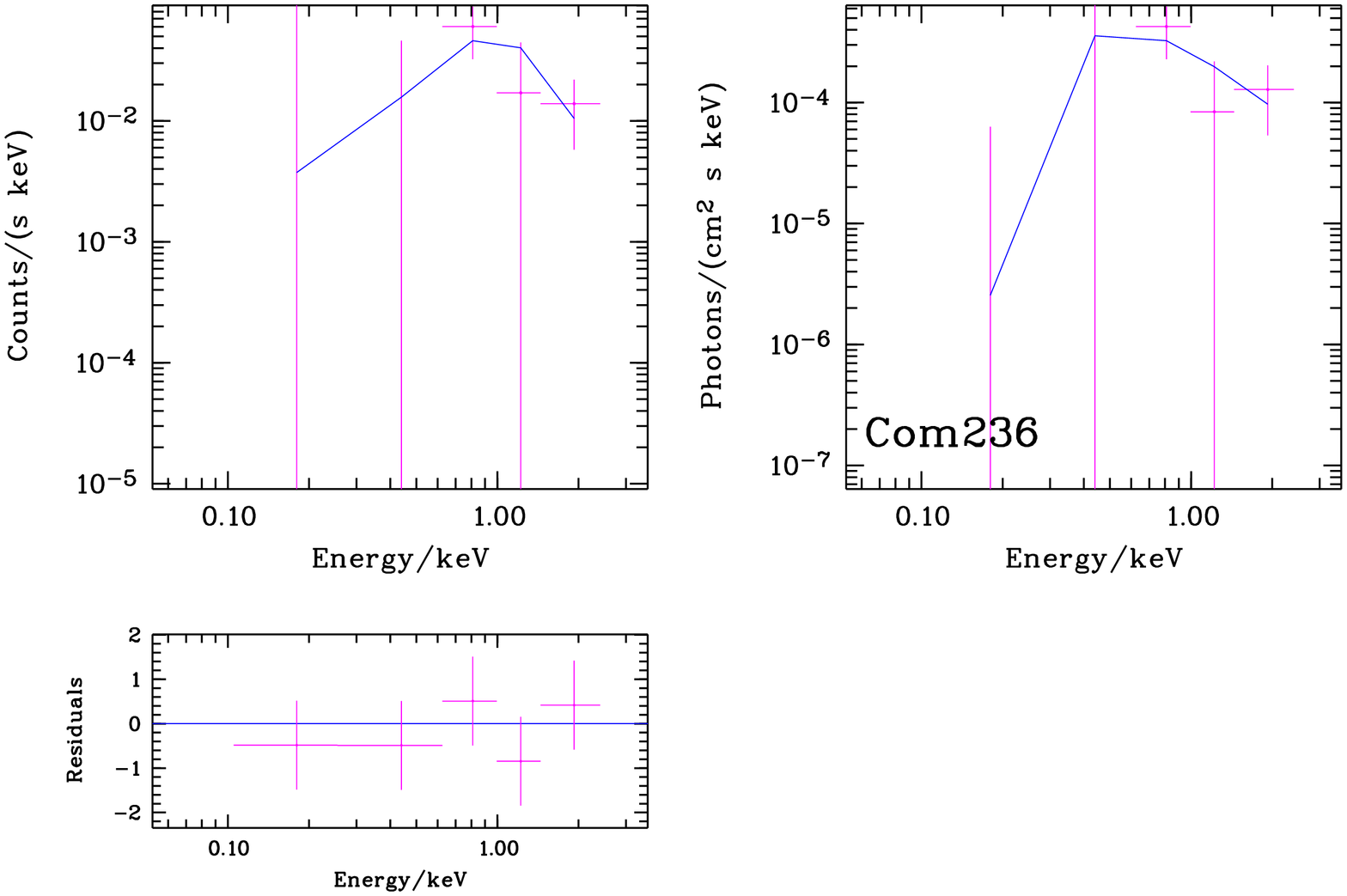}

 \caption[fsp]{{\bf contd.} ROSAT survey spectra of Com sources.}
 \end{figure*}

\setcounter{figure}{0}

\begin{figure*}[ht]

\includegraphics[width=3.9cm, bb=76 410 385 760, angle=-90,clip]{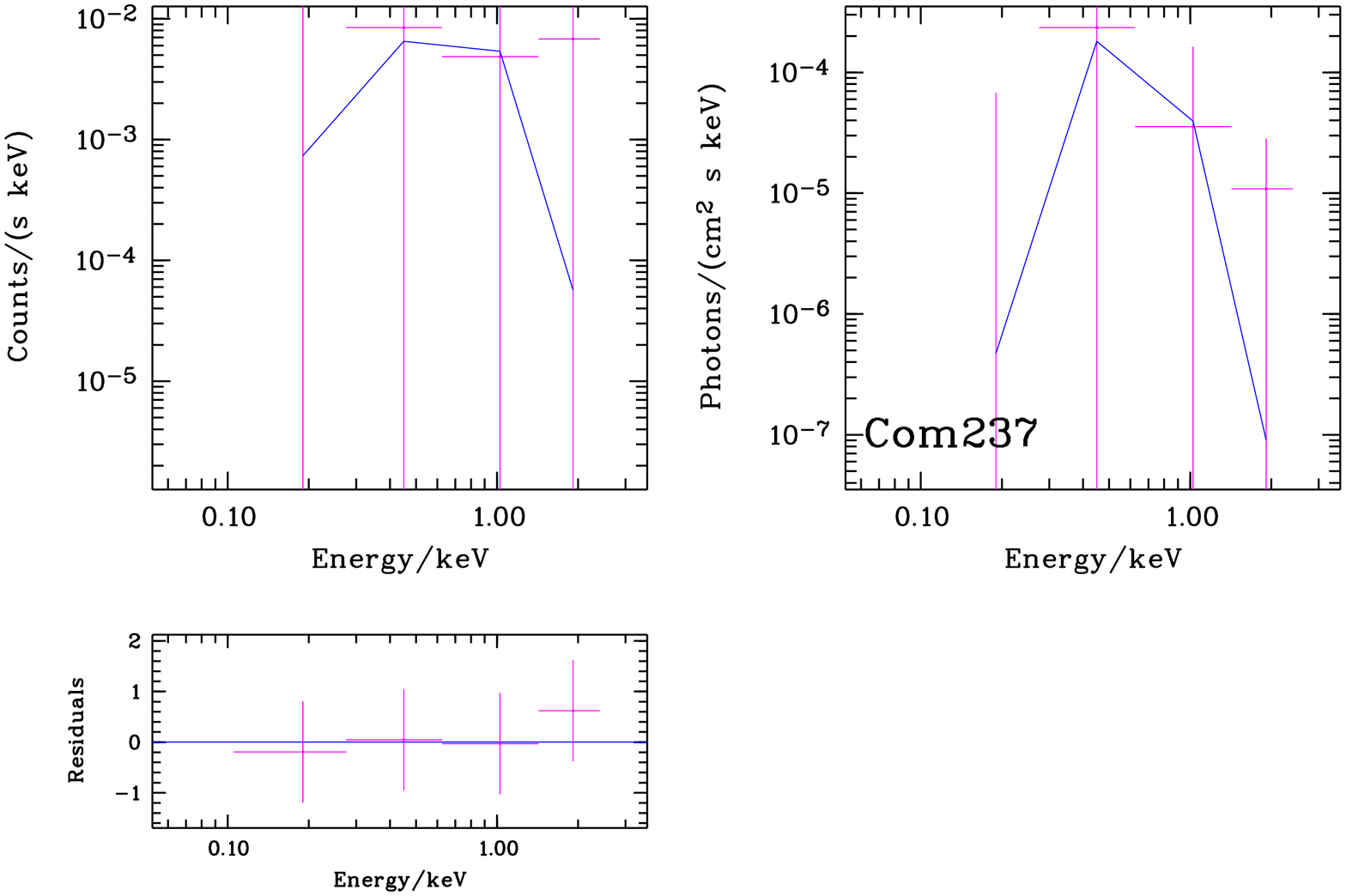}
\includegraphics[width=3.9cm, bb=76 410 385 760, angle=-90,clip]{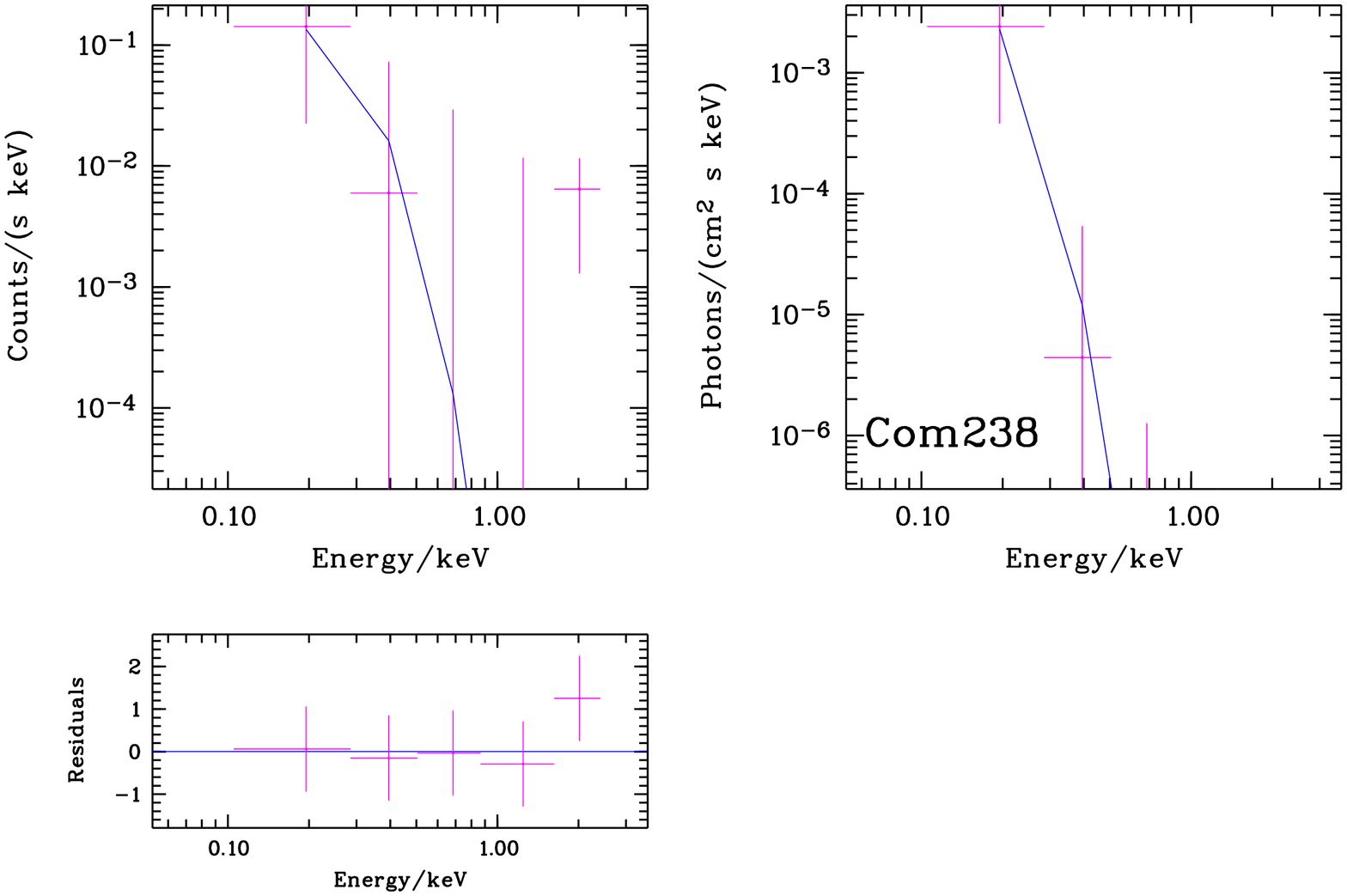}
 \caption[fsp]{{\bf contd.} ROSAT survey spectra of Com sources.}
 \end{figure*}

%%%
%%%%%% Sge Spektren
%%%

\begin{figure*}[ht]

\includegraphics[width=3.9cm, bb=76 410 385 760, angle=-90,clip]{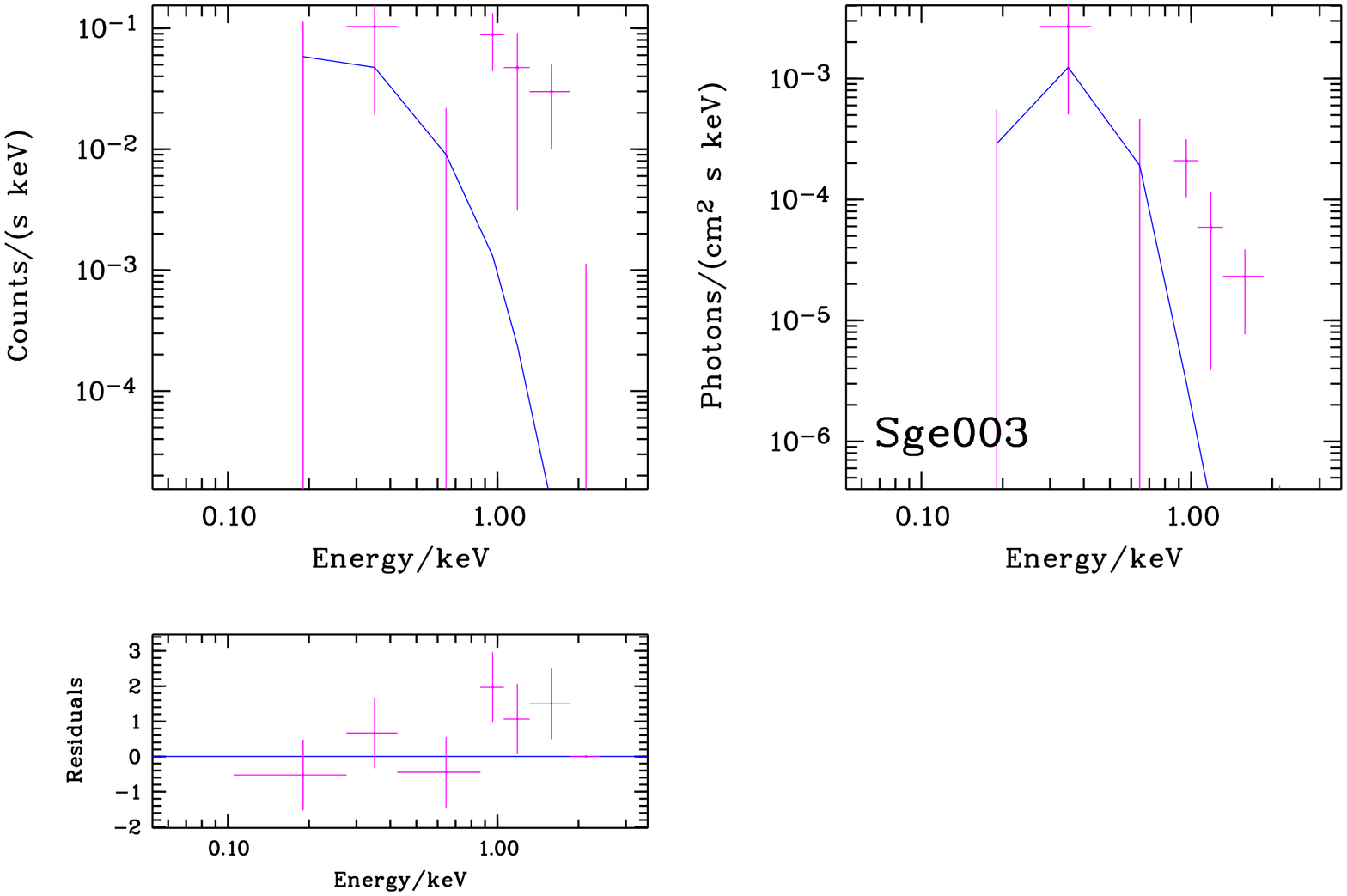}
\includegraphics[width=3.9cm, bb=76 410 385 760, angle=-90,clip]{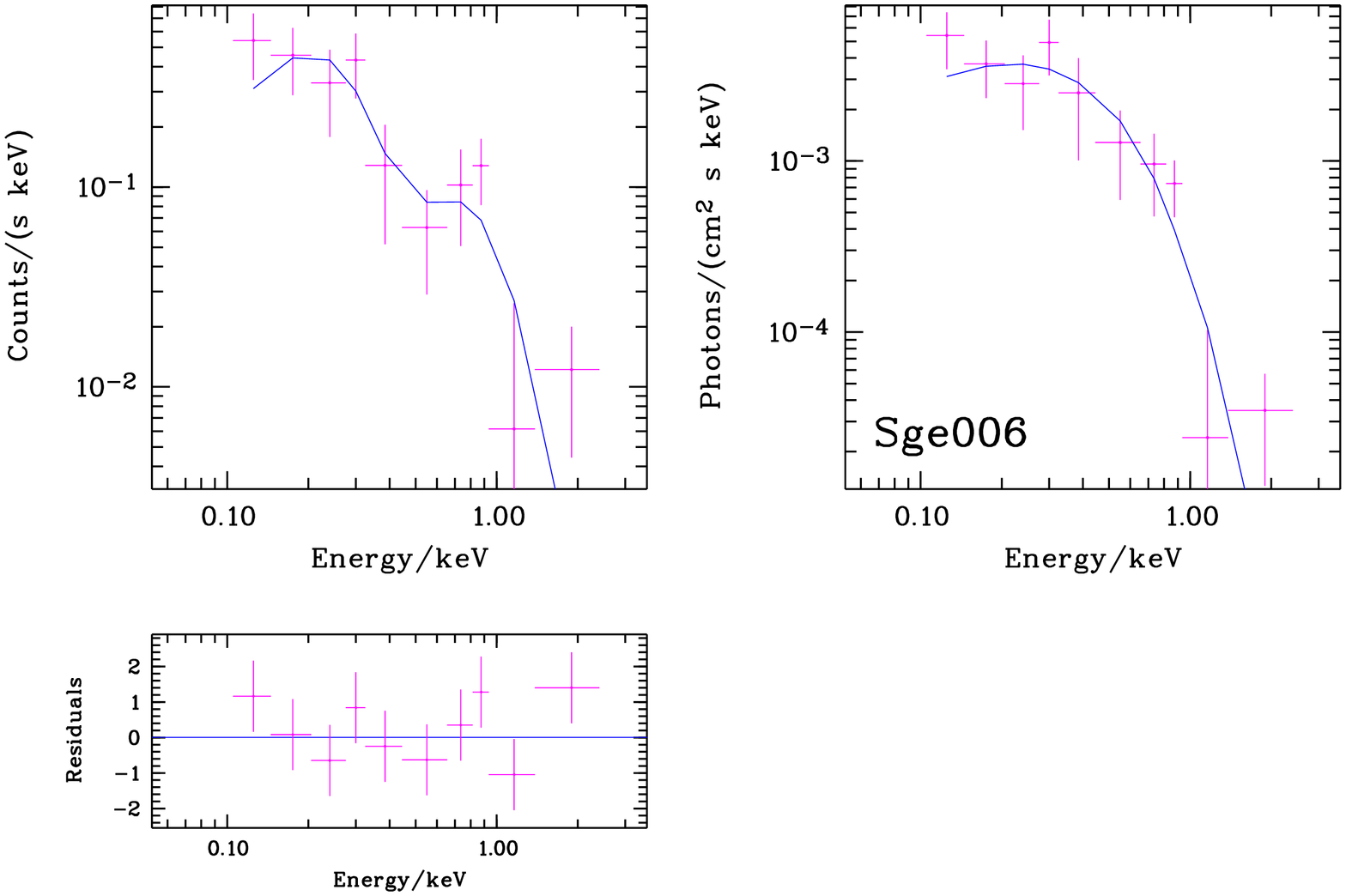}
\includegraphics[width=3.9cm, bb=76 410 385 760, angle=-90,clip]{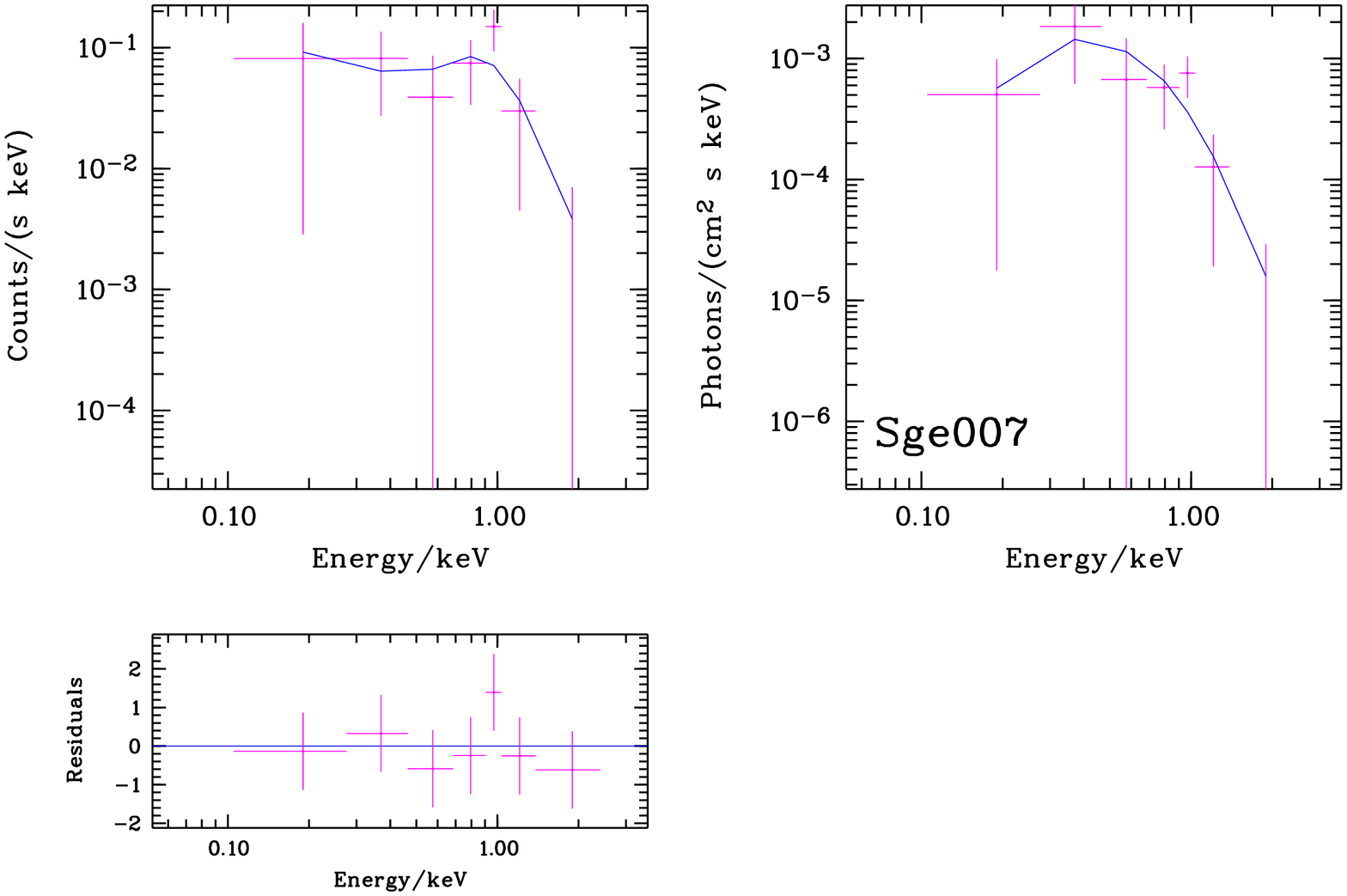}
\includegraphics[width=3.9cm, bb=76 410 385 760, angle=-90,clip]{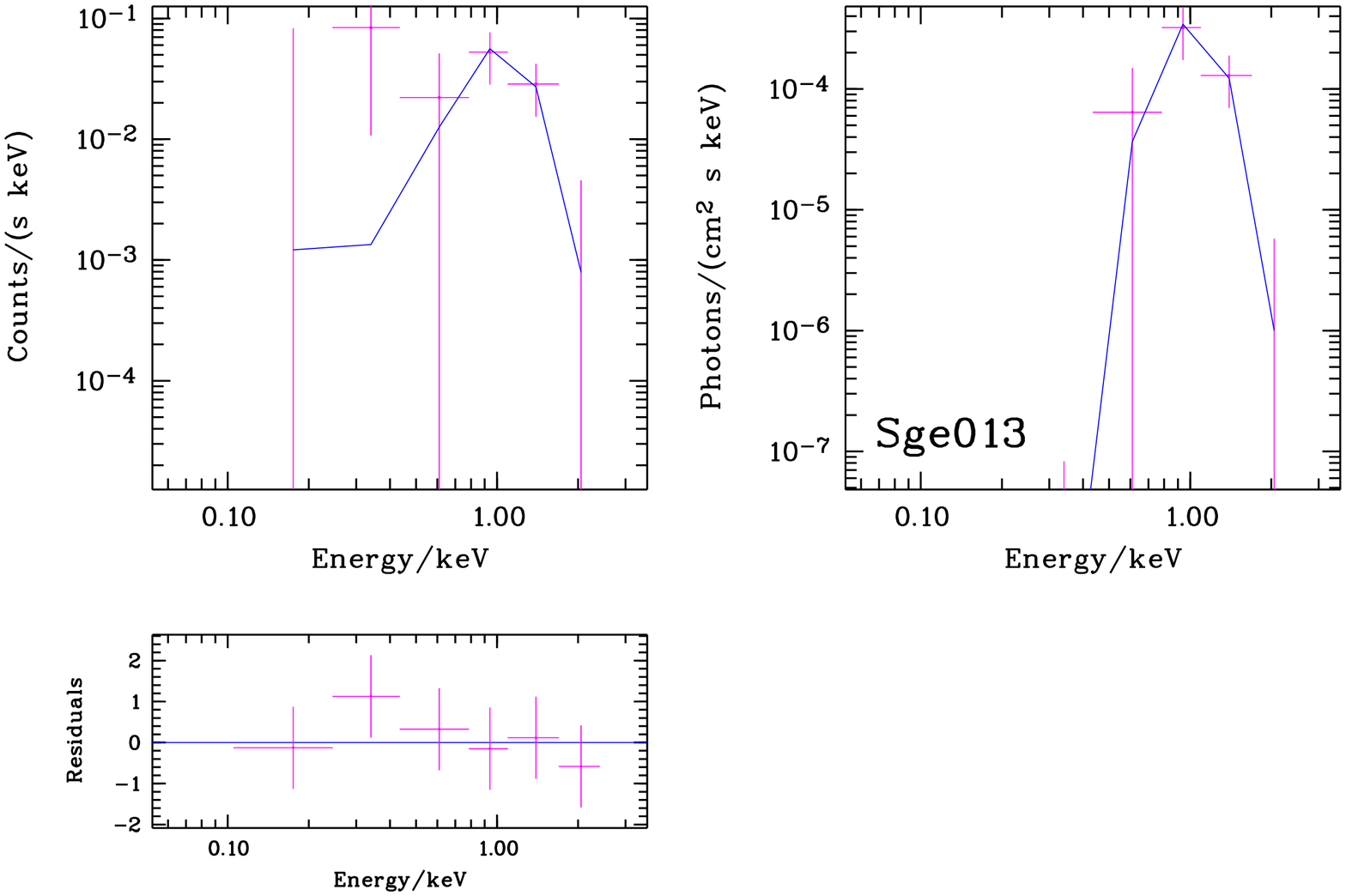}

\includegraphics[width=3.9cm, bb=76 410 385 760, angle=-90,clip]{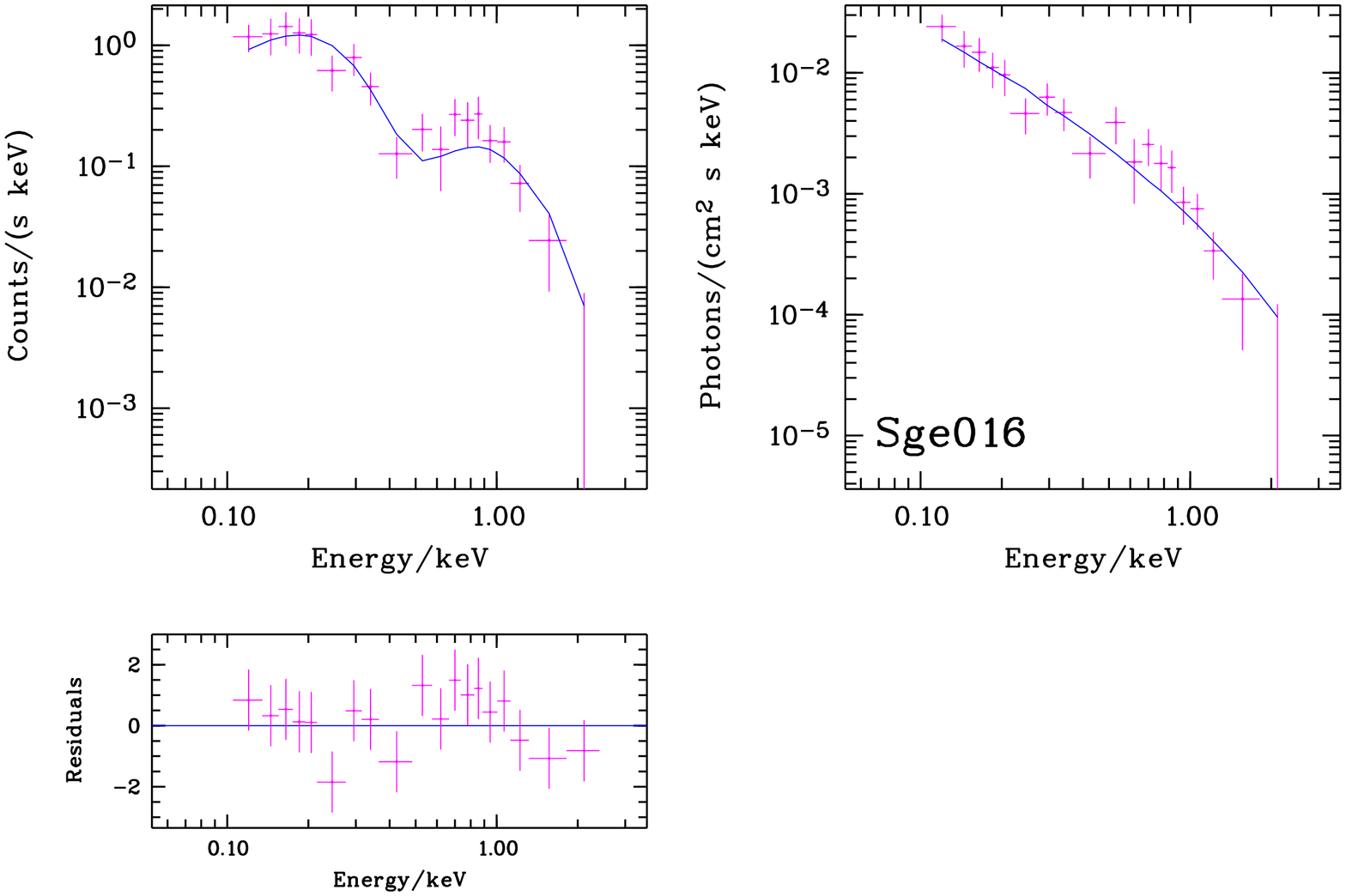}
\includegraphics[width=3.9cm, bb=76 410 385 760, angle=-90,clip]{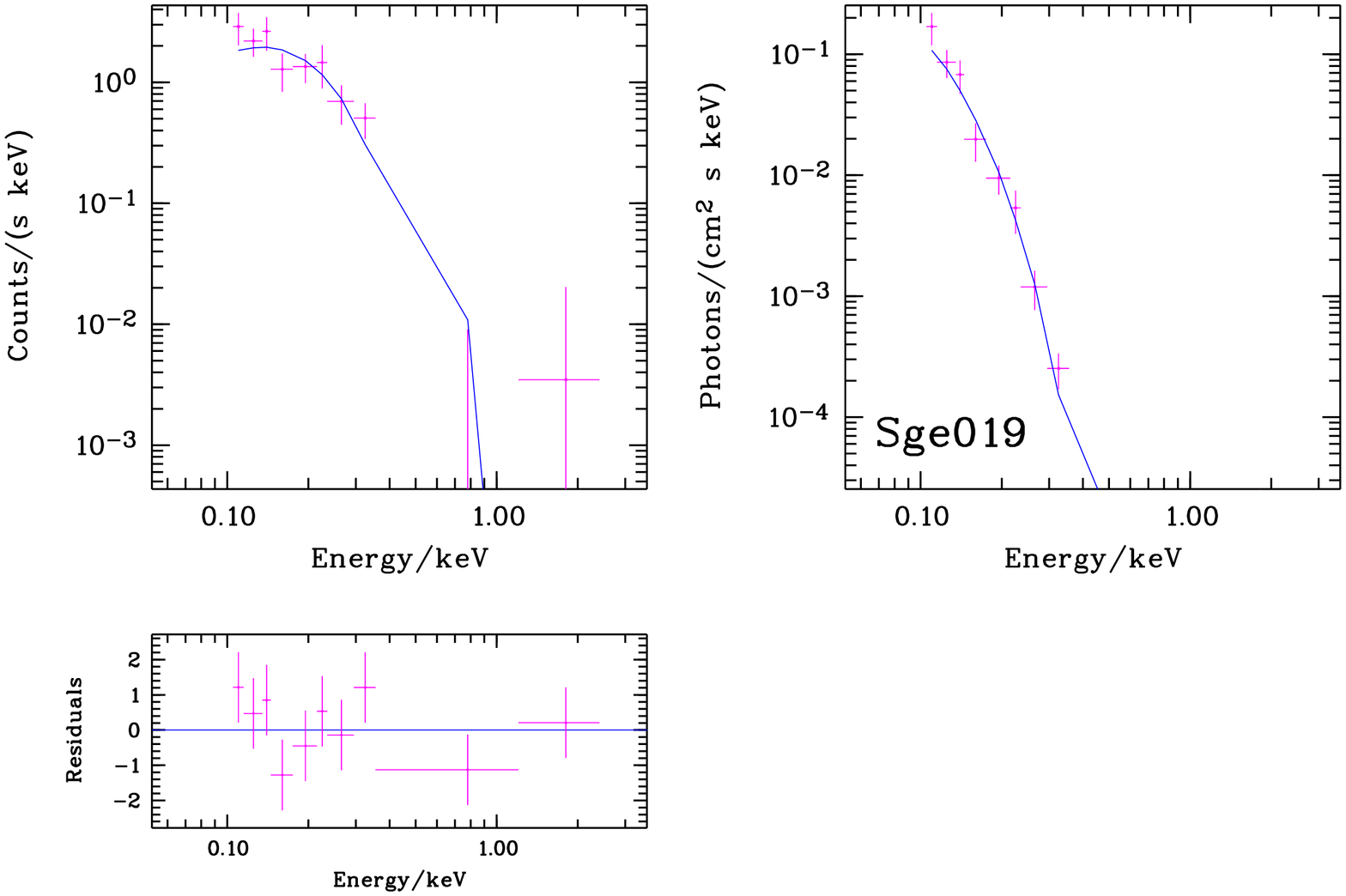}
\includegraphics[width=3.9cm, bb=76 410 385 760, angle=-90,clip]{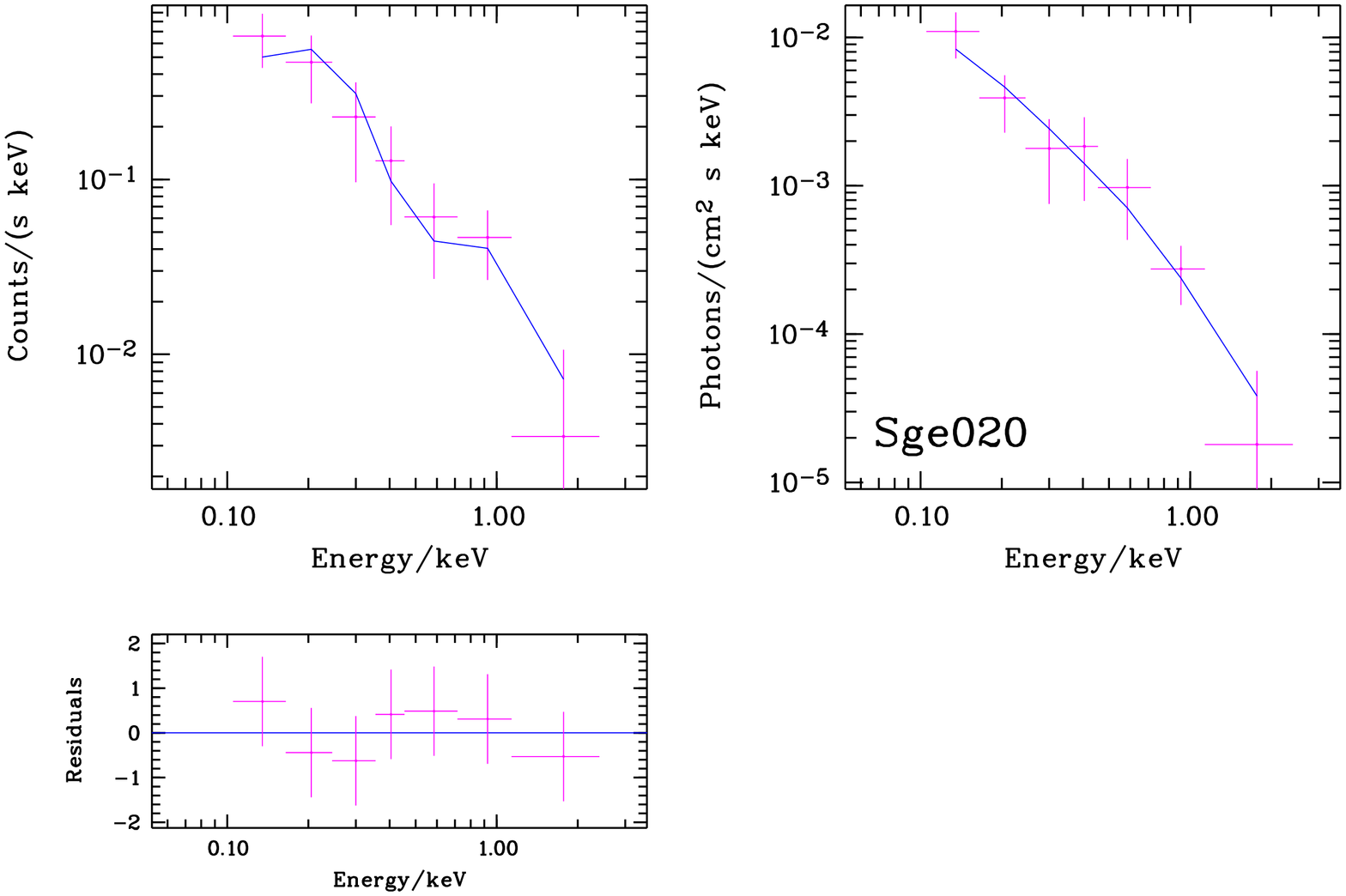}
\includegraphics[width=3.9cm, bb=76 410 385 760, angle=-90,clip]{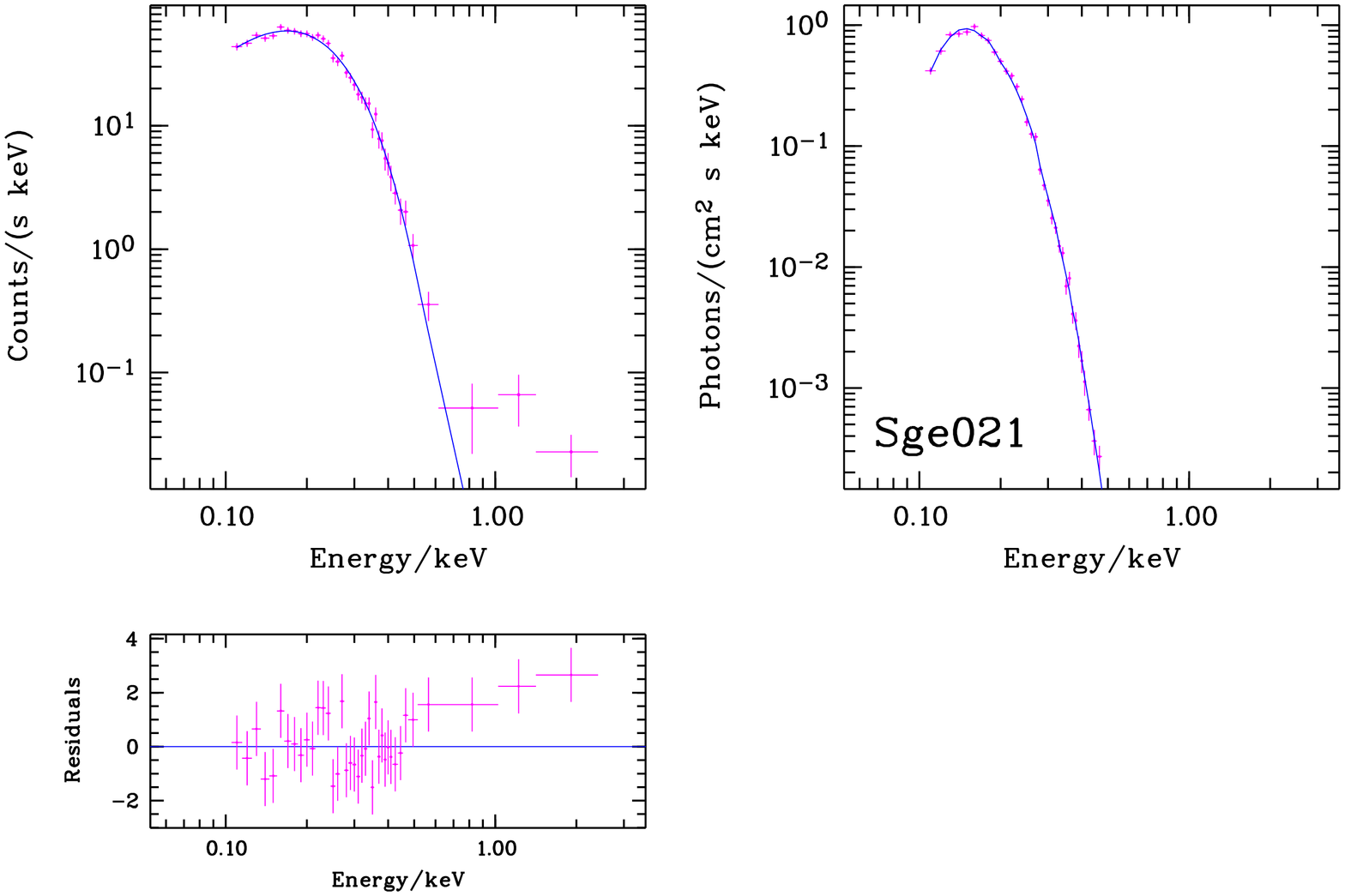}

\includegraphics[width=3.9cm, bb=76 410 385 760, angle=-90,clip]{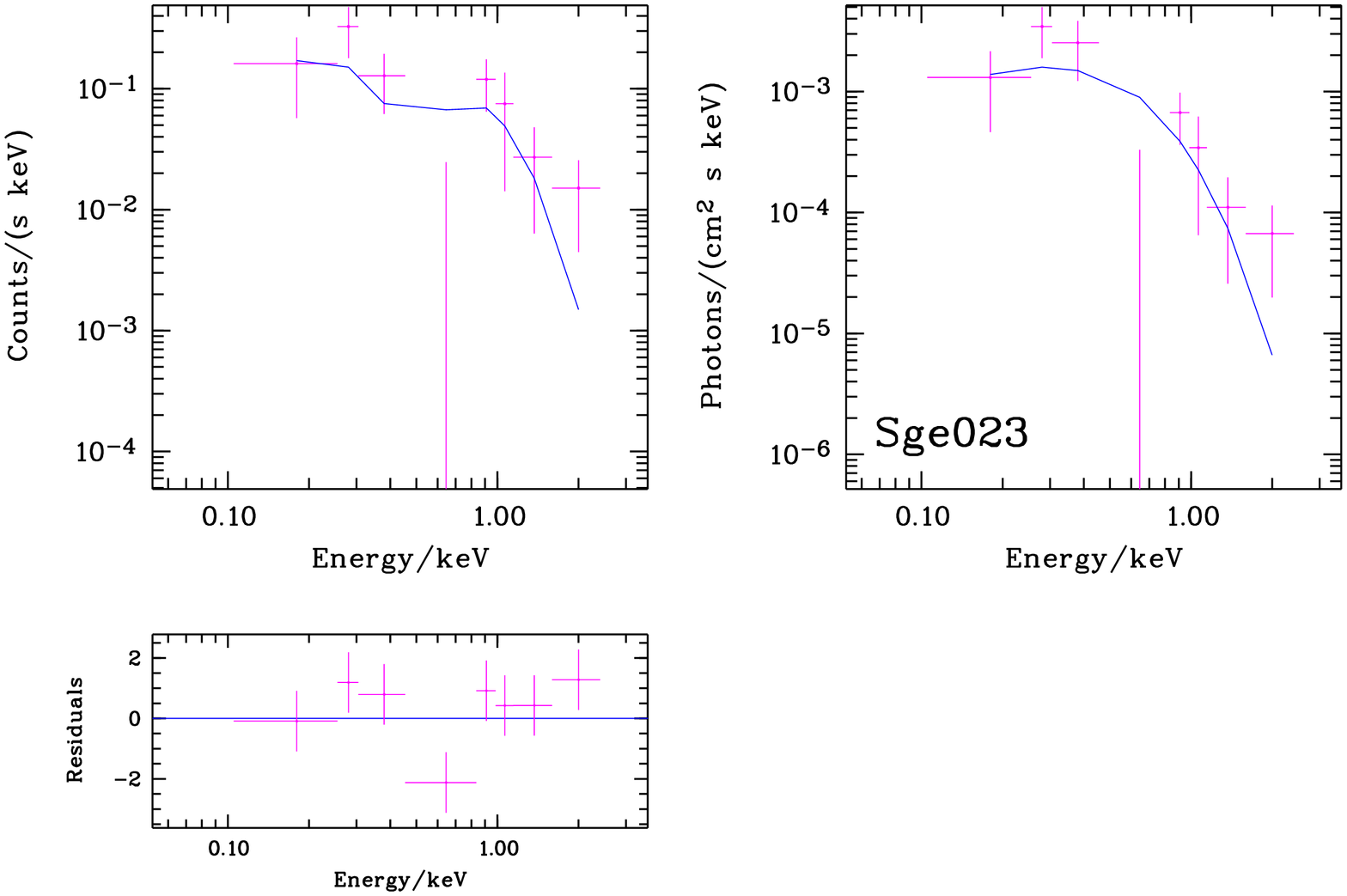}
\includegraphics[width=3.9cm, bb=76 410 385 760, angle=-90,clip]{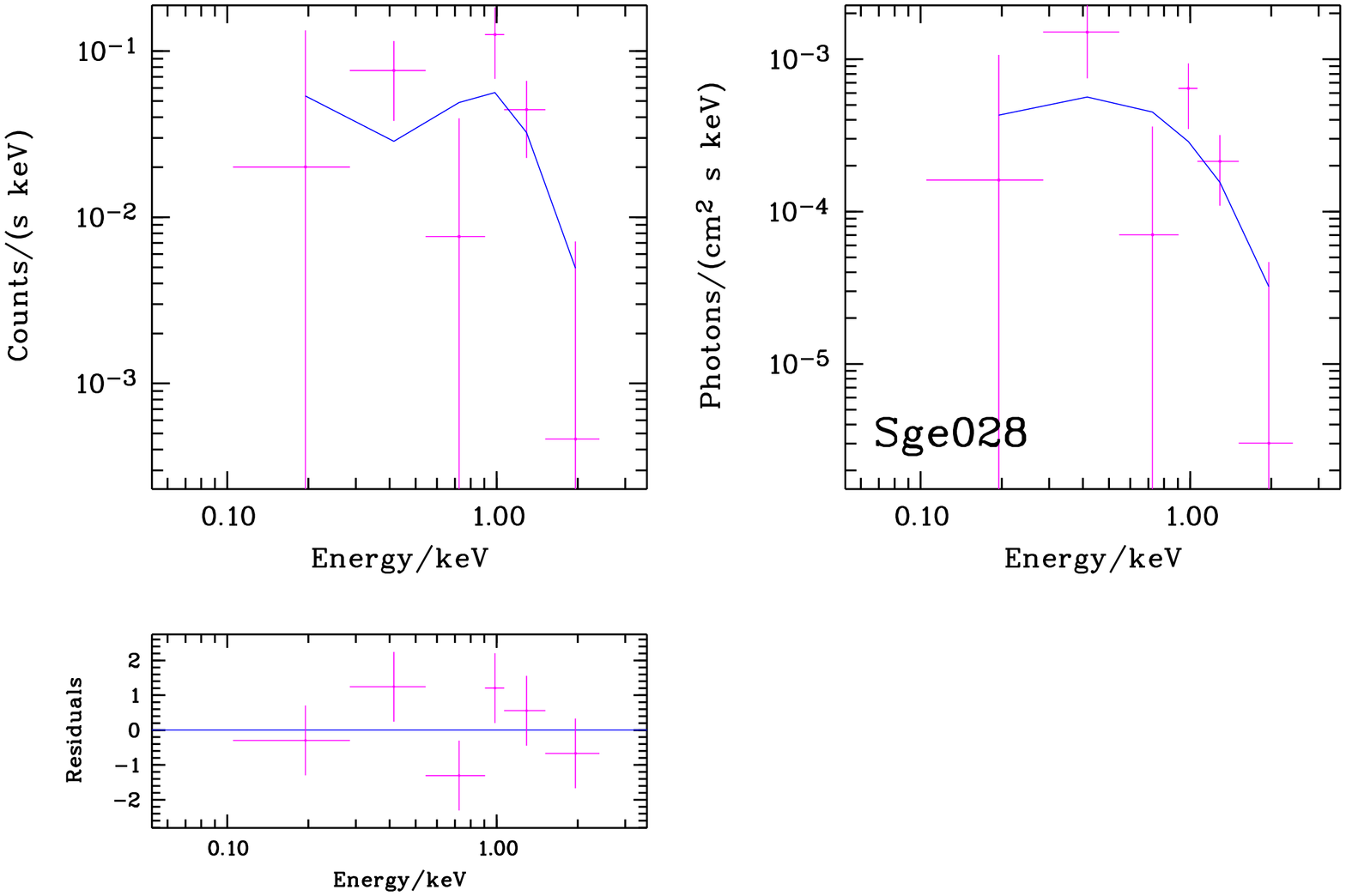}
\includegraphics[width=3.9cm, bb=76 410 385 760, angle=-90,clip]{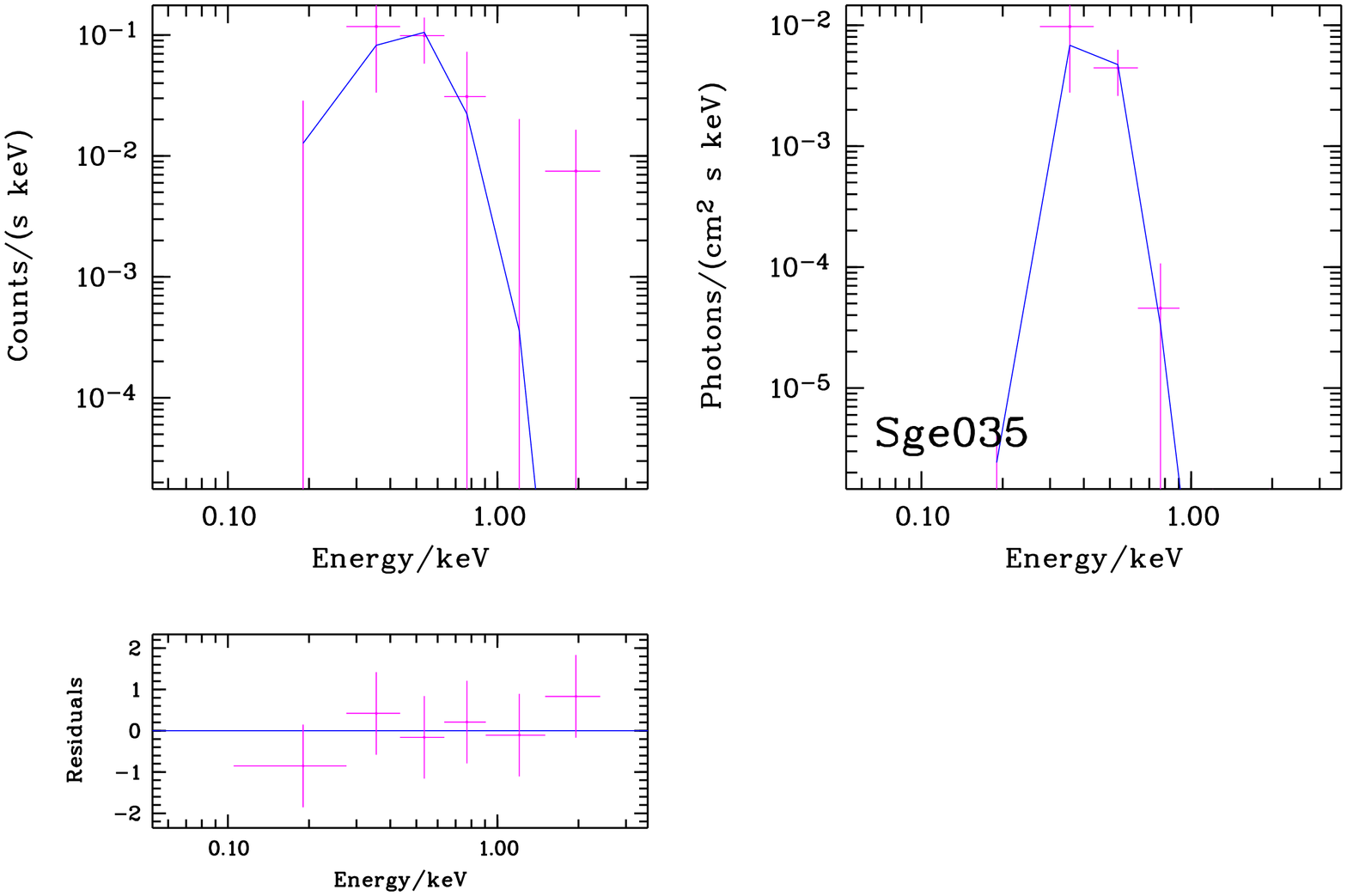}
\includegraphics[width=3.9cm, bb=76 410 385 760, angle=-90,clip]{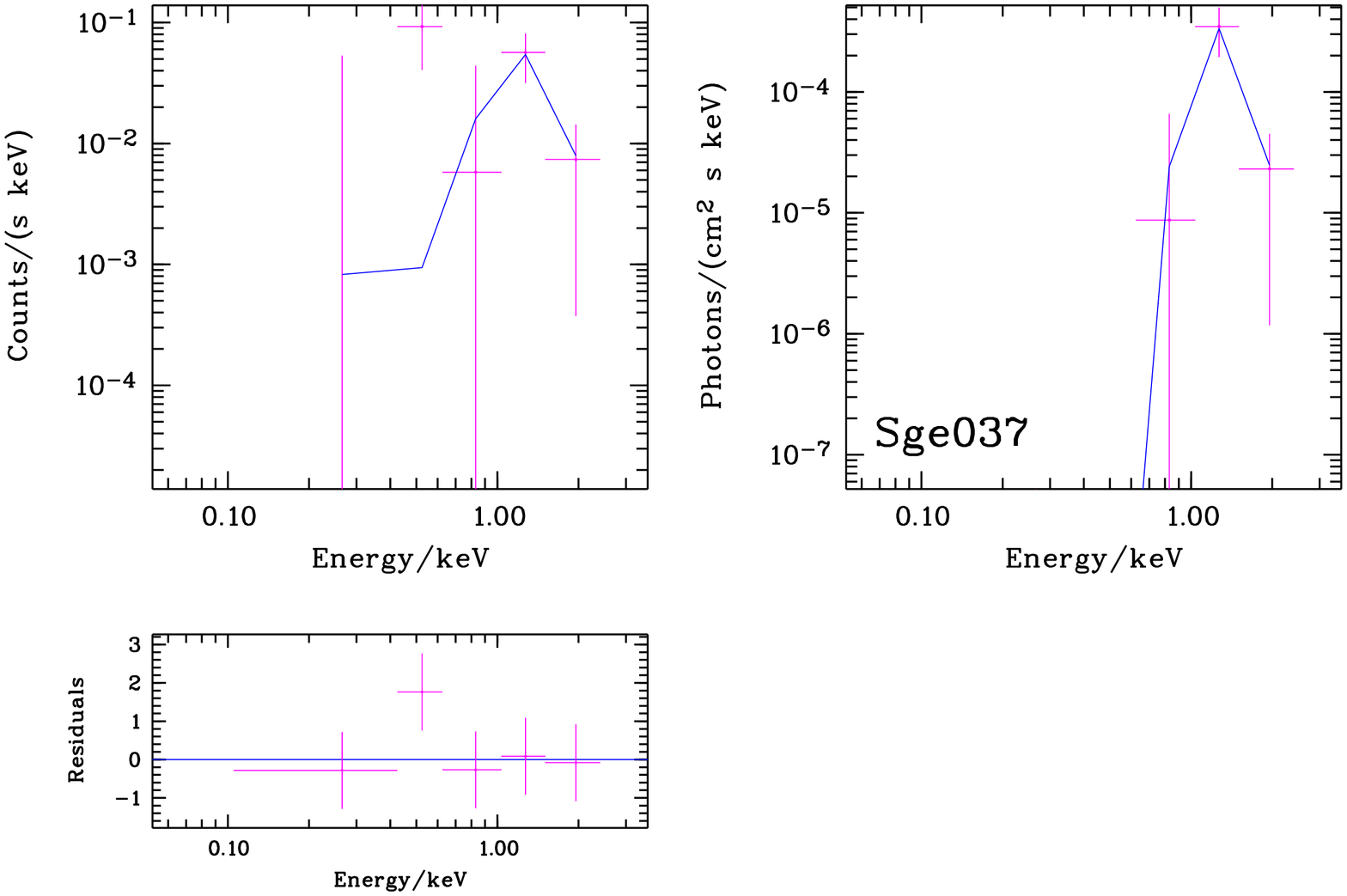}

\includegraphics[width=3.9cm, bb=76 410 385 760, angle=-90,clip]{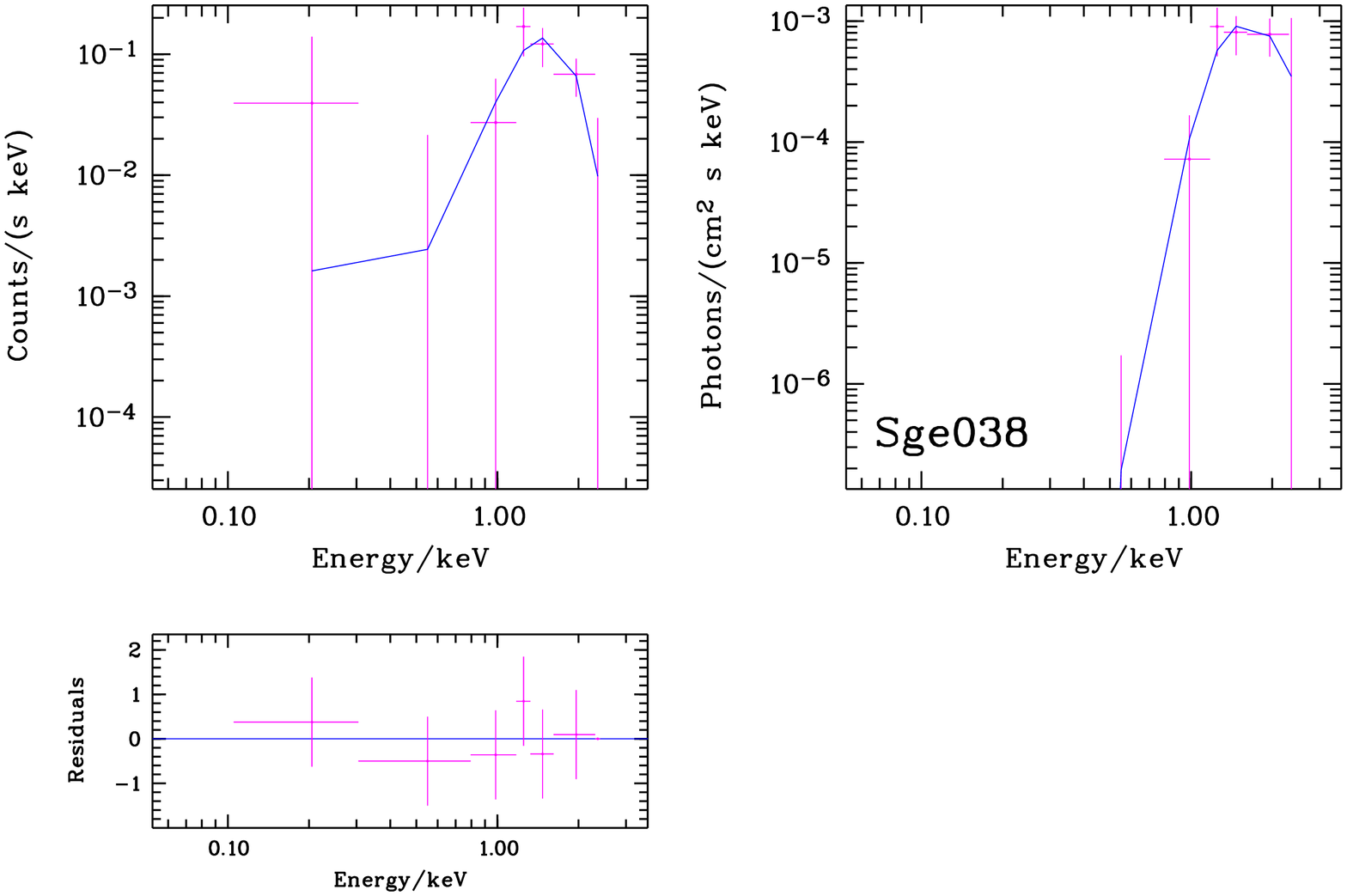}
\includegraphics[width=3.9cm, bb=76 410 385 760, angle=-90,clip]{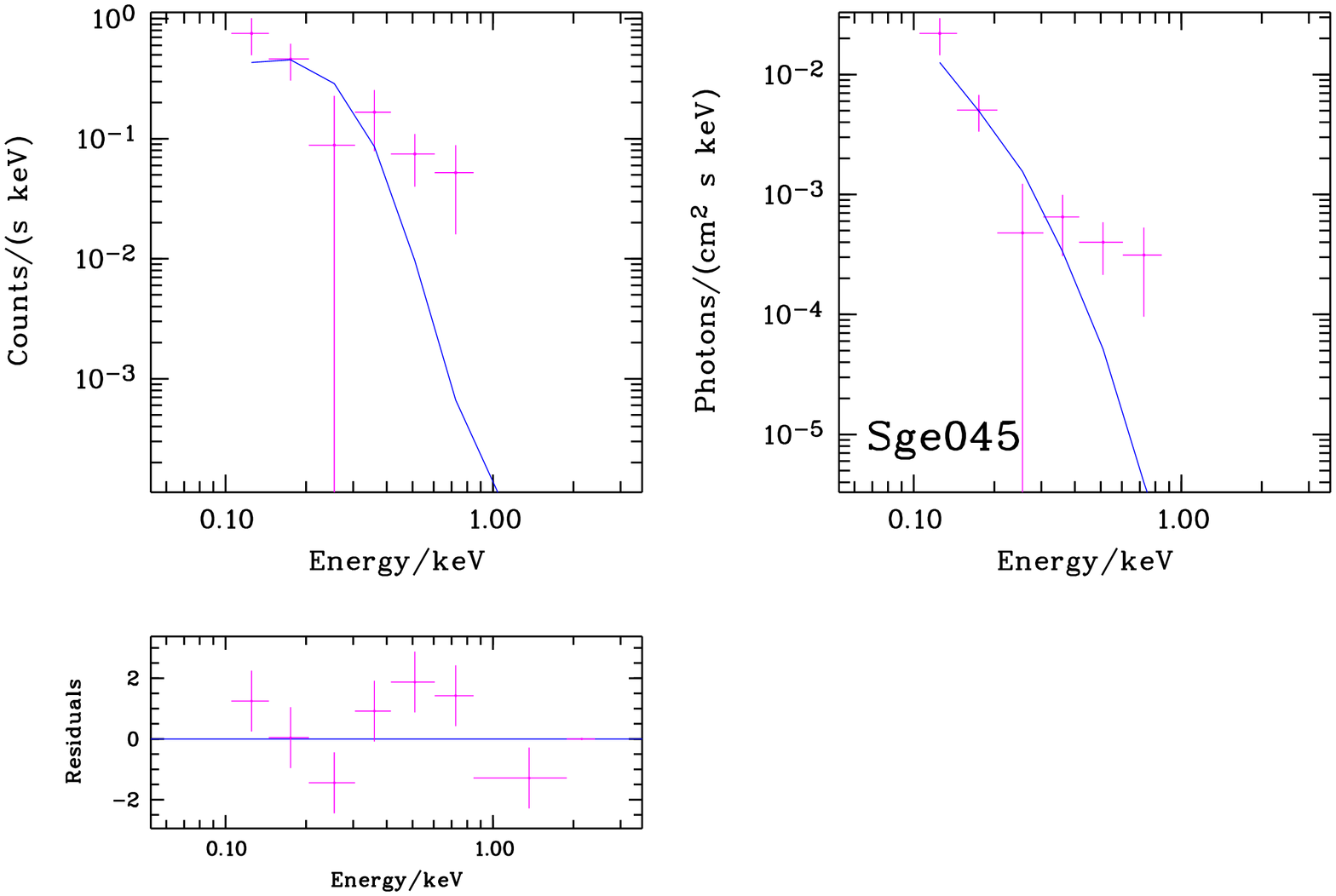}
\includegraphics[width=3.9cm, bb=76 410 385 760, angle=-90,clip]{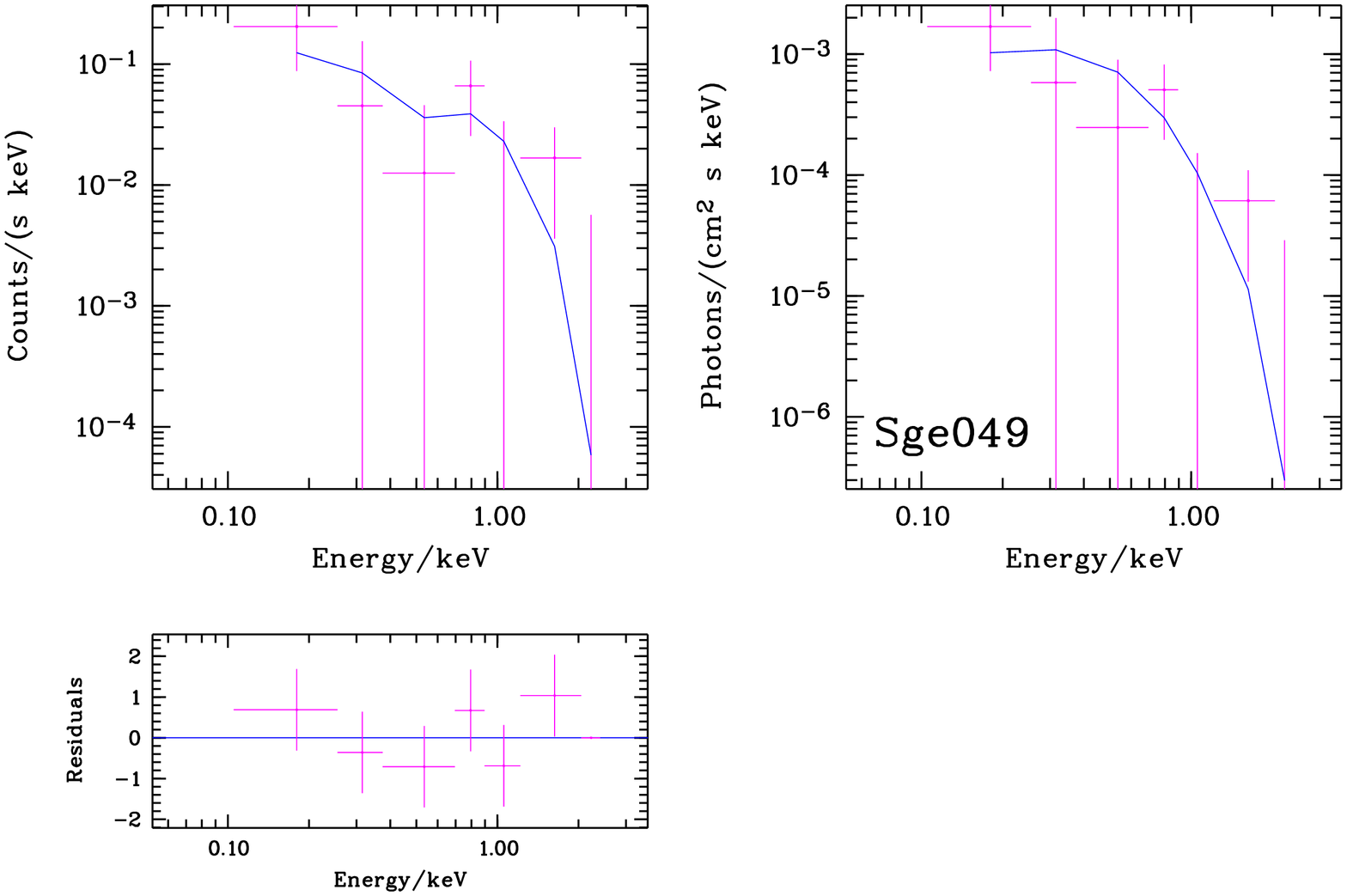}
\includegraphics[width=3.9cm, bb=76 410 385 760, angle=-90,clip]{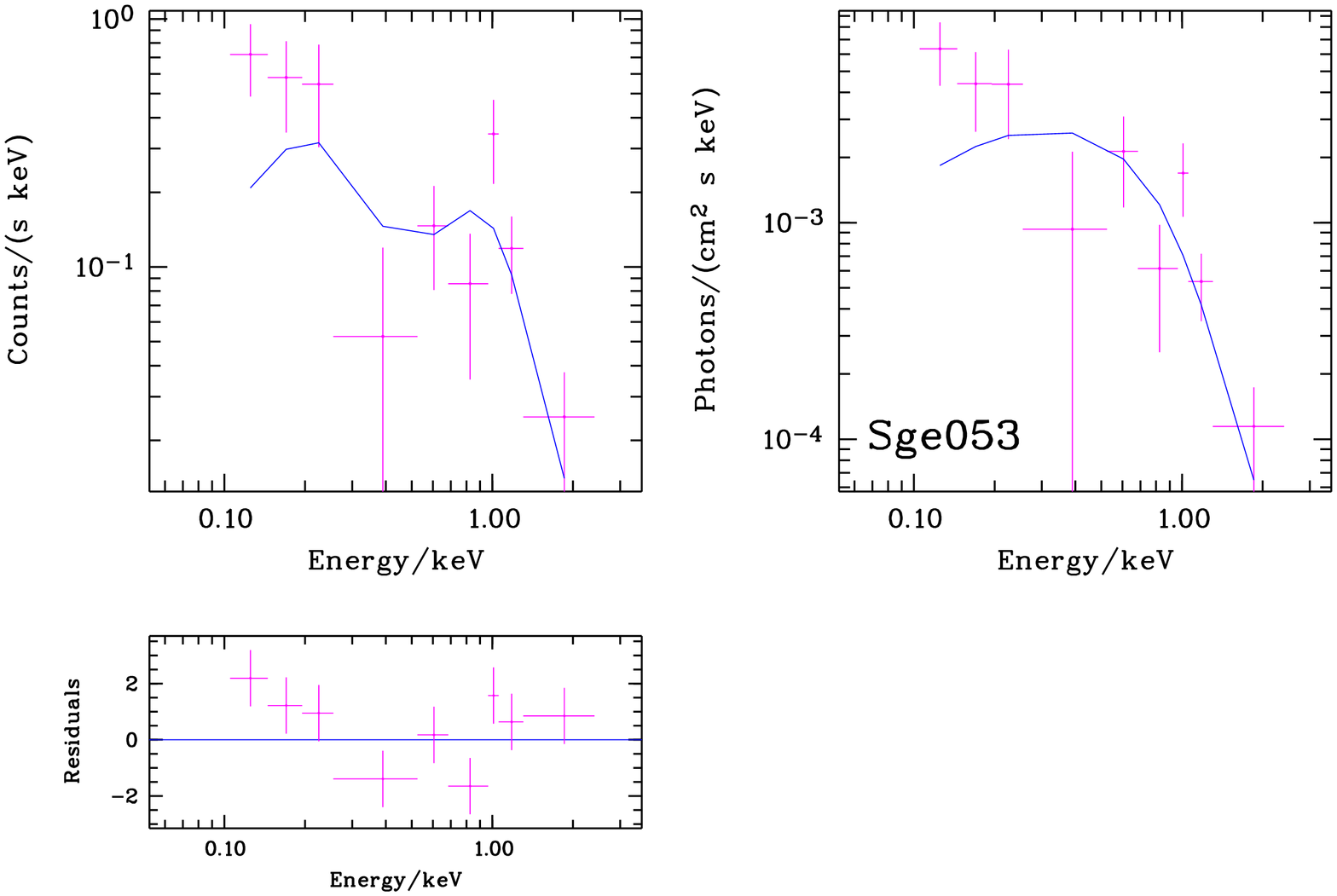}

\includegraphics[width=3.9cm, bb=76 410 385 760, angle=-90,clip]{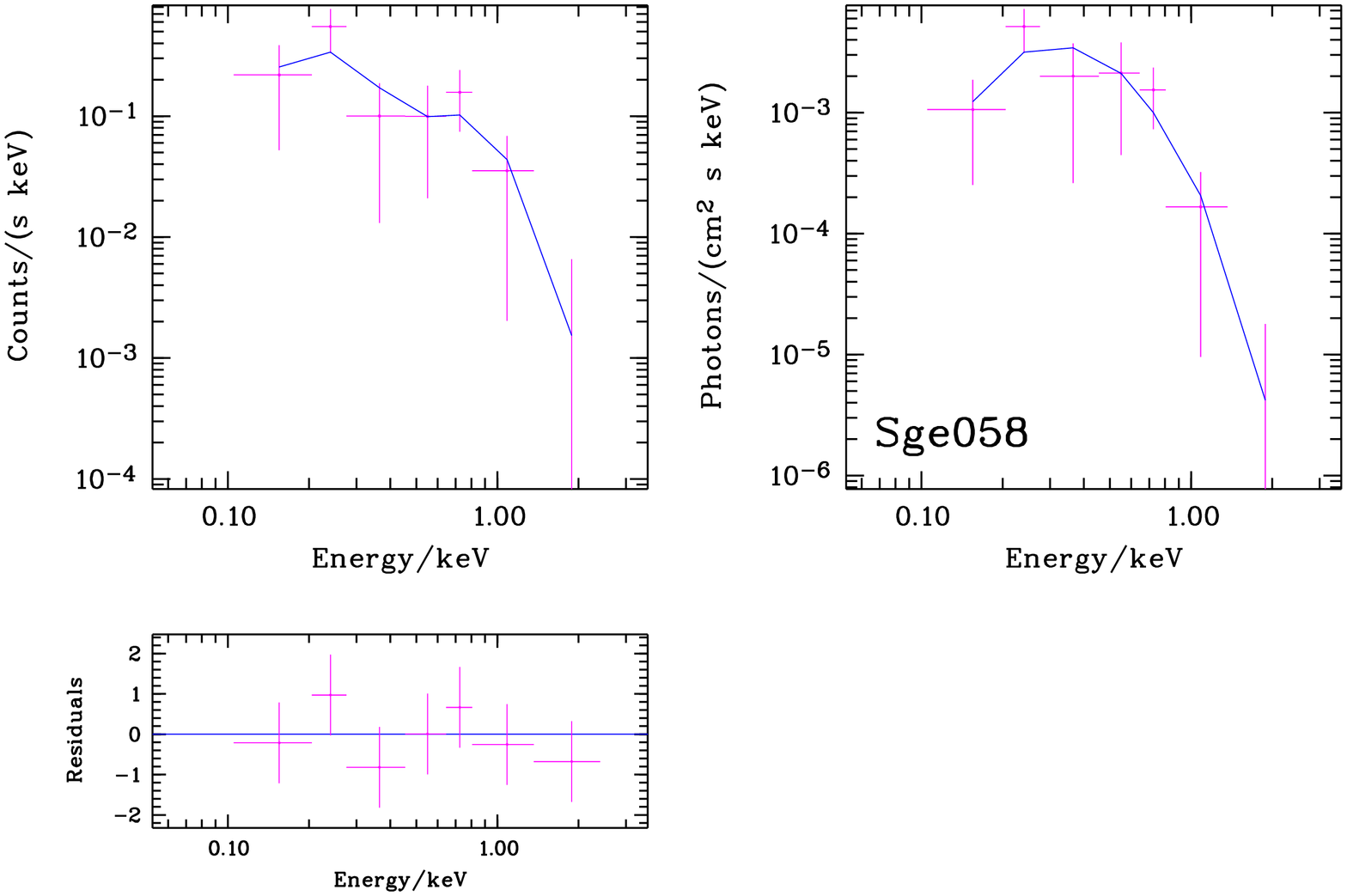}
\includegraphics[width=3.9cm, bb=76 410 385 760, angle=-90,clip]{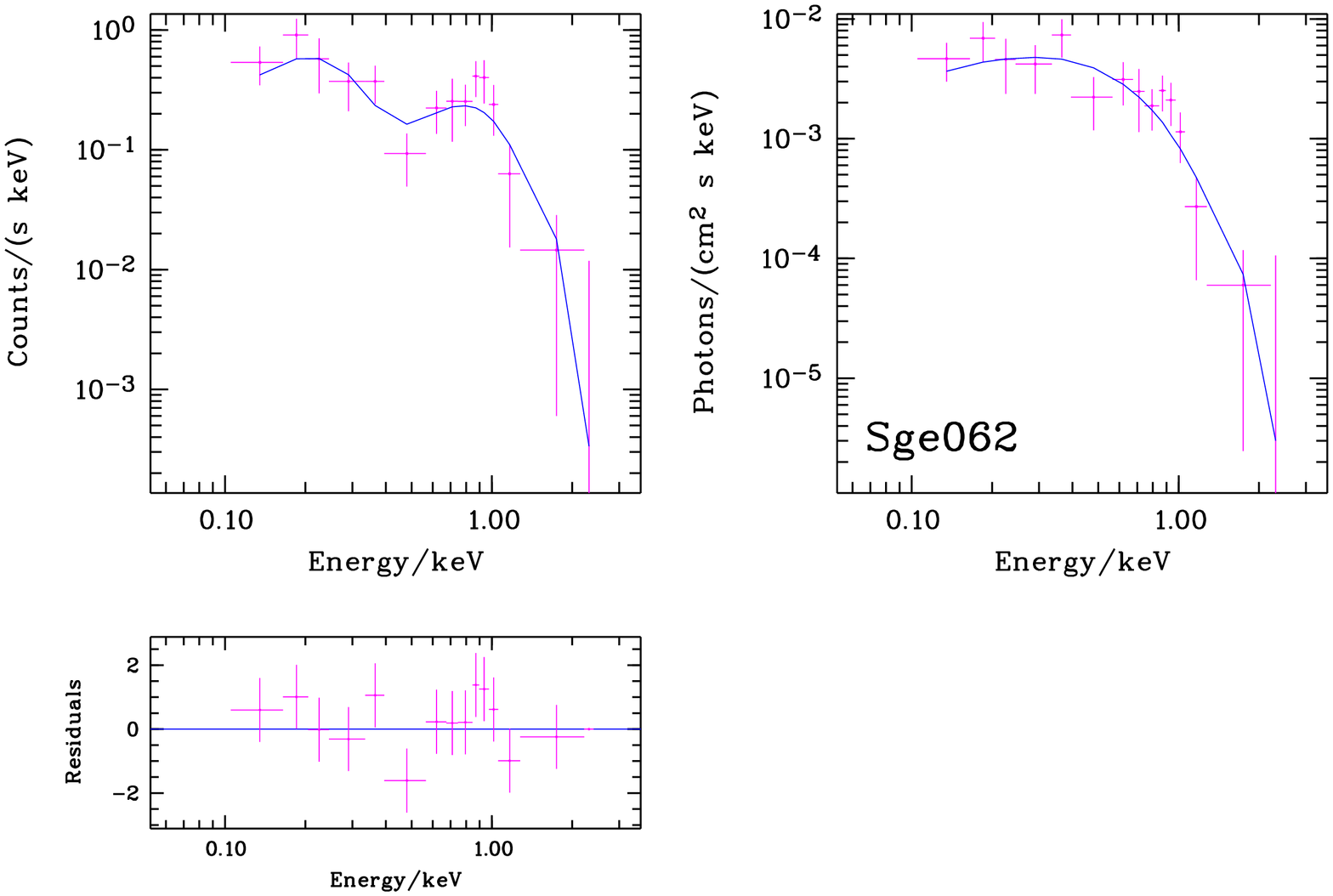}
\includegraphics[width=3.9cm, bb=76 410 385 760, angle=-90,clip]{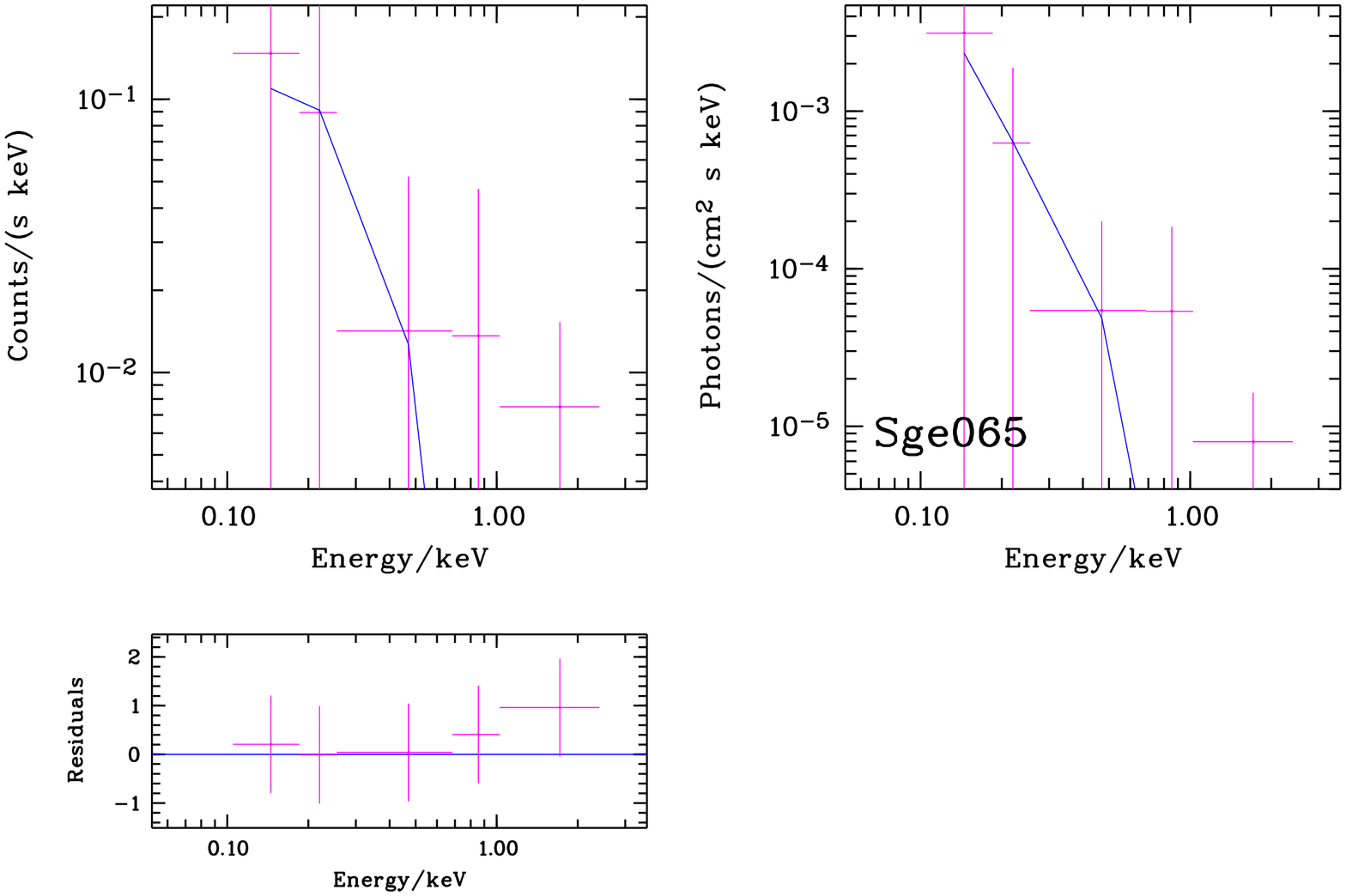}
\includegraphics[width=3.9cm, bb=76 410 385 760, angle=-90,clip]{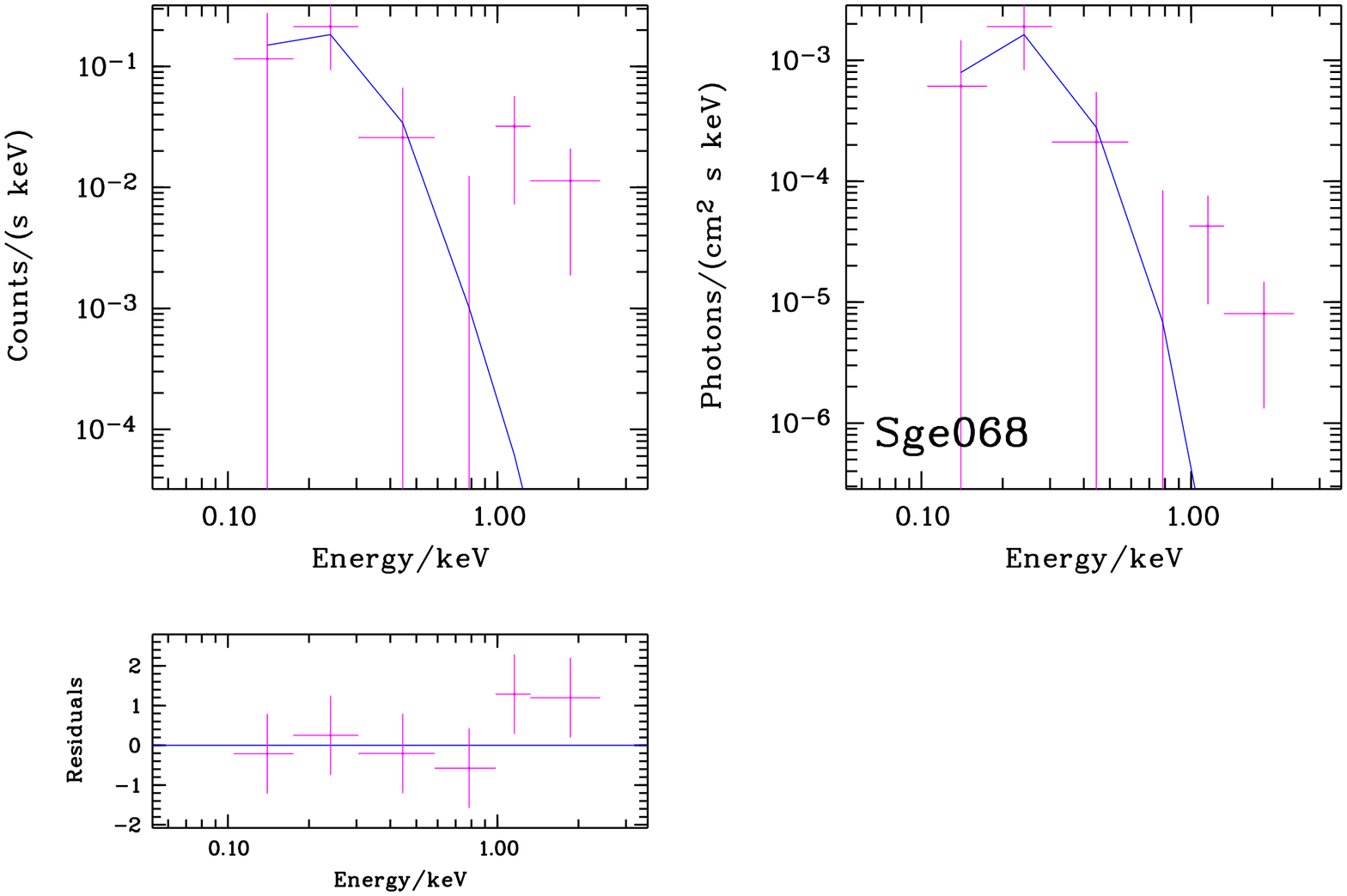}

\includegraphics[width=3.9cm, bb=76 410 385 760, angle=-90,clip]{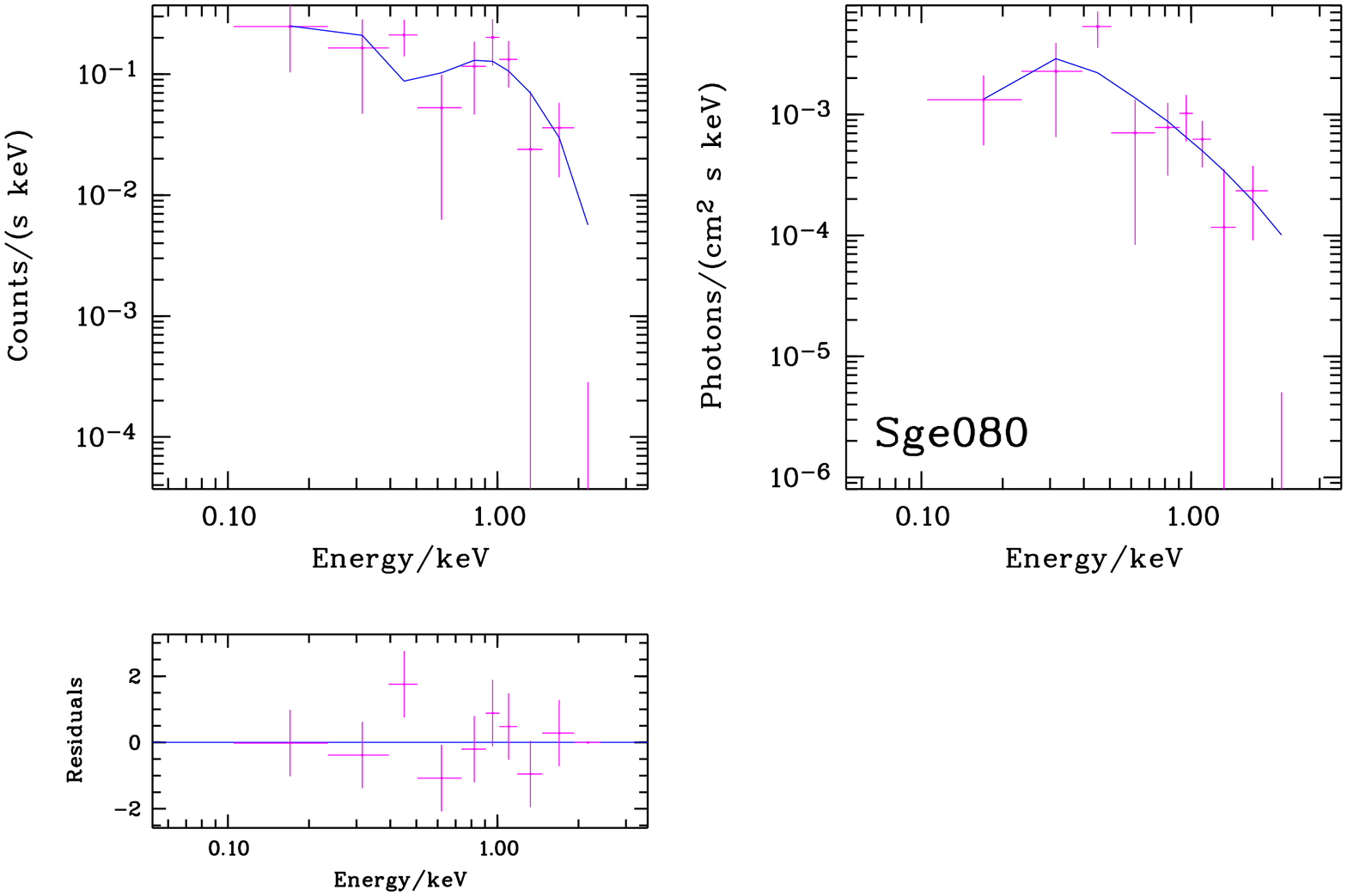}
\includegraphics[width=3.9cm, bb=76 410 385 760, angle=-90,clip]{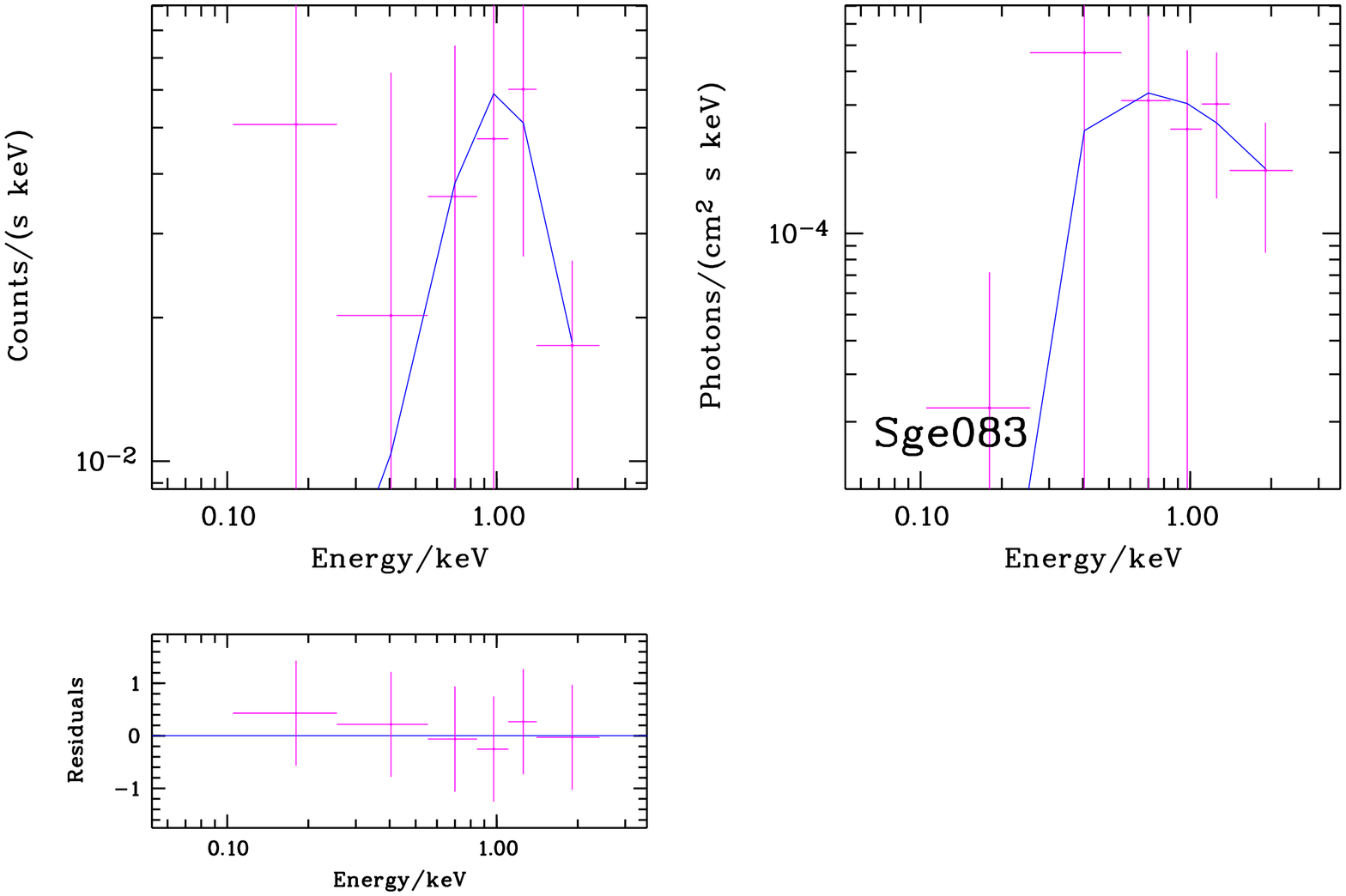}
\includegraphics[width=3.9cm, bb=76 410 385 760, angle=-90,clip]{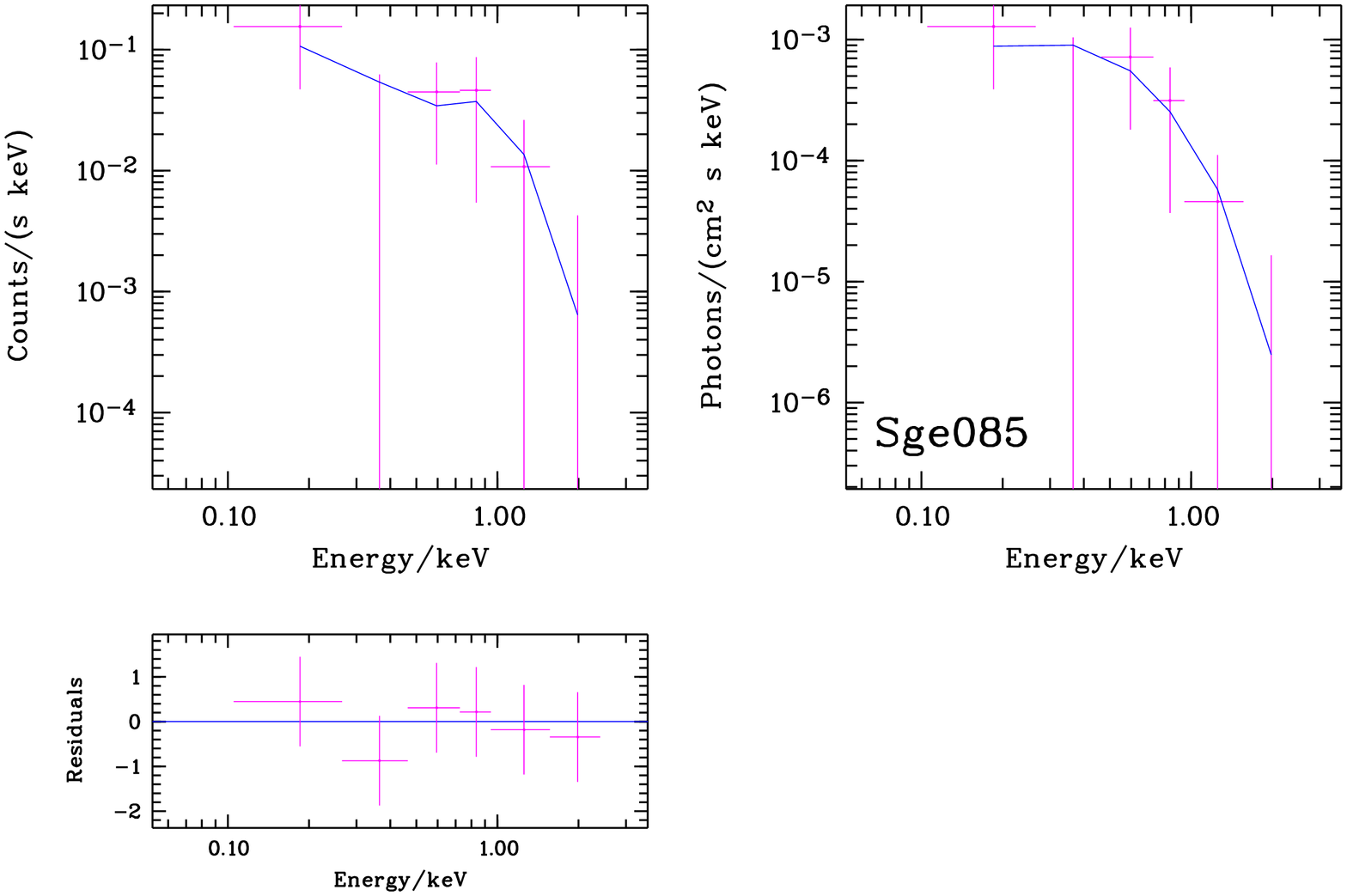}
\includegraphics[width=3.9cm, bb=76 410 385 760, angle=-90,clip]{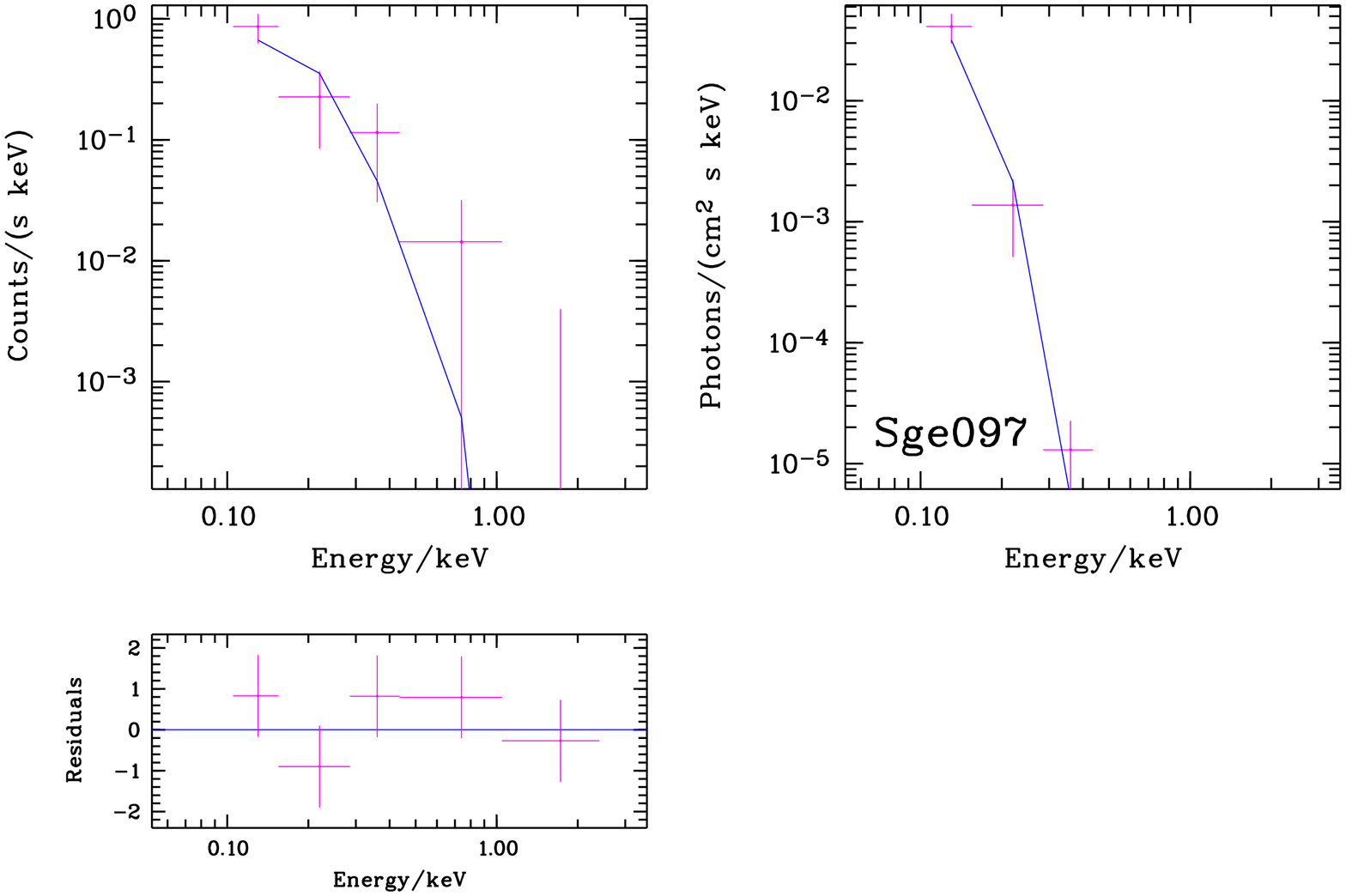}

\caption[fsp]{\label{sgespec} ROSAT survey spectra of Sge sources. 
For each source, the
photon spectrum is shown. The choice of the model for a given source
was made based on the lowest reduced $\chi^2$ and consistency with the
optical properties (see Table  \ref{specparsge} for model 
and spectral parameters).}
\end{figure*}

\setcounter{figure}{1}

\begin{figure*}[ht]

\includegraphics[width=3.9cm, bb=76 410 385 760, angle=-90,clip]{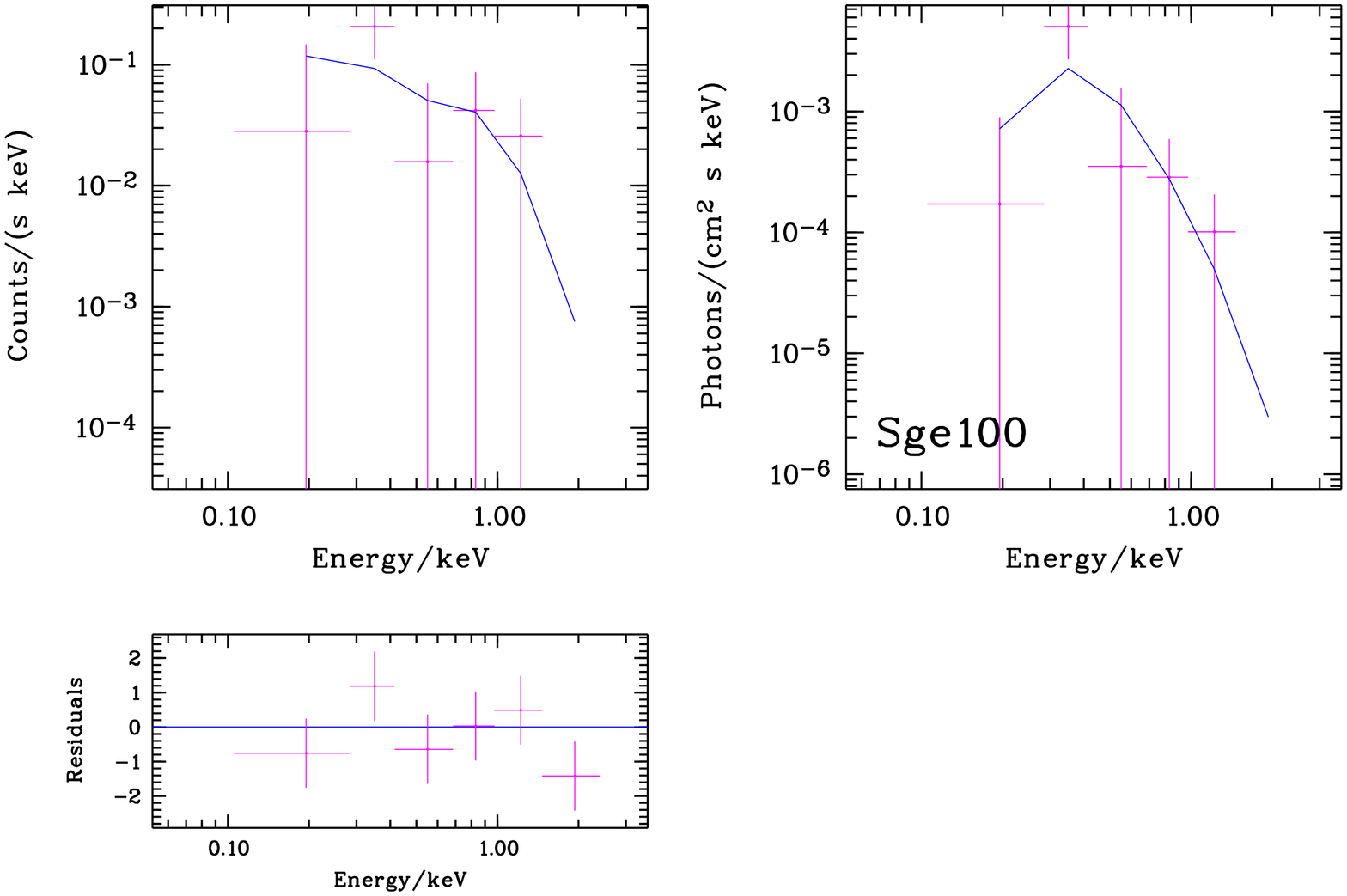}
\includegraphics[width=3.9cm, bb=76 410 385 760, angle=-90,clip]{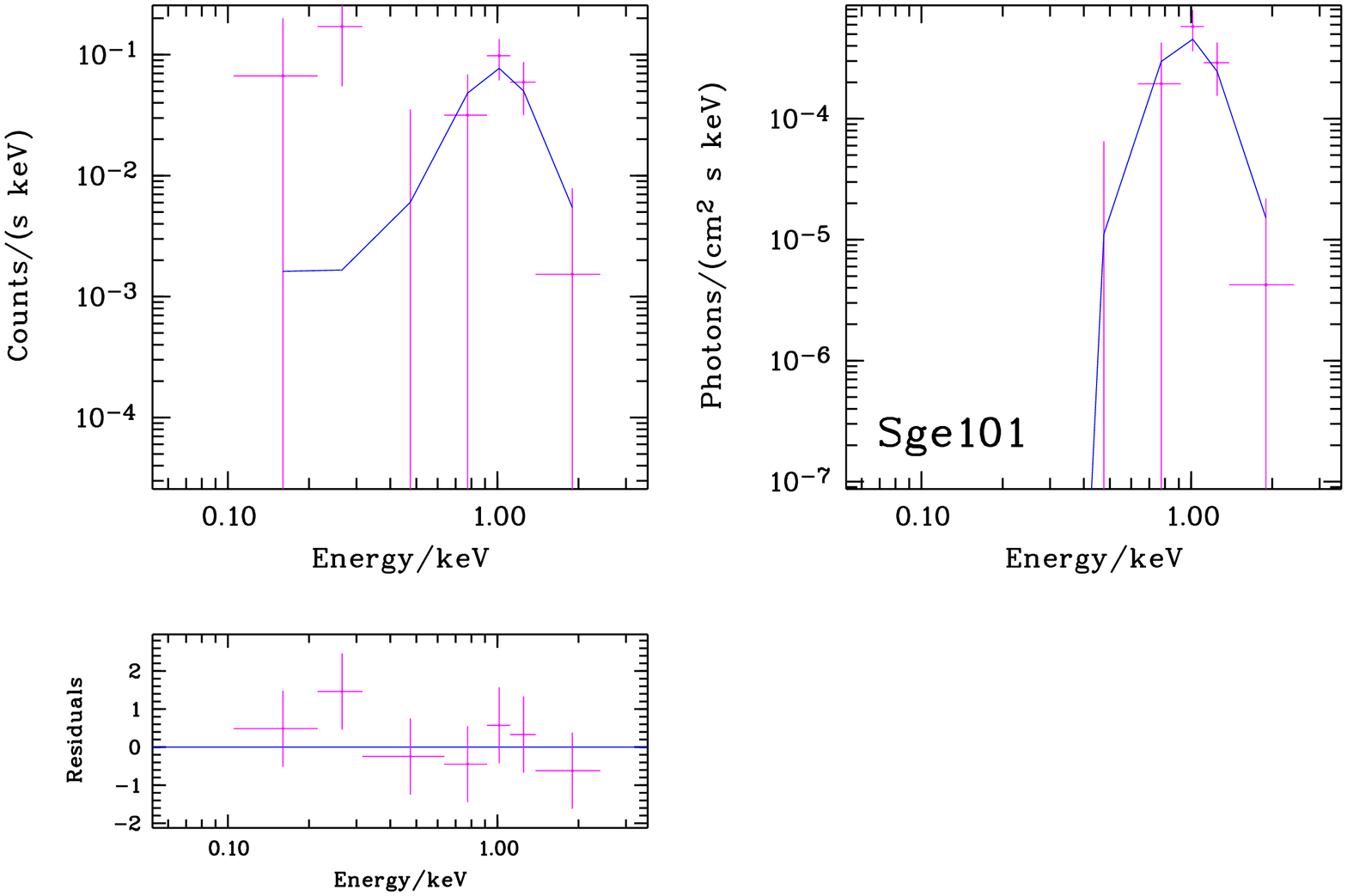}
\includegraphics[width=3.9cm, bb=76 410 385 760, angle=-90,clip]{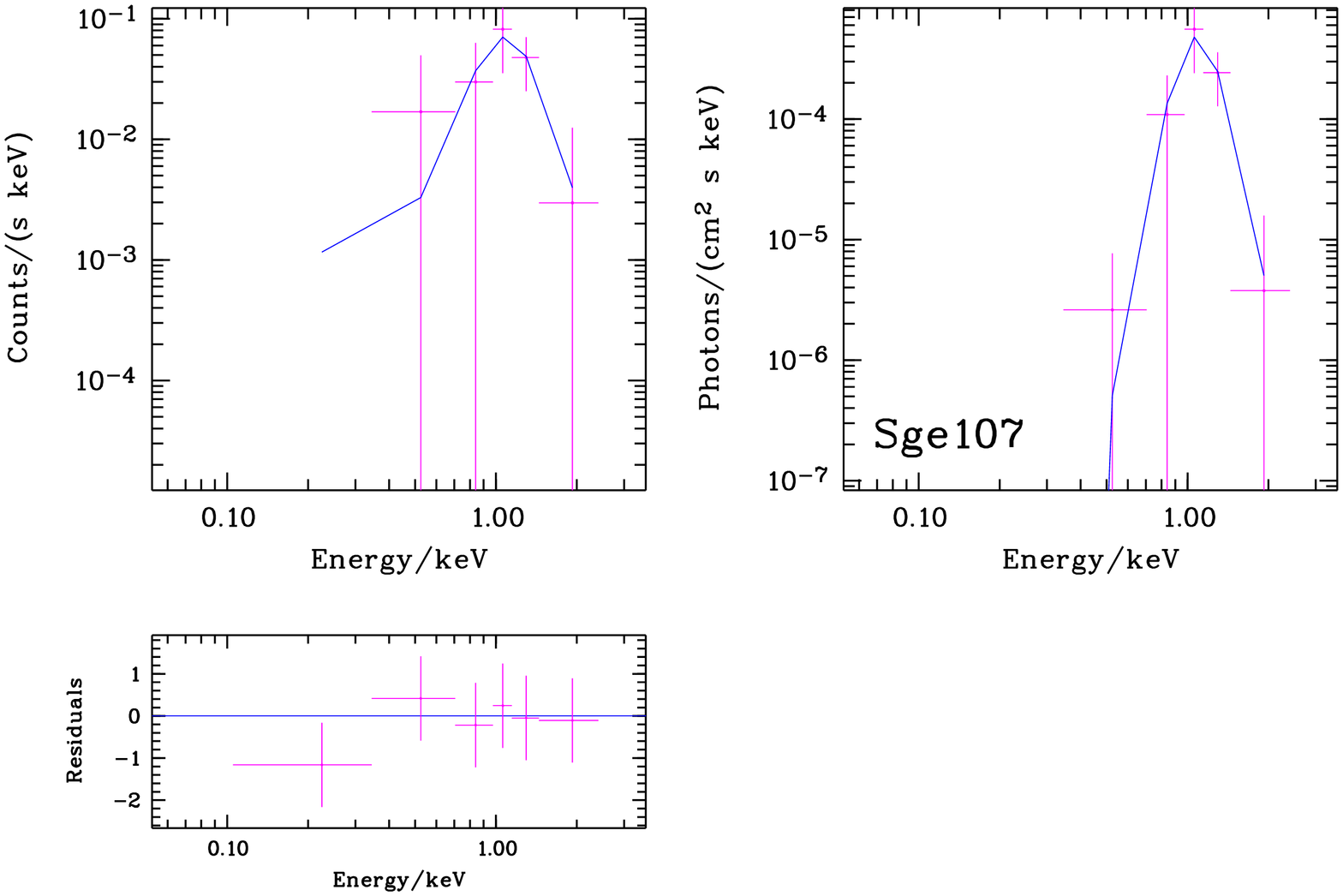}
\includegraphics[width=3.9cm, bb=76 410 385 760, angle=-90,clip]{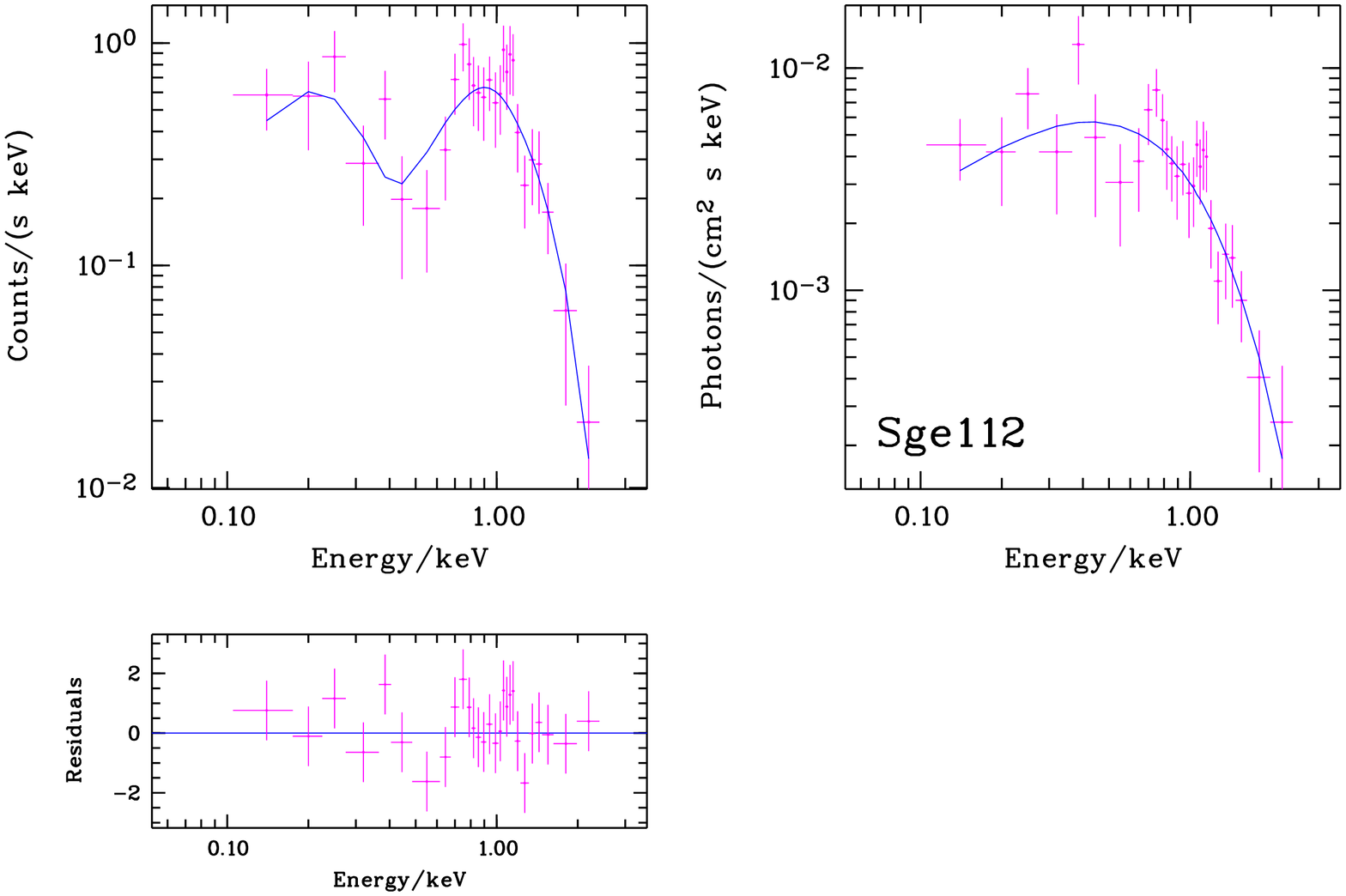}

\includegraphics[width=3.9cm, bb=76 410 385 760, angle=-90,clip]{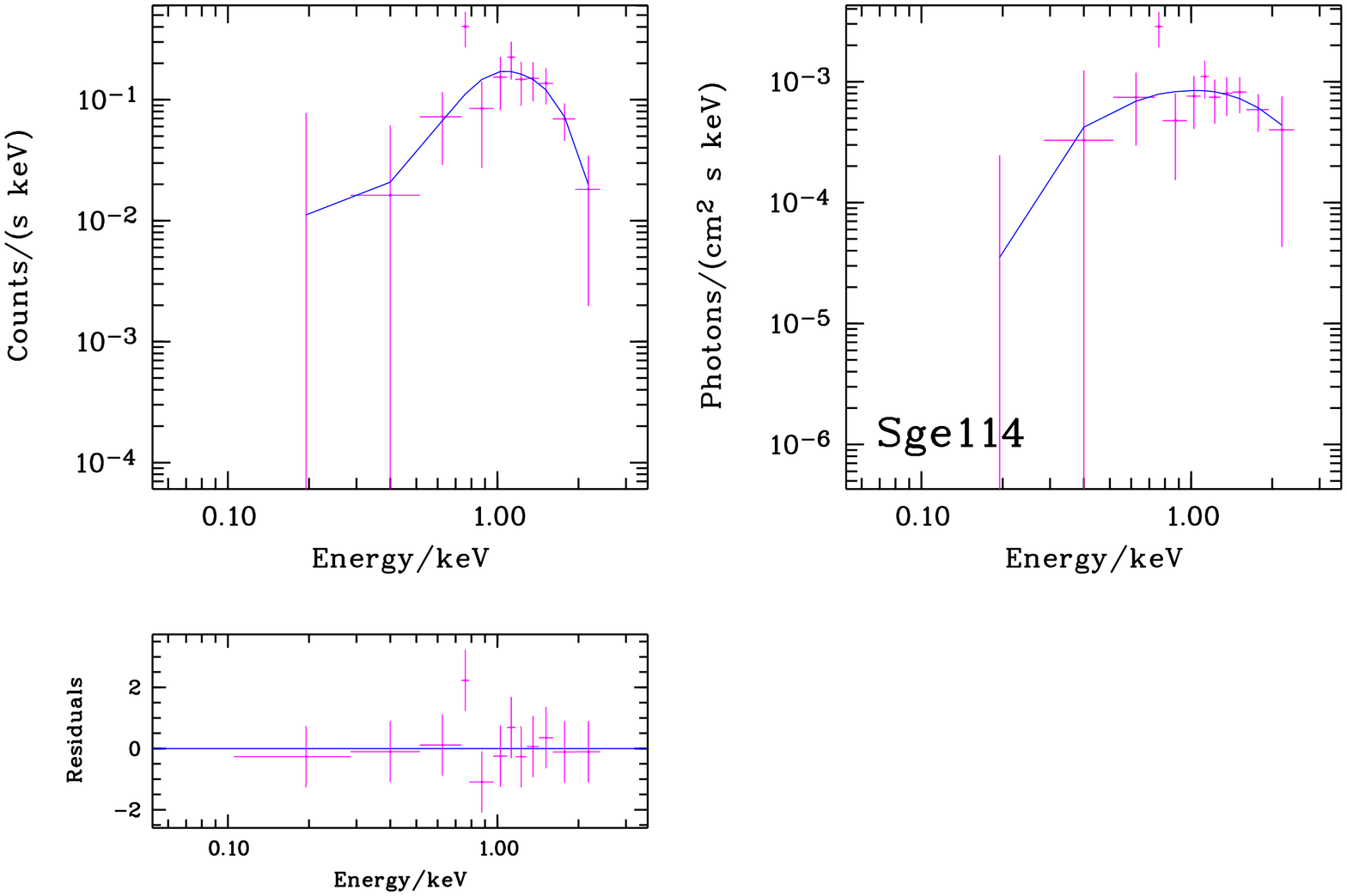}
\includegraphics[width=3.9cm, bb=76 410 385 760, angle=-90,clip]{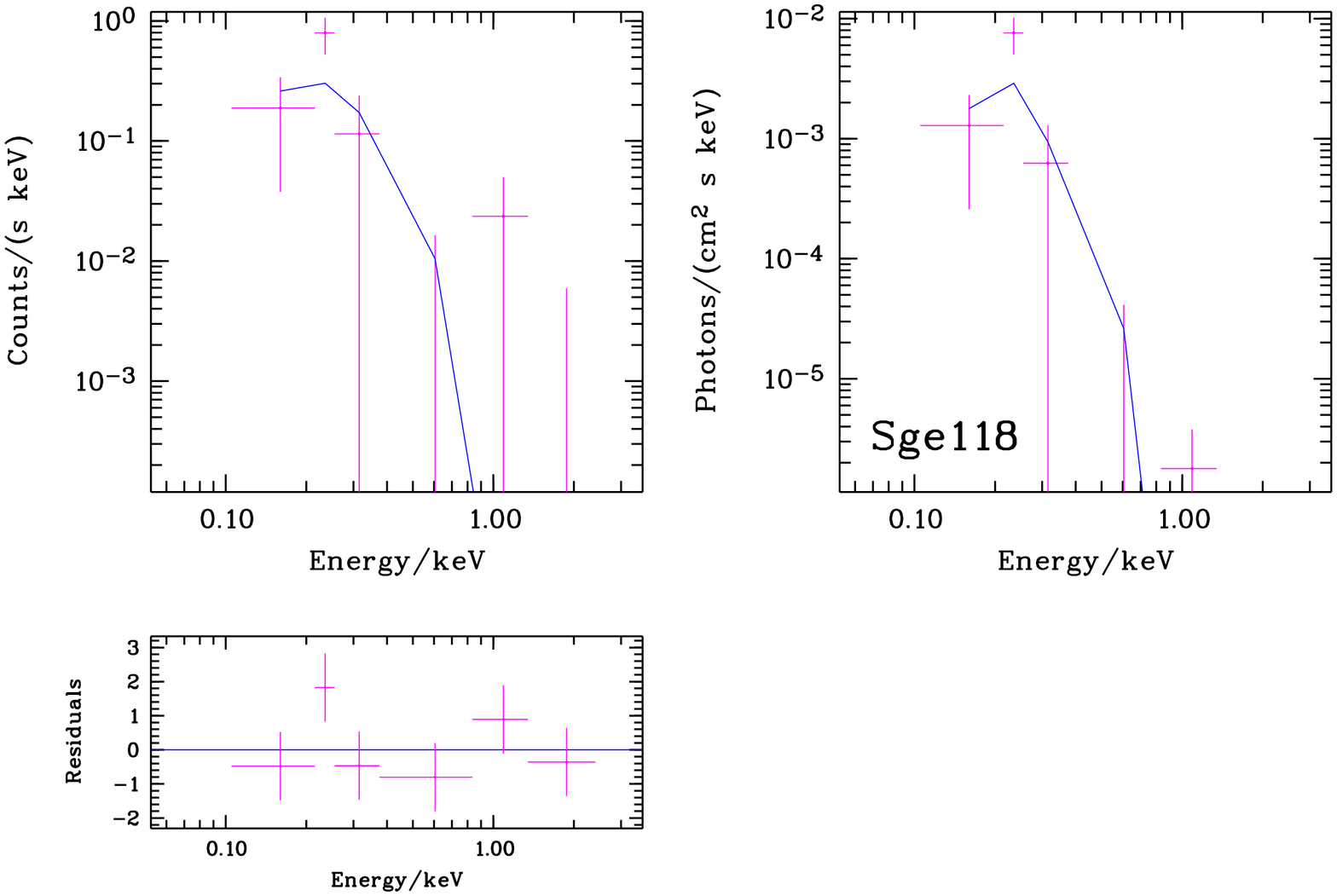}
\includegraphics[width=3.9cm, bb=76 410 385 760, angle=-90,clip]{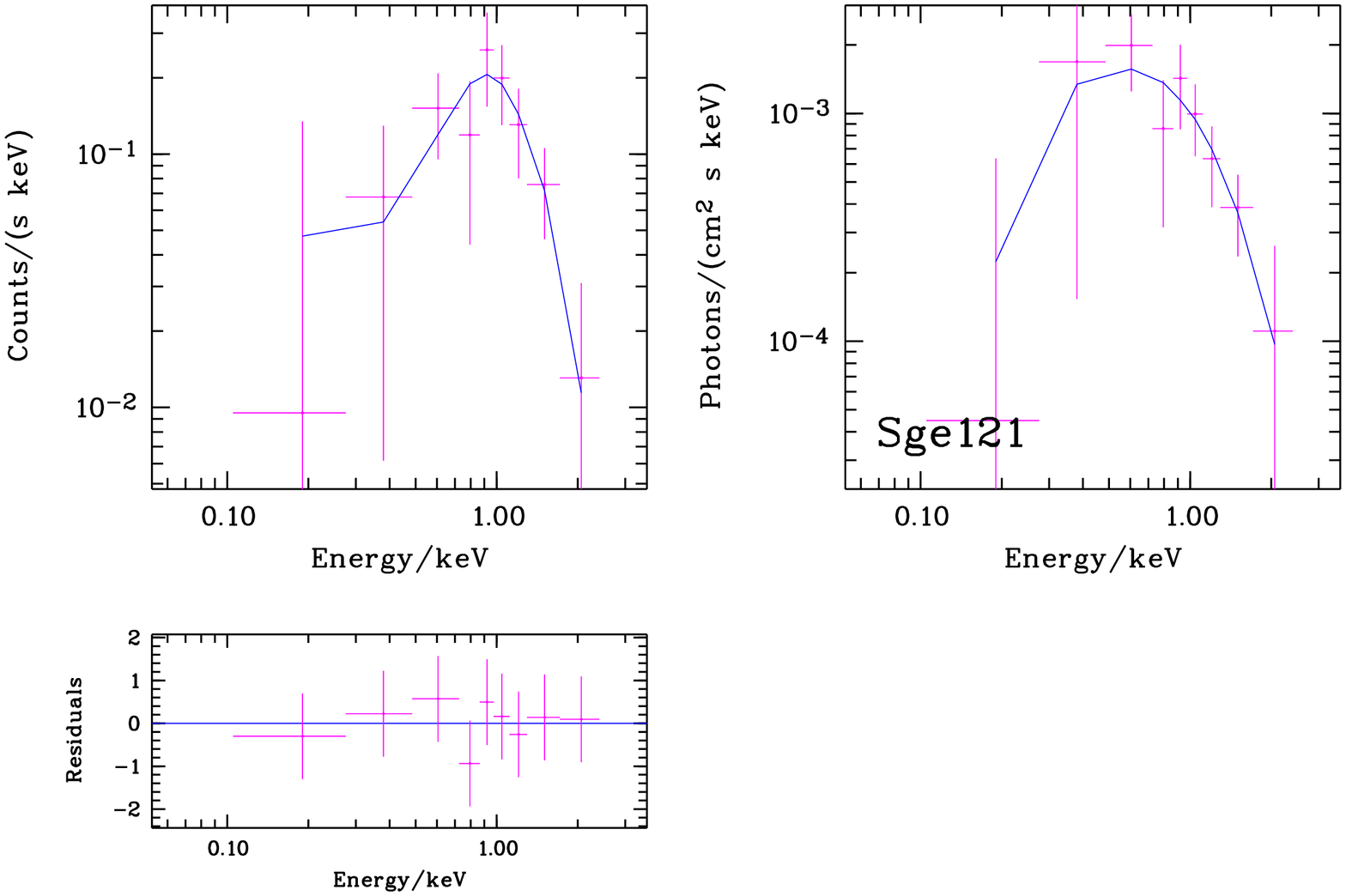}
\includegraphics[width=3.9cm, bb=76 410 385 760, angle=-90,clip]{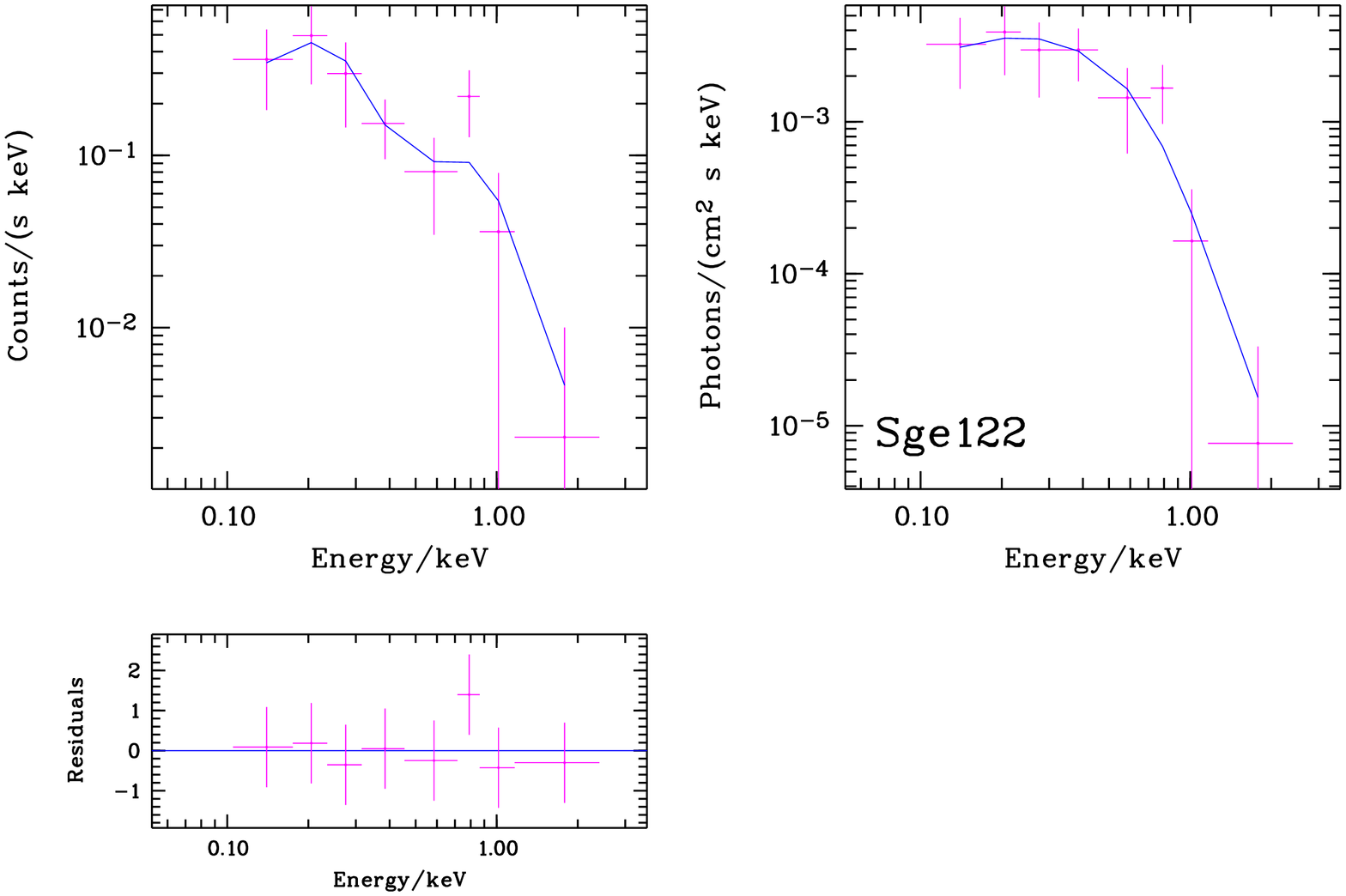}

\includegraphics[width=3.9cm, bb=76 410 385 760, angle=-90,clip]{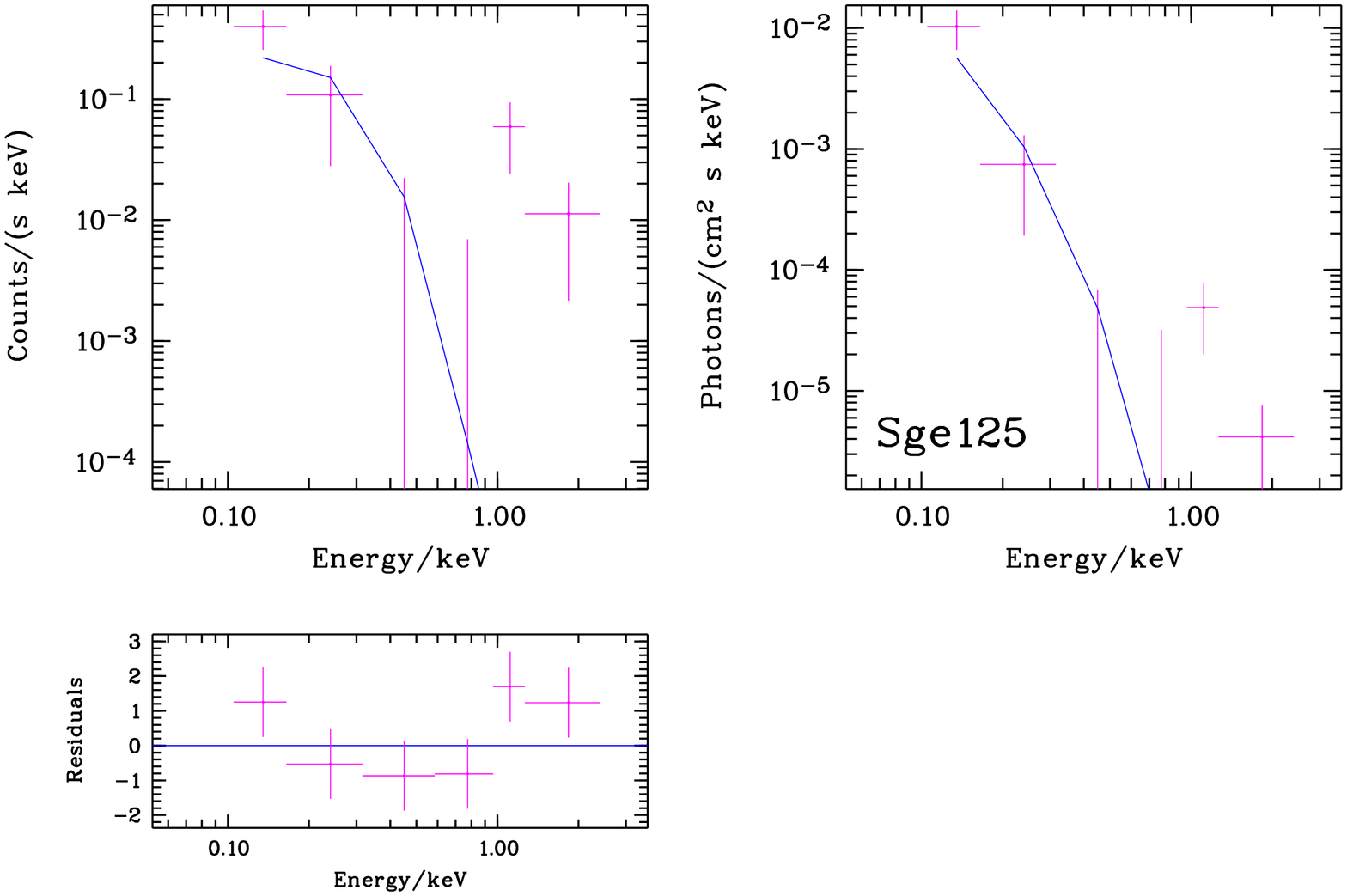}
\includegraphics[width=3.9cm, bb=76 410 385 760, angle=-90,clip]{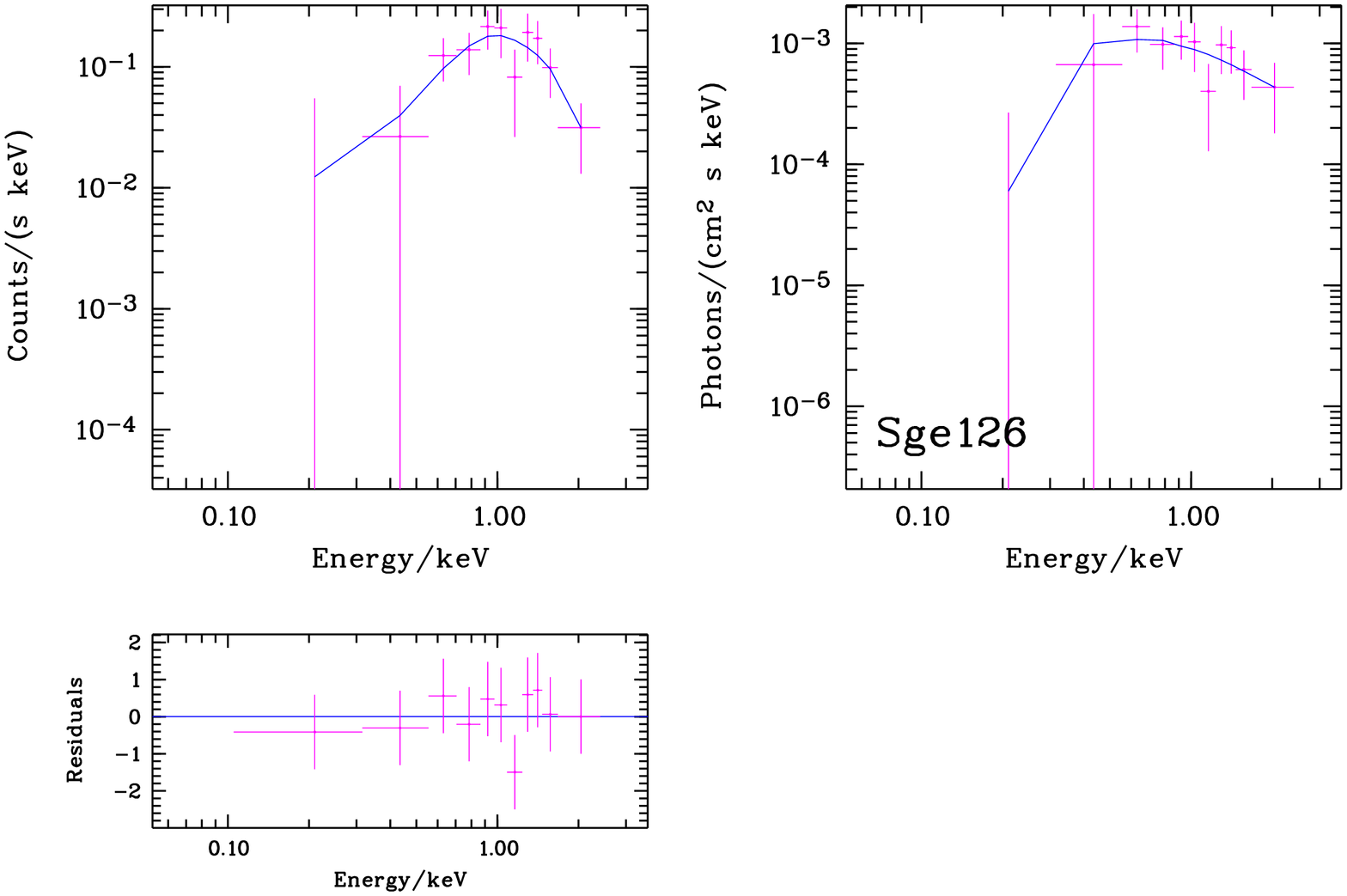}

 \caption[fsp]{{\bf contd.} ROSAT survey spectra of Sge sources.}
 \end{figure*}

\end{subfigures}

%% X-ray fit parameter

\begin{subtables}
\begin{table*}
 \caption{\label{specparcom} Spectral fit parameters for Com sources}
\begin{tabular}{cccccc}
   \hline
     \noalign{\smallskip}
   \noalign{\hspace{5.cm} Power law model}
     \noalign{\smallskip}
  No & $N_{\rm H}$         & Norm  & photon & $\chi_{\rm red}$ & Flux \\
     &(10$^{21}$ cm$^{-2}$) & (ph/cm$^2$/s/keV)  & index  &                 & ph/cm$^2$/s\\
   \noalign{\smallskip}
   \hline
   \noalign{\smallskip}
003  & 1.02E-01$\pm$2.65E-01 & 1.43E-04$\pm$2.05E-04 & -3.00E+00$\pm$1.89E+00 & 1.22 & 6.00E-04\\
007  & 1.00E-03$\pm$4.57E-01 & 4.87E-05$\pm$1.28E-04 & -2.26E+00$\pm$3.75E+00 & 0.04 & -2.91E-04\\
008  & 9.89E-02$\pm$8.95E-01 & 7.42E-05$\pm$2.01E-04 & -2.60E+00$\pm$4.78E+00 & 2.88 & 1.87E-05\\
010  & 1.62E-01$\pm$1.19E+00 & 4.76E-05$\pm$1.79E-04 & -3.24E+00$\pm$5.75E+00 & 2.20 & 8.78E-05\\
022  & 1.15E-01$\pm$2.41E-01 & 6.74E-04$\pm$2.31E-04 & -1.91E+00$\pm$1.04E+00 & 0.49 & 1.55E-03\\
027  & 1.00E-03$\pm$2.08E-01 & 3.51E-04$\pm$1.72E-04 & -1.37E+00$\pm$1.23E+00 & 0.64 & 8.15E-04\\
\hline
  \end{tabular}
  \end{table*}
\setcounter{table}{0}
\begin{table*}
 \caption{{\bf cont.:} Spectral fit parameters for Com sources}
   \begin{tabular}{cccccc}
   \hline
   \noalign{\smallskip}
   \noalign{\hspace{5.cm} Blackbody model}
     \noalign{\smallskip}
   No &  $N_{\rm H}$  & Norm &  kT   & $\chi_{\rm red}$ & Flux \\
      & (10$^{21}$ cm$^{-2}$) & (ph/cm$^2$/s) &  (keV)  &                  &    ph/cm$^2$/s \\
   \noalign{\smallskip}
   \hline
   \noalign{\smallskip}
004  & 5.72E+00$\pm$1.81E+01 & 1.66E-01$\pm$2.37E+00 & 9.99E-02$\pm$4.70E-01 & 0.88 & 7.00E-04\\
043  & 1.00E-03$\pm$1.60E-01 & 2.81E-03$\pm$2.42E-03 & 1.53E-01$\pm$2.56E-02 & 0.75 & 1.30E-03\\
062  & 1.00E-03$\pm$2.12E-01 & 2.16E-03$\pm$2.56E-03 & 1.48E-01$\pm$3.35E-02 & 0.79 & 8.89E-04\\
067  & 1.73E-01$\pm$3.22E+00 & 2.63E+00$\pm$3.04E+02 & 1.59E-02$\pm$4.95E-01 & 3.16 & 2.89E-04\\
113  & 1.00E-03$\pm$6.06E-02 & 2.34E+00$\pm$8.49E+00 & 1.78E-02$\pm$4.67E-03 & 2.66 & 1.75E-02\\
158  & 5.55E-02$\pm$9.94E-01 & 1.47E-03$\pm$1.04E-02 & 6.76E-02$\pm$4.61E-01 & 1.35 & 1.52E-04\\
160  & 2.84E+01$\pm$1.46E+02 & 5.26E+03$\pm$4.52E+05 & 6.04E-02$\pm$1.99E+00 & 0.53 & 3.47E-04\\
170  & 1.43E+01$\pm$1.28E+02 & 2.26E+04$\pm$2.91E+06 & 4.19E-02$\pm$2.49E+00 & 0.51 & 5.68E-04\\
192  & 1.00E-03$\pm$2.60E-01 & 1.41E-03$\pm$1.27E-03 & 2.18E-01$\pm$7.17E-02 & 1.19 & 8.81E-04\\
195  & 5.57E-02$\pm$1.56E-01 & 5.70E-03$\pm$4.36E-03 & 1.42E-01$\pm$1.78E-02 & 1.69 & 2.09E-03\\
221  & 8.70E-01$\pm$3.59E+00 & 1.44E-01$\pm$2.44E+00 & 4.99E-02$\pm$5.25E-01 & 0.44 & 6.85E-04\\
237  & 1.66E+00$\pm$4.25E+02 & 1.53E-03$\pm$7.74E-01 & 1.00E-01$\pm$9.56E+00 & 0.43 & 8.60E-05\\
238  & 1.00E-03$\pm$1.59E+02 & 1.96E-03$\pm$1.15E+01 & 2.67E-02$\pm$2.10E+01 & 0.84 & 8.49E-05\\
  \noalign{\smallskip}
   \hline
   \noalign{\smallskip}
   \hline
   \end{tabular}
\end{table*}

\begin{table*}
 \caption{Spectral fit parameters for Sge sources}
   \begin{tabular}{cccccc}
   \hline
     \noalign{\smallskip}
   \noalign{\hspace{5.cm} Bremsstrahlung model}
     \noalign{\smallskip}
   No &  $N_{\rm H}$   & Norm &  kT   & $\chi_{\rm red}$ & Flux \\
      & (10$^{21}$ cm$^{-2}$) & (ph/cm$^2$/s/keV) &  (keV)  &                  & ph/cm$^2$/s \\
   \noalign{\smallskip}
   \hline
   \noalign{\smallskip}
003  & 6.63E-01$\pm$1.16E+01 & 1.95E-06$\pm$1.04E-04 & 1.00E-01$\pm$2.04E+00 & 2.72 & 2.07E-04\\
013  & 1.98E+01$\pm$1.55E+02 & 4.80E-02$\pm$3.13E-01 & 1.00E-01$\pm$3.96E+00 & 0.59 & 2.08E-04\\
016  & 1.00E-03$\pm$5.43E-02 & 6.30E-04$\pm$3.94E-04 & 1.21E+00$\pm$7.22E-01 & 0.98 & 1.56E-03\\
020  & 1.00E-03$\pm$1.29E-01 & 1.79E-04$\pm$2.17E-04 & 6.79E-01$\pm$7.71E-01 & 0.47 & 5.82E-04\\
045  & 1.00E-03$\pm$5.91E-01 & 1.16E-07$\pm$1.02E-05 & 9.84E-02$\pm$2.84E-01 & 2.92 & 2.39E-04\\
065  & 1.00E-03$\pm$3.84E+00 & 2.47E-08$\pm$1.38E-05 & 9.68E-02$\pm$1.31E+00 & 0.57 & 9.15E-05\\
068  & 1.47E-01$\pm$2.09E+00 & 2.79E-07$\pm$2.71E-05 & 9.98E-02$\pm$1.24E+00 & 1.19 & 9.37E-05\\
080  & 1.52E-01$\pm$1.89E-01 & 6.23E-04$\pm$3.60E-04 & 1.46E+00$\pm$3.86E+00 & 1.07 & 7.48E-04\\
083  & 1.07E+00$\pm$3.62E+01 & 3.89E-04$\pm$1.63E-03 & 1.00E+02$\pm$1.72E+04 & 0.12 & 6.23E-04\\
100  & 4.08E-01$\pm$1.61E+00 & 1.24E-04$\pm$3.13E-04 & 2.71E-01$\pm$5.16E-01 & 1.55 & 2.48E-04\\
125  & 1.00E-03$\pm$1.90E+00 & 1.11E-08$\pm$3.72E-06 & 8.26E-02$\pm$4.33E-01 & 2.56 & 8.44E-05\\
126  & 9.74E-01$\pm$6.37E+00 & 1.15E-03$\pm$1.37E-03 & 1.93E+01$\pm$4.81E+02 & 0.5 & 1.76E-03\\
  \noalign{\smallskip}
   \hline
  \end{tabular}
   \begin{tabular}{cccccc}
   \hline
     \noalign{\smallskip}
   \noalign{\hspace{5.cm} Blackbody model}
     \noalign{\smallskip}
   No &  $N_{\rm H}$    & Norm &  kT   & $\chi_{\rm red}$ & Flux \\
      & (10$^{21}$ cm$^{-2}$) & (ph/cm$^2$/s) &  (keV) &           & ph/cm$^2$/s \\
   \noalign{\smallskip}
   \hline
   \noalign{\smallskip}
006  & 1.00E-03$\pm$1.59E-01 & 1.88E-03$\pm$1.89E-03 & 1.36E-01$\pm$4.07E-02 & 1.11 & 7.21E-04\\
007  & 1.07E-01$\pm$4.84E-01 & 1.24E-03$\pm$2.06E-03 & 1.77E-01$\pm$6.42E-02 & 0.73 & 4.69E-04\\
019  & 1.00E-03$\pm$4.15E-01 & 3.02E-02$\pm$5.67E-01 & 2.40E-02$\pm$4.75E-02 & 1.05 & 7.13E-04\\
021  & 7.03E-02$\pm$8.81E-02 & 2.07E+00$\pm$5.51E+00 & 2.63E-02$\pm$1.43E-02 & 1.29 & 1.61E-02\\
023  & 1.00E-03$\pm$3.14E-01 & 1.05E-03$\pm$1.71E-03 & 1.77E-01$\pm$7.69E-02 & 1.88 & 5.12E-04\\
028  & 1.00E-03$\pm$6.02E-01 & 5.51E-04$\pm$7.91E-04 & 2.57E-01$\pm$1.71E-01 & 1.85 & 4.17E-04\\
\hline
   \end{tabular}
   \label{specparsge}
\end{table*}
\end{subtables}

\begin{subtables}
\begin{table*}
\caption{Swift/XRT observations of ROSAT sources in the Com field}
\vspace{-0.3cm}
\begin{tabular}{rcrlcc}
   \hline
   \noalign{\smallskip}
 Name & Date of & Exposure & XRT coordinate (2000.0)       & Position &
   CR  \\
      & observation  & (sec)~~~ & ~~~~~~RA~~~~~~~~  Decl. & error (\asec) &
  (cts/s)  \\
  \noalign{\smallskip}
  \hline
  \noalign{\smallskip}
Com11  & 2007-07-17 & 3230 & --                       & --  & $<$0.002  \\
Com12  & 2008-12-01 & 1653 & 12 53 28.00 +25 34 08.9 & 6.1 & 0.015$\pm$0.003  \\
Com15  & 2008-02-22 &  747 & 12 59 54.90 +25 28 09.0 & 7.5 & 0.013$\pm$0.005  \\
Com15  & 2008-02-23 &  558 & 12 59 55.25 +25 28 09.2 & 13.5 & $<$0.006$\pm$0.005 \\
Com25  & 2008-04-20 & 1487 & 12 46 38.18 +25 13 24.6 & 5.8 & 0.010$\pm$0.003  \\
Com28  & 2007-07-19 & 1583 & --                      & -- & $<$0.006  \\
\hline
  \end{tabular}
  \label{comxrt}

  \noindent{$^{a)}$ diffuse emission.}
\end{table*}

\begin{table*}
\caption{Swift/XRT observations of ROSAT sources in the Sge field}
\vspace{-0.3cm}
\begin{tabular}{rcrlcc}
   \hline
   \noalign{\smallskip}
 Name & Date of & Exposure & XRT coordinate (2000.0)       & Position &
   CR  \\
      & observation  & (sec)~~~ & ~~~~~~RA~~~~~~~~  Decl. & error (\asec) &
  (cts/s)  \\
  \noalign{\smallskip}
  \hline
  \noalign{\smallskip}
Sge5   & 2007-07-10 & 3654 & 19 53 14.97 +23 50 43.7 & 6.0 & 0.004$\pm$0.001 \\
Sge8   & 2007-11-24 & 8318 & 19 57 26.30 +23 28 46.9 & 4.1 & 0.018$\pm$0.002 \\
Sge9   & 2008-02-17 & 3658 & -- & -- & $<$0.002  \\
Sge10  & 2008-02-17 & 6489 & 19 43 14.15 +23 20 15.9 & 5.3 & 0.003$\pm$0.001 \\
Sge11  & 2007-11-28 & 3514 & 20 13 51.56 +23 17 51.8 & 6.3 & 0.004$\pm$0.001 \\
Sge12  & 2007-08-07 &  666 & 20 15 34.24 +23 12 47.8 & 8.0 & 0.016$\pm$0.006 \\
\hline
  \end{tabular}
  
  \noindent{$^a$ A second, shorter observation did also not reveal any source.}
  \label{sgexrt}
\end{table*}
\end{subtables}

\subsection{SDSS spectra}
\label{sec:ID_SDSS}

All 313 objects in Table \ref{comopt} have been cross-correlated 
with the DR9 spectral data release, and 139 matches were found.
Interestingly, while it provided optical identifications for
hitherto 21 unclassified objects, this affected only 2 of our 
optical identifications of X-ray sources (see below). The more
important impact, however, was the provision of redshifts for
all these 139 objects which are listed in Table \ref{sdss}.

The two objects where SDSS-spectra helped in the identification
were: 
(i) {\it Com097}: without spectra, no distinction was possible
between {\it Com097a} and {\it Com097b}, while SDSS-III provided a spectrum
for both, thus solving the ambiguity.
(ii) {\it Com109}: similarly, the distinction between the 5 optical
objects was resolved by the SDSS-III spectrum of {\it Com109a}.

\begin{longtable}{cccrrc}
\caption{\label{sdss} Optical objects in the Com field with
existing SDSS-III spectra where the source ID refers to Table \ref{comopt}. }\\
\hline\hline
  Source ID & Plate & MJD & Fiber & Ident. & z \\
\hline
\endfirsthead
\caption{continued.}\\
\hline\hline
  Source ID & Plate & MJD & Fiber & Ident. & z \\
\hline
\endhead
\hline
\endfoot
003a & 2661 & 54505 & 550 &    QSO & 0.1957 \\ 
007a & 2660 & 54504 & 454 &    QSO & 0.6680 \\ 
008a & 2658 & 54502 & 617 & GALAXY & 0.0668 \\ 
010a & 2658 & 54502 & 464 &    QSO & 0.1598 \\ 
012a & 2661 & 54505 & 516 &    QSO & 0.1417 \\ 
014b & 2657 & 54502 & 543 &    QSO & 0.6789 \\ 
020a & 2660 & 54504 & 431 & GALAXY & 0.1834 \\ 
024a & 2659 & 54498 & 464 &    QSO & 0.3408 \\
\end{longtable}

%%%%%%%%%%%% optisch
%%%%%%%%%%%%

\begin{subtables}

\begin{longtable}{rcrllccrrrll}
\caption{\label{comopt} Optical objects in or near the ROSAT X-ray error 
 circle in the Com field. } \\
\hline\hline
  No  & Coordinates (2000.0) & D   & Designation & Name if & Type & 
   Spec. & R~~ & B~~ & V~~ & ~Ampl. & $\!\!$log(\fxv)\\
      &      R.A ~~~~~ Decl.     & \asec &  & variable&     &   
     & mag & mag & mag & ~~mag  &\\
\hline
\endfirsthead
\caption{continued.}\\
\hline\hline
  No  & Coordinates (2000.0) & D   & Designation & Name if & Type & 
   Spec. & R~~ & B~~ & V~~ & ~Ampl. & $\!\!$log(\fxv)\\
      &      R.A ~~~~~ Decl.     & \asec&  & variable&     &   
    & mag & mag & mag & ~~mag & \\
\hline
\endhead
\hline
\endfoot
001  a & 12 33 56.7 +26 02 33&    13&               &               & AGN?   &         & 18.6 & 18.7 & 18.6 &          & +0.31\\
002  a & 12 36 02.0 +25 59 20&   ~40& NGC 4565      & ULX-4        &  ULX  &         &      & 10.4 &      &           &--\\
003  a & 12 54 09.1  +25 55 24&    2&         &  S10944 Com  & QSO    &         & 16.9 & 16.7 & 16.8 & 16.6--18:  &+0.15\\
004  a & 12 23 08.5 +25 51 06&    14& HD 107793     &               &        & F8V     &  8.8 &  9.6 &  9.1 &         & -2.89 \\
005  a & 12 24 03.5 +25 51 06&    15& HD 107935     &               &        & Am      &    - &  7.0 &  6.7 &         & -4.84 \\
006  a & 12 22 30.3 +25 50 46&    23& HD 107700     &               &        & F2      &    - &  5.4 &  4.9 &         & -3.91 \\
007  a & 12 41 36.0 +25 52 51&     8&               &  S10945 Com  & QSO    &         & 17.2 & 17.3 & 17.2 & 16.9--18.5& -0.24\\
008  a & 12 28 12.8 +25 49 34&     5& GSC 1989 2197 &  S10966 Com  & AGN    &         & 15.6 & 16.4 & 15.9 & 16.6--18: & -0.97\\
\end{longtable}

\begin{longtable}{rcrllccrrrll}
\caption{\label{sgeopt} Optical objects in or near the ROSAT X-ray error 
 circle in the Sge field. } \\
\hline\hline
  No  & Coordinates (2000.0) & D   & Designation & Name if & Type & 
   Spec. & R~~ & B~~ & V~~ & ~Ampl. & $\!\!$log(\fxv) \\
      &      R.A ~~~~~ Decl.     & \asec&  & variable&     &   
    & mag & mag & mag & ~~mag & \\
\hline
\endfirsthead
\caption{continued.}\\
\hline\hline
  No  & Coordinates (2000.0) & D   & Designation & Name if & Type & 
   Spec. & R~~ & B~~ & V~~ & ~Ampl. & $\!\!$log(\fxv) \\
      &      R.A ~~~~~ Decl.     & \asec &  & variable&     &   
    & mag & mag & mag & ~~mag & \\
\hline
\endhead
\hline
\endfoot
001 a & 19 37 43.5 +24 15 40&  30& HD 185334    &              &         & A0&  7.8&  7.9&  7.8&           & -3.90\\
    b & 19 37 41.7 +24 14 57&  20& GSC 2138 966 &              &         &   & 13.4& 12.5& 13.1&           & -1.78\\
    c & 19 37 44.8 +24 15 27&  33&              &              &         &   & 12.4& 14.1& 13.0&           & -1.82\\
002 a & 19 56 29.0 +23 54 45&  13& HD 345287    &  B0 Vul      & EA      & F0& 10.8& 10.5& 10.7&  10.5-13.3& -3.15\\
    b & 19 56 29.7 +23 54 12&  47& GSC 2140 745 &              &         &   & 11.6& 12.6& 12.0&           & -2.63\\
003 a & 19 58 00.1 +23 54 15&  15& HD 189181    &              &         & A0&  8.1& 12.8& 9.9&            &-3.06\\
    b & 19 57 57.5 +23 53 50&  39& GSC 2141 123 &              &         &   & 11.2& 13.5& 12.1&           & -2.18\\
004 a & 20 07 02.6 +23 52 33&  14& GSC 2154 555&  S11007 Vul &CA &   & 11.9& 13.7& 12.6&  13.6-13.9 &-2.53\\
\end{longtable}

\end{subtables}

\twocolumn

\subsection{Notes to individual objects in the Com Field (Table \ref{comopt})}
\label{sec:IDComNotes}

{\it 1a:}  On one Tautenburg Schmidt plate (1944 Apr. 6) and two overlapping 
Palomar prints (1950 Apr. 10 and 1955 May 21) the object is visible with
nearly equal brightness. 

{\it 2a:}
The large distance from the ROSAT source indicates that the 
X-ray source may be near the periphery of the galaxy NGC 4565. 
Chandra and XMM positions confirm this. Because of its large luminosity
this source was classified as an ultra-luminous X-ray source (ULX-4;
Wu et al. 2002). Thus, the identification is not with the galaxy,
but with the ULX in that galaxy. The $B$ magnitude is for the
whole galaxy. Since the optical counterpart of the X-ray source is
not known, no \fxo\ is given. See also Brinkmann \etal\ (1995).

\begin{figure}[ht]
 \includegraphics[width=8.0cm]{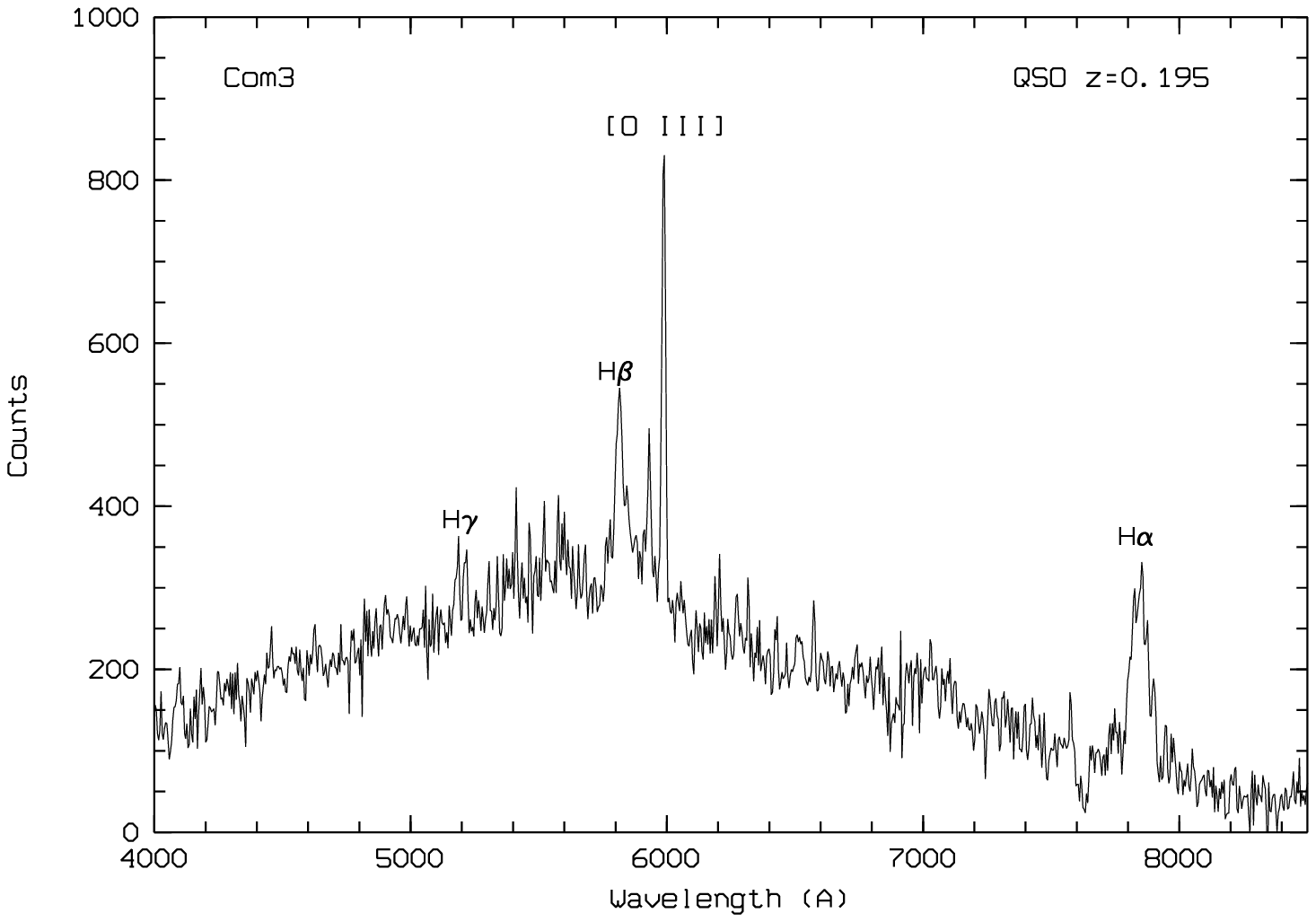}
\caption[v4]{Not flux-calibrated optical spectrum of object Com003, showing 
broad Balmer lines, and narrow [O III], suggesting a narrow-line Sy1 (z=0.195).
Courtesy S. Zharykov, ONAM Ensenada.
\label{com3ospec}}
 \end{figure}

{\it 3a:}  Because this faint object is near the border of the astrograph 
plates, it is faintly indicated only on some dozen plates. Variable within 
some weeks. 
Accidentally, this object is visible on 
3 overlapping fields of the POSS (1955 Apr. 15 17\fm3, 1955 May 21 17\fm0,
1956 May 17 16\fm6) and on 4 plates of the 2\,m Schmidt 
telescope of Tautenburg, which shows the object at different brightnesses. 
A low-S/N, not flux-calibrated optical spectrum (courtesy S. Zharykov, 
ONAM Ensenada, Mexico) reveals this object to be
an AGN at z = 0.195 (see Fig. \ref{com3ospec}), consistent with the
very soft X-ray spectrum. 
A 2-hr V-band monitoring on March 14/15 1996 (23:58--2:05 UT)
shows variability by 0.3 mag.
The SDSS-III spectrum identifies it as QSO at z = 0.1957 and 
a ``starburst broadline'' sub-class.

{\it 6a:} HR 4707 $\equiv$ 12 Com. 

{\it 7a:} Very soft X-ray spectrum.
A low-S/N, not flux-calibrated optical spectrum (courtesy S. Zharykov, 
ONAM Ensenada, Mexico) reveals only one emission line 
at 465 nm, so it is likely an AGN.
The SDSS-III spectrum identifies it as broadline QSO at z = 0.6680.

{\it 8a:} Faint on two Tautenburg plates (3999 from J.D. 2442094 and 
8586 from J.D. 244 9449).  
This variability, the \fxo\ ratio and the blue colour suggest an
  AGN or CV nature, rather than a stellar member of the Coma Berenices
  open cluster as proposed by Randich et al. (1996). 
The SDSS-III spectrum identifies it as a AGN at  z = 0.0668.

{\it 9a:} Fainter companion is totally blended. 

{\it  10a:} On POSS print and one Tautenburg Schmidt plate of equal
 brightness. CV Com is 4\amin\ outside the ROSAT position. 
The SDSS-III spectrum identifies it as a QSO at  z = 0.1598,
with ``starburst broadline'' sub-class.

{\it 11a,b:} 
Investigation on Sonneberg astrograph plates not possible due to blending.
The Swift XRT observation did not detect this source. The upper limit
of $<$0.002 cts/s is about a factor 5 below the ROSAT flux.

{\it 12a:} Probable active galaxy. The  published USNO-A2
$B$ magnitude of 17.2 is substantially different from that in USNO-B 
which gives B1 = 16.11 mag. This object is too faint for Sonneberg plates.
The red colour of the galaxy does not suggest an AGN nature, and the \fxo\
is too high for normal, inactive galaxies. 
Besides the all-sky survey, this sky area was observed at two more
occasions with the ROSAT PSPC: once for 12.01 ksec between Dec. 17, 1991
and Jan. 6, 1992, and again for 7.35 ksec on Jun 4, 1992. At the first
occasion, the measured count rate is 0.084$\pm$0.004 cts/s
while at the second it is 0.067$\pm$0.004 cts/s, thus this source is
clearly variable when compared to the all-sky survey rate of 
0.045$\pm$0.010 cts/s. 
The Swift XRT position confirms the association of the X-ray source
to this galaxy. Considering the soft spectrum, the XRT count rate is
similar to the intensity during the ROSAT all-sky survey.
This X-ray intensity pattern, in particular the rise between 1990 and 1992, 
is rather slow compared to what one would expect 
for a tidal disruption event (e.g. Komossa \& Greiner 1999).
The SDSS-III spectrum reveals broad emission lines, and identifies it
as QSO at z=0.1417.

{\it 13a:} $UBV$ photometry by Zeilik \etal\ (1982). See also Fleming \etal\ 
(1989).

{\it 14b:} Is quasar 4C 25.39 $\equiv$ PKS J1217+2529, and
is a blue object. The \fxo\ ratio favours the identification
of the ROSAT source with this quasar, rather than the brighter star {\it 14a}.
The SDSS-III spectrum reveals broad emission lines, and identifies it
as QSO at z=0.6789.

{\it  15a:} Triple blend on Sonneberg plates. The $B-V$ colour and \fxo\
are consistent with a K or M spectral type and the object being the
optical counterpart of the X-ray source. This is verified by
the Swift XRT data.
{\it 15b:} Object is too faint for Sonneberg plates. On POSS (243 5550) about
19.0 mag, and on 2 Tautenburg plates (244 1400 and 244 2074) about 19.5 mag.
Nevertheless, variability is uncertain.
The blue colour suggests an AGN nature (the X-ray hardness ratio argues against 
a white dwarf nature).

{\it  16a:} HD111813 is the obvious counterpart, which is also
 confirmed by the XMM position (2XMMp J125138.4+253033; it is plotted
with larger uncertainty for visibility reasons in Fig. \ref{comfc}).
The two pointings mentioned under Com012a also cover this source,
and the corresponding X-ray count rates are 0.040$\pm$0.002 and
0.035$\pm$0.003, respectively.

{\it  17a:} The ROSAT position coincides with the centre of the Sy2
galaxy NGC 4725, but XMM resolves this into 2 sources (2XMMp J125026.6+253003
and 2XMMp J125027.3+253026), and in addition finds three more nearby sources
which may add to the ROSAT source (2XMMp J125024.2+252947,
2XMMp J125024.3+252941, and 2XMMp J125024.9+253053).

{\it  18a:}  Flare star GJ 3739. No X-ray flares observed. 
 $B-V$ colour and \fxo\ are consistent with coronal emission from
an M star.

{\it 19a:} Probably galaxy pair; if interacting, or with one active member,
it could explain the X-ray emission.
 The two pointings mentioned under {\it Com012a} also cover this source,
and the corresponding X-ray count rates are 0.086$\pm$0.006 and
0.078$\pm$0.007, respectively. Both rates are about a factor 2 higher 
than the rate seen during the all-sky survey.

{\it 20a:} Invisible on Sonneberg plates. The red colour and
the fuzzy appearance on the digitized sky survey suggest a galaxy. 
This is supported by the SDSS-III spectrum which identifies this
object as galaxy at z=0.1834, though Kouzuma \etal\ (2010) identify 
it as AGN candidate due to the NIR colours from 2MASS.
However, the X-ray hardness ratio hints more towards a cluster identity.
The similar colours and fuzzy appearance of several nearby 
objects, most notably the other two sources North and north-west
of {\it 20a} within the RASS error circle, suggest galaxies at the same 
redshift.

{\it 21a:} Invisible on Sonneberg plates. 
The blue colour and large \fxo\ ratio suggests an AGN nature, 
in which case it could be the counterpart.

{\it 22a:} BL Lac object at z=0.135; see Brinkmann \etal\ (1995) and 
Sowards-Emmerd \etal\ (2005). Not tested for variability; on POSS
(243 5249) and on Tautenburg plate (244 9449) of equal brightness.

{\it 23a:}  The variability is not sure. Probably constant. 
If the FG spectral type is correct, then the \fxo\ ratio argues against
this object being the counterpart of the X-ray source.

{\it 24a:} On Tautenburg plate 8586 (TJD = 24\,49449) invisible.
The SDSS-III spectrum identifies it as QSO at z=0.3408 with
``starburst broadline'' sub-class.

{\it 25:} The Swift XRT position favours object {\it 25b}; being too
faint for the Sonneberg plates.
It is listed
as SDSS J124640.80+251149.5 and type ``Galaxy'' at z=0.084 in 
Rines et al. (2001), 
with the SDSS-III giving an identification as starforming galaxy.
The Swift XRT intensity is about the same as that seen with ROSAT.
 
{\it  26a:} {\it Einstein} X-ray source (Stocke \etal\ 1983).
 Not tested for variability. 
The SDSS-III spectrum identifies it as QSO at z=0.0637 with
``starburst broadline'' sub-class.

{\it  27a:}  Very difficult because near the brightness limit and 
near the edge of the plate, but the variability seems to be sure. 
Brightness varies irregularly.
The SDSS-III spectrum identifies {\it 27a} as broadline QSO at z=0.5447,
and {\it 27b} as star.

\begin{figure}[ht]
 \includegraphics[width=3.4cm, angle=270]{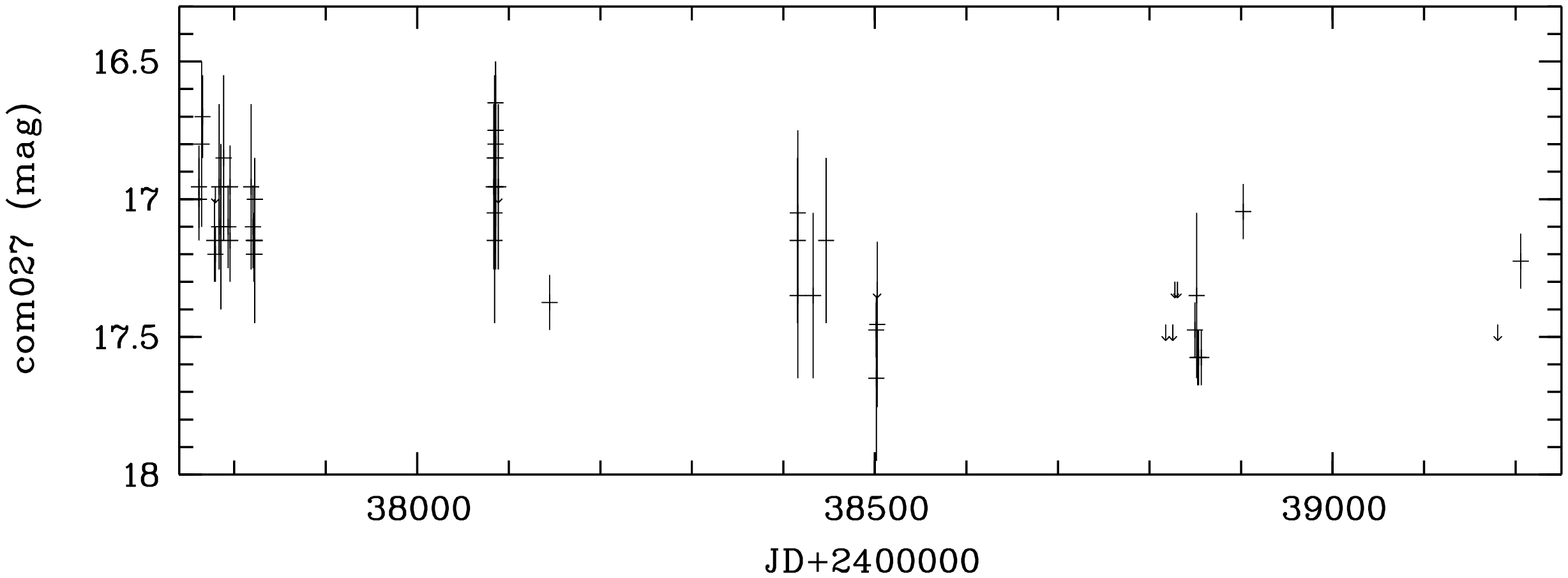}
\caption[]{Blow-up of the light curve of Com027a (Fig. \ref{comolc}) 
showing pronounced variability.}
\label{com27lc}
 \end{figure}

{\it  28a:} 
Not visible on Sonneberg plates. Blue colour suggests an AGN nature
(see also Chen \etal\ 2002).
The Swift XRT observation did not detect this source; the upper limit
is not constraining, given the soft X-ray spectrum.

{\it  29a:} Invisible on Sonneberg plates.
Due to the blue colour most likely a QSO, 
which is confirmed by the SDSS-III spectrum (broadline QSO at z=0.5865).
{\it 29b} is identified by Kouzuma \etal\ (2010) as AGN candidate 
due to the NIR colours from 2MASS.

{\it  30a:} Seyfert 1.5 galaxy (TON 0616, 4C 1834, FBQS J122539.5+245836)
 at z=0.268. On Tautenburg plates (244 2094 and 244 9449) faint.
See also the optical catalogue of QSOs by 
  Hewitt \& Burbidge (1987), and Brinkmann \etal\ (1995).
The SDSS-III spectrum classifies it as QSO at z=0.2679
with ``starburst broadline'' sub-class.

{\it  31:} Nothing visible on Sonneberg plates.
The Swift XRT position suggests {\it  31a} as counterpart. 

{\it  32a:} Within the error circle is one radio source identified
during the FIRST bright quasar survey:
FBQS J1249+2452 = FIRST J124958.8+245233 (White et al. 2000).
This coincides within 1\asec\ with object {\it 32a} which we therefore
classify as an AGN.
The SDSS-III spectrum supports this, and gives z=0.2465.

{\it  33:} No  optical counterpart candidate brighter than $B$ = 19.5 mag 
within the ROSAT error circle.  
The Swift XRT observation reveals clearly extended emission, centred
around 12\h45\m15\fss0 +24\degr53\amin35\asec\ and with a diameter of about
1\farcm5. The ROSAT and Swift X-ray fluxes are comparable. The hard
spectrum and the diffuse emission suggest a cluster origin. 
The ROSAT source
is likely the detection of a brighter blob at the south-eastern rim of
the extended emission.

\begin{figure}[ht]
 \includegraphics[width=3.4cm, angle=270]{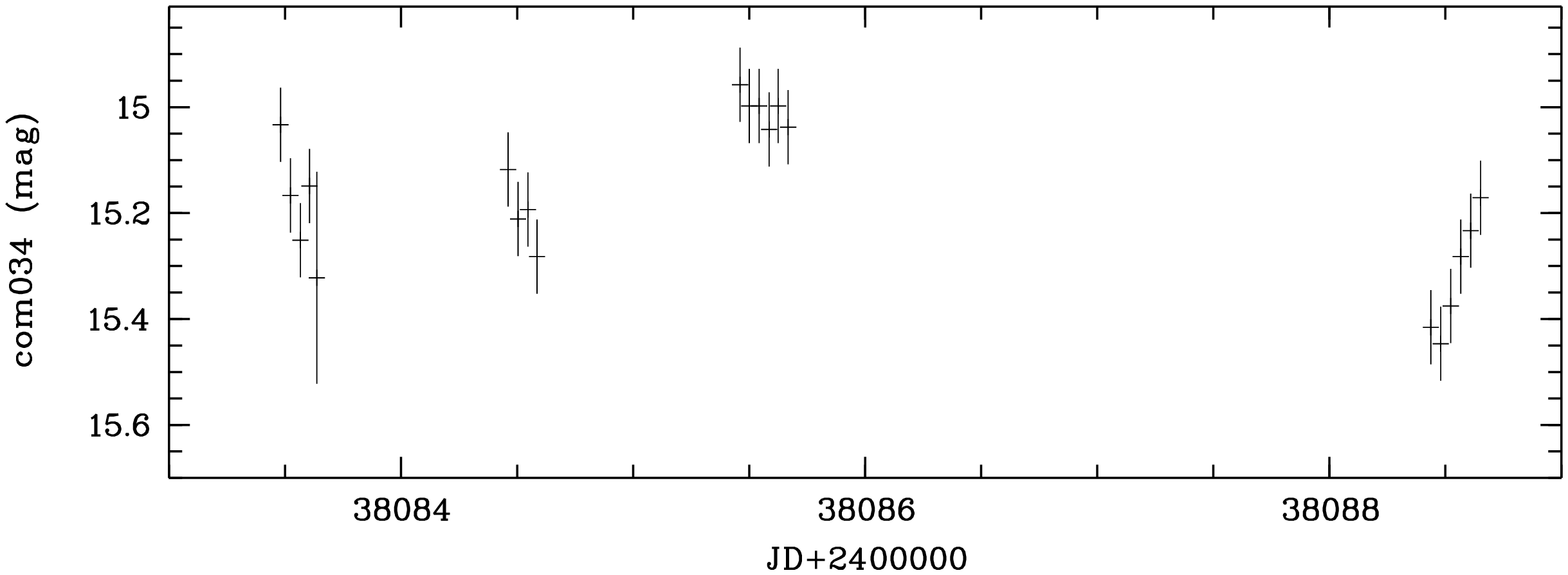}
\caption[]{Blow-up of the light curve of Com034a (Fig. \ref{comolc}) 
showing pronounced variability.}
\label{com34lc}
 \end{figure}

{\it  34a:} M star. Eclipse minima of short duration (1 day or shorter) seem
to exist, but because of irregularities in the light curve no secure period 
could be found. 
A period length of 0.35 days seems to be indicated, but there are
contradictions with bright states. Very probably chromospherically active
star of RS CVn type. Observed minima:
JD 2438\,083.64 (15.3 mag; descent, Fig. \ref{com34lc}),
         088.47 (15.4; ascent, Fig. \ref{com34lc}),
         089.54 (15.2),
         090.57 (15.2),
         091.63 (15.2),
         093.66 (15.3),
         116.38 (15.3),
         140.45 (15.3),
         142.35 (15.3),
         146.51 (15.3),
         473.32 (15.3),
         817.66 (15.2),
   2445\,812.51 (15.5:),
   2447\,205.55 (15.3:),
         262.47 (15.4),
   2448\,682.51 (15.3).

{\it  35a:}   Invisible on Sonneberg plates.
Due to the blue colour most likely a QSO.
The SDSS-III spectrum classifies it as QSO at z=0.2184
with ``starburst broadline'' sub-class.

{\it  36a:} Other name HD 111\,395.

{\it  37b:} Blue object, thus possibly AGN; but {\it 37a} is variable 
and is therefore the 
better candidate for the optical counterpart of the ROSAT source. 
This is confirmed with the Swift/XRT position.
The optical light 
curve shows waves with a length of about 20--30 days. May be a BY Dra star 
or a massive X-ray binary.  

\begin{figure}[ht]
 \includegraphics[width=3.4cm, angle=270]{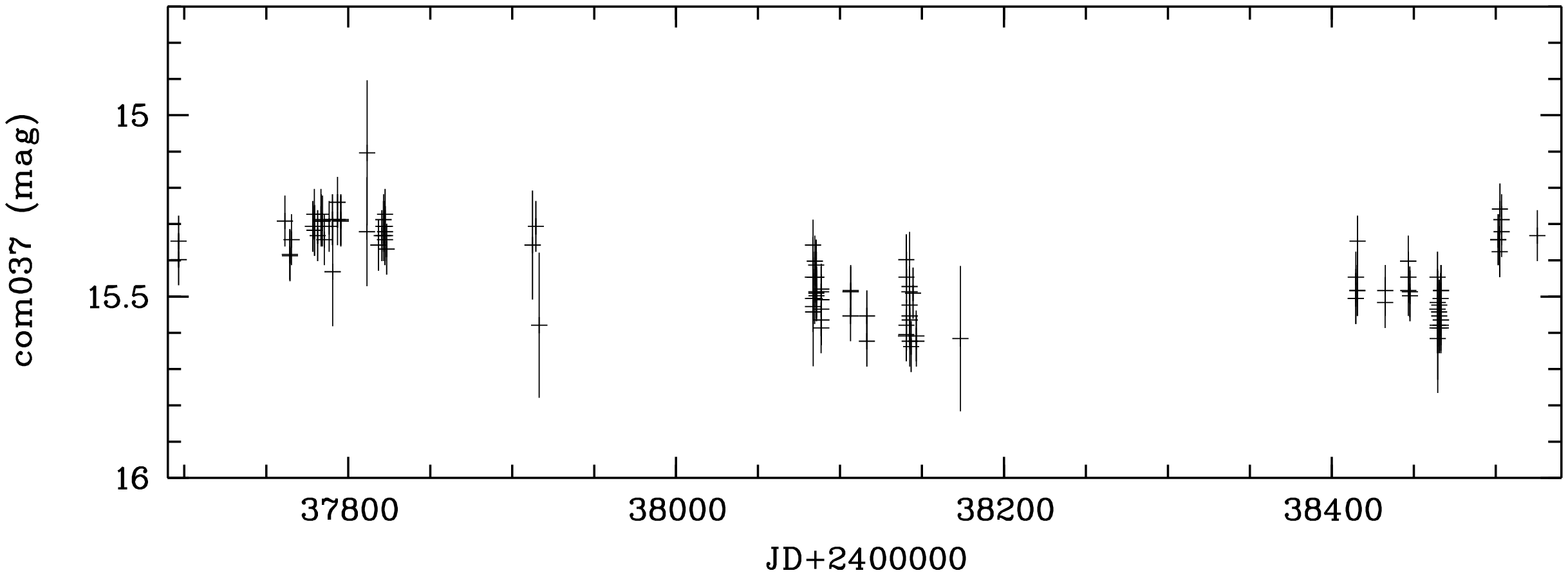}
\caption[]{Blow-up of the light curve of Com037a (Fig. \ref{comolc}) 
showing pronounced variability.}
\label{com37lc}
 \end{figure}

{\it  38a:} Classified as QSO in Zickgraf \etal\ (2003); too faint on
Sonneberg plates for variability assessment. 
The SDSS-III spectrum classifies it as broadline QSO at z=0.5471.

\begin{figure}[ht]
 \includegraphics[width=6.5cm, angle=270]{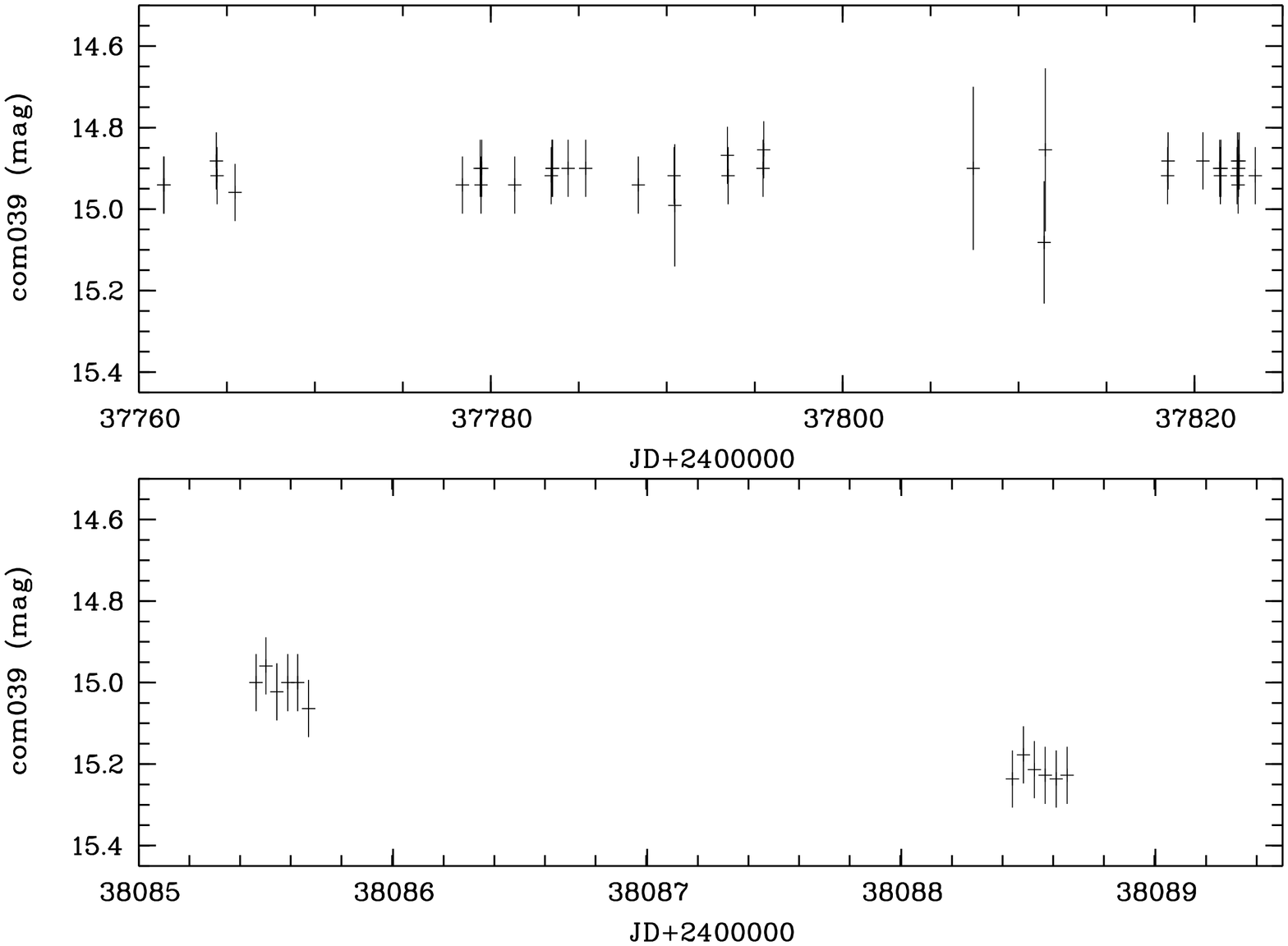}
\caption[]{Blow-up of the light curve of Com039a (Fig. \ref{comolc}) 
showing pronounced variability.}
\label{com39lc}
 \end{figure}

{\it  39a:} Classified as M star in Zickgraf \etal\ (2003); 
$B-V$ colour and \fxo\ are consistent with this. 
Most of the time at a bright (Fig. \ref{com39lc}, top panel), 
sometimes at mean magnitude level (Fig. \ref{com39lc}, left part of bottom 
panel), with only
small variations. Sometimes fadings with a duration of some days or 
several weeks (Fig. \ref{com39lc}, right part of bottom panel).

{\it  40a:} See also Brinkmann \etal\ (1995) 
for the BL Lac classification and redshift z=0.218.
On Sonneberg plates invisible. On POSS print 135 (J.D. 243 3291) 
at $B$ $\sim$ 16.6 mag.
 On Tautenburg Schmidt plates 3999 (J.D. 244 2094) $B$ = 16.6 mag, and 8586 
(J.D. 244 9449) $B$ = 17.8 mag.  Therefore variable. 
The SDSS-III spectrum classifies it as galaxy at z=0.2187,
despite its very blue continuum.

{\it  41a:} Faint object, on only 4 astrograph plates just visible. 
May therefore be sometimes brighter than 18 mag and probably variable.  
The blue colour and \fxo\ ratio suggest an AGN nature.

{\it  42a:} This object has been spectroscopically identified as QSO 
at z=0.438 (Chen \etal\ 2002).
 This very blue object is just visible on the 6 best plates. 
Nothing can be said about variability. Object
{\it 42b} also has a blue colour, suggesting also an AGN nature.
The SDSS-III spectrum classifies {\it 42b} as broadline QSO at z=1.2078.
The RASS position suggests that {\it 42a} is the optical counterpart of the
X-ray source.

{\it 43a:} Other name BD+25\degr2511. See also Raveendran (1984).

{\it 44a:} Invisible on Sonneberg plates.
The blue colour suggest an AGN nature.
The SDSS-III spectrum classifies it as broadline QSO at z=0.3710.

{\it 45a:} Already detected by the {\it Einstein} Observatory 
 (Stocke \etal\ 1983).
 Narrow-line Sy1 galaxy at z=0.186, with ROSAT data and optical
 variability based on Sonneberg data already reported earlier 
(Greiner \etal\ 1996). 
The SDSS-III spectrum classifies it as QSO at z=0.1856
with ``starburst broadline'' sub-class.

{\it 46:} Both objects invisible on Sonneberg plates.
The Swift XRT position points to {\it 46a} as the counterpart, and the
blue colour and \fxo\ ratio suggest an AGN nature.

{\it 47a:} Within the error circle is one radio source,
FBQS J122214.6+242420 at a spectroscopic distance of z=0.12 
(White \etal\ 2000).
This coincides within 1\asec\ with object {\it 47a} which we therefore
classify as an AGN. On Tautenburg plates invisible ($>$18 mag).
The Swift XRT observation finds this source at twice the 
ROSAT rate, and confirms this association through an accurate X-ray position.
The SDSS-III spectrum classifies it as QSO at z=0.1167
with ``AGN broadline'' sub-class.

{\it  48a:} Not tested for variability.  
The blue colour suggest an AGN nature.
The SDSS-III spectrum classifies it as broadline QSO at z=0.8282.

{\it  49a,b,c:} If the K spectral type for {\it 49c} is correct, then
it is excluded as counterpart.  {\it  49a,b} are possible counterpart
candidates, with {\it  49a} having higher likelihood.
The Swift XRT observation detects 3 photons from this source {\it  49a}; 
this low intensity is consistent with the moderately soft X-ray spectrum.
The SDSS-III spectrum classifies {\it 49a} as broadline QSO at z=1.0444.

{\it  50:} No optical counterpart candidate brighter than about 
$B$ = 20 mag inside the  ROSAT  error circle. 
The Swift XRT observation suggests diffuse X-ray emission,
with an extent of about 20-30\asec\ (likely biased due to counting
statistics). Also, the ROSAT and the Swift XRT X-ray flux measurements
are the same.

\begin{figure}[ht]
 \includegraphics[width=3.4cm, angle=270]{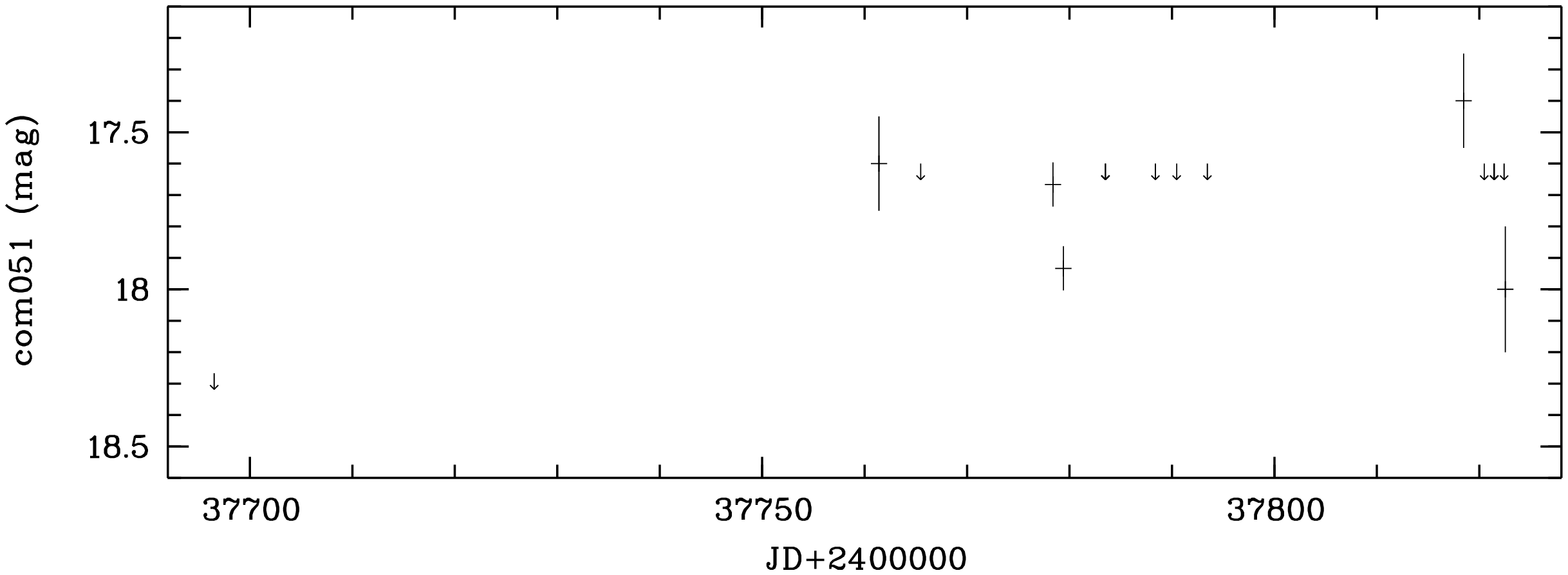}
\caption[]{Blow-up of the light curve of Com051a (Fig. \ref{comolc}) 
showing pronounced variability.}
\label{com51lc}
 \end{figure}

{\it  51a:} Only seen on the 8 best plates. On POSS print 1435 
(J.D. 243 5249) about 1 mag fainter than on Sonneberg plates
as can be seen in Figs. \ref{com51lc} \& \ref{comolc}. 
The SDSS-III spectrum classifies it as broadline QSO at z=0.465.

{\it  52b:} Already detected by the {\it Einstein} Observatory 
 (Stocke \etal\ 1983).
BL Lac object at z=0.14, with ROSAT data and optical
 variability based on Sonneberg data already reported earlier 
(Greiner \etal\ 1996). 
XMM detects three X-ray sources in the field (see Fig. \ref{comfc}),
one of which (2XMMp J125731.9+241240) coincides with {\it 52b}, thus 
confirming the earlier identification.
The SDSS-III spectrum classifies {\it 52b} as galaxy at z=0.1406.

{\it  53a:} Slightly outside the ROSAT error circle, and too faint for
 variability analysis on Sonneberg plates, but the XMM position
(2XMMp J125638.7+241252) confirms the identification.

{\it  54a:} Not tested for variability due to faintness. 
The SDSS-III spectrum classifies it as broadline QSO at z=0.5071.

{\it  55a:} Blue colour suggests an AGN nature. 
Not tested for variability, since too faint on Sonneberg plates.
On Tautenburg plate 3999 (244 2094) about 18.0 mag.
The SDSS-III spectrum classifies it as broadline QSO at z=0.3795.

\begin{figure}[ht]
 \includegraphics[width=3.4cm, angle=270]{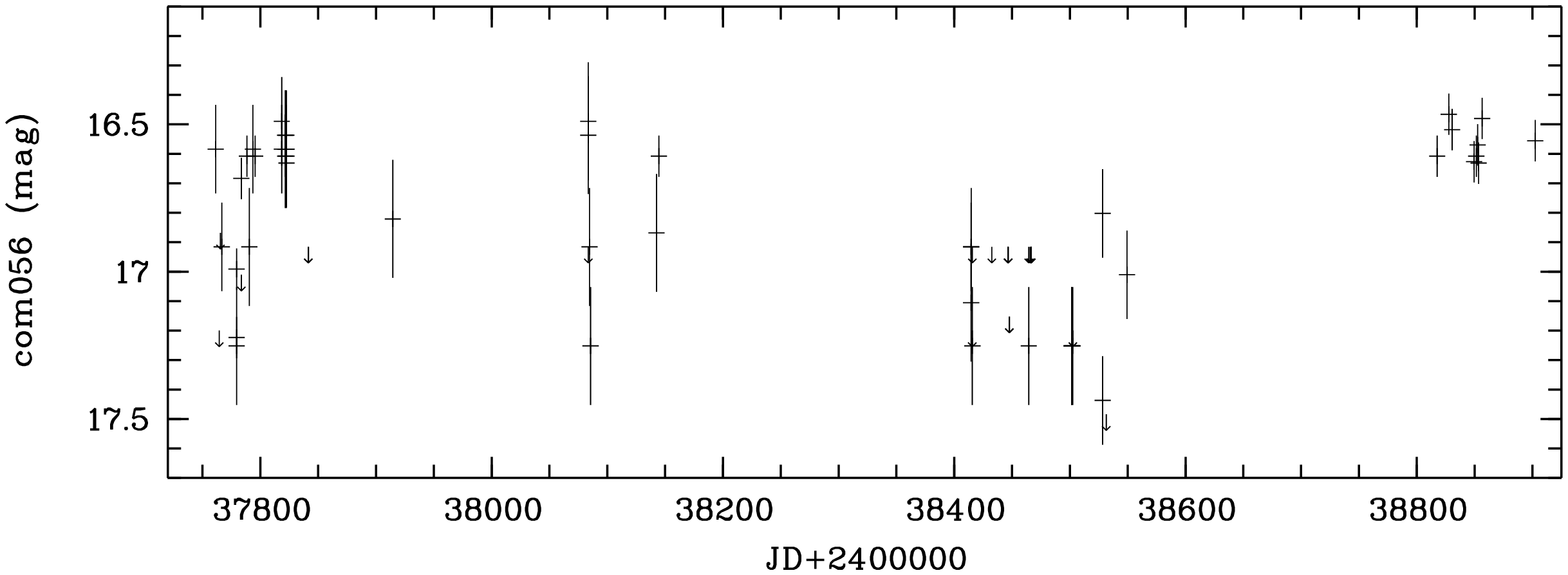}
\caption[]{Blow-up of the light curve of Com056a (Fig. \ref{comolc}) 
showing pronounced variability.}
\label{com56lc}
 \end{figure}

{\it  56a:} Spectroscopically classified as Seyfert galaxy at z=0.188
by Chen \etal\ (2002).
Difficult because of faintness. At the beginning of the
observations (J.D. 243 7696 - 7822) rather quick brightness changes 
between 16.4 and 17.2 mag, and the whole amplitude can be run through 
within 1 day. Short weakenings from a generally bright level in irregular 
time intervals. From J.D. 243 8085 to 8549 the brightness varies between 
16.8 and $>$17.5 mag. Faint observations preponderate, so that there are 
relatively short 
maxima. From J.D. 243 8817 to 8902 the object is obviously always bright 
between 16.4 mag and 16.6 mag (see Fig. \ref{com56lc}). 
The few observations from 243 9144 - 244 2540 
show the object between 16.6 mag and 17.0 mag. Since J.D. 244 4292 the object 
is nearly always invisible. An observation of B = 17.2 mag at 244 5818 must 
already be considered as a brightening. On the blue print of the Palomar 
survey (JD 243 5550) about 18 mag (Figs. \ref{com56lc} \& \ref{comolc}).  
The SDSS-III spectrum classifies it as QSO at z=0.1884,
with ``starburst broadline'' sub-class.
The RR Lyr star EL Com has a distance of about 
210\asec\ to the ROSAT position and is not considered a counterpart
candidate.

{\it  57a:} Not tested for variability.   

{\it  58a:} Not investigated on Sonneberg plates.

{\it  59:} Cluster of galaxies 
NSC J122158+235426 at z=0.149 (Gal \etal\ 2003).
{\it 59a:} 
Irregular galaxy 2MASX J12215975+2354425, seemingly not active.
The SDSS-III spectrum classifies {\it 59a} as galaxy at z=0.1318,
while Kouzuma \etal\ (2010) identify it as AGN candidate due to the
NIR colours from 2MASS.
The hard X-ray spectrum (hardness ratio) suggests that the cluster is
the counterpart, not the galaxy {\it 59a}.

{\it  60:} No optical counterpart candidate brighter than $B$ = 21 mag 
in the ROSAT error circle.  
The Swift XRT observation did not detect this source; the upper limit
is not constraining, given the very soft X-ray spectrum.
The brightest SDSS (DR9) objects are three galaxies with
g'=22.4 at 23\asec\ offset,
g'=22.6 mag at 25\asec\ offset, and  g'=21.7 at 31\asec\ offset.
The latter implies a lower limit on log(\fxo) \gax\ +1.2.

{\it  61a,b:} Both objects have blue colour, and are likely AGN,
see also Mickaelian \etal\ (2006). 
The Swift XRT observation suggests {\it  61a} as the counterpart.
The SDSS-III spectrum classifies {\it 61a} as galaxy at z=0.3984.

{\it  62a-e:} Several faint galaxies within the error circle; not tested
for variability. Remarkable is the high X-ray intensity.
A ROSAT HRI observation indicates that the X-ray source is possibly
extended (Gliozzi \etal\ 1999), 
and thus the cluster of galaxies at z=0.134 could be the counterpart.
However, the X-ray flux is centred on
the quasar {\it  62e}, and, more importantly,
 the flux seen in the HRI is a factor 4 lower 
than that from the survey, suggesting that the quasar dominantly contributes.
{\it  62a} is a galaxy (NGP9 F378-0239966) at z=0.132 (Gliozzi \etal\ 1999) 
which coincides
with the radio source FIRST J123438.6+235013, thus probably is an AGN.
{\it  62c} is a galaxy (NGP9 F378-0239961) at z=0.135 (Gliozzi \etal\ 1999) 
which coincides
with the radio source FIRST J123438.6+235013, thus probably also is an AGN.
{\it  62d} is a galaxy (NGP9 F378-0239941) at z=0.133 (Gliozzi \etal\ 1999) 
which coincides
with the radio source FIRST J123437.1+235016, thus probably also is an AGN.
{\it  62e} is a quasar at previously unknown redshift (Bade \etal\ 1998),
for which the SDSS-III spectrum provides z=0.1382 together with a 
classification as broadline QSO. Based on the X-ray variability, the
hardness ratio and \fxo\ we identify {\it 62e} as the counterpart.

{\it  63a:} Potentially a member of the Coma Berenices open cluster
(Randich et al. 1996).

{\it  64a:} This object is about 21 mag on the Palomar print. It is not 
contained in 
the USNO catalogues, just in the APM (McMahon \etal\ 2000) from 
which the magnitudes are taken. 
It is invisible on all Sonneberg plates with the exception 
of J.D. 243 8085.459 (B = 15.5 mag); on the contrary, at J.D. 243 8085.501 
it is invisible ($>$17 mag), implying a decline of $>$1.5 mag within
1 hr.
Reflected light microscopy (see e.g. Greiner \etal\ 1990 for
a similar application) gives no indication for a plate fault. 
Based on this, the object may be a flare star. 
Nevertheless, the optical variability remains to be confirmed.
The arrival times of the 13 X-ray photons are not compatible with
a single flare, but seem to be distributed equally. 

{\it  65:} Red star of 20 mag at a distance of 19\asec. Much too faint for 
Sonneberg plates.
The Swift XRT observation did not detect this source; the upper limit
is not constraining, given the moderately soft X-ray spectrum.
The SDSS-III spectrum classifies {\it 65a} as AGN at z=0.1377.

{\it  66a:} Within a cluster of galaxies. {\it 66b} is the heavily 
disturbed galaxy pair IC 3314+IC3312, which therefore is probably 
the optical counterpart of the ROSAT source. 
The Swift XRT observation did not detect a point source; the upper limit
suggests a factor of a few fading relative to the ROSAT measurement.

{\it  67a:} Stellar object with UV excess (Green \etal\ 1986).
The very soft X-ray spectrum (HR1=-1.0) clearly points
towards the white dwarf PG 1232+238 as the counterpart, 
consistent with the identification by
Zickgraf \etal\ (2003).

{\it 68a:}  The blue colour suggests an AGN nature, and the
the SDSS-III spectrum provides z=0.2583 together with a 
classification as broadline QSO.

{\it  69a/b:} Close pair of stars of equal brightness; the eastern component 
{\it 69a} is variable, and is identical with S 10937 $\equiv$ KX Com
(Richter \etal\ 1995).  

{\it 70:} The SDSS-III spectrum classifies {\it 70a} as starforming galaxy 
at z=0.1947, but the \fxo\ ratio suggests the QSO {\it 70b} as
the counterpart, for which the  SDSS-III spectrum provides z=1.1775.

{\it  71:} Some galaxies fainter than 21 mag are within the ROSAT error 
circle, but too faint for Sonneberg plates. The hard X-ray spectrum 
might suggest a cluster identity.

{\it  72:} While {\it 72a} is a galaxy based on its colours and too faint 
for Sonneberg plates, {\it 72b} is an A0 star according to SDSS-III,
and difficult on Sonneberg plates because of faintness.
Due to the inconsistent \fxo\ ratio, the A0 star cannot be the counterpart.

{\it  73a:} On 4 Tautenburg Schmidt plates and on POSS of equal brightness.  
The \fxo\ ratio excludes a stellar counterpart, but rather suggests
 a CV or AGN origin. However, the optical colours are not particularly blue. 

{\it  74:} Nothing visible on Sonneberg plates.
The Swift XRT position identifies {\it  74a} as  counterpart.
The blue colour and large \fxo\ suggest a AGN or CV identification.
Given the moderately soft X-ray spectrum, the ROSAT and Swift 
intensities are comparable.
The SDSS-III spectrum classifies {\it 74a} as QSO at z=0.2728,
with ``starburst broadline'' sub-class.

{\it  75a:} Considering the faintness of all optical objects in question, 
the high X-ray intensity is remarkable. Possibly an 
optically faint X-ray binary or an X-ray bright AGN, based on \fxo. 
Zickgraf \etal\ (2003) report ``BLUE-WK'', a moderately blue continuum,
 weak point-like object.
The SDSS-III spectrum classifies it as broadline QSO at z=0.1606.

{\it  76a:} Because of faintness variability not quite sure. 
The Tautenburg plate 3999 shows the object at 17.5 mag.
{\it 76a} and {\it 76b} are a narrow pair.   

{\it 77a:}  Because of faintness nothing can be said about variability.
The blue colour suggests an AGN (or CV) nature which would be consistent
with the \fxo\ ratio.
The SDSS-III spectrum classifies it as broadline QSO at z=0.6667.

{\it  78b:} Because of faintness only on a part of plates visible. 
Seemingly there are long flat waves 
(slowly oscillating brightness variations)
with a timescale of some 100 days, 
standstills seem also to occur. During the intervals J.D. 243 7695 - 7850, 
8415 - 8550, 8815 - 9100 about 
$B$ = 17.5 mag. Thereafter $B$ = 17.8 mag and fainter, but observations 
are only sporadic. J.D. 244 4700 again brighter, thereafter weakening. 
On one plate of Tautenburg Schmidt telescope at 244 2094 about $B$ = 18 mag. 
The SDSS-III spectrum classifies it as broadline QSO at z=1.3866.

{\it  79:} Both objects too faint for Sonneberg plates. 
The SDSS-III spectrum classifies {\it 79a} as broadline QSO at z=0.3412.

{\it 81a:} At first glance seems to be variable between about 17.5 and 
18.5 mag. But in spite of the seemingly large amplitude, the variability is
not quite sure because of being near the plate limit.
The SDSS-III spectrum classifies it as broadline QSO at z=0.9669.
{\it  81b:} Though beyond the plate limit, marginally visible on some 
Sonneberg plates. Nevertheless, variability questionable.

{\it  82:} No object brighter than 21 mag in the X-ray error circle. 
The SDSS-III spectrum classifies {\it 82a} as broadline QSO at z=0.7227
which we consider to be the counterpart despite its somewhat large
distance to the X-ray centroid position.

{\it  83a:} This faint object seems to be barely visible on some plates, 
but this may be artefacts. Nevertheless, a true variability cannot be 
fully excluded. 
The SDSS-III spectrum classifies it as broadline QSO at z=0.5299.

{\it  84a:} On POSS 1435 (1956 May 20) about $B$ = 19 mag, on Tautenburg 
Schmidt plate 4005 (1974 Feb. 15) about $B$ = 17.5: mag. Seems to be faintly 
indicated on some Sonneberg plates. Nevertheless, optical variability is not 
quite sure. The   
The Swift/XRT observation reveals no detection, with an upper limit
about a factor 4 below the RASS rate, so clearly X-ray variable. This and 
the \fxo\ ratio suggest an AGN nature.
The SDSS-III spectrum classifies it as QSO at z=0.2033, with
``starburst broadline'' sub-class. 

{\it  85:} No optical counterpart candidate brighter than about $B$ = 20 mag 
inside the ROSAT position. 
The brightest SDSS (DR9) objects are a galaxy 
(g'=21.0 mag, 35\asec\ offset) and a star (g'=21.35, 26\asec\ offset).
The former implies a lower limit on log(\fxo) \gax\ +1.1.

{\it  86a,b:} Both objects are outside the error circle. Inside the error 
circle no object brighter than 21 mag.   {\it  86b} is of equal brightness
on POSS (243 5249) and the Tautenburg plate (244 2094).

{\it 87a:} Variable with  a time 
scale of several days to months: possibly chromospherically active star.
Spectral classification from Garcia Lopez \etal\ (2000) who
  discard Coma Berenices cluster membership
which had been proposed by Randich et al. (1996), 
but find spectroscopic  evidence for chromospheric activity. 

\begin{figure}[ht]
 \includegraphics[width=3.4cm, angle=270]{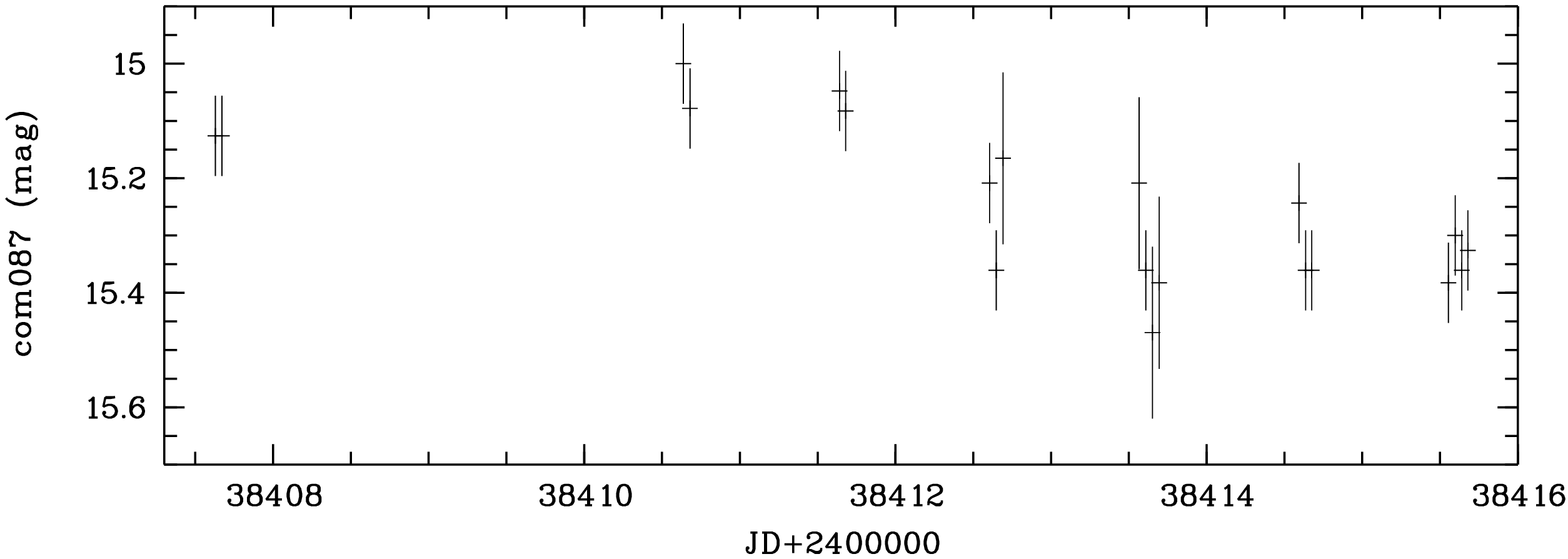}
\caption[]{Blow-up of the light curve of Com087a (Fig. \ref{comolc}) 
showing pronounced variability.}
\label{com87lc}
 \end{figure}

{\it 88a:} KZ Com = S 10939. See Richter \etal\ (1995).

{\it 89a+b:} The SDSS resolves this into 4 objects, and all
objects are consistent with the ROSAT X-ray position.
The \fxo\ ratio does not indicate an AGN nature for any of those, 
although Kouzuma \etal\ (2010) identify both {\it 89a+b:} 
as AGN candidates due to the NIR colours from 2MASS.
The hard X-ray spectrum
argues against a tidal disruption event in an inactive galaxy.
The SDSS-III spectrum classifies {\it 89b} as galaxy at z=0.1405.
A Swift/XRT observation does not reveal an obvious point source.
There is, however, a blob of diffuse emission slightly west
of {\it 89a+b}, and a further blob of diffuse emission is about 2\amin\ 
to the west.  
We thus propose a galaxy cluster identification.

{\it 90a:} 
Seems to be a spiral galaxy, 
for which the  SDSS-III spectrum provides a classification as
starforming galaxy,  at z=0.0760.
While this could be the counterpart of the ROSAT X-ray emission,
we note that (i) there is no point source in a Swift/XRT pointing
(the one centred on {\it 89}), but (ii) the ROSAT position
falls in the middle of two Swift/XRT sources at
RA(2000.0) = 12 18 25.6, Decl.(2000.0)= +22 50 22.5 and
RA(2000.0) = 12 17 42.1, Decl.(2000.0)= +22 48 43.7,
thus raising the possibility that {\it 89}  and {\it 90} 
form part of a cluster about 7\amin\ in diameter.

{\it 91:} No optical counterpart candidate brighter than $B$ = 21 mag in 
the ROSAT error circle.  
The Swift/XRT observation does not reveal the source, with an upper limit
a factor 5 below the RASS rate.

{\it 92a:} 
The SDSS-III spectrum classifies it as broadline QSO at z=0.13724.

\begin{figure}[ht]
 \includegraphics[width=3.4cm, angle=270]{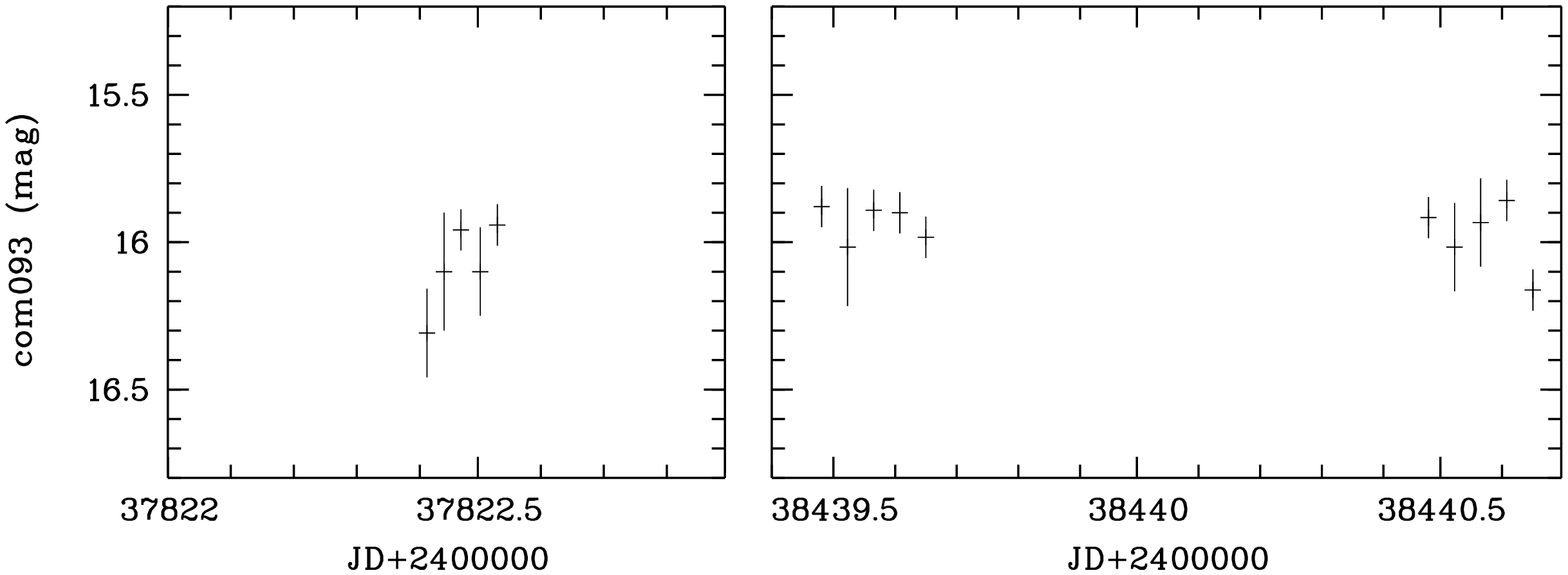}
\caption[]{Blow-up of the light curve of Com093a (Fig. \ref{comolc}) 
showing pronounced variability.}
\label{com93lc}
 \end{figure}

{\it  93a:} Observed minima: J.D. 243 7783.516 (16.2 mag), 7822.417 
(16.3: mag, ascent), 8440.654 (16.2 mag, descent), 244 5021.546 (16.3 mag)
(Fig. \ref{com93lc}).  

{\it  94a:} KW Com = S 10936.  See Richter \etal\ (1995).

{\it  95a:} Other name: NGP9 F378-0391299 (Odewahn \etal\ 1995).
Brightest member of a cluster of galaxies.   
The Swift XRT observation reveals a point source, so the
X-ray emission is unlikely from the cluster.
The  SDSS-III spectrum classifies it as broadline QSO  at z=0.1603.

{\it 96a:} Spectral classification from Garcia Lopez \etal\ (2000) who
  discard Coma Berenices cluster membership, 
which had been proposed by Randich et al. (1996),
   but find spectroscopic evidence for chromospheric activity.

{\it 97a,b:} Invisible on Sonneberg plates.
The  SDSS-III spectrum classifies {\it 97a} as broadline QSO  at z=0.5538,
and {\it 97b} as F5 star.

{\it 98a:} Because of the large distance to the ROSAT source, it is 
unlikely that this is the optical counterpart, despite its optical variability.
From a bright normal 
light we have short (about two hours) minima (in parentheses B magnitude): 
J.D. 243 7795.474 (16.1), 244 2452.491 (16.3), 2480.524 (16.1), 
2887.358 (16.2), 4701.522 (16.2:), 6173.434 (16,2), 6827.618 (16.1), 
6876.473 (16.2), 7206.516 (16.1), 9445.386 (16.2:). Strikingly, from 
1962--1975 only one minimum was observed, but thereafter
eleven minima until 1994. The duration of the minima is about 0.05 days.  
{\it ~98b:} This object coincides with a strong radio source 
  (NVSS 122401+223939; Condon \etal\ 1998), and has therefore been 
 proposed as a BL Lac
  candidate (Sowards-Emmerd \etal\ (2005). Thus, it is the more likely
 counterpart of the X-ray source.
The  SDSS-III spectrum classifies it (with a small $\chi^2$ warning)
as galaxy at z=0.4821.

{\it  99a:} Not tested for variability; too bright for astrograph plates. 
The very soft X-ray spectrum and the \fxo\ ratio are consistent with
a G star counterpart, despite the relatively large distance between optical
and X-ray position.

{\it 100a:} Invisible on Sonneberg plates.
Spectroscopically identified as Seyfert galaxy at z=0.086
by Chen \etal\ (2002). 

{\it  101a:} Clearly visible only on some plates. Mostly fainter than 18 mag. 
Gets as bright as 17.3 mag between J.D. 243 7696 and 7820 and between 
243 9557 and 9609. On Palomar print 1435 (1956 May 21) about $B$ = 18.5 mag. 
On Tautenburg Schmidt plates from 1974 Feb. 15 and 1991 Apr. 9 both about 
$B$ = 17.9 mag.  
The Swift XRT position clearly favours {\it  101b} as counterpart which
is also a radio source (FIRST J125232.6+223338).
The  SDSS-III spectrum classifies it as broadline QSO, at z=0.2113.

{\it 102:} Both optical objects too faint for Sonneberg plates.
The Swift XRT observation clearly detects this bright source, and
identifies {\it 102b} as the counterpart.
The  SDSS-III spectrum classifies {\it 102b} as broadline QSO, 
at z=0.1296.

{\it 103:} No optical counterpart candidate brighter than $B$ = 21 mag 
within either the ROSAT or Swift XRT error circle.   

{\it 104a:} Variability with amplitude \lax 0.2 mag is indicated,
 but is very uncertain.
The Swift XRT position confirms this source as the counterpart.

{\it  105a:} Difficult because of faintness. More frequently faint than 
bright. There are brightness changes up to 0.3 mag within some hours. 
Classified as AGN by Zickgraf \etal\ (2003).
The  SDSS-III spectrum classifies it as QSO at z=0.1019,
with ``starburst broadline'' sub-class.

{\it  106:} Within  X-ray error circle no object brighter than $B$ = 21 mag. 
{\it 106a} is possibly variable, but very uncertain.   
The Swift XRT observation did not detect this source, but due to the 
short exposure this is only mildly hinting at X-ray variability.

{\it 107a:} During the whole interval of observation the brightness declines. 
At the beginning (J.D. 243 7696 - 8530) 16.3--16.7 mag, then ( - 244 5100) 
16.5--17.0 mag, thereafter often fainter than 17.0 mag. Extreme value on a 
good plate (244 7945) at 18 mag. Light variations within one night seem 
questionable. One plate of Tautenburg Schmidt telescope (J.D. 244 2094) 
shows the object at 17.3 mag.  
The  SDSS-III spectrum classifies it as broadline QSO at z=0.5230.

{\it  108a:} Fainter than the Sonneberg plate limit, therefore not tested for 
optical variability.
Pair consists of a QSO at z=0.436 and a star (Oscoz \etal\ 1997).
The  SDSS-III spectrum classifies it as broadline QSO at z=0.4365.
{\it ~108b:} Spectral classification from Garcia Lopez \etal\ (2000) who
  discard Coma Berenices cluster membership
which had been proposed by Randich et al. (1996), 
and find no spectroscopic 
  evidence for chromospheric activity; this and \fxo\ make it unlikely to be 
the optical counterpart.

{\it 109a:} On plates of the Sonneberg astrographs always invisible. 
The  SDSS-III spectrum classifies it as QSO at z=0.3392,
with ``starburst broadline'' sub-class.
{\it ~109b:} No brightness variations. 
{\it ~109c,d:} Two galaxies, possibly intrinsic pair.
{\it ~109e} is invisible on all Sonneberg plates. Magnitude on POSS print from 
1950 Jan 09 about $B$ = 18.5, and from 1955 May 21 about $B$ = 20.5. 
Therefore obviously variable, but too far from the ROSAT position to be 
considered as counterpart of the X-ray source.
Thus, we consider {\it 109a} as the most likely counterpart.

{\it  110a:} 
The  SDSS-III spectrum classifies it as broadline QSO at z=1.004.

{\it  111a:} This very blue object is marginally seen on only a few 
plates, which show it at about 18.4 mag. On POSS 1435 blue print (1955 May 21) 
it is about 18.5 mag. On one Tautenburg Schmidt plate (J.D. 244 8356)
scarcely visible at about 20 mag. Therefore very probably variable.  

{\it  112a:} Not investigated on Sonneberg plates.

{\it  113a:} Single white dwarf PG 1254+223. 
Stellar object with UV excess (Green \etal\ 1986); identified by 
McCook \& Sion (1987). UBV measurements reported by Cheselka \etal\ (1993),
proper motion by Eggen et al. (1967). 
The XMM position (2XMMp J125702.3+220151) is plotted with somewhat
larger uncertainty radius for visibility reasons (Fig. \ref{comfc}). 
This and the very soft X-ray spectrum leaves no doubt about the 
identification.

{\it  114a:} The brightness varies irregularly within some days with 
amplitudes of about 0.15 mag. The mean value of magnitudes varies slowly: 
14.0--14.1 mag in 1962--1967, 14.2 mag in 1972--1978, then brightening, 
14.0 mag in 1991, thereafter slightly fading.  

{\it  115a-c:} Too faint for Sonneberg plates.
The  SDSS-III spectrum classifies {\it 115a} as M3 star. Could be a flare star,
though the spectrum shows only marginal H$\alpha$ emission. M star 
classification and \fxo\ is also marginally consistent with chromospheric
emission. {\it 115c} is a galaxy at z=0.2257 according to SDSS-III,
and unlikely the counterpart because of the large distance of 50\asec\
and the \fxo\ ratio.

{\it  116a:} Brightness usually at about 16.7 mag. From 
J.D. 243 7764 - 7806 about 16.9 mag. In the interval 243 8501 - 8525 fainter 
(about 17.1 mag). From 244 4342 - 4367 brighter (about 16.5 mag), from 
J.D. 244 6113 - 6121 still brighter (16.4 mag), then until 244 6177 fading 
(16.7 mag). Occasionally fading to nearly 17.5 mag.  
The  SDSS-III spectrum classifies it as QSO at z=0.1798,
with ``starburst broadline'' sub-class.

\begin{figure}[ht]
 \includegraphics[width=6.5cm, angle=270]{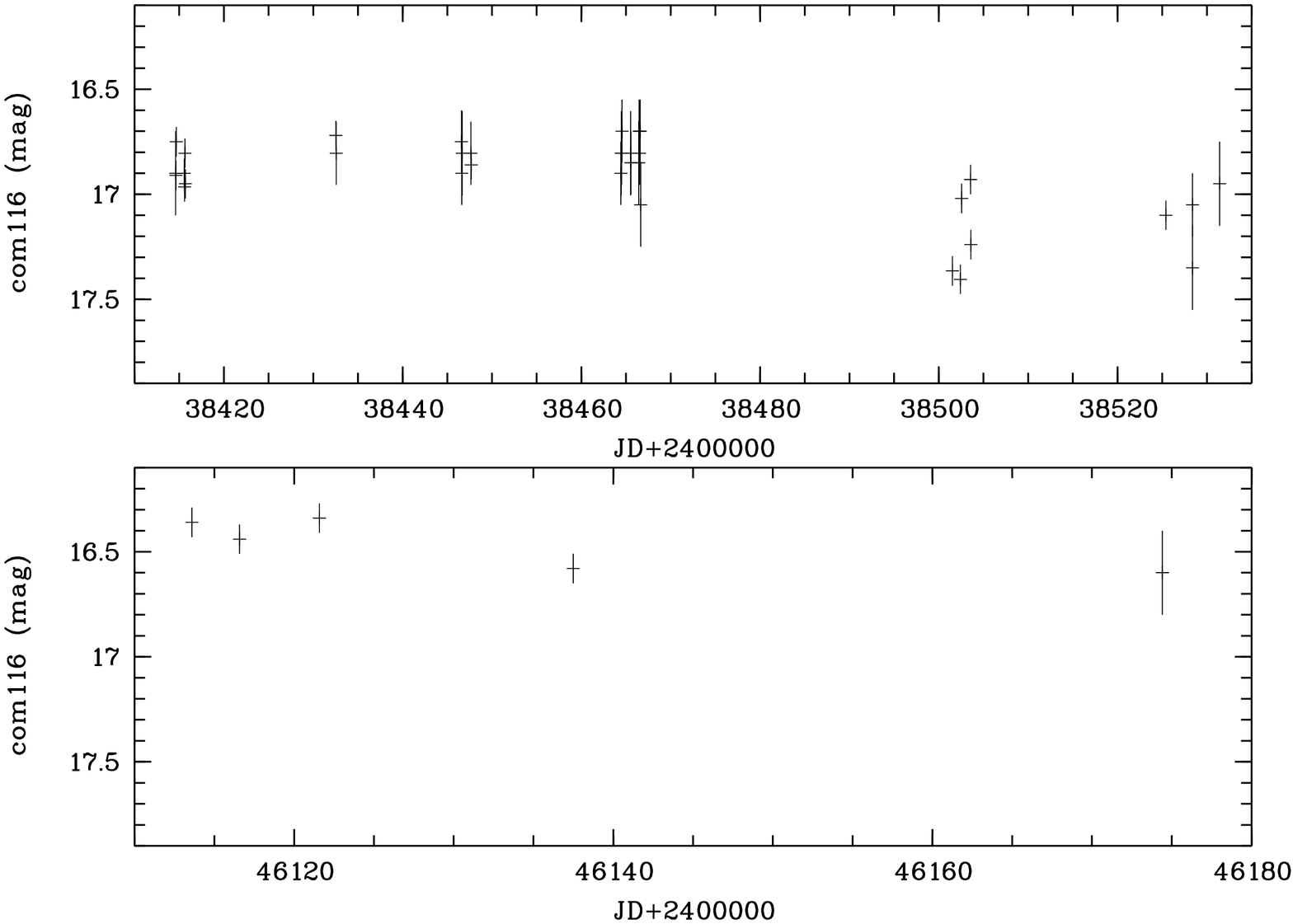}
\caption[]{Blow-up of the light curve of Com116a (Fig. \ref{comolc}) 
showing pronounced variability.}
\label{com116lc}
 \end{figure}

{\it 119a:} Sometimes the object seems to get slightly brighter than 18 mag. 
But this is questionable, because it is at the very plate limit.   
The blue colour suggests an AGN nature.
The  SDSS-III spectrum classifies it as QSO at z=0.4363.

{\it 120a:} Visible only on 20 of the best plates. Within the error circle 
no object brighter than about B = 21 mag. 
The Swift XRT observation did not detect this source, and the
upper limit implies a fading of about a factor three.

{\it  121a,b:} Both objects have blue colour.
In spite of the relatively large optical amplitude of {\it 121a} no 
exact statement
about the brightness variations can be given because of faintness. Small 
magnitudes are favoured. Brightness changes seem to occur within a single 
night. One Tautenburg plate (JD 244 2094) shows the object at $B$=17.7 mag.
The  SDSS-III spectrum classifies {\it 121b} as A0 star.
Due to the smaller distance to the X-ray position and the optical 
variability we prefer {\it 121a} as counterpart over {\it 121b}.

{\it 122a:} Galaxy; the NED reports a redshift of z=0.4, but without
reference. If the redshift is true, this should be an AGN, since X-ray emission
from non-active galaxies is not luminous enough to be seen in the ROSAT
all-sky survey up to z=0.4. 
The  SDSS-III spectrum classifies it as starforming galaxy at z=0.1114.

{\it 123a:} See also Brinkmann \etal\ (1995).

{\it  124a:} Too faint for Sonneberg plates. Blue magnitudes: On POSS 
print No. 135 (J.D. 243 3291): 16.4 mag, on Tautenburg Schmidt plate No. 4004
(J.D. 244 2094): 17.8 mag.  
The  SDSS-III spectrum classifies it as broadline QSO at z=0.3478.

{\it 126a:} Southern (blue) component of a double galaxy inside a cluster 
of faint galaxies (NSCS J122935+213714) at z=0.25 (Lopes \etal\ 2004).
The blue optical colour, X-ray hardness ratio and \fxo\ ratio suggest
an AGN nature.
The Swift XRT observation marginally detects this source, which is
consistent with the very soft X-ray spectrum. The Swift XRT image suggests
also large-scale diffuse emission.
The  SDSS-III spectrum classifies it as broadline QSO at z=0.3484.

{\it 127:} Inside a cluster of faint galaxies. Within the error circle there 
are 3 galaxies fainter than 18 mag. The NED lists {\it 127a} 
as QSO, but without reference. See also Mickaelian \etal\ (2006).
The  SDSS-III spectrum classifies {\it 127a} as broadline QSO at z=0.2336.

{\it 128:} A ROSAT HRI pointing improves the X-ray coordinate to
an error circle of 10\asec\ (1RXH J125627.3+213117) and falling on top 
of the galaxy {\it 128a}.
The  SDSS-III spectrum classifies it as QSO at z=0.0757,
with ``starburst broadline'' sub-class.

{\it 129:} No optical counterpart candidate brighter than $B$ = 20 mag 
within the ROSAT error circle.  

{\it 130a:} Only on the best astrograph plates visible at about $B$ = 18 mag. 
On some poor plates there seem to exist some brightenings to about 16 mag,
but this must be regarded as uncertain. One Tautenburg Schmidt plate 
7482 (J.D. 244 8356) shows the object at about $B$ = 18.5 mag.
The  SDSS-III spectrum classifies it as broadline QSO at z=0.2354.

{\it  131a:} 
Could not be tested for variability because of blend.
The  SDSS-III spectrum classifies it as starburst galaxy at z=0.1156.

{\it 132a:} Galaxy (by extent). Zickgraf \etal\ (2003) reports
``BLUE-WK'', a moderately blue continuum, weak point-like object.
The \fxo ratio suggests an AGN nature. Too faint on Sonneberg plates
for variability assessment. Even invisible on two Tautenburg plates.
The  SDSS-III spectrum classifies it as QSO at z=0.1896,
with ``starburst broadline'' sub-class.

{\it  133a:} Because of the blue colour probably an AGN (or less likely a CV),
and thus more likely the counterpart than the galaxy {\it 133b}.
The  SDSS-III spectrum classifies it as broadline QSO at z=1.4783.

{\it 134a:} 4C+21.35. Too faint for Sonneberg plates.
On POSS print 135 (J.D. 244 3291) about 1.5 mag brighter
  than on Tautenburg Schmidt plate 4004 (J.D. 244 2094).  
See also the optical catalogue of QSOs by Hewitt \& Burbidge (1987) and
 Brinkmann \etal\ (1995).
The  SDSS-III spectrum classifies it as broadline QSO at z=0.4338.

{\it 135a:} S 10938 Com (Richter \etal\ 1995). 

{\it 137a:} Very slow variation with a time scale of months to years.
At the beginning of the observations around JD 243\,7705:  $\sim$13.05 mag.
Later at 243\,7605--7825: $\sim$13.1 mag, 243\,8083--8173: $\sim$13.2 mag,
243\,8410--9205: $\sim$13.0--13.2 mag, 243\,9530--244\,4367: 
$\sim$13.0--13.1 mag,
244\,4634--4697: $\sim$12.9--13.0 mag, 244\,5000--9163: $\sim$12.8--13.0 mag,
244\,9410--9864: $\sim$12.8--12.9 mag. 
The colour indicates that it may be a coronal active object.

{\it 138a:} Non-stellar due to extent, coincident with near-infrared
source 2MASX J12380987+2114014. 
The  SDSS-III spectrum classifies it as galaxy at z=0.1089.
{\it 138b:} Not tested for variability. 
 Spectral type and \fxo\ do not match, so not a counterpart candidate.

{\it  139:} Zickgraf \etal\ (2003) identify the galaxy {\it 139a}
as the counterpart. 
The Swift observation clearly favours the galaxy as the counterpart.
The X-ray intensity is about a factor of three less than during the
ROSAT all-sky survey.
The  SDSS-III spectrum classifies {\it 139a} as galaxy at z=0.0514
with ``AGN broadline'' sub-class, and {\it 139b} as A0 star.

{\it 140a:} 
The  SDSS-III spectrum classifies it as QSO at z=0.1391,
with ``starburst broadline'' sub-class.

{\it 141a:} IR Com = S 10932. Atypical UG star with eclipses (see
Richter \& Greiner 1995b, Kroll \& Richter 1996, and Richter \etal\ 1997). 
The heavy absorption at X-ray wavelengths is likely an artefact of the
small photon number (24) statistics and/or an inappropriate X-ray spectral
model, since the optical colour is pretty blue.
The  SDSS-III spectrum classifies it as CV, and also shows
substantial He II emission. 

{\it  142a:} Other name: HIP 61204.
Variability discovered by {\it Hipparcos} (Hipparcos \& Tycho 
catalogues 1997). Two close components of nearly equal brightness.
Obviously a chromospherically active star.

{\it 143a:} It cannot be excluded that the object is slightly variable with an 
amplitude of about 0.1 mag. However, to verify such variability,
photoelectric observations are required.

{\it  144a:} Narrow-line Sy1 galaxy at z=0.335, with ROSAT data and 
optical variability based on Sonneberg data already reported earlier 
(Greiner \etal\ 1996). The XMM position (2XMMp J122541.9+205503) confirms
the identification beyond doubt.

{\it  145a:} Variability discovered by  {\it Hipparcos} (Hipparcos \& Tycho 
catalogues 1997) from which also the $B$ amplitude is extracted. 
XMM finds two sources one of which 
(2XMMp J123209.9+205508 = 1XMM J123210.0+205507) coincides with {\it  145a}.
The other (2XMMp J123209.4+205553) has no visible optical counterpart
on the POSS sky survey print.

{\it  146a:} Outside the error circle. Not tested for variability. 
Inside the error circle no object brighter than 21 mag.  
A Swift/XRT observation did not reveal this source; the upper limit
implies a fading by a factor of 5.
The  SDSS-III spectrum classifies {\it 146a} as starforming galaxy 
at z=0.0851.

{\it 147a:} Just outside the error circle, but spectral type and \fxo\
suggest this to be the counterpart.
Too bright for astrograph plates. 

{\it  148a:} Blue object. Because of faintness only on the best 25 
Sonneberg plates from 1962--1993 visible. The brightness varies 
irregularly between 17.2 mag and 18.0 mag. 
On some poor plates the object seems to be slightly brighter
than 17 mag, but this must be regarded as questionable.
Two Tautenburg Schmidt plates 
give similar brightness: 17.6 mag at J.D. 244 2094 and 18.0 mag at 244 2453. 
On the other hand, POSS print No. 1435 (1955 May 21) shows the object at 
about 18.5 mag. 
The  SDSS-III spectrum classifies it as broadline QSO at z=1.3540.
{\it 148b} is constant within the error limits.  

{\it 149a:} Galaxy (18.4 mag) just outside the error circle
which the SDSS-III spectrum classifies as starburst galaxy at z=0.1963.
Within the error circle are
 are two faint (about 21 mag) galaxies which are much too faint for Sonneberg 
plates.
The Swift/XRT observation reveals two blobs of diffuse emission about
2\farcm5 away towards the north-west and south-east, respectively,
but no point source at the ROSAT position.
This and the hard X-ray spectrum suggest a galaxy cluster nature.

{\it 150a:} Possibly AGN due to blue colour.
The SDSS-III spectrum classifies it as broadline QSO at z=0.5678.

{\it 151a-c:} Magnitudes of all three objects in Table \ref{comopt} 
are taken from the USNO-B1 catalogue.   Too faint for Sonneberg plates.
The Swift/XRT observation, despite only with 540 sec exposure, 
reveals a 2.5$\sigma$ X-ray source at a rate consistent with that of
the RASS rate, and the X-ray position clearly suggests {\it 151c} as 
counterpart.
The \fxo\ ratio and the relatively hard X-ray spectrum suggest an AGN
nature. The SDSS-III spectrum classifies it as broadline QSO at z=0.8401.

{\it 152a:} Very faint for Sonneberg plates.
Seems to be variable within the limits 17.0--17.3 mag, in 
some cases also fainter. But because near the plate limit, these variations 
are not quite sure. 
The SDSS-III spectrum classifies it as broadline QSO at z=0.6332.

{\it 153a:} On Sonneberg plates diffuse appearance (blend or galaxy).
Classified as AGN by Zickgraf \etal\ (2003).
The SDSS-III spectrum classifies it as QSO at z=0.0756
with ``starburst broadline'' sub-class.

{\it 155a:} Too faint for Sonneberg plates.
$B-V$ colour and \fxo\ ratio suggest an AGN nature, and
Swift XRT localization  proves this identification.
The SDSS-III spectrum classifies it as broadline QSO at z=0.4287.

{\it 156a:} Not tested for variability because diffuse appearance (galaxy).

{\it 157a:} Obviously variable, but with small amplitude (15.6-15.8 mag). 
On two plates of Tautenburg Schmidt telescope similar values. Brightness 
cycles with a duration of about 12 days possibly exist. Superposed are some 
fainter values up to 16.0 mag which, if real, could be the result of 
eclipses in an RS CVn system. Faint observations: J.D. 243 8085.459 
(15.95: mag), 243 8085.501 (15.9 mag), 243 8146.517 (16.0: mag), 244 4701.444
  (15.95: mag), 244 5384.628 (16.0: mag), 244 5403.540 (15.9 mag), 
244 6109.541 (15.9: mag), 244 6563.507 (16.0: mag), 244 7613.457 (16.0: mag); 
244 8683.507 (16.0 mag).  
The X-ray spectrum reveals substantial absorption, implying
an $A_{V} \sim 1.8$ mag, which is consistent with the very red optical colours.

\begin{figure}[ht]
 \includegraphics[width=3.4cm, angle=270]{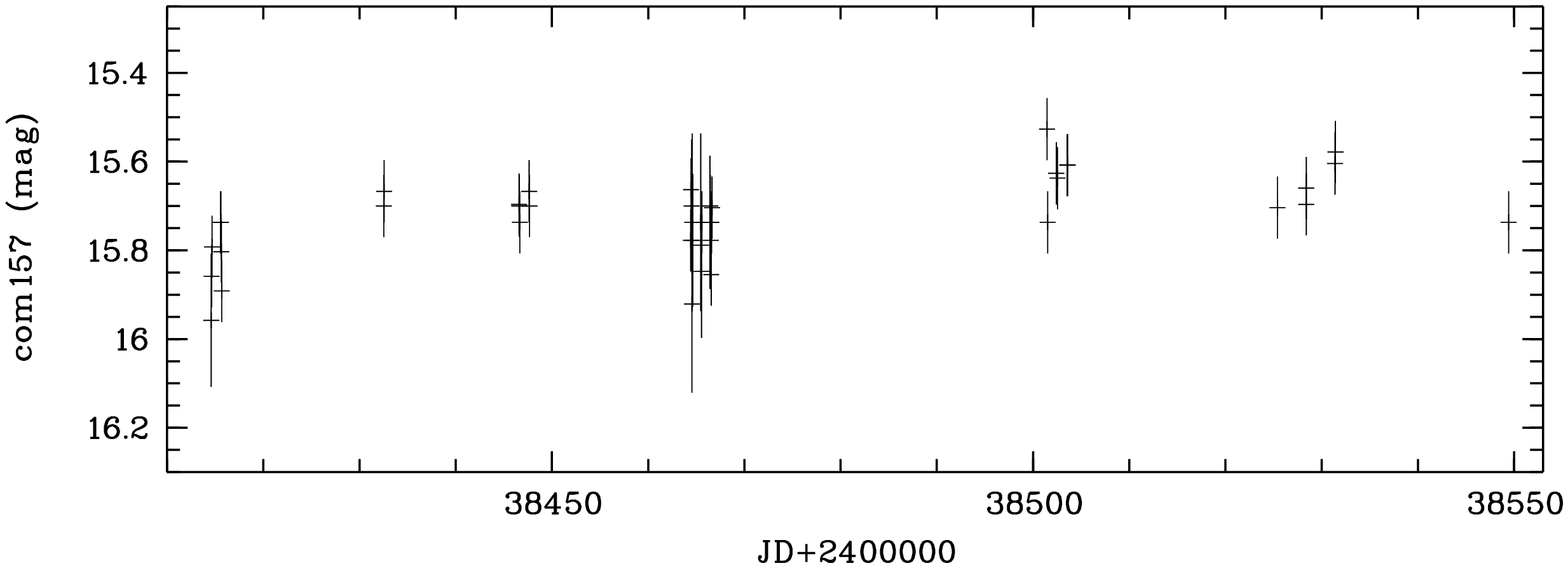}
\caption[]{Blow-up of the light curve of Com157a (Fig. \ref{comolc}) 
showing pronounced variability.}
\label{com157lc}
 \end{figure}

{\it  158a:} Not tested for variability, but possibly a coronal active star.

{\it  159a:} Sy1 galaxy Mrk 771 $\equiv$ TON 1542 $\equiv$ PGC 41532 at 
z=0.063. Stellar object with UV excess (Green \etal\ 1986), see also
optical catalogue of QSOs by Hewitt \& Burbidge (1987), and 
high-energy spectrum by Malaguti \etal\ (1994).
Diffuse appearance on Sonneberg plates, so not tested for variability.

{\it  162a:} S 10935 (see Richter \etal\ 1995). 
Mean  magnitude about 15.5 mag. 
According to  Bade (priv. comm.)  K spectral type.  

{\it  163a:} Irregular light changes. Coincidence with the radio source
7C 1237+2010 $\equiv$   87GB 123717.5+201037 (Brinkmann \etal\ 1997). 
This and the blue colour suggest a  QSO identification
which is proven by the SDSS-III spectrum, giving also z=0.2394.

\begin{figure}[ht]
 \includegraphics[width=3.4cm, angle=270]{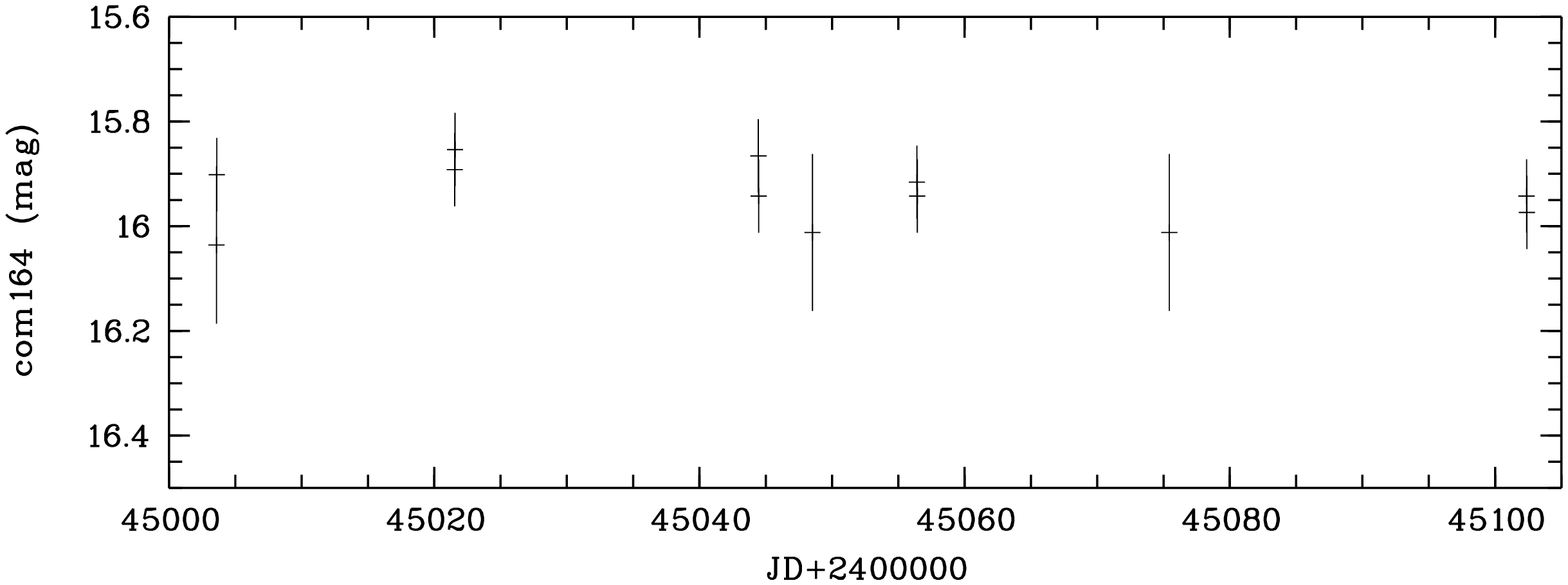}
\caption[]{Blow-up of the light curve of Com164a (Fig. \ref{comolc}) 
showing pronounced variability.}
\label{com164lc}
 \end{figure}

 {\it 164a:} Difficult because of the small amplitude. Irregular variable 
with waves from about some dozens to more than 100 days 
(Fig. \ref{com164lc}),  suggestive of chromospheric activity.
The SDSS-III spectrum classifies it as F9 star.
{\it 164b} is constant on Sonneberg plates.

{\it 165:}  In a 13.7 ksec observation with the ROSAT HRI 
in Jun 2-4, 1995 (Obs.-ID 800435), two nearby sources are detected:
the northern at 12\h26\m43\fss7 +19\degr50\amin46\asec\ (with 0.0018 cts/s),
the southern at 12\h26\m43\fss0 +19\degr50\amin11\asec\ (with 0.0007 cts/s).
The northern source is a factor 4 fainter than the all-sky survey
source, the sum of both sources is still fainter by a factor of 3
(accounting for the factor 3 difference in sensitivity between PSPC and HRI
for spectrally hard sources).
It is not clear whether one of the two HRI sources is related to
the all-sky survey source. Also, none of these two sources
coincides with object {\it 165a}
which the SDSS-III spectrum identifies as galaxy at z=0.2232.
There are several radio sources
within the all-sky survey source's error circle, but none coincides
with either {\it 165a} or the two HRI sources.
The Swift XRT observation reveals a blob of faint, diffuse emission
which is centred between the two HRI positions, but extends much 
further, to about 40 arcsec. The total flux is identical to that
measured during the ROSAT all-sky survey. This suggests diffuse emission,
either from the galaxy {\it 165a} or a larger structure. 

{\it 166:} Cluster of very faint (B $\sim$ 21 mag) galaxies.  The next
 named galaxy cluster is more than 3\arcmin\ away.
The Swift XRT observation reveals faint, diffuse emission which 
extends about 3 arcmin towards the south.

{\it  167a:} Galaxy with number 041 in cluster Abell 1570 (Flin \etal\ 1995)
for which the SDSS-III spectrum provides a redshift of z=0.2139.
The X-ray hardness ratio is large, so the ROSAT emission is rather cluster 
gas emission 
than X-rays from this individual galaxy. The optical cluster has an extent of
10\amin$\times$15\amin, and {\it 167a} is near its centre. 
The Swift XRT observation reveals large-scale, diffuse emission.

{\it  168a:} Classified as AGN by Zickgraf \etal\ (2003), and as 
counterpart of the X-ray source. This is supported by the
Swift XRT position.
The SDSS-III spectrum classifies it as QSO at z=0.1603
with ``starburst broadline'' sub-class.
{\it  168b:} Suspected variable, but too faint for confirmation. 
Light variations are not certain.   
{\it  168c:} The SDSS-III spectrum classifies it as F9 star.

{\it 169a:} On $\sim$70\% of the Sonneberg plates invisible.
Some observations on poor plates near 16 mag are questionable.  
On 4 Tautenburg 
Schmidt plates the $B$ magnitudes are: 17.4 mag (1974 Feb. 16; JD 244 2095), 
17.9 mag (1992 Mar. 1; JD 244 8683), 17.9 mag (also 1992 Mar. 1), and 17.9 mag 
(1992 Mar. 2; JD 244 8684). 
Zickgraf \etal\ (2003) report ``EBL-WK'', an extremely blue
continuum, weak point-like object.
Kouzuma \etal\ (2010) identify it as AGN candidate due to the
NIR colours from 2MASS.
The SDSS-III spectrum classifies it as QSO at z=0.22457 
with ``starburst broadline'' sub-class.

{\it  171a:} Classified as possible AGN by Mickaelian \etal\ (2006).
The Swift XRT position confirms this without doubt as the counterpart.
The SDSS-III spectrum classifies it as QSO at z=0.1313
with ``starburst broadline'' sub-class.

{\it  172a:} Galaxy in cluster ACO 1548 at z=0.16 (see Schneider \etal\ 1983).
The 0.1--2.4 keV X-ray luminosity of 1.3$\times$10$^{43}$ erg/s and 
spectral hardness
suggests cluster emission rather than the galaxy as counterpart.

{\it  173b:} This blue object is at the magnitude limit of Sonneberg plates. 
A variability of small amplitude near 18 mag therefore seemed initially 
doubtful. But the Tautenburg Schmidt plate from J.D. = 244 2453 shows 
the object distinctly fainter (B about 18.6) than on the POSS print 1576 
(J.D. 243 5548,  B about 18.0 mag). 
Therefore the object is probably variable. 
The SDSS-III spectrum classifies it as broadline QSO at z=0.5417.
{\it 173a} is likely also an AGN due to the blue colour, but based on
the optical variability we prefer {\it  173b} as counterpart.

{\it  174a:} Narrow-line Sy1 galaxy at z=0.28, with ROSAT data 
already reported earlier (Greiner \etal\ 1996). 

{\it  175a:} Not in the USNO-A2, but in USNO-B1 catalogue, from
which magnitudes are taken. Coincides within 1\asec\ with radio source 
NVSS 124538+192050 (Condon \etal\ 1998). We therefore identify this 
object as an AGN. 
This is confirmed by the SDSS-III spectrum: broadline QSO at z=0.6930.

{\it 176a:} Bright galaxy pair NGC 4561 $\equiv$ KPG 346 $\equiv$ LEDA 42020
at z=0.0047.

{\it  177:} No optical counterpart candidate brighter than $B$ = 21 mag in 
the ROSAT error circle.  

{\it  178a,b:} Both are viable counterpart candidates, and
both are too faint for Sonneberg plates. 
The SDSS-III spectrum identifies {\it 178a} as broadline QSO 
at z=0.4994, and {\it 178b} as broadline QSO at z=0.9963.

{\it  180a:} There seem to be brightness changes within 0.3 days between
17.0--17.2 mag. But considering the faintness and the small 
amplitude, this is very uncertain.  
The USNO as well as 2MASS colours suggest a K5 star suffering no
extinction, and thus the \fxo\ ratio is (marginally) consistent
with this star to be the counterpart.
The source is not detected in a Swift XRT observation, with an
upper limit about a factor 3 below the RASS rate, so possibly the RASS
detected the object during an active period, explaining the elevated
\fxo\ ratio.

{\it 182a:} Magnitudes taken from USNO-B1.
{\it 182b:} Variability cannot fully be excluded because of relatively 
high dispersion of the magnitudes. 
{\it 182c:} Coincides with the 365MHz radio source TXS 1223+192 
(Douglas \etal\ 1996) which most likely corresponds to GB6 J1225+1858.
Given the radio emission and \fxo, we identify this source as the
optical counterpart of the ROSAT source.

\begin{figure}[ht]
 \includegraphics[width=6.5cm, angle=270]{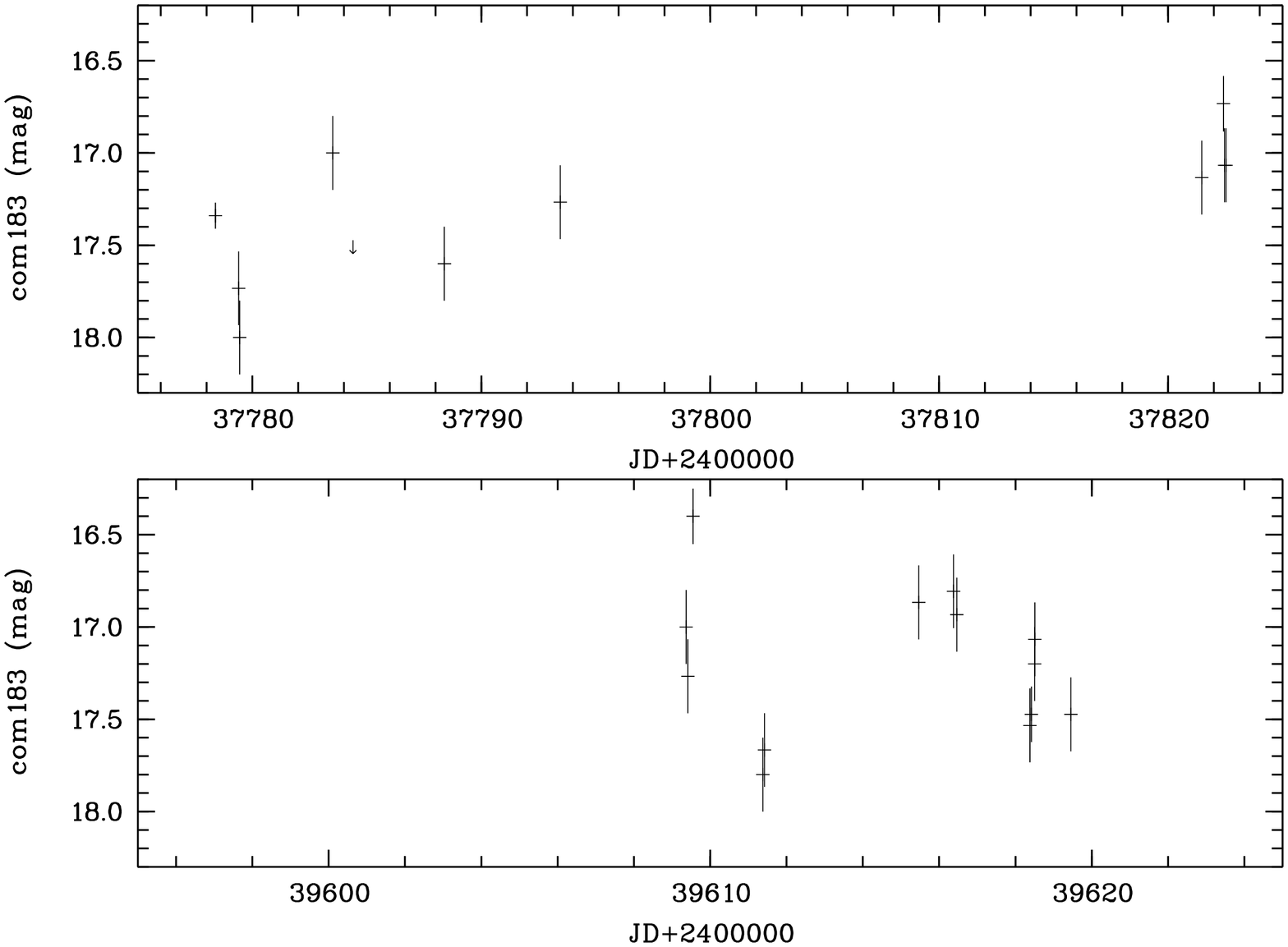}
\caption[]{Blow-up of the light curve of Com183a (Fig. \ref{comolc}) 
showing pronounced variability.}
\label{com183lc}
 \end{figure}

{\it 183a:} Difficult because of faintness. Only visible on good plates. 
The magnitude changes are characterized by a mean brightness level 
from which 
declines occur with a duration of several days. The mean brightness level is 
about 17.1--17.3 mag at the beginning of the observations 
J.D. = (243 7700 -- 8902)  and rises slowly to 16.8--17.0 mag 
(244 4663 - 8362), see light curve in Fig. \ref{com183lc}.  
AGN identification from Zickgraf \etal\ (2003).
The SDSS-III spectrum identifies it as broadline QSO 
at z=0.2223.

{\it 184a:} The brightness varies with a time scale of weeks and months. 
An isolated observation on Tautenburg 
Schmidt plate 4272 (J.D. 244 2453) shows the object at magnitude 17.4.   

\begin{figure}[ht]
 \includegraphics[width=6.5cm, angle=270]{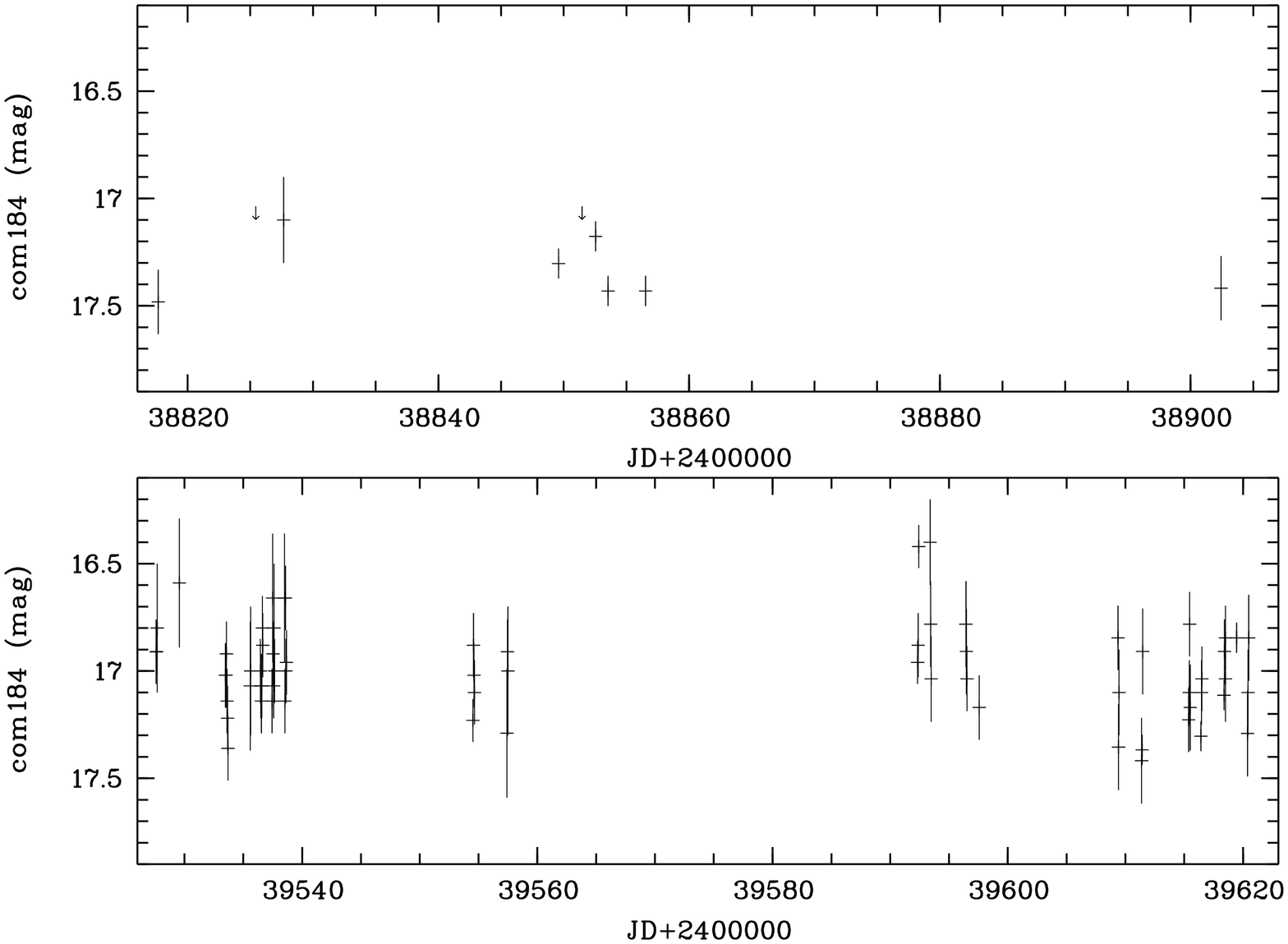}
\caption[]{Blow-up of the light curve of Com184a (Fig. \ref{comolc}) 
showing pronounced variability.}
\label{com184lc}
 \end{figure}

{\it 184a:} 
The SDSS-III spectrum identifies it as broadline QSO at z=0.5992.

{\it 185a:} Difficult because of faintness. There are typically waves with 
a length of about one week and 0.4 mag amplitude. Superposed are slow 
irregular changes. Variations of 0.2 mag within one night seem to occur, 
but this is unsure. On POSS print 89 (JD 243 3391) at 17.0 mag,
and on Tautenburg plates 1576 (JD 243 5548) at 17.2 mag and on 4011
(JD 244 2095) at 18.0: mag.
The SDSS-III spectrum identifies it as QSO at z=0.1970
with ``starburst broadline'' sub-class.

\begin{figure}[ht]
 \includegraphics[width=6.5cm, angle=270]{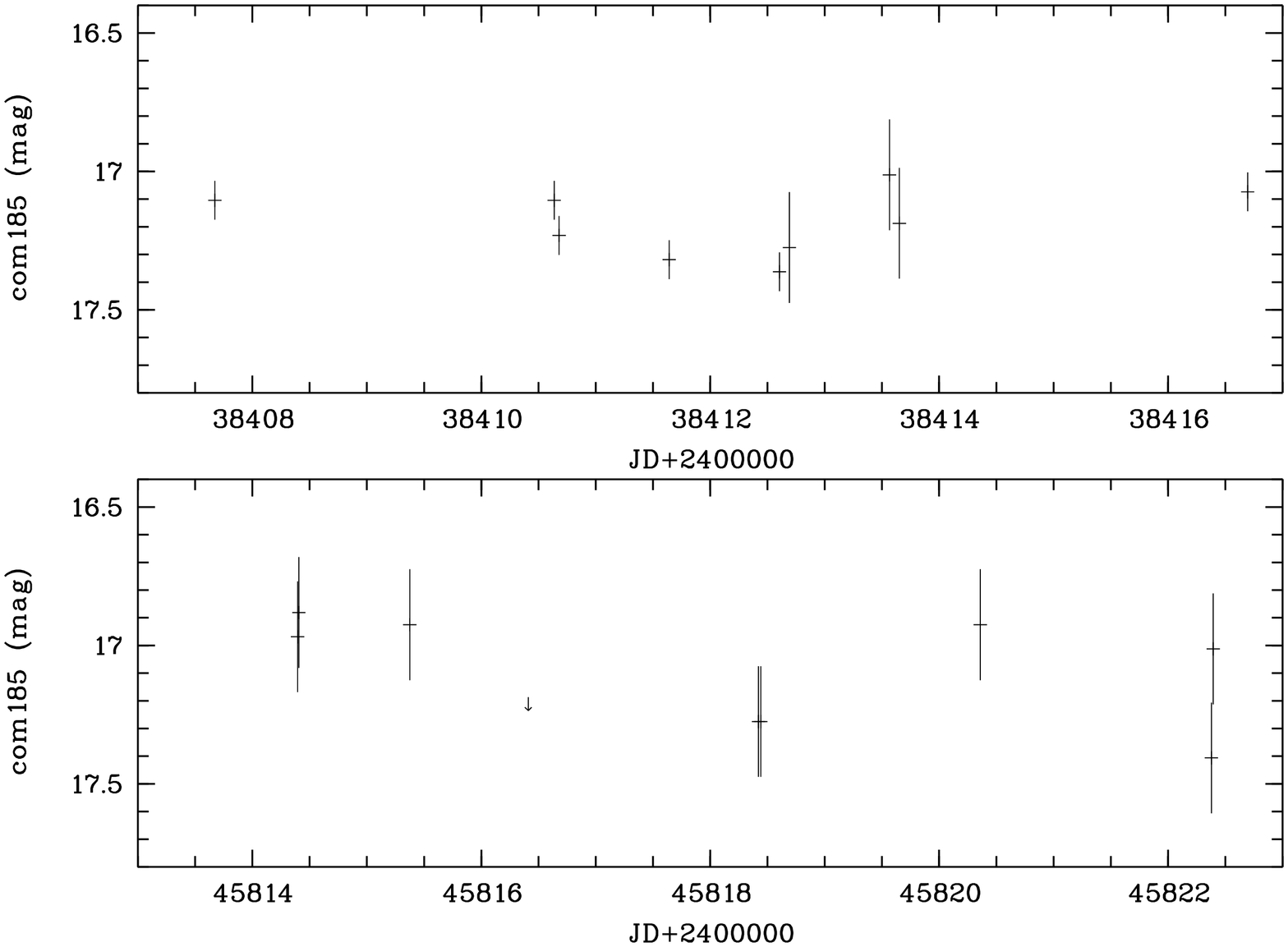}
\caption[]{Blow-up of the light curve of Com185a (Fig. \ref{comolc}) 
showing pronounced variability.}
\label{com185lc}
 \end{figure}

{\it 186a:} During some hours the whole brightness interval is passed 
through, but seemingly not quite regular. A possible relationship to eclipsing 
stars of the W UMa  type is therefore questionable. A relationship to 
RS CVn stars cannot be excluded. The EW star SS Com is 
6\amin\ apart from the ROSAT position, and not considered as counterpart.   

\begin{figure}[ht]
 \includegraphics[width=3.4cm, angle=270]{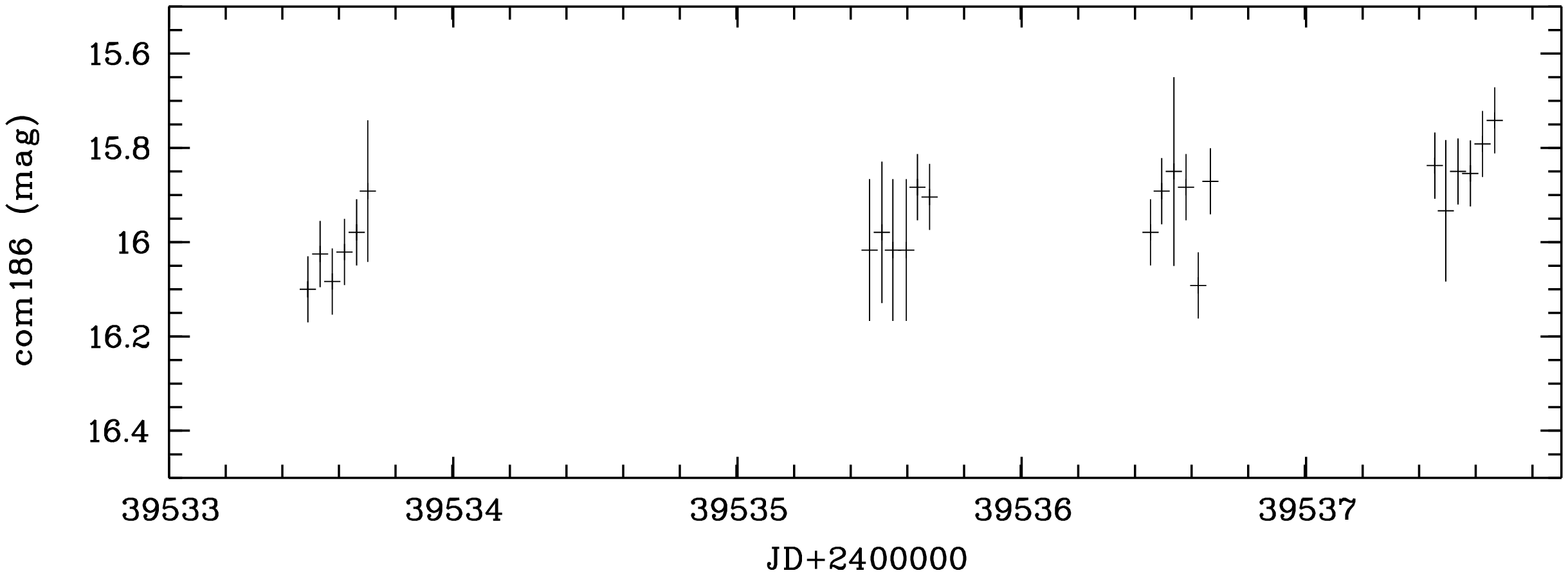}
\caption[]{Blow-up of the light curve of Com186a (Fig. \ref{comolc}) 
showing pronounced variability.}
\label{com186lc}
 \end{figure}

{\it 187a:} QSO at z=0.723 (Wills \& Wills 1976).
See optical catalogue of QSOs by Hewitt \& Burbidge (1987), 
and Brinkmann \etal\ (1995).
Difficult to investigate because near the plate limit. 
Sometimes fadings of short duration (about 1 hour).   

{\it 188b:} Galaxy MCG +03-32-083 at z=0.07,
brightest member of a cluster of galaxies. The X-ray emission
as seen by ROSAT as well as XMM (2XMMp J124117.3+183430)
is not only clearly extended, but elongated with the same orientation as the
galaxy, thus the galaxy itself rather than the cluster is the
most likely optical counterpart. This is supported by the
XMM position which aligns well with the galaxy centre.
Identification also proposed by Zickgraf \etal\ (2003).
The SDSS-III spectrum identifies {\it 188c} as galaxy at z=0.0711.

{\it 189a:} At the border of a cluster of galaxies. The hard X-ray 
spectrum indicates a cluster nature, rather than that of an AGN.

{\it 190:} No optical counterpart candidate brighter than $B$ = 20 mag 
in the error circle. 
HD 106972 is 84" distant from the ROSAT position, too far to be considered
the counterpart.

{\it 191:} No optical counterpart candidate brighter than $B$ = 20 mag 
in the error circle.

{\it 192a:} FN Com = S 8058 (Hoffmeister 1964). 
The observations of Wenzel and Ziegler 
(see Meinunger \& Wenzel 1968) in connection with the yellow colour 
and the X-radiation let us suppose that this is a star of the
BY Dra type.   

{\it 193:} No optical counterpart candidate brighter than $B$ = 20 mag 
within the error circle. The following objects all have a rather large distance
to the X-ray source.
{\it 193a:} The blue colour suggests an AGN classification.
{\it 193c:} QSO at z=1.401, see optical catalogue of 
QSOs by Hewitt \& Burbidge (1987).
It is variable according to Uomoto \etal\ (1976).
Variability is confirmed by observation on POSS (17.1 mag at JD 243 5548)
and Tautenburg plates ($>$17.6 mag at 244 2095, 17.9 mag at 244 8683.60,
%17.8 mag at 244 8683 
and 17.8 at 244 8683.66). On Sonneberg plates beyond the plate limit.

{\it 194:} No object brighter than 18.5 mag within the ROSAT error circle,
both objects {\it 194a} and {\it 194b} are too faint for Sonneberg plates.
{\it 194b} is nearly one magnitude fainter on POSS 1576 than on Tautenburg 
plates, but it is difficult to say whether this effect is caused by variability
or by the diffuse image of this object which may possibly be a galaxy.
The SDSS-III spectrum identifies {\it 194b} as QSO at z=0.2502
with ``starburst broadline'' subclass.

{\it 195:} No object  brighter than 21 mag within the error circle.
Remarkable is its high X-ray luminosity during the ROSAT all-sky survey,
which has declined by at least a factor of 8 to explain the non-detection
in the Swift XRT observation.
The SDSS-III spectrum identifies {\it 195a} as broadline 
QSO at z=0.1341. But due to its large distance to the X-ray centroid
the association remains open.
Turner et al. (2010) list the X-ray source as ``BL Lac-type object''.

{\it 196a:} Triple star. In addition to the second component of M-type 
(R = 13.9, B = 15.7, V = 14.6), there is a faint red third component 
(B about 18.4). The second component is possibly variable, but because 
of blending this is not sure. 
{\it 196a} is listed among the periodic
variables coincident with bright ROSAT sources (Kiraga 2012).

{\it 197a} is an AGN according to Zickgraf \etal\ (2003), despite not  
particularly blue colour in the USNO catalogue. It was
not investigated on Sonneberg plates due to blending with the bright star 
{\it 197b}.
{\it 197c} is a galaxy classified as ``Red-WK'' by Zickgraf \etal\ (2003), 
which is classified as starforming by the SDSS-III spectrum, at
z=0.0667.
It is constant on Sonneberg plates.

{\it 198a:} On two overlapping POSS prints and two Tautenburg plates of
nearly equal brightness.

{\it 199:} The RRc-type 
star AN Com is 96\asec\ away from the ROSAT position and thus not considered
as counterpart. 
The QSO {\it 199a} is the likely counterpart, 
and the SDSS-III spectrum provides z=0.2540.

{\it 200a:} 
Not tested for variability because of diffuse appearance on Sonneberg plates
(blend or galaxy).
The SDSS-III spectrum identifies it as QSO at z=0.0753,
with ``starburst broadline'' sub-class.

{\it 201:} Both objects too faint for Sonneberg plates. The
Swift XRT observation suggests object {\it 201b} as counterpart.

{\it 202a:} Slow light changes (within years) are superposed by waves of 
small amplitude (0.1--0.3 mag) and short duration (several days to 50 days). 
Small variations probably also within one night.   The mean brightness 
rises (with fluctuations) from 16.5 mag since the beginning of the 
observations (J.D. 243 7650) to 16.1 mag (J.D. 243 8080 - 9150) with 
a flat dip of 16.3 in J.D. 243 8500. From J.D. 243 9200 - 9620 variations 
between 16.2 and 16.4 mag. Sporadic observations between J.D. 244 2840 and 
3250 give 16.5--17.0 mag. From J.D. 244 4345 - 6530 bright again between 
16.0 and 16.5 mag. Thereafter (only few observations) 16.4--16.9 mag, only 
at J.D. 244 8350 a little brighter, about 16.3.
The SDSS-III spectrum identifies it as broadline QSO at z=0.3526.

{\it 203:} 
Object {\it 203a} is coincident with the radio source FIRST J124406.5+174831.
The SDSS-III spectrum identifies it as galaxy (no emission lines) 
at z=0.1670. Could possibly be a blazar. Alternatively a cluster of
galaxies, as {\it 203b} is another galaxy (though at unknown redshift), 
and the X-ray spectrum is pretty hard.

{\it  204a:} On POSS print 1572 (1956 Mar. 14) B = 16.6 mag, on 
Tautenburg Schmidt plates No. 4027 (1974 Feb. 17) B = 17.4, 
No. 7869 (1992 Feb 9) B = 18.3 mag, No. 7871 (1992 Feb. 9) B = 18.2 mag.   
On Sonneberg plates invisible.
The SDSS-III spectrum identifies it as QSO at z=0.2630
with ``starburst broadline'' sub-class.

{\it 205a:} Close pair of stars. Not tested for variability.

{\it 206a,b:} It is not easy to decide which of the two objects is the 
better candidate for the ROSAT source. 
While the FG star {\it 206a} is consistent with the soft X-ray spectrum,
its \fxo\ ratio is high, somewhat borderline (see Fig. \ref{stocke}).
On the contrary, the QSO {\it 206b}, 
being at z=0.5478 according to the SDSS-III spectrum
is clearly outside the ROSAT error circle.
We tend to think that {\it 206a} is the counterpart.

{\it 207a:} Too faint for Sonneberg plates. On POSS print and 4 Tautenburg 
Schmidt plates of equal brightness. 
The SDSS-III spectrum identifies it as broadline QSO at z=0.4067.

{\it 208c:} Brightness changes seem to be indicated, but not certain. 
The Swift XRT observation does not detect this X-ray source;
the upper limit is not constraining due to the very soft spectrum during
the ROSAT all-sky survey.

{\it  209a,b:} Blended on Sonneberg plates.
Not tested for variability, no variability found by Hewitt \&
Burbidge (1987). On POSS prints Nos. 1572 and 
1576 (1956 Mar. 14) and on Tautenburg Schmidt plate no. 4027 (1974 Feb. 17) 
of equal brightness.   
The SDSS-III spectrum identifies {\it 209a} as broadline QSO at z=0.1.2822.

{\it 210a:}
The SDSS-III spectrum identifies it as galaxy (no emission lines) 
at z=0.0720.

{\it 211a:} Not in USNO-A2 or USNO-B1, but in APM catalogue, from
which the magnitudes in Table \ref{comopt} are taken.
Only on a single Sonneberg plate barely indicated: 
J.D. 243 8853, B = 17.7: mag. Visible also on two POSS prints: 1956 Mar. 14,  
B = 21 mag and 1956 Mar. 15,  B = 18.4 mag. On three Tautenburg Schmidt 
plates invisible, fainter than 20.5 mag. Therefore variable. Type unknown, 
may be AGN or CV.

{\it 212a:} Blended on Sonneberg plates. 
The Swift XRT position clearly favours the galaxy
which according to the SDSS-III spectrum is a galaxy without emission
lines at z=0.0804.  The red colour, 
low \fxo\ and hard X-ray spectrum argue against an AGN. The Swift XRT
flux is about a factor 4 lower than the ROSAT measured flux.
The hard X-ray spectrum would also be unusual for a tidal-disruption event.

{\it  215a:} Within a cluster of faint galaxies. Remarkable 
is the high X-ray luminosity. Too faint for Sonneberg plates. 
On POSS print and on five Tautenburg plates similar magnitudes.   
Coincides with radio source NVSS 123305+170132 (Condon \etal\ 1998)
 $\equiv$ FIRST J123305.1+170132, therefore likely AGN/Blazar (see also
Laurent-Muehleisen \etal\ 1997). 
The SDSS-III spectrum identifies it as star with
``WD magnetic'' sub-class, but without any clear sign of either absorption
nor emission line this is doubtful. The featureless blue spectrum argues
more in favour of a BL Lac type object.

{\it 216a:} Coincides with the radio source FIRST J123528.8+170036.
The SDSS-III spectrum identifies it as galaxy at z=0.3806.
Likely a BL Lac object.

{\it 217:} Within a cluster of faint galaxies. No optical counterpart
candidate brighter than  $B$ = 21 mag within the ROSAT error circle. 
The Swift XRT observation did reveal extended emission,
suggesting emission from cluster gas.

{\it 218a:} 
QSO LBQS 1229+1711 at redshift 0.209 (Hewett \etal\ 1995).

{\it 219:} 
Invisible on Sonneberg plates.
The Swift XRT observation does not detect this source,
suggesting about a factor 10 variability.

{\it 220:} According to Gioia and Luppino (1994) rich loose cluster of 
galaxies. No dominant galaxy.  Hard X-ray emission (HR1=1) consistent with
cluster-emission. Distance to the {\it Einstein} source
MS 1241.5+1710 is 1\farcm5, which in turn had been
associated to the galaxy cluster at z=0.31 (Stocke \etal\ 1991).
Also ASCA source 1AXG J124359+1653 (Ueda \etal\ 2001).
The Swift XRT emission indicates large-scale diffuse emission, with the
maximum about 1\farcm7 north, at 12\h44\m01\fss2 +16\degr53\amin51\asec,
consistent with the ASCA position.

{\it 221:} Invisible on Sonneberg plates.
Source not detected with a Swift XRT observation.
Given the very soft spectrum of the ROSAT source, the upper limit
is not constraining, however.

{\it 222a:} Probably eclipsing binary. Difficult because of small amplitude. 
Supposed minima, in parentheses magnitudes: J.D. 243 7820.474 (14.6), 8085.668
 (14.6), 8142.454 (14.7:), 9611.380 (14.6), 9618.461 (14.6), 244 4702.540 
(14.7:), 6552.376  (14.6), 7206.536 (14.7), 7613.447 (14.6), 9482.403 (14.7). 
We did not succeed in finding a period. Some of the minima may be erroneous
because of small amplitude. Presumably chromospherically active star.

{\it 223a,b:} Possibly AGN due to very blue colour.
The SDSS-III spectrum identifies it  as broadline QSO at z=2.4468.
{\it 223c:}  Coincides with radio source NVSS 122944+164002
(Condon \etal\ 1998) $\equiv$
FIRST J122944.5+164003, thus most likely a QSO.

{\it 225:} Source not detected with a Swift XRT observation, indicating
a fading of about a factor of 10, or very soft spectrum. 
The SDSS-III spectrum identifies {\it 225b} as starforming galaxy
 at z=0.0763.

{\it 226a:} The elements given by  Meinunger \& Wenzel (1968) 
need only small corrections (Richter \& Greiner 1995a). 
They are: m = 243 7668.521 + 0.7354422 * E. 
The minima are given in Table \ref{cncom}.
The SDSS-III spectrum identifies {\it 226b} as a K7 star,
thus unlikely the counterpart due to the high \fxo.

\begin{table}
\caption{\label{cncom} Minima of CN Com $\equiv$ Com226a}
\begin{tabular}{rcrr}
\hline
\noalign{\smallskip}
 J.D. (2400000+) & B  & E~~ &  B-R (d) \\
\noalign{\smallskip}
\hline
\noalign{\smallskip}
   37668.503 &  13.85 &    0 &  -0.018  \\
   37737.649 &  14.0  &   94 &  -0.004  \\
   37749.416 &  13.9  &  110 &  -0.004  \\
   37752.377 &  14.0  &  114 &  +0.016  \\
   37821.502 &  13.9  &  208 &  +0.009  \\
   38090.665 &  13.7  &  574 &   0.000  \\
   38407.630 &  13.9  &  1005&  -0.010  \\
   38471.633 &  13.7  &  1092&  +0.009  \\
   39205.593 &  13.8  &  2090&  -0.002 \\
   44704.501 &  13.9  &  9567&  +0.004 \\
   45780.439 &  13.9  & 11030&  -0.009 \\
   45822.379 &  13.7  & 11087&  +0.010  \\
   47612.455 &  13.7  & 13521&  +0.020 \\
   48329.487 &  13.8  & 14496&  -0.004 \\
   48357.436 &  13.9  & 14534&  -0.002 \\
   48682.506 &  13.8  & 14976&  +0.003 \\
\noalign{\smallskip}
\hline
\end{tabular}
\end{table}

{\it 227a:} QSO at z=1.017, on Sonneberg plates small irregular brightness
changes, probably also within some nights.   
The SDSS-III spectrum identifies {\it 227b} as galaxy at z=0.0635.

{\it 228:} The Swift XRT observation reveals a bright, extended
source. Together with the hard X-ray spectrum this suggests a cluster origin.
Thus, {\it 228a} is likely not the counterpart.

{\it 229a:} Very close pair of stars with similar $B$ magnitude. 
A supposed variability of the northern component is uncertain because of 
faintness and blending. Identification as K star from Mickaelian \etal\ (2006).
{\it 229b} is probably a AGN due to its blue colour, and therefore a
good counterpart candidate. 

{\it 230:} See Brinkmann \etal\ (1995).
In the midst of a cluster of galaxies.  
The X-ray emission could also be associated to the Abell cluster 
ACO 1569 (see e.g. Gomez \etal\ 1997), rather than to the radio galaxy 
{\it 230a} at z=0.0684.

{\it 231a:} Other name 3C 275.1. Optical identification see 
Sandage \etal\ (1965). According to Uomoto \etal\ (1976) and
Hewitt \& Burbidge (1987) variable. See also Brinkmann \etal\ (1995).
The SDSS-III spectrum identifies it as broadline QSO at z=0.5552.

{\it 232:} Invisible on Sonneberg plates.
Not detected in a Swift XRT observation.

{\it 233:} Invisible on Sonneberg plates.
Marginally detected in a Swift XRT observation. Due to the
soft X-ray spectrum of the ROSAT source, the low count rate is
consistent with the ROSAT rate.
The \fxo\ ratio suggests an AGN nature.

{\it 234a:} Invisible on Sonneberg plates.
Likely  an AGN due to its blue colour on POSS sky survey prints.
The Swift XRT position identifies this object as counterpart.
The radio source FIRST J123439.3+160611 (Becker \etal\ 1997,
catalogue version of Apr 2003; positional error of 1\asec)
does not coincide with {\it 234a}. 

{\it 235a:} Other name: SDSS J123544.26+160536.3; QSO at z=0.07
(data release 5 as obtained Jun. 28, 2006;
http://www.sdss.org/dr5/products/spectra/getspectra.html).

{\it 236b:} Other name: SDSS J123942.92+160612.1;  galaxy at z=0.07.
(data release 5 as obtained Jun. 28, 2006;
http://www.sdss.org/dr5/products/spectra/getspectra.html). See also 
Mickaelian \etal\ (2006).
The SDSS-III spectrum identifies it as starforming galaxy at
z=0.0702.
{\it 236c:} slightly outside the error circle of ROSAT position, not tested 
for variability.  
{\it 236d:} Coincides with NVSS 123945+160600 (Condon \etal\ 1998)
$\equiv$ FIRST J123945.8+160558 (Becker \etal\ 1997; catalogue version of 
Apr 2003), thus likely AGN.
The SDSS-III spectrum identifies it as starforming galaxy at z=0.2808.
The Swift XRT position suggests this object as the counterpart of the
ROSAT source.

{\it 237:} No optical counterpart candidate brighter than $B$ = 21 mag 
in or near the error circle of the
ROSAT position. The X-ray source is located near the border of our
extracted X-ray data region, and its detection significance is 
near the detection limit, thus it could potentially be a spurious X-ray 
detection.
It is not detected in a Swift XRT observation.

{\it 238:} The soft X-ray spectrum and the \fxo\ ratio support the 
identification with the bright FG star.

\subsection{Notes on individual objects in the Sge field (Table \ref{sgeopt})}
\label{sec:IDSgeNotes}

{\it 1a:} A0 star HD 185334, coincident with the radio source 
  WSRTGP 1935+2408 (Taylor \etal\ 1996).
  The RXS position favours {\it 1b} as optical counterpart.

{\it 2a:} Algol star with variable period length (see GCVS). Whether it is 
a chromospheric active star is still to be proved.  

{\it 3a:} The USNO-A2 magnitudes are incompatible with the spectral type;
the $B$ magnitude must be wrong. 
{\it 3b:} GSC 2141 123 is a very close pair of stars of nearly equal 
$B$ magnitudes.

\begin{figure}[ht]
 \includegraphics[width=6.5cm, angle=270]{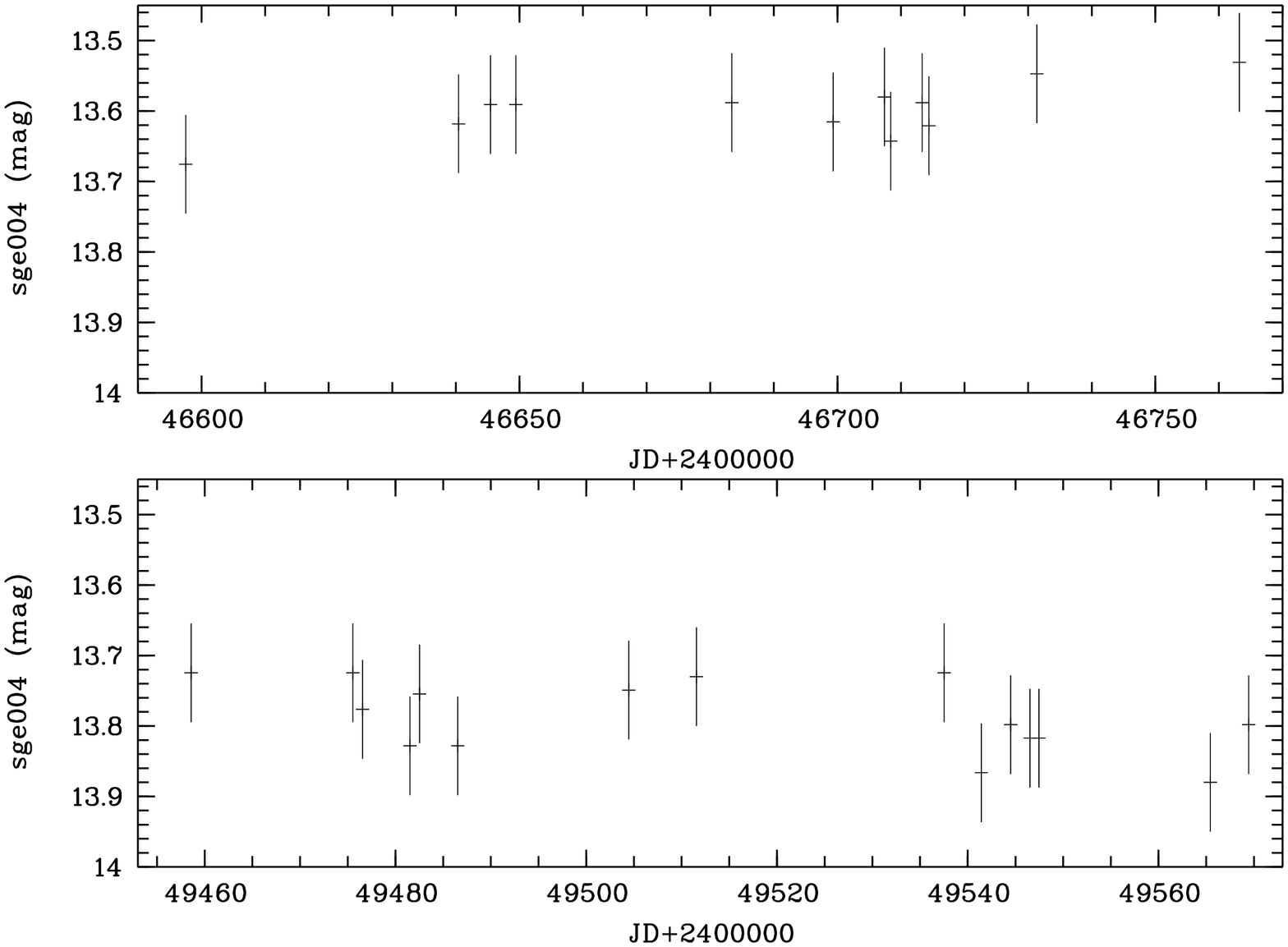}
\caption[]{Blow-up of the light curve of Sge004a (Fig. \ref{sgeolc}) 
showing pronounced variability.}
\label{sge004lc}
 \end{figure}

{\it  4a:} Optically variable with short waves (4--10 days) of small 
amplitude (0.2 mag) and long waves (years) with 0.3 mag amplitude
(Fig. \ref{sge004lc}).  

{\it  5a:} The Swift XRT position is marginally consistent with
this object, so we suggest it as counterpart of the ROSAT source though
an optical variability could not be established photographically.

{\it  6a:} Soft X-ray spectrum is consistent with G spectral type.
{\it  6b:} Close pair of stars. 
On Sonneberg astrograph plates totally blended by {\it 6a}.

\begin{figure}[ht]
 \includegraphics[width=6.5cm, angle=270]{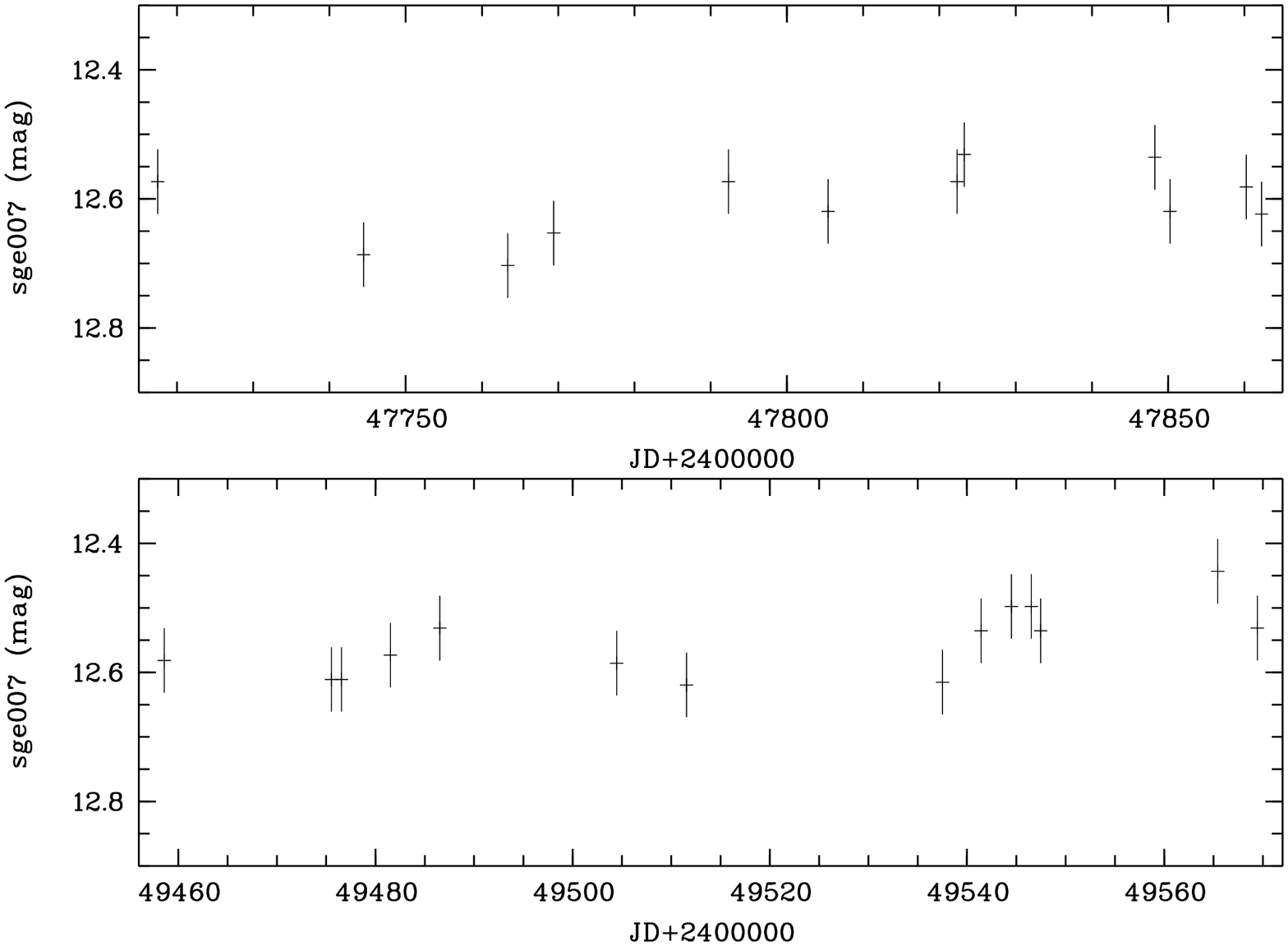}
\caption[]{Blow-up of the light curve of Sge007a (Fig. \ref{sgeolc}) 
showing pronounced variability.}
\label{sge007lc}
 \end{figure}

{\it  7a:} Variability near the detectability limit. Amplitude only 0.2 mag. 
Waves with a length of several days to some months. Standstills of several 
100 days at about 12.6 mag (Fig. \ref{sge007lc}).  

{\it  8a:} On Sonneberg plate unresolved blend of four stars. 
The Swift XRT position falls on the centre of the top three, and
  excludes the southern-most object.
Optical colours suggest late-spectral type, or counterpart is behind substantial
foreground extinction.

{\it 9a:} Brighter (eastern) component of a double star. 
Variability is difficult to assess because the object is near the edge of 
the plate. From time to time waves with a length of about one month. 
The X-ray emission is extremely soft, suggesting a small distance.
Not detected in a 3.6 ksec Swift XRT pointing; the upper limit is
consistent with the ROSAT flux, given the very soft spectrum.

 {\it 10:} Star cluster NGC 6823. {\it 10b} seems to be just visible 
on some plates though the USNO magnitude is about 1.5 mag fainter 
than the plate limit. Nevertheless, the variability could not 
be definitively confirmed. {\it 10g} is identical with Hoag 9, a B0.5V star 
(Massey \etal\ 1995) which is unlikely the counterpart due to the 
hard X-ray spectrum and too large \fxo\ for a B star and the large 
distance from the X-ray position.
The Swift XRT position suggests {\it 10e} to be the counterpart.

{\it 11a:} Not tested for variability. 
The G spectral type and the \fxo\ ratio are consistent with 
identification as counterpart. This is supported by the Swift XRT
source position.

{\it  12a,b:} On Sonneberg plates difficult because of blending. 
The components of the double star 12a cannot be clearly separated on Sonneberg
plates. Therefore, the investigation of the combined light curve can be
misleading. Thus, it cannot be ruled out that some observed flares of
about 0.5 mag amplitude are only spurious, and, if real, no assignment
can be made to one or the other component.
The Swift XRT position suggests the south-western component of 
{\it 12a} as the counterpart.

{\it  13:} There is little doubt on {\it  13a} being the counterpart 
which is also verified by the Swift/XRT position as this object
is contained in the pointing on Sge014.

{\it  14:} The Swift XRT position identifies 
{\it 14a} as the counterpart.

{\it  15a:}
The G spectral type and the \fxo\ ratio are consistent with 
identification as counterpart.
The X-ray source is not detected in two Swift/XRT pointings on
2007 Nov. 29 (2163 ksec) and 2008 Feb. 29 (1675 sec); the upper 
limits are consistent with the brightness of the ROSAT source,
given its soft spectrum.

\begin{figure}[ht]
 \includegraphics[width=6.5cm, angle=270]{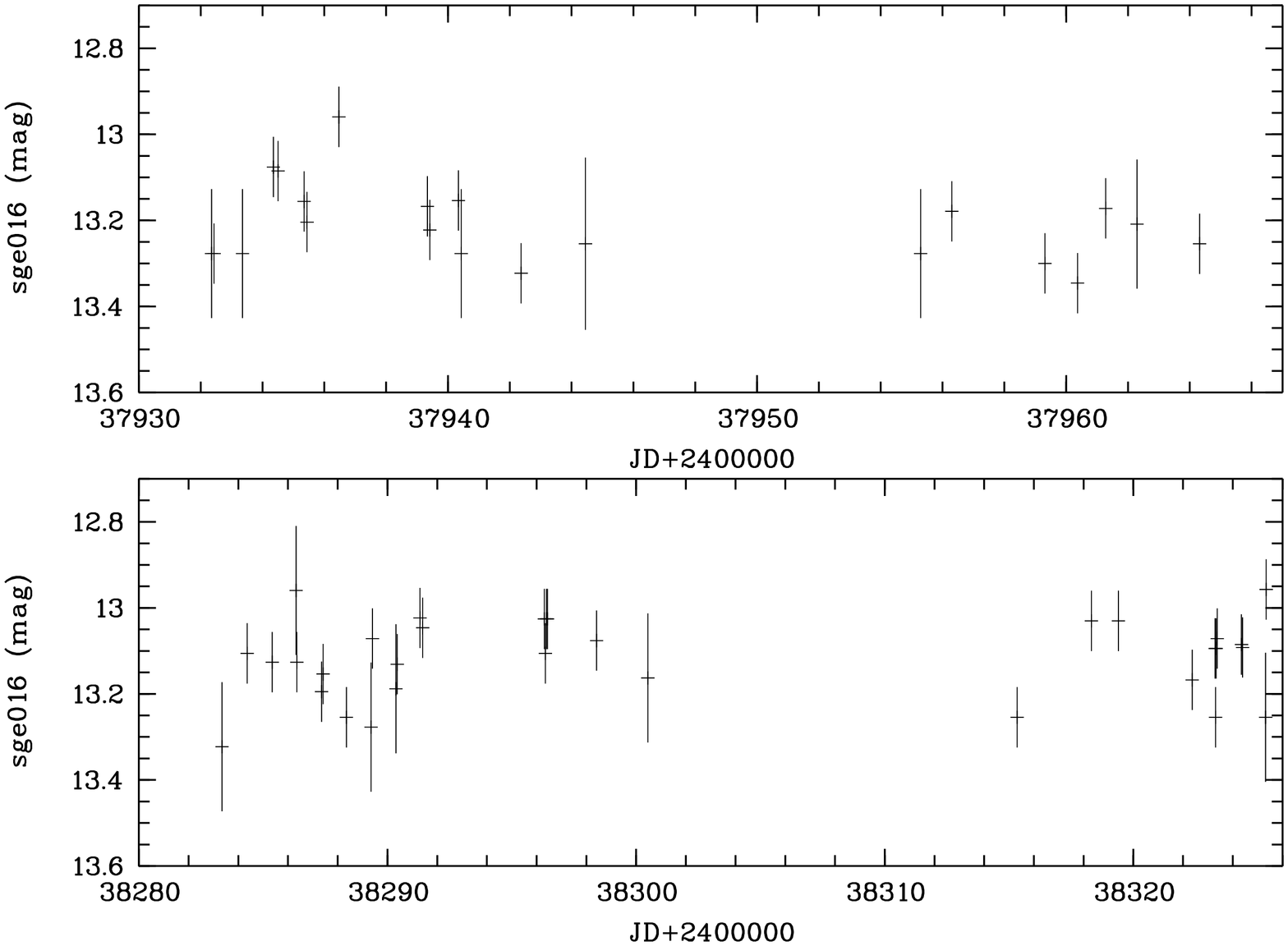}
\caption[]{Blow-up of the light curve of Sge016a (Fig. \ref{sgeolc}) 
showing pronounced variability.}
\label{sge016lc}
 \end{figure}

{\it 16a:} The optical brightness varies quickly with small amplitude between 
12.9 mag and 13.3 mag. There seem to exist also brightenings or flares
(see Fig. \ref{sge016lc} and \ref{sgeolc}): 
J.D. 243 7936.471 (25),  B = 12.9 mag, 8530.480 (22), 
B = 12.8 mag, 9238.585 (23), 12.2 mag (in parentheses: exposure time in min.).  
The X-ray source exhibits rather strong intensity, with practically no 
neutral hydrogen absorption, thus it must be at a rather small distance. 
The X-ray light curve shows a bright
(2.2$\pm$0.3 cts/s) flare during one scan, with levels of 0.25$\pm$0.13 cts/s
and 0.31$\pm$0.17 cts/s before and after (96 min apart).
This GSC star has already been
proposed as counterpart of this ROSAT source (Motch \etal\ 1998),
and is classified as high-proper motion star in Ivanov (2002).

{\it  17a,b:} These objects are at no time brighter than the plate 
limit 17.5 mag. The Swift XRT position identifies 
{\it 17c} as the counterpart which is difficult to measure on Sonneberg plates
due to blending.

{\it  18a:} Brighter component of a double star. Difficult to treat 
because of small amplitude and blend. The colour indicates that it may 
be a star of about F or G type. Minima of short duration seem to be 
indicated, which could be interpreted as eclipse minima of an RS CVn star, 
if real.  

\begin{figure}[ht]
 \includegraphics[width=6.5cm, angle=270]{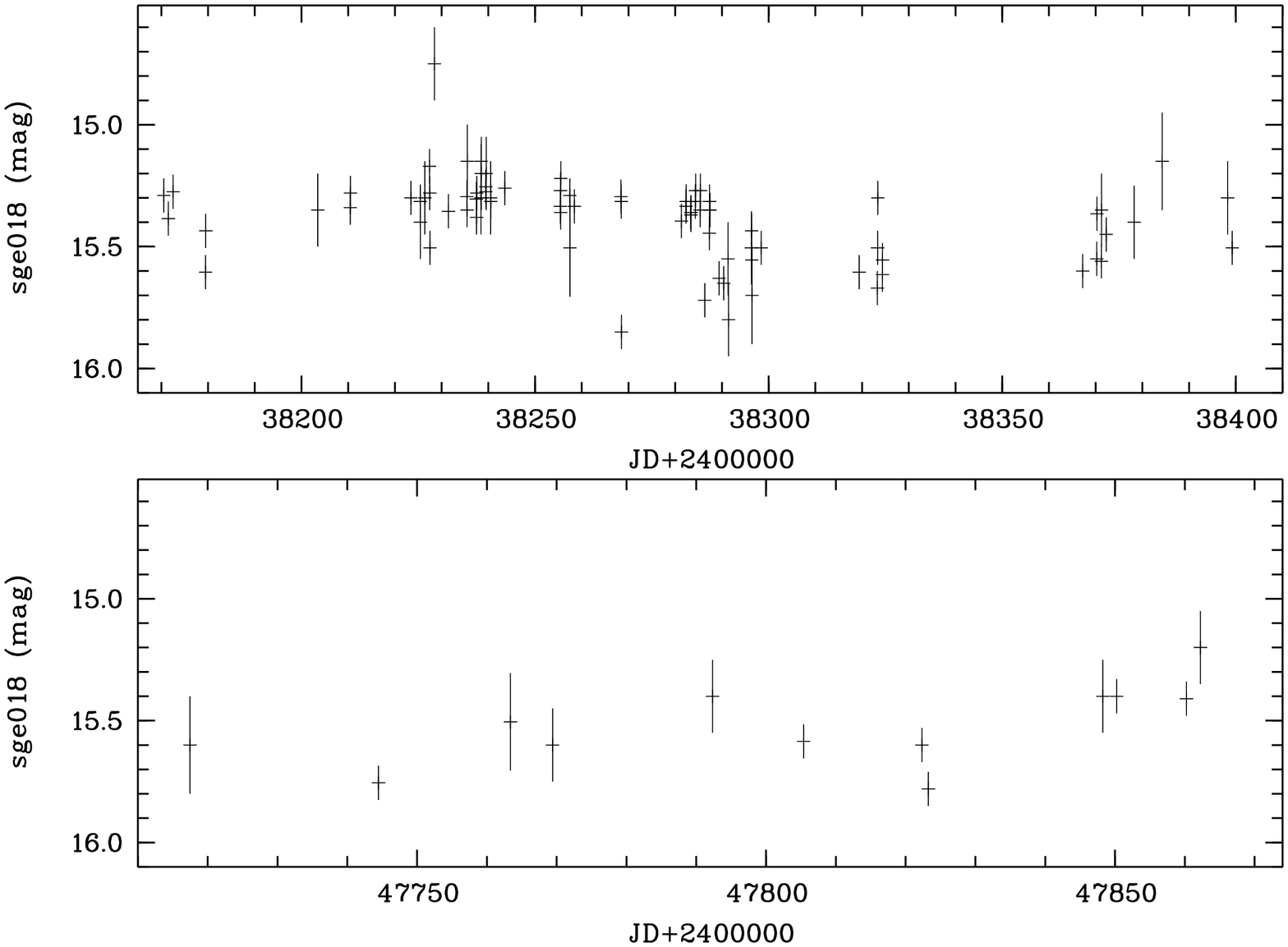}
\caption[]{Blow-up of the light curve of Sge018a (Fig. \ref{sgeolc}) 
showing pronounced variability.}
\label{sge018lc}
 \end{figure}

{\it 19:} Central star of planetary nebula M27 $\equiv$ NGC 6853 (Dumpbell 
nebula). Supersoft X-ray spectrum (Kreysing \etal\ 1992).

{\it 20a:} Soft X-ray spectrum consistent with identification
with this bright G star.

{\it 21a:}  Discovered already by {\it Einstein} Observatory observations,
and identified by Nousek \etal\ (1984). Lightcurve published by
Fuhrmann (1984).

\begin{figure}[ht]
 \includegraphics[width=3.5cm, angle=270]{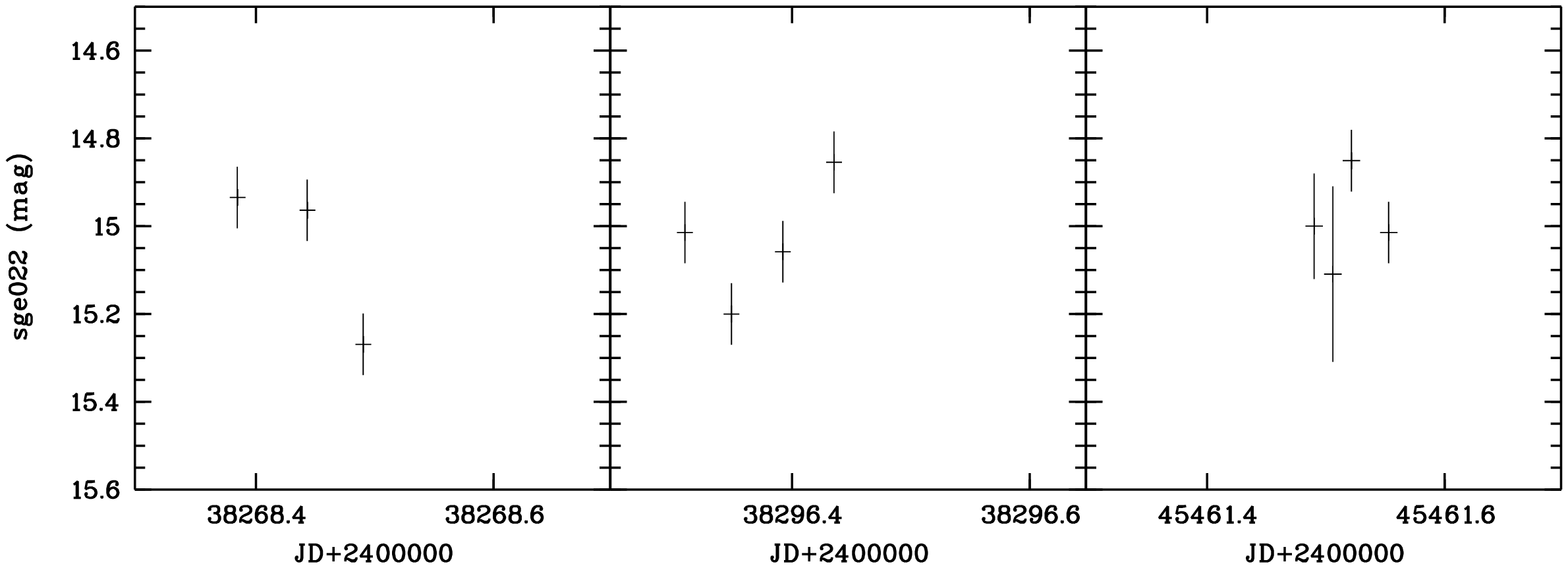}
\caption[]{Blow-up of the light curve of Sge022b (Fig. \ref{sgeolc}) 
showing pronounced variability.}
\label{sge022lc}
 \end{figure}

{\it 22a:} Double star.
{\it 22b:} Short period eclipsing star, may be 
of RS CVn type. Observed minima (in parentheses B magnitude): 
J.D. 243 0848.512 (15.3), 6807.367 (15.2), 8268.493 (15.3), 8296.353 (15.2), 
8323.332 (15.2), 244 4116.430 (15.2), 7848.262 (15.3), 8894.371 (15.2)
(Fig. \ref{sge022lc}). 
We did not succeed in finding a period. 

\begin{figure}[ht]
 \includegraphics[width=3.4cm, angle=270]{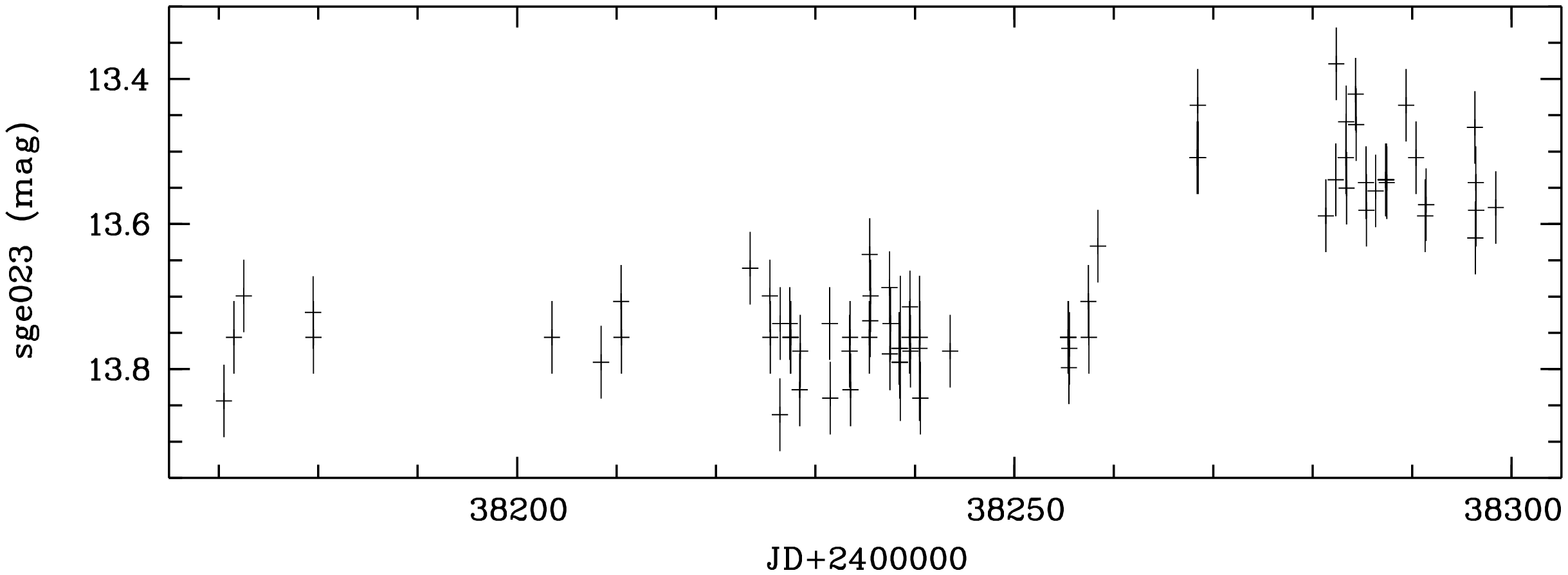}
\caption[]{Blow-up of the light curve of Sge023a (Fig. \ref{sgeolc}) 
showing pronounced variability.}
\label{sge023lc}
 \end{figure}

{\it  23a:} Rapid magnitude changes, mostly between 13.5 and 13.8 mag, 
occasionally a little brighter (13.4) (Fig. \ref{sge023lc}). 
There is a single deep minimum 
(14.0 mag, J.D. 243 3561). Sometimes brightness changes within some hours. 
Probably chromospherically active star.

{\it 24a:} 
Discovered in X-rays with {\it Einstein} Observatory observations, 
it has been originally interpreted by Takalo \& Nousek (1985) 
as an old nova,
but based on higher-quality optical spectra, Shara \etal\ (1990)
identifies it as starburst galaxy at z=0.029.
Light changes on Sonnerberg plates already reported by Andronov (1988),
with outbursts up to 1.7 mag amplitude at several occasions. 
IRAS data are given by Harrison \& Gehrz (1991).
The X-ray hardness ratio as well as the optical colours are consistent 
with a source experiencing the full galactic absorbing column.

{\it 25:} Not detected in a 6.5 ksec Swift XRT pointing on 2008 Feb. 28/29;
 the upper limit of $<$0.001 cts/s is not consistent with the ROSAT flux 
despite the very soft spectrum, but implies about a factor 3 variability.

{\it 26a:} Relatively hard X-ray spectrum is inconsistent with G star
identification unless foreground extinction is considerable.
However, Swift XRT position clearly favours {\it 26a} as counterpart.
Possibly interacting binary?

{\it 28:} For object {\it a}, the NSV  catalogue (Kholopov \etal\ 1982) 
indicates that the variability of this star is doubtful. But the coincidence
with a ROSAT source might suggest that it indeed 
may be variable with a small photoelectrically measurable amplitude, possibly 
a chromospherically active star. The position as derived from a pointed 
ROSAT HRI observation excludes the other two objects {\it 28 b,c} as 
counterpart candidates. {\it 28a} is in the list of double stars
by Jevers \& Vasilewskis (1978).

{\it  29a:} Possibly variable at the detection limit of photographic 
plates. Amplitude smaller than 0.2 mag. Chromospherically
 active star? The Swift XRT position clearly favours {\it 29a} 
as counterpart.

{\it 30a:} Relatively hard X-ray spectrum is inconsistent with G star
identification. Furthermore,
this G star is the brightest member of a blend of several fainter stars 
(see Fig. \ref{sgefc}), so other stars of this blend may be potential
counterpart candidates. The Swift XRT position uncertainty prohibits
solving this ambiguity.

\begin{figure}[ht]
 \includegraphics[width=3.4cm, angle=270]{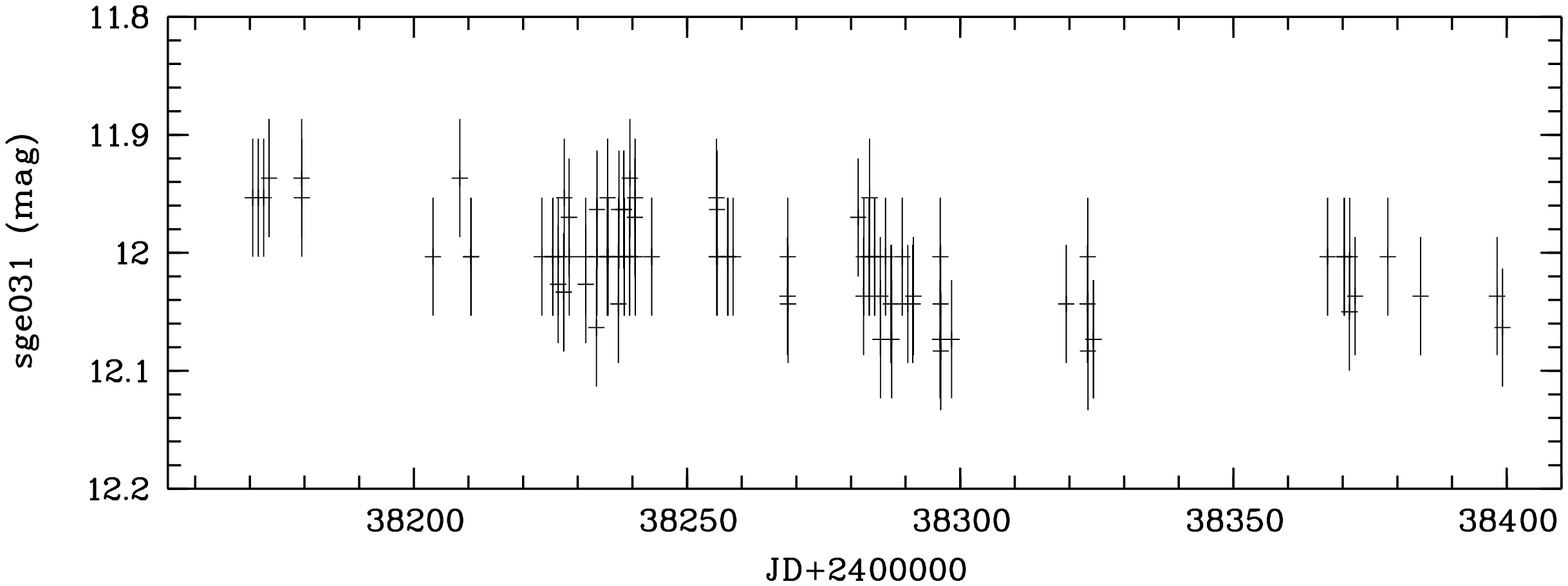}
\caption[]{Blow-up of the light curve of Sge031a (Fig. \ref{sgeolc}) 
showing pronounced variability.}
\label{sge031lc}
 \end{figure}

{\it  31a:} Irregular variable with an amplitude of only 0.2 mag, 
therefore very difficult. May be chromospherically active (see also
Fig. \ref{sge031lc}). 

{\it  32:} 
The 4.5 ksec Swift XRT observation on 2007 Dec. 28 does not detect
the ROSAT source. Given the hard (or absorbed) spectrum, this implies
an X-ray variability of a factor of $\approx$9.
The optical brightness dispersion of {\it  32b}  is relatively high, but is not 
sufficient to confirm variability.  However, since the other two optical
sources within the ROSAT error circle are constant, we tentatively
identify the ROSAT source with {\it  32b}.

\begin{figure}[ht]
 \includegraphics[width=3.4cm, angle=270]{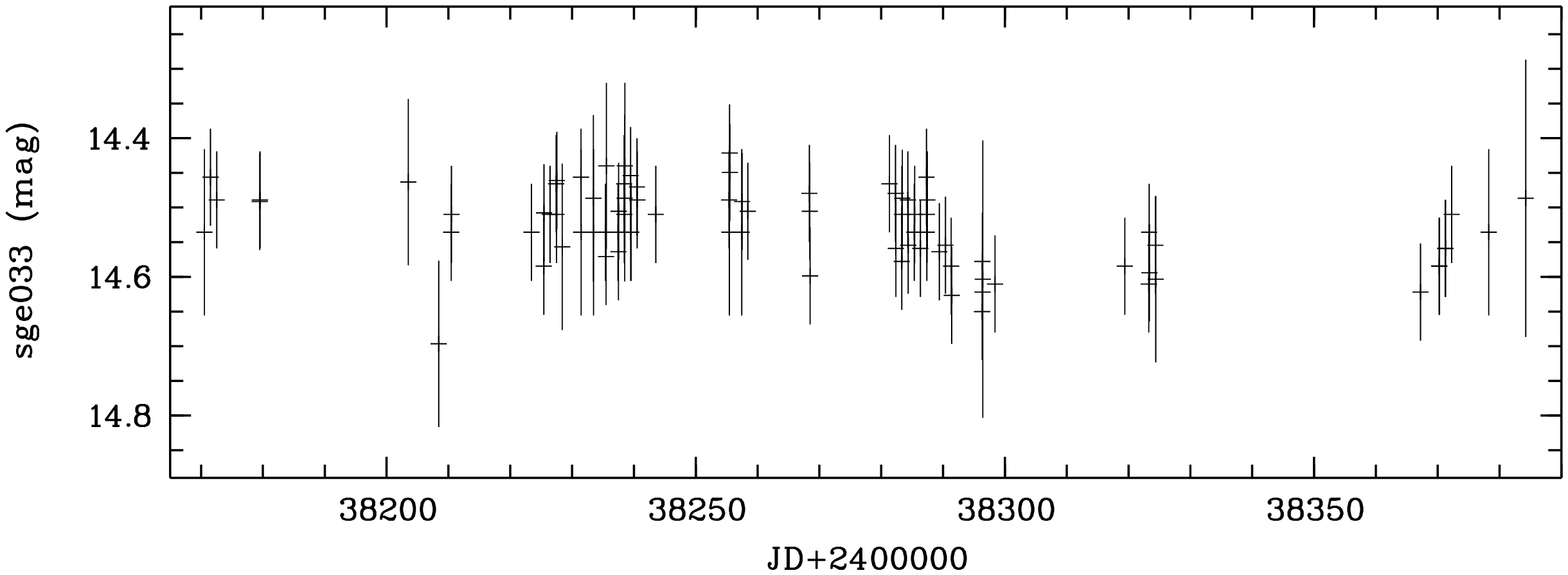}
\caption[]{Blow-up of the light curve of Sge033a (Fig. \ref{sgeolc}) 
showing pronounced variability.}
\label{sge033lc}
 \end{figure}

{\it 33a:} Flat waves with a length of about 50 days (Fig. \ref{sge033lc}). 
In addition slow 
changes within years. Ten exposures in different spectral regions of the
Tautenburg Schmidt telescope could be used: two B plates (JD 243\,8004
and 8005 show the object at 14.8 and 14.9 mag, respectively.
In the U band the object is about 0.2 mag brighter with one exception
at JD 243\,8302 (having 15.3 mag), thus implying 0.7 mag amplitude. 
Whether this is an eclipse minimum
(the only one observed) or a plate fault is difficult to decide. 

{\it  34a:} No brightness changes could be found - constant within 0.1 mag. 
Stars of spectral type A are usually very faint X-ray sources, thus the
\fxo\ ratio is somewhat high to claim the identification with certainty.
But the Swift XRT position supports this ID.

{\it 36:} The 5.4 ksec Swift XRT observation on 2007 Jul. 24 does not 
detect the ROSAT source. Given the hard (or absorbed) spectrum, this implies
an X-ray variability of a factor of $\approx$15.
{\it 36a} seems to be variable within 0.1 mag, but uncertain. 
The Mira star DD Vul is 5\farcm5 apart from the ROSAT position, thus
not considered as counterpart.

{\it  37a:} Difficult because of blending by {\it 37b}. Brightness changes of 
small amplitude. The few (23) observations during the time interval 
J.D. 242 9130 - 243 1320 show the object always bright (12.6--12.7 mag). 
In the interval J.D. 243 7570 - 244 1980 rapid variations between 12.7 
and 12.9 mag within only a few days with growing tendency to smaller 
magnitudes. The few (12) exposures between J.D. 244 2370 - 4520 show 
the object bright again (12.6--12.7 mag), since J.D. 244 5265 mostly 
weak (12.8--12.9 mag). May be a chromospherically active star according 
to the yellow colour. {\it 37c} is invisible on Sonneberg plates because of 
blending by {\it 37a} and {\it 37b}. 

{\it  38a:} The ROSAT HRI position favours object {\it 38a}
which is listed in USNO-B with R1 = 17.25, B2 = 19.41 and R2 = 16.48.
Besides suggesting variability, the colour is very red. 2MASS magnitudes
are J=12.85, H=11.85, K=11.45, suggesting substantial extinction.
No flares have been observed on Sonneberg plates. 
The radio source NVSS J194356+211826 (Condon \etal\ 1998)
is consistent with the HRI position, but not with object {\it 38a}. 
The X-ray spectrum is strongly absorbed,
(corresponding to $A_{\rm V} \sim$ several magnitudes 
depending on the spectral model), consistent with the optical colours.
This indicates a very distant source, possibly even an AGN shining 
through the Milky Way disk.

{\it  39:} The Swift XRT position is outside the ROSAT error circle,
so this may be a different source, though the X-ray intensity is similar
to that of the ROSAT source.
{\it  39c} may be variable with small amplitude (0.2 mag), but very uncertain. 

{\it 40:} A ROSAT HRI observation (6.3 ksec during Apr 22-25, 1998) 
improves the X-ray coordinates to
$\pm10$\asec\ (1RXH J195305.2+211449), thus supporting the identification
with the dwarf nova V405 Vul (= {\it 40a} $\equiv$ S 10943); 
Richter \etal\ (1998).

{\it  42a:} Apart from flat waves (15.8--16.0 mag) with a timescale of 
weeks,  occasionally fadings down to 16.4 mag of very short duration are 
observed, presumably about 1 hour. Because of the slightly blue colour of this 
object it may be more probably a cataclysmic object than an RS CVn star. 
Observed fadings (in parentheses B magnitude): J.D.243 1318.380  (16.2), 
243 8243.541 (16.5:), 243 8255.393 (16.3), 243 8286.325 (16.2), 
243 9686.474 (16.4), 244 1917.430 (16.2), 244 6355.344 ($>$16.4), 
244 7365.493 (16.4:), 244 7763.381 (16.3), 244 9840.560 (16.6:).  
The XMM Slew Survey source XMMSL1 J200624.8+210709 is consistent with
this object.

\begin{figure}[ht]
 \includegraphics[width=3.4cm, angle=270]{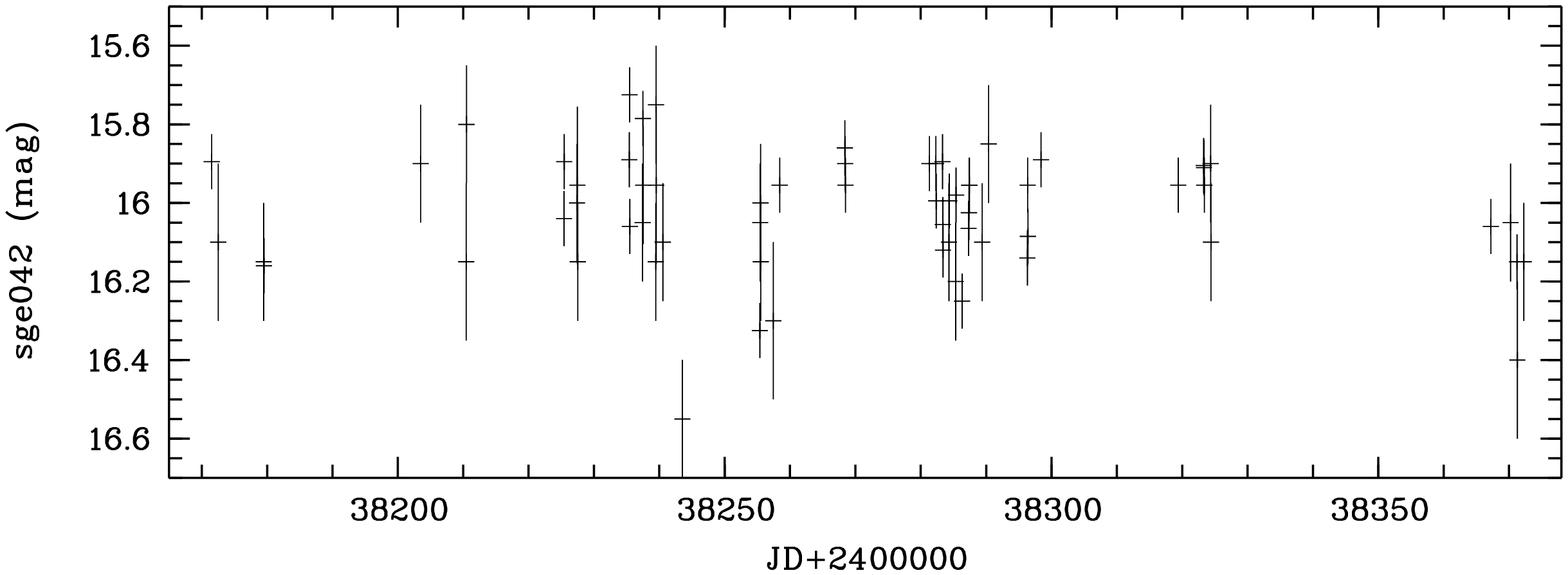}
\caption[]{Blow-up of the light curve of Sge042a (Fig. \ref{sgeolc}) 
showing pronounced variability.}
\label{sge042lc}
 \end{figure}

{\it 43a:} For the present time only analysed since 1983. From time to 
time sinusoidal waves with a length of about 40 days and an amplitude 
of 0.1-0.4 mag. Also long-time brightness variations. The colour is yellow. 
May be a chromospherically active star. 

{\it  44a:} Though normally far below the plate limit, it is faintly visible 
on several plates. Therefore probably variable. Type unknown. The reddish 
colour excludes CV type. On the other hand, the optical behaviour is 
not quite typical for a chromospherically active star. The following supposed 
brightenings can be seen (in parentheses approximate B magnitude): 
J.D. 242 9429.518 (16.5), 243 0848.512 (16.5), 243 8238.419 (17), 
243 9765.381 (17), 244 2369.235 (17), 244 3777.369 (16.5), 244 4116.445 (17), 
244 6640.430 (16.5), 244 7411.376 (16.5), 244 9163.470 (16.5), 244 9213 (17). 

{\it  45b:} On Sonneberg plates totally blended by {\it 45a}. 

{\it  46:}
The X-ray source has faded by a factor of 30 after the all-sky survey:
4 ROSAT HRI pointings in May 1991 (total exposure time of 13.4 ksec)
do not detect the source, and a 15.9 ksec
PSPC observation in Oct 20--23, 1991 just recovers 8 photons from the source,
corresponding to a vignetting-corrected count rate of 0.0005$\pm$0.0002 cts/s.
The error circle of 20\asec\ is fully contained within the all-sky survey 
positional error (see Fig. \ref{sgefc}), but does not help in distinguishing
between the counterpart candidates. The Swift/XRT observation in
Feb. 2008 also provides only an upper limit.
{\it  46b} possibly is slightly variable, but difficult because of blending. 

{\it 48:} The X-ray spectrum is very soft, so given the galactic 
latitude of b=-3\fdg4 the object must be nearby. A ROSAT HRI observation 
for 7.3 ksec during Apr 18-21, 1998 reveals 27 cts, thus allowing to 
improve the X-ray position to $\pm$10\asec: 
19\h53\m10\fss7 +20\degr43\amin53\asec. (The small count rate as compared 
to that during the all-sky survey is consistent with the 
factor 8 smaller sensitivity of the HRI for very soft X-ray spectra).
This improved X-ray position hints toward {\it 48h} as  counterpart.
The Swift XRT position with 5\asec\ accuracy supports this ID.
Moreover, the last of the three Swift observations finds the source
a factor 5 brighter in X-rays, while the two earlier Swift XRT flux
measurements are compatible with the ROSAT flux.
USNO-B1 reports $B$ magnitudes of $B1$=17.4 and $B2$=15.9, suggesting strong 
variability. All these properties are consistent with an AM Her variable.
{\it 48e:} Is not compatible with the ROSAT HRI position, but found
to be variable. Not in USNO-A catalogue. 
Marginally visible only on a few of the best plates. 
Between J.D. 243 8282 and 8287 rising from 19 mag to 
17.5 mag. At  J.D. 243 8323 still bright. Further bright at 
J.D. 243 9765 (18.4 mag), 243 9792 (17.5: mag), 244 1896 (17.3 mag), 
244 5489 (17.5 mag), 244 9213 (17.5 mag), 244 9504 (18.4 mag). 
On POSS print No. 190 at $B$=19.5 mag, on POSS print No. 302 at $B$=20.0 mag. 

 {\it 49a:} 
It is just outside of the 30\asec\ error circle, but 
consistent with the RXS X-ray position (Voges \etal\ 1999)
as well as the XMM slew survey source XMMSL1 J200648.6+204155.
Optical light curve see H\"aussler (1975).
Might be a symbiotic star.

{\it 50a} was originally favoured as counterpart due to its \fxo\ ratio.
 {\it 50b:} On Sonneberg plates fairly blended by {\it 50a}. Between 
J.D. 244 5460 and 5492, it may be a little brighter than at 
other times, but uncertain. The RXS X-ray position (Voges \etal\ 1999)
favours this star as counterpart which is supported by the
Swift XRT measurement.

 {\it 51a:} Seems to be variable with small amplitude, but because of 
faintness (on most plates invisible) not sure.
The Swift XRT observation does not detect this source; the upper
limit suggests a fading by about a factor of 8.

{\it  52:} Baade (1928) and Prager (1934) proposed NSV 12597 to be variable,
but Richter \& Greiner (1996) could not
decide whether or not NSV 12597 = AN 59.1928 is the optical counterpart of 
the ROSAT source, because the coordinates published in the NSV catalogue 
(Kholopov 1982) are questionable.  In the meantime, Skiff (1997) has 
identified the southwestern star of the pair of stars {\it 52b} as NSV 12597.
The coordinates in Table \ref{sgeopt} are for the unresolved pair from USNO-A2,
while Skiff (1997) reports RA = 19\h 57\m 01\fss2,
 Dec. = +20\degr 05\amin 43\asec\ for NSV 12597.
On Sonneberg plates no variability of {\it 52b} could be found.
However, the possibility cannot be excluded that a small variability,
if existing, will be masked because of blending by the northeastern
component of this pair of stars.
The Swift XRT observation identifies {\it 52b} as the counterpart of
the X-ray source. While the XRT position favours the northeastern
component, the southwestern component cannot be excluded.

{\it 53:} Both objects are constant. Based on the ROSAT data, the 
identification 
 with HD 350742 is not obvious since one would expect a softer X-ray spectrum
 for a nearby G0 star. Foreground X-ray absorption is negligible for all
  spectral models tested, and a power law model gives a significantly 
  better reduced $\chi^2$ than a bremsstrahlung (4.7) or blackbody model 
  (2.6). However, the Swift XRT position clearly favours {\it 53a} $\equiv$
   HD 350742 as the counterpart.

{\it  54a:} Possibly variable between 16.2 and 16.8 mag. 
Very uncertain because near the plate edge. 
The Swift XRT position is clearly outside the ROSAT error circle,
 and coincides with the bright star {\it 54c}.

{\it  55a:} Difficult because of small amplitude. Usually nearly constant 
for several years. From time to time fluctuations with small amplitudes
up to 0.2 mag and a cycle length of about 1 month. May be of BY Dra type. 

\begin{figure}[ht]
 \includegraphics[width=3.4cm, angle=270]{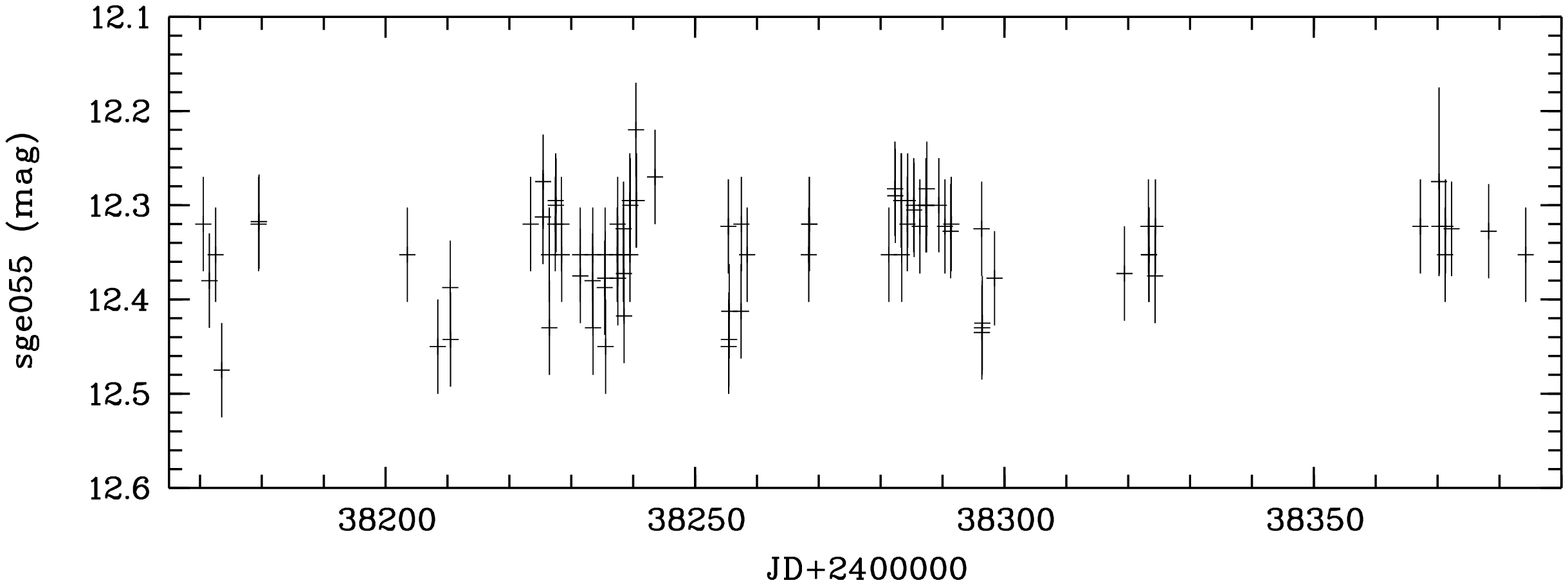}
\caption[]{Blow-up of the light curve of Sge055a (Fig. \ref{sgeolc}) 
showing pronounced variability.}
\label{sge055lc}
 \end{figure}

 {\it 56a:} May be variable with an amplitude of about 0.1 mag, but very 
uncertain. The Swift XRT position proves this object to be the 
counterpart.

{\it  57a:} Very close double star; the Table \ref{sgeopt} gives the brightness 
 and coordinates of the northern component.  The brightness of the 
southern component is $R=11.3$ mag, $B=12.6$ mag, $V=11.8$ mag.
The Mira star RR Sge (2\farcm5) and the suspected variable 
63.1928 = NSV 12599 (3\amin) are too far from the ROSAT position to be 
considered as counterpart candidates. 
The position of the only bright Swift XRT source coincides with the 
bright star {\it 57e} which is 58\asec\ away from the ROSAT position.

 {\it 58a:} On most of the astrograph plates heavily blended by {\it 58b}.
Nevertheless it cannot be excluded that it is variable with an amplitude 
of even about 1 mag, but very questionable.
If real, it could be an RS CVn star. 
{\it 58b} is a double star, even on POSS prints 
not fully separated.  The Swift XRT position suggests it to be
the counterpart.

{\it 59:} Not detected in the Swift XRT pointing, though the upper limit
is consistent with the ROSAT flux given the soft spectrum.

{\it  60:} 
The 1.8 ksec Swift XRT observation on 2008 Apr. 23 does not 
detect this ROSAT source.
{\it  60a} Seems to be variable, but the amplitude is too small to claim
this with certainty.

{\it 62b:} The USNO-A2 magnitudes/colours are likely incorrect, possibly
due to blending with {\it 62a}.

{\it 63:} While {\it 63a} is constant on Sonneberg plates, 
the Swift XRT position coincides with {\it 63b}.

{\it 64:} This X-ray source is positionally coincident with 
1E 1950.8+1844. 
Magnitudes of {\it 64e} and 
{\it 64f} are from the USNO-B1 catalogue. Though the $B$ magnitude of {\it 64f}
on the POSS is about 18\m, the object seems to be marginally visible (about
17 mag) on some Sonneberg plates. Nevertheless its  variability
 is questionable. 
The Swift/XRT position clearly coincides with {\it 64g} $\equiv$ 
TYC 1624-1644-1, with a count rate
which is compatible to that of the ROSAT source. 

{\it  65:} The Swift XRT position suggests {\it  65a} as counterpart.
Variability of {\it  65b} with an amplitude near 0.1 mag cannot be fully 
excluded.  
QY Sge is more than 5\amin\ off the ROSAT position,
and thus not considered to be the counterpart.

\begin{figure}[ht]
 \includegraphics[width=3.4cm, angle=270]{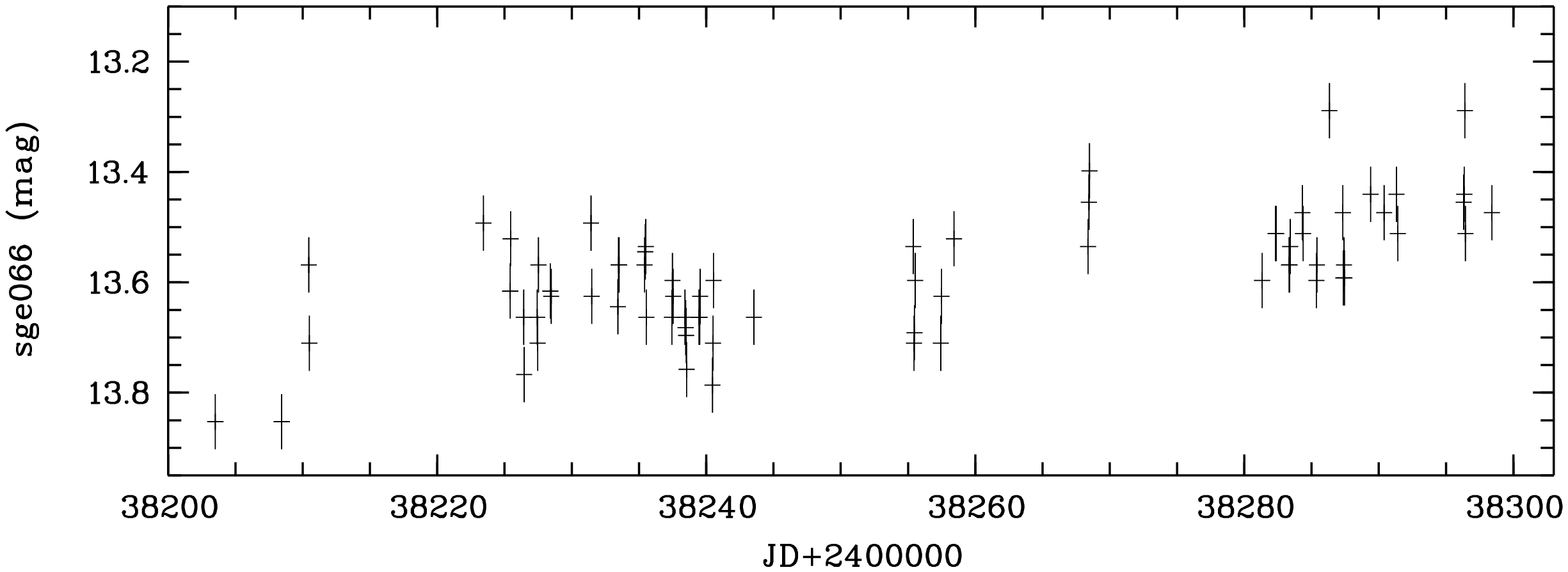}
\caption[]{Blow-up of the light curve of Sge066a (Fig. \ref{sgeolc}) 
showing pronounced variability.}
\label{sge066lc}
 \end{figure}

{\it  66a:} Difficult because of blending by {\it 66b} 
for which magnitudes
are taken from the USNO-B1 catalogue. Violent light changes, 
particularly before 1970 (see Fig. \ref{sge066lc}). 
From time to time constant over many years 
at about $B$ = 13.5.

{\it  67a:} See photoelectric catalogue of M stars by Neckel (1974).
May be the correct counterpart of  the ROSAT source in 
spite of the relatively large positional offset of 35\asec. 
Spectral type M2.5II-III + B9V according to GCVS. Probably a symbiotic star.

{\it  68:} The Swift XRT position clearly favours {\it  68f}
as counterpart.

{\it  69:}
The 4.1 ksec Swift XRT observation on 2007 Jul. 18 did not detect
this source; the upper limit is a factor 3 below the ROSAT flux.

{\it  70:} Though the USNO-A2
brightness of {\it 70c} is $B$ = 18.4, the object seems to be
sometimes just visible on Sonneberg plates. Therefore, variability is
possible. 
The Swift/XRT position clearly coincides with {\it 70d} with a count rate
which is compatible to that of the ROSAT source.

{\it  71:} Object
{\it  71b} is heavily blended by a star with $B$ = 18.0. 
It appears on Sonneberg plates as one object. At first glance it seems to be 
variable, but being at the very plate limit, this is questionable. 
The Swift XRT position clearly favours {\it  71c}
as counterpart. This object is listed as radio source MRC 2002+182
and candidate gravitational lens (King et al. 1999).

{\it  72:}  The Swift XRT position suggests {\it  72a}
as counterpart.

{\it  73:}
The 3.0 ksec Swift XRT observation on 2008 Jun. 17 did not detect
this source; the upper limit of $<$0.002 is a factor 15 below the ROSAT flux.

{\it  74a:} Extremely slow brightness changes. In 1938--1944 at $B$=12.80, 
during 1961--1967 at $B$=12.75, during 1973--1978 at $B$=12.70, 
during 1979--1993 $B$=12.65, and since 1994 
about $B$=12.65--12.70. About 165\asec\ away from the ROSAT position is the 
Mira star RT Sge, which surely has nothing to do with the X-ray source. 

{\it  75a:} Seems to be variable, but not quite sure because of faintness. 
The Swift XRT observations do not detect this source, implying a factor
  of about 4--6 variability.

{\it 76a:} Is the well known star $\alpha$ Sge.

{\it 77:} The Swift XRT observation does not detect this source,
implying variability by about a factor of 4.
The ROSAT spectrum as well as the \fxo\ ratio suggest {\it 77a}
as a chromospherically active K/M star counterpart, but other identifications
can not be excluded.

{\it  78:}
The 8.0 ksec Swift XRT observation on 2007 Dec. 20/21 did not detect
this source; the upper limit of $<$0.001 is a factor 15 below the ROSAT flux.
There is, however, a bright point source at about 1\farcm3 distance, at
RA = 20\h 15\m 21\fss64,  Decl.= +17\degr 59\amin 50\farcs1 ($\pm$4\farcs5),
in both pointings. However, its distance is substantially larger than
the ROSAT localisation error, so we do not consider this source to be
related to the ROSAT source.

{\it  79:} 
The 4.3 ksec Swift XRT observation on 2007 Jul. 23 did not detect
this source; the upper limit of $<$0.002 is a factor 10 below the ROSAT flux.
We detect three other sources within 3\amin\ of the ROSAT position, all
with count rates below the ROSAT all-sky survey limit.
{\it  79a} seems to be variable between 16.4 and 16.9 mag, but very 
uncertain, because the object is near the plate edge, and thus the 
plate limit. 

{\it 80a:} See for example Vogt \& Bateson (1982).
WZ Sge is the obvious counterpart which is nicely confirmed 
 by the XMM position (see small error circle in Fig. \ref{sgefc}).
(This is the only XMM observation in the Sge field.) 
IRAS data are given by Harrison \& Gehrz (1991).

{\it  81a:} In spite of the small amplitude obviously really variable. 
Long waves with a cycle length of several 100 days. Due to variability 
preferred as optical identification. The Swift XRT position supports
this ID. This is 38\asec\ from the RXS position.

\begin{figure}[ht]
 \includegraphics[width=3.4cm, angle=270]{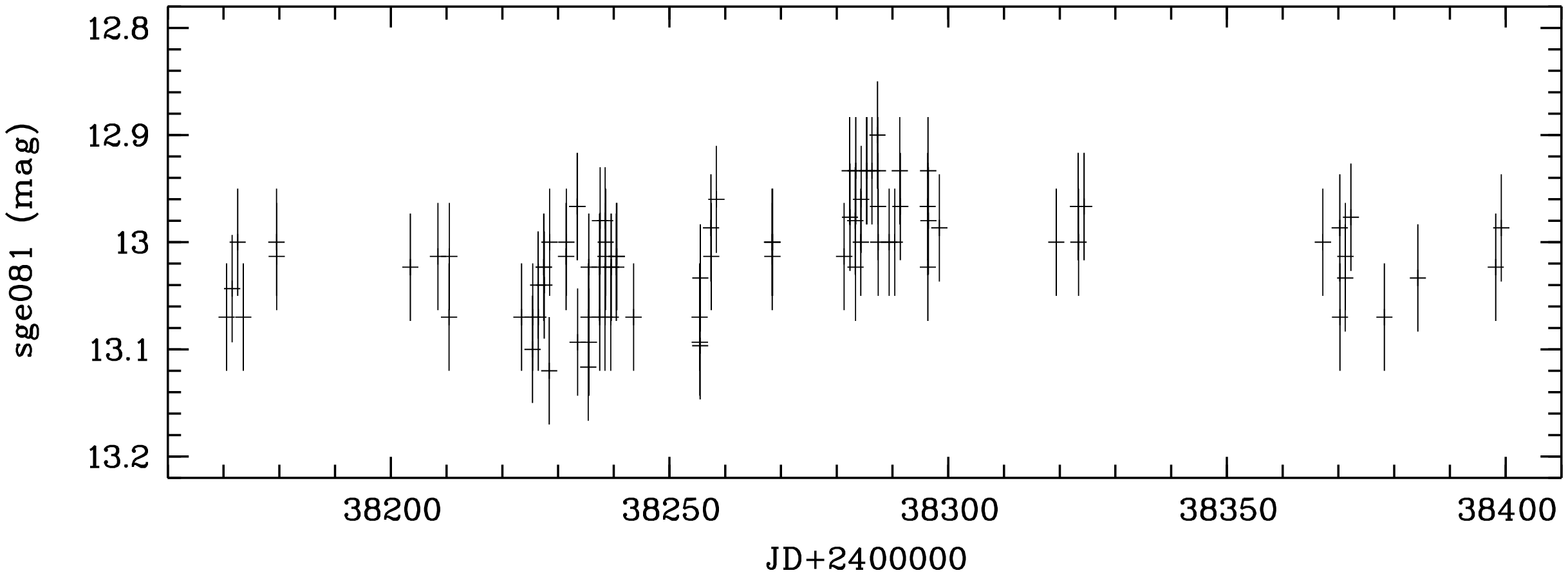}
\caption[]{Blow-up of the light curve of Sge081a (Fig. \ref{sgeolc}) 
showing pronounced variability.}
\label{sge081lc}
 \end{figure}
 
{\it  82c:} The dispersion of the magnitudes is relatively high, 
variability cannot be fully excluded.  
The Swift XRT position falls near to it, but is slightly off.
The probable Algol star 
DO Sge is about 3\amin\ apart from the ROSAT position, too far to
be considered a counterpart candidate. 

{\it 83a:} The published USNO-A2 $R,B,V$ magnitudes (non-simultaneous!) 
are marked as ``probably wrong''. 
The K2 spectrum as listed in Simbad can be traced back to the AGK3
catalogue, but Cannon \& Mayall (1949) list it as M0 star with $V$=7.7 mag!
The USNO-B1 magnitudes are $B1$=9.38, $R1$=6.90, $B2$=8.44, and $R2$=6.83.
With a photometric accuracy of better than 0.3 mag, this might
suggest variability (the object is too bright to check
variability on our astrograph plates). The
\fxo\ of HD 190720 is compatible with an K or M star to be the 
optical counterpart.
The Swift XRT position supports this ID.

{\it  84:} 
The two Swift XRT observations in 2007 did not detect
this source; the upper limit of $<$0.003 is a factor $\approx$2 below the 
ROSAT flux when accounting for the soft spectrum.
{\it  84f} is a double star.

{\it  85a:} The magnitudes show waves with a length of some 100 days and an 
amplitude smaller than 0.1 mag. The magnitude changed from 12.3 to 12.4 mag 
between J.D.242 7570 and 242 9630 to 12.4 to 12.5 later. More rapid changes 
within several days  were observed only from J.D. 243 8200 to 243 8400. 
Probably BY Dra type.  

{\it  86a:} Light changes in waves of some weeks or months with an 
amplitude up to 0.5 mag. Superposed are very slow variations. Mean magnitude: 
J.D.= 242 7500 - 244 0400: brightness fluctuates between 14.2 and 14.6 mag;
J.D.= 244 1900 - 244 7850: slowly fading to about 15.0 mag;
J.D.= 244 7850 - 255 0300: rising from 15.0 to 14.2 mag (Fig. \ref{sgeolc}).

{\it  87-91,93,94:} SNR 053.6-02.2 = 3C 400.2. This is an extended 
X-ray source (of order 20\amin), and our
source detection ``finds'' 7 spurious X-ray sources, as it was tailored 
for finding point sources. The stars in these regions were not tested for 
variability, and no finding charts are given. See Saken \etal\ (1995)
on the ROSAT data of 3C 400.2.

{\it 92:} Objects a--d constant on Sonneberg plates.
{\it 92e} is invisible on most of the plates, but appears occasionally
with clearly different magnitudes (though some of the brighter observations
might be artefacts). Because of the relatively blue colour it may be
a cataclysmic object. 
The Swift XRT position is consistent with this source, though we cannot
distinguish from {\it 92b}. The combined evidence of variability, 
X-ray emission and \fxo\ ratio suggest object {\it 92e} to be 
the counterpart.

{\it  95:}  
The 2.9 ksec Swift XRT observation on 2007 Dec. 17 did not detect
this source; the upper limit of $<$0.001 is a factor 20 below the ROSAT flux.
Both objects {\it  95a,b:} are only visible on the best plates. No variability 
could be found. 

{\it  96:} The Swift XRT observation suggests {\it  96b} as counterpart
which is much too faint to be visible on Sonneberg plates.

{\it  97a:} Brightness amplitude photographically not measurable. 
It may be a chromospherically active star. See also Blanco \etal\ (1972).
The very soft X-ray spectrum as well as \fxo\
are compatible with the G spectral type.

{\it 98a:} A pointed ROSAT PSPC observation yields a detection
(2RXP J200318.5+170253) with a position 
which is accurate to 5\asec, and thus leaves no doubt that RZ Sge
is the counterpart of this X-ray source. For optical properties
see for example Vogt \& Bateson (1982) and Khruzina \& Shugarov (1991).
Curiously, Kouzuma \etal\ (2010) identify this as AGN candidate due to 
the NIR colours, but in addition also list {\it 98c} and two fainter objects
just outside the 30\asec\ error circle as AGN candidates.

{\it 99:} This source is not detected by a  Swift XRT observation,
with an upper limit consistent with the ROSAT flux.

{\it 100a:} Blend of three reddish stars. Because of this blend and the
small expected amplitude a photographic search for variability seems to 
be without chance of success. Therefore not tested for variability.
The USNO coordinates and $BVR$ magnitudes are the combined values of this
blend.

{\it 101b:} Extremely blue. On plates of the Sonneberg astrograph 
blended by {\it 101a} which is considered to be a viable counterpart
due to the match of spectral type and \fxo. 

\begin{figure}[ht]
 \includegraphics[width=3.4cm, angle=270]{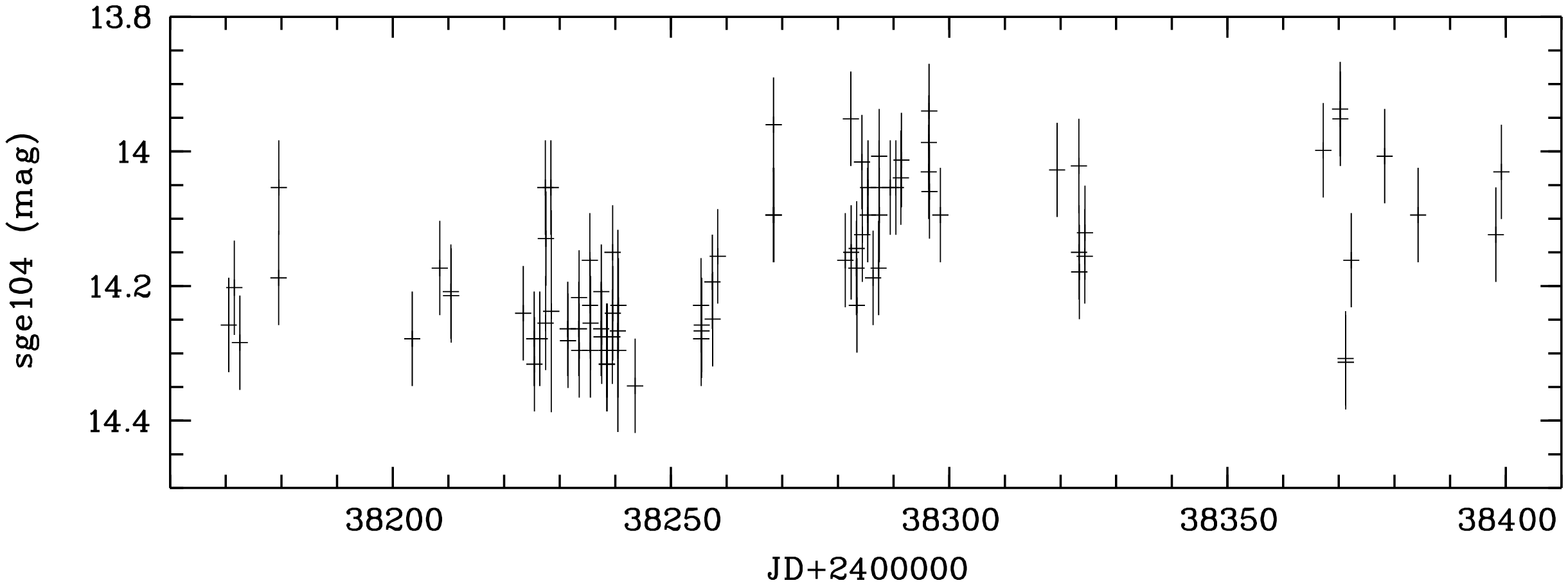}
\caption[]{Blow-up of the light curve of Sge104a (Fig. \ref{sgeolc}) 
showing pronounced variability.}
\label{sge104lc}
 \end{figure}

{\it 104a:} Brightness changes with small amplitude with a timescale 
of months or years (Fig. \ref{sge104lc}). 
Moreover, three distinct minima of short duration 
(some hours or perhaps one day) were found. These might be eclipse minima: 
J.D. 243 7584.308 (14.5 mag), 243 8371.224 (14.3 mag), 243 8371.266 
(14.3 mag), 244 7850.238 (14.4 mag). The star may be of RS CVn type. 

{\it 105a:} 
X-ray spectral hardness, \fxo\ and optical colour all suggest
this to be the counterpart.

{\it 109:}
The Swift XRT observation suggests {\it  109b} as counterpart.
No other optical object brighter than 18 mag within a distance of 40\asec\ 
from the ROSAT source is visible on Sonneberg plates except {\it 109a},
which is visible only on some of the best plates. 
No brightness fluctuations could be found. 
Object {\it  109b} coincides with the radio source 4C +16.67 
(Douglas et al. 1996), and 
has been identified by Kouzuma \etal\ (2010) as AGN candidate due to 
the NIR colours.

{\it 110:} The semi-regular variable (sub-type A) star DY Sge is 
about 6\amin\ 
away from the ROSAT position and not considered to be the counterpart.
The Swift XRT position excludes objects {\it 110a,c,d}, but is
not accurate enough to distinguish between {\it 110e} which is 
invisible on all Sonneberg plates, or the blend 
of four objects {\it 110b}.

{\it 111:} The Algol star DV Sge is nearly 4\amin\ away from the 
ROSAT position and not considered to be the counterpart. 
The \fxo\ ratio suggests {\it 111b} as counterpart, the 
brightness amplitude of which is constrained to less than 0.1 mag, 
if variable.

{\it 112b,c:} On astrograph plates totally blended by {\it 112a}. 

{\it 113a:} Difficult because of faintness. At the limits of detectability,
 but the variability seems to be sure. Long waves of small amplitude. 

{\it 114c:} Magnitudes are from the USNO-B catalogue. Very blue colour;
may be variable, but uncertain due to faintness.
{\it 114f:} Difficult because of faintness. On two overlapping POSS
plates 276 and 372 (JD 243 3835 and 243 3910) at nearly 17.3 mag.
 A CV nature is possible
because of the positive \fxo\ ratio.
Identification is ambiguous.

\begin{figure}[ht]
 \includegraphics[width=3.4cm, angle=270]{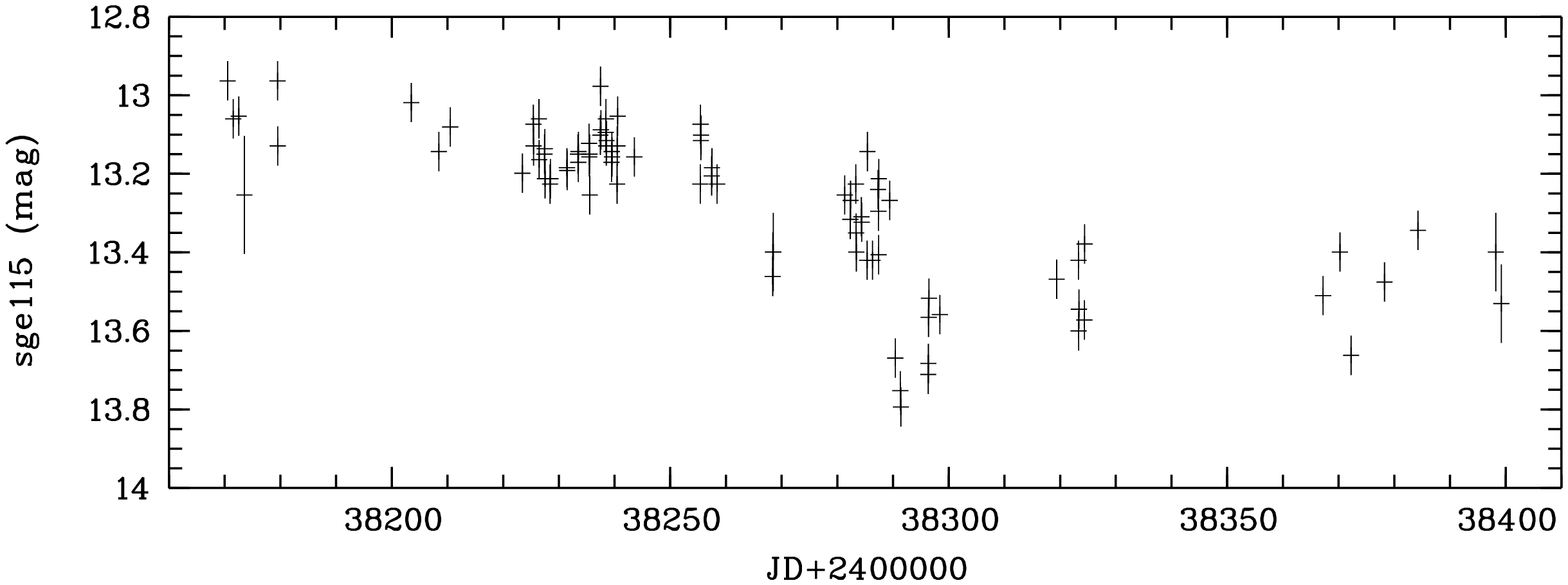}
\caption[]{Blow-up of the light curve of Sge115b (Fig. \ref{sgeolc}) 
showing pronounced variability.}
\label{sge115lc}
 \end{figure}

{\it 115a} is constant within $\pm$0.1 mag,  but the Swift XRT
position suggests this to be the counterpart.
{\it  115b} is disturbed by {\it 115a}. {\it 115b} itself consists of two
very close objects of similar brightness, on Sonneberg astrograph
plates only the combined light can be seen. Therefore it cannot be 
decided which of the two objects is variable. Magnitude changes with
a time scale of several weeks or even years (Fig. \ref{sge115lc}). 
Because of the blue colour
possibly CV type. The USNO magnitude is most likely
only for the western component.

{\it  116a:} See Raveendran (1984) and Richter \& Greiner (2000).

{\it  117:} Dispersion of magnitudes of object {\it  117d}
relatively high. Small amplitude 
variability cannot be excluded. The Swift XRT observation marginally
detects the source, and suggests {\it  117a} to be the counterpart,
consistent with the M2.5 spectral classification and the \fxo\ ratio.

{\it  119a:} Additional plates of the 140 mm triplet and of the neighbouring 
field $\gamma$ Aql could be used. Irregular light changes of small amplitude.

\begin{figure}[ht]
 \includegraphics[width=3.4cm, angle=270]{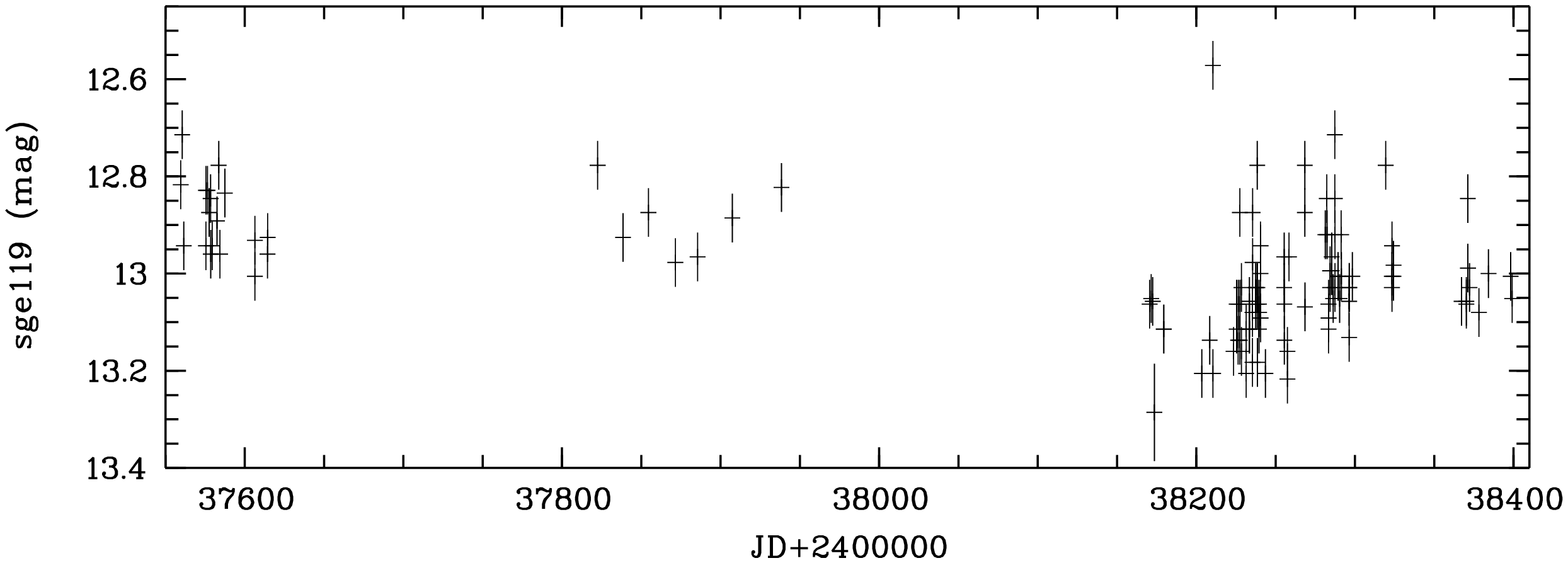}
\caption[]{Blow-up of the light curve of Sge119a (Fig. \ref{sgeolc}) 
showing pronounced variability.}
\label{sge119lc}
 \end{figure}

{\it  120a:} Though usually fainter than the plate limit, this very 
blue object can be seen on some plates: Variable between about 
16.5--18.0 mag. More frequently faint than bright.
The Swift XRT position clearly identifies this as counterpart of the 
ROSAT source. There is about 50\% X-ray variability between the two
Swift observations spaced by 10 days, where the second observation shows
the source about a factor 3 brighter than the ROSAT flux. 
There is also a XMM Slew survey source XMMSL1 J201359.1+154419
within 8\asec\ which is likely the same source.
The blue optical
colour, the hard X-ray spectrum and the clear optical and X-ray variability
suggest an intermediate polar type CV.
Alternatively, Kouzuma \etal\ (2010) identify it as AGN candidate due to 
the NIR colours from 2MASS. 
However, the blue optical colour together with a galactic foreground
$A_{\rm V} = 0.52$ mag make an AGN shining through the disk of our
Galaxy unlikely.

{\it  122a:} Spotted star according to Hall \& Henry (1992).

{\it  123a:} Variations within weeks or months. Also more rapid brightness 
changes within days. Plates of the 170 mm triplet  and of the neighbouring 
field $\gamma$ Aql could also be used. {\it 123b,c:} Just visible on 
good plates. No brightness variations could be found.  
The Swift/XRT observation targeted on {\it 124} clearly detects 
this X-ray source at 0.027 cts/s, about a factor 3 brighter than during
the RASS. 
The derived X-ray position favours {\it 123d or e}. 
Both objects are invisible on Sonneberg plates, so nothing can
be said about optical variability. Kouzuma \etal\ (2010) identify
{\it 123e} as AGN candidate.

\begin{figure}[ht]
 \includegraphics[width=3.4cm, angle=270]{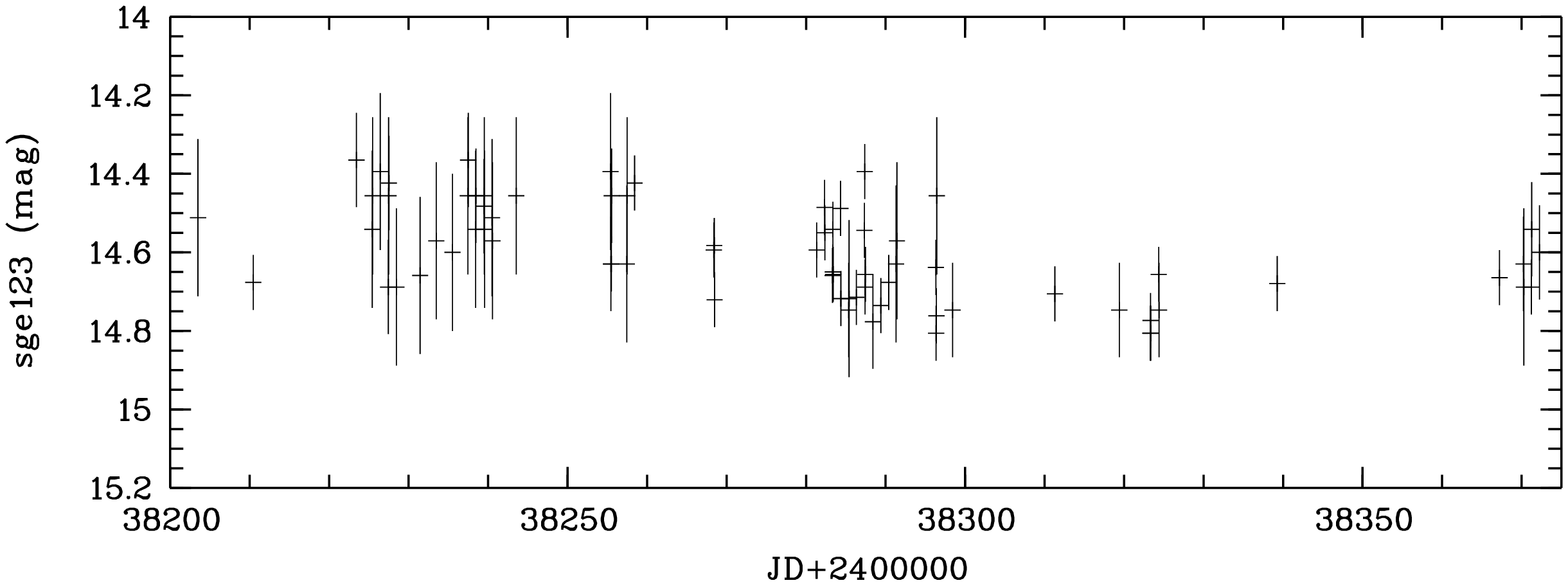}
\caption[]{Blow-up of the light curve of Sge123a (Fig. \ref{sgeolc}) 
showing pronounced variability.}
\label{sge123lc}
 \end{figure}

\begin{table*}
  \caption{Summary of identification statistics, separated in secure,
   suggested (sugg.) and no identification (ID), separately for genuine RASS 
   catalogue sources and our additional RASS sources (50\%/97\% more sources
  for the Com/Sge fields). The numbers in parentheses
   are the fractions for each group. 
   }
  \vspace{-0.2cm}
  \begin{center}
    \begin{tabular}{lrrrrrrrrr}
     \hline
     \noalign{\smallskip}
  Field &  \multicolumn{3}{c}{RASS catalogue} & \multicolumn{3}{c}{additional RASS source} & \multicolumn{3}{c}{Total (as in Tables \ref{comX}, \ref{sgeX})} \\
        & secure ID & sugg. ID & no ID~~ & secure ID & sugg. ID & no ID~~
         & secure ID & sugg. ID & no ID~~ \\
    \noalign{\smallskip}
    \hline
    \noalign{\smallskip}
% sum              159                     | 79
  Com & 122 (77\%) & 23 (14\%) & 14 ~~(9\%) & 54 (68\%) & 14 (18\%) & 11 (14\%) 
      & 176 (74\%) & 37 (15\%) & 25 (11\%) \\
  Sge &  45 (70\%) & 11 (17\%) &  ~~8 (13\%) & 35 (51\%) & 19 (28\%) & 14 (21\%)
      &  80 (61\%) & 30 (23\%) & 22 (17\%) \\
    \noalign{\smallskip}
    \hline
 \end{tabular}
 \end{center}
 \label{tab:IDstat}
\end{table*}

{\it 124a:} Additional plates of the 170 mm triplet and of the neighbouring 
field $\gamma$ Aql could also be used. Relatively high dispersion of 
magnitudes, nevertheless variability uncertain. The \fxo\ ratio supports it 
to be the counterpart, as does the Swift XRT position. 
The Mira star V 1019 Aql is more 
than 5\amin\ distant from the ROSAT position and not considered as
counterpart candidate. 

{\it  126a:} 
Identified by Kouzuma \etal\ (2010) as AGN candidate due to 
NIR colours.
{\it  126b:} Difficult because at the extreme edge of the plates. 
Some fadings seem to occur from a relatively constant brightness level. 
Eclipsing variable of RS CVn type? Observed minima: 
J.D. 243 7560.524 (15.8: mag), 8371.224 (15.9: mag), 9385.370 (15.8 mag), 
9792.287 (16.1 mag), 244 6683.403 (15.9: mag), 6763.219 (15.9: mag), 
9213.422 (15.9: mag).

 {\it 127b:} Distinctly beyond the plate limit, but on two occasions 
faintly visible: J.D. 242 9846.473 (15.6: mag), 244 2713.261 (16.7 mag). 
It is difficult to say whether these are real brightenings or artefacts. 
However, the Swift XRT position is consistent with {\it 127b} being
the counterpart, lending more credit to this optical behaviour. 

{\it 128:} Both, {\it 128a} and {\it 128b} are constant on Sonneberg plates,
and not considered as potential counterparts. The Swift XRT position
suggests {\it 128c} instead.

 {\it 129a:} Any light variations of this object are surely 
smaller than 0.03 mag.

{\it  130b:} Though the brightness is clearly beyond the plate limit, 
this object is indicated on five of about 250 plates: 
J.D. 242 9844.531 (16.9: mag), 242 9845.502 (16.6 mag), 
243 9702.411 (16.6: mag), 244 8096.490 (16.7), 244 8186.348 (16.3: mag). 
Therefore it is variable with an amplitude of about 1 mag. 
According to the blue colour it may be a cataclysmic variable. 

{\it 131:} The slowly variable star NSV 12500 is too far from the 
ROSAT position (11\amin) to be the optical counterpart. {\it 131a:} Small 
brightness variations within the limits 11.3--11.4 mag seem to be 
indicated. In single cases small flares (11.1 mag) which last about 
1--2 days. 
Since the \fxo\ ratio is inconsistent with a CV, this
is likely a chromospherically active star.

\begin{figure}[ht]
 \includegraphics[width=3.4cm, angle=270]{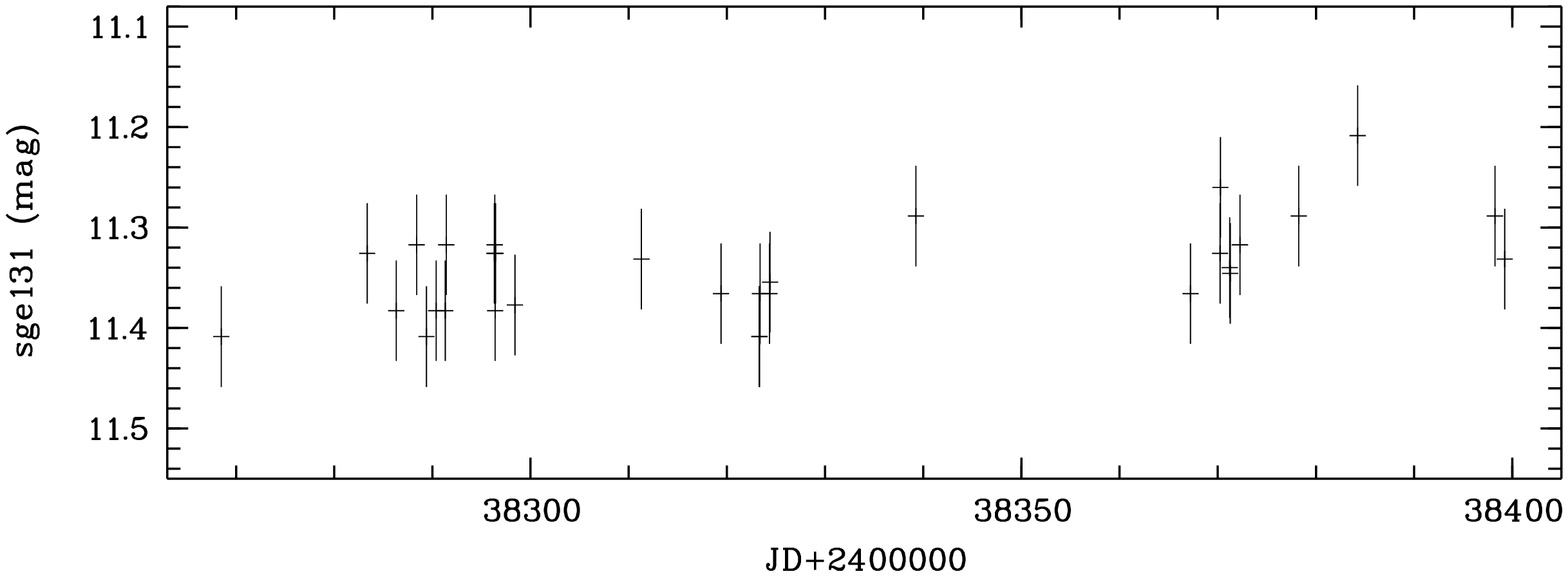}
\caption[]{Blow-up of the light curve of Sge131a (Fig. \ref{sgeolc}) 
showing pronounced variability.}
\label{sge131lc}
 \end{figure}

{\it 132:} The Swift XRT position clearly favours {\it 132b} as 
counterpart which is invisible on Sonneberg plates.

\section{Results and Discussion}

\subsection{Identification results and global statistics}

We are able to identify $\sim$70\% of the X-ray sources in the
two selected fields, and suggest likely counterparts for another 20\%
of objects (see Table \ref{tab:IDstat}). 
This is a surprisingly high number given that the expectation
for the optical brightness of our faintest X-ray sources through the
\fxo\ ratio is $B \sim 22^{\rm m}$ (see Fig. \ref{stocke}).

As mentioned in Sect. 2, our source lists for each field are larger 
than the published 1RXS catalogue (Voges \etal\ 1999)
due to primarily the slightly lower S/N ratio required for
the X-ray detection
(Voges \etal\ 1999). This results in 50\%/97\% more sources
in the Com/Sge field, respectively. The rate of source identification
is very similar for the official vs. this extended source list
(see columns 2--4 vs. 5--7 in Table \ref{tab:IDstat}), suggesting
that the number of spurious X-ray sources in our lists is still
negligible. 
This is also supported by the distribution of optical to X-ray
position offsets (Fig. \ref{posit}) which does not show any 
systematic errors (see below).

A breakdown of the identified sources is given in Table \ref{IDsum},
showing the dominance of AGN in the high-galactic field, and
bright and/or chromospherically active stars in the low-galactic field.
Most probably, the majority of optical counterparts in the "empty" fields 
(as well as some of the faint galaxies) are still undiscovered AGN. 
Therefore, nearly 2/3 of all the ROSAT sources in the 26 Com field may be
of extragalactic origin.

Somewhat surprisingly, we find 9 clusters of galaxies and another 
8 candidates, despite none of the X-ray sources was detected as extended.
If true, this implies a cluster content in the RASS about a factor 2
more than hitherto found (e.g. B\"ohringer \etal\ 2013).
This has interesting implications for the upcoming eROSITA survey
for which the expected number of clusters (100.000;  
Chon \& B\"ohringer 2013) is based on the RASS count.

\begin{table}[ht]
  \caption{Summary of the secure and probable optical counterparts of 
  ROSAT sources.}
  \vspace{-0.2cm}
  \begin{center}
    \begin{tabular}{lccc}
     \hline
     \noalign{\smallskip}
  Type &   26 Com   & $\gamma$ Sge  & remarks \\
    \noalign{\smallskip}
    \hline
    \noalign{\smallskip}
variable AGN and AGN? & 34 &  1 & 1 \\
constant AGN and AGN? & 88 &  3 & 2 \\
cluster of galaxies   & 17 &  0 &   \\
CA and CA?            & 18 & 25 & 3 \\
LB and SRB            &  2 &  1 & 4\\
bright stars          & 31 & 34 & 5\\
UV and UV?            &  6 & 1 & 6 \\
eruptive binaries     &  2 & 14 & 7 \\
magnetic stars (Sp. Am) & 1& 0  &  \\
DA                    & 2 & 0 & \\
PN                    & 0 & 1 & \\
supernova remnant     & 0 & 1 & 8 \\
other                 & 11 & 5 & \\
not identified        &  25 & 21 & 9 \\
not classified        &   1 & 19 & 10 \\
 \noalign{\smallskip}
 \hline
 \end{tabular}
 \end{center}
 \label{IDsum}
 \vspace{-.25cm}
 \noindent{\small
         1: Some of them may be eruptive binaries. \\
         2: Variability photographically not established (too faint, blends 
            or too small amplitudes). \\
         3: Chromospherically active stars (RS CVn and BY Dra). \\
         4: Slow irregular and semiregular variables of late spectral type
            (see GCVS, Kholopov 1985), some of which may possibly 
            also be related to the chromospherically active stars.\\
         5: Mostly overexposed on the photographic plates, and thus
              not investigated for variability. Because of the scarcity of
              bright stars as counterparts of ROSAT sources, these
              matches are likely real associations (see Fig. \ref{sbs}). 
              These stars are therefore
              with high probability chromospherically active stars. \\
         6: Flare stars of UV Cet type. \\
         7: Mainly cataclysmic variables (CVs), but also X-ray binaries of 
            AM Her type.\\
         8: 7 X-ray sources close together, since source detection algorithm
            was optimized for point sources.\\
         9: Out of the 25 unclassified objects in the Com field (dash in the
            last column of Table \ref{comX}), the error circle of 9 objects is
            empty on the DSS2 (about $B \sim 20$th mag), 
            and for another
            17 X-ray sources (totalling 26), no optical object is visible
            on the Sonneberg astrograph plates (about $B \sim 17$th mag).\\
        10: Positional coincidence implies secure correlation, but source
            type remains open.
}
   \end{table}

The Hamburg/RASS catalogue with nearly 4400 optical identifications
of northern hemisphere (Decl. $>$ 30\degr), 
high-galactic latitude ($>$30\degr) X-ray sources
yielded as largest source population $\approx$42\% AGN, followed by 
$\approx$31\% stellar 
counterparts, whereas galaxies and cluster of galaxies comprised only
$\approx$4\% and $\approx$5\%, respectively (Zickgraf et al. 2003).
Another large effort was the identification of a complete 
flux-limited sample of
northern RASS sources in six sub-fields covering a total of 685 
square degrees, which resulted in the spectroscopic identification
of 662 out of 674 sources (Zickgraf et al. 1997, 
Appenzeller et al. 1998, Krautter et al. 1999). These authors find
relative source class fractions of 42.1\% AGN, 40.7\% stars,
11.6\% cluster of galaxies, and 3.9\% galaxies. 
The ROSAT galactic plane survey (Motch \etal\ 1998) achieved 
identification of 76 out of 93 sources, among them 40 bright/active
stars, 25 AGN and 6 cataclysmic variables
(this survey selected according to X-ray brightness and
various hardness ratio criteria, so did not aim at completeness
in a given area).
These relative fractions of source types are largely consistent 
with our results (Table \ref{IDsum}), though some differences show up.
The low fraction of AGN in the two high-galactic latitude
identification programs  might be due
to the bias against fainter counterparts for which objective prism
spectra were either not available or inconclusive -- this would
rise the fraction of bright-vs.-faint counterparts, and thus the
ratio of stellar-vs.-extragalactic objects.
The nearly complete lack of AGN in our low-galactic latitude field 
as compared to the nearly 20\% fraction in Motch \etal\ (1998) 
is most likely due to their survey covering galactic latitudes
up to $\pm$20\degr, and our optical material not being deep enough
to recover absorbed AGN behind the Milky Way disk.

Fig. \ref{posit} shows the histogram of the distances of the 
secure optical counterparts to their respective centroid ROSAT
X-ray position for the Coma field. The distribution for the Sge field
looks very similar. The mean offset is 15\asec, and 90\% of 
all X-ray sources have their counterparts within 30\asec.

In the Coma field, 5 of our identifications have offsets relative to
the X-ray centroid of more than 40\asec, and constitute the very tail
of the offset distribution in Fig. \ref{posit}: these are
Com025b, 82a, 99a, 147a, and 212a. We have double-checked these
for possible mis-identifications, but find no systematic problem.
Two of these sources (99a, 147a) are identified with bright stars,
two others (25b, 212a) are driven by Swift/XRT positions, and one is 
a SDSS-III classified QSO. None of 82a, 99a, 147a have any other object
brighter than R$\sim$20 mag in their error circles. Only for two objects,
potential concerns could be risen: for 147a the \fxo\ ratio is on the
low side if the K0 stellar classification is correct, and for 212a
the Swift/XRT flux is a factor 4 below the RASS rate, so would have
been undetected in RASS, and thus being possibly a different source
if the RASS source has faded by a factor of 10.

\begin{figure}
  \hspace{-0.5cm}\includegraphics[width=6.7cm, angle=270]{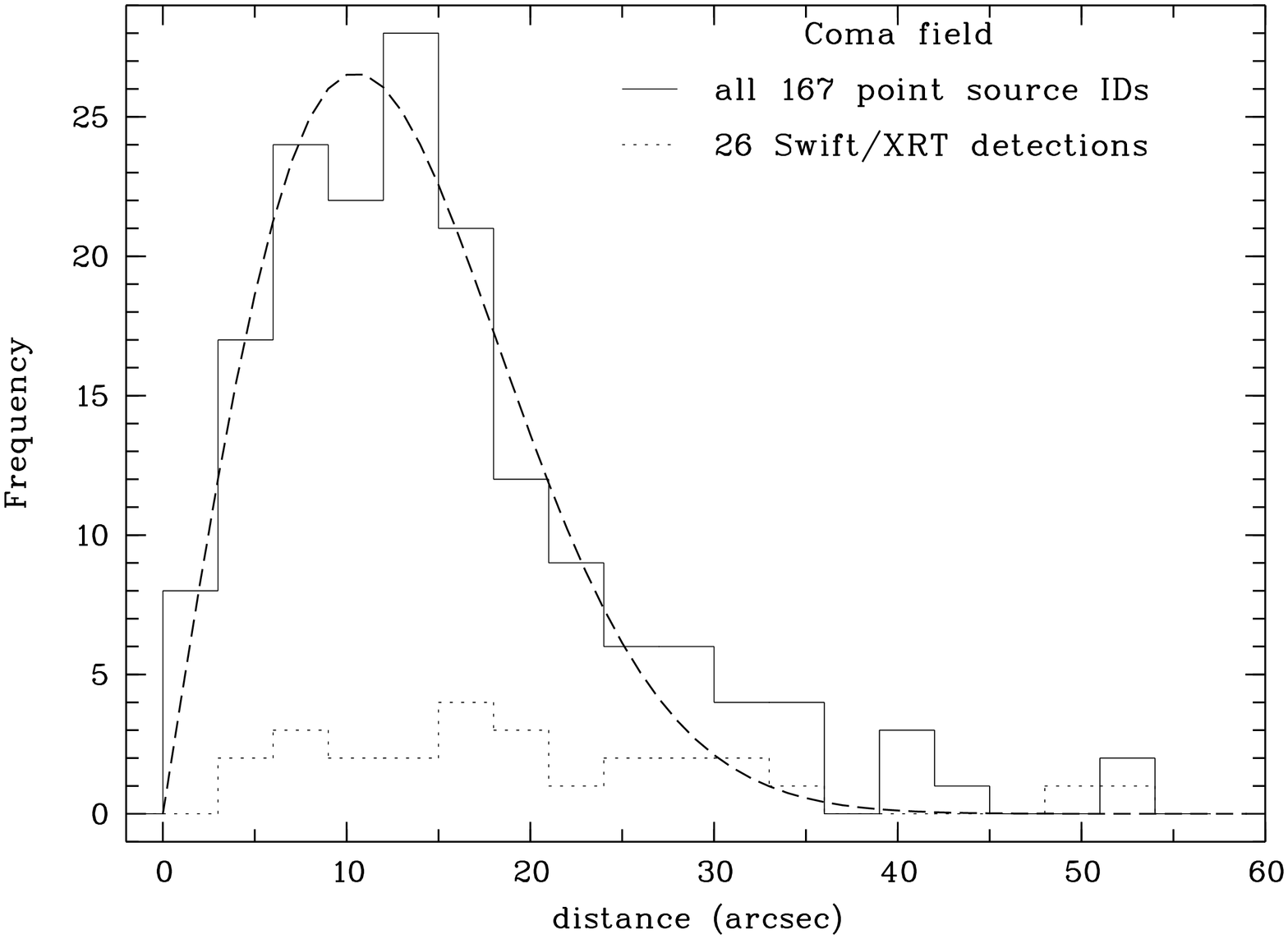}
  \caption[]{Distribution of positional offsets between centroid X-ray
    and optical position of the identified objects in the Com field (solid
    line), and the subset of 26 sources for which Swift/XRT positions have
    been obtained. The dashed line shows the expected Rayleigh 
    distribution, with a best-fit of $\sigma$ = 10\farcs5.
 \label{posit}}
\end{figure}

\subsection{Variability as a classification criterion}

In our study, variability has been used as an additional
criterion of an object 
to be an optical counterpart of a ROSAT point source, besides the classical
$F_{\rm x}/F_{\rm opt}$ ratio. Obviously, this variability criterion has not
been applied to extended sources, like supernova remnant SNR 053.6-02.2 
and clusters of galaxies. 

For the Sonneberg plate material used here, the best magnitude interval 
to detect variables on astrograph plates is about 13--15 mag. In this 
magnitude range we can discover brightness amplitudes of 0.2 mag or even 
smaller. At fainter magnitudes, near the plate limit at about 17 mag, 
no variations smaller than  about 0.5 mag can be found.

From Tables  \ref{comX},\ref{sgeX} and \ref{comopt},\ref{sgeopt} we can 
deduce the following (see Table \ref{IDsum}):
40 (30\%) of the supposed optical counterparts in $\gamma$ Sge 
and 61 (26\%) 
in 26 Com were found to be variable. Some of these variables are already 
known, most are newly detected, and have therefore been assigned a S number
(for new Sonneberg variables) in column 5 of Table \ref{comopt},\ref{sgeopt}. 
Moreover, during the search for optical variability a total of 5 variable
objects were found which seem not to be the counterparts of ROSAT sources,
either because there are better candidates for optical identification or
because of the relatively large spatial separation to the ROSAT source: 
Com 98a (E),
Com 101a (CV?),
Com 109e (AGN),
Sge 48e (CV?), and
Sge 115b (CV?).
Ignoring these, but considering that our sensitivity to detect
variability degrades rapidly beyond $B \sim 16.5$ mag, we find
that 50\% (Com) and 39\% (Sge) of the identified sources are variable, 
respectively.
If we also take into consideration that most HD stars correlating 
with ROSAT sources and have not been tested for 
variability may be variables of small amplitude, 
a good fraction of all of the optical counterparts could be variable. 
Exceptions are objects classified "C" 
(constant) which are either spurious counterparts, or their light amplitudes 
are too small to be detected photographically, or they are real 
counterparts without expected optical brightness changes 
(e.g. Seyfert-2 galaxies, 
hot white dwarfs, brown dwarfs, see e.g. Neuh\"auser \& Comeron 1998).

Optical variability is, not unexpectedly, tightly correlated with
X-ray emission, and thus our new method for optically identifying 
X-ray sources is demonstrated to be feasible. Given the large number of
optical plates used, this method was likely not more efficient
than e.g. optical spectroscopy. However, it required no proposal writing
and no telescope time,
but just access to publicly available archival data.
We have shown that an unequivocal optical identification of ROSAT sources 
is not possible in every case,  but for a large fraction of objects
(65--80\%) we can estimate for each object the 
reliability of a supposed identification 
by including optical variability as additional criterion.
It remains, however, true that optical spectroscopy will remain a
highly demanded asset, in particular to determine redshifts or sub-types
of the AGN-population: only then important physical parameters like the
luminosity, binarity or presence of an accretion disk  can be deduced. 

Yet, optical variability can serve as an
efficient tool in pre-selecting objects from specific source classes.
Together with the optical colour as e.g. provided
by the USNO catalogue, the \fxo\ ratio and details of the optical light curve
often provides already sufficiently accurate clues on the source type.
This is proven by the (even for us surprising) little impact which
the SDSS-III DR9 spectra had (due to their late availability relative to 
the completion of this work) on our identifications. Thus, 
optical variability can be an efficient substitute for optical
spectroscopy in the 
early identification process
of many thousands or million of objects. 
This is particularly true as long as no systematic spectroscopy of
all objects on the sky down to 20th mag is available -- which might
well apply even for the first years of {\it eROSITA} operation.

\subsection{Impact of variability as reliable identification criterion}

With the about 200.000 X-ray sources from the ROSAT mission, 
massive identification programs have been employed based on spatial
correlations with other large catalogues. A very early attempt was
a correlation with the Hamburg objective prism survey which provided
identifications for nearly 4400 ROSAT Bright Source Catalog sources
(Zickgraf \etal\ 2003). 
Another correlation was done with the IRAS Faint Source Survey,
the NRAO VLA Sky Survey, the FIRST Radio Survey, the Two Micron All-sky Survey
and the APM and COSMOS catalogues of digitized Schmidt plates, revealing
over 1500 sources with an identification reliability greater than 90\%
(Lonsdale \etal\ 2000). And finally the correlation with the
Sloan Digital Sky Survey (SDSS, data release 7) 
results in approximately 10.000 confirmed quasars and other active 
galactic nuclei that are likely RASS identifications (Binder \etal\ 2011).

A different identification method used multicolour CCD imaging with the goal
of a classification via the spectral energy distribution. For one
particular medium-deep ROSAT pointing, 149 of the 156 detected objects
within the error circles of 75 X-ray sources were classified, but despite
a 15-filter dataset per object, no identifications are proposed
(Zhang \etal\ 2004).

While these examples are far from exhaustive, they show the
size of the identifications of the various programs. With about
\lax 20.000 ROSAT sources identified, this also demonstrates the 
corresponding challenge in getting identifications for the remaining
$\approx$180.000 ROSAT sources.

As such, the herewith proposed identification of 256 ROSAT sources
is just another little step. However, our main goal was to prove
optical variability as a new, efficient classification criterion, 
and thus pave the way for larger projects of this kind, both
for ROSAT as well as the upcoming {\it eROSITA} all-sky survey 
(Predehl et al. 2006).

Despite the better positional accuracy of 1\asec--5\asec\ of the 
presently used X-ray missions {\it Chandra}, {\it XMM-Newton} and 
{\it Swift/XRT}, the ROSAT survey is still considered
a unique resource when it comes to large areas, or all-sky coverage.
Ignoring the $\sim$180.000 X-ray sources because of missing
identifications might be a superficial decision. Our suggestion
to use optical variability as an efficient criterium could be
used in the near-term future in a two-fold manner: 
(i) the digitization of large 
photographic plate collections, e.g. at Harvard (Grindlay et al. 2009,
Grindlay et al. 2013)
or Sonneberg Observatories (Kroll 2009)
 is well advanced, and
is going to provide a census of the optical variability of the sky
over the last 120 years down to limiting magnitudes of order 14--16 mag.
(ii) Digital sky surveys such as the Panoramic Survey Telescope 
and Rapid Response System (Pan-STARRS; Tonry et al. 2012),
the Palomar Transient Factory (PTF; Kulkarni 2012),
the Catalina Real-Time Transient Survey (CRTS; Drake \etal\ 2012),
the La Silla-QUEST variability Survey (LSQ; Hadjiyska \etal\ 2012),
or the Sloan Digitial Sky Survey (SDSS; strip 82, MacLeod et al. 2012)
already provide multi-epoch observations in various filter and cadence
combinations, and promise to be a great resource in a systematic
identification prgram of the RASS.
The recent cross-correlation of periodic variables from the ASAS 
survey with bright ROSAT sources (Kiraga \& Stepien 2013) is a 
promising first step.
In the long-term, the Large Synoptic Survey Telescope 
(LSST; Ivezic et al. 2008)
will provide the ultimate database
in 5 filters with 3-day cadence to a depth which is fully sufficient for
the identification of the majority of RASS sources.

The upcoming {\it eROSITA} all-sky survey (Predehl \etal\ 2006),
a 4-yr survey providing 8 consecutive all-sky surveys in the
0.3-10 keV band, will have source position uncertainties similar
to that of the ROSAT survey. While large (and expensive) multi-object
fiber spectrographs are being planned and build,
such as BigBOSS (Schlegel \etal\ 2010), 
MOONS (Cirasuolo \etal\ 2012), 4MOST (de Jong \etal\ 2012)
or WEAVE (Dalton \etal\ 2012),
it is well conceivable that these spectroscopic surveys will not
be available during the first years of the {\it eROSITA} operation.
Identification programs using optical variability
could be a valuable approach.

\subsection{Comments on specific source classes}

\subsubsection{Semiregular and irregular variables}

Red semiregular and irregular variables are usually not X-ray sources. 
On the other hand, three such objects (FN Com, HH Sge and $\delta$ Sge) 
have a very 
small distance to the ROSAT position. Possibly, they are members of X-ray 
radiating symbiotic binaries. This remains to be checked spectroscopically.

\begin{figure}[ht]
  \includegraphics[width=8.8cm]{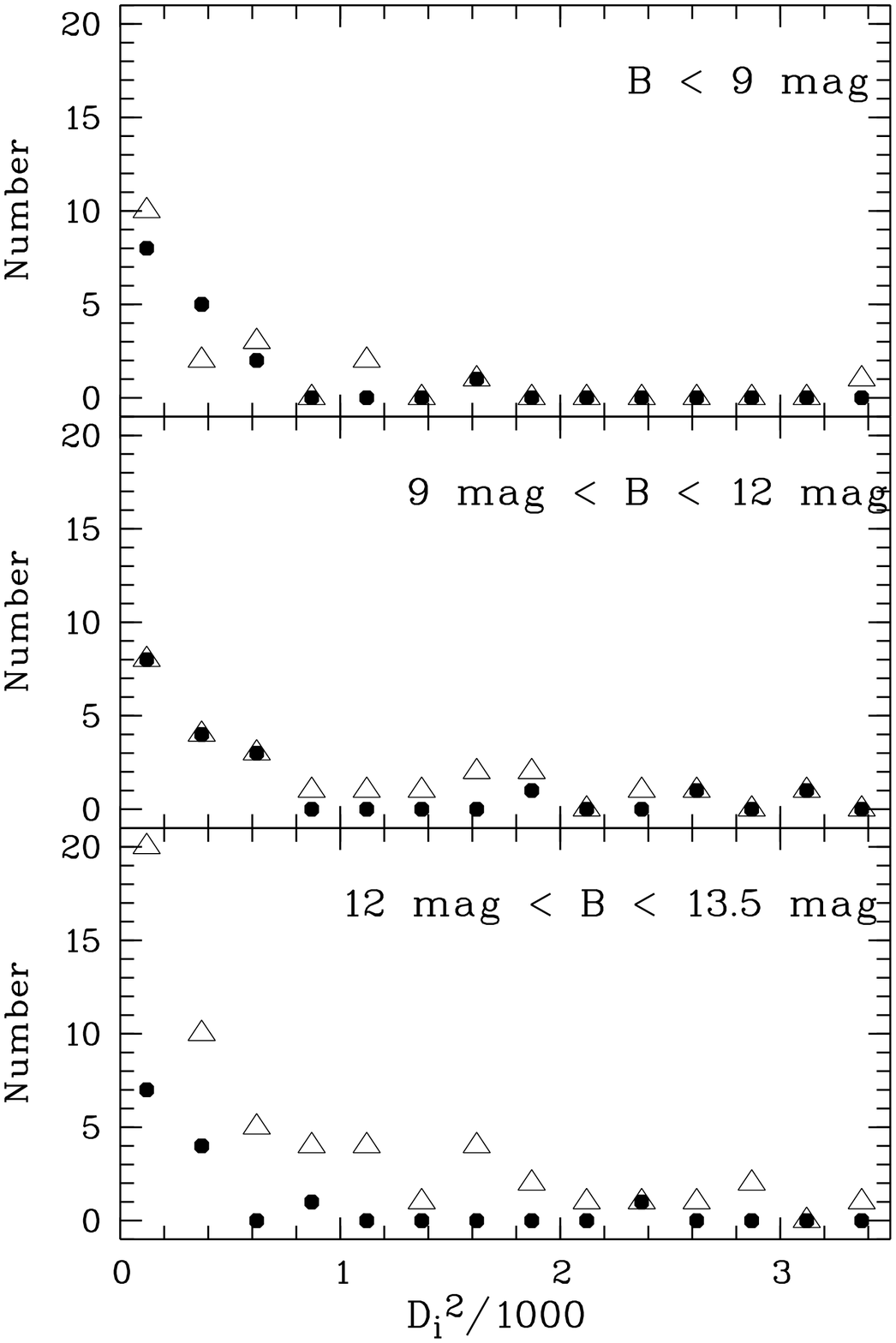}
  \caption[]{Number of bright stars in dependence of the offset from the
    X-ray centroid position (D$_i$ in \asec) for three different brightness 
    intervals
    as labeled in the three panels for the Com (filled hexagon) and 
    Sge (open triangles) fields, respectively. The overdensity within
    the 30\asec\ X-ray error circle. The percentage of real identifications
    (within 27\asec) is larger than 90\% (Coma) and 65\% (Sge) fields,
    respectively.
 \label{sbs}}
\end{figure}

\subsubsection{Chromospherically active and flare stars}

Chromospherically active (CA) stars and UV Ceti stars (flare stars) are 
known to exist not only isolated, but also in T 
associations and star 
clusters, representing a very young stellar population. According to 
Gershberg and Shakhovskaya (1974) and Rodono (1980), flare stars and 
chromospherically active stars are related and both should therefore 
be young. Thus, the large number of these objects in our Com field is 
not a feature of stellar population, but it reflects the fact that a large 
area of our Coma Berenices field is covered by the extended, young open 
cluster Melotte 111. 
Based on a literature comparison (Randich \etal\ 1996, 
Odenkirchen \etal\ 1998, Abad \etal\ 1999), we assume that some of our 
objects are cluster members. In order to verify this, proper motion or
radial velocity measurements are necessary.

\subsubsection{Bright stars}

The above tables \ref{comopt},\ref{sgeopt} contain many bright stars 
($B$ \lax 13.5 mag) within or near the ROSAT X-ray positions.
On plates of the Sonneberg field patrol those objects are overexposed.
In the few cases, in which variability measurements and/or optical
spectroscopy are available, these stars are mostly chromospherically
active stars. While an unequivocal identification is not possible in
every case, a statistical estimate of the correlation and of the
completeness and reliability of the samples can be made (Richter 1975).
Separately for the Coma and Sge fields, we have sub-divided our 
sample into three sub-groups of apparent magnitude $B<9$ mag, 
9 mag $<B<$12 mag, and 12 mag$<B<$13.5 mag, and determined the number
of bright stars in rings of increasing distance $D_{\rm i}$ (in \asec)
around the X-ray centroid. The resulting numbers (Fig. \ref{sbs}) 
clearly show an excess of bright stars for distances smaller than 
about 27\asec\ (corresponding to $D^2<$750). The corresponding total
number of bright star counterparts are 37 (Coma) and 54 (Sge).
This is surprisingly close to the number of proposed counterparts 
as given in Table \ref{IDsum}, the slight difference being caused by
the fact that some objects with already known variability are present
in the above statistics, but not in Table  \ref{IDsum}.

\subsubsection{Cataclysmic variables}

In the Coma field, one of the two known cataclysmic variables (CV) is
detected in the RASS, and one new possible CV is discovered. 
In the Sge field,
4 of the 7 known CVs are detected in RASS, and 9 new possible
CVs are discovered;
see Table \ref{tab:CV} for a summary of these sources.
The non-detections of known CVs are readily explained, as none is
expected to provide an X-ray flux large enough for the RASS sensitivity.

\begin{table}
  \caption{Summary of {known and candidate} cataclysmic variables in 
  our two fields.}
  \vspace{-0.2cm}
    \begin{tabular}{llccl}
     \hline
     \noalign{\smallskip}
  Name & Src-ID & Type & B (mag) & Comment \\
       &        &      & (USNO)    &         \\
     \noalign{\smallskip}
     \hline
     \noalign{\smallskip}
  IR Com     & Com141a & SU  & 13.4 &   \\
  J1238+1946 & --      & WD  & 17.4 & 1  \\
  S10980 Com & Com101a &     & 18.5 & 2 \\
  DO Vul     & --      & SU  & 20.7 &   \\
  J1953+1859 & --      & SU  & 20.4 &   \\
  V458 Vul   & --      & Nova&      & 3 \\
  QQ Vul     & Sge21a  & AM  & 14.4 &   \\
  V405 Vul   & Sge40a  & SU  & 19.0 &   \\
  WZ Sge     & Sge80a  & SU  & 15.0 &   \\
  HM Sge     & Sge102a & Nova&      &   \\
  S11015 Vul & Sge33a  &     & 14.8 &   \\
  S11017 Vul & Sge42a  &     & 15.5 &   \\
  S11020 Vul & Sge48e  &     & 18.6 & 2 \\
             & Sge48h  & AM  & 17.2 &   \\
  S11035 Vul & Sge92e  &     & 17.8 &   \\
  S11029 Vul & Sge114f &     & 16.3 & 4 \\
  S11030 Vul & Sge115b &     & 14.6 & 2 \\
  S11038 Aql & Sge120a & IP  & 17.2 &   \\
  S11033 Aql & Sge130b &     & 17.6 &   \\
     \noalign{\smallskip}
     \hline
  \end{tabular}

     $^{1)}$ WD = white dwarf. Double-degenerate binary (Brown et al. 2013). \\
     $^{2)}$ Is not the counterpart of the X-ray source.\\
     $^{3)}$ ROSAT observation was prior to outburst of 2007.\\
     $^{4)}$ Is possibly not the counterpart of the X-ray source.
  \label{tab:CV}
\end{table}

While we miss a more detailed sub-classification of these new possible
CVs, we can at least make statistical 
estimates of the population density of various classes of objects at
low and high galactic latitudes. 
In spite of their low luminosity ($M$ = +5...+12 mag at minimum), the 
high concentration of CVs towards low galactic latitudes is obvious, 
indicating a 
young (disk) stellar population. This nicely corresponds to earlier results of 
Richter (1968) who derived a logarithmic density gradient perpendicular to the 
galactic plane as large as $\delta log \nu / \delta z = 3.9$.

The identifications in our two fields, if all confirmed, would
double the number of cataclysmic variables in the fields.
Given the observed optical brightness of the new CV candidates, 
as well as their
optical colours and the X-ray hardness ratios, for none
of these sources a distance larger than $\sim$1 kpc would be required to
model the observed properties. This is particularly true for 
Sge48e$\equiv$S11020 Vul and Sge120a$\equiv$S11038 Aql. For the latter,
WISE photometry suggests a distance of about 350 pc, while for the
former an even smaller distance is mandatory due to the extremely soft,
i.e. unabsorbed, X-ray spectrum.
Thus, despite the lower X-ray luminosity both the optical magnitude
range as well as the covered distance/volume are very similar to
that used by Pretorius et al. (2013) when deriving the space density
of magnetic CVs. While we are missing the orbital periods and more
accurate distance estimates, the present results indicate that
the space density of CVs might be a factor 2 larger than derived
in Pretorius et al. (2013). This prospect may motivate more
detailed studies of these new CV candidates in the future.

\subsubsection{AGN variability} 

The field $\gamma$ Sge is basically free of AGN and other galaxies, 
except {\it Sge24a} and possibly
{\it Sge38a} $\equiv$ RXS J194356.1+211731, see text.
The following therefore is deduced exclusively from the high-galactic
latitude Coma field.

We start by noting that the X-ray and optical measurements were
not taken simultaneously, but years to decades apart. Long-term
flux variations of a factor of ten or even more are nothing unusual.
Optical variability studies have been moving from monitoring
a few dozen sources over weeks to years (e.g. Cristiani et al. 1990)
to the statistical study of thousands of sources over at most a 
few years (e.g. Bauer et al. 2009, Zuo et al. 2012). Here, we have
analysed the optical variability of two dozen AGN over timescales of 
50 years, which provides a less biased census of the more rare,
large-amplitude variations. 
These follow the anti-correlation with luminosity as
has been found in several previous studies (e.g. Cristiani et al. 1990,
Vanden Berk et al. 2004, Zuo et al. 2012). 

\section{Outlook - the quest for a massive identification program of RASS sources}

The X-ray source population of a galaxy provides important clues
  to the nature of its constituents and their evolution over the 
  lifetime of the galaxy, in particular also by tracing the endpoints
  of stellar evolution through accreting compact objects. The
  statistical analysis of X-ray source populations in nearby galaxies,
  e.g. number density, the X-ray luminosity function of various 
  source types as well as variability patterns, have led to
  substantial progress in our understanding of the evolution of
  important underlying physical parameters like metallicity and
  star formation rate (Gilfanov 2004, Grimm et al. 2003/2005,
  Townsley et al. 2006, Pietsch 2008, Fabbiano 2010, 
  Mineo et al. 2012). As we are moving from the study of just
  the most luminous sources (L$_{\rm X} > 10^{37}$ erg/s)
  to fainter thresholds, sometimes already as low as a few times
  10$^{35}$ erg/s, it becomes increasingly important to understand
  the variety of sources contributing in this luminosity range.
  This is where more detailed studies of X-ray sources in our
  galaxy come into play. Admittedly, luminosity estimates are
  much more uncertain for galactic sources due to limitations in the
  distance estimates and the clumpiness of foreground extinction.
  However, the diversity of spectral and temporal (variability at
  large, flares/outbursts, duty cycle, pulsed fractions) parameters
  are much easier to investigate for galactic sources. All the
  sources reported here will be among the bright sources of the 
  forthcoming eROSITA survey, allowing to do exactly those spectral 
  and timing studies. Only when having their optical identifications 
  at hand including the multi-wavelength properties can the full 
  power of these methods towards constraining evolutionary models 
  be realized.

\begin{acknowledgements}

We thank the 4pi Systeme - Gesellschaft f\"ur 
Astronomie und Informationstechnologie mbH, Sonneberg, Germany, 
especially P. Kroll and H.-J. Br\"auer,
for support of this project. We are indebted to the team of the 
Karl Schwarzschild Observatory Tautenburg for lending us plates from the 
2\,m Schmidt telescope. 
We are grateful to S. Nedyalkov (Uni. Bremen) and J. Brunschweiger (TU
M\"unchen) for their thorough
help in preparing the finding charts and part of tables \ref{comX},\ref{sgeX} 
and \ref{comopt},\ref{sgeopt}, to S. Zharykov (ONAM Ensenada, Mexico)
for the spectra of Com003 and Com007, and M. Salvato for discussion
about X-ray properties of AGN.
JG thanks N. Gehrels and the Swift team for accepting a substantial fraction
of the sources to be observed as ``fill-in'' targets with the Swift X-ray
telescope.
The early work (1993--1996) of this project has been supported by funds of the 
German Bundesministerium f\"ur Forschung und Technologie under grant 
05 2SO524 (GAR), and FKZ 50 OR 9201 (JG).
The ROSAT project was supported by the German 
Bundesministerium f\"ur Forschung und Technologie (BMBF/DLR) and 
the Max-Planck Society.
We appreciate the work of the XMM survey science centre, in particular
the regular updates of the EPIC Source  and XMM Slew Survey catalogues.
This research has made use of data obtained through the High Energy 
Astrophysics Science Archive Research Center Online Service, provided by the 
NASA/Goddard Space Flight Center.
This research has also made use of the VizieR catalogue access tool, CDS, 
Strasbourg, France, and the NASA/IPAC Extragalactic Database (NED) 
which is operated by the Jet Propulsion Laboratory, California Institute 
of Technology, under contract with the National Aeronautics and Space 
Administration.
Funding for SDSS-III has been provided by the Alfred P. Sloan Foundation, 
the Participating Institutions, the National Science Foundation, and 
the U.S. Department of Energy Office of Science. The SDSS-III web site 
is http://www.sdss3.org/.

SDSS-III is managed by the Astrophysical Research Consortium for the 
Participating Institutions of the SDSS-III Collaboration including the 
University of Arizona, the Brazilian Participation Group, Brookhaven National 
Laboratory, University of Cambridge, Carnegie Mellon University, University 
of Florida, the French Participation Group, the German Participation Group, 
Harvard University, the Instituto de Astrofisica de Canarias, the Michigan 
State/Notre Dame/JINA Participation Group, Johns Hopkins University, 
Lawrence Berkeley National Laboratory, Max Planck Institute for Astrophysics, 
Max Planck Institute for Extraterrestrial Physics, New Mexico State University,
New York University, Ohio State University, Pennsylvania State University, 
University of Portsmouth, Princeton University, the Spanish Participation 
Group, University of Tokyo, University of Utah, Vanderbilt University, 
University of Virginia, University of Washington, and Yale University. 
\end{acknowledgements}

\begin{subfigures}
\begin{figure*}[ht]
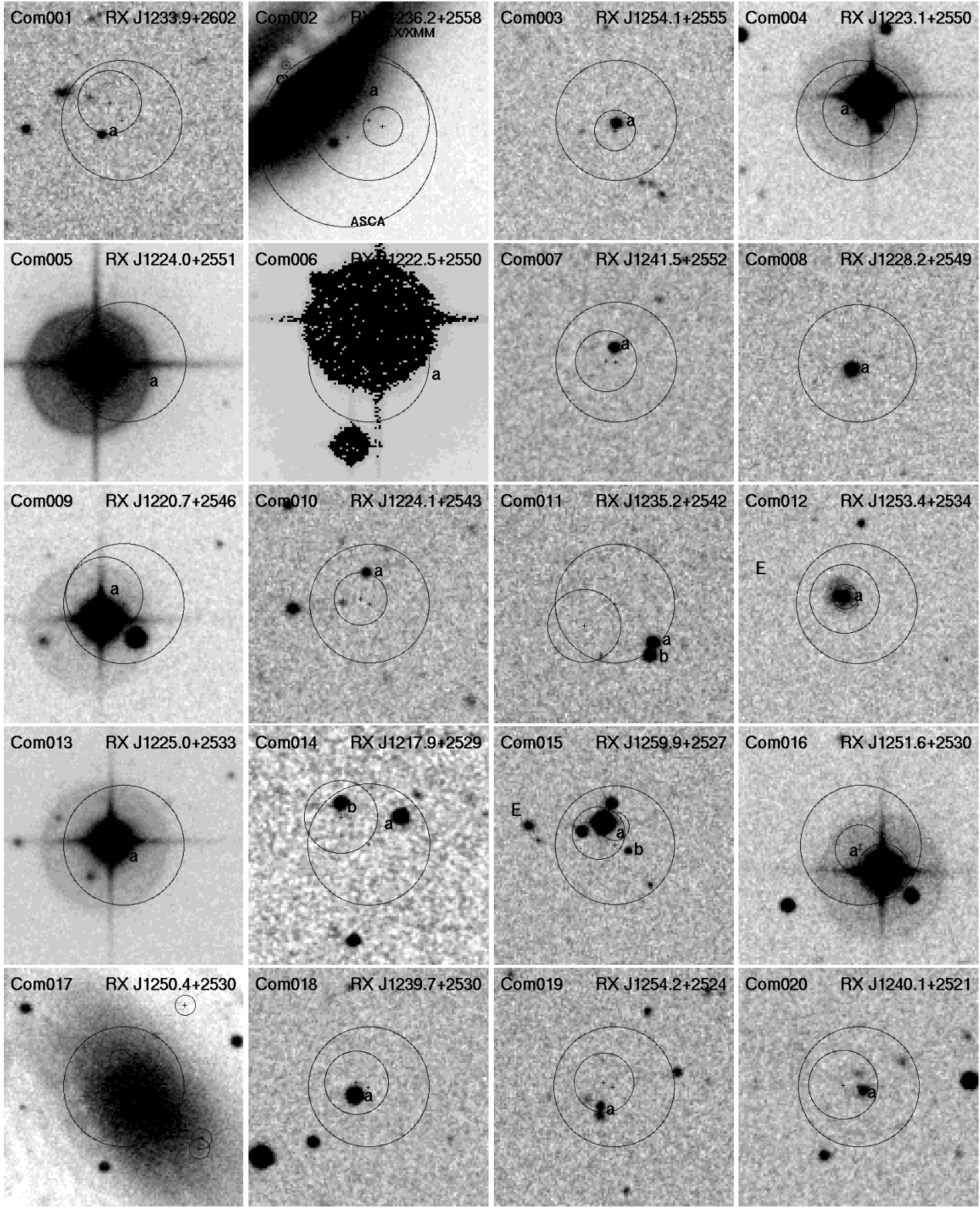

 \includegraphics[width=4.1cm]{dsscom/com001_d2b.ps}
 \includegraphics[width=4.1cm]{dsscom/com002_d2b.ps}
 \includegraphics[width=4.1cm]{dsscom/com003_d2b.ps}
 \includegraphics[width=4.1cm]{dsscom/com004_d2b.ps}

 \includegraphics[width=4.1cm]{dsscom/com005_d2b.ps}
 \includegraphics[width=4.1cm]{dsscom/com006_d2b.ps}
 \includegraphics[width=4.1cm]{dsscom/com007_d2b.ps}
 \includegraphics[width=4.1cm]{dsscom/com008_d2b.ps}

 \includegraphics[width=4.1cm]{dsscom/com009_d2b.ps}
 \includegraphics[width=4.1cm]{dsscom/com010_d2b.ps}
 \includegraphics[width=4.1cm]{dsscom/com011_d2b.ps}
 \includegraphics[width=4.1cm]{dsscom/com012_d2b.ps}

 \includegraphics[width=4.1cm]{dsscom/com013_d2b.ps}
 \includegraphics[width=4.1cm]{dsscom/com014_d2b.ps}
 \includegraphics[width=4.1cm]{dsscom/com015_d2b.ps}
 \includegraphics[width=4.1cm]{dsscom/com016_d2b.ps}

 \includegraphics[width=4.1cm]{dsscom/com017_d2b.ps}
 \includegraphics[width=4.1cm]{dsscom/com018_d2b.ps}
 \includegraphics[width=4.1cm]{dsscom/com019_d2b.ps}
 \includegraphics[width=4.1cm]{dsscom/com020_d2b.ps}

\caption[fcc]{\label{comfc} DSS blue finding chart of Com sources 
      Com001 to Com020.
      The size is 2\farcm5 $\times$  2\farcm5, North is up and East 
      to the left.
      The central circle is the 30\asec\ error circle as derived from
      our source detection on the RASS data. Shifted circles of similar (or 
      slightly smaller) size are ROSAT source position errors taken from the 
      1RXS catalogue (see 4th column in Tables \ref{comX}, \ref{sgeX}), 
      and very small shifted circles are source positions taken from the 
      ROSAT HRI (1RXH) catalogue or XRT positions from our Swift follow-up
      (see the notes on individual objects for details). 
      Many of the alphabetically labelled 
      objects have been investigated for variability, and details are
      given in Table \ref{comopt},\ref{sgeopt}.}
\end{figure*}

\setcounter{figure}{0}

\begin{figure*}[ht]
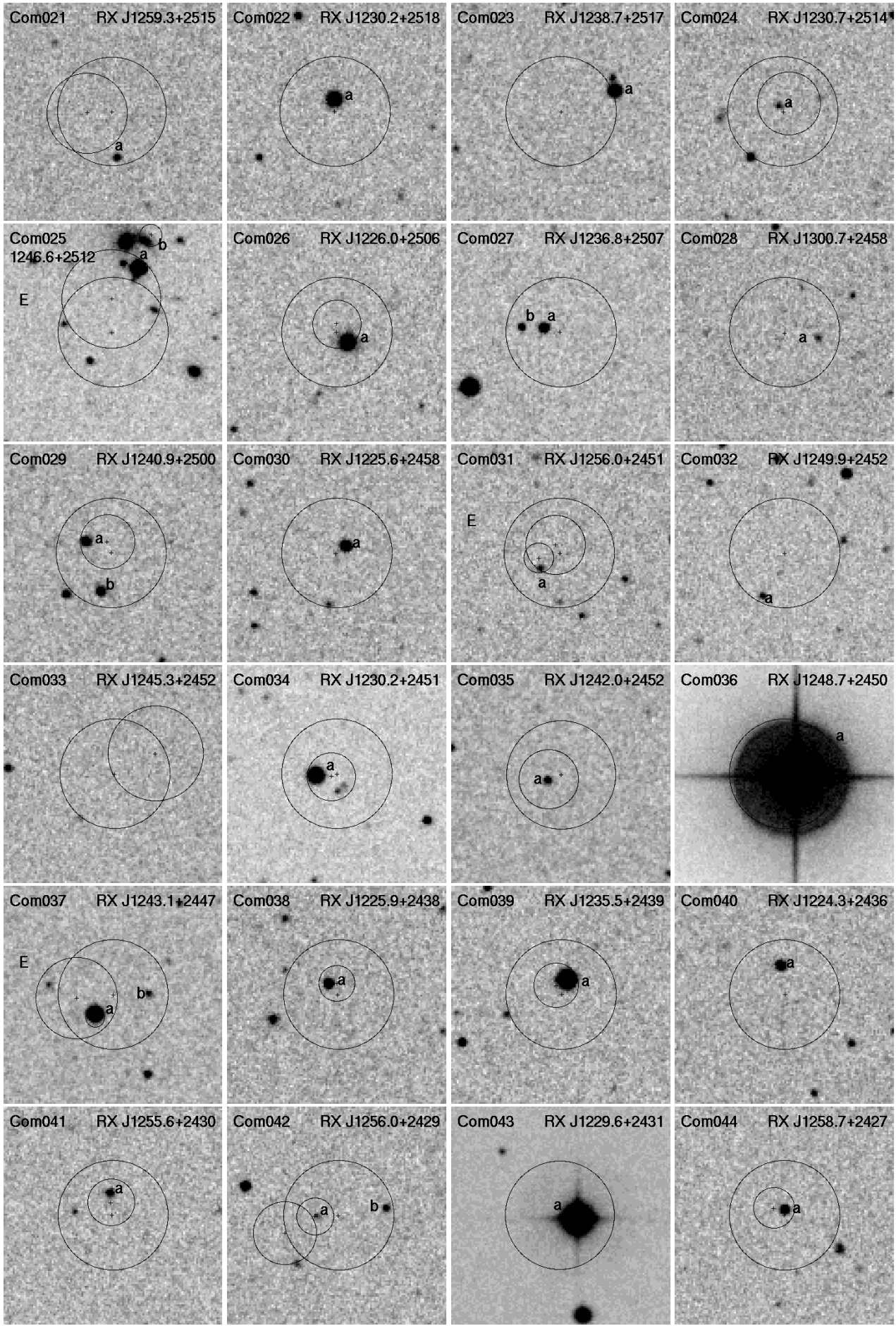

 \includegraphics[width=4.1cm]{dsscom/com021_d2b.ps}
 \includegraphics[width=4.1cm]{dsscom/com022_d2b.ps}
 \includegraphics[width=4.1cm]{dsscom/com023_d2b.ps}
 \includegraphics[width=4.1cm]{dsscom/com024_d2b.ps}

 \includegraphics[width=4.1cm]{dsscom/com025_d2b.ps}
 \includegraphics[width=4.1cm]{dsscom/com026_d2b.ps}
 \includegraphics[width=4.1cm]{dsscom/com027_d2b.ps}
 \includegraphics[width=4.1cm]{dsscom/com028_d2b.ps}

 \includegraphics[width=4.1cm]{dsscom/com029_d2b.ps}
 \includegraphics[width=4.1cm]{dsscom/com030_d2b.ps}
 \includegraphics[width=4.1cm]{dsscom/com031_d2b.ps}
 \includegraphics[width=4.1cm]{dsscom/com032_d2b.ps}

 \includegraphics[width=4.1cm]{dsscom/com033_d2b.ps}
 \includegraphics[width=4.1cm]{dsscom/com034_d2b.ps}
 \includegraphics[width=4.1cm]{dsscom/com035_d2b.ps}
 \includegraphics[width=4.1cm]{dsscom/com036_d2b.ps}

 \includegraphics[width=4.1cm]{dsscom/com037_d2b.ps}
 \includegraphics[width=4.1cm]{dsscom/com038_d2b.ps}
 \includegraphics[width=4.1cm]{dsscom/com039_d2b.ps}
 \includegraphics[width=4.1cm]{dsscom/com040_d2b.ps}

 \includegraphics[width=4.1cm]{dsscom/com041_d2b.ps}
 \includegraphics[width=4.1cm]{dsscom/com042_d2b.ps}
 \includegraphics[width=4.1cm]{dsscom/com043_d2b.ps}
 \includegraphics[width=4.1cm]{dsscom/com044_d2b.ps}
 \caption[fcc]{{\bf contd.} Com sources Com021 to Com044.}
 \end{figure*}

\setcounter{figure}{0}

\begin{figure*}[ht]
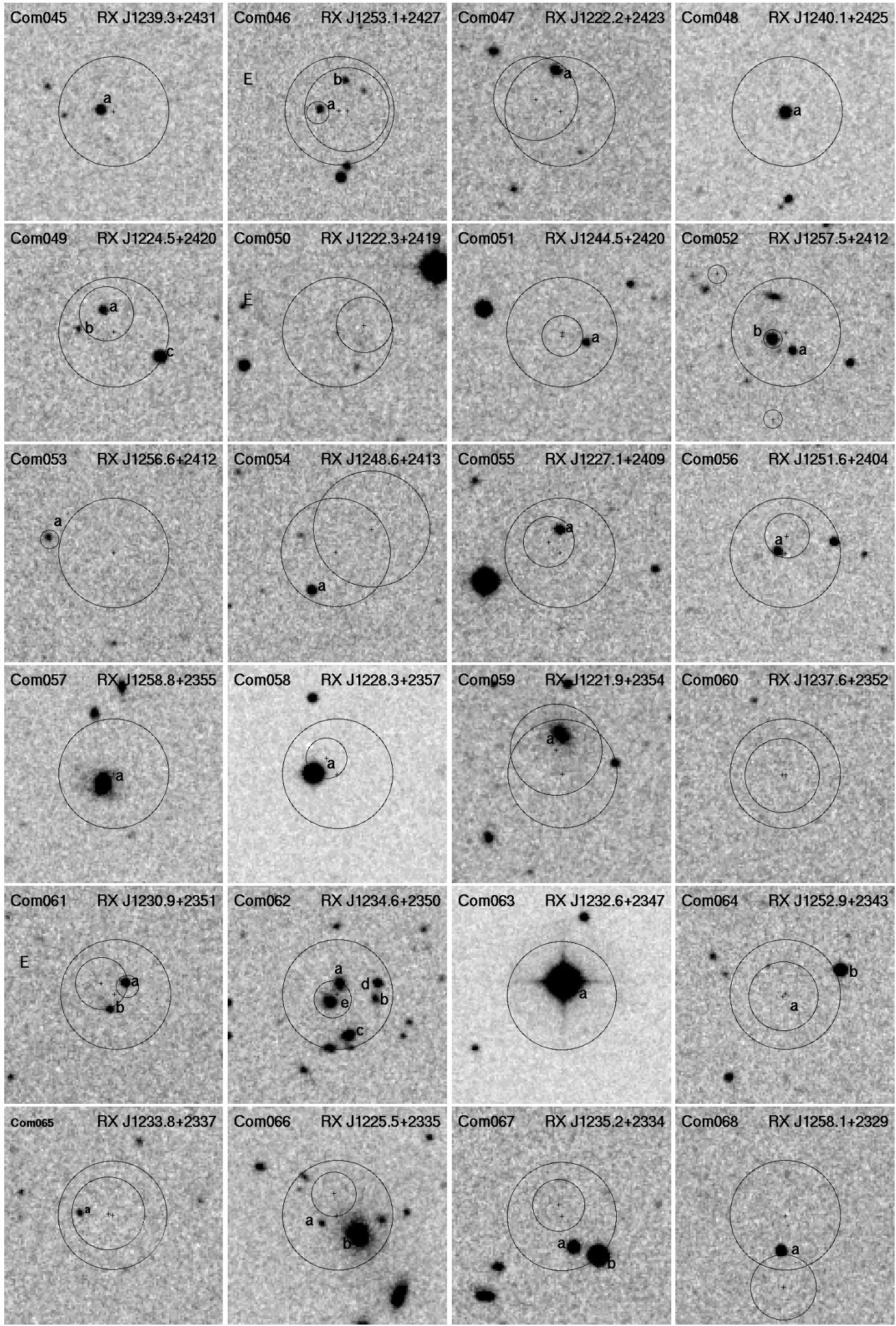

 \includegraphics[width=4.1cm]{dsscom/com045_d2b.ps}
 \includegraphics[width=4.1cm]{dsscom/com046_d2b.ps}
 \includegraphics[width=4.1cm]{dsscom/com047_d2b.ps}
 \includegraphics[width=4.1cm]{dsscom/com048_d2b.ps}

 \includegraphics[width=4.1cm]{dsscom/com049_d2b.ps}
 \includegraphics[width=4.1cm]{dsscom/com050_d2b.ps}
 \includegraphics[width=4.1cm]{dsscom/com051_d2b.ps}
 \includegraphics[width=4.1cm]{dsscom/com052_d2b.ps}

 \includegraphics[width=4.1cm]{dsscom/com053_d2b.ps}
 \includegraphics[width=4.1cm]{dsscom/com054_d2b.ps}
 \includegraphics[width=4.1cm]{dsscom/com055_d2b.ps}
 \includegraphics[width=4.1cm]{dsscom/com056_d2b.ps}

 \includegraphics[width=4.1cm]{dsscom/com057_d2b.ps}
 \includegraphics[width=4.1cm]{dsscom/com058_d2b.ps}
 \includegraphics[width=4.1cm]{dsscom/com059_d2b.ps}
 \includegraphics[width=4.1cm]{dsscom/com060_d2b.ps}

 \includegraphics[width=4.1cm]{dsscom/com061_d2b.ps}
 \includegraphics[width=4.1cm]{dsscom/com062_d2b.ps}
 \includegraphics[width=4.1cm]{dsscom/com063_d2b.ps}
 \includegraphics[width=4.1cm]{dsscom/com064_d2b.ps}

 \includegraphics[width=4.1cm]{dsscom/com065_d2b.ps}
 \includegraphics[width=4.1cm]{dsscom/com066_d2b.ps}
 \includegraphics[width=4.1cm]{dsscom/com067_d2b.ps}
 \includegraphics[width=4.1cm]{dsscom/com068_d2b.ps}
 \caption[fcc]{{\bf contd.} Com sources Com045 to Com068.
   Object {\it a} for Com064 = RX J1252.9+2343
   is either strongly variable or very red, and
   not visible on the DSS plate.}
 \end{figure*}

\setcounter{figure}{0}

\begin{figure*}[ht]
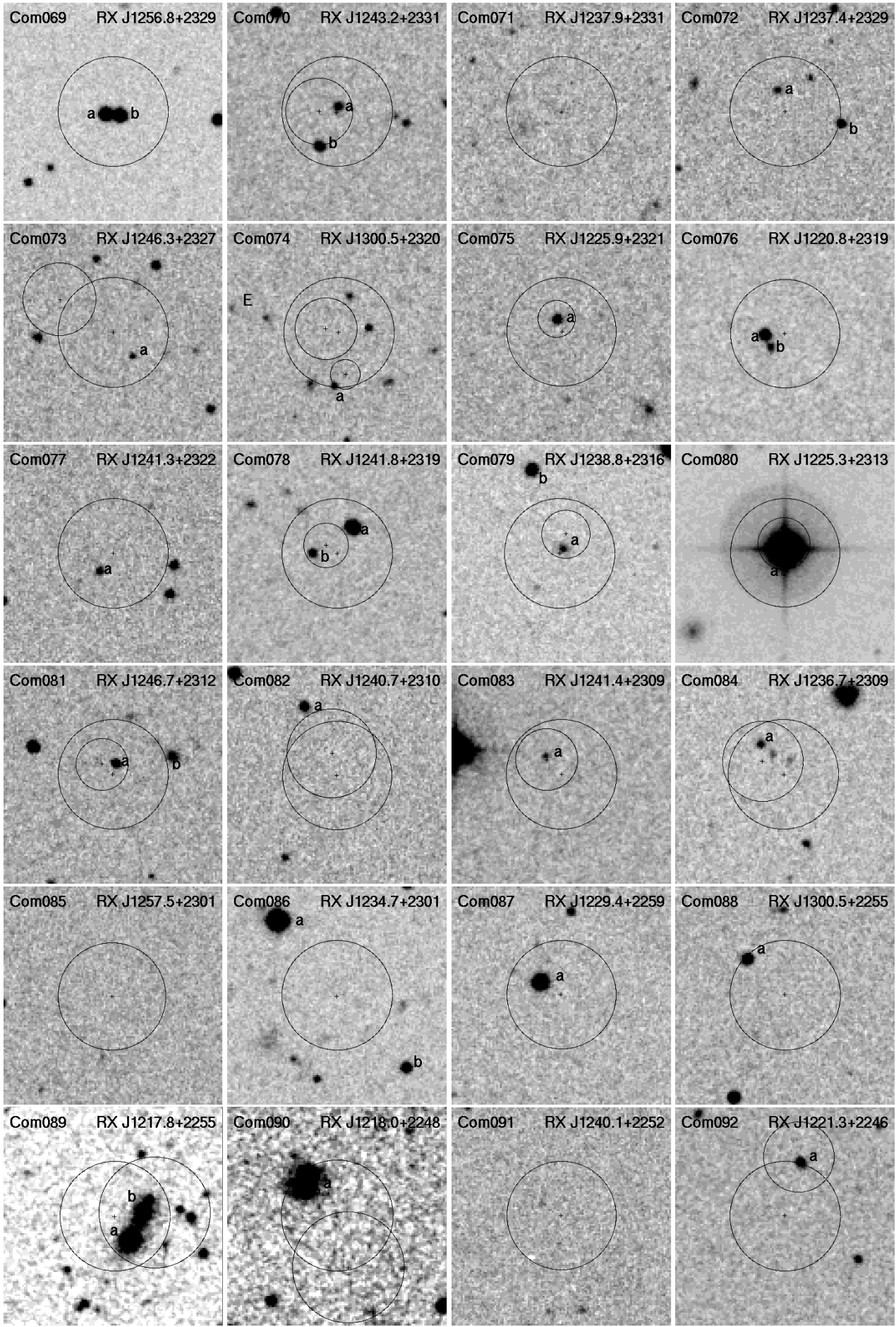

 \includegraphics[width=4.1cm]{dsscom/com069_d2b.ps}
 \includegraphics[width=4.1cm]{dsscom/com070_d2b.ps}
 \includegraphics[width=4.1cm]{dsscom/com071_d2b.ps}
 \includegraphics[width=4.1cm]{dsscom/com072_d2b.ps}

 \includegraphics[width=4.1cm]{dsscom/com073_d2b.ps}
 \includegraphics[width=4.1cm]{dsscom/com074_d2b.ps}
 \includegraphics[width=4.1cm]{dsscom/com075_d2b.ps}
 \includegraphics[width=4.1cm]{dsscom/com076_d2b.ps}

 \includegraphics[width=4.1cm]{dsscom/com077_d2b.ps}
 \includegraphics[width=4.1cm]{dsscom/com078_d2b.ps}
 \includegraphics[width=4.1cm]{dsscom/com079_d2b.ps}
 \includegraphics[width=4.1cm]{dsscom/com080_d2b.ps}

 \includegraphics[width=4.1cm]{dsscom/com081_d2b.ps}
 \includegraphics[width=4.1cm]{dsscom/com082_d2b.ps}
 \includegraphics[width=4.1cm]{dsscom/com083_d2b.ps}
 \includegraphics[width=4.1cm]{dsscom/com084_d2b.ps}

 \includegraphics[width=4.1cm]{dsscom/com085_d2b.ps}
 \includegraphics[width=4.1cm]{dsscom/com086_d2b.ps}
 \includegraphics[width=4.1cm]{dsscom/com087_d2b.ps}
 \includegraphics[width=4.1cm]{dsscom/com088_d2b.ps}

 \includegraphics[width=4.1cm]{dsscom/com089_d2b.ps}
 \includegraphics[width=4.1cm]{dsscom/com090_d2b.ps}
 \includegraphics[width=4.1cm]{dsscom/com091_d2b.ps}
 \includegraphics[width=4.1cm]{dsscom/com092_d2b.ps}
 \caption[fcc]{{\bf contd.} Com sources Com069 to Com092.}
 \end{figure*}

\setcounter{figure}{0}

\begin{figure*}[ht]
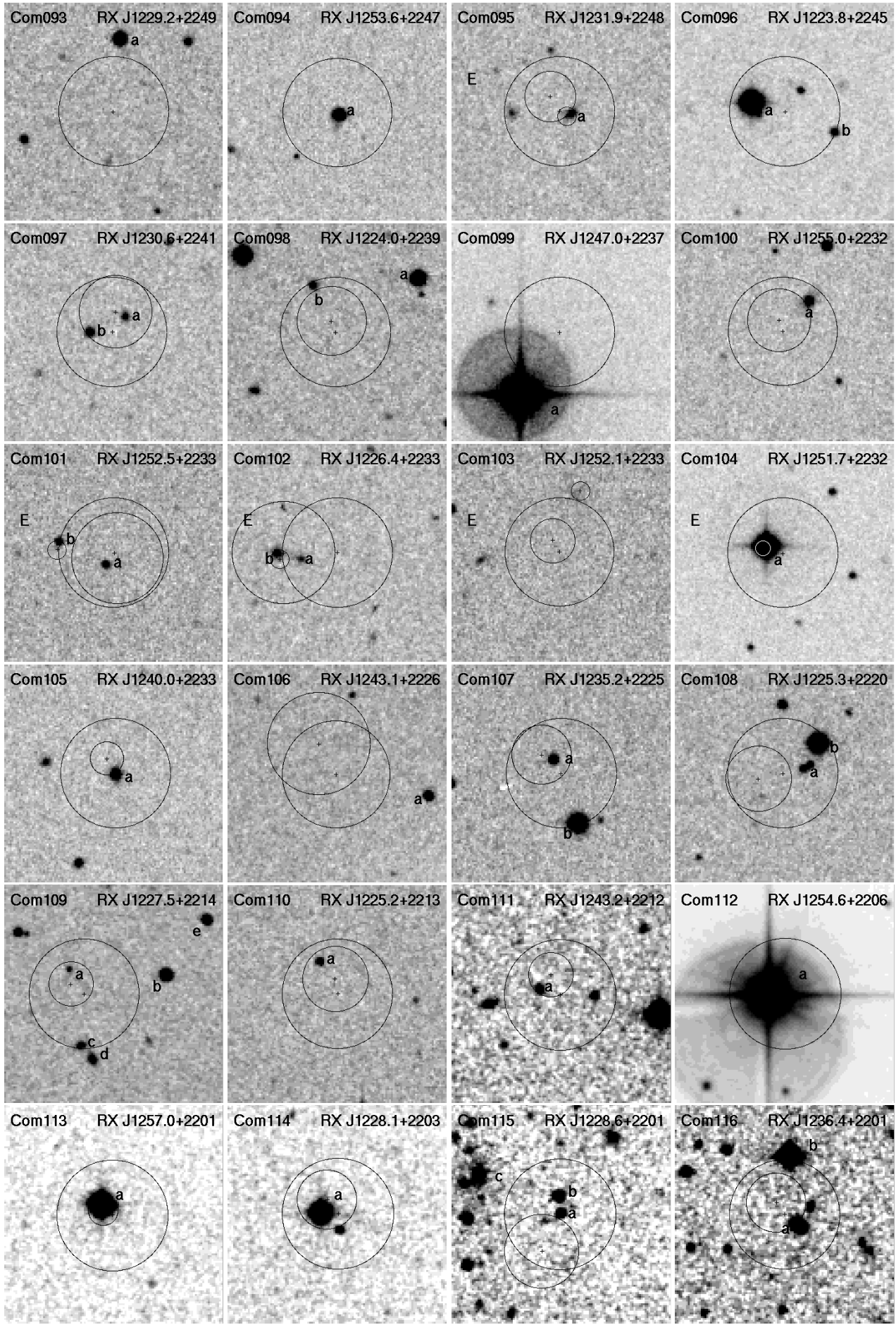

 \includegraphics[width=4.1cm]{dsscom/com093_d2b.ps}
 \includegraphics[width=4.1cm]{dsscom/com094_d2b.ps}
 \includegraphics[width=4.1cm]{dsscom/com095_d2b.ps}
 \includegraphics[width=4.1cm]{dsscom/com096_d2b.ps}

 \includegraphics[width=4.1cm]{dsscom/com097_d2b.ps}
 \includegraphics[width=4.1cm]{dsscom/com098_d2b.ps}
 \includegraphics[width=4.1cm]{dsscom/com099_d2b.ps}
 \includegraphics[width=4.1cm]{dsscom/com100_d2b.ps}

 \includegraphics[width=4.1cm]{dsscom/com101_d2b.ps}
 \includegraphics[width=4.1cm]{dsscom/com102_d2b.ps}
 \includegraphics[width=4.1cm]{dsscom/com103_d2b.ps}
 \includegraphics[width=4.1cm]{dsscom/com104_d2b.ps}

 \includegraphics[width=4.1cm]{dsscom/com105_d2b.ps}
 \includegraphics[width=4.1cm]{dsscom/com106_d2b.ps}
 \includegraphics[width=4.1cm]{dsscom/com107_d2b.ps}
 \includegraphics[width=4.1cm]{dsscom/com108_d2b.ps}

 \includegraphics[width=4.1cm]{dsscom/com109_d2b.ps}
 \includegraphics[width=4.1cm]{dsscom/com110_d2b.ps}
 \includegraphics[width=4.1cm]{dsscom/com111_d2b.ps}
 \includegraphics[width=4.1cm]{dsscom/com112_d2b.ps}

 \includegraphics[width=4.1cm]{dsscom/com113_d2b.ps}
 \includegraphics[width=4.1cm]{dsscom/com114_d2b.ps}
 \includegraphics[width=4.1cm]{dsscom/com115_d2b.ps}
 \includegraphics[width=4.1cm]{dsscom/com116_d2b.ps}
 \caption[fcc]{{\bf contd.} Com sources Com093 to Com116.}
 \end{figure*}

\setcounter{figure}{0}

\begin{figure*}[ht]
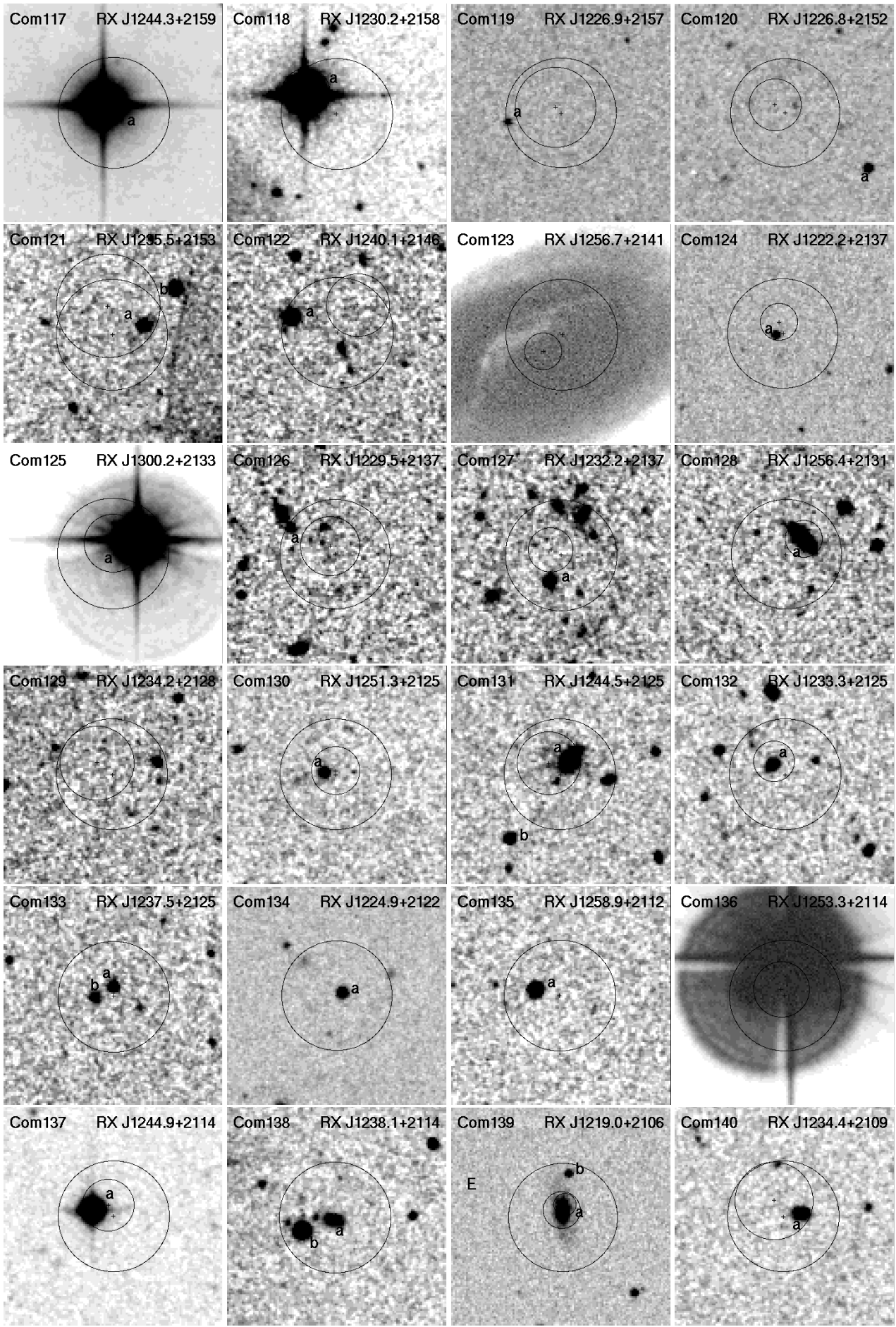

 \includegraphics[width=4.1cm]{dsscom/com117_d2b.ps}
 \includegraphics[width=4.1cm]{dsscom/com118_d2b.ps}
 \includegraphics[width=4.1cm]{dsscom/com119_d2b.ps}
 \includegraphics[width=4.1cm]{dsscom/com120_d2b.ps}

 \includegraphics[width=4.1cm]{dsscom/com121_d2b.ps}
 \includegraphics[width=4.1cm]{dsscom/com122_d2b.ps}
 \includegraphics[width=4.1cm]{dsscom/com123_d2b.ps}
 \includegraphics[width=4.1cm]{dsscom/com124_d2b.ps}

 \includegraphics[width=4.1cm]{dsscom/com125_d2b.ps}
 \includegraphics[width=4.1cm]{dsscom/com126_d2b.ps}
 \includegraphics[width=4.1cm]{dsscom/com127_d2b.ps}
 \includegraphics[width=4.1cm]{dsscom/com128_d2b.ps}

 \includegraphics[width=4.1cm]{dsscom/com129_d2b.ps}
 \includegraphics[width=4.1cm]{dsscom/com130_d2b.ps}
 \includegraphics[width=4.1cm]{dsscom/com131_d2b.ps}
 \includegraphics[width=4.1cm]{dsscom/com132_d2b.ps}

 \includegraphics[width=4.1cm]{dsscom/com133_d2b.ps}
 \includegraphics[width=4.1cm]{dsscom/com134_d2b.ps}
 \includegraphics[width=4.1cm]{dsscom/com135_d2b.ps}
 \includegraphics[width=4.1cm]{dsscom/com136_d2b.ps}

 \includegraphics[width=4.1cm]{dsscom/com137_d2b.ps}
 \includegraphics[width=4.1cm]{dsscom/com138_d2b.ps}
 \includegraphics[width=4.1cm]{dsscom/com139_d2b.ps}
 \includegraphics[width=4.1cm]{dsscom/com140_d2b.ps}
 \caption[fcc]{{\bf contd.} Com sources Com117 to Com140.}
 \end{figure*}

\setcounter{figure}{0}

\begin{figure*}[ht]
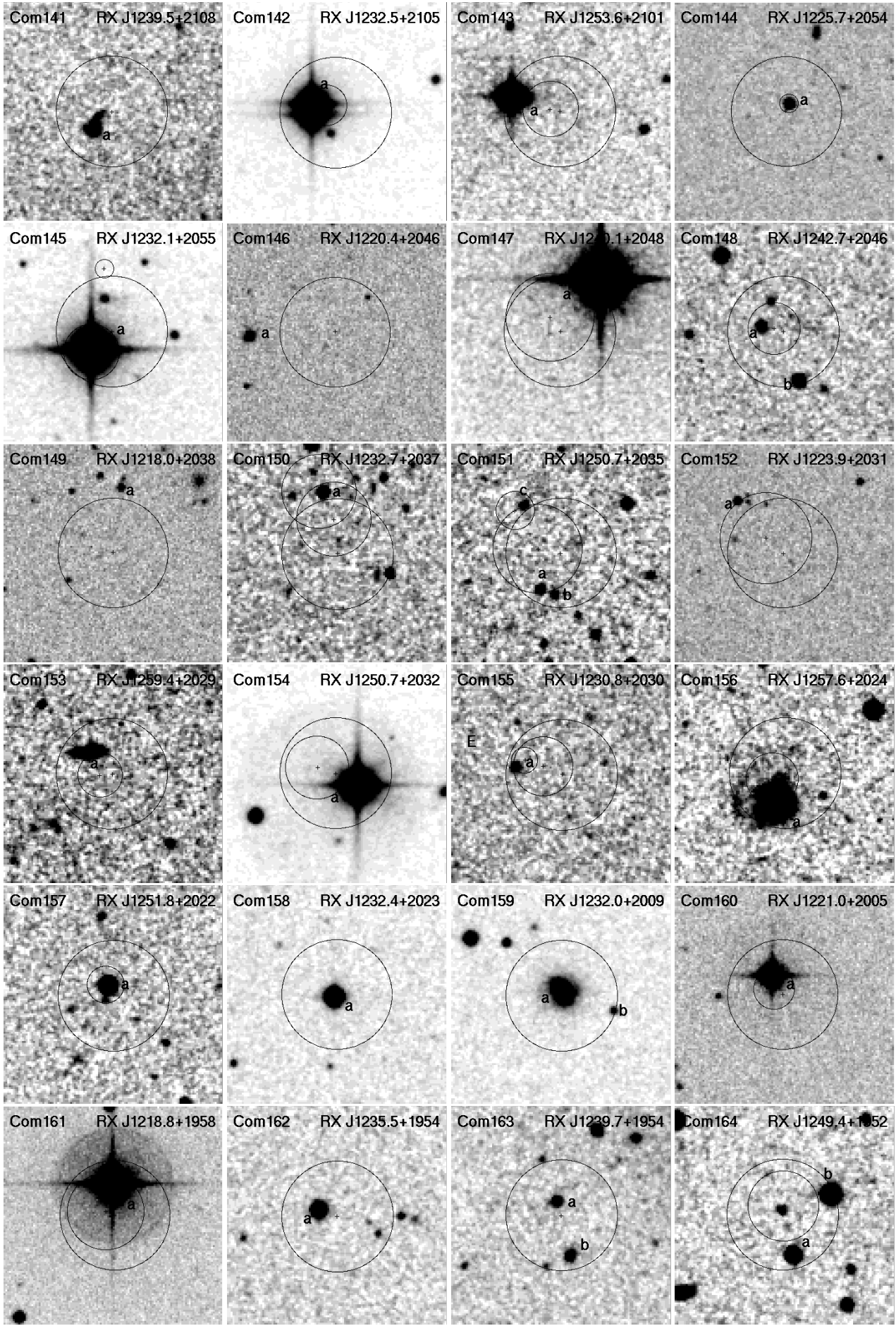

 \includegraphics[width=4.1cm]{dsscom/com141_d2b.ps}
 \includegraphics[width=4.1cm]{dsscom/com142_d2b.ps}
 \includegraphics[width=4.1cm]{dsscom/com143_d2b.ps}
 \includegraphics[width=4.1cm]{dsscom/com144_d2b.ps}

 \includegraphics[width=4.1cm]{dsscom/com145_d2b.ps}
 \includegraphics[width=4.1cm]{dsscom/com146_d2b.ps}
 \includegraphics[width=4.1cm]{dsscom/com147_d2b.ps}
 \includegraphics[width=4.1cm]{dsscom/com148_d2b.ps}

 \includegraphics[width=4.1cm]{dsscom/com149_d2b.ps}
 \includegraphics[width=4.1cm]{dsscom/com150_d2b.ps}
 \includegraphics[width=4.1cm]{dsscom/com151_d2b.ps}
 \includegraphics[width=4.1cm]{dsscom/com152_d2b.ps}

 \includegraphics[width=4.1cm]{dsscom/com153_d2b.ps}
 \includegraphics[width=4.1cm]{dsscom/com154_d2b.ps}
 \includegraphics[width=4.1cm]{dsscom/com155_d2b.ps}
 \includegraphics[width=4.1cm]{dsscom/com156_d2b.ps}

 \includegraphics[width=4.1cm]{dsscom/com157_d2b.ps}
 \includegraphics[width=4.1cm]{dsscom/com158_d2b.ps}
 \includegraphics[width=4.1cm]{dsscom/com159_d2b.ps}
 \includegraphics[width=4.1cm]{dsscom/com160_d2b.ps}

 \includegraphics[width=4.1cm]{dsscom/com161_d2b.ps}
 \includegraphics[width=4.1cm]{dsscom/com162_d2b.ps}
 \includegraphics[width=4.1cm]{dsscom/com163_d2b.ps}
 \includegraphics[width=4.1cm]{dsscom/com164_d2b.ps}
 \caption[fcc]{{\bf contd.} Com sources Com141 to Com164.}
 \end{figure*}

\setcounter{figure}{0}

\begin{figure*}[ht]
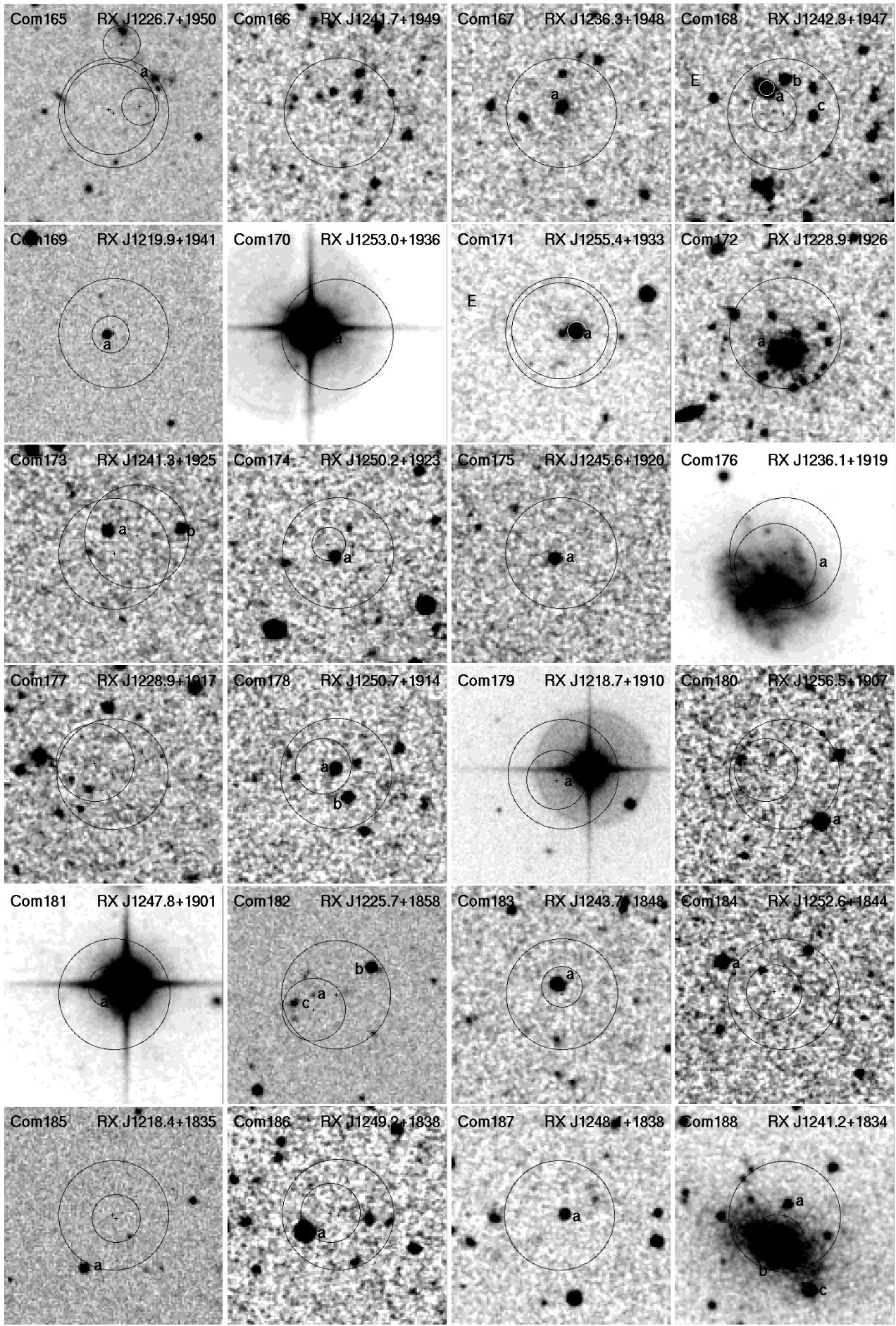

 \includegraphics[width=4.1cm]{dsscom/com165_d2b.ps}
 \includegraphics[width=4.1cm]{dsscom/com166_d2b.ps}
 \includegraphics[width=4.1cm]{dsscom/com167_d2b.ps}
 \includegraphics[width=4.1cm]{dsscom/com168_d2b.ps}

 \includegraphics[width=4.1cm]{dsscom/com169_d2b.ps}
 \includegraphics[width=4.1cm]{dsscom/com170_d2b.ps}
 \includegraphics[width=4.1cm]{dsscom/com171_d2b.ps}
 \includegraphics[width=4.1cm]{dsscom/com172_d2b.ps}

 \includegraphics[width=4.1cm]{dsscom/com173_d2b.ps}
 \includegraphics[width=4.1cm]{dsscom/com174_d2b.ps}
 \includegraphics[width=4.1cm]{dsscom/com175_d2b.ps}
 \includegraphics[width=4.1cm]{dsscom/com176_d2b.ps}

 \includegraphics[width=4.1cm]{dsscom/com177_d2b.ps}
 \includegraphics[width=4.1cm]{dsscom/com178_d2b.ps}
 \includegraphics[width=4.1cm]{dsscom/com179_d2b.ps}
 \includegraphics[width=4.1cm]{dsscom/com180_d2b.ps}

 \includegraphics[width=4.1cm]{dsscom/com181_d2b.ps}
 \includegraphics[width=4.1cm]{dsscom/com182_d2b.ps}
 \includegraphics[width=4.1cm]{dsscom/com183_d2b.ps}
 \includegraphics[width=4.1cm]{dsscom/com184_d2b.ps}

 \includegraphics[width=4.1cm]{dsscom/com185_d2b.ps}
 \includegraphics[width=4.1cm]{dsscom/com186_d2b.ps}
 \includegraphics[width=4.1cm]{dsscom/com187_d2b.ps}
 \includegraphics[width=4.1cm]{dsscom/com188_d2b.ps}
 \caption[fcc]{{\bf contd.} Com sources Com165 to Com188.}
 \end{figure*}

\setcounter{figure}{0}

\begin{figure*}[ht]
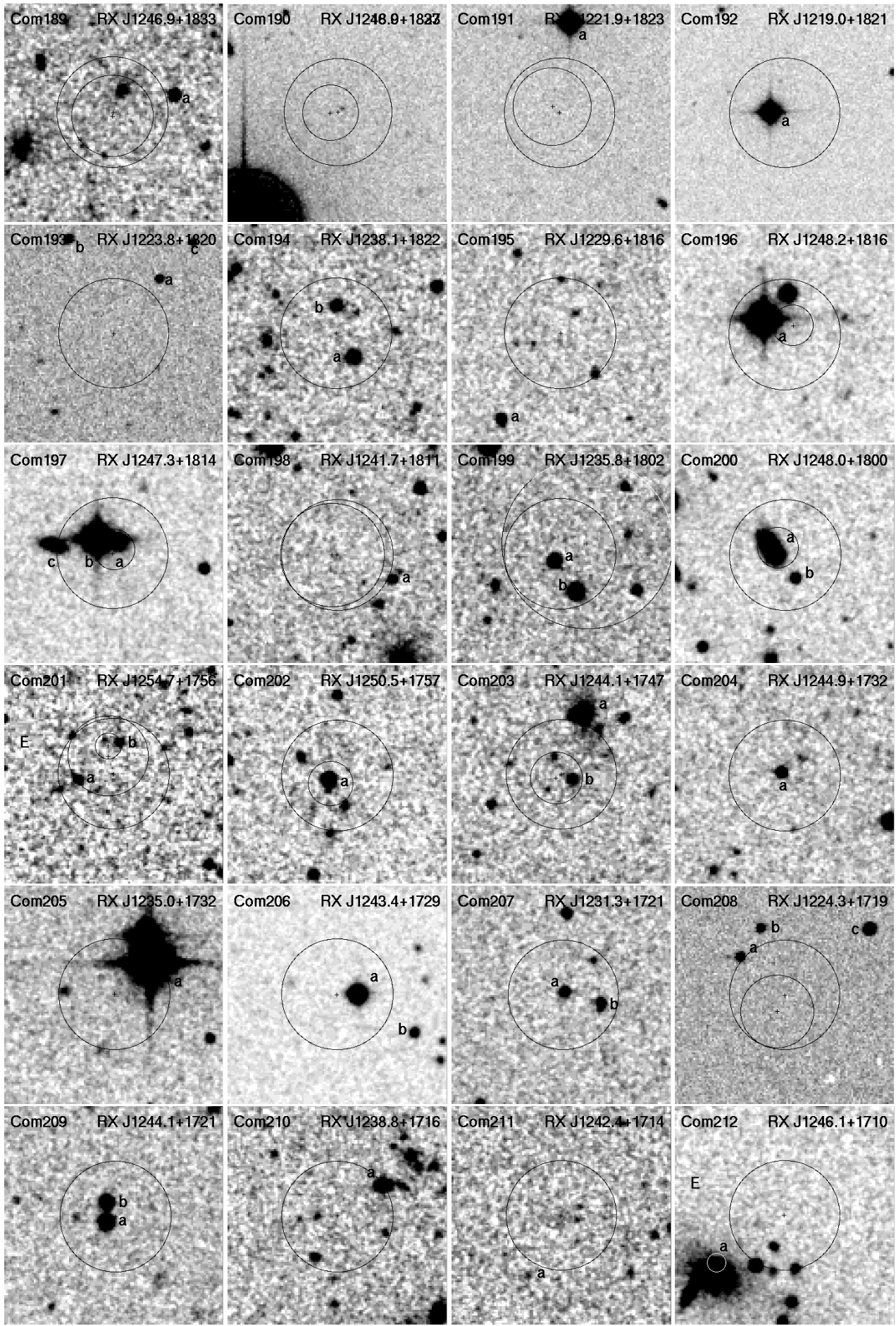

 \includegraphics[width=4.1cm]{dsscom/com189_d2b.ps}
 \includegraphics[width=4.1cm]{dsscom/com190_d2b.ps}
 \includegraphics[width=4.1cm]{dsscom/com191_d2b.ps}
 \includegraphics[width=4.1cm]{dsscom/com192_d2b.ps}

 \includegraphics[width=4.1cm]{dsscom/com193_d2b.ps}
 \includegraphics[width=4.1cm]{dsscom/com194_d2b.ps}
 \includegraphics[width=4.1cm]{dsscom/com195_d2b.ps}
 \includegraphics[width=4.1cm]{dsscom/com196_d2b.ps}

 \includegraphics[width=4.1cm]{dsscom/com197_d2b.ps}
 \includegraphics[width=4.1cm]{dsscom/com198_d2b.ps}
 \includegraphics[width=4.1cm]{dsscom/com199_d2b.ps}
 \includegraphics[width=4.1cm]{dsscom/com200_d2b.ps}

 \includegraphics[width=4.1cm]{dsscom/com201_d2b.ps}
 \includegraphics[width=4.1cm]{dsscom/com202_d2b.ps}
 \includegraphics[width=4.1cm]{dsscom/com203_d2b.ps}
 \includegraphics[width=4.1cm]{dsscom/com204_d2b.ps}

 \includegraphics[width=4.1cm]{dsscom/com205_d2b.ps}
 \includegraphics[width=4.1cm]{dsscom/com206_d2b.ps}
 \includegraphics[width=4.1cm]{dsscom/com207_d2b.ps}
 \includegraphics[width=4.1cm]{dsscom/com208_d2b.ps}

 \includegraphics[width=4.1cm]{dsscom/com209_d2b.ps}
 \includegraphics[width=4.1cm]{dsscom/com210_d2b.ps}
 \includegraphics[width=4.1cm]{dsscom/com211_d2b.ps}
 \includegraphics[width=4.1cm]{dsscom/com212_d2b.ps}
 \caption[fcc]{{\bf contd.} Com sources Com189 to Com212.}
 \end{figure*}

\setcounter{figure}{0}

\begin{figure*}[ht]
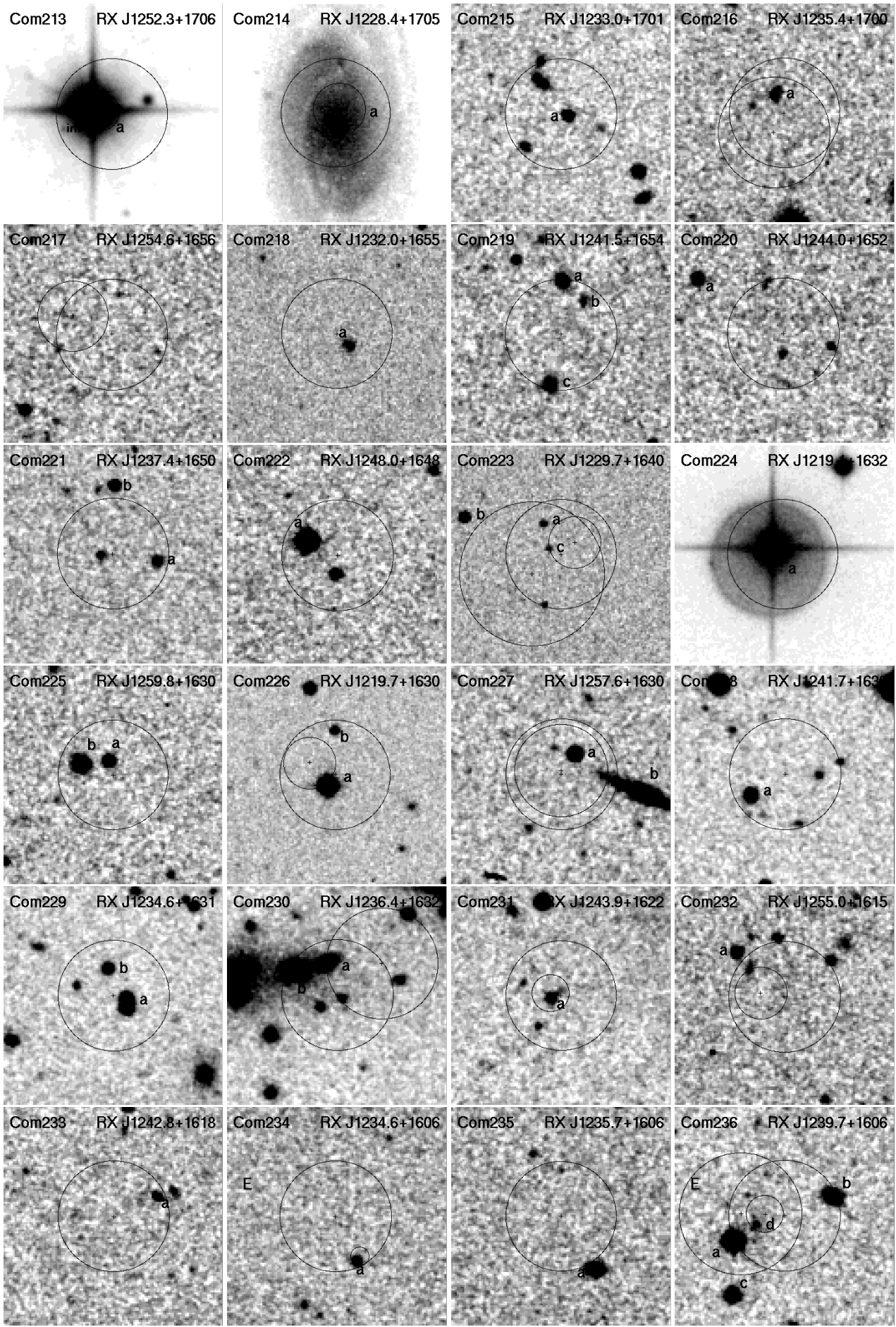

 \includegraphics[width=4.1cm]{dsscom/com213_d2b.ps}
 \includegraphics[width=4.1cm]{dsscom/com214_d2b.ps}
 \includegraphics[width=4.1cm]{dsscom/com215_d2b.ps}
 \includegraphics[width=4.1cm]{dsscom/com216_d2b.ps}

 \includegraphics[width=4.1cm]{dsscom/com217_d2b.ps}
 \includegraphics[width=4.1cm]{dsscom/com218_d2b.ps}
 \includegraphics[width=4.1cm]{dsscom/com219_d2b.ps}
 \includegraphics[width=4.1cm]{dsscom/com220_d2b.ps}

 \includegraphics[width=4.1cm]{dsscom/com221_d2b.ps}
 \includegraphics[width=4.1cm]{dsscom/com222_d2b.ps}
 \includegraphics[width=4.1cm]{dsscom/com223_d2b.ps}
 \includegraphics[width=4.1cm]{dsscom/com224_d2b.ps}

 \includegraphics[width=4.1cm]{dsscom/com225_d2b.ps}
 \includegraphics[width=4.1cm]{dsscom/com226_d2b.ps}
 \includegraphics[width=4.1cm]{dsscom/com227_d2b.ps}
 \includegraphics[width=4.1cm]{dsscom/com228_d2b.ps}

 \includegraphics[width=4.1cm]{dsscom/com229_d2b.ps}
 \includegraphics[width=4.1cm]{dsscom/com230_d2b.ps}
 \includegraphics[width=4.1cm]{dsscom/com231_d2b.ps}
 \includegraphics[width=4.1cm]{dsscom/com232_d2b.ps}

 \includegraphics[width=4.1cm]{dsscom/com233_d2b.ps}
 \includegraphics[width=4.1cm]{dsscom/com234_d2b.ps}
 \includegraphics[width=4.1cm]{dsscom/com235_d2b.ps}
 \includegraphics[width=4.1cm]{dsscom/com236_d2b.ps}
 \caption[fcc]{{\bf contd.} Com sources Com213 to Com236.}
\end{figure*}

\setcounter{figure}{0}

\begin{figure*}[ht]
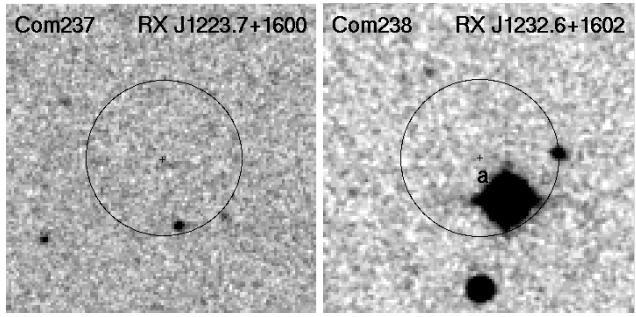

 \includegraphics[width=4.1cm]{dsscom/com237_d2b.ps}
 \includegraphics[width=4.1cm]{dsscom/com238_d2b.ps}
 \caption[fcc]{{\bf contd.} Com sources Com237 to Com238.}

\end{figure*}

\newpage

\begin{figure*}[ht]
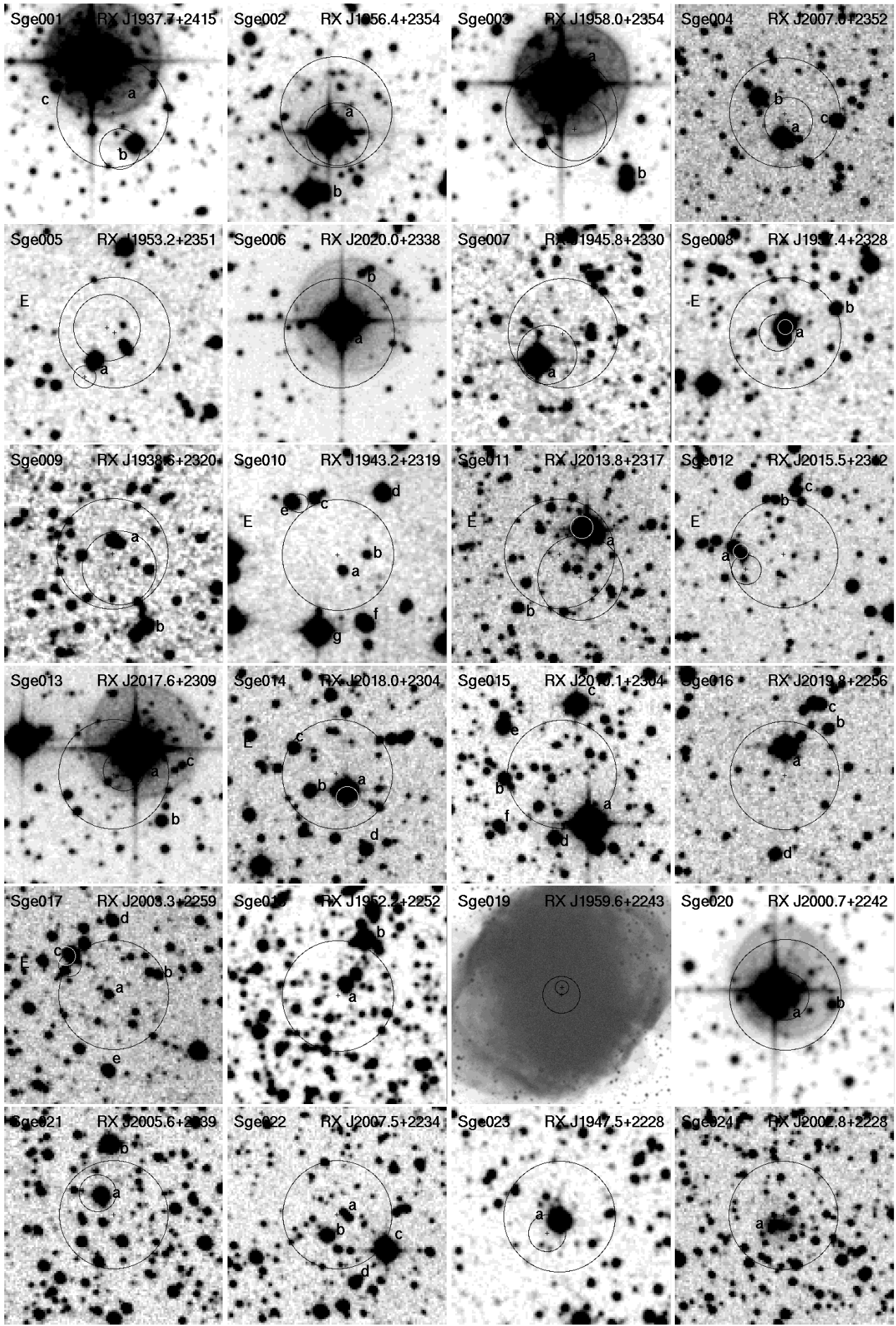

 \includegraphics[width=4.1cm]{dsssge/sge001_d2b.ps}
 \includegraphics[width=4.1cm]{dsssge/sge002_d2b.ps}
 \includegraphics[width=4.1cm]{dsssge/sge003_d2b.ps}
 \includegraphics[width=4.1cm]{dsssge/sge004_d2b.ps}

 \includegraphics[width=4.1cm]{dsssge/sge005_d2b.ps}
 \includegraphics[width=4.1cm]{dsssge/sge006_d2b.ps}
 \includegraphics[width=4.1cm]{dsssge/sge007_d2b.ps}
 \includegraphics[width=4.1cm]{dsssge/sge008_d2b.ps}

 \includegraphics[width=4.1cm]{dsssge/sge009_d2b.ps}
 \includegraphics[width=4.1cm]{dsssge/sge010_d2b.ps}
 \includegraphics[width=4.1cm]{dsssge/sge011_d2b.ps}
 \includegraphics[width=4.1cm]{dsssge/sge012_d2b.ps}

 \includegraphics[width=4.1cm]{dsssge/sge013_d2b.ps}
 \includegraphics[width=4.1cm]{dsssge/sge014_d2b.ps}
 \includegraphics[width=4.1cm]{dsssge/sge015_d2b.ps}
 \includegraphics[width=4.1cm]{dsssge/sge016_d2b.ps}

 \includegraphics[width=4.1cm]{dsssge/sge017_d2b.ps}
 \includegraphics[width=4.1cm]{dsssge/sge018_d2b.ps}
 \includegraphics[width=4.1cm]{dsssge/sge019_d2b.ps}
 \includegraphics[width=4.1cm]{dsssge/sge020_d2b.ps}

 \includegraphics[width=4.1cm]{dsssge/sge021_d2b.ps}
 \includegraphics[width=4.1cm]{dsssge/sge022_d2b.ps}
 \includegraphics[width=4.1cm]{dsssge/sge023_d2b.ps}
 \includegraphics[width=4.1cm]{dsssge/sge024_d2b.ps}
\caption[fcs]{\label{sgefc} DSS blue finding charts of Sge sources Sge001 to 
       Sge024. The size is 2\farcm5 $\times$  2\farcm5 (except for 
      Sge019 = RX J1959.6+2243 which is three times as large to show
      the full extent of the underlying nebula), North is up and East 
      to the left.
}
\end{figure*}

\setcounter{figure}{1}

\begin{figure*}[ht]
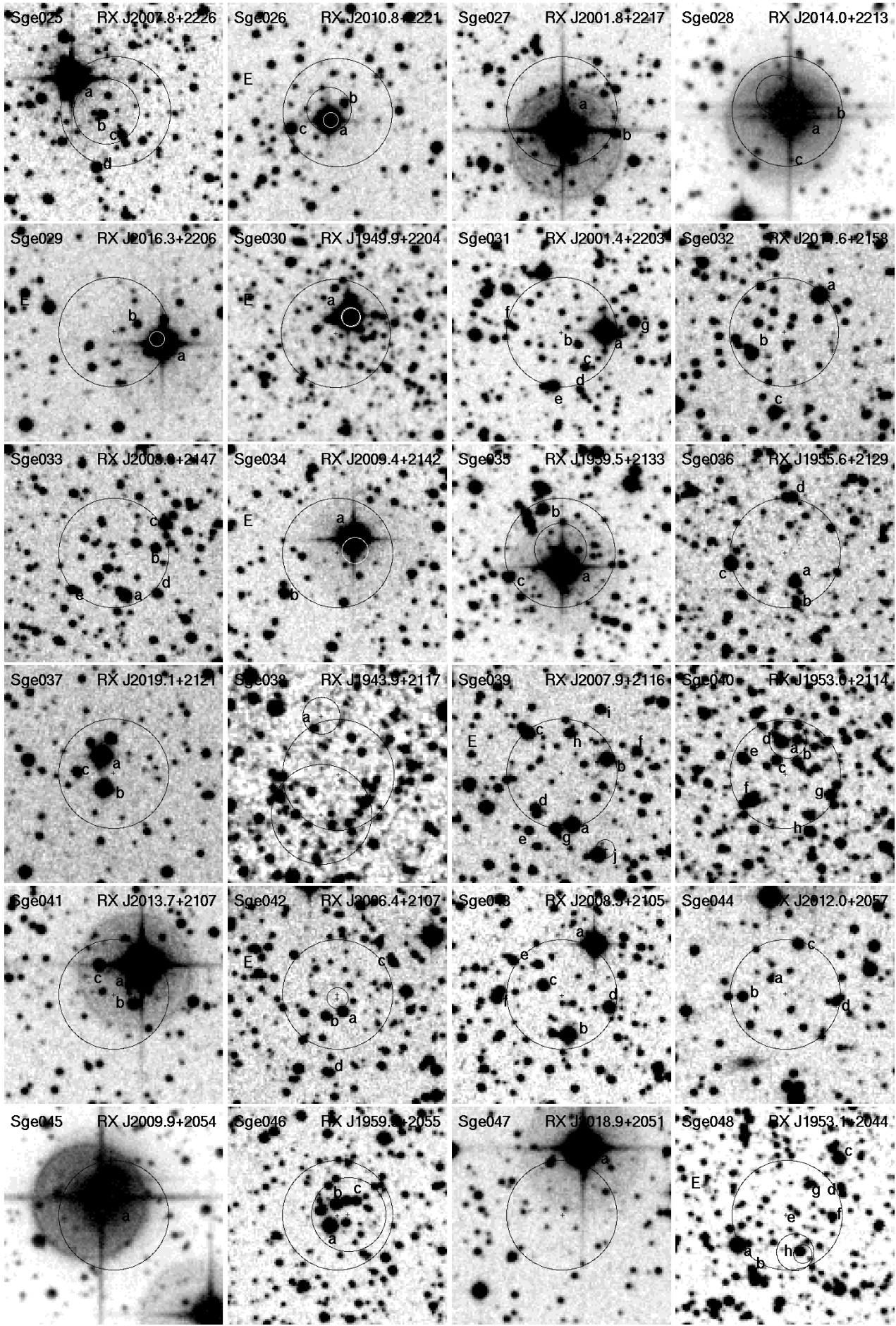

 \includegraphics[width=4.1cm]{dsssge/sge025_d2b.ps}
 \includegraphics[width=4.1cm]{dsssge/sge026_d2b.ps}
 \includegraphics[width=4.1cm]{dsssge/sge027_d2b.ps}
 \includegraphics[width=4.1cm]{dsssge/sge028_d2b.ps}

 \includegraphics[width=4.1cm]{dsssge/sge029_d2b.ps}
 \includegraphics[width=4.1cm]{dsssge/sge030_d2b.ps}
 \includegraphics[width=4.1cm]{dsssge/sge031_d2b.ps}
 \includegraphics[width=4.1cm]{dsssge/sge032_d2b.ps}

 \includegraphics[width=4.1cm]{dsssge/sge033_d2b.ps}
 \includegraphics[width=4.1cm]{dsssge/sge034_d2b.ps}
 \includegraphics[width=4.1cm]{dsssge/sge035_d2b.ps}
 \includegraphics[width=4.1cm]{dsssge/sge036_d2b.ps}

 \includegraphics[width=4.1cm]{dsssge/sge037_d2b.ps}
 \includegraphics[width=4.1cm]{dsssge/sge038_d2b.ps}
 \includegraphics[width=4.1cm]{dsssge/sge039_d2b.ps}
 \includegraphics[width=4.1cm]{dsssge/sge040_d2b.ps}

 \includegraphics[width=4.1cm]{dsssge/sge041_d2b.ps}
 \includegraphics[width=4.1cm]{dsssge/sge042_d2b.ps}
 \includegraphics[width=4.1cm]{dsssge/sge043_d2b.ps}
 \includegraphics[width=4.1cm]{dsssge/sge044_d2b.ps}

 \includegraphics[width=4.1cm]{dsssge/sge045_d2b.ps}
 \includegraphics[width=4.1cm]{dsssge/sge046_d2b.ps}
 \includegraphics[width=4.1cm]{dsssge/sge047_d2b.ps}
 \includegraphics[width=4.1cm]{dsssge/sge048_d2b.ps}
 \caption[fcs]{{\bf contd.} Sge sources Sge025 to Sge048.}
 \end{figure*}

\setcounter{figure}{1}

\begin{figure*}[ht]
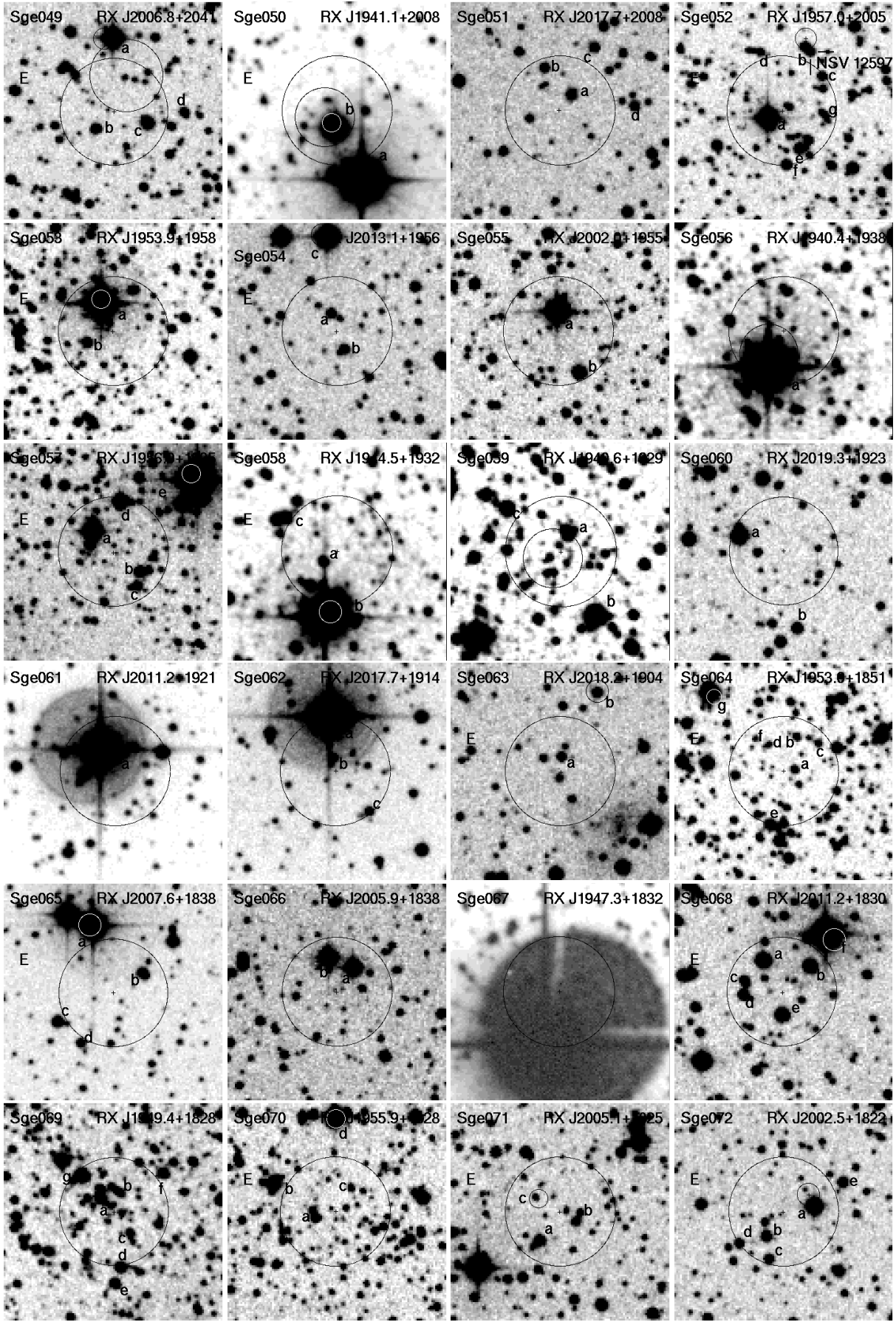

 \includegraphics[width=3.9cm]{dsssge/sge049_d2b.ps}
 \includegraphics[width=3.9cm]{dsssge/sge050_d2b.ps}
 \includegraphics[width=3.9cm]{dsssge/sge051_d2b.ps}
 \includegraphics[width=3.9cm]{dsssge/sge052_d2b.ps}

 \includegraphics[width=3.9cm]{dsssge/sge053_d2b.ps}
 \includegraphics[width=3.9cm]{dsssge/sge054_d2b.ps}
 \includegraphics[width=3.9cm]{dsssge/sge055_d2b.ps}
 \includegraphics[width=3.9cm]{dsssge/sge056_d2b.ps}

 \includegraphics[width=3.9cm]{dsssge/sge057_d2b.ps}
 \includegraphics[width=3.9cm]{dsssge/sge058_d2b.ps}
 \includegraphics[width=3.9cm]{dsssge/sge059_d2b.ps}
 \includegraphics[width=3.9cm]{dsssge/sge060_d2b.ps}

 \includegraphics[width=3.9cm]{dsssge/sge061_d2b.ps}
 \includegraphics[width=3.9cm]{dsssge/sge062_d2b.ps}
 \includegraphics[width=3.9cm]{dsssge/sge063_d2b.ps}
 \includegraphics[width=3.9cm]{dsssge/sge064_d2b.ps}

 \includegraphics[width=3.9cm]{dsssge/sge065_d2b.ps}
 \includegraphics[width=3.9cm]{dsssge/sge066_d2b.ps}
 \includegraphics[width=3.9cm]{dsssge/sge067_d2b.ps}
 \includegraphics[width=3.9cm]{dsssge/sge068_d2b.ps}

 \includegraphics[width=3.9cm]{dsssge/sge069_d2b.ps}
 \includegraphics[width=3.9cm]{dsssge/sge070_d2b.ps}
 \includegraphics[width=3.9cm]{dsssge/sge071_d2b.ps}
 \includegraphics[width=3.9cm]{dsssge/sge072_d2b.ps}
 \caption[fcs]{{\bf contd.} Sge sources Sge049 to Sge072.}
 \end{figure*}

\setcounter{figure}{1}

\begin{figure*}[ht]
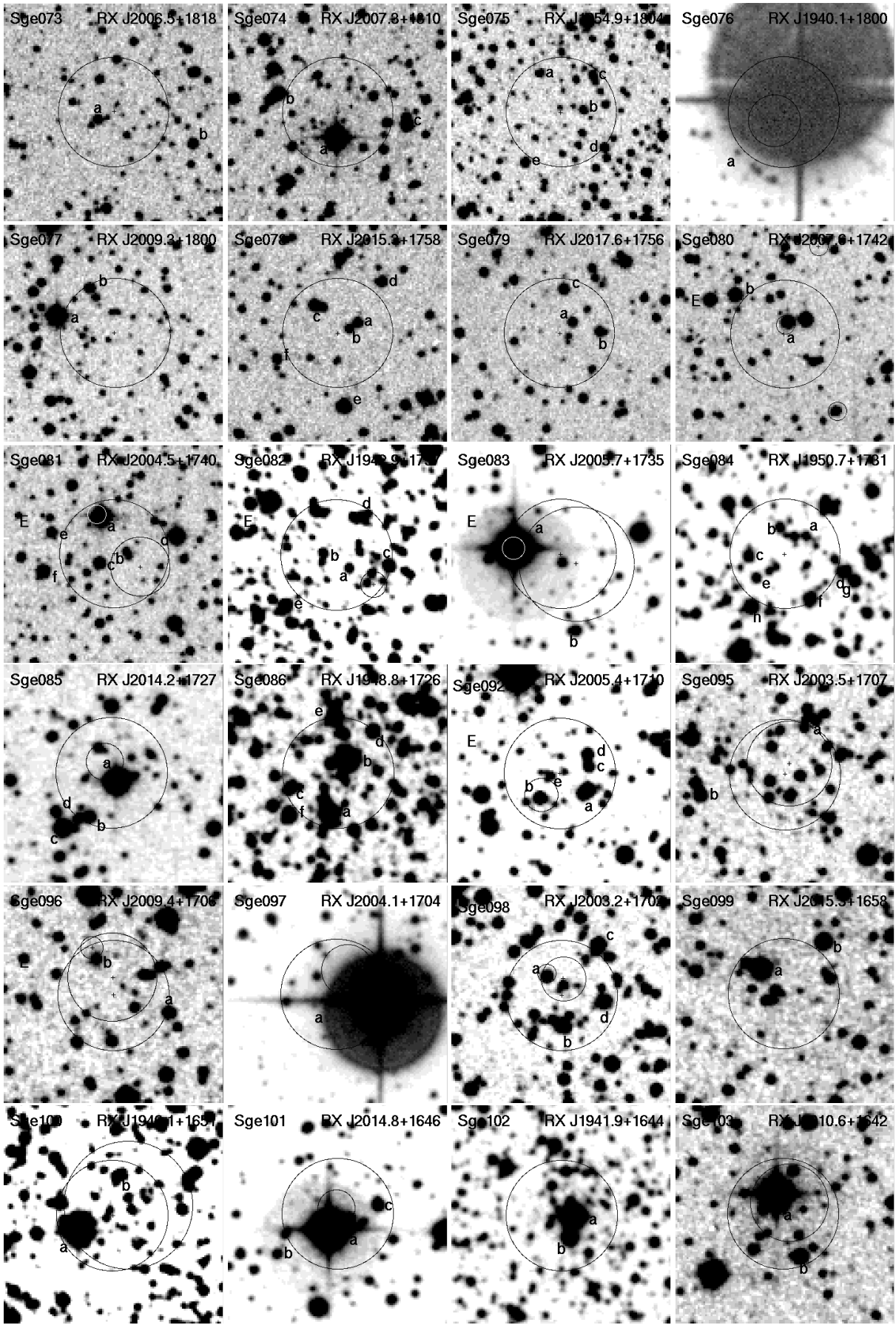

 \includegraphics[width=3.9cm]{dsssge/sge073_d2b.ps}
 \includegraphics[width=3.9cm]{dsssge/sge074_d2b.ps}
 \includegraphics[width=3.9cm]{dsssge/sge075_d2b.ps}
 \includegraphics[width=3.9cm]{dsssge/sge076_d2b.ps}

 \includegraphics[width=3.9cm]{dsssge/sge077_d2b.ps}
 \includegraphics[width=3.9cm]{dsssge/sge078_d2b.ps}
 \includegraphics[width=3.9cm]{dsssge/sge079_d2b.ps}
 \includegraphics[width=3.9cm]{dsssge/sge080_d2b.ps}

 \includegraphics[width=3.9cm]{dsssge/sge081_d2b.ps}
 \includegraphics[width=3.9cm]{dsssge/sge082_d2b.ps}
 \includegraphics[width=3.9cm]{dsssge/sge083_d2b.ps}
 \includegraphics[width=3.9cm]{dsssge/sge084_d2b.ps}

 \includegraphics[width=3.9cm]{dsssge/sge085_d2b.ps}
 \includegraphics[width=3.9cm]{dsssge/sge086_d2b.ps}
 \includegraphics[width=3.9cm]{dsssge/sge092_d2b.ps}
 \includegraphics[width=3.9cm]{dsssge/sge095_d2b.ps}

 \includegraphics[width=3.9cm]{dsssge/sge096_d2b.ps}
 \includegraphics[width=3.9cm]{dsssge/sge097_d2b.ps}
 \includegraphics[width=3.9cm]{dsssge/sge098_d2b.ps}
 \includegraphics[width=3.9cm]{dsssge/sge099_d2b.ps}

 \includegraphics[width=3.9cm]{dsssge/sge100_d2b.ps}
 \includegraphics[width=3.9cm]{dsssge/sge101_d2b.ps}
 \includegraphics[width=3.9cm]{dsssge/sge102_d2b.ps}
 \includegraphics[width=3.9cm]{dsssge/sge103_d2b.ps}

 \caption[fcs]{{\bf contd.} Sge sources Sge073 to Sge103.}
 \end{figure*}

\setcounter{figure}{1}

\begin{figure*}[ht]
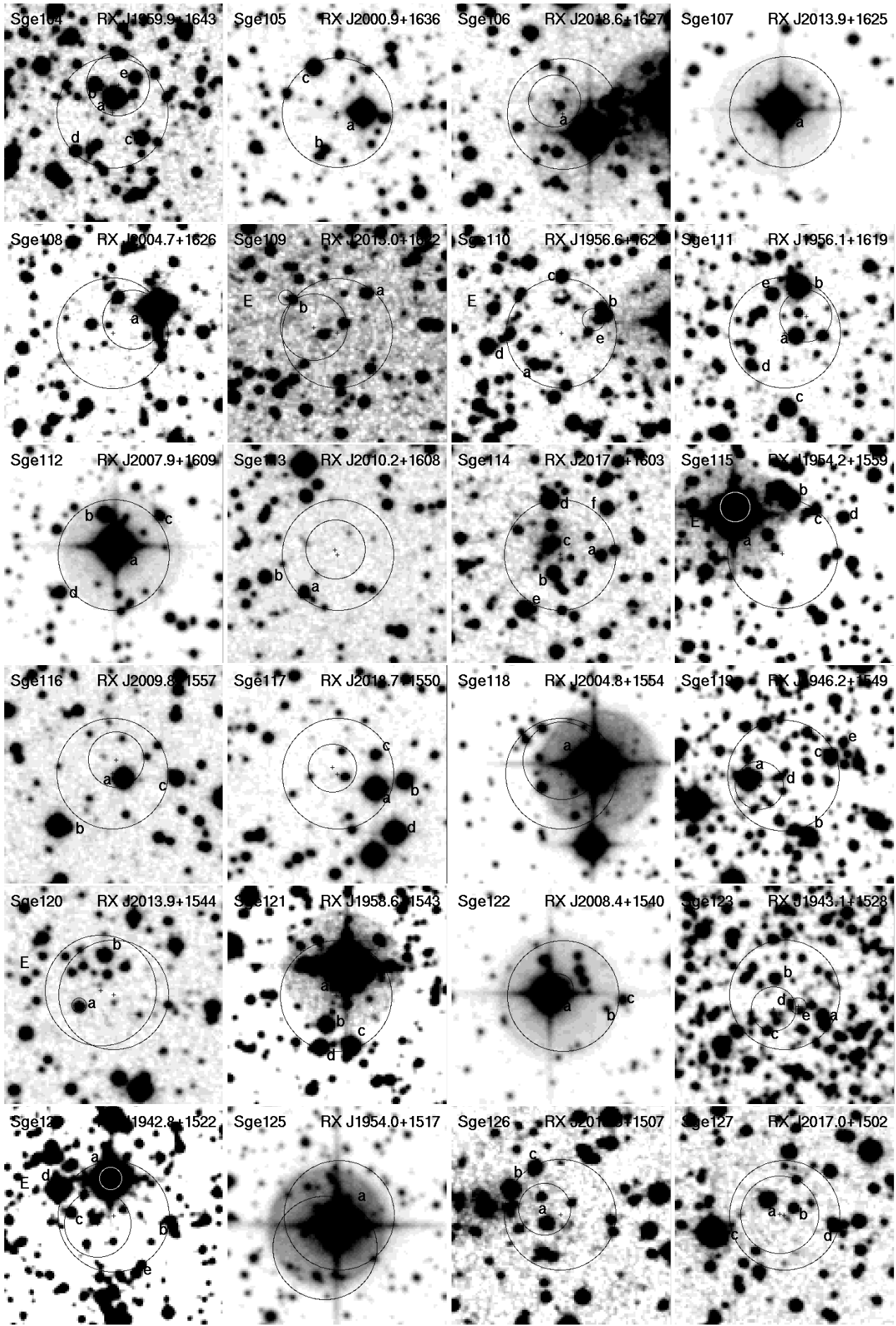

 \includegraphics[width=4.1cm]{dsssge/sge104_d2b.ps}
 \includegraphics[width=4.1cm]{dsssge/sge105_d2b.ps}
 \includegraphics[width=4.1cm]{dsssge/sge106_d2b.ps}
 \includegraphics[width=4.1cm]{dsssge/sge107_d2b.ps}

 \includegraphics[width=4.1cm]{dsssge/sge108_d2b.ps}
 \includegraphics[width=4.1cm]{dsssge/sge109_d2b.ps}
 \includegraphics[width=4.1cm]{dsssge/sge110_d2b.ps}
 \includegraphics[width=4.1cm]{dsssge/sge111_d2b.ps}

 \includegraphics[width=4.1cm]{dsssge/sge112_d2b.ps}
 \includegraphics[width=4.1cm]{dsssge/sge113_d2b.ps}
 \includegraphics[width=4.1cm]{dsssge/sge114_d2b.ps}
 \includegraphics[width=4.1cm]{dsssge/sge115_d2b.ps}

 \includegraphics[width=4.1cm]{dsssge/sge116_d2b.ps}
 \includegraphics[width=4.1cm]{dsssge/sge117_d2b.ps}
 \includegraphics[width=4.1cm]{dsssge/sge118_d2b.ps}
 \includegraphics[width=4.1cm]{dsssge/sge119_d2b.ps}

 \includegraphics[width=4.1cm]{dsssge/sge120_d2b.ps}
 \includegraphics[width=4.1cm]{dsssge/sge121_d2b.ps}
 \includegraphics[width=4.1cm]{dsssge/sge122_d2b.ps}
 \includegraphics[width=4.1cm]{dsssge/sge123_d2b.ps}

 \includegraphics[width=4.1cm]{dsssge/sge124_d2b.ps}
 \includegraphics[width=4.1cm]{dsssge/sge125_d2b.ps}
 \includegraphics[width=4.1cm]{dsssge/sge126_d2b.ps}
 \includegraphics[width=4.1cm]{dsssge/sge127_d2b.ps}

 \caption[fcs]{{\bf contd.} Sge sources Sge104 to Sge127.}
 \end{figure*}

\setcounter{figure}{1}

\begin{figure*}[ht]
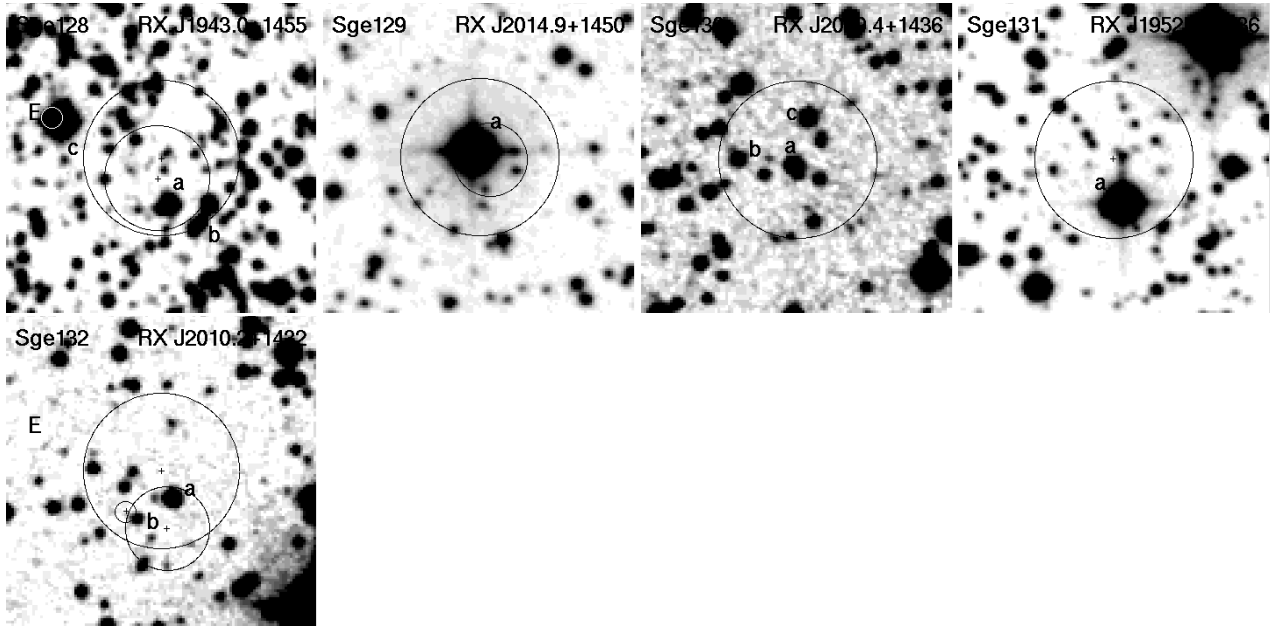

 \includegraphics[width=4.1cm]{dsssge/sge128_d2b.ps}
 \includegraphics[width=4.1cm]{dsssge/sge129_d2b.ps}
 \includegraphics[width=4.1cm]{dsssge/sge130_d2b.ps}
 \includegraphics[width=4.1cm]{dsssge/sge131_d2b.ps}

 \includegraphics[width=4.1cm]{dsssge/sge132_d2b.ps}
 \caption[fcs]{{\bf contd.} Sge source Sge132.}
 \end{figure*}

\end{subfigures}

\clearpage

% Beginn Lichtkurven

\begin{subfigures}

\begin{figure*}[ht]
\includegraphics[width=18.0cm,clip]{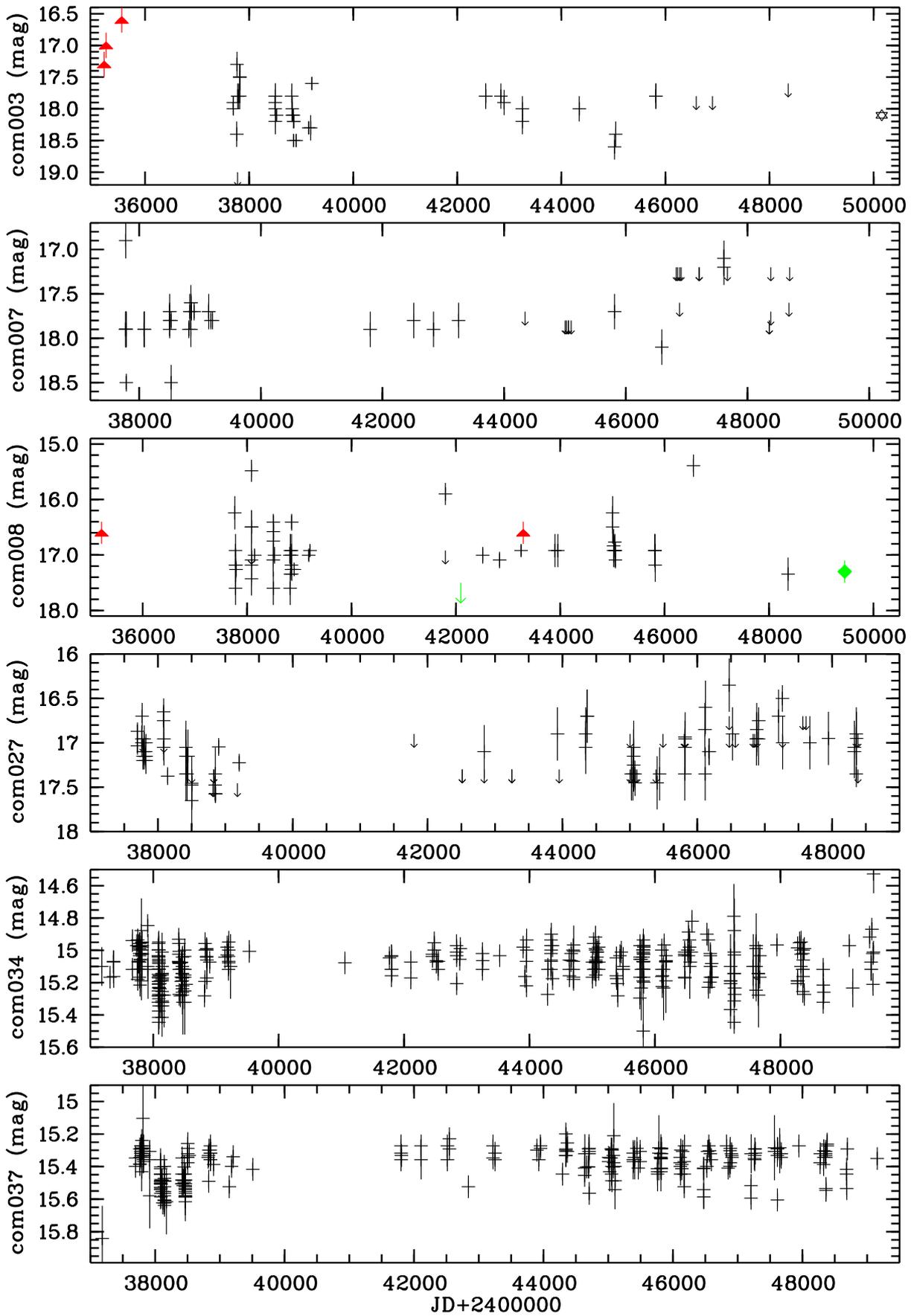}
\vspace{-0.5cm}
\caption[colc1]{\label{comolc} Optical light curves  of Com sources. 
For each source, the measurements are shown by crosses, and upper limits 
as arrows. Filled triangles (red) and hexagons (green) are measurements of 
the POSS and the Tautenburg Schmidt plates, respectively, and open stars
refer to CCD observations at SAO Russia.
}
\end{figure*}

\setcounter{figure}{0}

\begin{figure*}[ht]
\includegraphics[width=18.0cm,clip]{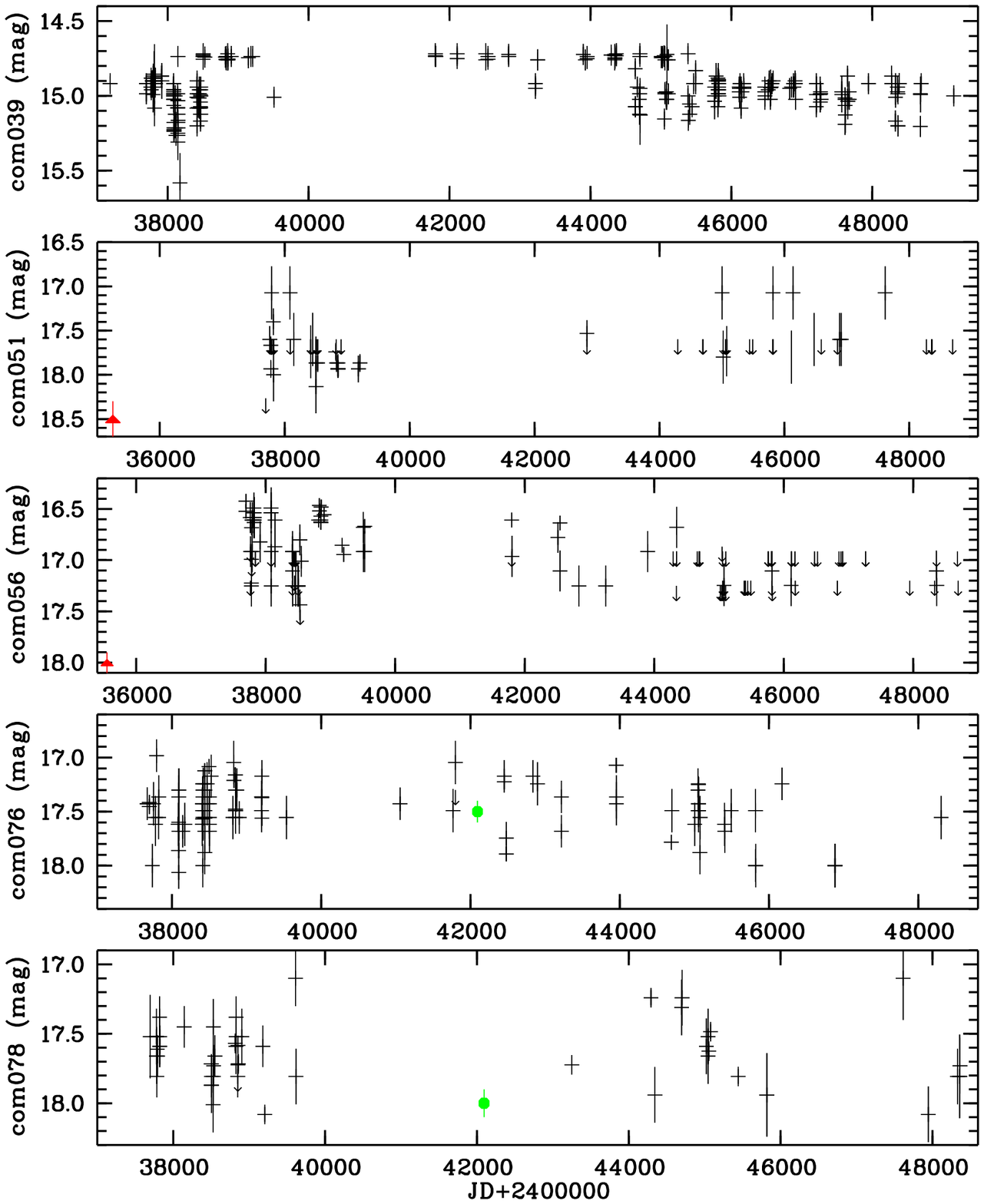}
 \caption[colc2]{{\bf contd.} Optical light curves of Com sources. }
 \end{figure*}

\setcounter{figure}{0}

\begin{figure*}[ht]
\includegraphics[width=18.0cm, clip]{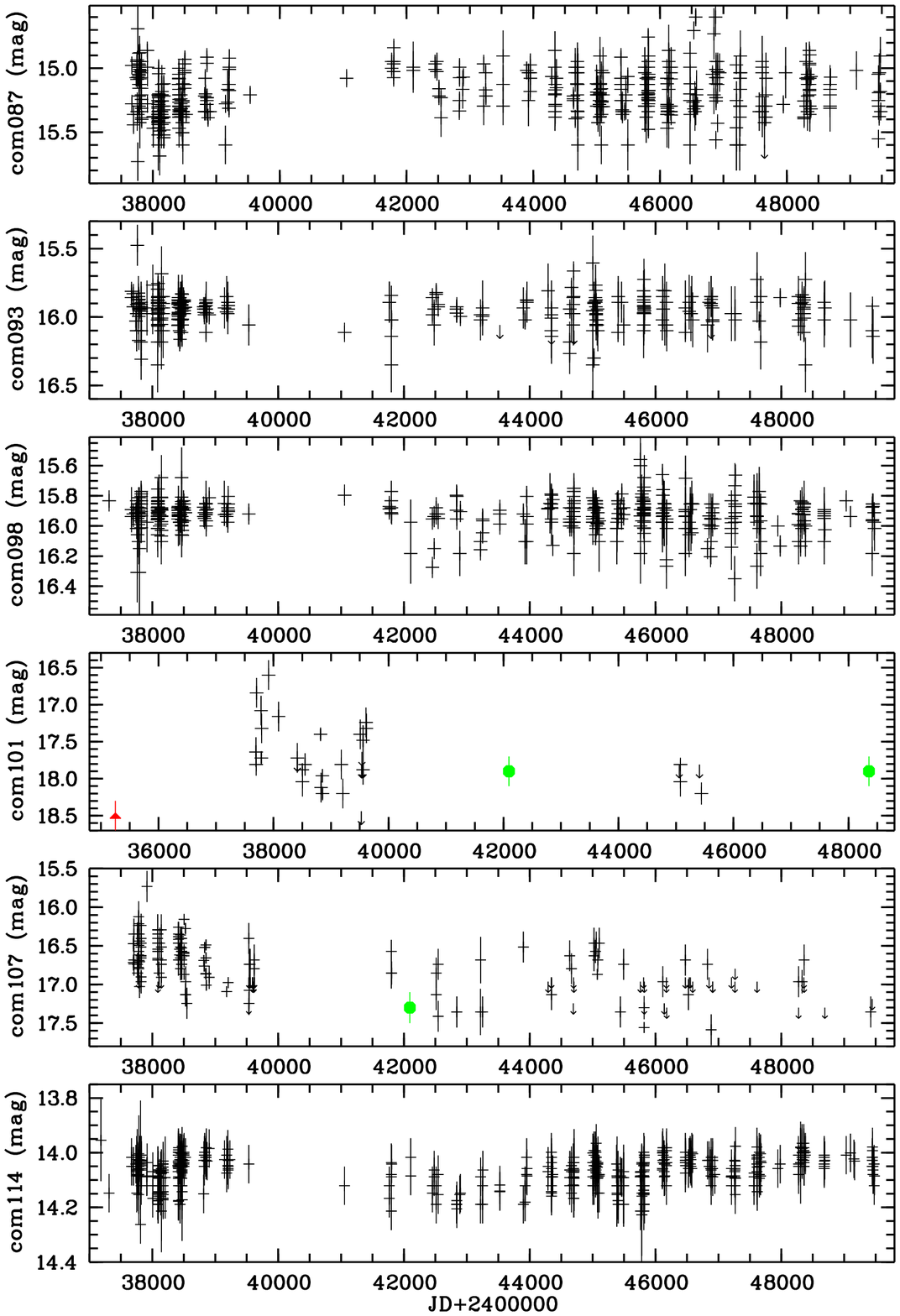}
 \caption[colc3]{{\bf contd.} Optical light curves of Com sources. }
 \end{figure*}

\setcounter{figure}{0}

\begin{figure*}[ht]
\includegraphics[width=18.0cm, clip]{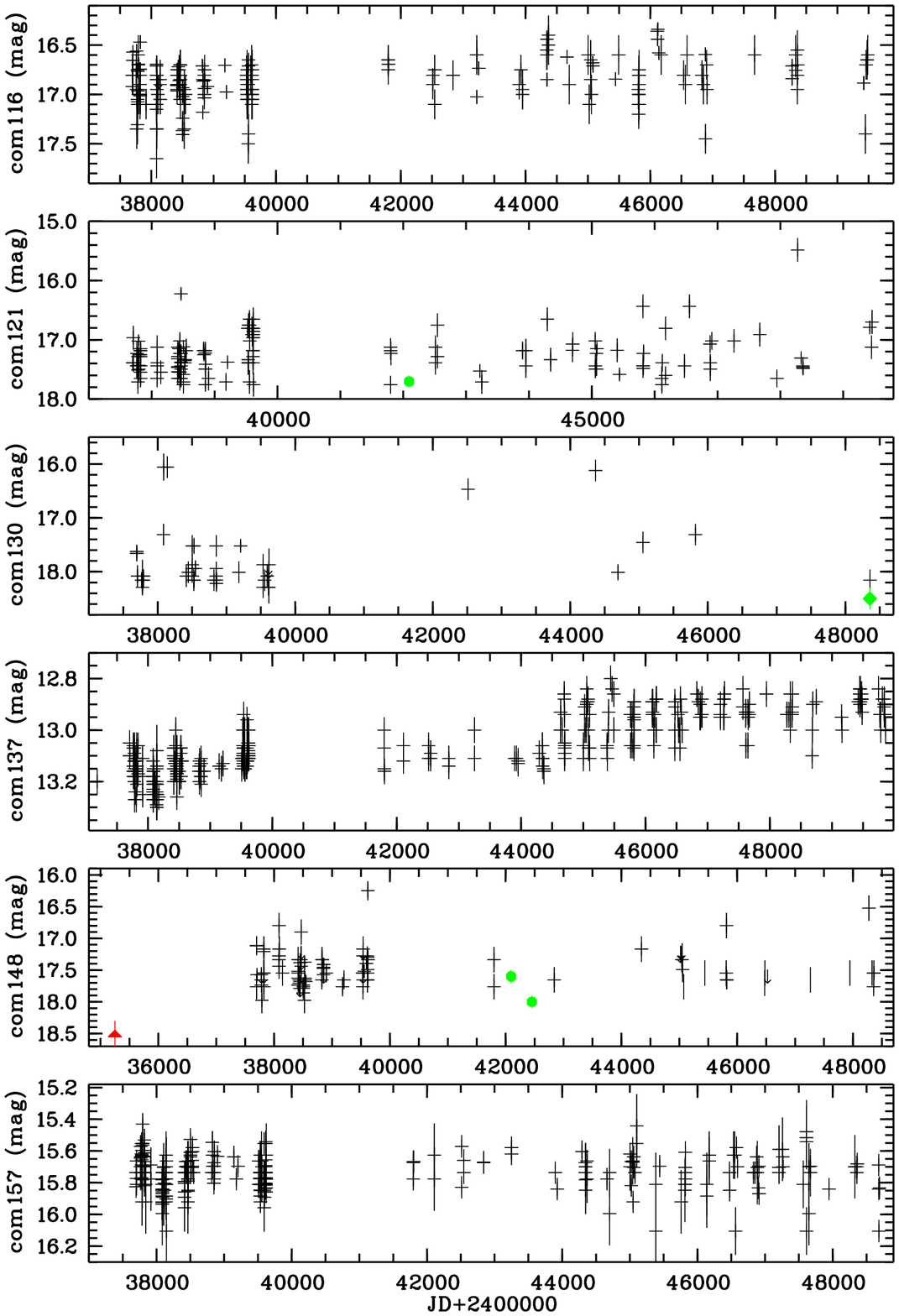}
 \caption[colc4]{{\bf contd.} Optical light curves of Com sources. }
 \end{figure*}

\setcounter{figure}{0}

\begin{figure*}[ht]
\includegraphics[width=18.0cm, clip]{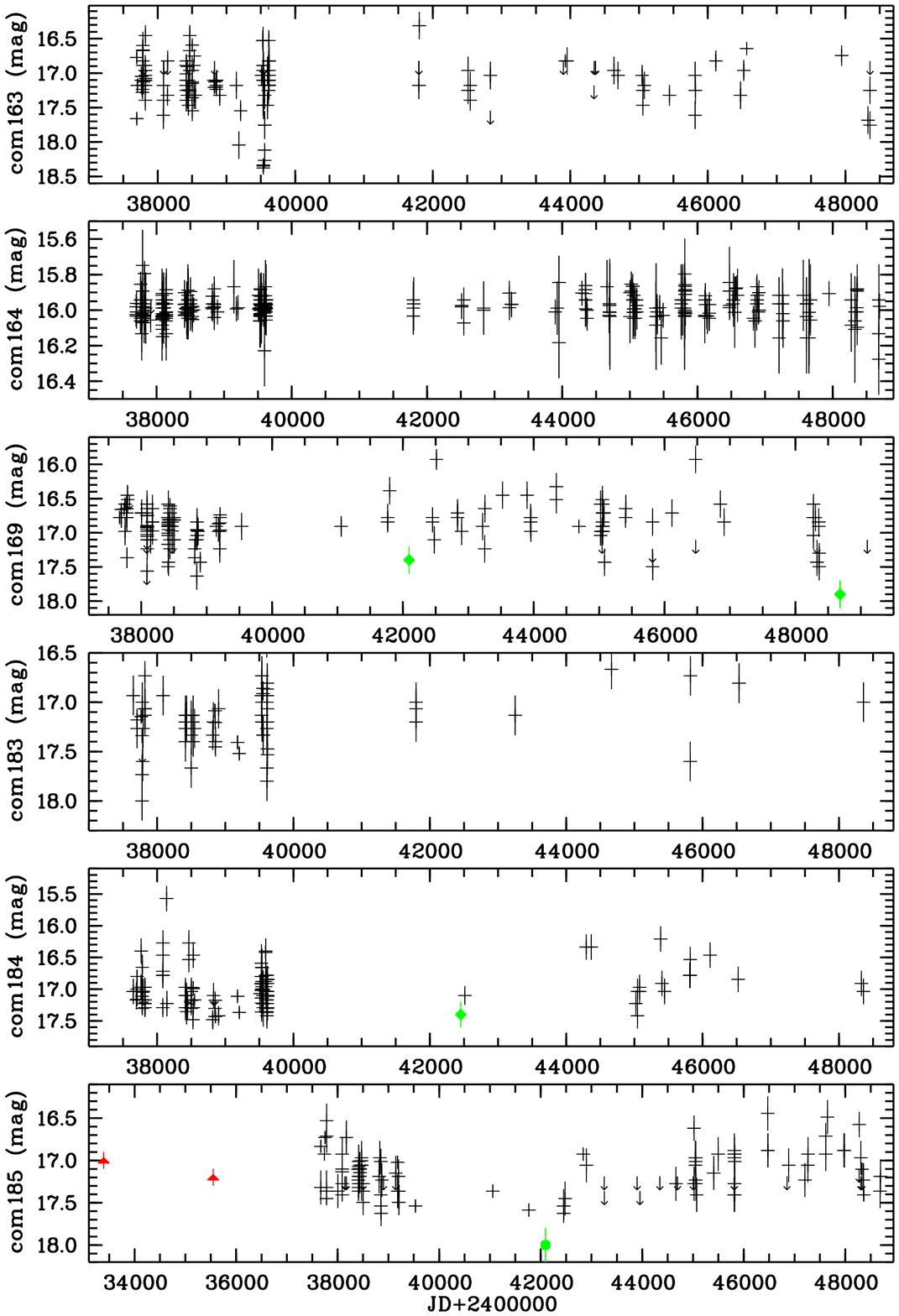}
 \caption[colc5]{{\bf contd.} Optical light curves of Com sources. }
 \end{figure*}

\setcounter{figure}{0}

\begin{figure*}[ht]
\includegraphics[width=18.0cm, clip]{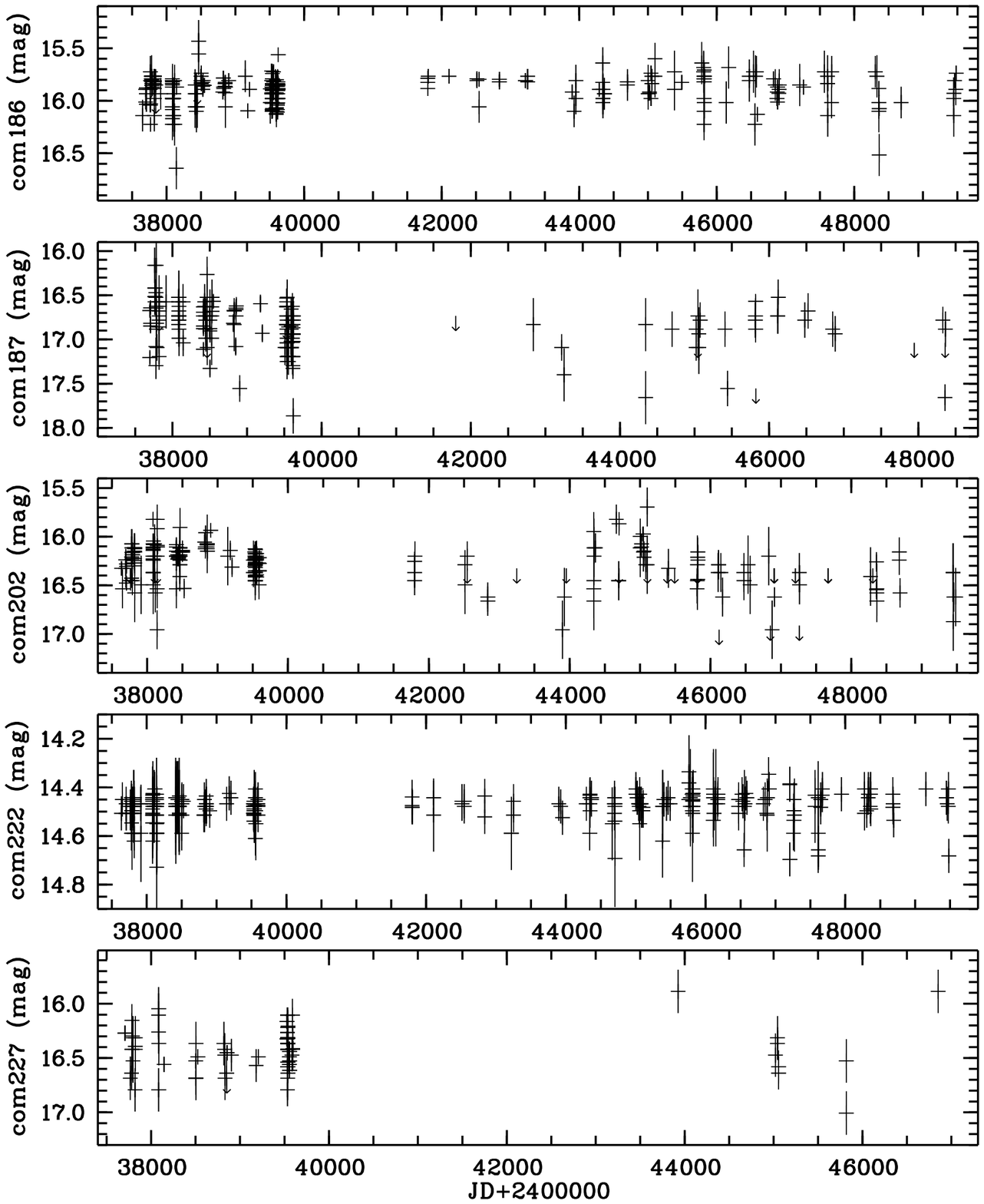}
 \caption[colc6]{{\bf contd.} Optical light curves of Com sources. }
 \end{figure*}

%%%
%%%%%% Sge Lichtkurven
%%%

\begin{figure*}[ht]
\includegraphics[width=18.0cm,clip]{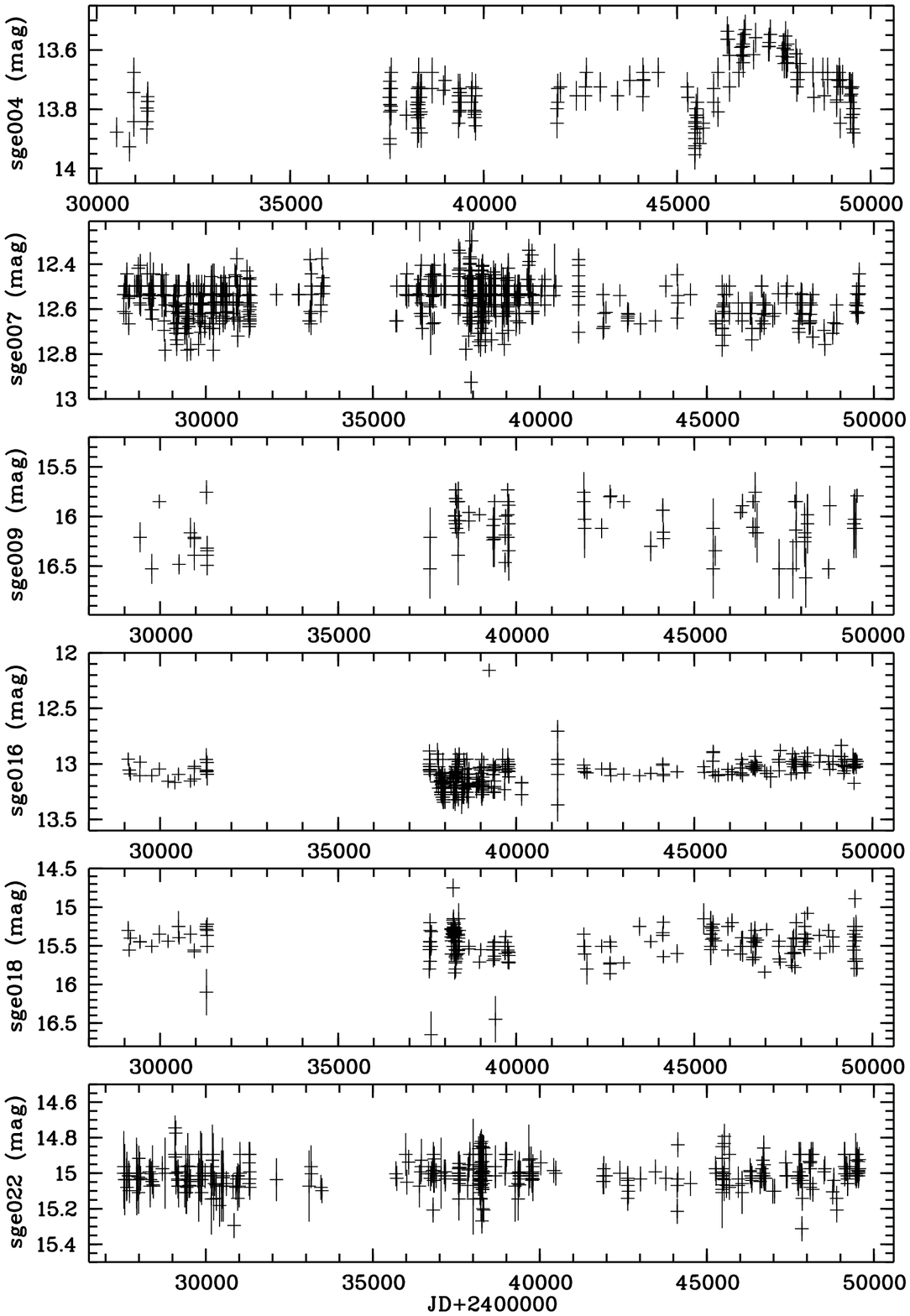}
\caption[solc1]{\label{sgeolc} Optical light curves of Sge sources. 
 Labels as in \ref{comolc}.}
\end{figure*}

\setcounter{figure}{1}

\begin{figure*}[ht]
\includegraphics[width=18.0cm, clip]{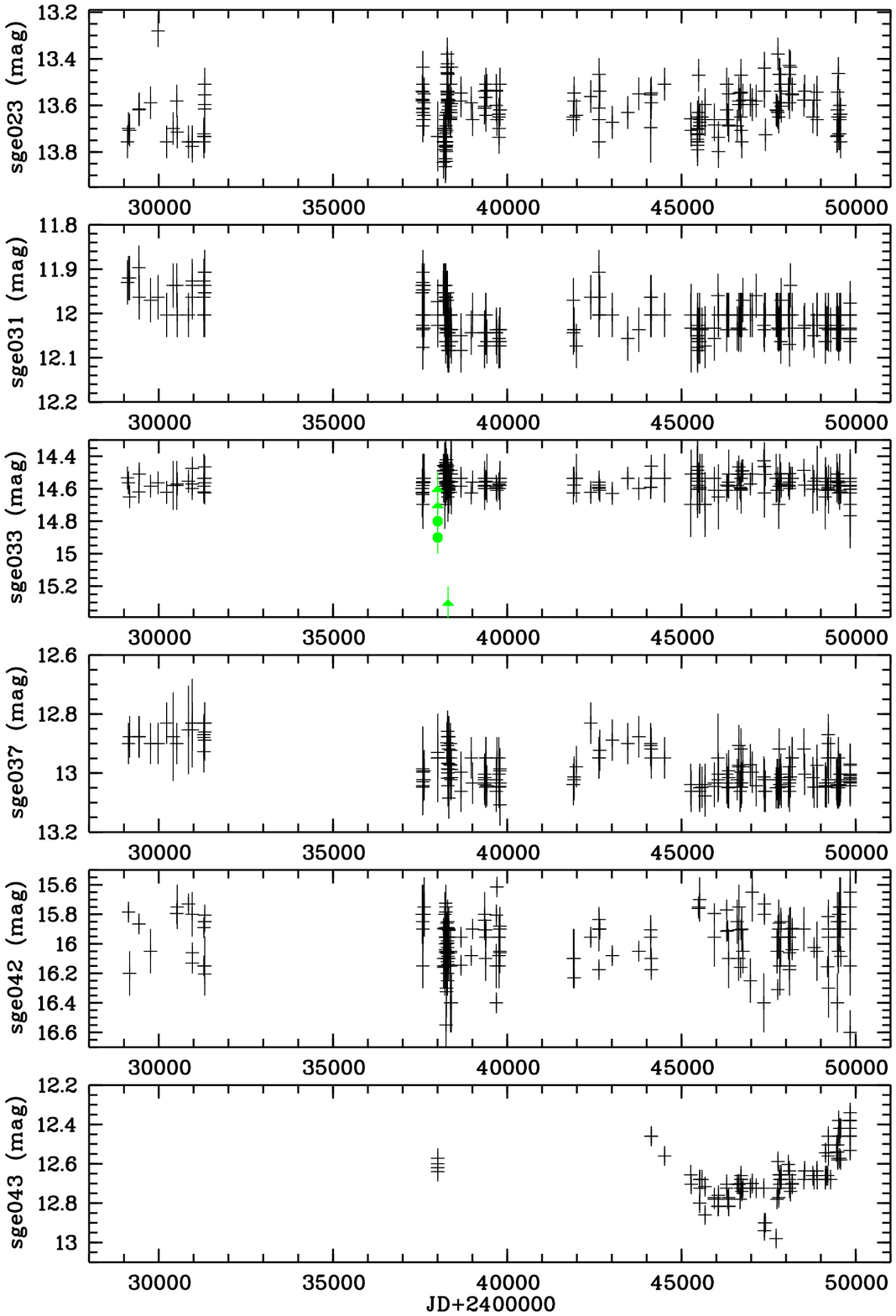}
 \caption[solc2]{{\bf contd.} Optical light curves of Sge sources.}
 \end{figure*}

\setcounter{figure}{1}

\begin{figure*}[ht]
\includegraphics[width=18.0cm, clip]{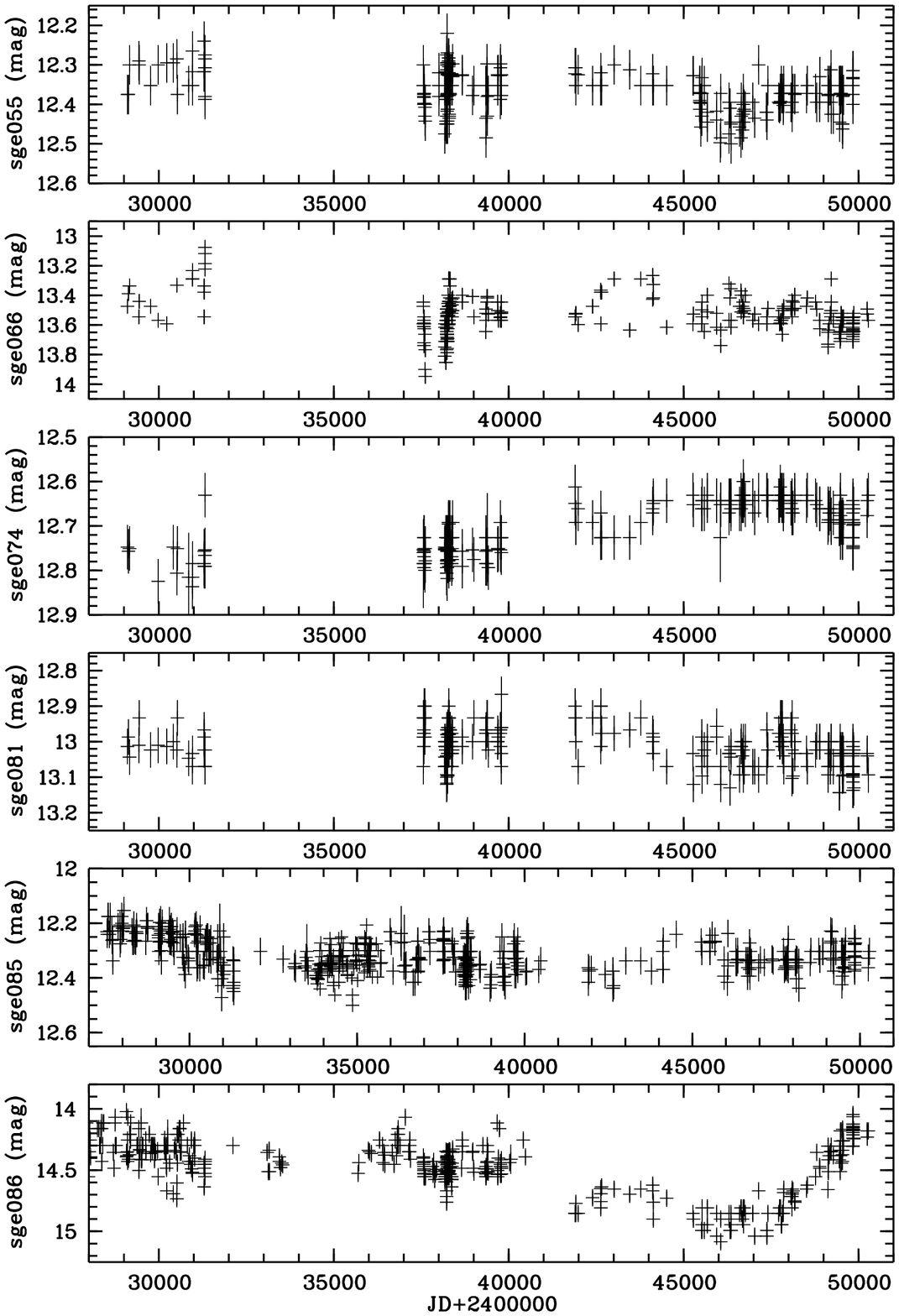}
 \caption[solc3]{{\bf contd.} Optical light curves of Sge sources.}
 \end{figure*}

\setcounter{figure}{1}

\begin{figure*}[ht]
\includegraphics[width=18.0cm, clip]{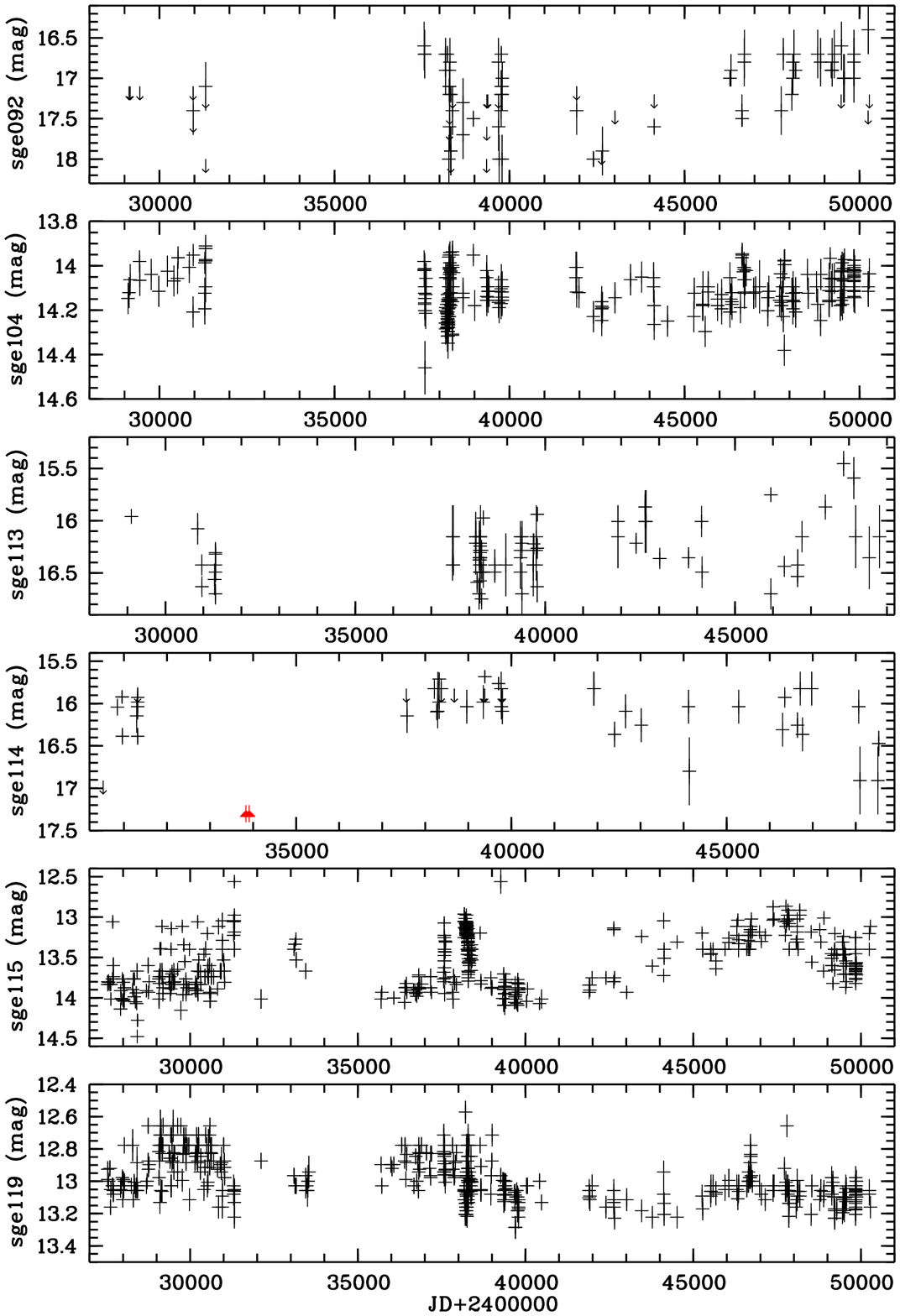}
 \caption[solc4]{{\bf contd.} Optical light curves of Sge sources.}
 \end{figure*}

\setcounter{figure}{1}

\begin{figure*}[ht]
\includegraphics[width=18.0cm, clip]{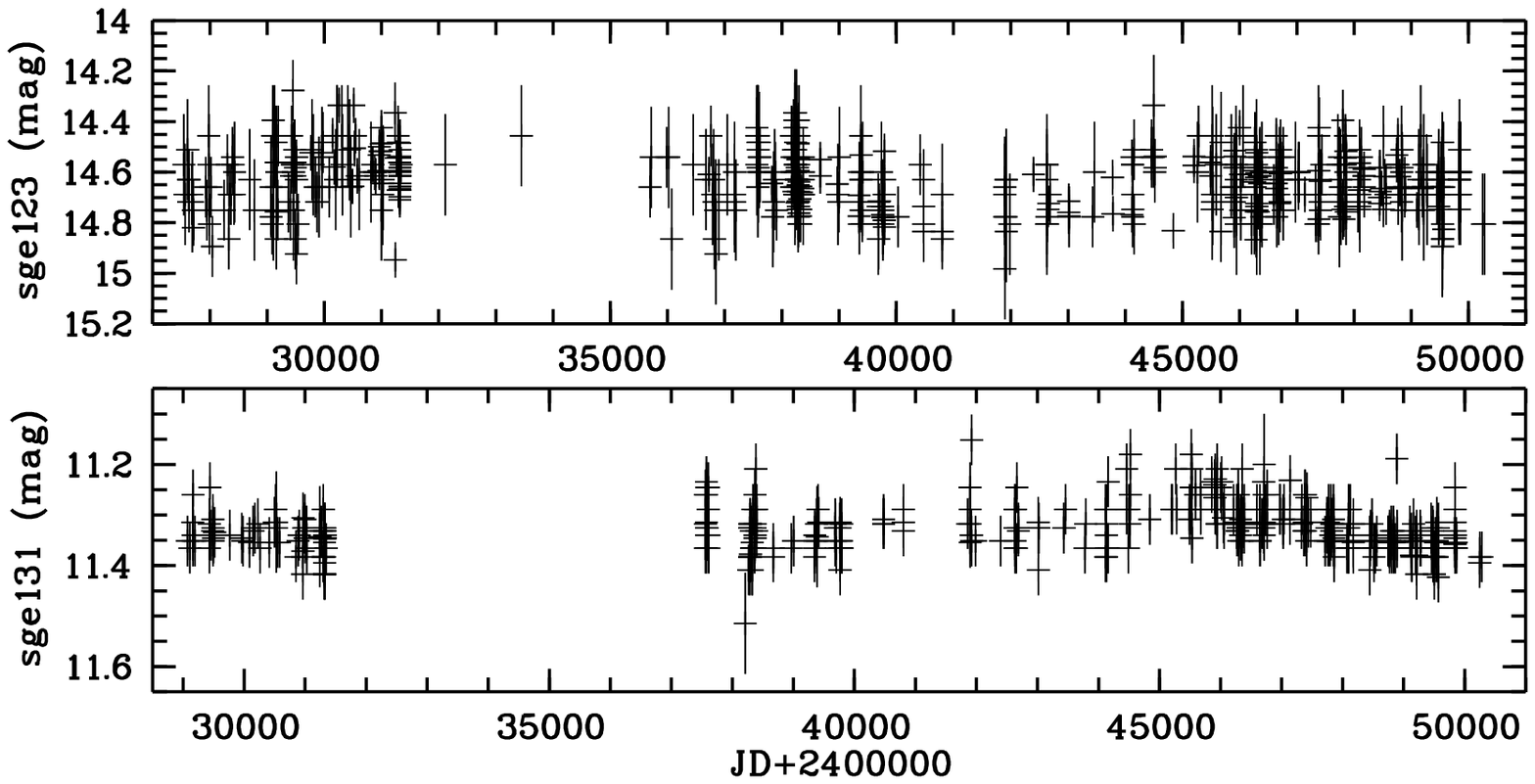}
 \caption[solc5]{{\bf contd.} Optical light curves of Sge sources.}
 \end{figure*}

\end{subfigures}

\end{document}